\PassOptionsToPackage{colorlinks=true,linkcolor=blue,urlcolor=blue}{hyperref}

\documentclass[11pt]{article}

\usepackage{amsmath,amssymb,amsthm,bm}   % TeX4ht relies on these
\usepackage{mathtools}            % optional but harmless
\usepackage{rotating}

\usepackage[margin=1cm, paperwidth=8.5in, paperheight=11in]{geometry}
\usepackage[
  hypertexnames=false
]{hyperref}
\usepackage{kotex}
\usepackage{etex}
\usepackage{fullpage}
\usepackage{mathtools}
\usepackage{amsfonts}
\usepackage{graphicx}
\usepackage{amsthm}
\usepackage[utf8]{inputenc}
\usepackage{afterpage}
\usepackage{adjustbox}
\usepackage{placeins}
\usepackage{fixltx2e}
\usepackage{supertabular} 
\usepackage{titlesec}
\usepackage{array}
\usepackage{tabularx}
\usepackage{color}
\usepackage{algorithm}
\usepackage{algpseudocode}
\usepackage{subcaption}
\usepackage{hyperref}
\usepackage{pgfplots}
\usepackage{relsize}

\usepackage{enumitem}
\usepackage{fancyhdr}
\usepackage{alltt}
\usepackage{soul}
\usepackage{fancyvrb}
\usepackage{xcolor}
\usepackage[margin=0.1cm]{subcaption}
\usepackage{multicol}
\usepackage{refcount}

\usepackage{breakurl} % Break URL
\usepackage{multirow}
\usepackage[most]{tcolorbox}

\tcbset{breakable}
\usepackage{esvect}
\usepackage{hyperref}
\usepackage{mathdots}
\usepackage{pifont} 
\usepackage{booktabs}
\usepackage{arydshln}
\usepackage{xcolor}

\usepackage{mathtools}  
\usepackage{array}
\usepackage{hyperref}
\usepackage{hyperref}
\usepackage{mathtools, nccmath}
\usepackage{titling}

\usepackage{tocloft}
\usepackage{pdflscape}
\usepackage[all=normal, floats, bibnotes, wordspacing, charwidths, indent]{savetrees}
\makeatletter%
\usepackage{enumitem}
\setlist{nolistsep}
\setlist[itemize]{leftmargin=*}
\setlist[enumerate]{leftmargin=*}

\makeatletter
\newcommand\bunderline{}% check for being undefined
\DeclareRobustCommand\bunderline[1]{\mathord{\mathpalette\b@underline{#1}}}
\newcommand{\b@underline}[2]{%
  \sbox\z@{$\m@th#1\underline{#2}$}%
  \raisebox{-\dp\z@}{\scalebox{.5}[.25]{$\m@th#1[$}}%
  \copy\z@
  \raisebox{-\dp\z@}{\scalebox{.5}[.25]{$\m@th#1]$}}%
}

\makeatother

\usepackage{graphicx}
\def\rotatecharone#1{\rotatebox[origin=c]{180}{#1}}

\usepackage{hyperref}
\skip\footins 0.5\baselineskip %8.1pt plus 4pt minus 2pt
\everypar{\looseness=-1 }

\newenvironment{myproof}[1][\proofname]{%
  \begin{proof}[#1]$ $\par\nobreak\ignorespaces
}{%
  \end{proof}
}
\makeatletter
\renewcommand{\sectionautorefname}{\S\@gobble}
\renewcommand{\subsectionautorefname}{\S\@gobble}
\renewcommand{\subsubsectionautorefname}{\S\@gobble}
\renewcommand{\appendixautorefname}{\S\@gobble}
\renewcommand{\appendixautorefname}{\S\@gobble}

\makeatother
\newcommand{\ignore}[1]{}

\makeatletter%
\usepackage{stackengine,xcolor}
\input pdf-trans
\newbox\qbox
\def\usecolor#1{\csname\string\color@#1\endcsname\space}
\newcommand\bordercolor[1]{\colsplit{1}{#1}}
\newcommand\fillcolor[1]{\colsplit{0}{#1}}
\newcommand\colsplit[2]{\colorlet{tmpcolor}{#2}\edef\tmp{\usecolor{tmpcolor}}%  
  \def\tmpB{}\expandafter\colsplithelp\tmp\relax%
  \ifnum0=#1\relax\edef\fillcol{\tmpB}\else\edef\bordercol{\tmpC}\fi}
\def\colsplithelp#1#2 #3\relax{%
  \edef\tmpB{\tmpB#1#2 }%
  \ifnum `#1>`9\relax\def\tmpC{#3}\else\colsplithelp#3\relax\fi
}
\newcommand\outline[1]{\leavevmode%
  \def\maltext{#1}%
  \setbox\qbox=\hbox{\maltext}%
  \boxgs{Q q 2 Tr \thickness\space w \fillcol\space \bordercol\space}{}%
  \copy\qbox%
}
\newcommand\mathcalbb[2][1]{%
  \stackengine{0pt}{\def\thickness{.55}\outline{$\mathcal{#2}$}}{\kern.1pt\outline{$\mathcal{#2}$}}{O}{l}{F}{F}{L}}
\bordercolor{black}
\fillcolor{white}
\def\thickness{.1}% TO CHANGE THICKNESS OF SHADOW
\usepackage{lmodern}
\usepackage{lipsum}

\makeatletter
\DeclareRobustCommand{\ccong}{\mathrel{\mathpalette\@verequiv\sim}}
\newcommand{\@verequiv}[2]{%
  \lower.5\p@\vbox{
    \lineskiplimit\maxdimen
    \lineskip-.5\p@
    \ialign{%
      $\m@th#1\hfil##\hfil$\crcr
      #2\crcr
      \equiv\crcr
    }%
  }%
}
\renewcommand*\env@matrix[1][c]{\hskip -\arraycolsep
  \let\@ifnextchar\new@ifnextchar
  \array{*\c@MaxMatrixCols #1}}

\newcommand{\hathat}{\tilde}

\makeatother

\begin{document}
\title{\Huge{\textbf{The Beginner's Textbook}}\\ \Huge{\textbf{for Fully Homomorphic Encryption}}}
\author{
        \textbf{Ronny Ko} \thanks{\large \textit{\text{Acknowledgments}:}\\  
        \textbf{Robin Geelen} (KU Leuven)\\  
        \textbf{Tianjian Yang} (Peking University)\\ \textbf{Yongwoo Lee} (Inha University) \\ \textbf{Nolan Carouge} (Grenoble INP Ensimag) \\
        \textbf{Navid Abapour} (University of Surrey)
        }
        \\{LG Electronics Inc.}}%\\\texttt{\small{hajoon.ko@lge.com}}}
\date{}

\begin{titlingpage}
\maketitle
\end{titlingpage}

\clearpage

\section*{Preface}
Fully Homomorphic Encryption (FHE) is a cryptographic scheme that enables computations to be performed directly on encrypted data, as if the data were in plaintext. After all computations are performed on the encrypted data, it can be decrypted to reveal the result. The decrypted value matches the result that would have been obtained if the same computations had been applied to the plaintext data.

FHE supports basic operations such as addition and multiplication on encrypted numbers. Using these fundamental operations, more complex computations can be constructed, including subtraction, division, logic gates (e.g., AND, OR, XOR, NAND, MUX), and even advanced mathematical functions such as ReLU, sigmoid, and trigonometric functions (e.g., sin, cos). These functions can be implemented either as exact formulas or as approximations, depending on the trade-off between computational efficiency and accuracy. 

FHE enables privacy-preserving machine learning by allowing a server to process the client’s data in its encrypted form through an ML model. With FHE, the server learns neither the plaintext version of the input features nor the inference results. Only the client, using their secret key, can decrypt and access the results at the end of the service protocol.
FHE can also be applied to confidential blockchain services, ensuring that sensitive data in smart contracts remains encrypted and confidential while maintaining the transparency and integrity of the execution process.
Other applications of FHE include secure outsourcing of data analytics, encrypted database queries, privacy-preserving searches, efficient multi-party computation for digital signatures, and more.

This book is designed to help the reader understand how FHE works at the mathematical level. The book comprises the following four parts: 

$ $

\begin{itemize}
\item \textbf{\autoref{part:basic-math}:~\nameref{part:basic-math}} explains necessary background concepts for FHE, such as groups, fields, orders, polynomial rings, cyclotomic polynomials, vectors and matrices, the Chinese Remainder Theorem, Taylor series, polynomial interpolation, and the Fast Fourier Transform.

\item \textbf{\autoref{part:pqc}:~\nameref{part:pqc}} explains well-known lattice-based cryptographic schemes, which include LWE, RLWE, GLWE, GLev, and GGSW cryptosystems.

\item \textbf{\textbf{\autoref{part:generic-fhe}:~\nameref{part:generic-fhe}}} explains the generic techniques of FHE adopted by many existing schemes, such as homomorphic addition, multiplication, modulus switching, and key switching.

\item \textbf{\textbf{\autoref{part:fhe-schemes}:~\nameref{part:fhe-schemes}}} explains four widely used FHE schemes: TFHE, BFV, CKKS, and BGV, as well as their RNS-variant versions.
\end{itemize}

$ $

%These parts are designed in an incremental manner, and therefore understanding each part requires the understanding of its prior part(s). 

%$ $

This book is available both as an \href{https://arxiv.org/abs/2503.05136}{\textbf{arXiv PDF}} and on an \href{https://fhetextbook.github.io} {\textbf{auto-generated dynamic website}} (powered by \href{https://www.kodymirus.cz/overleaf-html-sample/main.html}{make4ht}). We provide a Python Demo FHE library (TFHE, BFV, BGV, CKKS) for educational purposes, which is available at \href{https://github.com/fhetextbook/fhe-textbook}{\texttt{https://github.com/fhetextbook/fhe-textbook}}.
Please report any errors regarding the book draft on the \href{https://github.com/fhetextbook/fhe-textbook/issues}{\textbf{Issues Board}}.

\subsubsection*{Acknowledgments}
Special thanks go to the following researchers: Robin Geelen (KU Leuven, \href{mailto:robin.geelen@esat.kuleuven.be}{robin.geelen@esat.kuleuven.be}), for his thoughtful and dedicated feedback; Yongwoo Lee (Inha University, \href{mailto:yongwoo@inha.ac.kr}{yongwoo@inha.ac.kr}), for his general advice; Tianjian Yang (Peking University, \href{mailto:robin.geelen@esat.kuleuven.be}{2300012738@stu.pku.edu.cn}), for correcting numerous typos and other errors throughout the draft; Nolan Carouge (Grenoble INP Ensimag, \href{mailto:nolan.carouge.pro@gmail.com}{nolan.carouge.pro@gmail.com}), for proof-reading Part 1; and Navid Abapour (University of Surrey, \href{mailto:n.abapour@surrey.ac.uk}{n.abapour@surrey.ac.uk}), for proof-reading Part 3 and 4.

\thispagestyle{empty}

\newpage

\tableofcontents

\titleformat*{\section}{\LARGE\bfseries\scshape}
\titleformat*{\subsection}{\Large\bfseries}
\titleformat*{\subsubsection}{\bfseries}
\titleformat*{\paragraph}{\itshape\subsubsectionfont}
\titleformat*{\subparagraph}{\large\bfseries}

% page header foot 
%\usepackage{fancyhdr}
%\pagestyle{fancy}
%\lhead{Security and Privacy in Cyber-Physical Systems: Foundations and Applications}
%\rfoot{Copyright \textcopyright 2016 by Wiley}
% \thispagestyle{fancy}, after \maketitle
\newcommand{\para}[1]{\vspace{0.05in}\noindent{\bf{#1}}}

\newcommand{\white}[1]{{\textcolor{white}{#1}}} % phantom upgrade

\newcommand{\gap}[1]{\text{ } \text{#1} \text{ }}

\newcommand{\tboxlabel}[1]{$\bm{\langle}$Summary~{#1}$\bm{\rangle}$}

\newcommand{\tboxdef}[1]{$\bm{\langle}$Definition~{#1}$\bm{\rangle}$}

\newcommand{\tboxtheorem}[1]{$\bm{\langle}$Theorem~{#1}$\bm{\rangle}$}

\newtheorem{proposition}{Proposition}

\newtcolorbox[blend into=tables]{mytable}[2][]{float=htb, halign=center,  title={#2}, every float=\centering, #1}

% $\hat Y = \frac{1}{1 + e^{-Z}}$.

% $Z = {w_1 \cdot X_1 + w_2 \cdot X_2 + \dots + w_n \cdot X_n + b}$

\clearpage

%\section{Background}

\part{Basic Math}
\label{part:basic-math}

\renewcommand{\thesection}{A-\arabic{section}}
\setcounter{section}{0}

This chapter explains the basic mathematical components of number theory: group, field, order, roots of unity, cyclotomic polynomial, polynomial ring, and decomposition. These are essential building blocks for post-quantum cryptography.

\clearpage

\section{Modulo Arithmetic}
\label{sec:modulo}

\noindent \textbf{- Reference} 

\href{https://www.youtube.com/watch?v=fz1vxq5ts5I}{YouTube -- Extended Euclidean Algorithm Tutorial}.

\subsection{Overview}
\label{subsec:modulo-overview} 

\begin{tcolorbox}[title={\textbf{\tboxdef{\ref*{subsec:modulo-overview}} Integer Modulo}}]

\begin{itemize}

\item \textbf{Modulo} is the operation of computing the remainder obtained when one number is divided by another. \textbf{modulo} is often abbreviated as \textbf{mod}.

$ $

\item \textbf{$\bm{a}$ mod $\bm{q}$ (i.e., $\bm{a} \bm{\text{ modulo } q}$)} is the remainder after dividing $a$ by $q$, which is always an element of $\{0, 1, 2, 3, \cdots, q-1\}$. For example, $7 \bmod 5 = 2$, because the remainder of dividing 7 by 5 is 2. 

$ $

\item \textbf{Modulus:} Given $\bm{a}$ mod $\bm{q}$, we call the divisor $q$ the modulus, whereas \emph{modulo} refers to the operation.

$ $

\item \textbf{Modulo Congruence ($\bm{\equiv}$):} $a$ is congruent to $b$ modulo $q$ (i.e., $a \equiv b \textbf{ mod } q$) if they have the same remainder when divided by $q$. For example, $5 \equiv 12 \bmod 7$, because $5 \bmod 7 = 5$ and $12 \bmod 7 = 5$. In mathematics, the notation $a \equiv b \bmod q$ is identical to $a = b \pmod q$, meaning that the remainder of $a$ divided by $q$ is the same as the remainder of $b$ divided by $q$. Note that this notation differs from $a = b \bmod q$, which states that $a$ equals the remainder of $b$ divided by $q$.

$ $

\item \textbf{Congruence \textit{vs.} Equality:} 

$a \equiv b \bmod q \Longleftrightarrow a = b + k\cdot q$ \text{ } (for some integer $k$)

$ $

This means that $a$ and $b$ are congruent modulo $q$ if and only if $a$ and $b$ differ by some multiple of $q$. For example, $5 \equiv 12 \bmod 7 \Longleftrightarrow 5 = 12 + (-1)\cdot 7$

\end{itemize}

\end{tcolorbox}

\subsection{Modulo Arithmetic}
\label{subsec:modulo-arithmetic}

The supported modulo operations are addition, subtraction, and multiplication. The properties of these modulo operations are as follows:

\begin{tcolorbox}[title={\textbf{\tboxtheorem{\ref*{subsec:modulo-arithmetic}.1} Properties of Modulo Operations}}]
For any integer $x$, the following is true:

\begin{enumerate}
\item \textbf{Addition:} $a \equiv b \bmod q \Longleftrightarrow a + x\equiv b + x\bmod q$

\item \textbf{Subtraction:} $a \equiv b \bmod q \Longleftrightarrow a - x\equiv b - x\bmod q$

\item \textbf{Multiplication:} $a \equiv b \bmod q \Longleftrightarrow a \cdot x \equiv b \cdot x \bmod q$. 
This equivalence holds provided that $\gcd(x,q)=1$. Without this assumption, only the implication $a \equiv b \bmod q \Rightarrow a \cdot x \equiv b \cdot x \bmod q$ is guaranteed.

\end{enumerate}

\end{tcolorbox}

\begin{proof}

$ $

For any integer $x$,

\begin{enumerate}

\item \textbf{Addition:} $a \equiv b \bmod q \Longleftrightarrow a = b + k q $ (for some integer $k$) \textcolor{red}{ \text{ } \# $a$ and $b$ differ by some multiple of $q$}

$\Longleftrightarrow a + x = b + k\cdot q + x$

$\Longleftrightarrow a + x = b + x + k\cdot q$ \textcolor{red}{ $\rhd$ $a+x$ and $b+x$ differ by some multiple of $q$}

$\Longleftrightarrow a + x \equiv b + x \bmod q$

$ $

\item \textbf{Subtraction:} $a \equiv b \bmod q \Longleftrightarrow a = b + k q $ (for some integer $k$)

$\Longleftrightarrow a - x = b + k\cdot q - x$

$\Longleftrightarrow a - x = b - x + k\cdot q$  \textcolor{red}{ $\rhd$ $a-x$ and $b-x$ differ by some multiple of $q$}

$\Longleftrightarrow a - x \equiv b - x \bmod q$

$ $

\item \textbf{Multiplication:} $a \equiv b \bmod q \Longleftrightarrow a = b + k q $ (for some integer $k$)

$\Longrightarrow a \cdot x = b \cdot x + k\cdot q \cdot x$

$\Longrightarrow a \cdot x = b \cdot x + k_x\cdot q$ (where $k_x = k \cdot x$)  \textcolor{red}{ $\rhd$ $a\cdot x$ and $b\cdot x$ differ by some multiple of $q$}

$\Longrightarrow a \cdot x \equiv b \cdot x \bmod q$

Conversely, if $x$ and $q$ are coprime (i.e., $\gcd(x,q)=1$), then $x$ has a multiplicative inverse $x^{-1}$ modulo $q$. From $a \cdot x \equiv b \cdot x \bmod q$

$\Longrightarrow a \cdot x \cdot x^{-1} \equiv b \cdot x \cdot x^{-1} \bmod q$

$\Longrightarrow a \equiv b \bmod q$

\end{enumerate}
\end{proof}

Based on the modulo operations in Theorem~\ref*{subsec:modulo-arithmetic}.1, we can also derive the following properties of modulo arithmetic:

\begin{tcolorbox}[title={\textbf{\tboxtheorem{\ref*{subsec:modulo-arithmetic}.2} Properties of Modulo Arithmetic}}]

\begin{enumerate}
\item \textbf{Associative:} $(a \cdot b) \cdot c \equiv a \cdot (b \cdot c) \bmod q$

\item \textbf{Commutative:} $(a \cdot b) \equiv (b \cdot a) \bmod q$

\item \textbf{Distributive:} $(a \cdot (b + c)) \equiv ((a \cdot b) + (a \cdot c))  \bmod q$

\item \textbf{Interchangeable:} Congruent values are interchangeable in modulo arithmetic. 

For example, suppose $(a \equiv b \bmod q)$ and $(c \equiv d \bmod q)$. Then, $a$ and $b$ are interchangeable, and $c$ and $d$ are interchangeable in modulo arithmetic as follows:

$(a + c) \equiv (b + d) \equiv (a + d) \equiv (b + c) \bmod q$

$(a - c) \equiv (b - d) \equiv (a - d) \equiv (b - c) \bmod q$

$(a \cdot c) \equiv (b \cdot d) \equiv (a \cdot d) \equiv (b \cdot c) \bmod q$

\end{enumerate}

\end{tcolorbox}

The proof of Theorem~\ref*{subsec:modulo-arithmetic}.2 is similar to that of Theorem~\ref*{subsec:modulo-arithmetic}.1, which we leave as an exercise for the reader.

\subsection{Inverse}
\label{subsec:modulo-inverse}

\begin{tcolorbox}[title={\textbf{\tboxdef{\ref*{subsec:modulo-inverse}} Inverse in Modulo Arithmetic}}]

In modulo $q$ (i.e., in the world of remainders where all numbers have been divided by $q$), for each $a \in \{0, 1, 2, \cdots, q-1\}$:

\begin{itemize}

\item \textbf{Additive Inverse} of $a$ is denoted as $a_+^{-1}$ which satisfies $a + a_+^{-1} \equiv 0 \bmod q$. For example, in modulo 11, $3_+^{-1} = 8$, because $3 + 8 \equiv 0 \bmod 11$.

\item \textbf{Multiplicative Inverse} of $a$ is denoted as $a_*^{-1}$ which satisfies $a \cdot a_*^{-1} \equiv 1 \bmod{q}$. Such an inverse exists if and only if $\gcd(a,q)=1$. For example, modulo 11, $3_*^{-1} = 4$, because $3 \cdot 4 \equiv 1 \bmod{11}$.

\end{itemize}

\end{tcolorbox}

\subsection{Modulo Division}
\label{subsec:modulo-division}

In modulo arithmetic, \textit{modulo division} is different from regular numeric division. Strictly speaking, there is no separate operation called “modulo division”, because the modulo operation already returns only the remainder of a division. In practice, one uses “modulo division” to mean multiplying by a modular inverse when it exists, i.e., when $\gcd(a,q)=1$. \textit{Modulo division} of $b$ by $a$ modulo $q$ is equivalent to computing the \textit{modular} multiplication $b \cdot a^{-1} \bmod q$. The result of \textit{modulo division} is different from that of numeric division because \textit{modulo division} always gives an integer (a residue modulo $q$) (as it multiplies two integers modulo $q$), whereas numeric division gives a real number. The inverse of an integer modulo $q$ can be computed using the extended Euclidean algorithm (\href{https://www.youtube.com/watch?v=fz1vxq5ts5I}{YouTube tutorial}).

\subsection{Centered Residue Representation}
\label{subsec:modulo-centered}

Throughout this section, we have assumed that the residues are positive integers. For example, the possible residues modulo $q$ are assumed to be $\{0, 1, \cdots, q-1\}$. This system is called the canonical (i.e., unsigned) residue representation. On the other hand, there is also a counterpart system that assumes signed (i.e., centered) residues $\left\{-\dfrac{q}{2}, -\dfrac{q}{2} + 1, \cdots, 0, \cdots, \dfrac{q}{2} - 2, \dfrac{q}{2} - 1\right\}$\footnote{Here, we assume $q$ is an even number. In the case where $q$ is an odd number, the residues are $\left\{ -\dfrac{q-1}{2}, -\dfrac{q-3}{2}, \cdots, 0, \cdots, \dfrac{q-3}{2}, \dfrac{q-1}{2}\right\}$}, with the residues centered around $0$, and the total number of residues is the same, namely $q$. In both systems, a modulo operation changes a given value to another value within the system's residue range such that: (1) if the given value is greater than the upper bound of the residue range, the value is subtracted by the modulus $q$; (2) if the value is less than the lower bound of the residue range, the value is increased by the modulus $q$. The only difference between these two (canonical and centered) systems is their upper bounds and lower bounds: $0$ and $q-1$ in the canonical residue system, whereas $-\dfrac{q}{2}$ and $\dfrac{q}{2} - 1$ in the centered residue system. The canonical residue representation assumes that $\mathbb{Z}_q = \{0, 1, \cdots, q-1\}$, whereas the centered residue system assumes that $\mathbb{Z}_q = \left\{-\dfrac{q}{2}, -\dfrac{q}{2} + 1, \cdots, 0, \cdots, \dfrac{q}{2} - 2, \dfrac{q}{2} - 1\right\}$. 

In both systems, the same properties hold for addition, subtraction, multiplication, and division (when the divisor is invertible). This can be proved using the reasoning from \autoref{subsec:modulo-arithmetic}: any two congruent residues differ by an integer multiple of $q$ in either representation.

Also, the same property holds for an inverse: an inverse of $a$ modulo $q$ is $a^{-1}$ such that $a \cdot a^{-1} \equiv 1 \bmod q$. 

Using a signed residue representation is useful in certain cases. In an example of canonical (i.e., unsigned) residue representation, suppose we have the relation $a + b \bmod q$ and we know that in a given application, $a + b$ is guaranteed to be within the $[0, q-1]$ range (i.e., $0 \leq a + b \leq q-1$). Then, $(a + b \bmod q)$ = $a + b$, and thus we can remove the modulo operation, simplifying the relation. Now, suppose a different example of centered (i.e., signed) residue representation where we have the relation $a - b \bmod q$, and we know that in a given application, $a - b$ is guaranteed to be within the range $\left[-\dfrac{q}{2}, \dfrac{q}{2} - 1\right]$. Then, $(a - b \bmod q) = a - b$. However, notice that if the relation $a - b \bmod q$ were in a canonical residue representation, then we cannot remove the modulo operation because if $a - b$ is negative, then this becomes smaller than the lower bound of the canonical residue system (i.e., $0$), and thus a modulo reduction (i.e., addition by one or more $q$) is needed. 

In \autoref{subsec:rns-fastbconvex}, we design the \textsf{FastBConvEx} operation based on this beneficial property of centered residue representation: in this algorithm design, we can simplify $(\mu + u \bmod b_\alpha)$ to $\mu + u$ because we know that $-\dfrac{b_\alpha}{2} \leq \mu + u < \dfrac{b_\alpha}{2}$.

\clearpage

\section{Group}
\label{sec:group}
%\textbf{- First Read:} 
%\href{https://e.math.cornell.edu/people/belk/numbertheory/CyclotomicPolynomials.pdf}{Fields and Cyclotomic Polynomials}

\subsection{Definitions}
\label{subsec:group-def}

\begin{tcolorbox}[title={\textbf{\tboxdef{\ref*{subsec:group-def}} Group}}]
\noindent \textbf{\underline{Set Elements}}
\begin{itemize}
\item \textbf{Set ($\mathbb{S}$):} An unordered collection of elements: $\mathbb{S} = \{a, b, c, \ldots\}$
\item \textbf{Set Operations $\bm{(+, \cdot)}$:} We consider two binary operations on $\mathbb{S}$: addition $(+)$ and multiplication $(\cdot)$.
\item \textbf{Additive Identity ($0_{(+)}$ often written $0$):} An element $i \in \mathbb{S}$ is an additive identity if for all $a \in \mathbb{S}$, $i + a = a = a + i$.
\item \textbf{Multiplicative Identity ($1_{(\cdot)}$ often written $1$):} An element $i \in \mathbb{S}$ is a multiplicative identity if for all $a \in \mathbb{S}$, $i \cdot a = a = a \cdot i$
\item \textbf{Additive Inverse ($a^{-1}_{(+)}$):} For each $a \in \mathbb{S}$, its additive inverse $a^{-1}_{(+)}$, often written $-a$, is defined as an element such that $a + a^{-1}_{(+)} = 0_{(+)} = a^{-1}_{(+)} + a$ (i.e., additive identity)
\item \textbf{Multiplicative Inverse ($a^{-1}_{(\cdot)}$):} For each $a \in \mathbb{S}$ that is invertible with respect to $(\cdot)$, its multiplicative inverse $a^{-1}_{(\cdot)}$, often written $a^{-1}$, is defined as an element such that $a \cdot a^{-1}_{(\cdot)} = 1_{(\cdot)} = a^{-1}_{(\cdot)} \cdot a$ (i.e., multiplicative identity)
\end{itemize}

$ $

\noindent \textbf{\underline{Element Operation Features}}
\begin{itemize}
\item \textbf{Closed:} A set $\mathbb{S}$ is closed under the $(+)$ operation if for every $a, b \in \mathbb{S}$, it is the case that $a + b \in \mathbb{S}$. Likewise, a set $\mathbb{S}$ is closed under the $(\cdot)$ operation if for every $a, b \in \mathbb{S}$, it is the case that $a \cdot b \in \mathbb{S}$. 
\item \textbf{Associative:} For any $a,b,c \in \mathbb{S}$, $(a + b) + c = a + (b + c)$
\item \textbf{Commutative:} For any $a,b \in \mathbb{S}$, $a + b = b + a$
\item \textbf{Distributive:} If both $(+)$ and $(\cdot)$ are defined (e.g. in a ring), then $a \cdot (b + c) = (a \cdot b) + (a \cdot c)$, and $(a + b) \cdot c = a\cdot c + b\cdot c$.
\end{itemize}

$ $

\noindent \textbf{\underline{Group Types}}
\begin{itemize}
\item \textbf{Semigroup:} A semigroup is a set of elements which is closed and associative on a single operation ($+$ or $\cdot$)
\item \textbf{Monoid:} A monoid is a semigroup with an identity element $e$ (a neutral element that leaves any other element unchanged under the operation).

(e.g., $0$ is the identity element for $+$ operator, $1$ is the identity element for the $\cdot$ operator)
\item \textbf{Group:} A group is a monoid, and every element has an inverse with respect to the operation.
\item \textbf{Abelian Group:} An abelian group is a group, plus its operation is commutative.
\end{itemize}
\end{tcolorbox}

\subsection{Examples}
\label{subsec:group-ex}

$\mathbb{Z}$ (i.e., the set of all integers) is an abelian group under addition ($+$), because:
\begin{itemize}
\item \textbf{Closed:} For any integer $a, b \in \mathbb{Z}$, $a + b = c$ is also an integer (i.e. $a+b \in \mathbb{Z}$).
\item \textbf{Associative:} For any integer $a, b, c \in \mathbb{Z}$, $(a + b) + c = a + (b + c)$.
\item \textbf{Identity:} The additive identity is 0 because, for any $a \in \mathbb{Z}$, $a + 0 = a$.
\item \textbf{Inverse:} For each $a \in \mathbb{Z}$, its additive inverse is $-a$, as $a + (-a) = 0$.
\item \textbf{Commutative: } For any integer $a, b \in \mathbb{Z}$, $a + b = b + a$.
\end{itemize}

$ $

\noindent $\mathbb{Z}$ is a monoid under multiplication ($\cdot$) because:
\begin{itemize}
\item \textbf{Closed:} For any integer $a, b \in \mathbb{Z}$, $a \cdot b = c$ is also an integer (i.e., $a\cdot b \in \mathbb{Z}$).
\item \textbf{Associative:} For any integer $a, b, c \in \mathbb{Z}$, $(a \cdot b) \cdot c = a \cdot (b \cdot c)$.
\item \textbf{Identity:} The multiplicative identity is 1, because for any $a \in \mathbb{Z}$, $a \cdot 1 = a$.
\item \textbf{NO Inverse:} For an integer $a \in \mathbb{Z}$, its multiplicative inverse is $\dfrac{1}{a}$, but this is not necessarily an integer ($\notin \mathbb{Z}$); therefore, not every element has a multiplicative inverse. Thus, $(\mathbb{Z},\cdot)$ is not a group (though it is a monoid).
\end{itemize}
$ $

\noindent $\mathbb{R}^\times$ (i.e., the set of all nonzero real numbers) is an abelian group under multiplication ($\cdot$), because:
\begin{itemize}
\item \textbf{Closed:} For any real number $a, b \in \mathbb{R}^\times$, $a \cdot b = c$ is also a real number (and remains in $\mathbb{R}^\times$).
\item \textbf{Associative:} For any real number $a, b, c \in \mathbb{R}^\times$, $(a \cdot b) \cdot c = a \cdot (b \cdot c)$.
\item \textbf{Identity:} The multiplicative identity is 1, as for any real number $a \in \mathbb{R}^\times$, $a \cdot 1 = a$.
\item \textbf{Inverse:} For each real number $a \in \mathbb{R}^\times$, its multiplicative inverse is $\dfrac{1}{a}$, which is a non-zero real number ($\in \mathbb{R}^{\times}$).
\end{itemize}

\clearpage

\section{Field}
\label{sec:field}
\textbf{- Reference:} 
\href{https://e.math.cornell.edu/people/belk/numbertheory/CyclotomicPolynomials.pdf}{Fields and Cyclotomic Polynomials}~\cite{cyclotomic-polynomial}

\subsection{Definitions}
\label{subsec:field-def}

\begin{tcolorbox}[title={\textbf{\tboxdef{\ref*{subsec:field-def}} Field Definitions}}]
\begin{itemize}
\item \textbf{Ring:} A set $R$ that is an abelian group under addition $(+)$, equipped with a multiplication $(\cdot)$ that is closed and associative, and such that multiplication distributes over addition on both sides: $a\cdot(b+c)=a\cdot b+a\cdot c$ and $(a+b)\cdot c=a\cdot c+b\cdot c$ for all $a,b,c\in R$. (Multiplication is not necessarily commutative, e.g., a matrix multiplication, and an identity element for $(\cdot)$ is optional unless stated “ring with unity”.)
\item \textbf{Field:} A set $F$ that is an abelian group under $(+)$, whose nonzero elements $F^\times=F\setminus\{0\}$ form an abelian group under $(\cdot)$, with multiplication distributing over addition.
\item \textbf{Galois Field ($\mathrm{GF}(p^n)$):} A field with $p^n$ elements for some prime $p$ and positive integer $n$.
\item \textbf{$\mathbb{Z}_p$ ($\mathbb{Z}/p\mathbb{Z}$):} For a prime $p$, the set $\{0,1,\ldots,p-1\}$ with addition and multiplication modulo $p$ forms a finite field. More generally, for any integer $m\ge 2$, $\mathbb{Z}_m$ is a commutative ring, and it is a field iff $m$ is prime.\end{itemize}
\end{tcolorbox}

$ $

\subsection{Examples}
\label{subsec:field-ex}

$\mathbb{Z}$ (the set of all integers) is a ring but not a field, because not all of its elements have a multiplicative inverse (as shown in \autoref{subsec:group-ex}). 

$ $

\noindent $\mathbb{R}$ (the set of all real numbers) is a field. As shown in \autoref{subsec:group-ex}, it is an abelian group under $(+)$; its nonzero elements form an abelian group under $(\cdot)$, and multiplication distributes over addition.

$ $

\noindent $\mathbb{Z}_7 = \{0, 1, 2, 3, 4, 5, 6\}$ is a finite field because:
\begin{itemize}
\item \textbf{Closed:} For any $a,b \in \mathbb{Z}_7$, there exist $c_1,c_2 \in \mathbb{Z}_7$ such that $a+b \equiv c_1 \pmod{7}$ and $a\cdot b \equiv c_2 \pmod{7}$.
\item \textbf{Associative:} For any $a,b,c \in \mathbb{Z}_7$, $(a+b)+c=a+(b+c)$, and $(a\cdot b)\cdot c=a\cdot(b\cdot c)$.
\item \textbf{Commutative:} For any $a,b \in \mathbb{Z}_7$, $a+b=b+a$, and $a\cdot b=b\cdot a$.
\item \textbf{Distributive:} For any $a,b,c \in \mathbb{Z}_7$, $(a+b)\cdot c=a\cdot c+b\cdot c$, and $a\cdot(b+c)=a\cdot b+a\cdot c$.
\item \textbf{Identity:} The additive identity is $0$, and the multiplicative identity is $1$.
\item \textbf{Inverse:} For any $a \in \mathbb{Z}_7$, there exists $a' \in \mathbb{Z}_7$ such that $a+a' \equiv 0 \pmod{7}$ (e.g., the additive inverse of $3$ is $4$ since $3+4\equiv0 \pmod{7}$). For any $a \in \mathbb{Z}_7^\times=\{1,\dots,6\}$, there exists $b \in \mathbb{Z}_7^{\times}$ such that $ab \equiv 1 \pmod{7}$ (e.g., $3\cdot5=15\equiv1 \pmod{7}$).
\end{itemize}

\subsection{Theorems}
\label{subsec:field-theorem}

\begin{tcolorbox}[title={\textbf{\tboxtheorem{\ref*{subsec:field-theorem}} Field Theorems}}]
\begin{enumerate}
\item \textbf{Size of Finite Field:} Every finite field (also called a Galois Field) has $p^n$ elements for some prime $p$ and positive integer $n$, conversely, for each such prime power $p^n$, there exists a finite field of order $p^n$ (unique up to isomorphism).
\item \textbf{Isomorphic Fields:} Any two finite fields $\mathbb{F}_1$ and $\mathbb{F}_2$ with the same number of elements are isomorphic, i.e., there exists a bijection $f:\mathbb{F}_1\to\mathbb{F}_2$ such that for all $a,b\in\mathbb{F}_1$, $f(a+b)=f(a)+f(b)$ and $f(ab)=f(a)f(b)$.
\end{enumerate}
\end{tcolorbox}

\clearpage

\section{Order}
\label{sec:order}
\textbf{- Reference:} 
\href{https://e.math.cornell.edu/people/belk/numbertheory/CyclotomicPolynomials.pdf}{Fields and Cyclotomic Polynomials}~\cite{cyclotomic-polynomial}

\subsection{Definitions}
\label{subsec:order-def}

\begin{tcolorbox}[title={\textbf{\tboxdef{\ref*{subsec:order-def}} Order Definition}}]
$\bm{\textsf{ord}_{\mathbb{F}}(a)}$: For $a \in \mathbb{F}^{\times}$ (a finite field, \autoref{subsec:field-def}), $a$'s order is the smallest positive integer $k$ such that $a^k = 1$. 
\end{tcolorbox}

\subsection{Theorems}
\label{subsec:order-theorem}

\begin{tcolorbox}[title={\textbf{\tboxtheorem{\ref*{subsec:order-theorem}.1} Order Property (I)}}]
For $a \in \mathbb{F}^{\times}$, and $n \geq 1$, $a^n = 1$ if and only if \textbf{\textsf{ord}}$_{\mathbb{F}}(a) \text{ } \mid \text{ } n$

(i.e., $\textsf{ord}_{\mathbb{F}}(a)$ divides $n$).
\end{tcolorbox}

\begin{myproof}
    \begin{enumerate}
    \item \textit{Forward Proof:} If $\textsf{ord}_{\mathbb{F}}(a) \text{ } | \text{ } n$, then for $\textsf{ord}_{\mathbb{F}}(a) = k$ where $k$ is $a$'s order, and $n = lk$ for some integer $l$. 
    
    Then, $a^n = a^{lk} = (a^k)^l = 1^l = 1$.
    \item \textit{Backward Proof:} If $a^n = 1$ and $\textsf{ord}_{\mathbb{F}}(a)=k$, write $n=qk+r$ with $0 \le r < k$. Then $1=a^n=a^{qk+r}=(a^k)^q a^r=a^r$. By minimality of $k$, we must have $r=0$, hence $k \mid n$.
    \end{enumerate}
\end{myproof}

\begin{tcolorbox}[title={\textbf{\tboxtheorem{\ref*{subsec:order-theorem}.2} Order Property (II)}}]
If $\textsf{ord}_{\mathbb{F}}(a) = k$, then for any $n \geq 1$, $\textsf{ord}_{\mathbb{F}}(a^n) = \dfrac{k} {\gcd(k, n)}$.
\end{tcolorbox}
\begin{myproof}
    \begin{enumerate}
    \item $a^k, a^{2k}, a^{3k}, \ldots = 1$. 
    \item Given $\textsf{ord}_{\mathbb{F}}(a^n) = m$, $(a^n)^m, (a^n)^{2m}, (a^n)^{3m}, \ldots = 1$ 
    \item Note that by definition of order, $x=k$ is the smallest value that satisfies $a^x$ = 1. Thus, given $\textsf{ord}_{\mathbb{F}}(a^n) = m$, then $m$ is the smallest integer that makes $(a^n)^m = 1$. Note that $(a^n)^m$ should be a multiple of $a^k$, which means $mn$ should be a multiple of $k$. The smallest possible integer $m$ that makes $mn$ a multiple of $k$ is $m = \dfrac{k}{\gcd(k, n)}$. 
    \end{enumerate}
\end{myproof}

\begin{tcolorbox}[title={\textbf{\tboxtheorem{\ref*{subsec:order-theorem}.3} Order Property (III)}}]
Suppose $k$ divides $n$.  Then, $\textsf{ord}_{\mathbb{F}}(a) = kn$ if and only if $\textsf{ord}_{\mathbb{F}}(a^k) = n$.
\end{tcolorbox}
\begin{myproof}
\begin{enumerate}
    \item \textit{Forward Proof:} Given $\textsf{ord}_{\mathbb{F}}(a) = kn$, and given Theorem~\ref*{subsec:order-theorem}.2, $\textsf{ord}_{\mathbb{F}}(a^k) = \dfrac{nk}{\gcd(k, nk)} = \dfrac{nk}{k} = n$.
    \item \textit{Backward Proof:} Given $\textsf{ord}_{\mathbb{F}}(a^k)=n$ and letting $\textsf{ord}_{\mathbb{F}}(a)=m$, Theorem~\ref*{subsec:order-theorem}.2 gives $\textsf{ord}_{\mathbb{F}}(a^k)=\dfrac{m}{\gcd(m,k)}=n$, so $m=n\cdot\gcd(m,k)$ (i.e., $m$ is some multiple of $n$). But since $k$ divides $n$, $k$ also divides $m$. This means that $\gcd(m,k) = k$. Hence, $\textsf{ord}_{\mathbb{F}}(a)=m=n\cdot\gcd(m,k)=nk$.
\end{enumerate}
\end{myproof}

\begin{tcolorbox}[title={\textbf{\tboxtheorem{\ref*{subsec:order-theorem}.4} Fermat's Little Theorem}}]
Given $|\mathbb{F}| = p$ (a prime) and $a \in \mathbb{F}$, $a^p = a$.
\end{tcolorbox}
\begin{myproof}
    \begin{enumerate}
    \item If $a=0$, then $a^p=a=0$.
    \item If $a\ne 0$, then $a \in \mathbb{F}^{\times}$, the multiplicative group of the field, which has size $|\mathbb{F}^{\times}|=p-1$.
    By Lagrange's theorem (in a finite group $G$, the order of any element divides $|G|$), the order of $a$ divides $p-1$, hence $a^{p-1}=1$. Therefore $a^p=a$.
    \end{enumerate}
\end{myproof}

\clearpage

\section{Polynomial Ring}
\label{sec:polynomial-ring}
\textbf{- Reference 1:} 
\href{https://en.wikipedia.org/wiki/Polynomial_ring}{Polynomial Ring (Wikipedia)}~\cite{polynomial-ring}

\noindent \textbf{- Reference 2:} 
\href{https://math.libretexts.org/Bookshelves/Combinatorics_and_Discrete_Mathematics/Applied_Discrete_Structures_(Doerr_and_Levasseur)/16%3A_An_Introduction_to_Rings_and_Fields/16.03%3A_Polynomial_Rings}{Polynomial Rings (LibreTexts)}~\cite{polynomial-rings}

\subsection{Overview}
\label{subsec:poly-ring-overview}

\textbf{A polynomial ring} is a set of polynomials where polynomial computations over the $(+, \cdot)$ operators (e.g., $f_1 + f_2$, $f_1 \cdot (f_2 - f_3)$, $f_1 + f_2 + f_4$) are closed, associative, commutative, and distributive.

A polynomial ring ${\mathbb{Z}_p[x] / (x^n + 1)}$ is the set of all polynomials $f_i$ that have the following properties:

\begin{tcolorbox}[title={\textbf{\tboxlabel{\ref*{subsec:poly-ring-overview}} Ring}}]

For a polynomial $f \in \mathbb{Z}_p[x] / (x^n + 1)$ where $f = c_0 + c_1x^1 + \cdots + c_{n-1}x^{n-1}$:

\begin{itemize}
\item \textbf{Coefficient Ring:} each coefficient $c_j \in \mathbb{Z}_p$.

$ $

\item \textbf{Degree Bound:} Any $f' \in \mathbb{Z}_p[x]$ can be written as:

$ $

$f'=(x^n+1)f_q+f_r,\qquad \deg f_r < n,$

$ $

so in the quotient ring $\mathbb{Z}_p[x]/(x^n+1)$ we have $f' \equiv f_r \pmod{x^n+1}$. $f_q$ is called a quotient polynomial and $f_r$ is called a remainder polynomial resulting from the polynomial division of $f'$ divided by $x^n + 1$.

$ $

\item \textbf{Polynomial Congruence:} If two polynomials are congruent, they belong to the same equivalence class, in which case they are interchangeable in the polynomial operations ($+, \cdot$) in the polynomial ring. For example, if: 

$ $

$f' \equiv f_{r1} \in \mathbb{Z}_p[x] / (x^n + 1)$

$f{''} \equiv f_{r2} \in \mathbb{Z}_p[x] / (x^n + 1)$

$f_{r1} + f_{r2} \equiv f_{r3} \in \mathbb{Z}_p[x] / (x^n + 1)$

$ $

Then the polynomial operation result of $f' + f''$ is in the same equivalence class as: 

$f' + f'' \equiv f_{r1} + f_{r2} \equiv f_{r3} \in \mathbb{Z}_p[x] / (x^n + 1)$

\end{itemize}

To make the notation simple, we denote the polynomial ring $\mathbb{Z}_p[x] / (x^n + 1)$ as $\mathcal{R}_{\langle n, p \rangle}$
\end{tcolorbox}

Recall that in $\mathbb{Z}_p$, any $b$ writes $b = mp + r$ with $0\le r<p$, hence $b \equiv r \pmod{p}$ (the quotient $m$ disappears). Similarly, in a polynomial ring $\mathcal{R}_{\langle n, p \rangle}$, a high-degree polynomial $f_{big}$ can be divided by the polynomial modulo $x^n + 1$, which yields:

$f_{big} = (x^n + 1)\cdot(f_q) + f_r \equiv f_r \in \mathcal{R}_{\langle n, p \rangle}$

\noindent, whereas $f_q$ is a quotient polynomial, and $f_r$ is a remainder polynomial. In this case, $f_{big}$ is congruent to (i.e., it is in the same equivalence class as) $f_r$. Thus, $f_q$ can be eliminated, and $f_r$ (i.e., the simplified version of $f_{big}$) can be used interchangeably for polynomial operations $(+, \cdot)$ in the polynomial ring. 
Polynomial simplification (i.e., reduction) in a polynomial ring is done by substituting $x^n \equiv -1$ into $f_{big}$ because $x^n + 1 \equiv 0$ in the polynomial ring (this is the same as the case of a number ring modulo $p$, where we reduce a number by substituting $0$ for $p$). This way, a high-degree polynomial $f_{big}$ can be recursively simplified to a polynomial of degree less than $n$ by recursively substituting $x^n \equiv -1$ into $f_{big}$.

For a polynomial modulo, we normally choose a cyclotomic polynomial $x^n + 1$ (where $n$ is $2^p$ for some integer $p$) as the divisor, as it provides computational efficiency. 

\subsubsection{Example}
\label{subsubsec:poly-ring-ex}
Given $f \in \mathbb{Z}_7[x] / (x^2 + 1)$, suppose $f = x^4 + 3x^3 + 11x^2 + 6x + 10$. Then, 

$ $

$f = (x^2)\cdot(x^2) + 3x\cdot(x^2) + 11x^2 + 6x + 10$ 

$ \equiv (-1)(-1) + 3x(-1) +  (11 \bmod 7)(-1) + 6x + (10 \bmod 7)$

$ = 3x \in \mathbb{Z}_7[x] / (x^2 + 1)$

$ $

\noindent Thus, $f(x) = x^4 + 3x^3 + 11x^2 + 6x + 10$ is equivalent to ($\equiv$) $3x$ in the polynomial ring $\mathbb{Z}_7[x] / (x^2 + 1)$.

\subsubsection{Discussion}
\label{subsubsec:polynomial-ring-discuss}

\begin{table}[h]
\centering
\footnotesize
%\noindent\adjustbox{max width=\columnwidth}{
\begin{tabular}{|c||c|c|} % left align
\hline
&\textbf{Ring} & \textbf {Polynomial Ring} \\
\hline
\hline
\textbf{{Set Elements}} & number & polynomial \\
\hline
\textbf{{Ring Notation}} & $\mathbb{Z}_p = \{0, 1, \ldots, p - 1\}$  & $\mathbb{Z}_p[x] / (x^n + 1)$ \\
\textbf{{\& Definition}} & The set of all integers between $0$ and $p$ & The set of all polynomials $f$ such that\\
&& $f= c_0 + c_1x^1 + c_2x^2 \cdots + c_{n-1}x^{n-1}$ \\
&& where each coefficient $c_i \in \mathbb{Z}_p$ \\
&& and $f$'s degree is less than $n$ \\
\hline
\textbf{{Supported}}&$(+, \cdot)$& $(+, \cdot)$ \\
\textbf{{Operations}}&(Addition, Multiplication)& (Addition, Multiplication) \\
\hline
\textbf{{($+$) Operation}} & We know how to add numbers & $f_a = a_0 + a_1x^1 + a_2x^2 \cdots + a_{d_a-1}x^{d_a-1}$ \\
 & & $f_b = b_0 + b_1x^1 + b_2x^2 \cdots + b_{d_b-1}x^{d_b-1}$ \\
 & & Then, $f_a + f_b$ is computed as: \\
 & & $f_c = \sum\limits_{i=0}^{\textsf{max}(d_a,d_b)}(a_i+b_i)x^i$ \\
\hline
\textbf{{($\cdot$) Operation}} & We know how to multiply numbers & $f_a = a_0 + a_1x^1 + a_2x^2 \cdots + a_{d_a-1}x^{d_a-1}$ \\
 & & $f_b = b_0 + b_1x^1 + b_2x^2 \cdots + b_{d_b-1}x^{d_b-1}$ \\
 & & Then, $f_a \cdot f_b$ is computed as: \\
 & & $f_c = \sum\limits_{i=0}^{d_a+d_b}\sum\limits_{j=0}^{i}a_jb_{i-j}x^i$ \\
\hline
& For $a, b \in \mathbb{Z}_p$ & For $f_a, f_b \in \mathbb{Z}_p[x]/(x^n + 1)$,\\
\textbf{{Commutative}}  & $ a + b = b + a$ & $f_a + f_b = f_b + f_a$\\
\textbf{{Associative}} & $(a + b) + c = a + (b + c)$ & $(f_a + f_b) + f_c = f_a + (f_b + f_c)$\\
\textbf{{Distributive}} & $a \cdot (b + c) = a\cdot b + a\cdot c$ & $f_a \cdot (f_b + f_c) = f_a\cdot f_b + f_a\cdot f_c$\\
\textbf{{Closed}}& $a + b \equiv c \in \mathbb{Z}_p$, $a \cdot b \equiv d \in \mathbb{Z}_p$ & $f_a + f_b \equiv f_c \in \mathcal{R}_{\langle n, p \rangle}$, \\
&&$f_a \cdot f_b \equiv f_d \in \mathcal{R}_{\langle n, p \rangle}$\\
\hline
\textbf{{Congruency ($\equiv$)}} & Two numbers $a \equiv b$ in modulo $p$ if: & Two polynomials $f_a \equiv f_b$ in $\mathbb{Z}_p[x] / (x^n + 1)$ if:    \\
 & $(a \bmod p) = (b \bmod p)$   & $f'_a = f_a \bmod (x^n + 1) = \sum\limits_{i=0}^{d_a} a_ix^i$ \\
 & & $f'_b = f_b \bmod (x^n + 1) = \sum\limits_{i=0}^{d_b} b_ix^i$, \\
&& $d_a = d_b$ and $a_i \equiv b_i$ in modulo $p$\\
&&for all $0 \leq i \leq d_a$\\ 
\hline
\end{tabular}%}
\centering
\caption{Comparison between a number ring and a polynomial ring.
}
\label{tab:ring-comparison}
\end{table}

\para{Congruency:} If two numbers are congruent, they belong to the same \textit{congruence class}. The same is true for two congruent polynomials. If the computation results of two mathematical formulas belong to the same congruency class, then their computations wrap around within the modulus of their congruency. This is a useful property for cryptographic schemes where encryption \& decryption computations wrap around their values within a limited set of binary bits. Congruency is useful for simplifying computations. For example, a large number or a complex polynomial can be \textit{normalized} to a simpler number or polynomial by using the congruency rule, which reduces the computational overhead. 

\para{Polynomial Evaluation:} Note that two numbers that belong to the same congruence class are not necessarily the same number. For example, $5 \equiv 10$ modulo 5, but these two numbers are not the same. Likewise, two congruent polynomials are not the same. While two congruent polynomials in a polynomial ring can be interchangeably used for polynomial operations supported in the ring (i.e., $(+, \cdot)$), such as $f_1 + f_2$ or $f_1 \cdot (f_2 - f_3)$, two congruent polynomials do not necessarily yield the same result when evaluated for a certain variable value $x = a$. For example, in the previous example of \autoref{subsubsec:poly-ring-ex}, the two polynomials $x^4 + 3x^3 + 11x^2 + 6x + 10$ and $3x$ are congruent in the polynomial ring $\mathbb{Z}_7[x] / (x^2 + 1)$. However, these two polynomials do not give the same evaluation results for $x = 0$: the original polynomial gives 10, whereas the reduced (i.e., simplified) polynomial gives 0. This discrepancy in evaluation occurs because we defined two polynomials $M_1$ and $M_2$ to be congruent over $x^n + 1$ (i.e., $M_1 \equiv M_2$) if their remainder is the same after being divided by $x^n + 1$ (i.e., $M_1 = Q \cdot (x^n + 1) + M_2$ for some polynomial $Q$). Therefore, $M_1$ and $M_2$ will be evaluated to the same polynomial $M_2$ if they are evaluated at the $x$ values such that $x^n = -1$, which makes the $x^n +1$ term 0. The solutions for $x^n = -1$ are called the $n$-th roots of unity, which we will learn in \autoref{sec:roots} and \autoref{sec:cyclotomic-polynomial-integer-ring}. We summarize as follows: 

\begin{tcolorbox}[title={\textbf{\tboxlabel{\ref*{subsubsec:polynomial-ring-discuss}} Polynomial Evaluation over Polynomial Ring}}]

Suppose polynomials $M_1$ and $M_2$ are congruent over the polynomial ring $x^n + 1$. This is equivalent to the following relation: $M_1 = Q \cdot (x^n + 1) + M_2$ for some polynomial $Q$. Then, $M_1(X)$ and $M_2(X)$ are guaranteed to be evaluated to the same value if $X=x_i$ is the solution for $x^n + 1$ (i.e., $X$ is the $n$-th root of unity). That is , $M_1(x_i) = M_2(x_i)$. 

\end{tcolorbox}

\para{Number Ring \& Polynomial Ring:} These two rings share many common characteristics, which are summarized in \autoref{tab:ring-comparison}.

\subsection{Coefficient Rotation}
\label{subsec:coeff-rotation}

Coefficient rotation is a process of shifting the entire coefficients of a polynomial (either to the left or right) in a polynomial ring. In order to rotate the entire coefficients of a polynomial by $h$ positions to the left, we multiply $x^{-h}$ with the polynomial. 

For example, suppose we have a polynomial as follows:

$ $

$f(x) = c_0 + c_1x^1 + c_2x^2 + \cdots + c_hx^h + \cdots + c_{n-1}x^{n-1} \in \mathcal{R}_{\langle n, p \rangle}$

$ $

To shift the entire coefficients of $f$ to the left by $h$ positions (i.e., shift $f$'s $h$-th coefficient to the constant term), we compute $f \cdot x^{-h}$, which is:

$ $

\begin{tcolorbox}[title={\textbf{\tboxlabel{\ref*{subsec:coeff-rotation}} Polynomial Rotation}}]

Given the $(n-1)$-degree polynomial: 

$ $

$f(x) = c_0 + c_1x^1 + c_2x^2 + \cdots + c_hx^h + \cdots + c_{n-1}x^{n-1} \in \mathcal{R}_{\langle n, p \rangle}$

$ $

The coefficients of $f(x)$ can be rotated to the left by $h$ positions by multiplying to $f(x)$ by $x^{-h}$ as follows:

$ $

$f(x) \cdot x^{-h} = c_0\cdot x^{-h} + c_1x^1 \cdot x^{-h} + c_2x^2 \cdot x^{-h} + \cdots + c_hx^h \cdot x^{-h} + \cdots + c_{n-1}x^{n-1} \cdot x^{-h}$
$\equiv c_h + c_{h+1}x + c_{h+2}x^2 + \cdots + c_{n-1}x^{n-1-h} - c_0x^{n - h} - \cdots - c_{h-1}x^{n-1} \in \mathcal{R}_{\langle n, p \rangle}$
\end{tcolorbox}

$ $

Note that multiplying the two polynomials $f$ and $x^{-h}$ will yield a congruent polynomial in $\mathcal{R}_{\langle n, p \rangle}$. Therefore, the rotated polynomial, which is the result of $f \cdot x^{-h}$, will also have a congruent polynomial in $\mathcal{R}_{\langle n, p \rangle}$. 

Note that the coefficient signs change when they rotate around the boundary of $x^n (= -1)$, as the computation is conducted in the polynomial ring $\mathbb{Z}_p[x] / (x^n + 1)$.

\subsubsection{Example}
\label{subsec:coeff-rotation-ex}

Suppose we have a polynomial $f \in \mathbb{Z}_8[x] / (x^4 + 1)$ as follows:

We use the centered residue system for $\mathbb{Z}_8$, i.e., $\{-4,-3,-2,-1,0,1,2,3\}$.

$f = 2 + 3x - 4x^2 -x^3$

$ $

The polynomial ring $\mathbb{Z}_8[x] / (x^4 + 1)$ has the following 4 congruence relationships: 

{\boldmath{$x^4 \equiv -1$}}

$x^4 \cdot x^{-1} \equiv -1 \cdot x^{-1}$

{\boldmath{$x^3 \equiv -x^{-1}$}}

$ $

$x^4 \equiv -1$

$x^4 \cdot x^{-3} \equiv -1 \cdot x^{-3}$

{\boldmath{$x \equiv -x^{-3}$}}

$ $

$x^4 \equiv -1$

$x^4 \cdot x^{-2} \equiv -1 \cdot x^{-2}$

{\boldmath{$x^2 \equiv -x^{-2}$}}

$ $

Then, based on the coefficient rotation technique in Summary~\ref*{subsec:coeff-rotation}, rotating 1 position to the left is equivalent to computing $f \cdot x^{-1}$ as follows:

$f\cdot x^{-1} = 2\cdot(x^{-1}) + 3x\cdot(x^{-1}) - 4x^2\cdot(x^{-1}) - x^3\cdot(x^{-1})$

$\equiv -2x^{3} + 3 - 4x^1 - x^2$

$= 3 - 4x^1 - x^2 -2x^{3}$

$ $

As another example, rotating 3 positions to the left is equivalent to computing $f \cdot x^{-3}$ as follows:

$f\cdot x^{-3} = 2\cdot(x^{-3}) + 3x\cdot(x^{-3}) - 4x^2\cdot(x^{-3}) - x^3\cdot(x^{-3})$

$\equiv -2x - 3x^2 + 4x^3 - 1$

$= -1 - 2x - 3x^2 + 4x^3$

$= -1 - 2x - 3x^2 + (4 \equiv -4 \bmod 8) x^3$

$\equiv -1 - 2x - 3x^2 -4x^3$

\clearpage

\section{Decomposition}
\label{sec:decomp}
Decomposition is a mathematical technique used to represent a number in a smaller base (radix) while preserving its value. This section will explain number decomposition and polynomial decomposition. 

\subsection{Number Decomposition}
\label{subsec:number-decomp}
We fix a modulus $q\ge 2$ and write $\mathbb{Z}_q=\mathbb{Z}/q\mathbb{Z}$. Let $\gamma\in\mathbb{Z}_q$. Number decomposition expresses $\gamma$ as a sum of multiple numbers in base $\beta$ as follows: 

$ $

$\gamma = \gamma_1 \dfrac{q}{\beta^1} + \gamma_2 \dfrac{q}{\beta^2} + \cdots + \gamma_\ell \dfrac{q}{\beta^\ell}  $

\begin{figure}[h!]
    \centering
  \includegraphics[width=0.8\linewidth]{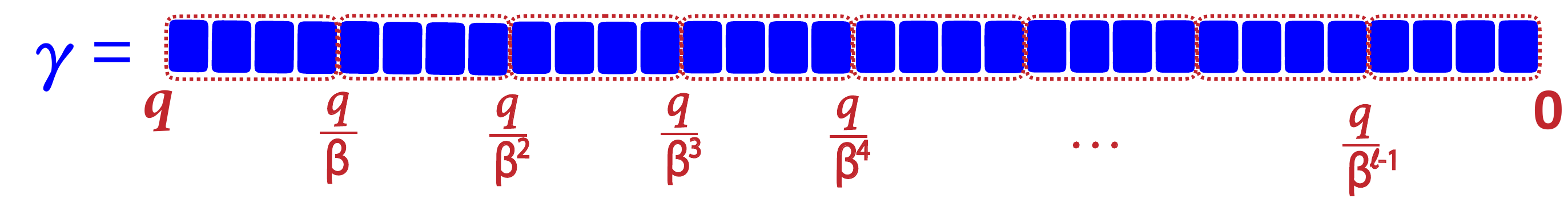}
  \caption{An illustration of number decomposition.}
  \label{fig:decomp}
\end{figure}

$ $

\noindent where $\beta\ge 2$ is a base and $\ell\ge 1$ is the decomposition level. We assume $\beta^\ell \mid q$ and take digits $\gamma_i\in\{0,1,\ldots,\beta-1\}$; under these conditions, the decomposition is unique. This is visually shown in \autoref{fig:decomp}. (If $\beta^\ell\nmid q$, see \autoref{subsec:approx-decomp}.) 
Each $\gamma_i$ is a digit in the base-$\beta$ representation of $\gamma$, where $i=1$ is the most significant digit. When $q$ is a power of two, this corresponds to a shift by $i\cdot \log_2\beta$ bits.

We define the decomposition of the number $\gamma$ into base $\beta$ with level $\ell$ as follows:

$ $

$\textsf{Decomp}^{\beta, \ell}(\gamma) = (\gamma_1, \gamma_2, \text{ } \cdots , \text{ } \gamma_\ell)$.
 
$ $

Number decomposition is also called radix decomposition, where the base $\beta$ is referred to as a radix.

\subsubsection{Example}

Suppose we take $\gamma=13$ in $\mathbb{Z}_{16}$. Suppose we want to decompose 13 with the base $\beta = 2$ and level $\ell = 4$. Then, 13 is decomposed as follows:

$ $

$13 = 1 \cdot \dfrac{16}{2^1} + 1 \cdot \dfrac{16}{2^2} + 0 \cdot \dfrac{16}{2^3} + 1 \cdot \dfrac{16}{2^4}$

$ $

$\textsf{Decomp}^{2, 4}(13) = (1, 1, 0, 1)$

\subsection{Polynomial Decomposition}
\label{subsec:poly-decomp}

This time, suppose we have a polynomial $f$ in the polynomial ring ${\mathbb{Z}_q[x] / (x^n + 1)}$. Therefore, each coefficient $c_i$ of $f$ is an integer modulo $q$. Polynomial decomposition expresses $f$ as a sum of multiple polynomials in base $\beta$ and level $\ell$ as follows:

\begin{tcolorbox}[title={\textbf{\tboxlabel{\ref*{subsec:poly-decomp}} Polynomial Decomposition}}]

Given $f\in \mathbb{Z}_q[x]/(x^n+1)$, fix $\beta\ge 2$ and $\ell\ge 1$ with $\beta^\ell\mid q$. We write

$ $

$
f=\sum\limits_{i=1}^{\ell} f_i\,\dfrac{q}{\beta^i}, \qquad f_i\in \mathbb{Z}_q[x]/(x^n+1)
$

$ $

where each $f_i$ is obtained by decomposing every coefficient of $f$ in base $\beta$. If $f=\sum\limits_j c_j x^j$ with $c_j\in\mathbb{Z}_q$, then
$c_j=\sum\limits_{i=1}^{\ell} c_{j,i}\,\dfrac{q}{\beta^i}$ with $c_{j,i}\in\{0,\ldots,\beta-1\}$,
and $f_i=\sum_j c_{j,i} x^j$.
We denote the decomposition of the polynomial $f$ into the base $\beta$ with the level $\ell$ as follows:

$ $

$\textsf{Decomp}^{\beta, \ell}(f) = (f_1, f_2, \text{ } \cdots , \text{ } f_\ell)$
 $ $
\end{tcolorbox}

\subsubsection{Example}

Suppose we have the following polynomial in the polynomial ring $\mathbb{Z}_{16}[x] / (x^4 + 1)$:

$ $

$f = 6x^3 + 3x^2 + 14x + 7 \in \mathbb{Z}_{16}[x] / (x^4 + 1)$

$ $

Suppose we want to decompose the above polynomial with base $\beta = 4$ and level $\ell = 2$. Then, each degree's coefficient of the polynomial $f$ is decomposed as follows:

$ $

${\bm{x}^{\bm{3}}}$: $6 = \color{blue}{1 \cdot \dfrac{16}{4^1}} \color{black}+ \color{red}{2 \cdot \dfrac{16}{4^2}}$

${\bm{x}^{\bm{2}}}$: $3 = \color{blue}{0 \cdot \dfrac{16}{4^1}} \color{black}+ \color{red}{3 \cdot \dfrac{16}{4^2}}$

${\bm{x}^{\bm{1}}}$: $14 = \color{blue}{3 \cdot \dfrac{16}{4^1}} \color{black}+ \color{red}{2 \cdot \dfrac{16}{4^2}}$

${\bm{x}^{\bm{0}}}$: $7 = \color{blue}{1 \cdot \dfrac{16}{4^1}} \color{black}+ \color{red}{3 \cdot \dfrac{16}{4^2}}$

$ $

The decomposed polynomial is as follows:

$f = 6x^3 + 3x^2 + 14x + 7 = \color{blue}{(1x^3 + 0x^2 + 3x + 1) \cdot \dfrac{16}{4^1}} \color{black}+ \color{red}{(2x^3 + 3x^2 + 2x + 3) \cdot \dfrac{16}{4^2}} \color{black}$

$ $

$\textsf{Decomp}^{4, 2}(6x^3 + 3x^2 + 14x + 7) = (1x^3 + 0x^2 + 3x + 1, 2x^3 + 3x^2 + 2x + 3)$

\subsubsection{Discussion}

Note that after decomposition, the original coefficients of the polynomial have been reduced to smaller numbers. This characteristic is importantly used in the multiplication of polynomials in FHE ciphertexts to reduce the growth rate of the noise. Normally, the polynomial coefficients of ciphertexts are large because they are uniformly random numbers. Reducing such large constants is important for reducing the noise growth during homomorphic multiplication. We will discuss this in more detail in \autoref{subsec:tfhe-mult-cipher}.

\subsection{Approximate Decomposition}
\label{subsec:approx-decomp}

\begin{figure}[h!]
    \centering
  \includegraphics[width=0.7\linewidth]{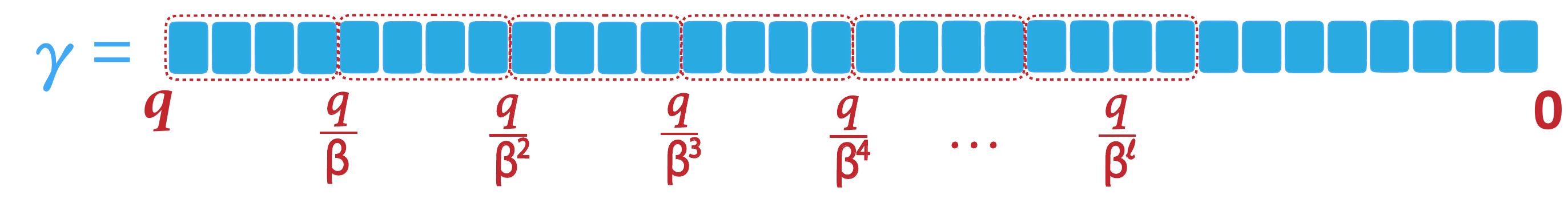}
  \caption{An illustration of approximate decomposition}
  \label{fig:decomp3}
\end{figure}

If no level $\ell$ satisfies $\beta^\ell \mid q$, then some lower-order digits of $q$ (in base $\beta$) must be discarded during decomposition, as shown in \autoref{fig:decomp3}. Such lower bits can be rounded to the nearest multiple of $\dfrac{q}{\beta^\ell}$ during decomposition. In such a case, the decomposition is an approximate decomposition. 
Formally, when $\beta^\ell\nmid q$ we can write
$$
\gamma=\sum_{i=1}^{\ell}\gamma_i\,\frac{q}{\beta^{i}}+\varepsilon,\qquad
\gamma_i\in\{0,\ldots,\beta-1\},\quad
|\varepsilon|\le \frac{q}{2\beta^{\ell}}
$$
(using nearest-integer rounding and identifying $\gamma$ with its integer representative)The polynomial case is analogous, coefficient-wise.

\subsection{Gadget Decomposition}
\label{subsec:gadget-decomposition}

Gadget decomposition is a generalized form of number decomposition (\autoref{subsec:number-decomp}). In number decomposition, a number $\gamma$ is decomposed as follows: 

$\gamma = \gamma_1 \dfrac{q}{\beta^1} + \gamma_2 \dfrac{q}{\beta^2} + \cdots + \gamma_\ell \dfrac{q}{\beta^\ell} $

$ $

In gadget decomposition, we decompose $\gamma$ as follows: 

$\gamma = \gamma_1 g_1 + \gamma_2 g_2 + \cdots + \gamma_\ell g_\ell $

$ $

We denote $\vec{g} = (g_1, g_2, \cdots, g_\ell)$ as a gadget vector, and $\textsf{Decomp}^{\vec{g}}(\gamma) = (\gamma_1, \gamma_2, \text{ } \cdots , \text{ } \gamma_\ell) $

$ $

Then, $\gamma = \langle \textsf{Decomp}^{\vec{g}}(\gamma), \vec{g} \rangle $

$ $

In the case of number decomposition (\autoref{subsec:number-decomp}), its gadget vector is $\vec{g} = \Bigg(\dfrac{q}{\beta}, \dfrac{q}{\beta^2}, \cdots, \dfrac{q}{\beta^\ell}\Bigg)$.

\clearpage

\section{Roots of Unity}
\label{sec:roots}
\textbf{- Reference:} 
\href{https://e.math.cornell.edu/people/belk/numbertheory/CyclotomicPolynomials.pdf}{Fields and Cyclotomic Polynomials}~\cite{cyclotomic-polynomial}

\subsection{Definitions}
\label{subsec:roots-def}
\begin{tcolorbox}[title={\textbf{\tboxdef{\ref*{subsec:roots-def}} Definitions for Roots of Unity}}]
\begin{itemize}
\item \textbf{$\bm{n}$-th root of Unity:} A complex number $\zeta$ that satisfies the equation $\zeta^n = 1$
\item \textbf{Primitive $\bm{n}$-th Root of Unity:} Any $n$-th root of unity $\zeta$ such that $\textsf{ord}_{\mathbb{C}}(\zeta) = n$. We denote $P(n)$ as a set of primitive $n$-th root of unity. 
\end{itemize}

$ $

A primitive $n$-th root of unity is considered a \textit{generator} of all $n$ $n$-th roots of unity.
\end{tcolorbox}

\subsection{Theorems}
\label{subsec:roots-theorem}

\begin{figure}[h!]
    \centering
  \includegraphics[width=0.4\linewidth]{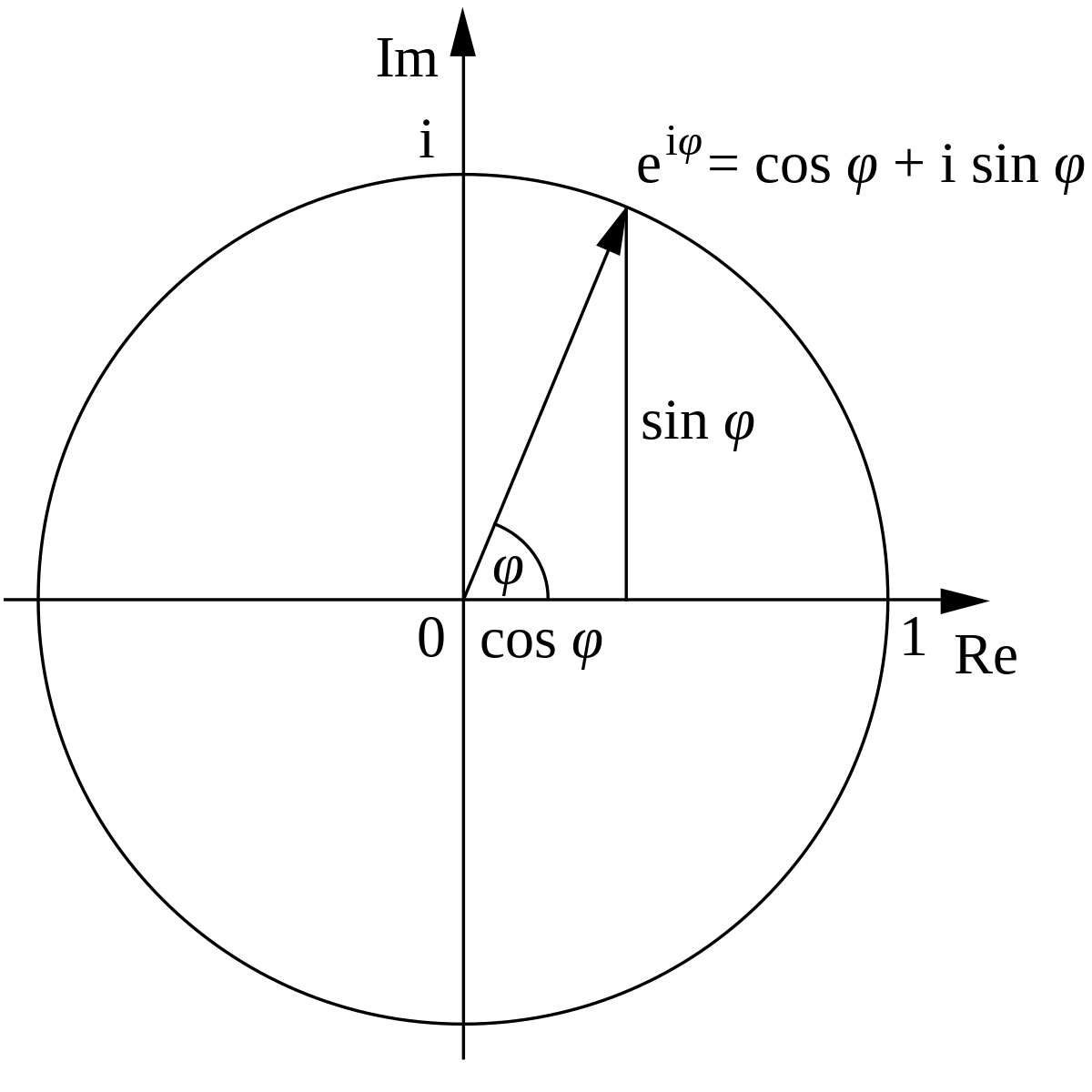}
  \caption{The figure illustrates a circle of Euler's formula in the complex plane \href{https://en.wikipedia.org/wiki/Euler's_formula}{(Source)}}
  \label{fig:complex-plane}
\end{figure}

\begin{tcolorbox}[title={\textbf{\tboxtheorem{\ref*{subsec:roots-theorem}.1} Formula for $\bm{n}$-th Root of Unity}}]
Given $\zeta^n = 1$, there exist exactly $n$ different $n$-th roots of unity: 

$ $

$\zeta = e^{2k\pi i/n} = \cos\left(\dfrac{2k\pi}{n}\right) + i\cdot\sin\left(\dfrac{2k\pi}{n}\right)$, 

$ $

\noindent for $n$ different $k$ values, where $k = \{0, 1, \cdots, n-1\}$.
\end{tcolorbox}
\begin{myproof} 
    \begin{enumerate}
    \item Suppose $\zeta = e^{2k\pi i/n}$. Then, $\zeta^n = (e^{2k\pi i/n})^n = e^{2k\pi i}$, and since $\zeta^n = 1$, we need to find the $k$ values such that $e^{2k\pi i} = 1$
    \item Euler's formula states that $e^{i\cdot x} = \text{cos}(x) + i \cdot \text{sin}(x)$. Therefore, if $x = 2k\pi$, then $e^{2k\pi i} = \text{cos}(2k\pi) + i \cdot \text{sin}(2k\pi)$. This formula becomes 1 if $k = 0, 1, 2, ...$. Thus, $e^{2k\pi i} = 1$ for any integer $k \geq 0$. 
    \item If $\zeta = e^{2k\pi i/n} = \text{cos}(\frac{2k\pi}{n}) + i\cdot\text{sin}(\frac{2k\pi}{n})$, then the first $n$ roots for $k = 0, 1, ... \text{ } n-1$ are all distinct values, because they lie on the circle in the complex plane (where $x$-axis is a real value and $y$-axis is a complex value coefficient) at each angle $2k\pi/n$ for $k = \{0, 1, \cdots, n-1\}$.
    \item Note $\zeta^n = 1$ is an $n$-th polynomial, so it can have at most $n$ roots. Thus, we can consider the first $n$ roots $e^{2k\pi i/n}$ for $k = \{0, 1, \cdots, n-1\}$ as the $n$ distinct roots and ignore the rest of roots (i.e., $k \geq n$), considering them to be repetitions of the first $n$ roots on a circle in the complex plane (see \autoref{fig:complex-plane}). 
    \end{enumerate}
\end{myproof}

\begin{tcolorbox}[title={\textbf{\tboxtheorem{\ref*{subsec:roots-theorem}.2} Order of the Root of Unity}}]
Given $\zeta \in \mathbb{C}$ (the complex number domain) and $\zeta^n = 1$ where $n \geq 1$, $\zeta$ is an $n$-th root of unity if and only if $\textsf{ord}_{\mathbb{C}}(\zeta) \text{ } | \text{ } n$.
\end{tcolorbox}
\begin{proof}
    We use 
    Theorem~\ref*{subsec:order-theorem}.1:
    \begin{enumerate}
    \item \textit{Forward Proof:} Since $\textsf{ord}_{\mathbb{C}}(\zeta)=k$ is the smallest integer such that $\zeta^k = 1$, for any $n$ that satisfies $\zeta^n = 1$, $n$ must be a multiple of $k$. This means that $k \mid n$.
    \item \textit{Backward Proof:} If $k \mid n$, then $n$ is a multiple of $k$, which means that $\zeta^n = 1$. 
    \end{enumerate}
    %$(Here, note that our \b\bmodulo computation wraps around a circle of Euler's formula on the complex plane instead of a finite integer field).
\end{proof}

\begin{tcolorbox}[title={\textbf{\tboxtheorem{\ref*{subsec:roots-theorem}.3} Set of All $\bm{n}$-th Roots of Unity}}]
The set of all $n$-th roots of unity is the union $\bigcup_{d|n} P(d)$ (i.e., the union of all primitive $d$-th roots of unity where $d \text{ } | \text{ } n$).
\end{tcolorbox}
\begin{myproof}
    \begin{enumerate}
    \item Let $\omega = e^{2\pi i/n}$.
    Given $\zeta^n = 1$, for each $n$-th root of unity $\zeta$ is, $\zeta = \omega^{k_i}$ for $k_i = \{0, 1, \cdots, n-1\}$. Note that according to Theorem~\ref*{subsec:order-theorem}.1, $(\omega^{k_i})^n = 1$ if and only if $\textsf{ord}_{\mathbb{C}}(\omega^{k_i}) \text{ } | \text{ } n$. 
    \item Let $\textsf{ord}_{\mathbb{C}}(\omega^{k_i}) = d_i$. Then, $(\omega^{k_i})^{d_i} = 1$. Combining these two facts, each $n$-th root of unity $\omega^{k_i}$ is also the primitive $d_i$-th root of unity (i.e., a solution for $\zeta^{d_i} = 1$), that is, $\omega^{k_i} \in P(d_i)$. 
    \item Remember that for each $\textsf{ord}_{\mathbb{C}}(\omega^{k_i}) = d_i$, $d_i \text{ } | \text{ } n$. For every $d_i$ that divides $n$, all the (primitive) $d_i$-th roots of unity are also the $n$-th root of unity. This is because the (primitive) $d_i$-th root of unity that satisfies $\zeta^{d_i} = 1$ also satisfies $\zeta^{n} = 1$ (as $n$ is a multiple of $d_i$).
    \item Step 2 concludes that each $n$-th root of unity is a primitive $d_i$-th root of unity for some $d_i$ that divides $n$. Step 3 concludes that each $d_i$-th root of unity, where $d_i$ divides $n$, is also the $n$-th root of unity. Combining these two conclusions, the set of all $n$-th root of unity is equivalent to the union of all primitive $d_i$-th roots of unity where $d_i$ divides $n$ (i.e., $\bigcup_{d|n} P(d)$). 
    \end{enumerate}
\end{myproof}

\begin{tcolorbox}[title={\textbf{\tboxtheorem{\ref*{subsec:roots-theorem}.4} Condition for Primitive $\bm{n}$-th Roots of Unity}}]
Given an $n$-th root of unity $\zeta = \omega^k$ for $k = \{0, 1, \cdots, n-1\}$ where $\omega = e^{2\pi i/n}$, $\zeta$ is a primitive $n$-th root of unity if and only if $\text{gcd}(n, k) = 1$ (i.e., $k$ is co-prime to $n$).
\end{tcolorbox}
\begin{myproof}
    \begin{enumerate}
    \item Note that $\zeta^n = 1$ and $\zeta = \omega$ for $k = 1$. Thus, $\textsf{ord}_{\mathbb{C}}(\omega) = n$.
    \item Theorem~\ref*{subsec:order-theorem}.2 states that if $ord_{\mathbb{F}}(a) = k$, then for any $n \geq 1$, $ord_{\mathbb{F}}(a^n) = \dfrac{k}{\text{gcd}(k,n)}$. Similarly, if $\textsf{ord}_{\mathbb{C}}(\omega) = n$, then for any $k \geq 1$, $\textsf{ord}_{\mathbb{C}}(\omega^k) = \dfrac{n}{\text{gcd}(k, n)}$.
    \item Step 2 implies that $\textsf{ord}_{\mathbb{C}}(\omega^k) = n$ (i.e., $\omega^k$ is a primitive $n$-th root of unity) if and only if $\text{gcd}(k, n) = 1$.
    \end{enumerate}
\end{myproof}

\begin{tcolorbox}[title={\textbf{\tboxtheorem{\ref*{subsec:roots-theorem}.5} The number of Primitive $\bm{n}$-th Roots of Unity}}]
The number of primitive $n$-th roots of unity is $\phi(n)$ (i.e., the number of elements in $\{1, \cdots, n-1 \}$ that are coprime to $n$).
\end{tcolorbox}

\begin{proof}
$ $
\begin{enumerate}
\item Given $\zeta^n = 1$, the roots of unity are $\zeta = \omega^k$ where $\omega = e^{2\pi i/n}$ and $k = \{0, 1, \cdots, n-1\}$ 
\item By definition, $\omega^k$ is a primitive $n$-th root of unity if and only if $\textsf{ord}_{\mathbb{C}}(\omega^k) = n$. 
\item $\omega$ is a primitive $n$-th root of unity because $\textsf{ord}_{\mathbb{C}}(\omega) = n$. 
\item According to Theorem~\ref*{subsec:order-theorem}.2, if $\textsf{ord}_{\mathbb{C}}(\omega) = n$, then $\textsf{ord}_{\mathbb{C}}(\omega^k) = \dfrac{n}{\text{gcd}(k,n)}$. Therefore, in order for $\textsf{ord}_{\mathbb{C}}(\omega^k) = n$, $\text{gcd}(k,n)$ has to be 1. In other words, $k$ and $n$ have to be co-prime.
The total number of such co-primes between $n$ and $k = \{1, 2, \cdots, n-1\}$ (excluding 0 because $\text{gcd}(0, n)= n$ and also $\textsf{ord}_{\mathbb{C}}(\omega^0) = \textsf{ord}_{\mathbb{C}}(1) = 1 \neq n$) is $\phi(n)$, which corresponds to the total number of the primitive $n$-th root of unity.
\end{enumerate}
\end{proof}

\clearpage

\section{Cyclotomic Polynomial}
\label{sec:cyclotomic}
\textbf{- Reference:} 
\href{https://e.math.cornell.edu/people/belk/numbertheory/CyclotomicPolynomials.pdf}{Fields and Cyclotomic Polynomials}~\cite{cyclotomic-polynomial}

\subsection{Definitions}
\label{subsec:cyclotomic-def}
\begin{tcolorbox}[title={\textbf{\tboxdef{\ref*{subsec:cyclotomic-def}} Cyclotomic Polynomial}}]
 \textbf{The $\bm{n}$-th Cyclotomic Polynomial:} is a polynomial whose roots are the primitive $n$-th roots of unity, that is: 
 
 \[ \Phi_n(x) = \prod_{\zeta \in P(n)} (x - \zeta)  = \prod_{\substack{0 \leq k \leq n-1,\\ 
 \text{gcd}(k, n) = 1}} (x - \omega^k) \textcolor{white}{......} \text{, where } \omega = e^{2\pi i/n} \]

 Remember the Euler's formula: $e^{2k\pi i/n} = \cos\left(\dfrac{2k\pi}{n}\right) + i\cdot\sin\left(\dfrac{2k\pi}{n}\right)$

$ $

A few pre-computed cyclotomic polynomials are as follows:
 
\begin{multicols}{2}
$\Phi_1(x) = x - 1$
\newline $\Phi_2(x) = x + 1$
\newline $\Phi_3(x) = x^2 + x + 1$
\newline $\Phi_4(x) = x^2 + 1$
\newline $\Phi_5(x) = x^4 + x^3 + x^2 + x + 1$
\newline $\Phi_6(x) = x^2 - x + 1$
\newline $\Phi_7(x) = x^6 + x^5 + x^4 + x^3 + x^2 + x + 1$
\newline $\Phi_8(x) = x^4 + 1$
\newline $\Phi_9(x) = x^6 + x^3 + 1$
\newline $\Phi_{10}(x) = x^4 - x^3 + x^2 - x + 1$
\end{multicols}
\end{tcolorbox}

As one example, 

\[\Phi_4(x) = \prod_{\substack{0 \leq k \leq 3,\\ 
 \text{gcd}(k, 4) = 1}} (x - \omega^k) = (x - \omega^1)(x - \omega^3) = (x - e^{2\pi i / 4})(x - e^{2\cdot3\pi i / 4}) = (x - e^{\pi i / 2})(x - e^{3\pi i / 2})\]

$= \left(x - \left(\cos\left(\dfrac{\pi}{2}\right) + i \cdot \sin\left(\dfrac{\pi}{2}\right) \right) \right) \cdot \left(x - \left(\cos\left(\dfrac{3\pi}{2}\right) + i \cdot \sin\left(\dfrac{3\pi}{2}\right) \right) \right)$

$= (x - i)(x + i) = x^2 + 1$

\subsection{Theorems}
\label{subsec:cyclotomic-theorem}

\begin{tcolorbox}[title={\textbf{\tboxtheorem{\ref*{subsec:cyclotomic-theorem}.1} Roots of the $M$-th Cyclotomic Polynomial}}]

Suppose that $M$ is a power of 2 and the $M$-th cyclotomic polynomial $\Phi_M(x) = x^n + 1$ (where $M = 2n$). Then, the roots of the $M$-th cyclotomic polynomial are $\omega, \omega^3, \omega^5, \cdots, \omega^{2n-1}$, where $\omega = e^{i\pi/n}$

\end{tcolorbox}

\begin{proof}

According to Definition~\ref*{subsec:cyclotomic-def} in \autoref{subsec:cyclotomic-def}, the roots of $\Phi_M(x)$ are $e^{2k\pi i/M} = e^{2k\pi i/(2n)} = e^{k\pi i/n}$ where $0 \leq k < M = 2n$ and $\textsf{gcd}(k, M = 2n) = 1$, thus $k = \{1, 3, 5, \cdots, 2n-1\}$. If we let $\omega = e^{i\pi/n}$, then the roots of $\Phi_M(x)$ are $\omega, \omega^3, \omega^5, \cdots, \omega^{2n-1}$.

\end{proof}

\begin{tcolorbox}[title={\textbf{\tboxtheorem{\ref*{subsec:cyclotomic-theorem}.2} Polynomial Decomposition into Cyclotomic Polynomials}}]
For any positive integer $n$, \[ x^n - 1 = \prod_{d \mid n} \Phi_d(x) \]
\end{tcolorbox}
\begin{myproof}
\begin{enumerate}
    \item The roots of $x^n - 1$ are all the $n$-th roots of unity. Thus, $x^n - 1 = (x - \omega^0)(x - \omega^1)...(x - \omega^{n-1})$, where $\zeta = \omega^k$.
    \item Theorem~\ref*{subsec:roots-theorem}.3 states that each $n$-th root of unity ($\omega^k$) is a primitive $d$-th root of unity for some $d$ that divides $n$. In other words, each $n$-th root of unity belongs to some $P(d)$ where $d \mid n$. Meanwhile, by definition, $\Phi_d(x) = \Pi_{\zeta \in P(d)} (x - \zeta)$. Therefore, $x^n - 1$ is the product of all $\Phi_d(x)$ such that $d \mid n$. 
\end{enumerate}
\end{myproof}
\begin{tcolorbox}[title={\textbf{\tboxtheorem{\ref*{subsec:cyclotomic-theorem}.3} Integer Coefficients of Cyclotomic Polynomials}}]
A cyclotomic polynomial has only integer coefficients.
\end{tcolorbox}
\begin{myproof}
\begin{enumerate}
    \item We prove by induction. When $n=1$, $\Phi_1(x) = x - 1$, where each coefficient is an integer.
    \item Let $x^n - 1 = f(x) \cdot g(x) = (\Sigma_{i=0}^{p}a_ix^i)(\Sigma_{j=0}^{q}b_jx^j)$. As an induction hypothesis 1, we will prove that if $f(x)$ has only integer coefficients, then $g(x)$ will also have only integer coefficients. Given our target equation is $x^n - 1$, we know that $a_px^p \cdot b_qx^q = x^n$, and thus $a_pb_q = 1$, which means $a_p = \pm 1$ (as we hypothesized that $f(x)$ has only integer coefficients). We also know that $a_0b_0 = -1$. All the other coefficients should be 0. Thus, for any $r < q$, the coefficients are either: (i) $a_pb_{r} + a_{p-1}b_{r + 1} + ... + a_{p-q+r}b_{q} = 0$; or (ii) $a_pb_{r} + a_{p-1}b_{r + 1} + ... + a_{0}b_{r+p} = 0$. Both case (i) and (ii) represent $f(x)\cdot g(x)$'s computed coefficient of some $x^i$ where $0 < i < n$. Now, we propose another induction hypothesis 2, which is that $b_{q}, ... \text{ } b_{r+1}$ are all integers.
    \item In the case of (i), $a_pb_{r} = -(a_{p-1}b_{r + 1} + ... + a_{p-q+r}b_{q})$, and dividing both sides by $a_p$ (which is either $1$ or $-1$), $b_{r} = \pm(a_{p-1}b_{r + 1} + ... + a_{p-q+r}b_{q})$, as every $a_i$ is an integer based on our hypothesis. By induction hypothesis 1 and 2, $b_r$ is an integer. The same is true in the case of (ii). 
    \item We set $b_q$ (an integer coefficient) as the starting point for induction hypothesis 2. Then, according to induction proof 2, all of $b_j$ for $0 \leq j \leq q$ are integers.
    \item Now, we set $\Phi_1(x)$ (an integer coefficient polynomial) as the starting point for induction hypothesis 1. Let $x^n - 1 = \Phi_{d_1}(x)\Phi_{d_2}(x)...\Phi_{d_k}(x)\Phi_{n}(x)$, where each $d_i \mid n$ (Theorem~\ref*{subsec:cyclotomic-theorem}.2). We know that $\Phi_{d_1}(x)\Phi_{d_2}(x)\cdots\Phi_{d_k}(x)$ forms an integer coefficient polynomial. We treat $\Phi_{d_1}(x)\Phi_{d_2}(x)\cdots\Phi_{d_k}(x)$ as $f(x)$, and $\Phi_n(x)$ as $g(x)$.
    Then, according to step 4's induction proof, $\Phi_{n}(x)$ is an integer coefficient polynomial (also note that $\Phi_n(x)$ is monic, whose the highest degree's coefficient is 1). 
    \item As we marginally increase $n$ to $n+1$ to compute $x^{n+1} - 1 = \Phi_{d'_1}(x)\Phi_{d'_2}(x)...\Phi_{d'_k}(x)\Phi_{n+1}(x)$ (where each $d'_i \mid (n+1)$), we know that $\Phi_{d'_1}(x)\Phi_{d'_2}(x)\cdots\Phi_{d'_k}(x)$ is a monic polynomial, as proved by the previous induction step. Thus, $\Phi_{n+1}(x)$ is also monic. 
\end{enumerate}
\end{myproof}
\begin{tcolorbox}[title={\textbf{\tboxtheorem{\ref*{subsec:cyclotomic-theorem}.4} Formula for $\bm{\Phi_{nk}(x)}$}}]
If $k \mid n$, then $\Phi_{nk}(x) = \Phi_n(x^k)$.
\end{tcolorbox}
\begin{myproof}
\begin{enumerate}
    \item Theorem~\ref*{subsec:order-theorem}.3 states that given $k \mid n$, $\text{ord}_{\mathbb{F}}(a) = kn$ if and only if $\text{ord}_{\mathbb{F}}(a^k) = n$. This means that for $\zeta \in \mathbb{C}$, $\text{ord}_{\mathbb{C}}(\zeta) = nk$ if and only if $\text{ord}_{\mathbb{C}}(\zeta^k) = n$. In other words, $\zeta$ is a primitive $nk$-th root of unity if and only if $\zeta^k$ is the primitive $n$-th root of unity. This implies that $\zeta$ is a root of $\Phi_{nk}(x)$ if and only if $\zeta^k$ is a root of $\Phi_{n}(x)$. 
    \item Let $\Phi_{nk}(x) = (x - \zeta_1)(x - \zeta_2)...(x - \zeta_p)$, where $P(nk)$ has $p$ primitive $nk$-th roots of unity.
    \item $\Phi_{n}(x) = (x - \zeta_1^k)(x - \zeta_2^k)...(x - \zeta_p^k)$. 
    %Note that $P(n)$ should also have $p$ primitive $n$-th roots of unity, because elements of $P(nk)$ are isomorphic to the elements of $P(n)$ (as they preserved if and only if relationships in step 2). 
    Note that raising each element of $P(nk)$ to the $k$-th power yields an element of $P(n)$, thus mapping $P(nk)$ onto $P(n)$ (by the result of step 2).
    Now, it's also true that $\Phi_{n}(y) = (y - \zeta_1^k)(y - \zeta_2^k)...(y - \zeta_p^k)$, where $y = x^k$. In this case, $x = \{\zeta_1, \zeta_2, ... \zeta_p\}$.
    \item $\Phi_{nk}(x)$ and $\Phi_{n}(y) = \Phi_{n}(x^k)$ have the same roots with the same coefficients. Therefore, $\Phi_{nk}(x) = \Phi_{n}(y) = \Phi_{n}(x^k)$. 
\end{enumerate}
\end{myproof}

\clearpage

\section{Roots of Unity and Cyclotomic Polynomial over Ring}
\label{sec:cyclotomic-polynomial-integer-ring}
In \autoref{sec:roots} and \autoref{sec:cyclotomic}, we learned about the definition and properties of the $\mu$-th roots of unity and the $\mu$-th cyclotomic polynomial over complex numbers (i.e., $X \in \mathbb{C}$) as follows: 

\begin{itemize}
\item \textbf{The $\bm \mu$-th roots of unity} are the solutions in $\mathbb{C}$ to the equation $X^\mu = 1$. In other words, all $\mu$-th roots of unity can be written as $X = e^{2 \pi i k / \mu}$ for integers $k$ with $0 \le k < \mu$.
\item \textbf{The primitive $\bm \mu$-th roots of unity (denoted as $\bm \omega$)} are those $\mu$-th roots of unity whose order (\autoref{subsec:order-def}) is $\mu$ (i.e., $\omega^{\mu} = 1$ and $\omega^d \neq 1$ for any $1 \le d < \mu$ with $d \mid \mu$).
\item Given any primitive $\mu$-th roots of unity $\omega$, one can generate all primitive $\mu$-th roots of unity by computing $\omega^{k'}$ such that $k'$ is an integer $0 < k' < \mu$ and $\textsf{gcd}(k', \mu) = 1$ (Theorem~\ref*{subsec:roots-theorem}.4 in \autoref{subsec:order-theorem}). 
\item \textbf{The $\bm \mu$-th cyclotomic polynomial} is defined as a polynomial whose roots are the primitive $\mu$-th roots of unity. That is, \[ \Phi_{\mu}(x) = \prod_{\omega \in P({\mu})} (x - \omega) = \prod_{\substack{0 \leq k \leq {\mu}-1,\\ \textsf{gcd}(k, {\mu}) = 1}} (x - \omega^k) \]
\end{itemize}

In this section, we will explain the $\mu$-th cyclotomic polynomial over $\mathbb{Z}_p$ (with $p$ prime), which is structured as follows:

\begin{tcolorbox}[title={\textbf{\tboxdef{\ref*{sec:cyclotomic-polynomial-integer-ring}} Roots of Unity and Cyclotomic Polynomial over Ring $\mathbb{Z}_p$}}]

\begin{itemize}
\item \textbf{The $\bm \mu$-th roots of unity (denoted as $\bm \omega$)} are the solutions for $X^\mu \equiv 1 \bmod p$. Note that in contrast to the complex case, these solutions cannot be expressed as $X = e^{2 \pi i k / \mu}$. 
\item \textbf{The primitive $\bm \mu$-th roots of unity} are defined as those $\mu$-th roots of unity whose order is $\mu$ (i.e., $\omega^{\mu} \equiv 1 \bmod p$, and $\omega^d \not\equiv 1 \bmod p$ for any $1 \le d < \mu$ with $d \mid \mu$). 
\item Given any primitive $\mu$-th roots of unity $\omega$, it can generate all primitive $\mu$-th roots of unity by computing $\omega^{k'}$ such that $k'$ is an integer $0 < k' < \mu$ and $\textsf{gcd}(k', \mu) = 1$.
\item \textbf{The $\bm \mu$-th cyclotomic polynomial} is defined as a polynomial whose roots are the primitive $\mu$-th roots of unity. That is, \[ \Phi_{\mu}(x) = \prod_{\omega \in P({\mu})} (x - \omega) = \prod_{\substack{0 \leq k \leq {\mu}-1,\\ \textsf{gcd}(k, {\mu}) = 1}} (x - \omega^k) \]
\end{itemize}

\end{tcolorbox}

\begin{table}[h] %usepackage{array} 
\begin{tabular}{|c||c||c|}
\hline \hline
& \textbf{Polynomial over $\bm{\mathbb{C}}$} & \textbf{Polynomial over $\bm{\mathbb{Z}}_{\bm{p}}$} \\ 
& \textbf{(Complex Number)} & \textbf{(Ring)} \\ \hline \hline
\textbf{Definition}&All $X \in \mathbb{C}$ such that $X^\mu = 1$, (which are&All $X \in \mathbb{Z}_p$ such that $X^\mu \equiv 1 \bmod p$\\
\textbf{of the}&computed as $X = e^{2 \pi i k / \mu}$ for integer $k$&\\
\textbf{$\bm \mu$-th}&where $0 \leq k \leq \mu - 1$)&\\
\textbf{Root of Unity}&&\\\hline
\textbf{Definition}&Those $\mu$-th roots of unity $\omega$ such that&Those $\mu$-th roots of unity $\omega$ such that\\
\textbf{of the}&$\omega^{\mu} = 1$, and $\omega^{d} \neq 1$ &$\omega^{\mu} \equiv 1 \bmod p$, and $\omega^{d} \not\equiv 1 \bmod p$\\
\textbf{Primitive}&for any $1 \le d < \mu$ with $d \mid \mu$&for any $1 \le d < \mu$ with $d \mid \mu$\\
\textbf{$\bm \mu$-th}&&\\
\textbf{Root of}&&\\
\textbf{Unity}&&\\\hline
\textbf{Definition}&\multicolumn{2}{|c|}{The polynomial whose roots are the $\mu$-th primitive roots of unity as follows:}\\
\textbf{of the}&\multicolumn{2}{|c|}{$ \Phi_{\mu}(x) = \prod_{\omega \in P(\mu)} (x - \omega) $  \text{ } (see Definition~\ref*{subsec:cyclotomic-def} in \autoref{subsec:cyclotomic-def})}\\
\textbf{$\bm \mu$-th}&\multicolumn{2}{|c|}{}\\
\textbf{Cyclotomic}&\multicolumn{2}{|c|}{}\\
\textbf{Polynomial}&\multicolumn{2}{|c|}{}\\\hline
\textbf{Finding}&For $\omega = e^{2 \pi i/ \mu}$, compute all ${\omega}^k$ such that&Find one primitive $\omega$ that is a root of\\
\textbf{Primitive}&$0 < k < \mu $ and $\textsf{gcd}(k, \mu) = 1$&the $\mu$-th cyclotomic polynomial, and\\
\textbf{$\bm \mu$-th}&(Theorem~\ref*{subsec:roots-theorem}.4 in \autoref{subsec:roots-theorem})&compute all $\omega^k \bmod p$ such that\\
\textbf{Roots of}&&$0 < k < \mu $ and $\textsf{gcd}(k, \mu) = 1$\\
\textbf{Unity}&&\\\hline\hline
\end{tabular}
\caption{The roots of unity and cyclotomic polynomials over $X \in \mathbb{C}$ vs. over $X \in \mathbb{Z}_p$}
\label{tab:cyclotomic-polynomial-comparison}
\end{table}

Note that in the $\mu$-th cyclotomic polynomial, in both cases of over $X \in \mathbb{C}$ and over $X \in \mathbb{Z}_p$, each of their roots $\omega$ (i.e., the primitive $\mu$-th root of unity) has the order $\mu$ (i.e., $\omega^{\mu} = 1$ over $X \in \mathbb{C}$, and $\omega^{\mu} \equiv 1 \bmod p$ over $X \in \mathbb{Z}_p$). Also note that each root $\omega$ can generate all roots of the $\mu$-th cyclotomic polynomial by computing $\omega^{k'}$ such that $\textsf{gcd}(k', \mu) = 1$.

\autoref{tab:cyclotomic-polynomial-comparison} compares the properties of the roots of unity and the $\mu$-th cyclotomic polynomial over $\mathbb{C}$ (the complex numbers) and over $\mathbb{Z}_p$ (the ring).

\clearpage

\section{Lagrange's Polynomial Interpolation}
\label{sec:polynomial-interpolation}
Suppose we are given $n+1$ two-dimensional coordinates $(x_0, y_0), (x_1, y_1), \cdots, (x_n, y_n)$, where all $x$ values are distinct, but $y$ values are not necessarily distinct. Lagrange's polynomial interpolation is a technique to find a unique polynomial of degree at most $n$ that passes through such $n+1$ coordinates. The given points $(x_i,y_i)$ may lie either in $\mathbb{C}^2$ (the complex plane, which includes the real numbers) or in $\mathbb{Z}_p^2$ for some prime $p$.

\begin{tcolorbox}[title={\textbf{\tboxtheorem{\ref*{sec:polynomial-interpolation}} Lagrange's Polynomial Interpolation}}]

Suppose we are given $n+1$ two-dimensional coordinates $(x_0, y_0), (x_1, y_1), \cdots, (x_n, y_n)$, whereas all $X$ values are distinct but the $Y$ values don't need to be distinct. The domain of $(X, Y)$ can be either: $(x_i, y_i) \in \mathbb{C}^2$ (which includes the real domain) or $(x_i, y_i) \in \mathbb{Z}_p^2$ (where $p$ is a prime). Then, there exists a unique polynomial $f(X)$ of degree at most $n$ that passes through these $n+1$ coordinates. Such a polynomial $f(X)$ is computed as follows:

$f(X) = \sum\limits_{j=0}^{n}\dfrac{(X-x_0)\cdot(X-x_1)\cdots(X-x_{j-1})\cdot(X-x_{j+1})\cdots(X-x_{n})}{(x_j-x_0)\cdot(x_j-x_1)\cdots(x_j-x_{j-1})\cdot(x_j-x_{j+1})\cdots(x_j-x_{n})}\cdot y_j$

$\textcolor{white}{f(X) }= \sum\limits_{j=0}^{n} \left( \prod\limits_{\substack{0 \le k \le n\\ k \ne j}} \dfrac{X - x_k}{x_j - x_k} \cdot y_j\right)$

\end{tcolorbox}

\begin{myproof}
\begin{enumerate}

\item First, we will show that there exists an $n$-degree (or lesser degree) polynomial $f(X)$ that passes through the $n+1$ distinct coordinates: $(x_0, y_0), (x_1, y_1), \ldots, (x_n, y_n)$. Such a polynomial $f(X)$ is designed as follows:

$f(X) = \sum\limits_{j=0}^{n}\dfrac{(X-x_0)\cdot(X-x_1)\cdots(X-x_{j-1})\cdot(X-x_{j+1})\cdots(X-x_{n})}{(x_j-x_0)\cdot(x_j-x_1)\cdots(x_j-x_{j-1})\cdot(x_j-x_{j+1})\cdots(x_j-x_{n})}\cdot y_j$

$ \textcolor{white}{f(X) }= \sum\limits_{j=0}^{n} \left( \prod\limits_{\substack{0 \le k \le n\\ k \ne j}} \dfrac{X - x_k}{x_j - x_k} \cdot y_j\right)$

$ \textcolor{white}{f(X) }= \sum\limits_{j=0}^{n} \ell_j(X) \cdot y_j$ \textcolor{red}{ $\rhd$ where $\ell_j(X) =  \prod\limits_{\substack{0 \le k \le n\\ k \ne j}} \dfrac{X - x_k}{x_j - x_k}$}

$ $

We call $\{\ell_0(X), \ell_1(X), \ldots, \ell_{n}(X) \}$ the Lagrange basis for polynomials of degree $\leq n$. Given this design of $f(X)$, notice that for each of $(x_i, y_i) \in \{(x_0, y_0), (x_1, y_1), \ldots, (x_n, y_n)\}$, \text{ } $\ell_i(x_{i'}) = 1$ for $i' = i$, and $\ell_i(x_{i'}) = 0$ for $i' \neq i$. Therefore, $f(x_i) = \sum\limits_{j=0}^{n} \ell_j(x_i) \cdot y_j = 1 \cdot y_i = y_i$ for $0 \leq i \leq n$. In other words, $f(X)$ passes through the $n+1$ distinct coordinates: $\{(x_0, y_0), (x_1, y_1), \ldots, (x_n, y_n)\}$. Such a satisfactory $f(X)$ can be computed in the case where the domain of $(X, Y$) is either: $(x_i, y_i) \in \mathbb{C}^2$ (i.e., real and complex numbers), or $(x_i, y_i) \in \mathbb{Z}_p^2$ (where $p$ is a prime). Especially, a valid $f(X)$ can be computed also in the $\bmod \, p$ domain, because as we learned from Fermat's Little Theorem in Theorem~\ref*{subsec:order-theorem}.4 (\autoref{subsec:order-theorem}), $a^{p - 1} \equiv 1 \bmod p$ if and only if $a$ and $p$ are co-prime, and this means that if $p$ is a prime, then $a^{p - 1} \equiv 1 \bmod p$ for all $a \in \mathbb{Z}_p^{\times}$ (i.e., $\mathbb{Z}_p$ without $\{0\}$). Since every value in $\mathbb{Z}_p^{\times}$ has an inverse, 
we can perform each division in the formula for $f(X) = \sum\limits_{j=0}^{n}\dfrac{(X-x_0)\cdot(X-x_1)\cdots(X-x_{j-1})\cdot(X-x_{j+1})\cdots(X-x_{n})}{(x_j-x_0)\cdot(x_j-x_1)\cdots(x_j-x_{j-1})\cdot(x_j-x_{j+1})\cdots(x_j-x_{n})}\cdot y_j$ by multiplying with the corresponding inverses of those denominators.
 
\item Next, we will prove that no two distinct $n$-degree (or lesser degree) polynomials $f_1(X)$ and $f_2(X)$ can pass through the same $n+1$ distinct $(X, Y)$ coordinates. Suppose there exist such two polynomials $f_1(X)$ and $f_2(X)$. Then $f_{1-2}(X) = f_1(X) - f_2(X)$ will be a new $n$-degree (or lesser degree) polynomial that passes through $(x_0, 0), (x_1, 0), \cdots, (x_n, 0)$. This means that $f_{1-2}(X)$ has $n+1$ distinct roots. In other words, $f_{1-2}(X)$ is an $n+1$-degree (or higher-degree) polynomial. However, this contradicts the assumption that $f_{1-2}(X)$ is an $n$-degree (or lesser degree) polynomial. Therefore, there exist no two polynomials $f_1(X)$ and $f_2(X)$ that pass through the same $n+1$ distinct $(X, Y)$ coordinates. 

\item We have shown that there exists some $n$-degree (or lesser degree) polynomial $f(X)$ that passes through $n+1$ distinct $(X, Y)$ coordinates, and no such two or more distinct polynomials exists. Therefore, there exists only a unique polynomial that satisfies this requirement. 

\end{enumerate}
\end{myproof}

\subsection{The Field Condition for Unique Interpolation}
\label{subsec:polynomial-interpolation-field-condition}

\begin{tcolorbox}[title={\textbf{\tboxtheorem{\ref*{subsec:polynomial-interpolation-field-condition}} The Field Condition for Unique Interpolation}}]

If a polynomial is defined over a field $\mathbb{F}$
 (i.e., its $x$ and $y$
coordinates and its coefficients lie in $\mathbb{F}$), then for any $n+1$ distinct $x$ values and any $n+1$ arbitrary (i.e., can be overlapped) $y$ values in $\mathbb{F}$, there exists a unique polynomial of degree at most $n$ that interpolates them.

$ $

Examples of such fields are $\mathbb{R}$ (real number domain), $\mathbb{C}$ (complex number domain), and any finite field $\mathbb{Z}_t$ (where $t$ is a prime). 

\end{tcolorbox}

\begin{myproof}
\begin{enumerate}

\item Remember from Theorem~\ref*{sec:polynomial-interpolation} that Lagrange's polynomial interpolation is defined as: 

$f(X) = \sum\limits_{j=0}^{n} \left( \prod\limits_{\substack{0 \le k \le n\\ k \ne j}} \dfrac{X - x_k}{x_j - x_k} \cdot y_j\right)$

$ $

If every $x$, $y$ values and coefficients of the polynomial $f(X)$ lie in a field, then $(x_j - x_k)^{-1}$ is guaranteed to exist within the same field, because every non-zero element of a field has an inverse, and $(x_j - x_k)$ is not zero (since $x_j \neq x_k$). This implies that $\prod\limits_{\substack{0 \le k \le n\\ k \ne j}} \dfrac{X - x_k}{x_j - x_k} \cdot y_j$ also lies in the same field. Hence, a well-defined $f(X)$ is guaranteed to exist. And the degree of $f(X)$ is at most $n$, because its highest possible degree term is $X^{n}$.

\item For uniqueness of $f(X)$, suppose another polynomial $g(X)$ of degree at most $n$ also passes through the same $n+1$ points. Let $h(X) = f(X) - g(X)$. Then the degree of $h(X)$ is at most $n$, and $h(x_i) = 0$ for all $(n+1)$ distinct nodes $x_0,\dots,x_{n}$, so $h$ has $(n+1)$ distinct roots. Over a field, a non-zero polynomial of degree $d$ has at most $d$ roots; a nonzero $h(X)$ of degree at most $n$ could have at most $n$ roots. Since $h(X)$ has $(n+1)$ roots, $h(X)$ must be the zero polynomial, i.e. $f(X) = g(X)$. Therefore $f(X)$ is unique.

\end{enumerate}
\end{myproof}

\clearpage

\section{Vector and Matrix}
\label{sec:matrix}

\subsection{Vector Arithmetic}
\label{subsec:vector-arithmetic}

This section explains the basic arithmetic of vector and matrix computations, as well as advanced concepts such as vector/plane projection and the basis of planes (or spaces).

\begin{tcolorbox}[title={\textbf{\tboxdef{\ref*{subsec:vector-arithmetic}} Vector Arithmetic}}]

\begin{itemize}
\item \textbf{Addition:} Given two $n \times 1$ vectors (i.e., $n$-dimensional vectors) composed of $n$ numbers each:

$\vec{a} = (a_0, a_1,\cdots, a_{n-1})$, $\vec{b} = (b_0, b_1,\cdots, b_{n-1})$

$ $

Vector addition is defined as: 

$ \vec{a} + \vec{b} = (a_0 + b_0, \text{ } a_1  +  b_1, \text{ }\cdots, \text{ } a_{n-1}  +  b_{n-1})$

$ $

\item \textbf{Dot Product:} Given two $n$-dimensional vectors:

$\vec{a} = (a_0, a_1,\ldots, a_{n-1})$, $\vec{b} = (b_0, b_1,\ldots, b_{n-1})$

$ $

Dot product is defined as: 

$\langle \vec{a}, \vec{b} \rangle =  \vec{a} \cdot \vec{b} = a_0 b_0 + a_1 b_1 +\cdots + a_{n-1} b_{n-1}$.

$\vec{a} \cdot \vec{b}$ can also be expressed as $\|\vec{a}\|\,\|\vec{b}\|\cos\theta$, where $\theta$ is the angle between $\vec{a}$ and $\vec{b}$. If $\vec{a}$ and $\vec{b}$ point in the same direction, $\vec{a}\cdot\vec{b}$ attains its maximum value ($\|\vec{a}\|\,\|\vec{b}\|$). If $\vec{a}$ and $\vec{b}$ are orthogonal, then $\vec{a}\cdot\vec{b} = 0$.

$ $

\item \textbf{Hadamard Product:} Given two $n$-dimensional vectors:

$\vec{a} = (a_0, a_1,\ldots, a_{n-1})$, $\vec{b} = (b_0, b_1,\ldots, b_{n-1})$

$ $

Hadamard product is defined as: 

$\vec{a} \odot \vec{b} = (a_0 b_0, \text{ } a_1 b_1, \text{ }\cdots, \text{ } a_{n-1} b_{n-1})$

$ $

\item \textbf{Hermitian Product:} Given two $n$-dimensional complex vectors:

$\vec{a} = (a_0 + i \cdot a'_0, \text{ } a_1+ i\cdot a'_1, \text{ }\ldots,\text{ }  a_{n-1}+ i\cdot a'_{n-1})$ 

$\vec{b} = (b_0 + i\cdot b'_0, \text{ } b_1 + i\cdot b'_1, \text{ }\ldots, \text{ } b_{n-1} + i\cdot b'_{n-1})$

$ $

Hermitian product is a dot product with the 2nd operand as a conjugate (\autoref{subsec:imaginary}): 

$\langle\langle \vec{a}, \vec{b} \rangle\rangle =  \vec{a} \cdot \overline{\vec{b}} $

$ = (a_0 + i \cdot a'_0, \text{ } a_1+ i\cdot a'_1, \text{ }\cdots,\text{ } a_{n-1}+ i\cdot a'_{n-1})  \cdot (b_0 - i\cdot b'_0, \text{ } b_1 - i\cdot b'_1, \text{ }\cdots, \text{ } b_{n-1} - i\cdot b'_{n-1})$

\end{itemize}
\end{tcolorbox}

\subsection{Various Types of Matrix}
\label{subsec:vandermonde}

\begin{tcolorbox}[title={\textbf{\tboxdef{\ref*{subsec:vandermonde}} Matrices}}]

\begin{itemize}

\item An $n \times n$ \textbf{identity matrix} and a \textbf{reverse identity matrix} are defined as:

$I_n = \begin{bmatrix}
1 & 0 & 0 & \cdots & 0\\
0 & 1 & 0 & \cdots & 0 \\
0 & 0 & 1 & \cdots & 0 \\
\vdots & \vdots & \vdots & \ddots & \vdots \\
0 & 0 & 0 & \cdots & 1 \\
\end{bmatrix}$, $I^R_n = \begin{bmatrix}
0 & \cdots & 0 & 0 & 1\\
0 & \cdots & 0 & 1 & 0 \\
0 & \cdots & 1 & 0 & 0 \\
\vdots & \iddots & \vdots & \vdots & \vdots \\
1 & 0 & 0 & \cdots & 0 \\
\end{bmatrix}$

$ $

\item The \textbf{transpose of a matrix} $X$ is defined as element-wise swapping along the diagonal line, denoted as $X^T$, which is:

$X = \begin{bmatrix}
a_1 & a_2 & a_3 & \cdots & a_n\\
b_1 & b_2 & b_3 & \cdots & b_n \\
c_1 & c_2 & c_3 & \cdots & c_n \\
\vdots & \vdots & \vdots & \ddots & \vdots \\
m_1 & m_2 & m_3 & \cdots & m_n \\
\end{bmatrix}$, $X^T = \begin{bmatrix}
a_1 & b_1 & c_1 & \cdots & m_1\\
a_2 & b_2 & c_2 & \cdots & m_2\\
a_3 & b_3 & c_3 & \cdots & m_3\\
\vdots & \vdots & \vdots & \ddots & \vdots \\
a_n & b_n & c_n & \cdots & m_n\\
\end{bmatrix}$

$ $

\item A \textbf{Vandermonde matrix} is an $(m + 1) \times (n + 1)$ matrix defined as:

$\mathit{Vander}(x_0, x_1, \cdots, x_{m}) = \begin{bmatrix}
1 & x_0 & x_0^2 & \cdots & x_0^n\\
1 & x_1 & x_1^2 & \cdots & x_1^n \\
1 & x_2 & x_2^2 & \cdots & x_2^n \\
\vdots & \vdots & \vdots & \ddots & \vdots \\
1 & x_m & x_m^2 & \cdots & x_m^{n} \\
\end{bmatrix}$

%A Vandermonde matrix is often used for representing $m$ different $n$-degree polynomials.

\end{itemize}

\end{tcolorbox}

\subsection{Matrix Arithmetic}
\label{subsec:matrix-arithmetic}

Matrix-to-vector multiplication and matrix-to-matrix multiplication are defined as follows:

\begin{tcolorbox}[title={\textbf{\tboxdef{\ref*{subsec:matrix-arithmetic}} Matrix Arithmetic}}]

\begin{itemize}
\item \textbf{Matrix-to-Vector Multiplication:} Given a $m \times n$ matrix $A$ and a $n$-dimensional vector $x$:

$A = \begin{bmatrix}
 a_{1, 1} & a_{1, 2} & a_{1, 3} & \cdots & a_{1, n}\\
 a_{2, 1} & a_{2, 2} & a_{2, 3} & \cdots & a_{2, n} \\
 a_{3, 1} & a_{3, 2} & a_{3, 3} & \cdots & a_{3, n} \\
\vdots & \vdots & \vdots & \ddots & \vdots \\
 a_{m, 1} & a_{m, 2} & a_{m, 3} & \cdots & a_{m, n} \\
\end{bmatrix} = \begin{bmatrix} 
\vec{a}_{1, * } \\ 
\vec{a}_{2, * } \\ 
\vec{a}_{3, * } \\ 
\vdots\\
\vec{a}_{m, * } \\ 
\end{bmatrix}, \text{ } \vec{x} = (x_1, x_2, \cdots, x_{n})$

$ $

The result of $A \cdot \vec{x}$ is an $m$-dimensional vector computed as:

$A \cdot \vec{x} = \Big(\vec{a}_{1, * } \cdot \vec{x}, \text{ } \vec{a}_{2, * } \cdot \vec{x}, \text{ }\cdots,\text{ } \vec{a}_{m, * } \cdot \vec{x} \Big) = \left(\sum\limits_{i=1}^{n} a_{1, i} \cdot  x_i, \text{ } \sum\limits_{i=1}^{n}  a_{2, i} \cdot x_i,\cdots, \text{ } \sum\limits_{i=1}^{n} a_{m, i} \cdot x_i \right)$

$ $

\item \textbf{Matrix-to-Matrix Multiplication:} Given a $m \times n$ matrix $A$ and a $n \times k$ matrix $B$:

$A = \begin{bmatrix}
 a_{1, 1} & a_{1, 2} & a_{1, 3} & \cdots & a_{1, n}\\
 a_{2, 1} & a_{2, 2} & a_{2, 3} & \cdots & a_{2, n} \\
 a_{3, 1} & a_{3, 2} & a_{3, 3} & \cdots & a_{3, n} \\
\vdots & \vdots & \vdots & \ddots & \vdots \\
 a_{m, 1} & a_{m, 2} & a_{m, 3} & \cdots & a_{m, n} \\
\end{bmatrix}$, $B = \begin{bmatrix}
 b_{1, 1} & b_{1, 2} & b_{1, 3} & \cdots & b_{1, k}\\
 b_{2, 1} & b_{2, 2} & b_{2, 3} & \cdots & b_{2, k} \\
 b_{3, 1} & b_{3, 2} & b_{3, 3} & \cdots & b_{3, k} \\
\vdots & \vdots & \vdots & \ddots & \vdots \\
 b_{n, 1} & b_{n, 2} & b_{n, 3} & \cdots & b_{n, k} \\
\end{bmatrix}$

$ $

The result of $A \cdot B$ is a $m \times k$ matrix computed as:

$A \cdot B = \begin{bmatrix}
\sum\limits_{i=1}^{n}a_{1, i}b_{i, 1} & \sum\limits_{i=1}^{n}a_{1, i}b_{i, 2} & \sum\limits_{i=1}^{n}a_{1, i}b_{i, 3} & \cdots & \sum\limits_{i=1}^{n}a_{1, i}b_{i, k} \\
\sum\limits_{i=1}^{n}a_{2, i}b_{i, 1} & \sum\limits_{i=1}^{n}a_{2, i}b_{i, 2} & \sum\limits_{i=1}^{n}a_{2, i}b_{i, 3} & \cdots & \sum\limits_{i=1}^{n}a_{2, i}b_{i, k} \\
\sum\limits_{i=1}^{n}a_{3, i}b_{i, 1} & \sum\limits_{i=1}^{n}a_{3, i}b_{i, 2} & \sum\limits_{i=1}^{n}a_{3, i}b_{i, 3} & \cdots & \sum\limits_{i=1}^{n}a_{3, i}b_{i, k} \\
\vdots & \vdots & \vdots & \ddots & \vdots \\
\sum\limits_{i=1}^{n}a_{m, i}b_{i, 1}
& \sum\limits_{i=1}^{n}a_{m, i}b_{i, 2} & \sum\limits_{i=1}^{n}a_{m, i}b_{i, 3} & \cdots & \sum\limits_{i=1}^{n}a_{m, i}b_{i, k} \\
\end{bmatrix}$

\end{itemize}
\end{tcolorbox}

Given the above definitions of matrix and vector arithmetic, the following algebraic properties can be derived:

\begin{tcolorbox}[title={\textbf{\tboxtheorem{\ref*{subsec:matrix-arithmetic}} Matrix Arithmetic Properties}}]

\begin{itemize}

\item \textbf{Associative:} 

$(AB)C = A(BC)$

$A(Bx) = (AB)x$

$ $

\item \textbf{Distributive:} 

$A(x + y) = Ax + Ay$

%$A(x \odot y) = Ax \odot Ay $

%$A\langle x, y\rangle = \langle Ax, Ay\rangle$

%$A\langle \langle x, y\rangle\rangle = \langle\langle Ax, Ay\rangle\rangle$

$ $

However, $Ax \cdot Ay \neq A(x \cdot y)$, because the resulting dimensions do not match. Also, $A(x \odot y) \neq Ax \odot Ay $. Further, note that $A\langle x, y\rangle \neq \langle Ax, Ay\rangle$, and $A\langle \langle x, y\rangle\rangle \neq \langle\langle Ax, Ay\rangle\rangle$.

$ $

\item \textbf{NOT Commutative:} 

$Ax \neq xA$

$AB \neq BA$

\end{itemize}
\end{tcolorbox}

\begin{proof}

$ $

The properties described in \ref*{subsec:matrix-arithmetic} can be demonstrated by expanding the formulas on both sides of each equation using a variable representation for each element in the vectors/matrices and comparing the resulting formulas. We leave this expansion as an exercise for the reader.

\end{proof}

\subsection{Projection}
\label{subsec:projection}

There are two types of projections: a vector projection and an orthogonal (i.e., plane) projection.

\begin{figure}[!h]
    \centering
    \subfloat[Vector Projection \href{https://upload.wikimedia.org/wikipedia/commons/thumb/9/98/Projection_and_rejection.png/200px-Projection_and_rejection.png}{(Source)}]{
  \includegraphics[width=0.25\linewidth]{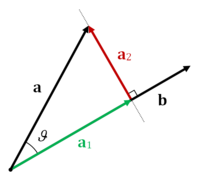}

  \label{fig:vector-projection}
}
    \subfloat[Orthogonal (Plane/Subspace) Projection \href{https://www.researchgate.net/publication/333616378/figure/fig2/AS:766235117617152@1559696096437/Geometric-illustration-of-the-orthogonal-projection-operator-P-A-vector-x-R-m-is.png}{(Source)}]{
  \includegraphics[width=0.4\linewidth]{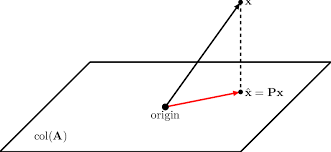}

  \label{fig:orthogonal-projection}
}
  \caption{}
\end{figure}

\para{Vector Projection:} Given two vectors $\vec{a}$ and $\vec{b}$ in the same $n$-dimensional vector space, the vector projection $\textsf{Proj}_{\vec{b}}(\vec{a})$ measures the component of $\vec{a}$ in the direction of $\vec{b}$ (i.e., the part of $\vec{a}$ that is parallel to $\vec{b}$). In the example of \autoref{fig:vector-projection}, $\vec{a}$'s projection on $\vec{b}$ is $\vec{a}_1$, where the length of $\vec{a}_1$ is geometrically $||a_1|| = ||a|| \cos \theta = ||a||\dfrac{a \cdot b}{||a||\cdot||b||} = \dfrac{a \cdot b}{||b||}$. Let $\vec{b'}$ be a unit vector of $\vec{b}$, that is $\vec{b'} = \dfrac{\vec{b}}{||b||}$. Then, $\vec{a}_1 = ||a_1||\cdot\vec{b'} = \dfrac{a \cdot b}{||b||} \cdot \dfrac{\vec{b}}{||b||} = \dfrac{a \cdot b}{||b||^2}\vec{b}$. Thus, $\textsf{Proj}_{\vec{b}}(\vec{a}) = \dfrac{a \cdot b}{||b||^2}\vec{b}$.

\para{Orthogonal Projection:} Given the vector $\vec{x}$ and a set of mutually orthogonal vectors $\vec{p}_0, \vec{p}_1, \cdots, \vec{p}_{n-1}$ that span the subspace $P$, the orthogonal projection $\textsf{Proj}_P(\vec{x})$ measures how much of $\vec{x}$ lies in $P$ (i.e., the component of $\vec{x}$ within $P$). In the example, $\vec{x}$’s projection onto $P$ (red arrow) equals the sum of the projections of $\vec{x}$ onto each orthogonal basis vector $\vec{p}_i$ spanning $P$: $\textsf{Proj}_P(\vec{x}) = \sum\limits_{i=0}^{n-1}\textsf{Proj}_{\vec{p}_i}(\vec{x})$. This computation can be viewed as expressing $\vec{x}$ in a coordinate system defined by the $n$ orthogonal vectors.

\begin{tcolorbox}[title={\textbf{\tboxdef{\ref*{subsec:projection}} Vector and Orthogonal Projections}}]

\begin{itemize}

\item \textbf{Vector Projection:} Given two vectors $\vec{a}$ and $\vec{b}$ in the same vector space, the vector projection of $\vec{a}$ on $\vec{b}$ is:

$\textsf{Proj}_{\vec{b}}(\vec{a}) = \vec{a}_p  = \dfrac{\vec{a} \cdot \vec{b}}{||b||^2} \vec{b}$ \textcolor{red}{ $\rhd$ where $||\vec{a}_p|| = ||\vec{a}|| \cos \theta$}

$ $

\item \textbf{Orthogonal Basis:} If the $n$-dimensional plane (or subspace) $P$ is spanned by the mutually orthogonal $n$-dimensional vectors $\vec{p}_0, \vec{p}_1, \cdots, \vec{p}_{n-1}$, 

then the matrix $P = \begin{bmatrix} 
\vec{p}_0\\
\vec{p}_1\\
\vdots\\
\vec{p}_{n-1}\\
\end{bmatrix}
$
is defined to be an orthogonal basis of plane $P$.

\item \textbf{Orthogonal Projection:} Given the orthogonal basis matrix $P = \begin{bmatrix} 
\vec{p}_0\\
\vec{p}_1\\
\vdots\\
\vec{p}_{n-1}\\
\end{bmatrix}
$, 

vector $\vec{a}$'s orthogonal projection on $P$ is:

$\textsf{Proj}_P(\vec{a}) = \sum\limits_{i=0}^{n-1} \textsf{Proj}_{\vec{p}_i}(\vec{a})$

\end{itemize}

\end{tcolorbox}

Based on the definition of orthogonal projection, the following properties are derived:  

\para{Orthogonal Basis:} In an $n$-dimensional vector space, any mutually orthogonal $n$ vectors in the vector space span the subspace $P$ that is identical to the entire vector space. Further, the orthogonal projection of any vector in the vector space on $P$ is guaranteed to be a unique vector.

\para{Non-orthogonal Basis:} In an $n$-dimensional vector space, suppose some $n$ non-orthogonal vectors satisfy the following two conditions: (i) they span the entire vector space; (ii) they are linearly independent (i.e., one vector cannot be expressed as a linear combination of the other vectors). Then, the $n \times n$ matrix $P$ comprised of these $n$ vectors forms a basis for the entire vector space $V$, and the matrix-to-vector multiplication $P\vec{v}$ for each $\vec{v}$ in the vector space is guaranteed to yield a unique vector. However, the formula $\textsf{Proj}_P(\vec{v})$ is not a valid geometric projection of the vector $\vec{v}$ on $P$, because the $n$ basis vectors are non-orthogonal. Yet, the computation of $P\vec{v}$ can be thought of as uniquely transforming $\vec{v}$ into a different coordinate system that expresses the vector space with respect to $n$ non-orthogonal vectors in $P$. 

\begin{tcolorbox}[title={\textbf{\tboxtheorem{\ref*{subsec:projection}} Uniqueness of Transformed Vectors}}]

\begin{itemize}
\item  \textbf{Orthogonal Basis:} If some $n$ vectors are a orthogonal basis of the plane $P$ in the $n$-dimensional vector space, then $P$ is the same as the entire vector space, and $\textsf{Proj}_P(v)$ for every vector $\vec{v}$ in the vector space is guaranteed to be a unique vector.
\item \textbf{Non-orthogonal Basis:} If some $n$ vectors are a non-orthogonal basis of the plane $P$ in the $n$-dimensional vector space (i.e., each vector is linearly independent and they span $P$), then $P$ is the same as the entire vector space, and $P\vec{v}$ is guaranteed to result in a unique vector.  
\end{itemize}
\end{tcolorbox}

\subsection{Basis of a Polynomial Ring}
\label{subsec:polynomial-ring-basis}

Given an $(n-1)$-degree polynomial ring $\mathbb{Z}[X] / (X^n + 1)$, a basis of the polynomial ring is defined as a set of polynomials that satisfies the following two requirements:

\begin{itemize}

\item \textbf{Linear Independence}: Each polynomial in the basis set cannot be expressed as a linear combination of the other polynomials in the same set
\item \textbf{Spanning the Polynomial Ring:} A linear combination of the polynomials in the basis set can express any polynomial in the polynomial ring
\end{itemize}

Note that for a $(n-1)$-degree polynomial ring, the number of polynomials that form a basis of the polynomial ring is exactly $n$.

\subsection{Isomorphism between Polynomials and Vectors over Integers (and Ring)}
\label{subsec:poly-vector-transformation}

Now, let's define a mapping $\sigma$ from the $(n-1)$-degree polynomial ring to the $n$-dimensional vector space, such that an input polynomial's list of $y$ values evaluated at $n$ distinct $X \in \mathbb{Z}_t$ coordinates (e.g., $x_0, x_1, \cdots, x_{n-1}$) forms the mapping's output vector, where $\mathbb{Z}_t$ is a field (i.e., $t$ is prime), so that the interpolated polynomial $f(X)$ over the $(n+1)$ distinct $X$ points is well-defined (as explained in \autoref{subsec:polynomial-interpolation-field-condition}). Technically, $\sigma$ is defined as:

$\sigma: f(X) \in \mathbb{Z}_t[X] / (X^n + 1) \text{ } \longrightarrow \text{ } (f(x_0), f(x_1), f(x_2), \cdots, f(x_{n-1})) \in \mathbb{Z}^n_t$

$ $

Now, we will explain why the mapping $\sigma$ is isomorphic, which means that $\sigma$ is a bijective one-to-one mapping from $\mathbb{Z}_t[X] / (X^n + 1)$ to $\mathbb{Z}^n_t$, and it preserves the algebraic operations $(+, \cdot)$ (i.e., $\sigma$ is a homomorphism for addition and multiplication).

$ $

\para{Bijective:} In the $(n-1)$-degree polynomial ring, a list of $y$ values evaluated at some statically chosen $n$ distinct $x$ coordinates defines a unique polynomial because, algebraically, there exists only one $(n-1)$-degree (or a lesser degree) polynomial that passes through each given set of $n$ distinct $(x, y)$ coordinates. We proved this in Lagrange Polynomial Interpolation (Theorem~\ref*{sec:polynomial-interpolation} in \autoref{sec:polynomial-interpolation}). 

$ $

\para{Homomorphic:} The homomorphism of the mapping $\sigma$ on the $(+, \cdot)$ operations means that the following two relationships hold: 

$\sigma(f_a(X) + f_b(X)) = \sigma(f_a(X)) + \sigma(f_b(X))$

$\sigma(f_a(X) \cdot f_b(X)) = \sigma(f_a(X)) \odot \sigma(f_b(X))$ \textcolor{red}{\text{ } \# $\odot$ is Hadamard vector multiplication (Summary~\ref*{subsec:vector-arithmetic})}

$ $

To prove our $\sigma$ mapping's homomorphism, let's denote the input polynomials $f_a(X)$, $f_b(X)$, and their $\sigma$-mapped output vectors as follows:

$f_a(X) = a_0 + a_1X + a_2X^2 + \cdots + a_{n-1}X^{n-1}$

$ \sigma(f_a(X)) = \left({f_a(x_0)}, {f_a(x_1)}, {f_a(x_2)}, \cdots, {f_a(x_{n-1}}) \right)  = \left(\sum\limits_{i=0}^{n-1}a_i(x_0)^i, \sum\limits_{i=0}^{n-1}a_i(x_1)^i, \sum\limits_{i=0}^{n-1}a_i(x_2)^i, \cdots, \sum\limits_{i=0}^{n-1}a_i(x_{n-1})^i\right)$

$ $

$f_b(X) = b_0 + b_1X + b_2X^2 + \cdots + b_{n-1}X^{n-1}$

$ \sigma(f_b(X)) = \left({f_b(x_0)}, {f_b(x_1)}, {f_b(x_2)}, \cdots, {f_b(x_{n-1}}) \right) = \left(\sum\limits_{i=0}^{n-1}b_i(x_0)^i, \sum\limits_{i=0}^{n-1}b_i(x_1)^i, \sum\limits_{i=0}^{n-1}b_i(x_2)^i, \cdots, \sum\limits_{i=0}^{n-1}b_i(x_{n-1})^i\right)$

$ $

Given the above setup, we can see that $\sigma$ preserves homomorphism on the $(+)$ operation as follows:

$\bm{\sigma}\bm{(}f_a(X) + f_b(X)\bm{)} = \bm{\sigma\bigg(}(a_0 + b_0) + (a_1 + b_1)X + (a_2 + b_2)X^2 + \cdots + (a_{n-1} + b_{n-1})X^{n-1}\bm{\bigg)}$

$= \left(\sum\limits_{i=0}^{n-1}(a_i + b_i)(x_0)^i, \text{ }\sum\limits_{i=0}^{n-1}(a_i + b_i)(x_1)^i, \text{ }\sum\limits_{i=0}^{n-1}(a_i + b_i)(x_2)^i, \cdots, \text{ }\sum\limits_{i=0}^{n-1}(a_i + b_i)(x_{n-1})^i\right)$

$= \left(\sum\limits_{i=0}^{n-1}a_i(x_0)^i, \text{ }\sum\limits_{i=0}^{n-1}a_i(x_1)^i, \text{ }\sum\limits_{i=0}^{n-1}a_i(x_2)^i, \cdots, \text{ } \sum\limits_{i=0}^{n-1}a_i(x_{n-1})^i\right)$

\text{ } $+ \left(\sum\limits_{i=0}^{n-1}b_i(x_0)^i, \text{ } \sum\limits_{i=0}^{n-1}b_i(x_1)^i, \text{ } \sum\limits_{i=0}^{n-1}b_i(x_2)^i, \cdots, \text{ } \sum\limits_{i=0}^{n-1}b_i(x_{n-1})^i\right)$

$= \bm{\sigma(}f_a(X)\bm{)} + \bm{\sigma(}f_b(X)\bm{)}$

$ $

Also, we can see that $\sigma$ preserves homomorphism on the $(\cdot)$ operation as follows:

$\bm{\sigma(}f_a(X)\cdot f_b(X)\bm{)} = \bm{\sigma\bigg(}\left(\sum\limits_{i=0}^{n-1}a_iX^i\right)\cdot\left( \sum\limits_{i=0}^{n-1}b_iX^i\right)\bm{\bigg)}$

$ $

$= \Bigg( \left(\sum\limits_{i=0}^{n-1}a_ix_0^i\right)\left( \sum\limits_{i=0}^{n-1}b_ix_0^i\right), \text{ }
\left(\sum\limits_{i=0}^{n-1}a_ix_1^i\right)\left( \sum\limits_{i=0}^{n-1}b_ix_1^i\right), \text{ }
\left(\sum\limits_{i=0}^{n-1}a_ix_2^i\right)\left( \sum\limits_{i=0}^{n-1}b_ix_2^i\right),$

$  \text{ } \text{ } \text{ } \text{ } \text{ } \text{ }
\cdots,
\left(\sum\limits_{i=0}^{n-1}a_ix_{n-1}^i\right)\left( \sum\limits_{i=0}^{n-1}b_ix_{n-1}^i\right)
\Bigg)$

$ $

$ = \Bigg(\sum\limits_{i=0}^{n-1}a_i(x_0)^i, \text{ } \sum\limits_{i=0}^{n-1}a_i(x_1)^i, \text{ } \cdots, \text{ } \sum\limits_{i=0}^{n-1}a_i(x_{n-1})^i\Bigg)$

$  \text{ } \text{ } \text{ } \text{ }  \odot  \Bigg(\sum\limits_{i=0}^{n-1}b_i(x_0)^i, \text{ } 
\sum\limits_{i=0}^{n-1}b_i(x_1)^i, \text{ }
\cdots, \text{ } \sum\limits_{i=0}^{n-1}b_i(x_{n-1})^i\Bigg)$

$ $

$= \bm{\sigma(}f_a(X)\bm{)} \odot \bm{\sigma(}f_b(X)\bm{)}$

$ $

In summary, $\sigma$ preserves the following homomorphism:

$\sigma(f_a(X) + f_b(X)) = \sigma(f_a(X)) + \sigma(f_b(X))$

$\sigma(f_a(X) \cdot f_b(X)) = \sigma(f_a(X)) \odot \sigma(f_b(X))$

$ $

However, for $\sigma(f_a(X) \cdot f_b(X)) = \sigma(f_a(X)) \odot \sigma(f_b(X))$, we need further reasoning to justify that this relation holds in polynomial rings, which is explained below.

\para{Polynomial Ring Reduction:} Suppose that we did not have the polynomial ring setup $X^n + 1$. Then, if we multiply $f_a(X)$ and $f_b(X)$, then $f_a(X) \cdot f_b(X)$ may become a new polynomial whose degree is higher than $n-1$. This higher-degree polynomial would still decode into the expected correct vector. Suppose the following:

$\sigma(f_a(X)) = (f_a(x_0), f_a(x_1), \cdots, f_a(x_{n-1})) = (v_0, v_1, \cdots, v_{n-1})$

$\sigma(f_b(X)) = (f_b(x_0), f_b(x_1), \cdots, f_b(x_{n-1})) = (u_0, u_1, \cdots, u_{n-1})$

$ $

Then, the following is true: 

$\sigma(f_a(X) \cdot f_b(X)) = (f_a(x_0)\cdot f_b(x_0), f_a(x_1)\cdot f_b(x_1), \cdots, f_a(x_{n-1})\cdot f_b(x_{n-1})) = (v_0u_0, v_1u_1, \cdots, v_{n-1}u_{n-1})$

$= (v_0, v_1, \cdots, v_{n-1}) \odot (u_0, u_1, \cdots, u_{n-1}) $

$ $

As shown above, even if $f_a(X) \cdot f_b(X)$ results in a polynomial with a degree higher than $n-1$, it can be decoded into the expected correct vector. However, the $\sigma$ mapping loses the property of isomorphism between a polynomial and a vector because if a polynomial's degree is higher than $n-1$, then there can be more than 1 polynomial that passes through the given $n$ distinct $X$ coordinates: $\{x_0, x_1, \cdots, x_{n-1}\}$. This is a problem because, if the $\sigma$ mapping supports only polynomial-to-vector mappings and not vector-to-polynomial mappings, then we cannot convert vectors into polynomials in the first place and do isomorphic computations. Another minor issue is that if the polynomial degree term is higher than $n-1$, then the computational overhead of decoding (i.e., polynomial evaluation) becomes larger than before. 

To resolve these two minor issues, we let the $n$ distinct $X$ coordinates of evaluation be the solutions of the polynomial ring modulo $X^n + 1$ (where $n$ is some power of 2) and reduce $f_a(X) \cdot f_b(X)$ to a new polynomial modulo $X^n + 1$ whose degree is at most $n - 1$. Let $f_{ab}(X) = f_a(X) \cdot f_b(X)$, and $f'_{ab}(X)$ be the reduced polynomial such that $f_{ab}(X) = Q(X)\cdot (X^n + 1) + f'_{ab}(X)$ for some quotient polynomial $Q(X)$. Then, as illustrated in Summary \ref*{subsubsec:polynomial-ring-discuss} (\autoref{subsubsec:polynomial-ring-discuss}), $f_{ab}(X)$ and $f'_{ab}(X)$ evaluate to the same value if they are evaluated at the roots of $X^n + 1$ (by zeroing out the $Q(X)$ term). Therefore, if we let the $n$ distinct evaluating points $\{x_0, x_1, \cdots, x_{n-1}\}$ be the roots of $X^n + 1$, then we can ensure that the decoded vector of $f'_{ab}(X)$ is identical to that of $f_{ab}(X)$, which we expect. Therefore, we can replace the higher-degree polynomial $f_{ab}(X)$ with the reduced polynomial $f'_{ab}(X)$ and continue with any further polynomial additions or multiplications using $f'_{ab}(X)$. Also, by applying polynomial ring reduction, we can enhance the computational efficiency of polynomial addition and multiplication, as well as preserve the isomorphism of the $\sigma$ mapping. Therefore, we can freely convert between vectors \& polynomials and perform additions and multiplications. 

For applying this polynomial ring reduction, the polynomial modulus can be any polynomial as long as it has at least $n$ distinct roots. In practice, we often choose $X^n + 1$ as the polynomial ring modulus, which is the $(\mu=2n)$-th cyclotomic polynomial (\autoref{subsec:cyclotomic-def}). The reason we let the polynomial ring modulus be a cyclotomic polynomial (especially the $(\mu=2n)$-th cyclotomic polynomial, $X^n + 1$) is that its $n$ distinct roots are well-defined (i.e., primitive $(\mu=2n)$-th roots of unity) and thus can be quickly identified even when $n$ is large.

$ $

\para{Polynomial Coefficient Modulo Reduction:} In addition, we often reduce the polynomial coefficients based on some modulus $t$ to keep the size of the coefficients lower than a certain limit for the purpose of computational efficiency. Suppose two polynomials $f_c(X)$ and $f_d(X)$ have congruent coefficients modulo $t$ as follows:

$f_c(X) = w_0 + w_1\cdot x_i + w_2\cdot x_i^2 + \cdots + w_{n-1}\cdot x_i^{n-1}$

$f_d(X) = w'_0 + w'_1\cdot x_i + w'_2\cdot x_i^2 + \cdots + w'_{n-1}\cdot x_i^{n-1} $

$w_i \equiv w'_i \bmod t$

$ $

Then, their evaluated value $f_c(x_i)$ and $f_d(x_i)$ for any $x_i$ is guaranteed to be congruent modulo $t$, as shown below:

$f_c(x_i) = w_0 + w_1\cdot x_i + w_2\cdot x_i^2 + \cdots + w_{n-1}\cdot x_i^{n-1}$

$\equiv (w_0 + w_1\cdot x_i + w_2\cdot x_i^2 + \cdots + w_{n-1}\cdot x_i^{n-1}) \bmod t $

$\equiv (w_0  \bmod t) + (w_1  \bmod t)\cdot x_i + (w_2  \bmod t)\cdot x_i^2 + \cdots + (w_{n-1} \bmod t)\cdot x_i^{n-1}) $

$\equiv w'_0 + w'_1\cdot x_i + w'_2\cdot x_i^2 + \cdots + w'_{n-1}\cdot x_i^{n-1} $

$= f_d(x_i)$

\para{Summary:} Since $\sigma$ is bijective and homomorphic, $\sigma$ is an isomorphic mapping between the $(n-1)$-degree polynomial ring $\mathbb{Z}_t[X]/X^n + 1$ and the $n$-dimensional vector space $\mathbb{Z}_t^n$.

$ $

\subsubsection{Finding Appropriate Modulus $t$} 
\label{subsubsec:poly-vector-transformation-modulus}

To isomorphically evaluate a polynomial in $\mathbb{Z}_t[X]/X^n + 1$ into an $n$-dimensional vector, we need to evaluate the polynomial at $n$ distinct roots of $X^n + 1 \bmod t$. However, $X^n + 1 \bmod t$ does not have $n$ distinct roots for all combinations of (degree, modulus) $ = (n, t)$. For example, if $n = 2$ and $t = 3$, then $X^2+ 1 \not\equiv 0 \bmod 3$ for any possible values of $X = \{0, 1, 2\}$. Therefore, our goal is to find a satisfactory $t$ given a fixed $n$ such that $n$ distinct roots of $X^n + 1 \pmod t$ exist in order to use the isomorphic $\sigma$ mapping. 

We start with two constraints: (1) choose $t$ to be a prime number; (2) ensure $t-1$ is a multiple of $2n$.

We learned from Fermat's Little Theorem in Theorem~\ref*{subsec:order-theorem}.4 (\autoref{subsec:order-theorem}) the following: $a^{t - 1} \equiv 1 \bmod t$ if and only if $a$ and $t$ are co-prime. This means that if $t$ is a prime, then $a^{t - 1} \equiv 1 \bmod t$ for all $a \in \mathbb{Z}_t^{\times}$ (i.e., $\mathbb{Z}_t$ without $\{0\}$). Suppose $g$ is the generator of $\mathbb{Z}_t^{\times}$ whose powered values generate all elements of $\mathbb{Z}_t^{\times}$. Then, $\textsf{Ord}_{\mathbb{Z}_t}(g) = t - 1$ and $g^{t - 1} \equiv 1 \bmod t$. Since $t - 1 = k \cdot 2n$ for some $k$, $g^{k\cdot2n} \equiv (g^{k})^{2n} \equiv 1 \bmod t$. Then, $\textsf{Ord}_{\mathbb{Z}_t}(g^{k}) \leq 2n$. However, since $\textsf{Ord}_{\mathbb{Z}_t}(g) = t - 1$, for all $a$ such that $a < t - 1 = k \cdot 2n$, $g^a \not\equiv 1 \bmod t$. In other words, for all $b$ such that $b < 2n$, $(g^k)^b \not\equiv 1 \bmod t$. Thus, $\textsf{Ord}_{\mathbb{Z}_t}(g^{k}) = 2n$. 

Let $c = g^k$. Since $\textsf{Ord}_{\mathbb{Z}_t}(c) = 2n$, $c^{2n} \equiv 1 \bmod t$. In other words, $(c^n)^2 \equiv 1 \bmod t$. 
Now, $c^n$ can be only 1 or -1. The reason is that in the relation $X^2 \equiv 1 \bmod t$, $X$ can be mathematically only $1$ or $-1 \equiv t - 1 \bmod t$. If we substitute $X = c^n$, then $c^n$ can be only $1$ or $-1 \equiv t - 1 \bmod t$. But $\textsf{Ord}_{\mathbb{Z}_t}(c) = 2n$, thus $c^n$ cannot be 1 (because $1^1 = 1$ and 1 is smaller than the order of $c$: $2n > 1$). Thus, $c^n$ can be only $-1 \equiv t - 1 \bmod t$. If $c^n = -1 \equiv t - 1 \bmod t$, then $c$ is the root of $X^n + 1 \bmod t$, because $X^n + 1 = c^n + 1 \equiv (t - 1) + 1 \equiv 0 \bmod t$. 

In conclusion, given a cyclotomic polynomial $X^n + 1$, if we choose a prime $t$ such that $t - 1 = k \cdot 2n$ for some integer $k$, then one root of $X^n + 1 \bmod t$ is: $X = c = g^k$. 

Once we have found one root of $X^n + 1$, then we can derive all $n$ distinct roots of $X^n + 1$. Suppose $\omega$ is one root of $X^n + 1 \bmod t$. Then we derive the following:

$\omega^n + 1 \equiv 0 \bmod t$

$\omega^n \equiv t - 1 \bmod t$

$\omega^{2n} \equiv (t - 1)\cdot(t - 1) \equiv  t^2 - 2t + 1 \equiv   1 \bmod t$

While $\omega^{2n} \equiv 1 \bmod t$, $\omega^{n} \not\equiv 1 \bmod t$, because if so, $X^n + 1 = \omega^n + 1 = 1 + 1 = 2 \neq 0$, which contradicts the fact that $\omega$ is a root of $X^n + 1$. Therefore, $\textsf{Ord}_{\mathbb{Z}_t}(c) = 2n$. 

Now, we derive the remaining $n-1$ distinct roots of $X^n + 1 \bmod t$ as follows:

$(\omega^3)^n + 1 \equiv (\omega^{2n})\cdot \omega^n + 1 \equiv \omega^n + 1 \equiv t - 1 + 1 \equiv 0 \bmod t$

$(\omega^5)^n + 1 \equiv (\omega^{4n})\cdot \omega^n + 1 \equiv \omega^n + 1 \equiv t - 1 + 1 \equiv 0 \bmod t$

$\vdots$

$(\omega^{2n - 1})^n + 1 \equiv 0 \bmod t$ \\ \textcolor{red}{ $\rhd$ for any odd $k = 2j+1$, $(\omega^{2j+1})^n = (\omega^{2n})^j \cdot \omega^n = 1^j \cdot (-1) = -1$, so $(\omega^k)^n + 1 = 0$}

$ $

Note that $\omega, \omega^3, \omega^5, \cdots, \omega^{2n - 1}$ are all distinct values in $\mathbb{Z}_t^{\times}$, because $\textsf{Ord}_{\mathbb{Z}_t}(c) = 2n$. Thus, $\{\omega^{2i + 1}\}_{i=0}^{n-1}$ are $n$ distinct roots of $X^n + 1$. At the same time, since these are the roots of the cyclotomic polynomial $X^n + 1$, these are $n$ distinct primitive $(\mu=2n)$-th roots of unity.

$ $

We summarize our findings as follows:

\begin{tcolorbox}[title={\textbf{\tboxtheorem{\ref*{subsubsec:poly-vector-transformation-modulus}} Isomorphism between Polynomials and Vectors over Integers (Ring)}}]

\begin{itemize}
\item Suppose we have an $(n-1)$-degree polynomial ring $\mathbb{Z}_t[X] / F(X)$ where $F(X)$ is an $n$-degree polynomial having $n$ distinct roots $\{x_0, \cdots, x_{n-1}\}$, and an $n$-dimensional vector space $\mathbb{Z}_t^n$ (integers mod $t$) with vector $\vec{v}$.

$\sigma: f(X) \in \mathbb{Z}_t[X] / F(X) \text{ } \longrightarrow \text{ } (f(x_0), f(x_1), f(x_2), \cdots, f(x_{n-1})) \in \mathbb{Z}_t^n$

$ $

Then, $\sigma$ preserves isomorphism over the $(+, \cdot)$ operations:

$\sigma(f_a(X) + f_b(X)) = \sigma(f_a(X)) + \sigma(f_b(X))$

$\sigma(f_a(X) \cdot f_b(X)) = \sigma(f_a(X)) \odot \sigma(f_b(X))$

$ $

%\item If the $n$-degree polynomial $F(x)$ has a fewer number of roots than $n$, then isomorphism between the vectors and polynomials over the $(+, \cdot)$ operations holds without modulo $t$ as follows:

%$\sigma': f(X) \in \mathbb{Z}[X] / F(X) \text{ } \longrightarrow \text{ } (f(x_0), f(x_1), f(x_2), \cdots, f(x_{n-1})) \in \mathbb{Z}^n$

%$ $

\item Suppose we have the $(\mu=2n)$-th cyclotomic polynomial $X^n + 1 \bmod t$ such that $t$ is a prime and $t - 1$ is some multiple of $2n$, and $g$ is a generator of $\mathbb{Z}_t^{\times}$. Then, $n$ distinct roots of $X^n + 1$ (i.e., primitive $(\mu=2n)$-th roots of unity) can be efficiently computed as: $\{\omega^{2i + 1}\}_{i=0}^{n-1}$ where $\omega = g^{\frac{t-1}{2n}} \bmod t$.

\end{itemize}
\end{tcolorbox}

\subsection{Isomorphism between Polynomials and Vectors over Complex Numbers}
\label{subsec:poly-vector-transformation-complex}

In Theorem~\ref*{subsec:polynomial-ring-basis} (\autoref{subsec:polynomial-ring-basis}), we learned the isomorphic mapping $\sigma: f(X) \in \mathbb{Z}_t[X]/(X^n + 1) \longrightarrow (f(x_0),f(x_1),f(x_2), \cdots, f(x_{n-1})) \in \mathbb{Z}_t^n$, where $x_0, x_1, x_2, \cdots, x_{n-1} \in \mathbb{Z}$ are the ($(\mu=2n)$-th primitive) roots of the cyclotomic polynomial $X^n + 1$, which are $\omega, \omega^3, \omega^5, \cdots, \omega^{2n-1}$, where $\omega$ can be any root of $X^n + 1$ (i.e., since each $\omega$ is a generator of all roots). In this subsection, we will demonstrate the isomorphism between a vector space and a polynomial ring over complex numbers as follows: 

$\sigma_c: f(X) \in \mathbb{R}[X]/(X^n + 1) \longrightarrow (f(\omega),f(\omega^3),f(\omega^5), \cdots, f(\omega^{2n-1})) \in \mathbb{\hat{C}}^{n}$

$ $

, where $X \in \mathbb{C}$, and $\omega = e^{i\pi/n}$, the root (i.e., the primitive $(\mu=2n)$-th root) of the cyclotomic polynomial $X^n + 1$ over complex numbers (Theorem~\ref*{subsec:cyclotomic-theorem}.1 in \autoref{subsec:cyclotomic-theorem}). We define $\mathbb{\hat{C}}^{n}$ to be an $n$-dimensional \textit{special} vector space whose second-half elements of each vector are reverse-ordered conjugates of the first-half elements (e.g., $(v_0, v_1, \cdots, v_{\frac{n}{2}-1}, \overline{v}_{\frac{n}{2}-1}, \cdots, \overline{v}_1, \overline{v}_0 )$). 

\subsubsection{Isomorphism between $\mathbb{C}^{\frac{n}{2}}$ and $\hat{\mathbb{C}}^{n}$}
\label{subsec:poly-vector-transformation-complex-isomorphism1}

In this section, we treat both $\mathbb{C}^{\frac{n}{2}}$ and $\hat{\mathbb{C}}^{n}$ as vector spaces of complex vectors, where the mapping between them is defined as $\phi: \mathbb{C}^{\frac{n}{2}} \rightarrow \hat{\mathbb{C}}^{n}$ as follows:

$\phi(z_0, z_1, \dots, z_{\frac{n}{2}-1}) = (z_0, z_1, \dots, z_{\frac{n}{2}-1}, \overline{z_{\frac{n}{2}-1}}, \dots, \overline{z_1}, \overline{z_0})$

$ $

In other words, $\hat{\mathbb{C}}^{n}$ is a conjugate extension of $\mathbb{C}^{\frac{n}{2}}$ such that the first-half elements of $\hat{\mathbb{C}}^{n}$ are identical to those of $\mathbb{C}^{\frac{n}{2}}$, and the second-half elements of $\hat{\mathbb{C}}^{n}$ are reverse-ordered conjugates of $\mathbb{C}^{\frac{n}{2}}$. 

Now, we will demonstrate that the mapping $\phi$ between $\mathbb{C}^{\frac{n}{2}}$ and $\hat{\mathbb{C}}^{n}$ is an isomorphism.

$ $

\para{Bijective:} The mapping $\phi$ is trivially bijective. Every vector in $\mathbb{C}^{\frac{n}{2}}$ uniquely maps to a vector in $\hat{\mathbb{C}}^{n}$ by appending its reverse-ordered complex conjugates. Conversely, any vector in $\hat{\mathbb{C}}^{n}$ can uniquely map back to $\mathbb{C}^{\frac{n}{2}}$ by simply dropping the second half of its elements, establishing a perfect one-to-one correspondence.

\para{Homomorphic:} We will demonstrate that $\phi$ preserves homomorphism over addition $(+)$ and element-wise vector multiplication $(\odot)$. Given two vectors $\vec{u}, \vec{v} \in \mathbb{C}^{\frac{n}{2}}$:
\begin{itemize}
    \item \textbf{Addition:} $\phi(\vec{u} + \vec{v}) = \phi(\vec{u}) + \phi(\vec{v})$ holds because complex conjugation distributes over addition (i.e., $\overline{u + v} = \overline{u} + \overline{v}$).
    \item \textbf{Element-wise Multiplication:} $\phi(\vec{u} \odot \vec{v}) = \phi(\vec{u}) \odot \phi(\vec{v})$ holds because the conjugate of a product is the product of the conjugates (i.e., $\overline{u \cdot v} = \overline{u} \cdot \overline{v}$).
\end{itemize}

$ $

Therefore, the mapping $\phi: \mathbb{C}^{\frac{n}{2}} \rightarrow \hat{\mathbb{C}}^{n}$ is an isomorphism.

\begin{tcolorbox}[title={\textbf{Vector Space of Complex Vectors with either Real or Complex Scalars}}]

By mathematical definition, a vector space must satisfy \textit{closure under scalar multiplication}: for any vector $\vec{v}$ in the space and any scalar $c$ from the scalar field, the product $c \cdot \vec{v}$ must also lie in the same vector space.  Under this requirement, $\mathbb{C}^{\frac{n}{2}}$ is naturally a vector space of complex vectors with complex scalars (though the same is true with real scalars, since complex includes real). On the other hand, $\hat{\mathbb{C}}^{n}$ is a vector space of complex vectors \textit{strictly with real scalars}.

$ $

To see why, recall that a vector $\vec{v} = (v_0, v_1, \dots, v_{n-1}) \in \hat{\mathbb{C}}^n$ must satisfy the Hermitian-symmetry constraint: $v_{n-1-k} = \overline{v_k}$ for all $k$. Consider an example with $n = 4$:

$$\vec{v} = (1+i,\ 2+2i,\ 2-2i,\ 1-i) \in \hat{\mathbb{C}}^4$$

Now, let us multiply $\vec{v}$ by the complex scalar $c = i$:

$$i \cdot \vec{v} = (-1+i,\ -2+2i,\ 2+2i,\ 1+i)$$

For this result to still lie in $\hat{\mathbb{C}}^4$, the third entry must be the conjugate of the second entry. But $\overline{-2+2i} = -2-2i$, which is not equal to $2+2i$. So $i \cdot \vec{v} \notin \hat{\mathbb{C}}^4$, and closure fails.

$ $

More generally, for any $\vec{v} \in \hat{\mathbb{C}}^n$ and any scalar $c$, the product $c \cdot \vec{v}$ preserves Hermitian symmetry if and only if $\overline{c} = c$; that is, if and only if $c$ is a real number. Therefore, $\hat{\mathbb{C}}^n$ is a valid vector space only when restricted to real scalars, and this is why $\hat{\mathbb{C}}^n$ is a vector space of complex vectors strictly with real scalars.

$ $

Throughout this book, we will regard both $\mathbb{C}^{\frac{n}{2}}$ and $\hat{\mathbb{C}}^{n}$ as vector spaces of complex vectors with real scalars.

$ $

\textbf{Isomorphism of Vector Spaces:} For a mapping between vector spaces to be an isomorphism, it has to be not only bijective and homomorphic over the $(+, \cdot)$ operations, but also homomorphic over scalar multiplication. Therefore, we need to prove that $\phi(c \cdot \vec{u}) = c \cdot \phi(\vec{u})$. This is always true in our case because the scalar $c$ is a real number, meaning it is unaffected by complex conjugation (i.e., $\overline{c} = c$, and thus $\overline{c \cdot u_k} = c \cdot \overline{u_k}$). 

$ $

\textbf{Dimension of Vector Spaces:} The dimension of a vector space is defined by how many scalars it takes to fully describe it. Therefore, the dimension of $\mathbb{C}^{\frac{n}{2}}$ is $n$ because it has $\dfrac{n}{2}$ elements where each element requires 2 scalars to describe (one real and one imaginary part). The dimension of $\hat{\mathbb{C}}^{n}$ is also $n$ because the first $\dfrac{n}{2}$ elements requires $\dfrac{n}{2}\cdot 2$ scalars to describe, and the latter $\dfrac{n}{2}$ elements are simply reverse-ordered conjugates of the former $\dfrac{n}{2}$ elements.

\end{tcolorbox}

$ $

\subsubsection{Isomorphism between $\hat{\mathbb{C}}^{n}$ and $\mathbb{R}[X] / (X^n + 1)$}
\label{subsec:poly-vector-transformation-complex-isomorphism2}

Now, we will demonstrate that $\sigma_c$ is an isomorphism (i.e., bijective and homomorphic) between $\hat{\mathbb{C}}^{n}$ and $\mathbb{R}[X] / (X^n + 1)$ by applying the same reasoning as described in the beginning of \autoref{subsec:poly-vector-transformation}. 

\begin{figure}[h!]
    \centering
  \includegraphics[width=0.3\linewidth]{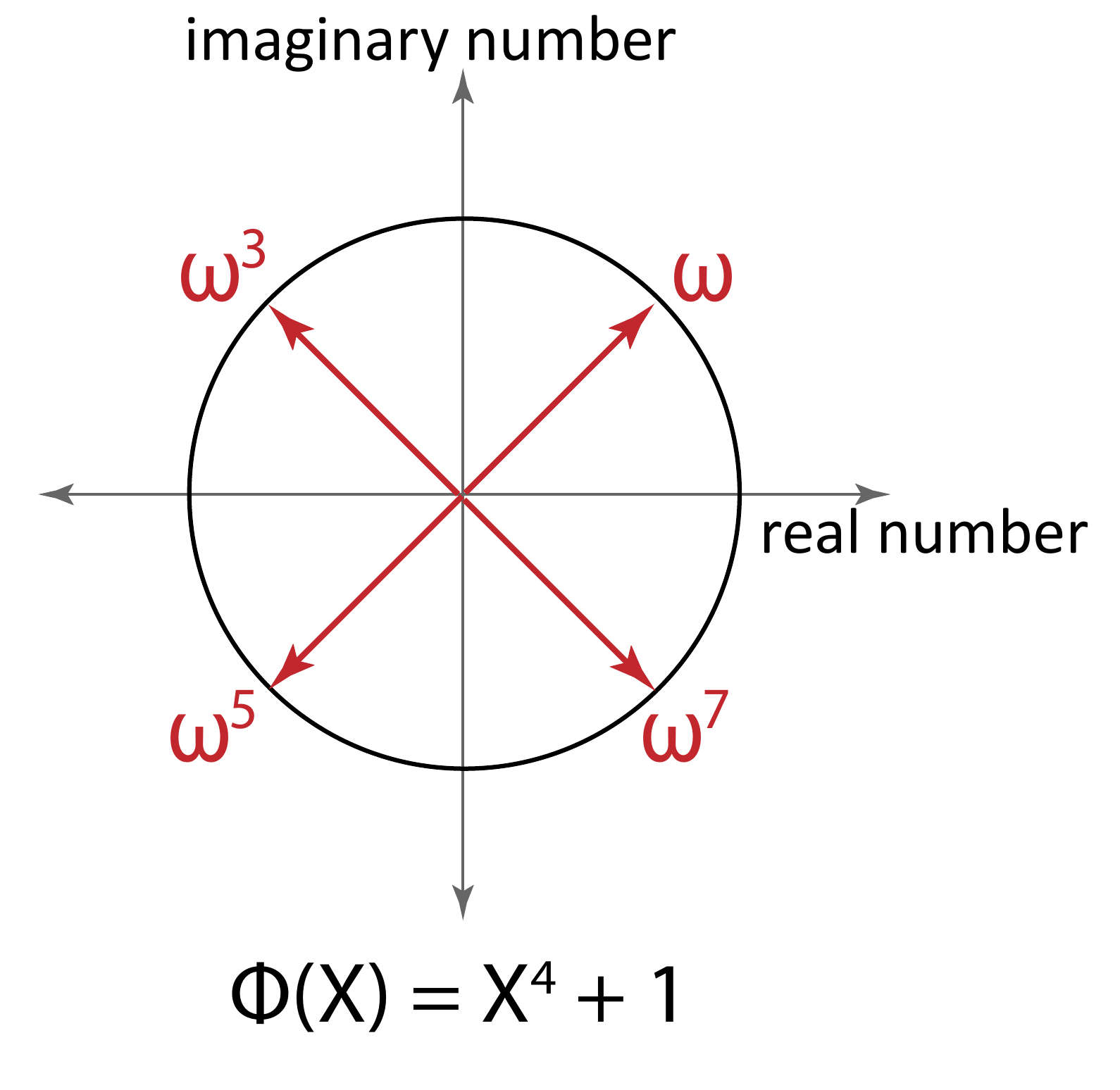}
  \caption{An illustration of the four roots of the 8th cyclotomic polynomial $x^4 + 1$}
  \label{fig:cyclotomic-polynomial}
\end{figure}

$ $

\para{Bijective:} Based on Euler's formula $e^{i\theta} = \cos\theta + i\cdot\sin\theta$ (\autoref{subsec:euler}), we can derive the following arithmetic relations: $\omega = \overline{\omega^{2n-1}}, \text{ } \omega^3 = \overline{\omega^{2n-3}}, \cdots, \omega^{n-1} = \overline{\omega^{n+1}}$. In other words, the one-half roots are conjugates of the other-half roots. This can also be pictorially understood based on a complex plane in \autoref{fig:cyclotomic-polynomial}, where red arrows represent the roots of the 8th cyclotomic polynomial $X^4 + 1$, comprising imaginary number and real number components. As shown in this figure, one half of the red arrows (i.e., roots) are a reflection of the other half on the $x$-axis (i.e., real number axis). This means that we can express these roots as an $n$-dimensional vector whose elements are the roots of $X^n + 1$, such that its second-half elements are a reverse-ordered conjugate of the first-half elements. Based on this vector design, the $\sigma_c$ mapping can be re-written as follows:

$\sigma_c(f(X)) = (f(\omega),f(\omega^3),f(\omega^5), \cdots, f(\omega^{n-1}), f(\overline{\omega^{n-1}}), f(\overline{\omega^{n-3}}), \cdots, f(\overline{\omega^3}), f(\overline{\omega}))$

$ $

Since $f(\overline{X}) = \overline{f(X)}$ (because $f(X)$ has strictly real coefficients), we can rewrite $\sigma_c$ as: 

$\sigma_c(f(X)) = (f(\omega),f(\omega^3),f(\omega^5), \cdots, f(\omega^{n-3}), f(\omega^{n-1}), \overline{f(\omega^{n-1})}, \overline{f(\omega^{n-3})}, \cdots, \overline{f(\omega^3)}, \overline{f(\omega)})$

$ $

This structure of vector exactly aligns with the definition of $\hat{\mathbb{C}}^n$: the second half of the elements of the $n$-dimensional vector $\vec{\hat v}$ is a reverse-ordered conjugate of the first half.

For bijectiveness, we also need to demonstrate that every $f(X) \in \mathbb{R}[X] / (X^n + 1)$ is mapped to some $\vec{\hat v} \in \hat{\mathbb{C}}^{n}$, and no two different $f_1(X), f_2(X) \in \mathbb{R}[X] / (X^n + 1)$ map to the same $\vec{\hat v} \in \hat{\mathbb{C}}^{n}$. The first requirement is satisfied because each polynomial $f(X) \in \mathbb{R}[X] / (X^n + 1)$ can be evaluated at the $n$ distinct roots of $X^n + 1$ to a valid number. The second requirement is also satisfied because in the $(n-1)$-degree polynomial ring, each list of $n$ distinct $(x, y)$ coordinates (where we fix the $X$ values as the $n$ distinct roots of $X^n + 1$ as $\{\omega, \omega^3, \cdots, \omega^{2n - 1}\}$) can be mapped only to a single polynomial within the $(n-1)$-degree polynomial ring, as proved by Lagrange Polynomial Interpolation (Theorem~\ref*{sec:polynomial-interpolation} in \autoref{sec:polynomial-interpolation}).

$ $

\para{Homomorphic:} $\sigma_c$ is homomorphic, because based on the reasoning shown in \autoref{subsec:poly-vector-transformation}, the relations $\sigma_c(f_a(X) + f_b(X)) = \sigma_c(f_a(X)) + \sigma_c(f_b(X))$ and $\sigma_c(f_a(X) \cdot f_b(X)) = \sigma_c(f_a(X)) \odot \sigma_c(f_b(X))$ mathematically hold regardless of whether the type of $X$ is a modulo integer or a complex number. Additionally, for any real scalar $c \in \mathbb{R}$, $\sigma_c(c \cdot f(X)) = c \cdot \sigma_c(f(X))$ holds because scaling the polynomial and then evaluating it is the same as evaluating it and then scaling the result.

$ $

Since $\sigma_c$ is both bijective and homomorphic over the $(+, \cdot)$ operations, it is an isomorphism.

\begin{tcolorbox}[title={\textbf{\tboxtheorem{\ref*{subsec:poly-vector-transformation-complex}} Isomorphism between Polynomials and Vectors over Complex Numbers}}]

The following mapping $\sigma_c$ between polynomials and vectors over complex numbers is an isomorphism:

$\sigma_c: f(X) \in \mathbb{R}[X]/(X^n + 1) \longrightarrow (f(\omega),f(\omega^3),f(\omega^5), \cdots, f(\omega^{2n-1})) \in \hat{\mathbb{C}}^{n} \text{ } (\cong \mathbb{C}^{\frac{n}{2}})$

$ $

, where $\omega = e^{i\pi/n}$, the root (i.e., the primitive $(\mu=2n)$-th root) of the cyclotomic polynomial $X^n + 1$ over complex numbers, and $\hat{\mathbb{C}}^{n}$ is an $n$-dimensional vector space of complex vectors (with real scalars) whose second-half elements are reverse-ordered conjugates of the first-half elements. 

\end{tcolorbox}

\subsection{Transforming Basis between Polynomial Ring and Vector Space of Complex Vectors with Real Scalars}
\label{subsec:poly-vector-basis-transfer}

Suppose some polynomials $f_0(X), f_1(X), \cdots, f_{n-1}(X)$ form a basis of the $(n-1)$-degree polynomial ring and $\sigma$ is an isomorphic mapping from the $(n-1)$-degree polynomial ring to the $n$-dimensional vector space. Then, $(\sigma(f_0(X)), \sigma(f_1(X)), \cdots, \sigma(f_{n-1}(X)))$ form a basis of the $n$-dimensional vector space. This is because the $\sigma$-mapped output vectors homomorphically preserve the same algebraic relationships on the $(+, \cdot)$ operations and the basis relationship between basis vectors and a subspace can be expressed as a linear algebraic formula consisting of the $(+, \cdot)$ operations (i.e., linear independence and spanning of the space). Therefore, if a set of polynomials satisfies a basis relationship, their $\sigma$-mapped vectors also preserve a basis relationship.

The same principle holds between a polynomial ring and vector space of complex numbers (with real scalars). Given the polynomial ring $\mathbb{R}[X]/(x^n + 1)$, the most intuitive way to set up a basis of $\mathbb{R}[X]/(x^n + 1)$ is as follows:

$f_0(X) = 1$

$f_1(X) = X$

$f_2(X) = X^2$

$\vdots$

$f_{n-1}(X) = X^{n-1}$

$ $

These $n$ polynomials are linearly independent, because each polynomial exclusively has its own unique exponent term, whereas one term cannot be expressed by a linear combination of the other terms. Also, these $n$ polynomials span the polynomial ring $\mathbb{R}[X]/(x^n + 1)$, because each polynomial's scalar multiplication can express any coefficient value of its own exponent term, and summing all such polynomials can express any polynomial in the polynomial ring $\mathbb{R}[X]/(X^n + 1)$.

Now, we will apply the $\sigma_c$ mapping to the above $n$ polynomials that are a basis of the $(n-1)$-degree polynomial ring $\mathbb{R}[X]/(x^n + 1)$. Then, according to the principle of polynomial-to-vector basis transfer (explained in Theorem~\ref*{subsec:polynomial-ring-basis} in \autoref{subsec:polynomial-ring-basis}), we can use these $n$ polynomials (i.e., the basis of the $(n-1)$-degree polynomial ring) and the isomorphic polynomial-to-vector mapping $\sigma_c$ to compute the basis of the $n$-dimensional special vector space $\hat{\mathbb C}^{n}$ as follows: 

$W = \begin{bmatrix}
\sigma_c(f_0(X)) \\
\sigma_c(f_1(X)) \\
\sigma_c(f_2(X)) \\
\vdots \\
\sigma_c(f_{n-1}(X)) \\
\end{bmatrix}
= \begin{bmatrix}
\sigma_c(1) \\
\sigma_c(X) \\
\sigma_c(X^2) \\
\vdots \\
\sigma_c(X^{n-1}) \\
\end{bmatrix}$
$= \begin{bmatrix}
1 & 1 & \cdots & 1 & 1\\
(\omega) & (\omega^3) & (\omega^5) & \cdots & (\omega^{2n-1})\\
(\omega)^2 & (\omega^3)^2 & (\omega^5)^2 & \cdots & (\omega^{2n-1})^2\\
\vdots & \vdots & \vdots & \ddots & \vdots \\
(\omega)^{n-1} & (\omega^3)^{n-1} & (\omega^5)^{n-1} & \cdots & (\omega^{2n-1})^{n-1}\\
\end{bmatrix}$

\text{ } \text{ }  $= \begin{bmatrix}
1 & 1 & 1 & \cdots & 1\\
(\omega) & (\omega^3) & \cdots & (\overline{\omega})^3 & (\overline{\omega})\\
(\omega)^2 & (\omega^3)^2 & \cdots & (\overline{\omega}^3)^2 & (\overline{\omega})^2\\
\vdots & \vdots & \vdots & \ddots & \vdots \\
(\omega)^{n-1} & (\omega^3)^{n-1} & \cdots & (\overline{\omega}^3)^{n-1} & (\overline{\omega})^{n-1}\\
\end{bmatrix}$

$ $

$W$ is a valid basis of the $n$-dimensional special vector space $\hat{\mathbb C}^{n}$.

\begin{tcolorbox}[title={\textbf{\tboxtheorem{\ref*{subsec:poly-vector-transformation-complex}} Transforming Basis between Polynomial Ring and Vector Space}}]

If $n$ polynomials form a basis of an $(n-1)$-degree polynomial ring and they are converted into $n$ distinct vectors via an isomorphic mapping $\sigma$ (or $\sigma_c$ in the case of the complex number domain) from the $(n-1)$-degree polynomial ring to the $n$-dimensional vector space, then those converted $n$ vectors form a basis of the $n$-dimensional vector space. 

\end{tcolorbox}

\clearpage

\section{Euler's Formula}
\label{sec:euler}

\subsection{Pythagorean Identity}
\label{subsec:pythagorean}

\begin{figure}[!h]
    \centering
    \subfloat[Pythagorean Theorem \href{https://socratic.org/questions/58a20042b72cff17846a6feb}{(Source)}]{
  \includegraphics[width=0.3\linewidth]{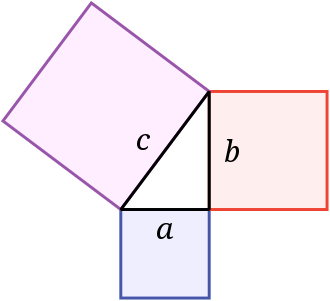}

  \label{fig:pythagorean-theorem}
}
    \subfloat[Pythagorean Identity \href{https://socratic.org/questions/58a20042b72cff17846a6feb}{(Source)}]{
  \includegraphics[width=0.3\linewidth]{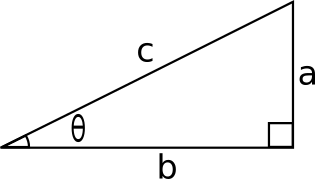}

  \label{fig:pythagorean-identity}
}
  \caption{}
\end{figure}

\begin{tcolorbox}[title={\textbf{\tboxtheorem{\ref*{subsec:pythagorean}} Pythagorean Identity}}]
$\sin^2 \theta + \cos^2 \theta = 1$
\end{tcolorbox}

\begin{proof}
$ $
\begin{enumerate}

\item In \autoref{fig:pythagorean-theorem}, by Pythagorean theorem, $a^2 + b^2 = c^2$.

\item In \autoref{fig:pythagorean-identity}, by definition of sin and cosine,

$\sin\theta = \dfrac{a}{c}, \text{ } \cos\theta = \dfrac{b}{c}$

\item Combining step 1 and 2:

$\sin^2 \theta + \cos^2 \theta = \dfrac{a^2}{c^2} + \dfrac{b^2}{c^2} = \dfrac{a^2 + b^2}{c^2} = 1$

\end{enumerate}
\end{proof}

\subsection{Imaginary Number}n
\label{subsec:imaginary}

\begin{tcolorbox}[title={\textbf{\tboxdef{\ref*{subsec:imaginary}} Imaginary Number}}]

\begin{itemize}

\item \boldmath{$i$} is defined to be an imaginary number that has the property: $i^2 =-1$
\item \textbf{Complex Number} is a number of the form $a + b\,i$, where $a$ and $b$ are real numbers (e.g., $a = 5.6 + 4.3i)$

\item \textbf{Real Number} is a number that does not involve any imaginary number (e.g., $a = 13.4$)

\item $\overline{a}$ is a \textbf{Conjugate} of $a$ if $a$ and $\overline{a}$ have the same real number part and an opposite-signed imaginary number part

(e.g., $a = 3 + i\cdot3.4$, \text{ } $\overline{a} = 3 - i\cdot3.4$)

\item \textbf{Hermitian Vector} is a vector where the 2nd half of its elements is the complex conjugate of the 1st half in reverse order, as illustrated by the $n$-dimensional vector below:

$\vec{v} = (v_1, v_2, v_3, \cdots, v_{\frac{n}{2}-1}, v_{\frac{n}{2}}, \overline{v}_{\frac{n}{2}}, \overline{v}_{\frac{n}{2} - 1}, \cdots, \overline{v}_3, \overline{v}_2, \overline{v}_1)$

\end{itemize}
\end{tcolorbox}

\subsection{Euler's Formula}
\label{subsec:euler}

\begin{tcolorbox}[title={\textbf{\tboxdef{\ref*{subsec:euler}} Euler's Formula}}]

$e^{i\theta} = \cos\theta + i\cdot\sin\theta$

\end{tcolorbox}

\begin{figure}[!h]
    \centering
  \includegraphics[width=0.4\linewidth]{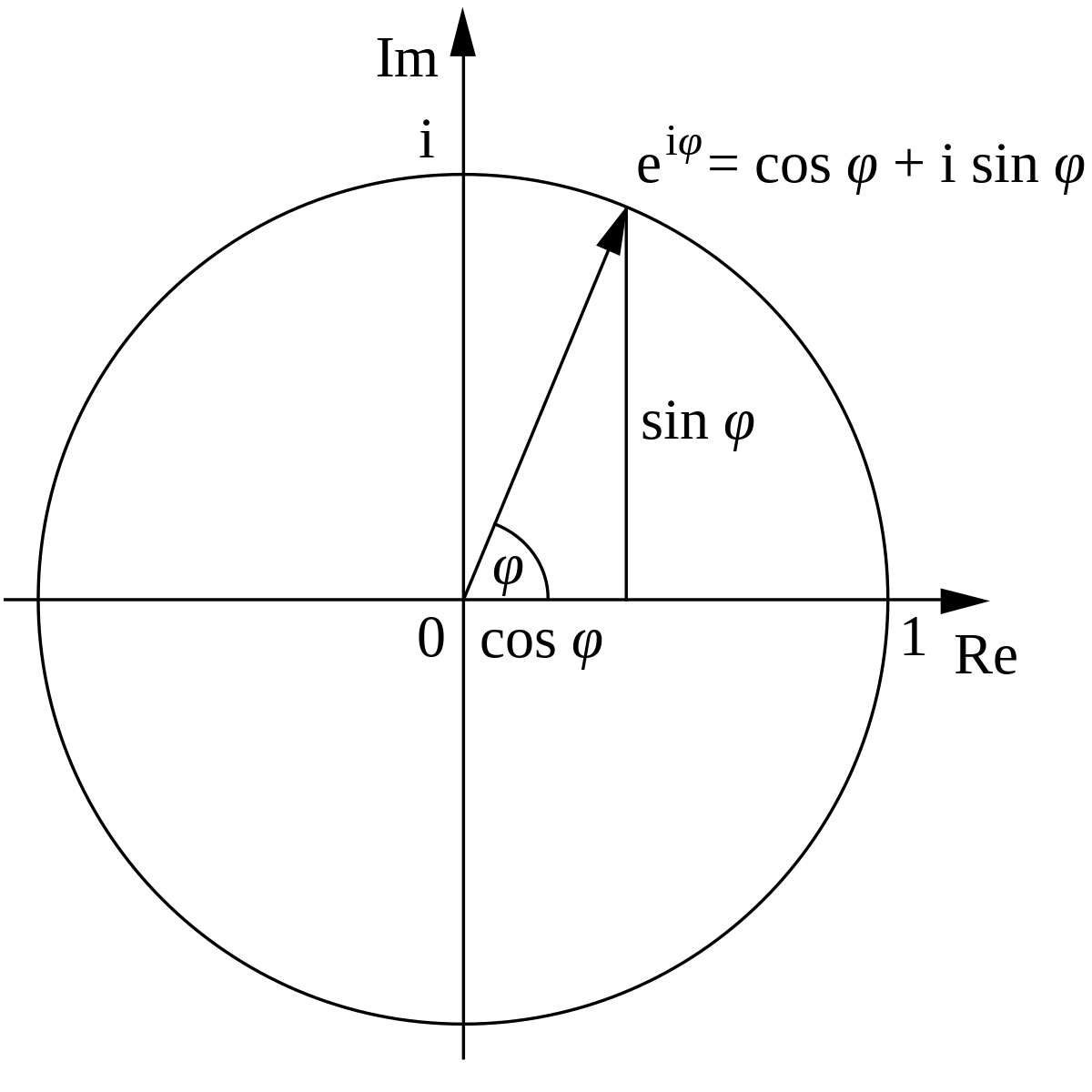}
  \caption{The figure illustrates Euler's formula on the unit circle in the complex plane \href{https://en.wikipedia.org/wiki/Euler's_formula}{(Source)}}
\end{figure}

The value of $e^{i\theta}$ is represented as a coordinate on a circle in the complex plane in \autoref{fig:complex-plane}, where the $x$-axis encodes the value's real number part and the $y$-axis encodes the value's imaginary number part. Note that as $\theta$ increases, the imaginary part ($\sin\theta$) oscillates between $1$ and $-1$ (reaching $i$ and $-i$ on the imaginary axis), and the real part ($\cos\theta$) oscillates between $1$ and $-1$, with period $2\pi$.

\subsection{Vandermonde Matrix with Roots of Cyclotomic Polynomial over Complex Numbers}
\label{subsec:vandermonde-euler}

In this subsection, we will build a Vandermonde matrix (\autoref{subsec:vandermonde}) with the $n$ distinct roots of the $\mu$-th cyclotomic polynomial over complex numbers (where $\mu$ is a power of 2) as follows:

\begin{tcolorbox}[title={\textbf{\tboxtheorem{\ref*{subsec:vandermonde-euler}} Vandermonde Matrix with the Roots of  \text{(power-of-2)}-th Cyclotomic Polynomial over Complex Numbers}}]

Suppose we have an $n \times n$ (where $n$ is a power of 2) Vandermonde matrix comprised of $n$ distinct roots of the $\mu$-th cyclotomic polynomial (explained in Theorem~\ref*{subsec:cyclotomic-theorem}.1 in \autoref{subsec:cyclotomic-theorem}), where $\mu$ is a power of 2 and $n = \dfrac{\mu}{2}$. In other words, $V = \mathit{Vander}(x_0, x_1, \cdots, x_{n-1})$, where each $x_j = (e^{i\pi/n})^{2j-1}$ for $1 \leq j \leq n$ (i.e., the primitive $\mu$-th roots of unity). Then, the following holds:

$V \cdot V^T = \begin{bmatrix}
0 & \cdots & 0 & 0 & n\\
0 & \cdots & 0 & n & 0\\
0 & \cdots & n & 0 & 0\\
\vdots & \iddots & \vdots & \vdots & \vdots \\
n & 0 & 0 & \cdots & 0\\
\end{bmatrix} = n \cdot I^R_n$

$ $

And $V^{-1} = \dfrac{V^T \cdot I_n^R}{n}$

\end{tcolorbox}

\begin{proof}

$ $

\begin{enumerate}

\item Given $\omega = e^{i\pi/n}$, each $x_j = (\omega)^{2j-1}$. Thus, we can expand as follows:

$V \cdot V^T = \begin{bmatrix}
1 & (\omega) & (\omega)^2 & \cdots & (\omega)^{n-1}\\
1 & (\omega^3) & (\omega^3)^2 & \cdots & (\omega^3)^{n-1}\\
1 & (\omega^5) & (\omega^5)^2 & \cdots & (\omega^5)^{n-1}\\
\vdots & \vdots & \vdots & \ddots & \vdots \\
1 & (\omega^{2n-1}) & (\omega^{2n-1})^2 & \cdots & (\omega^{2n-1})^{n-1}\\
\end{bmatrix} 
\cdot 
\begin{bmatrix}
1 & 1 & 1 & \cdots & 1\\
(\omega) & (\omega^3) & (\omega^5) & \cdots & (\omega^{2n-1})\\
(\omega)^2 & (\omega^3)^2 & (\omega^5)^2 & \cdots & (\omega^{2n-1})^2\\
\vdots & \vdots & \vdots & \ddots & \vdots \\
(\omega)^{n-1} & (\omega^3)^{n-1} & (\omega^5)^{n-1} & \cdots & (\omega^{2n-1})^{n-1}\\
\end{bmatrix} $

$ $

$=
\begin{bmatrix}
\sum\limits_{k=0}^{n-1} \omega^{2k} & \sum\limits_{k=0}^{n-1} \omega^{4k}  & \sum\limits_{k=0}^{n-1} \omega^{6k} & \cdots & \sum\limits_{k=0}^{n-1} \omega^{2nk} \\

\sum\limits_{k=0}^{n-1} \omega^{4k} & \sum\limits_{k=0}^{n-1} \omega^{6k}  & \sum\limits_{k=0}^{n-1} \omega^{8k} & \cdots & \sum\limits_{k=0}^{n-1} \omega^{2k(n+1)} \\

\sum\limits_{k=0}^{n-1} \omega^{6k} & \sum\limits_{k=0}^{n-1} \omega^{8k} & \sum\limits_{k=0}^{n-1} \omega^{10k} & \cdots & \sum\limits_{k=0}^{n-1} \omega^{2k(n+2)} \\

\vdots & \vdots & \vdots & \ddots & \vdots \\
\sum\limits_{k=0}^{n-1} \omega^{2nk} & \sum\limits_{k=0}^{n-1} \omega^{2(n+1)k} & \sum\limits_{k=0}^{n-1} \omega^{2(n+2)k} & \cdots & \sum\limits_{k=0}^{n-1} \omega^{2(n+n-1)k} \\

\end{bmatrix}$

$ $

\item The $V \cdot V^T$ matrix's anti-diagonal elements are $\sum\limits_{k=0}^{n-1} \omega^{2nk}$. We can derive the following:

$\sum\limits_{k=0}^{n-1} \omega^{2nk} = \sum\limits_{k=0}^{n-1} (e^{i\pi/n})^{2nk} = \sum\limits_{k=0}^{n-1}e^{2\pi k i} = \sum\limits_{k=0}^{n-1}(\cos(2\pi k) + i\sin(2\pi k)) = \sum\limits_{k=0}^{n-1} (1 + 0) = n$

This means that The $V \cdot V^T$ matrix's anti-diagonal elements are $n$. 

$ $

\item Next, we will prove that the $V \cdot V^T$ matrix has 0 for all positions except for the anti-diagonal ones. In other words, we will prove the following: 

$\sum\limits_{k=0}^{n-1} \omega^{2k} = \sum\limits_{k=0}^{n-1} \omega^{4k} = \sum\limits_{k=0}^{n-1} \omega^{6k} = \cdots = \sum\limits_{k=0}^{n-1} \omega^{2(n-1)k} = \sum\limits_{k=0}^{n-1} \omega^{2(n+1)k} = \cdots = \sum\limits_{k=0}^{n-1} \omega^{2(2n-1)k} = 0$

For this proof, we will leverage the Geometric Sum formula $\sum\limits_{i=0}^{n-1}x^i = \dfrac{x^n - 1}{x - 1}$:

\begin{tcolorbox}[title={\textbf{\tboxtheorem{\ref*{subsec:vandermonde-euler}.1} Geometric Sum Formula}}]

Let the geometric sum $S_n = 1 + x + x^2 + \cdots + x^{n - 1}$

Then, $x \cdot S_n = x + x^2 + x^3 + \cdots + x^{n}$

$x \cdot S_n - S_n = (x + x^2 + x^3 + \cdots + x^{n}) - (1 + x + x^2 + \cdots + x^{n - 1}) = x^n - 1$

$S_n\cdot (x - 1) = x^n - 1$

$S_n = \dfrac{x^n - 1}{x - 1}$ \textcolor{red}{ $\rhd$ with the constraint that $x \neq 1$}

\end{tcolorbox}

Leveraging the Geometric Sum formula $\sum\limits_{i=0}^{n-1}x^i = \dfrac{x^n - 1}{x - 1}$, 

$\sum\limits_{k=0}^{n-1} \omega^{2mk} = \dfrac{(\omega^{2m})^n - 1}{\omega^{2m} - 1} = \dfrac{(\omega^{2n})^{m} - 1}{\omega^{2m} - 1} = \dfrac{1 - 1}{\omega^{2m} - 1} = 0$ for $1 \leq m \leq (2n - 1)$  \textcolor{red}{ $\rhd$ since $\textsf{Ord}(\omega) = 2n$}

$ $

Therefore,

$\sum\limits_{k=0}^{n-1} \omega^{2k} = \sum\limits_{k=0}^{n-1} \omega^{4k} = \sum\limits_{k=0}^{n-1} \omega^{6k} = \cdots = \sum\limits_{k=0}^{n-1} \omega^{2(n-1)k} = \sum\limits_{k=0}^{n-1} \omega^{2(n+1)k} = \cdots = \sum\limits_{k=0}^{n-1} \omega^{2(2n-1)k} = 0$

$ $

\item Based on the proof of step 2 and 3, $V \cdot V^T = 
\begin{bmatrix}
0 & \cdots & 0 & 0 & n\\
0 & \cdots & 0 & n & 0\\
0 & \cdots & n & 0 & 0\\
\vdots & \iddots & \vdots & \vdots & \vdots \\
n & 0 & 0 & \cdots & 0\\
\end{bmatrix} = n \cdot I_n^R$

\item Given $V \cdot V^T = n \cdot I_n^R$, 

$V^{-1} \cdot V \cdot V^T = V^{-1} \cdot n \cdot I_n^R$

$V^T = V^{-1} \cdot n \cdot I_n^R$

$V^T \cdot I_n^R = V^{-1} \cdot n \cdot I_n^R  \cdot I_n^R$

$V^T \cdot I_n^R = V^{-1} \cdot n$ \textcolor{red}{\text{ } \# since $I_n^R  \cdot I_n^R = I_n$}

$V^{-1} = \dfrac{V^T \cdot I_n^R}{n}$

\end{enumerate}
\end{proof}

Later in the CKKS scheme (\autoref{sec:ckks}), we will use $V^{-1}$ to encode a complex vector into a real number vector, and $V^T$ to decode a real number vector into a complex vector (\autoref{subsec:ckks-enc-dec}).

$ $

%\para{Condition for $\bm \mu$:} It's worthwhile to note that the property $V\cdot V^T = n\cdot I_n^R$ does not hold if $\mu$ (denoting the $\mu$-th cyclotomic polynomial) is not a power of 2. In particular, step 3 of the proof does not hold anymore if $\mu$ is not a power of 2:

%$\sum\limits_{k=0}^{n-1} \omega^{2k} \neq \sum\limits_{k=0}^{n-1} \omega^{4k} \neq \sum\limits_{k=0}^{n-1} \omega^{6k} \neq \cdots \neq \sum\limits_{k=0}^{n-1} \omega^{2(n-1)k} \neq \sum\limits_{k=0}^{n-1} \omega^{2(n+1)k} \neq \cdots \neq \sum\limits_{k=0}^{n-1} \omega^{2(2n-1)k} \neq 0$

\para{Condition for $\bm \mu$:} The identity $V \cdot V^T = n \cdot I_n^R$ relies only on
$\textsf{Ord}(\omega) = 2n$ (step 3), so it holds for any $\mu = 2n$ (i.e., any even number).
The nodes $\omega, \omega^3, \ldots, \omega^{2n-1}$ are likewise the $n$ roots of $X^n + 1$
for any $n$, since $(\omega^{2j-1})^n = \omega^{2nj}\cdot\omega^{-n} = 1\cdot(-1) = -1$.

Nevertheless, the reason we take $\mu$ to be a power of 2 does not come from this theorem, but
from the FHE schemes themselves. In practice, CKKS, BFV, and BGV are most commonly instantiated
over the power-of-2 cyclotomic ring $\mathbb{Z}[X]/(X^n + 1)$ with $n$ a power of 2, because this choice makes $X^n + 1 = \Phi_{2n}(X)$ irreducible over rational numbers (which holds
precisely when $n$ is a power of 2), and also enables efficient negacyclic-NTT multiplication
and a clean Ring-LWE security reduction. Once
$n$ (hence $\mu = 2n$) is fixed as a power of 2, the nodes are automatically the primitive
$\mu$-th roots of unity: by Theorem~\ref*{subsec:roots-theorem}.4 (\autoref{subsec:roots-theorem}),
$\omega^k$ is a primitive $\mu$-th root iff $\gcd(\mu, k) = 1$, and every odd exponent
$1, 3, \ldots, 2n-1$ is coprime to $\mu = 2n$ iff $2n$ has no odd prime factor---i.e.\ iff $2n$
is a power of 2. So $\{\omega, \omega^3, \ldots, \omega^{2n-1}\}$ being exactly the primitive
$(\mu=2n)$-th roots is a consequence of the power-of-2 ring choice. Meanwhile, as far as $\mu$ is an even number, $V \cdot V^T = n \cdot I_n^R$ still holds.

\subsection{Vandermonde Matrix with Roots of Cyclotomic Polynomial over Ring ($\mathbb{Z}_p$)}
\label{subsec:vandermonde-euler-integer-ring}

Theorem~\ref*{subsec:vandermonde-euler} (in \autoref{subsec:vandermonde-euler}) showed that $V \cdot V^T = n \cdot I^R_n$, where $V$ is the Vandermonde matrix $V = \mathit{Vander}(x_0, x_1, \cdots, x_{n-1})$, where each $x_i$ is the primitive $\mu$-th root of unity over $X \in \mathbb{C}$ (i.e., complex number) and $\mu$ is a power of 2. In this subsection, we will show that the relation $V \cdot V^T = n \cdot I^R_n$ holds even if each $x_i$ is the primitive $\mu$-th root of unity over $X \in \mathbb{Z}_p$ (i.e., ring). In particular, we will prove Theorem~\ref*{subsec:vandermonde-euler}:

\begin{tcolorbox}[title={\textbf{\tboxtheorem{\ref*{subsec:vandermonde-euler-integer-ring}} Vandermonde Matrix with Roots of  \text{(power-of-2)}-th Cyclotomic Polynomial over Ring ($\mathbb{Z}_p$)}}]

The proof takes the same format as that of Theorem~\ref*{subsec:vandermonde-euler} (in \autoref{subsec:vandermonde-euler}). Suppose we have an $n \times n$ (where $n$ is a power of 2) Vandermonde matrix comprising $n$ distinct roots of the $\mu$-th cyclotomic polynomial over $X \in \mathbb{Z}_p$ (ring), where $\mu$ is a power of 2 and $n = \dfrac{\mu}{2}$. In other words, $V = \mathit{Vander}(x_0, x_1, \cdots, x_{n-1})$, where each $x_i$ is the root of $X^n + 1$ (i.e., the primitive $(\mu=2n)$-th roots of unity). Then, the following holds:

$V \cdot V^T = \begin{bmatrix}
0 & \cdots & 0 & 0 & n\\
0 & \cdots & 0 & n & 0\\
0 & \cdots & n & 0 & 0\\
\vdots & \iddots & \vdots & \vdots & \vdots \\
n & 0 & 0 & \cdots & 0\\
\end{bmatrix} = n \cdot I^R_n$

$ $

And $V^{-1} = n^{-1}\cdot V^T \cdot I_n^R$

\end{tcolorbox}
\begin{proof}
$ $
\begin{enumerate}
\item $V \cdot V^T$ is expanded as follows:

$V \cdot V^T = \begin{bmatrix}
1 & (\omega) & (\omega)^2 & \cdots & (\omega)^{n-1}\\
1 & (\omega^3) & (\omega^3)^2 & \cdots & (\omega^3)^{n-1}\\
1 & (\omega^5) & (\omega^5)^2 & \cdots & (\omega^5)^{n-1}\\
\vdots & \vdots & \vdots & \ddots & \vdots \\
1 & (\omega^{2n-1}) & (\omega^{2n-1})^2 & \cdots & (\omega^{2n-1})^{n-1}\\
\end{bmatrix} 
\cdot 
\begin{bmatrix}
1 & 1 & 1 & \cdots & 1\\
(\omega) & (\omega^3) & (\omega^5) & \cdots & (\omega^{2n-1})\\
(\omega)^2 & (\omega^3)^2 & (\omega^5)^2 & \cdots & (\omega^{2n-1})^2\\
\vdots & \vdots & \vdots & \ddots & \vdots \\
(\omega)^{n-1} & (\omega^3)^{n-1} & (\omega^5)^{n-1} & \cdots & (\omega^{2n-1})^{n-1}\\
\end{bmatrix} $

$ $

$=
\begin{bmatrix}
\sum\limits_{k=0}^{n-1} \omega^{2k} & \sum\limits_{k=0}^{n-1} \omega^{4k}  & \sum\limits_{k=0}^{n-1} \omega^{6k} & \cdots & \sum\limits_{k=0}^{n-1} \omega^{2nk} \\

\sum\limits_{k=0}^{n-1} \omega^{4k} & \sum\limits_{k=0}^{n-1} \omega^{6k}  & \sum\limits_{k=0}^{n-1} \omega^{8k} & \cdots & \sum\limits_{k=0}^{n-1} \omega^{2k(n+1)} \\

\sum\limits_{k=0}^{n-1} \omega^{6k} & \sum\limits_{k=0}^{n-1} \omega^{8k} & \sum\limits_{k=0}^{n-1} \omega^{10k} & \cdots & \sum\limits_{k=0}^{n-1} \omega^{2k(n+2)} \\

\vdots & \vdots & \vdots & \ddots & \vdots \\
\sum\limits_{k=0}^{n-1} \omega^{2nk} & \sum\limits_{k=0}^{n-1} \omega^{2(n+1)k} & \sum\limits_{k=0}^{n-1} \omega^{2(n+2)k} & \cdots & \sum\limits_{k=0}^{n-1} \omega^{2(n+n-1)k} \\

\end{bmatrix}$

$ $

, where $\omega$ (i.e., the primitive $(\mu=2n)$-th root of unity) has the order $2n$. 

$ $

\item Note that the $V \cdot V^T$ matrix's anti-diagonal elements are $\sum\limits_{k=0}^{n-1} \omega^{2nk}$. It can be seen that $\omega^{2n} \equiv 1 \bmod p$, because $\textsf{Ord}_p(\omega) = 2n$. Thus, the $V \cdot V^T$ matrix's every anti-diagonal element is $\sum\limits_{k=0}^{n-1} 1 = n$.

$ $

\item Next, we will prove that the $V \cdot V^T$ matrix has $0$ for all other positions than the anti-diagonal ones. In other words, we will prove the following: 

$\sum\limits_{k=0}^{n-1} \omega^{2k} = \sum\limits_{k=0}^{n-1} \omega^{4k} = \sum\limits_{k=0}^{n-1} \omega^{6k} = \cdots = \sum\limits_{k=0}^{n-1} \omega^{2(n-1)k} = \sum\limits_{k=0}^{n-1} \omega^{2(n+1)k} = \cdots = \sum\limits_{k=0}^{n-1} \omega^{2(2n-1)k} = 0$

$ $

The above is true by the Geometric Sum formula (Theorem~\ref*{subsec:vandermonde-euler}.1). As shown in the proof step 3 of Theorem~\ref*{subsec:vandermonde-euler}, each element is of the form $\sum_{k=0}^{n-1} (\omega^{2m})^k$ for some integer $m$ (where $2m$ is not a multiple of $2n$). Let $r = \omega^{2m}$. Then:

$\sum\limits_{k=0}^{n-1} r^k = \dfrac{r^n - 1}{r - 1} = \dfrac{(\omega^{2m})^n - 1}{\omega^{2m} - 1} = \dfrac{(\omega^n)^{2m} - 1}{\omega^{2m} - 1}$

$ $

Since $\omega$ is a root of $X^n + 1$, we know $\omega^n \equiv -1 \bmod p$. Thus:

$\dfrac{(-1)^{2m} - 1}{\omega^{2m} - 1} = \dfrac{1 - 1}{\omega^{2m} - 1} = 0$

$ $

Therefore, the sum is 0 for all off-anti-diagonal positions.

$ $

\item According to step 1 and 2, the $V \cdot V^T$ matrix has $n$ on its anti-diagonal positions and $0$ for all other positions.

\item Now we will derive the formula for $V^{-1}$. Given $V \cdot V^T = n \cdot I_n^R$, 

$V^{-1} \cdot V \cdot V^T = V^{-1} \cdot n \cdot I_n^R$

$V^T = V^{-1} \cdot n \cdot I_n^R$

$V^T \cdot I_n^R = V^{-1} \cdot n \cdot I_n^R  \cdot I_n^R$

$V^T \cdot I_n^R = V^{-1} \cdot n$ \textcolor{red}{\text{ } \# since $I_n^R  \cdot I_n^R = I_n$}

$ $

Now, there is one caveat: modulo operation does not support direct number division (as explained in \autoref{subsec:modulo-division}). This means that the formula $V^{-1} = \dfrac{V^T \cdot I_n^R}{n}$ in Theorem~\ref*{subsec:vandermonde-euler} (in \autoref{subsec:vandermonde-euler}) is inapplicable in our case, because our modulo $p$ arithmetic does not allow direct division of $V^T \cdot I_n^R$ by $n$. Therefore, we instead multiply $V^T \cdot I_n^R$ by the inverse of $n$ (i.e., $n^{-1}$). We continue as follows:

$V^T \cdot I_n^R = V^{-1} \cdot n$

$V^T \cdot I_n^R \cdot n^{-1}= V^{-1} \cdot n \cdot n^{-1}$

$V^{-1} = n^{-1}\cdot V^T \cdot I_n^R$

\end{enumerate}
\end{proof}

We finally proved that $V\cdot V^T = n \cdot I_n^R$, and $V^{-1} = n^{-1}\cdot V^T \cdot I_n^R$. Later in the BFV scheme (\autoref{sec:bfv}), we will use $V^{-1}$ to encode an integer vector into a vector of polynomial coefficients, and $V^T$ to decode it back to the integer vector (\autoref{subsec:bfv-batch-encoding}).

$ $

\para{Condition for $\bm \mu$:} Like in CKKS, it's worthwhile to note that the property $V\cdot V^T = n\cdot I_n^R$ does not hold if $\mu$ (denoting the $\mu$-th cyclotomic polynomial) is not a power of 2. In particular, step 3 of the proof does not hold anymore if $\mu$ is not a power of 2:

$\sum\limits_{k=0}^{n-1} \omega^{2k} \neq \sum\limits_{k=0}^{n-1} \omega^{4k} \neq \sum\limits_{k=0}^{n-1} \omega^{6k} \neq \cdots \neq \sum\limits_{k=0}^{n-1} \omega^{2(n-1)k} \neq \sum\limits_{k=0}^{n-1} \omega^{2(n+1)k} \neq \cdots \neq \sum\limits_{k=0}^{n-1} \omega^{2(2n-1)k} \neq 0$

\clearpage

\section{Modulo Rescaling}
\label{sec:modulus-rescaling}

\subsection{Rescaling Modulo of Congruence Relations}
\label{subsec:modulo-rescaling}

Remember from \autoref{sec:modulo} that $a \bmod q$ is the remainder of $a$ divided by $q$, and the congruence relation $a \equiv b \bmod q$ means that the remainder of $a$ divided by $q$ is the same as the remainder of $b$ divided by $q$. Its equivalent numeric equation is  $a = b + k\cdot q$, meaning that $a$ and $b$ differ by some multiple of $q$. The congruence and equation are two different ways of describing the relationship between two numbers $a$ and $b$. 

In this section, we introduce another way of describing the relationship between numbers. We will describe two numbers $a$ and $b$ in terms of a different modulo $q'$ instead of the original modulo $q$. Such a change of modulo in a congruence relation is called modulo scaling. When we rescale the modulo of a congruence relation, we also need to rescale the numbers involved in the congruence relation. 

Suppose we have the following congruence relations: 

$a \equiv b \bmod q$

$a + c \equiv b + d\bmod q$

$a \cdot c \equiv b \cdot d\bmod q$

Now, suppose we want to rescale the modulo of the above congruence relations from $q \rightarrow q'$, where $q' \mid q$ (meaning $q$ is a multiple of $q'$). Then, the accordingly updated congruence relations are as shown in \autoref{tab:rescaling}.

\begin{table}[h!]
{
\centering
\begin{tabular}{|l|l|l|}
\hline
\textbf{Congruence}  & \textbf{Rescaled Congruence Relation} & \textbf{Rescaled Congruence Relation}\\
\textbf{Relation} & \textbf{-- Exact} & \textbf{-- Approximate} \\
\hline\hline
$a \equiv b \bmod q$ & $\Bigg\lceil a\dfrac{q'}{q}\Bigg\rfloor \equiv \Bigg\lceil b\dfrac{q'}{q}\Bigg\rfloor \bmod q'$ & $\Bigg\lceil a\dfrac{q'}{q}\Bigg\rfloor \cong \Bigg\lceil b\dfrac{q'}{q}\Bigg\rfloor \bmod q'$\\
&(if $q$ divides both $aq'$ and $bq'$)&(if $q$ does not divide either $aq' \text{ or } bq'$)\\
\hline
$a + c \equiv b + d \bmod q$ & $\Bigg\lceil a\dfrac{q'}{q}\Bigg\rfloor + \Bigg\lceil c\dfrac{q'}{q}\Bigg\rfloor \equiv \Bigg\lceil b\dfrac{q'}{q}\Bigg\rfloor + \Bigg\lceil d\dfrac{q'}{q}\Bigg\rfloor \bmod q'$ & $\Bigg\lceil a\dfrac{q'}{q}\Bigg\rfloor + \Bigg\lceil c\dfrac{q'}{q}\Bigg\rfloor \cong \Bigg\lceil b\dfrac{q'}{q}\Bigg\rfloor + \Bigg\lceil d\dfrac{q'}{q}\Bigg\rfloor \bmod q'$\\
&(if $q$ divides all of $aq', bq', cq'$ and $dq'$)&(if $q$ does not divide:  $aq', bq', cq',$ or $dq'$)\\
\hline
$a \cdot c \equiv b \cdot d \bmod q$ & $\Bigg\lceil ac\dfrac{q'}{q}\Bigg\rfloor  \equiv \Bigg\lceil bd\dfrac{q'}{q}\Bigg\rfloor \bmod q'$ & $\Bigg\lceil ac\dfrac{q'}{q}\Bigg\rfloor  \cong \Bigg\lceil bd\dfrac{q'}{q}\Bigg\rfloor  \bmod q'$\\
&(if $q$ divides both $acq'$ and $bdq'$)&(if $q$ does not divide either  $acq'$ or $bdq'$)\\
\hline
\end{tabular} \par
}
\caption{Rescaling the congruence relations  from modulo $q\rightarrow q'$ (where $\lceil \rfloor$ denotes rounding to the nearest integer)}
\label{tab:rescaling}
\end{table}

\begin{proof}

$ $

\begin{enumerate}
%\item Let $\hat{a} = \Bigg\lceil a\dfrac{q'}{q}\Bigg\rfloor$, $\hat{b} = \Bigg\lceil b\dfrac{q'}{q}\Bigg\rfloor$, $\hat{c} = \Bigg\lceil a\dfrac{q'}{q}\Bigg\rfloor$, $\hat{d} = \Bigg\lceil a\dfrac{q'}{q}\Bigg\rfloor$, where 
\item $a \equiv b \bmod q$ $\Longleftrightarrow$ $a = b + q\cdot k$ (for some integer $k$)

$\Longleftrightarrow a \cdot \dfrac{q'}{q} = b \cdot \dfrac{q'}{q} + q\cdot k \cdot \dfrac{q'}{q}$

$\Longleftrightarrow a \cdot \dfrac{q'}{q} = b \cdot \dfrac{q'}{q} + k\cdot q'$

\begin{enumerate}
\item If $q$ divides both $aq'$ and $bq'$, then $a \cdot \dfrac{q'}{q} = \Bigg\lceil a\dfrac{q'}{q}\Bigg\rfloor$, and $b \cdot \dfrac{q'}{q} = \Bigg\lceil b\dfrac{q'}{q}\Bigg\rfloor$. Therefore:

$a \cdot \dfrac{q'}{q} = b \cdot \dfrac{q'}{q} + k\cdot q'$

$\Longleftrightarrow \Bigg\lceil a\dfrac{q'}{q}\Bigg\rfloor = \Bigg\lceil b\dfrac{q'}{q}\Bigg\rfloor + k\cdot q'$

$\Longleftrightarrow \Bigg\lceil a\dfrac{q'}{q}\Bigg\rfloor \equiv \Bigg\lceil b\dfrac{q'}{q}\Bigg\rfloor \bmod q'$ \text{ } $(\Longleftrightarrow a \equiv b \bmod q)$

$ $

\item If $q$ does not divide either $aq'$ or $bq'$, then $a \cdot \dfrac{q'}{q} \approx \Bigg\lceil a\dfrac{q'}{q}\Bigg\rfloor$, \text{ } $b \cdot \dfrac{q'}{q} \approx \Bigg\lceil b\dfrac{q'}{q}\Bigg\rfloor$. Therefore:

$a \cdot \dfrac{q'}{q} = b \cdot \dfrac{q'}{q} + k\cdot q'$

$\Longleftrightarrow \Bigg\lceil a\dfrac{q'}{q}\Bigg\rfloor \approx \Bigg\lceil b\dfrac{q'}{q}\Bigg\rfloor + k\cdot q'$

$\Longleftrightarrow \Bigg\lceil a\dfrac{q'}{q}\Bigg\rfloor \cong \Bigg\lceil b\dfrac{q'}{q}\Bigg\rfloor \bmod q'$ \text{ } $(\Longleftrightarrow a \equiv b \bmod q)$

\end{enumerate}

$ $

\item $a + c\equiv b + d\bmod q$ $\Longleftrightarrow$ $a + c = b + d + k\cdot q$ (for some integer $k$)

$\Longleftrightarrow a \cdot \dfrac{q'}{q} + c \cdot \dfrac{q'}{q} = b \cdot \dfrac{q'}{q} + d \cdot \dfrac{q'}{q}  + q\cdot k \cdot \dfrac{q'}{q}$

$\Longleftrightarrow a \cdot \dfrac{q'}{q} + c \cdot \dfrac{q'}{q}  = b \cdot \dfrac{q'}{q} + d \cdot \dfrac{q'}{q} + k\cdot q'$

\begin{enumerate}

\item If $q$ divides all of $aq'$, $bq'$, $cq'$, and $dq'$, then 

$a \dfrac{q'}{q} + c  \dfrac{q'}{q}  = \Bigg\lceil a\dfrac{q'}{q}\Bigg\rfloor + \Bigg\lceil c\dfrac{q'}{q}\Bigg\rfloor$, \text{ } $b \dfrac{q'}{q} + d \dfrac{q'}{q} = \Bigg\lceil b\dfrac{q'}{q}\Bigg\rfloor + \Bigg\lceil d\dfrac{q'}{q}\Bigg\rfloor$

Therefore:

$a \cdot \dfrac{q'}{q} + c \cdot \dfrac{q'}{q} = b \cdot \dfrac{q'}{q} + d \cdot \dfrac{q'}{q} + k\cdot q'$

$\Longleftrightarrow \Bigg\lceil a\dfrac{q'}{q}\Bigg\rfloor + \Bigg\lceil c\dfrac{q'}{q}\Bigg\rfloor = \Bigg\lceil b\dfrac{q'}{q}\Bigg\rfloor + \Bigg\lceil d\dfrac{q'}{q}\Bigg\rfloor + k\cdot q'$

$\Longleftrightarrow \Bigg\lceil a\dfrac{q'}{q}\Bigg\rfloor + \Bigg\lceil c\dfrac{q'}{q}\Bigg\rfloor \equiv \Bigg\lceil b\dfrac{q'}{q}\Bigg\rfloor + \Bigg\lceil d\dfrac{q'}{q}\Bigg\rfloor \bmod q'$ \text{ } $(\Longleftrightarrow a + c \equiv b + d \bmod q)$

$ $

\item If $q$ does not divide at least one of $aq'$, $bq'$, $cq'$, and $dq'$, then

$a \dfrac{q'}{q} + c  \dfrac{q'}{q}  \approx \Bigg\lceil a\dfrac{q'}{q}\Bigg\rfloor + \Bigg\lceil c\dfrac{q'}{q}\Bigg\rfloor$, \text{ } $b \dfrac{q'}{q} + d \dfrac{q'}{q} \approx \Bigg\lceil b\dfrac{q'}{q}\Bigg\rfloor + \Bigg\lceil d\dfrac{q'}{q}\Bigg\rfloor$

Therefore:

$a \cdot \dfrac{q'}{q} + c \cdot \dfrac{q'}{q} = b \cdot \dfrac{q'}{q} + d \cdot \dfrac{q'}{q}  + k\cdot q'$

$\Longleftrightarrow \Bigg\lceil a\dfrac{q'}{q}\Bigg\rfloor + \Bigg\lceil c\dfrac{q'}{q}\Bigg\rfloor \approx \Bigg\lceil b\dfrac{q'}{q}\Bigg\rfloor + \Bigg\lceil d\dfrac{q'}{q}\Bigg\rfloor + k\cdot q'$

$\Longleftrightarrow \Bigg\lceil a\dfrac{q'}{q}\Bigg\rfloor + \Bigg\lceil c\dfrac{q'}{q}\Bigg\rfloor \cong \Bigg\lceil b\dfrac{q'}{q}\Bigg\rfloor + \Bigg\lceil d\dfrac{q'}{q}\Bigg\rfloor \bmod q'$ \text{ } $(\Longleftrightarrow a + c \equiv b + d \bmod q)$

\end{enumerate}

$ $

\item $a \cdot c\equiv b \cdot d\bmod q$ $\Longleftrightarrow$ $a \cdot c = b \cdot d + k\cdot q$ (for some integer $k$)

$\Longleftrightarrow ac \cdot \dfrac{q'}{q}  = bd \cdot \dfrac{q'}{q} + q\cdot k \cdot \dfrac{q'}{q}$

$\Longleftrightarrow ac \cdot \dfrac{q'}{q}  = bd \cdot \dfrac{q'}{q} + k\cdot q'$

\begin{enumerate}

\item If $q$ divides all of $aq'$, $bq'$, $cq'$, and $dq'$, then 

$ac \cdot \dfrac{q'}{q} = \Bigg\lceil ac\dfrac{q'}{q}\Bigg\rfloor$, \text{ } $bd \cdot \dfrac{q'}{q} = \Bigg\lceil bd\dfrac{q'}{q}\Bigg\rfloor$

Therefore:

$ac \cdot \dfrac{q'}{q} = bd \cdot \dfrac{q'}{q} + k\cdot q'$

$\Longleftrightarrow \Bigg\lceil ac\dfrac{q'}{q}\Bigg\rfloor = \Bigg\lceil bd\dfrac{q'}{q}\Bigg\rfloor + k\cdot q'$

$\Longleftrightarrow \Bigg\lceil ac\dfrac{q'}{q}\Bigg\rfloor \equiv \Bigg\lceil bd\dfrac{q'}{q}\Bigg\rfloor \bmod q'$ \text{ } $(\Longleftrightarrow a\cdot c \equiv b\cdot d \bmod q)$

$ $

\item If $q$ does not divide any of $aq'$, $bq'$, $cq'$, or $dq'$, then

$ac \cdot \dfrac{q'}{q} \approx \Bigg\lceil ac\dfrac{q'}{q}\Bigg\rfloor$, \text{ } $bd \cdot \dfrac{q'}{q} \approx \Bigg\lceil bd\dfrac{q'}{q}\Bigg\rfloor$

Therefore:

$ac \cdot \dfrac{q'}{q} = bd \cdot \dfrac{q'}{q} + k\cdot q'$

$\Longleftrightarrow \Bigg\lceil ac\dfrac{q'}{q}\Bigg\rfloor \approx \Bigg\lceil bd\dfrac{q'}{q}\Bigg\rfloor + k\cdot q'$

$\Longleftrightarrow \Bigg\lceil ac\dfrac{q'}{q}\Bigg\rfloor \cong \Bigg\lceil bd\dfrac{q'}{q}\Bigg\rfloor \bmod q'$ \text{ } $(\Longleftrightarrow a\cdot c \equiv b\cdot d \bmod q)$

\end{enumerate}

$ $

\end{enumerate}
\end{proof}

As shown in the proof, if all numbers in the congruence relations are exactly divisible by the rescaling factor during the modulo rescaling, then the rescaled result gives exact congruence relations in the new modulo. On the other hand, if any numbers in the congruence relations are not divisible by the rescaling factor during the modulo rescaling (i.e., we need to round some decimals), then the rescaled result gives approximate congruence relations in the new modulo.

In a more complicated congruence relation that contains many $(+, -, \cdot)$ operations, the same principle of modulo rescaling explained above can be recursively applied to each pair of operands surrounding each operator. 

\subsubsection{Example}
\label{subsec:modulo-rescaling-ex}

Suppose we have the following congruence relation:

$b \equiv a\cdot s + \Delta \cdot m + e \bmod q$, \text{ } where: $q = 30$, \text{ } $s = 5$, \text{ } $a = 10$, \text{ } $\Delta = 10$, \text{ } $m = 1$, \text{ } $e = 10$, \text{ } $b = 40$

$ $

First, we can test if the above congruence relation is true by plugging in the given example values as follows: 

$b \equiv a\cdot s + \Delta \cdot m + e \bmod 30$

$40 \equiv 10 \cdot 5 + 10 \cdot 1 + 10 \bmod 30$

$40 \equiv 70 \bmod 30$ 

$ $

This congruence relation is true. 

$ $

Now, suppose we want to rescale the modulo from $30 \rightarrow 3$. Then, based on the rescaling principles described in \autoref{tab:rescaling}, we compute the rescaled values as follows: 

$q'= 3$, \text{ } $s = 5$, \text{ } $m = 1$

$\hat{a} = \Bigg\lceil a\cdot\dfrac{3}{30} \Bigg\rfloor = \Bigg\lceil 10\cdot\dfrac{3}{30} \Bigg\rfloor = 1$

$\hat{\Delta} = \Bigg\lceil \Delta\cdot\dfrac{3}{30} \Bigg\rfloor = \Bigg\lceil 10\cdot\dfrac{3}{30} \Bigg\rfloor = 1$

$\hat{e} = \Bigg\lceil e\cdot\dfrac{3}{30} \Bigg\rfloor = \Bigg\lceil 10\cdot\dfrac{3}{30} \Bigg\rfloor = 1$

$\hat{b} = \Bigg\lceil b\cdot\dfrac{3}{30} \Bigg\rfloor = \Bigg\lceil 40\cdot\dfrac{3}{30} \Bigg\rfloor = 4$

$ $

The rescaled congruence relation from modulo $30 \rightarrow 3$ is derived as follows:

$\Bigg\lceil b\dfrac{3}{30} \Bigg\rfloor \equiv \Bigg\lceil s\cdot a \dfrac{3}{30} \Bigg\rfloor + \Bigg\lceil m \cdot \Delta \dfrac{3}{30} \Bigg\rfloor + \Bigg\lceil e \dfrac{3}{30} \Bigg\rfloor \bmod 3$

$\hat{b} \equiv \hat{a} \cdot s + \hat{\Delta} \cdot m + \hat{e}  \bmod 3$
 \text{ } (an exact congruence relation, as all rescaled values have no decimals)

$4 \equiv 1 \cdot 5 + 1 \cdot 1 + 1 \bmod 3$

$4 \equiv 7 \bmod 3$

$ $

As shown above, the rescaled congruence relation preserves correctness, because all rescaled values are divisible by the rescaling factor. By contrast, if $\dfrac{q}{q'} = \dfrac{30}{3} = 10$ did not divide at least one of $a\cdot s$, $\Delta m$, or $e$, then the rescaled congruence relation would be an approximate (i.e., $\cong$) congruence relation.

\clearpage

\section{Chinese Remainder Theorem}
\label{sec:chinese-remainder}
\textbf{- Reference 1:} 
\href{https://brilliant.org/wiki/chinese-remainder-theorem/}{Brilliant -- Chinese Remainder Theorem}~\cite{crt}

\noindent \textbf{- Reference 2:} 
\href{https://www.youtube.com/watch?v=fz1vxq5ts5I}{YouTube -- Extended Euclidean Algorithm Tutorial}

\begin{tcolorbox}[title={\textbf{\tboxtheorem{\ref*{sec:chinese-remainder}.1} Chinese Remainder Theorem}}]

Suppose we have positive coprime integers $n_0, n_1, n_2, \cdots, n_k$. Let $N = n_0 n_1 \cdots n_k$. We sample $k + 1$ random integers $a_0, a_1, a_2, \cdots, a_k$ from each modulus $n_0, n_1, n_2, \dots, n_k$ (i.e., $a_0 \in \mathbb{Z}_{n_0}$, $a_1 \in \mathbb{Z}_{n_1}$, $\cdots$, $a_k \in \mathbb{Z}_{n_k}$). Then, there exists one and only one solution $x \bmod N$ such that $x \equiv a_i \pmod{n_i}$ for each $0 \le i \le k$. That is: 

\text{ } $x \equiv a_0 \bmod n_0$

\text{ } $x \equiv a_1 \bmod n_1$

\text{ } $x \equiv a_2 \bmod n_2$

\text{ } \text{ } $\vdots$

\text{ } $x \equiv a_k \bmod n_k$

$ $

To compute $x$, we first compute each $y_i$ and $z_i$ (for $0 \leq i \leq k$) as follows:

$y_i = \dfrac{N}{n_i}, \text{ } z_i = y_i^{-1} \bmod n_i$

$ $

Note that each $y_i$'s inverse (i.e., $y_i^{-1}$) can be computed by using the Extended Euclidean algorithm (watch the \href{https://www.youtube.com/watch?v=fz1vxq5ts5I}{YouTube tutorial}). Then, the unique solution $x$ can be computed as follows:

$x = \sum\limits_{i=0}^k a_i y_i z_i $ \textcolor{red}{ $\rhd$ Alternatively, we can compute $ x = \sum\limits_{i=0}^k |a_i z_i|_{n_i} y_i$ (where $|a_i z_i|_{n_i} = a_i z_i \bmod n_i$)}

$ $

Since such $x$ is unique in $\bmod \text{ } N$, there are isomorphic mappings between $x \bmod N$ and $(a_0, a_1, a_2, \cdots, a_k)$.

$ $

Also, $y_iz_i \equiv (y_iz_i)^2 \bmod N$  \text{ } for all $0 \leq i \leq k$ 
\end{tcolorbox}

\begin{proof}

$ $

\begin{enumerate}
\item Given $x = \sum\limits_{i=0}^k a_i y_i z_i$, let's compute $x \bmod n_i$ for each $i$ where $0 \leq i \leq k$: 

$x \bmod n_i = \sum\limits_{j=0}^k a_j y_j z_j \bmod n_i$

$= a_0 y_0 z_0 + a_1 y_1 z_1 + a_2 y_2 z_2 + \cdots + a_k y_k z_k \bmod n_i$

$= a_i y_i z_i \bmod n_i$ \textcolor{red}{ $\rhd$ because $y_j \equiv 0 \bmod n_i$ for all $j \neq i$, as they are a multiple of $n_i$}

$= a_i$ \textcolor{red}{ $\rhd$ because $y_i z_i \equiv y_i y_i^{-1} \equiv 1 \bmod n_i$}

$ $

Thus, the value of $x$ in each modulo $n_0, n_1, n_2, \cdots, n_k$ is congruent with $a_0, a_1, a_2, \cdots, a_k$.

$ $

Alternatively, note that the following is also true:

$x \bmod n_i = \sum\limits_{j=0}^k |a_j z_j|_{n_j} y_i \text{ } (\bmod n_i)$

$= |a_0 z_0|_{n_0} y_0 + |a_1 z_1|_{n_1} y_1 + |a_2 z_2|_{n_2} y_2 + \cdots + |a_k z_k|_{n_k} y_k \bmod n_i$

$= |a_i z_i|_{n_i} y_i \bmod n_i$ 

$= a_i$ 

$ $

\item Now, we prove that $x$ is a unique solution modulo $N$. Suppose there were two solutions: $x$ and $x'$ such that:

$x \equiv x' \equiv a_0 \bmod n_0$

$x \equiv x' \equiv a_1 \bmod n_1$

$x \equiv x' \equiv a_2 \bmod n_2$

\text{ } $\vdots$

$x \equiv x' \equiv a_k \bmod n_k$

$ $

Then, by definition of modulo congruence, $ n_0 \mid (x - x') , n_1 \mid (x - x') , \text{ } n_2 \mid (x - x'), \cdots, \text{ } n_k \mid (x - x')$.   

Also, since $n_0, n_1, n_2, \cdots, n_k$ are coprime, it must be the case that $n_0n_1n_2n_3\cdots n_k \mid (x - x')$, or $N \mid (x - x')$. This means that $x \equiv x' \bmod N$. Therefore, $x$ is a unique solution in modulo $N$.

$ $

\item Now, we will prove that $y_iz_i \equiv (y_iz_i)^2 \bmod N$  \text{ } for all $0 \leq i \leq k$. 

$ $

In the case of modulo $n_i$, $y_iz_i \equiv 1 \bmod n_i$, since $z_i$ is an inverse of $y_i$ modulo $n_i$. In the case of all other modulo $n_j$ where $i \neq j$, $y_iz_i \equiv 0 \bmod n_j$, because $y_i = \dfrac{N}{n_i}$ and thus $n_j$ divides $y_i$. 

$ $

By squaring both sides of $(y_iz_i) \equiv 1 \bmod n_i$, we get $(y_iz_i)^2 \equiv 1 \bmod n_i$. Similarly, by squaring both sides of $(y_iz_i) \equiv 0 \bmod n_j$, we get $(y_iz_i)^2 \equiv 0 \bmod n_j$.

$ $

Therefore, $y_iz_i - (y_iz_i)^2 \equiv 0 \bmod n_i$, and $y_iz_i - (y_iz_i)^2 \equiv 0 \bmod n_j$. In other words, $y_iz_i - (y_iz_i)^2 \equiv 0 \bmod n_j$ for all $0 \leq j \leq k$. 

$ $

Then, we do the similar reasoning as step 2: since every co-prime $n_j$ divides $y_iz_i - (y_iz_i)^2$, $n_0n_1\cdots n_k = N$ divides $y_iz_i - (y_iz_i)^2$. Thus, $y_iz_i - (y_iz_i)^2 \equiv 0 \bmod N$, which is $y_iz_i \equiv (y_iz_i)^2 \bmod N$. This is true for all $0 \leq i \leq k$. 

\end{enumerate}

\end{proof}

\subsection{Application: Residue Number System (RNS)} 
\label{subsec:crt-application}

In a modern processor, each data size is a maximum of 64 bits. If the data size exceeds 64 bits, its computations can be handled efficiently by using the Chinese remainder theorem, ensuring that each co-prime modulus $n_i$ satisfies $\log_2n_i \leq 64$ (where $N = n_0\cdot n_1\cdots\cdot n_k$), so that we can represent a large value $a \bmod N$ as $\vec{a}_{\mathit{crt}} = (a_0, a_1, \cdots , a_k)$, where $a \equiv a_i \bmod n_i$. Then, for any pair of big numbers $a$ and $b \bmod N$, we can compute $a + b \bmod N$ and $a \cdot b \bmod N$ as follows: 

\begin{itemize}
\item $a + b \equiv \sum\limits_{i=0}^k a_i y_i z_i + \sum\limits_{i=0}^k b_i y_i z_i \equiv \sum\limits_{i=0}^k (a_i y_i z_i + b_i y_i z_i) \equiv \sum\limits_{i=0}^k (a_i + b_i) y_i z_i \bmod N $

$ $

\item $a \cdot b \equiv \sum\limits_{i=0}^k a_i y_i z_i \cdot \sum\limits_{i=0}^k b_i y_i z_i \equiv \sum\limits_{i=0}^k (a_i \cdot b_i) (y_i z_i)^2  + \sum\limits_{i\neq j}^k (a_i \cdot b_j) y_i z_iy_j z_j \equiv \sum\limits_{i=0}^k (a_i \cdot b_i) (y_i z_i)^2$

\textcolor{red}{ $\rhd$ Note that all terms $y_iz_iy_jz_j$ where $i \neq j$ are 0 modulo $N$, because $y_iy_j \bmod N \equiv 0$. \\
This is because $y_i = n_0n_1\cdots n_{i-1}n_{i+1}\cdots$ and $y_j = n_0n_1\cdots n_{j-1}n_{j+1}\cdots$. \\
Thus $y_iy_j$ is a multiple of $N$.}

$ $

$\equiv \sum\limits_{i=0}^k (a_i \cdot b_i) (y_i z_i) \bmod N$

\textcolor{red}{ $\rhd$ This is because $(y_i z_i) \equiv (y_i z_i)^2$ as shown in step 3 in the proof of Theorem~\ref*{sec:chinese-remainder}.1}

\end{itemize}

$ $

Thus, the Chinese remainder theorem gives us the following useful formula: 

\begin{tcolorbox}[title={\textbf{\tboxtheorem{\ref*{sec:chinese-remainder}.2} Application of the Chinese Remainder Theorem}}]

Suppose there are two big numbers $a = \sum\limits_{i=0}^k a_i y_i z_i \bmod N$ and $b = \sum\limits_{i=0}^k b_i y_i z_i \bmod N$  where $N$ is a the product of co-prime moduli $n_0 \cdot n_1 \cdots n_k$, we have an isomorphism as follows:

$a \xrightarrow{\sigma} \vec{a}_{\mathit{crt}} = (a_0, a_1, \cdots, a_k)$

$b \xrightarrow{\sigma} \vec{b}_{\mathit{crt}} = (b_0, b_1, \cdots, b_k)$

$ $

Based on the above isomorphism, the following is true:

\begin{itemize}

\item $a + b \equiv \sum\limits_{i=0}^k (a_i + b_i) y_i z_i \bmod N \Longleftrightarrow \vec{a}_{\mathit{crt}} + \vec{b}_{\mathit{crt}} \equiv (a_0 + b_0, \text{ } a_1 + b_1, \cdots, \text{ } a_k + b_k) \bmod N$ 

\item $a \cdot b \equiv \sum\limits_{i=0}^k (a_i \cdot b_i) y_i z_i \bmod N \Longleftrightarrow \vec{a}_{\mathit{crt}} \odot \vec{b}_{\mathit{crt}} \equiv (a_0b_0, a_1 b_1, \cdots, a_k b_k)  \bmod N$

\end{itemize}

, where each element-wise addition/multiplication can be independently done modulo $n_i$

\end{tcolorbox}

\clearpage

\section{Taylor Series}
\label{sec:taylor-series}
The Taylor series is a way to represent an analytic function near a point by an infinite power series (assuming the series converges to the function in that neighborhood). Truncating this series gives polynomial approximations of the function. Formally speaking, the Taylor series of a function is an infinite sum of the evaluations of the function's derivatives at a single point. Given function $f(X)$, its Taylor series centered at $a$ is expressed as follows:

$
f(a) + \dfrac{f'(a)}{1!}(X - a) + \dfrac{f''(a)}{2!}(X - a)^2 + \dfrac{f'''(a)}{3!}(X - a)^3 + \cdots = \sum\limits_{d=0}^{\infty}\dfrac{f^{(d)}(a)}{d!}(X - a)^d
$

For a given function, one can truncate the Taylor series to a finite number of terms (up to degree $D$ instead of an infinite number of terms). Such a $D$-degree polynomial is also called the $D$-th Taylor polynomial approximating $f(X)$. Generally, the larger the degree $D$ (i.e. the more terms we include), the more accurate the approximation of $f(X)$ becomes. The accuracy of the approximation is higher for those coordinates nearby $X=a$, and lower for those coordinates away from $X=a$. To increase the accuracy for farther coordinates, we need to increase $D$.

\clearpage

\section{Efficient Polynomial Multiplication by FFT and NTT}
\label{sec:ntt}
\textbf{- Reference:} 
\href{http://web.cecs.pdx.edu/~maier/cs584/Lectures/lect07b-11-MG.pdf}{Polynomials and the Fast Fourier Transform (FFT)}~\cite{ntt}

\subsection{Background and Motivation}
\label{subsec:ntt-motivation}

Given two $(n-1)$-degree polynomials:

$A(X) = \sum\limits_{i=0}^{n-1}a_iX^i$, \textcolor{white}{...} $B(X) = \sum\limits_{i=0}^{n-1}b_iX^i$

, the polynomial multiplication $C(X) = A(X)\cdot B(X)$ is computed as follows:

$C(X) = \sum\limits_{i=0}^{2n-2}c_iX^{i}$, where $c_i = \sum\limits_{k=0}^{i}a_kb_{i-k}$

This operation of computing $\vec{c} = (c_0, c_1, \cdots, c_{2n-1})$ is also called the convolution of $\vec a$ and $\vec b$, denoted as $\vec{c} = \vec{a} \otimes \vec{b}$. The time complexity of this operation (i.e., the total number of multiplications between two numbers) is $O(n^2)$. 

Another way of multiplying two polynomials is based on \textbf{point-value representation}. The point-value representation of an $(n-1)$-degree (or lesser degree) polynomial $A(X)$ is a set of $n$ coordinates $\{(x_0, y_0), (x_1, y_1), \cdots (x_{n-1}, y_{n-1})\}$, where each $x_i$ is a distinct $X$ coordinate (whereas each $y_i$ is not necessarily a distinct $Y$ coordinate). Given a point-value representation of an $(n-1)$-degree (or lesser degree) polynomial, we can use polynomial interpolation (\autoref{sec:polynomial-interpolation}) to derive the polynomial. 
Let's denote the point-value representation of $(n-1)$-degree (or lesser degree) polynomial $A(X)$ and $B(X)$ as follows:

$A(X)$ : $\bm ( ({x}_0, {y}_0^{\langle a \rangle}), ({x}_1, {y}_1^{\langle a \rangle}), \cdots ({x}_{n-1}, {y}_{n-1}^{\langle a \rangle}) \bm )$

$B(X)$ : $\bm ( ({x}_0, {y}_0^{\langle b \rangle}), ({x}_1, {y}_1^{\langle b \rangle}), \cdots ({x}_{n-1}, {y}_{n-1}^{\langle b \rangle}) \bm )$

Then, the point-value representation of the polynomial $C(X) = A(X) \cdot B(X)$ can be computed as a Hadamard product (Definition~\ref*{subsec:vector-arithmetic} in \autoref{subsec:vector-arithmetic}) of the $y$ values of the point-value representation of $A(X)$ and $B(X)$ as follows:

$C(X)$ : $\bm ( ({x}_0, {y}_0^{\langle c \rangle}), ({x}_1, {y}_1^{\langle c \rangle}), \cdots ({x}_{n-1}, {y}_{n-1}^{\langle c \rangle}) \bm )$, where ${y}_i^{\langle c \rangle} = {y}_i^{\langle a \rangle} \cdot {y}_i^{\langle b \rangle}$

However, we cannot derive polynomial $C(X)$ based on these $n$ coordinates because the degree of $C(X)$ is $2n-2$ (or less than $2n-2$). But if we regard all polynomials (including $A(X), B(X)$ and $C(X)$) to be in the polynomial ring $\mathbb{R}[X]/ (X^n + 1)$ (or $\mathbb{Z}_p[X]/(X^n + 1)$), then we can reduce the $(2n-2)$-degree polynomial $C(X)$ to a congruent $(n-1)$-degree (or lesser degree) polynomial in the ring. Then, the $n$ coordinates of $C(X)$ are sufficient to derive $C(X)$. 

However, the time complexity of this new method is still $O(n^2)$. The Hadamard product between two polynomials' point-value representations takes $O(n)$, but evaluating a polynomial at $n$ distinct $x$ values takes $O(n^2)$ (because each polynomial has $n$ terms, and we have to compute each term for $n$ distinct $x$ values). The polynomial interpolation for deriving $C(X)$ also takes $O(n^2)$. 

To solve this efficiency problem, this section will explain an efficient technique for polynomial evaluation, which can evaluate a polynomial at $n$ distinct roots of unity in $O(n \log n)$. This technique is classified into 2 types: Fast Fourier Transform (FFT) and Number-theoretic Transform (NTT). These two types are technically almost the same, with the only difference that the FFT assumes a polynomial ring over complex numbers (\autoref{sec:roots}), whereas the NTT assumes a polynomial ring over a finite field (e.g., integers modulo a prime) (\autoref{sec:cyclotomic-polynomial-integer-ring}). Polynomial multiplication based on FFT (or NTT) comprises 3 steps: (1) forward FFT (or NTT); (2) point-value multiplication; and (3) inverse FFT (or NTT). 

\subsection{Forward FFT (or NTT)}
\label{subsec:ntt-forward}

We assume a polynomial ring of $\mathbb{R}[X]/ (X^n + 1)$ for FFT, and $\mathbb{Z}_p[X]/ (X^n + 1)$ for NTT (where $X^n + 1$ is a cyclotomic polynomial). The $x$ coordinates to evaluate the target polynomial are the $n$ distinct roots of $X^n + 1$, which are $\omega^1$, $\omega^3, \ldots, \omega^{2n-1}$, where $\omega$ is the primitive $2n$-th root of unity. Then, the point-value representation of the polynomial $A(X)$ is $\bm ( ({x}_0, {y}_0^{\langle a \rangle}), ({x}_1, {y}_1^{\langle a \rangle}), \cdots ({x}_{n-1}, {y}_{n-1}^{\langle a \rangle}) \bm )$, where: 

${y}_i^{\langle a \rangle} = A(\omega^{2i + 1}) = \sum\limits_{j=0}^{n-1}a_{j}\cdot (\omega^{2i + 1})^j = \sum\limits_{j=0}^{n-1}a_{j}\cdot \omega^{(2i + 1)\cdot j}$

$ $

We call the vector $\vec{y}^{\langle a \rangle} = (y_0^{\langle a \rangle}, y_1^{\langle a \rangle}, \cdots, y_{n-1}^{\langle a \rangle})$ the Discrete Fourier Transform (DFT) of the coefficient vector $\vec{a} = (a_0, a_1, \cdots, a_{n-1})$. We write this as $\vec{y}^{\langle a \rangle} = \textsf{DFT}(\vec{a})$. As explained in \autoref{subsec:ntt-motivation}, the computation of the DFT takes $O(n^2)$, because we have to evaluate $n$ distinct $X$ values for a polynomial that has $n$ terms.

\subsubsection{High-level Idea}
\label{subsec:ntt-forward-overview}

FFT (or NTT) is an improved method for computing the DFT, which reduces the time complexity from $O(n^2)$ to $O(n \log n)$. The high-level idea of FFT is to split the $(n-1)$-degree (or lesser degree) target polynomial $A(X)$ to evaluate into 2 half-degree polynomials $A_0(X)$ and $A_1(X)$ as follows:

$A(X) = a_0 + a_1X + a_2X^2 + \cdots + a_{n-1}X^{n - 1}$

$\textcolor{white}{A(X) }= A_0(X^2) + X \cdot A_1(X^2)$
$A_0(X) = a_0 + a_2X + a_4X^2 + \cdots + a_{n-2}X^{\frac{n}{2} - 1}$

$A_1(X) = a_1 + a_3X + a_5X^2 + \cdots + a_{n-1}X^{\frac{n}{2} - 1}$

The above method of splitting a polynomial into two half-degree polynomials is called the Cooley-Tukey step. As we split $A(X)$ into two smaller-degree polynomials $A_0(X)$ and $A_1(X)$, evaluating $A(X)$ at the odd-powered primitive $2n$-th roots of unity $\{\omega^1, \omega^3, \omega^5, \cdots, \omega^{2n-1}\}$ is equivalent to evaluating $A_0(X)$ and $A_1(X)$ at $n$ distinct \textit{squared} $n$-th roots of unity $\{(\omega^2)^1, (\omega^2)^3, (\omega^2)^5, \ldots, (\omega^2)^{2n-1}\}$ and computing $A_0(X^2) + X\cdot A_1(X^2)$. However, remember that the primitive $2n$-th root of unity $\omega$ has order $2n$ (i.e., $\omega^{2n} = 1$ and $\omega^m \neq 1$ for all $m < 2n$). Therefore, the second half of $\{(\omega^2)^1, (\omega^2)^3, (\omega^2)^5, \ldots, (\omega^2)^{2n-1}\}$ is a repetition of the first half. This implies that we only need to evaluate $A_0(X)$ and $A_1(X)$ at $\dfrac{n}{2}$ distinct $x$ coordinates each, instead of $n$ distinct coordinates, because the polynomial evaluation results for the other half are the same as those of the first half (as their input $x$ to the polynomial is the same). 

We recursively split $A_0(X)$ and $A_1(X)$ into half-degree polynomials and evaluate them at half-counted (i.e., $n/2$) $n$-th roots of unity. Then, the total number of rounds of splitting is $\log n$, and the maximum number of root-to-coefficient multiplications in each round is $n$, which aggregates to $O(n \log n)$. 

\subsubsection{Details}
\label{subsec:ntt-forward-details}

Suppose we have a polynomial ring that is either $\mathbb{Z}_{p}[X] / (X^8 + 1)$ (i.e., over a finite field with prime $p$) or $\mathbb{R}[X] / (X^8 + 1)$ (over complex numbers). 
We denote by $\omega$ a primitive $(2n=16)$-th root of unity, and 
the 8 distinct roots of $X^8 + 1$ are: $\{\omega^1, \omega^3, \omega^5, \omega^7, \omega^9, \omega^{11}, \omega^{13}, \omega^{15}\}$.

$ $

Now, we define our target polynomial to evaluate as follows: 

$A(X) = a_0 + a_1X + a_2X^2 + a_3X^3 + a_4X^4 + a_5X^5 + a_6X^6 + a_7X^7$

$ $

We split this 7-degree polynomial into the following two 3-degree polynomials (using the Cooley-Tukey step): 

$A_0(X) = a_0 + a_2X + a_4X^2 + a_6X^3$

$A_1(X) = a_1 + a_3X + a_5X^2 + a_7X^3$

$A(X) = A_0(X^2) + X \cdot A_1(X^2)$

$ $

We recursively split the two 3-degree polynomials above into 1-degree polynomials as follows:

$A_{0,0}(X) = a_0 + a_4X$, \textcolor{white}{...} $A_{0,1}(X) = a_2 + a_6X$

$A_0(X) = A_{0,0}(X^2) + X\cdot A_{0,1}(X^2)$

$ $

$A_{1,0}(X) = a_1 + a_5X$, \textcolor{white}{...} $A_{1,1}(X) = a_3 + a_7X$

$A_1(X) = A_{1,0}(X^2) + X\cdot A_{1,1}(X^2)$

$ $

$A(X) = A_0(X^2) + X \cdot A_1(X^2)$

$\mathcolor{white}{A(X)} = \underbrace{\underbrace{(\underbrace{A_{0,0}(X^4)}_{\text{FFT Level 1}} + X^2\cdot \underbrace{A_{0,1}(X^4)}_{\text{FFT Level 1}})}_{\text{FFT Level 2}} + X \cdot \underbrace{(\underbrace{A_{1,0}(X^4)}_{\text{FFT Level 1}} + X^2\cdot \underbrace{A_{1,1}(X^4)}_{\text{FFT Level 1}})}_{\text{FFT Level 2}}}_{\text{FFT Level 3}}$

$ $

To evaluate $A(X)$ at the $n$ distinct roots $X = \{\omega^1, \omega^3, \ldots, \omega^{15}\}$, we evaluate each FFT level of the above formula at $X = \{\omega^1, \omega^3, \cdots, \omega^{15}\}$, starting from level $1 \leq l \leq 3$. 

$ $

\para{FFT Level $\bm{l = 1}$:} We evaluate $A_{0,0}(X^4)$, $A_{0,1}(X^4)$, $A_{1,0}(X^4)$, and $A_{1,1}(X^4)$ at $X = \{\omega^1, \omega^3, \cdots, \omega^{15}\}$. However, notice that plugging in $X = \{\omega^1, \omega^3, \cdots, \omega^{15}\}$ to $X^4$ results in only 2 distinct values: $\omega^4$ and $\omega^{12}$ (which correspond to roots of $X^2+1$). This is because the order of $\omega$ is $2n$ (i.e., $\omega^{2n} = 1$), and thus $(\omega^1)^4 = (\omega^5)^4 = (\omega^9)^4 = (\omega^{13})^4 = \omega^4$, and $(\omega^3)^4 = (\omega^7)^4 = (\omega^{11})^4 = (\omega^{15})^4 = \omega^{12}$.
Therefore, we only need to evaluate $A_{0,0}(X^4)$, $A_{0,1}(X^4)$, $A_{1,0}(X^4)$, and $A_{1,1}(X^4)$ at 2 distinct $x$ values instead of 8, where each evaluation requires a constant number of arithmetic operations: computing 1 multiplication and 1 addition. As there are a total of 4 polynomials to evaluate (i.e., $A_{0, 0}(X^4), A_{0, 1}(X^4), A_{1, 0}(X^4), A_{1, 1}(X^4)$), we compute the FFT a total of $4 \cdot 2 = 8$ times. 

$ $

\para{FFT Level $\bm{l = 2}$:} Based on the evaluation results from FFT Level 1 as building blocks, we evaluate $A_{0}(X^2)$ and $A_{1}(X^2)$ at $X = \{\omega^1, \omega^3, \cdots, \omega^{15}\}$. However, notice that plugging in $X = \{\omega^1, \omega^3, \cdots, \omega^{15}\}$ to $X^2$ results in only 4 distinct values: $\omega^2$, $\omega^6$, $\omega^{10}$, and $\omega^{14}$ (which correspond to roots of $X^4+1$). This is because the order of $\omega$ is $2n$ (i.e., $\omega^{2n} = 1$), and thus $(\omega^1)^2 = (\omega^9)^2$, $(\omega^3)^2 = (\omega^{11})^2$, $(\omega^5)^2 = (\omega^{13})^2$, and $(\omega^7)^2 = (\omega^{15})^2$. Therefore, we only need to evaluate $A_{0}(X^2)$ and $A_{1}(X^2)$ at 4 distinct $x$ values instead of 8, where each evaluation requires a constant number of arithmetic operations: computing 1 multiplication and 1 addition (where we use the results from FFT Level 1 as building blocks, and the computational structure of FFT Level 2 is the same as that of FFT Level 1). There are a total of 2 polynomials to evaluate (i.e., $A_{0}(X^2), A_{1}(X^2)$); thus, we compute the FFT a total of $2 \cdot 4 = 8$ times. 

$ $

\para{FFT Level $\bm{l = 3}$:} Based on the evaluation results from FFT Level 2 as building blocks, we evaluate $A(X)$ at $X = \{\omega^1, \omega^3, \cdots, \omega^{15}\}$. For this last level of computation, we need to evaluate all 8 distinct $X$ values, since they are all unique values, and each evaluation requires a constant number of arithmetic operations: computing 1 multiplication and 1 addition. There is a total of 1 polynomial to evaluate (i.e., $A(X)$); thus, we compute the FFT a total of $1 \cdot 8 = 8$ times. 

$ $

\para{Generalization:} Suppose that the degree of the target polynomial to evaluate is at most $n-1$ (with $n$ terms), and we define $L = \log n$ (i.e., the total number of FFT levels). Then, the forward FFT operation requires a total of $L$ FFT levels, where each $l$-th level requires the evaluation of $2^{L - l}$ polynomials at $2^l$ distinct $X$ values. Therefore, the total number of FFT computations for the forward FFT is: $\sum_{l=1}^{L}(2^{L-l} \cdot 2^l) = L \cdot 2^L = n \log n$
%$\log (n) \cdot (2^{L - l} \cdot 2^l) = 2^L \log n = n \log n$
. Therefore, the time complexity of the forward FFT is $O(n \log n)$.

Using the FFT technique, we reduce the number of $x$ points to evaluate to half as the level decreases (while the number of polynomials to evaluate doubles), and their growth and reduction cancel each other, resulting in $O(n)$ for each level. Since there are $\log n$ such levels, the total time complexity is $O(n \log n)$. The core enabler of this optimization is the special property of the $x$ evaluation coordinates: its power (i.e., $\omega^i$) is cyclic. To enforce this cyclic property, FFT requires the evaluation points of $x$ to be the odd-powered primitive $2n$-th roots of unity.

\subsection{Point-wise Multiplication}
\label{subsec:pointwise-multiplication}

Once we have applied the forward FFT operation (\autoref{subsec:ntt-forward}) to polynomial $A(X)$ and $B(X)$ as $\vec{y}^{\langle a \rangle}$ and $\vec{y}^{\langle b \rangle}$, computing the point-value representation of $C(X) = A(X) \cdot B(X)$ can be done in $O(n)$ using the Hadamard product $ \vec{y}^{\langle c \rangle} = \vec{y}^{\langle a \rangle} \odot \vec{y}^{\langle b \rangle}$ (as explained in \autoref{subsec:ntt-motivation}).

\subsection{Inverse FFT (or NTT)}
\label{subsec:ntt-backward}

So far, we have computed:

$C(X)$ : $\bm ( ({x}_0, {y}_0^{\langle c \rangle}), ({x}_1, {y}_1^{\langle c \rangle}), \cdots ({x}_{n-1}, {y}_{n-1}^{\langle c \rangle}) \bm )$, where ${y}_i^{\langle c \rangle} = {y}_i^{\langle a \rangle} \cdot {y}_i^{\langle b \rangle}$

$ $

Our final step is to convert ${y}_i^{\langle c \rangle}$ back to $\vec{c}$, the polynomial coefficients of $C(X)$. We call this reversing operation the inverse FFT. 

Given an $(n-1)$-degree polynomial 
$C(X) = \sum\limits_{i=0}^{n-1}c_iX^i$, the forward FFT process is computationally equivalent to evaluating the polynomial at $n$ distinct $n$-th roots of $X^n + 1$ as follows:

${y}_i^{\langle c \rangle} = C(\omega^{2i+1}) = \sum\limits_{j=0}^{n-1}c_{j}\cdot (\omega^{2i + 1})^j = \sum\limits_{j=0}^{n-1}c_{j}\cdot \omega^{(2i+1)j}$

The above evaluation is equivalent to computing the following matrix-vector multiplication: 

$ $

${y}_i^{\langle c \rangle} = W \cdot \vec{c}$, \text{ } where $W = \begin{bmatrix}
(\omega^1)^0 & (\omega^1)^1 & \cdots & (\omega^1)^{n-1}\\
(\omega^3)^0 & (\omega^3)^1 & \cdots & (\omega^3)^{n-1}\\
\vdots & \vdots & \ddots & \vdots \\
(\omega^{2n-1})^0 & (\omega^{2n-1})^1 & \cdots & (\omega^{2n-1})^{n-1}\\
\end{bmatrix} $, \text{ } $\vec{c} = (c_0, c_1, \cdots, c_{n-1})$

$ $

We denote each element of $W$ as: $(W)_{i,j} = \omega^{(2i+1)j}$. The inverse FFT is a process of reversing the above computation. For this inversion, our goal is to find an inverse matrix $W^{-1}$ such that $W^{-1} \cdot {y}_i^{\langle c \rangle} = W^{-1} \cdot (W \cdot \vec{c}) = (W^{-1} \cdot W) \cdot \vec{c} = I_n \cdot \vec{c} =  \vec{c}$. 
As a solution, we propose the inverse matrix $W^{-1}$ as follows: 

$(W^{-1})_{j,k} = n^{-1}\cdot \omega^{-(2k+1)j}$

$ $

Now, we will show why $W^{-1} \cdot W = I_n$. Each element of $W^{-1} \cdot W$ is computed as: 

$(W^{-1} \cdot W)_{j, i} = \sum\limits_{k=0}^{n-1}(n^{-1}\cdot \omega^{-(2k+1)j} \cdot \omega^{(2k+1)i}) = n^{-1}\cdot\sum\limits_{k=0}^{n-1}\omega^{(2k+1)(i-j)}$

In order for $W^{-1} \cdot W$ to be $I_n$, the following should hold:

\[
(W^{-1} \cdot W)_{j, i} = \begin{cases}
\text{1 if } j = i \\
\text{0 if } j \neq i \\
\end{cases}
\]

If $j = i$ holds, the above condition is satisfied because: 

$(W^{-1} \cdot W)_{j, i} = n^{-1}\cdot\sum\limits_{k=0}^{n-1}\omega^{(2k+1)(0)} =  n^{-1}\cdot\sum\limits_{k=0}^{n-1}1 = n^{-1} \cdot n = 1$

$ $

In the case of $j \neq i$, we will leverage the Geometric Sum formula $\sum\limits_{i=0}^{n-1}x^i = \dfrac{x^n - 1}{x - 1}$ (the proof is provided below):

\begin{tcolorbox}[title={\textbf{\tboxtheorem{\ref*{subsec:ntt-backward}} Geometric Sum Formula}}]

Let the geometric sum $S_n = 1 + x + x^2 + \cdots + x^{n - 1}$

Then, $x \cdot S_n = x + x^2 + x^3 + \cdots + x^{n}$

$x \cdot S_n - S_n = (x + x^2 + x^3 + \cdots + x^{n}) - (1 + x + x^2 + \cdots + x^{n - 1}) = x^n - 1$

$S_n\cdot (x - 1) = x^n - 1$

$S_n = \dfrac{x^n - 1}{x - 1}$ \textcolor{red}{ $\rhd$ with the constraint that $x \neq 1$}

\end{tcolorbox}

Our goal is to compute $\sum\limits_{k=0}^{n-1}\omega^{(2k+1)(i-j)} = \omega^{i-j} \sum\limits_{k=0}^{n-1}(\omega^{2(i-j)})^k$. Leveraging the Geometric Sum formula with $x = \omega^{2(i-j)}$:

$\omega^{i-j} \sum\limits_{k=0}^{n-1}(\omega^{2(i-j)})^k = \omega^{i-j} \dfrac{(\omega^{2(i-j)})^n - 1}{\omega^{2(i-j)} - 1}$ 

$= \omega^{i-j} \dfrac{(\omega^{2n})^{i - j} - 1}{\omega^{2(i-j)} - 1}$

$= \omega^{i-j} \dfrac{(1)^{i - j} - 1}{\omega^{2(i-j)} - 1}$ \textcolor{red}{ $\rhd$ since the order of $\omega$ is $2n$, $\omega^{2n}=1$}

\textcolor{red}{ $\rhd$ Here, the denominator can't be 0. Since $i \neq j$ and $|i-j| < n$, the exponent $2(i-j)$ is not a multiple of $2n$.}

$= 0$

$ $

Thus, $(W^{-1} \cdot W)_{j, i}$ is 1 if $j = i$, and 0 if $j \neq i$. Therefore, the inverse FFT can be computed as $\vec{c} = W^{-1} \cdot y^{\langle c \rangle}$, where: 

$c_i = \sum_{j=0}^{n-1} (W^{-1})_{i,j} \cdot y_j^{\langle c \rangle}$

$ \sum\limits_{j=0}^{n-1} \left( n^{-1} \cdot \omega^{-(2j+1)i} \right) \cdot y_j^{\langle c \rangle}$ \textcolor{red}{ $\rhd$ since $(W^{-1})_{j,k} = n^{-1} \cdot \omega^{-(2k+1)j}$, and $(W^{-1})_{i,j} = n^{-1} \cdot \omega^{-(2j+1)i}$}

$ = n^{-1}\cdot \sum\limits_{j=0}^{n-1}{y}_j^{\langle c \rangle} \cdot \omega^{-(2j+1)i}$

$ = n^{-1} \sum\limits_{j=0}^{n-1} y_j^{\langle c \rangle} \cdot (\omega^{-2ji} \cdot \omega^{-i}) $ \textcolor{red}{ $\rhd$ since $\omega^{-(2j+1)i} = \omega^{-2ji - i} = \omega^{-2ji} \cdot \omega^{-i}$}

$= n^{-1} \omega^{-i} \sum\limits_{j=0}^{n-1} {y}_j^{\langle c \rangle} (\omega^{-2})^{ji}$

$ $

By reusing the recursive splitting technique explained in the forward process (\autoref{subsec:ntt-forward-details}), this inverse operation also achieves a time complexity of $O(n \log n)$.

\begin{comment}
\begin{algorithm}[t]
\algrenewcommand\algorithmicindent{1em} % Adjust indentation herei.e., 
\begin{algorithmic}[1]
\State \textsf{FFT}$(\vec{a}):$ %\textcolor{red}{ $\rhd$ an RLWE ciphertext storing $k$ zeros and non-zero random numbers, whose ratio varies}
\State $n' \gets$ length of $\vec{a}$
\If{$n' = 1$}
    \State \textbf{return} $a$
\EndIf
\State $\omega = e^{\frac{2\pi}{k}}$ \textcolor{red}{ $\rhd$ or $\omega = g^{\frac{p-1}{n}}$ in the case of NTT}
\State $\omega' = 1$
\State $\vec{a}^{\langle 0 \rangle} = (a_0, a_2, \cdots, a_{n'-2})$
\State $\vec{a}^{\langle 1 \rangle} = (a_1, a_3, \cdots, a_{n'-1})$
\State $\vec{y}^{\langle 0 \rangle} = \textsf{FFT}(\vec{a}^{\langle 0 \rangle})$
\State $\vec{y}^{\langle 1 \rangle} = \textsf{FFT}(\vec{a}^{\langle 1 \rangle})$
\For{$i \in \{0, 1, \cdots \frac{n'}{2} - 1\}$}
    \State $y_i \gets y_i^{\langle 0 \rangle} + \omega \cdot y_i^{\langle 1 \rangle}$
    \State $y_{i+\frac{n}{2}} \gets y_i^{\langle 0 \rangle} + \omega \cdot y_i^{\langle 1 \rangle}$
    \State $\omega' = \omega' \cdot \omega$
\EndFor
\State \textbf{return} $\vec{y}$ 
\end{algorithmic}
\caption{Fast Fourier Transform (FFT)}
\label{alg:fft}
\end{algorithm}
\end{comment}

\clearpage

\part{Post-quantum Cryptography}
\label{part:pqc}

\renewcommand{\thesection}{B-\arabic{section}}
\setcounter{section}{0}

This chapter explains lattice-based cryptographic schemes: LWE cryptosystem, RLWE cryptosystem, GLWE cryptosystem, GLev cryptosystem, and GGSW cryptosystem. These are the essential building blocks for FHE schemes.

$ $

\begin{tcolorbox}[
    title = \textbf{Required Background},    % box title
    colback = white,    % light background; tweak to taste
    colframe = black,  % frame colour
    boxrule = 0.8pt,     % line thickness
    left = 1mm, right = 1mm, top = 1mm, bottom = 1mm % inner padding
]

\begin{itemize}
\item \autoref{sec:group}: \nameref{sec:group}
\item \autoref{sec:field}: \nameref{sec:field}
\item \autoref{sec:polynomial-ring}: \nameref{sec:polynomial-ring}
\end{itemize}

\end{tcolorbox}

$ $

\clearpage

\section{Lattice-based Cryptography}
\label{sec:lattice}
\textbf{- Reference:} 
\href{https://mysite.science.uottawa.ca/mnevins/papers/StephenHarrigan2017LWE.pdf}{Lattice-Based Cryptography and the Learning with
Errors Problem}~\cite{lattice-crypto}

$ $

Lattice-based cryptography is often considered as  post-quantum cryptography, resistant against quantum computer attacks. This section describes the mathematical hard problem that is the basis of the lattice-based cryptosystems we will explore: LWE (Learning with Error) cryptosystem, RLWE (Ring Learning with Error) cryptosystem, GLWE (General Learning with Error) cryptosystem, GLev cryptosystem, and GGSW cryptosystem.

\subsection{Overview}
\label{subsec:lattice-overview}

Suppose we have a single unknown $k$-dimensional vector $\vec{s}$ as a secret key, many publicly known $k$-dimensional vectors $\vec{a}^{\langle i \rangle}$.

And suppose we have a large set of the following dot products $\vec{s} \cdot \vec{a}^{\langle i \rangle}$: 

$\vec{s} \cdot \vec{a}^{\langle 0 \rangle} = s_0\cdot a_{0}^{\langle 0 \rangle} + s_1\cdot a_{1}^{\langle 0 \rangle} + \cdots + s_{k-1}\cdot a_{k-1}^{\langle 0 \rangle} = b^{\langle 0 \rangle}$

$\vec{s} \cdot \vec{a}^{\langle 1 \rangle} = s_0\cdot a_{0}^{\langle 1 \rangle} + s_1\cdot a_{1}^{\langle 1 \rangle} + \cdots + s_{k-1}\cdot a_{k-1}^{\langle 1 \rangle} = b^{\langle 1 \rangle}$

$\vec{s} \cdot \vec{a}^{\langle 2 \rangle} = s_0\cdot a_{0}^{\langle 2 \rangle} + s_1\cdot a_{1}^{\langle 2 \rangle} + \cdots + s_{k-1}\cdot a_{k-1}^{\langle 2 \rangle} = b^{\langle 2 \rangle}$

\text{ } $\vdots$

$ $

Suppose that all $(\vec{a}^{\langle i \rangle}, b^{\langle i \rangle})$ tuples are publicly known. An attacker only needs $k$ such tuples to derive the secret vector $\vec{s}$. Specifically, as there are $k$ unknown variables (i.e., $s_0, s_1, \cdots, s_{k-1}$), the attacker can solve for those $k$ variables with $k$ equations by using linear algebra. 

However, suppose that in each equation above, we randomly add an unknown small noise $e^{\langle i \rangle}$ (i.e., error) as follows: 

$\vec{s} \cdot \vec{a}^{\langle 0 \rangle} = s_0\cdot a_{0}^{\langle 0 \rangle} + s_1\cdot a_{1}^{\langle 0 \rangle} + \cdots + s_{k-1}\cdot a_{k-1}^{\langle 0 \rangle} + e^{\langle 0 \rangle} \approx b^{\langle 0 \rangle}$

$\vec{s} \cdot \vec{a}^{\langle 1 \rangle} = s_0\cdot a_{0}^{\langle 1 \rangle} + s_1\cdot a_{1}^{\langle 1 \rangle} + \cdots + s_{k-1}\cdot a_{k-1}^{\langle 1 \rangle} + e^{\langle 1 \rangle} \approx b^{\langle 1 \rangle}$

$\vec{s} \cdot \vec{a}^{\langle 2 \rangle} = s_0\cdot a_{0}^{\langle 2 \rangle} + s_1\cdot a_{1}^{\langle 2 \rangle} + \cdots + s_{k-1}\cdot a_{k-1}^{\langle 2 \rangle} + e^{\langle 2 \rangle} \approx b^{\langle 2 \rangle}$

\text{ } $\vdots$

$ $

Then, even if the attacker has a sufficient number of $(\vec{a}^{\langle i \rangle}, b^{\langle i \rangle})$ tuples, it is not feasible to derive $s_0, s_1, \cdots, s_{k-1}$, because even a small amount of noise added to each equation prevents the linear-algebra-based direct derivation of the unknown variables. For each of the above equations, the attacker has to consider as many possibilities as there are possible values of $e^{\langle i \rangle}$. For example, if there are $r$ possible values for each noise $e^{\langle i \rangle}$, the attacker's brute-force search space for applying linear algebra to those $k$ equations is: $\overbrace{r \times r \times r \times \cdots \times r}^{\text{k times}} = r^k$. Thus, the number of noisy equations grows, and the aggregate possibilities of $e^{\langle i \rangle}$s grow exponentially, which means that the attacker's cost of attack grows exponentially. 

Based on this intuition, the mathematical hard problem that constitutes lattice-based cryptography is as follows:

$ $

\begin{tcolorbox}[title={\textbf{\tboxlabel{\ref*{subsec:lattice-overview}} The LWE (Learning with Errors) and RLWE Problems}}]
\textbf{\underline{LWE Problem}}

Consider samples of the form: $b = \vec{s} \cdot\vec{a} + e$ (where $e$ is a small noise to be explained later). 

For each encryption, a random $k$-dimensional vector $\vec{a} \in \mathbb{Z}_q^k$ and a small noise value $e \in \mathbb{Z}_q$ are newly sampled from $\{0, 1, \cdots, q - 1\}$, where $q$ is the ciphertext domain size. On the other hand, the $k$-dimensional secret vector $\vec{s}$ is the same for all encryptions. Suppose we have a sufficient number of ciphertext tuples: 

$(\vec{a}^{\langle 1 \rangle}, b^{\langle 1 \rangle})$, where $b^{\langle 1 \rangle} = \vec{s} \cdot \vec{a}^{\langle 1 \rangle} + e^{\langle 1 \rangle}$ 

$(\vec{a}^{\langle 2 \rangle}, b^{\langle 2 \rangle})$, where $b^{\langle 2 \rangle} = \vec{s} \cdot \vec{a}^{\langle 2 \rangle} + e^{\langle 2 \rangle}$ 

$(\vec{a}^{\langle 3 \rangle}, b^{\langle 3 \rangle})$, where $b^{\langle 3 \rangle} = \vec{s} \cdot \vec{a}^{\langle 3 \rangle} + e^{\langle 3 \rangle}$ 

\text{ } $\vdots$

Suppose that the attacker has a sufficiently large number of $(\vec{a}^{\langle i \rangle}, b^{\langle i \rangle})$ tuples. Given this setup, the following hard problems constitute the defense mechanism of the LWE (Learning with Errors) cryptosystem:

$ $

\begin{itemize}
\item \textbf{Search-Hard Problem:} There is no efficient algorithm for the attacker to find out the secret key vector $\vec{s}$.
\item \textbf{Decision-Hard Problem:} We create a black box system which can be configured to one of the following two modes: (1) all $b^{\langle i\rangle}$ values are purely randomly generated; (2) all $b^{\langle i\rangle}$ values are computed as the results of $\vec{s} \cdot \vec{a}^{\langle i\rangle} + e^{\langle i\rangle}$ based on the randomly picked known public (symmetric) keys $\vec{a}^{\langle i \rangle}$, randomly picked unknown noises $e^{\langle i\rangle}$, and a constant unknown secret vector $\vec{s}$. Given a sufficient number of $(\vec{a}^{\langle i \rangle}, b^{\langle i \rangle})$ tuples generated by this black box system, the attacker has no efficient algorithm to determine which mode this black box system is configured to.
\end{itemize}

$ $

These two problems are interchangeable.

$ $

\textbf{\underline{RLWE Problem}}

In the case of the RLWE (Ring Learning with Errors) problem, the only difference is that $\vec{a}$, $b$, $\vec{s}$, and $e$ are replaced by polynomials $(n-1)$-degree polynomials $A$, $B$, $S$, and $E$ in $\mathbb{Z}_q[X] / (x^n + 1)$, and its search-hard problem is finding the unknown $n$ coefficients of the secret polynomial $S$. 

\end{tcolorbox}

\subsection{LWE Cryptosystem}
\label{subsec:lattice-scheme}

The LWE cryptosystem uses the following encryption formula: $b = \vec{s} \cdot \vec{a} + \Delta \cdot m + e$ (where $\vec{s}$ is a secret key, $\vec{a}$ is a publicly known random vector picked per encryption, $m$ is a plaintext, $e$ is small noise randomly picked per encryption from a normal distribution, and $b$ is a ciphertext). $\Delta$ is a scaling factor of the plaintext $M$ (shifting $m$ by $\text{log}_2\Delta$ bits to the left). Before encrypting the plaintext, we left-shift the plaintext several bits (i.e., $\text{log}_2\Delta$ bits) to secure sufficient space to store the error in the lower bits.

\begin{figure}[h!]
    \centering
  \includegraphics[width=0.8\linewidth]{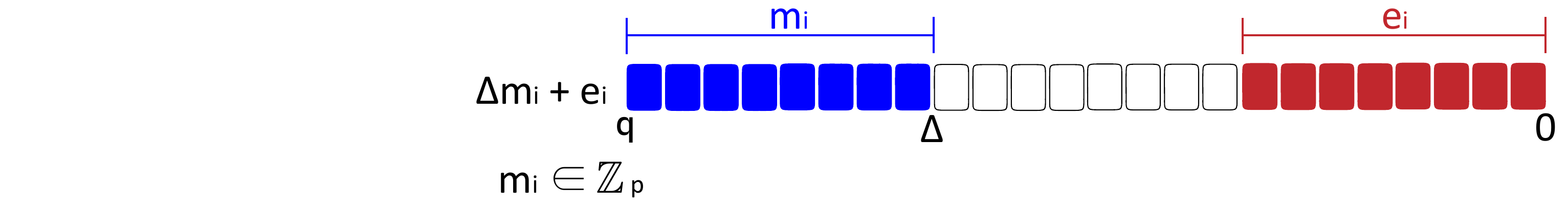}
  \caption{An illustration of LWE's plaintext scaling and adding a noise: $\Delta \cdot m + e \in \mathbb{Z}_q$}
  \label{fig:scaling}
\end{figure}

\autoref{fig:scaling} visually illustrates the term $\Delta \cdot m + e$, where the plaintext $m$ left-shifted by $\text{log}_2\Delta$ bits and noised by the noise $e$. The actual encryption and decryption formulas are as follows:

\begin{tcolorbox}[title={\textbf{\tboxlabel{\ref*{subsec:lattice-scheme}} Lattice-based LWE Cryptosystem}}]
\begin{itemize}
    \item \textbf{\underline{Encryption}: } $b^{\langle i \rangle} = \vec{s} \cdot \vec{a}^{\langle i \rangle} + \Delta \cdot m^{\langle i \rangle} + e^{\langle i \rangle}$, where $b^{\langle i \rangle}$ and $\vec{a}^{\langle i \rangle}$ are publicly known also to the attacker, while $\vec{s}, m^{\langle i \rangle}, e^{\langle i \rangle}$ are unknown (only known by the secret key owner). 

    $ $
    
    \item \textbf{\underline{Decryption}: } $ \dfrac{\lceil b^{\langle i \rangle} - \vec{s} \cdot \vec{a}^{\langle i \rangle} \rfloor_{\Delta}}{\Delta} = \dfrac{\lceil \Delta m^{\langle i \rangle} + e^{\langle i \rangle} \rfloor_{\Delta}}{\Delta} = m^{\langle i \rangle}$ $\Big($ provided $|e^{\langle i \rangle}| < \dfrac{\Delta}{2}\Big)$   
\end{itemize}
\end{tcolorbox}

$\lfloor \rceil_\Delta$ means rounding the number to the nearest multiple of $\Delta$. For example, $\lfloor 16 \rceil_{10} = 20$, which is rounding 16 to the nearest multiple of 10. As another example, $\lfloor 17 \rceil_{8} = 16$, which rounds 17 to the nearest multiple of 8 (note that 17 is closer to 16 than to 24; thus, it is rounded to 16).

$ $

\noindent \textbf{Correctness: } In the decryption scheme, computing $b^{\langle i \rangle} - \vec{s} \cdot \vec{a}^{\langle i \rangle}$ gives $\Delta \cdot m^{\langle i \rangle}+ e^{\langle i \rangle}$, which is \autoref{fig:scaling}. Then, $\lceil \Delta \cdot m^{\langle i \rangle} + e^{\langle i \rangle} \rfloor_{\Delta}$ (i.e., rounding the value to the nearest multiple of $\Delta$) gives $\Delta \cdot m^{\langle i \rangle}$, provided the added noise $|e^{\langle i \rangle}| < \frac{\Delta}{2}$. That is, if the noise is less than $\frac{\Delta}{2}$, it will disappear during the rounding. Finally, right-shifting $\Delta \cdot m^{\langle i \rangle}$ by $\text{log}_2 \Delta$ bits gives $m^{\langle i \rangle}$. To summarize, if we ensure $|e^{\langle i \rangle}| < \frac{\Delta}{2}$ (which is why the noise $e^{\langle i \rangle}$ should be smaller than this threshold), then we can eliminate $e^{\langle i \rangle}$ during the decryption's rounding process and retrieve the original $\Delta \cdot m^{\langle i \rangle}$. The reason we scaled $m^{\langle i \rangle}$ by $\Delta$ is to: (i) create space for storing $e^{\langle i \rangle}$ in the lower bits during encryption such that the noise bits do not interfere with the plaintext bits (to avoid corrupting the plaintext bits); and (ii) blow away the noise $e^{\langle i \rangle}$ stored in the lower bits during decryption without corrupting the plaintext $m^{\langle i \rangle}$. 

$ $

\noindent \textbf{Security: } Given that an attacker has a large list of $(\vec{a}^{\langle i \rangle}, b^{\langle i \rangle})$ (i.e., many ciphertexts), it is almost impossible for them to derive $\vec{s}$, due to the random noise $e^{\langle i \rangle}$ added in each encryption (which is a search-hard problem described in \autoref{subsec:lattice-overview}). This is because even small added unknown noises $e^{\langle i \rangle}$ greatly change the mathematical solution for $\vec{s}$ that satisfies all the $b^{\langle i \rangle} = \vec{s} \cdot \vec{a}^{\langle i \rangle} + \Delta \cdot m^{\langle i \rangle} + e^{\langle i \rangle}$ equations. 

%$Y = S \cdot X$

Even in the case that the attacker has a large list of $(\vec{a}^{(j)}, b^{(j)})$ generated for the same ciphertext $m^{\langle i \rangle}$ (where each ciphertext used different $\vec{a}^{(j)}$ and $e^{(j)}$ to encrypt the same $m^{\langle i \rangle}$), he still cannot derive $m^{\langle i \rangle}$, because a randomly picked different noise $e^{(j)}$ is used for every $(\vec{a}^{(j)}, b^{(j)})$ and is accumulated over ciphertexts, which exponentially complicates the difficulty of the linear algebra involved in solving $\vec{s}$. Also, in the actual cryptosystem (\autoref{sec:lwe}), the publicly known random vector $\vec{a}^{\langle i \rangle}$ and the secret key $\vec{s}$ are not a single number but a long vector comprising many random numbers. Thus, adding $\vec{a}^{\langle i \rangle }\cdot \vec{s}$ to $\Delta m^{\langle i \rangle} + e^{\langle i \rangle}$ increases the entropy of randomness against the attack. 

$ $

To summarize, lattice-based cryptography hides plaintext by adding the encryption component $\vec{s} \cdot \vec{a}$ to it, along with a small random noise $e$. During decryption, the secret key owner re-creates this encryption component $\vec{a} \cdot \vec{s}$ by using her $\vec{s}$ and removes it. She then removes the noise $e$ using the rounding technique and finally right-shifts the remaining $\Delta m$ by $\text{log}_2 \Delta$ bits to get $m$.

\subsection{RLWE Cryptosystem}
\label{subsec:lattice-scheme2}

In the RLWE cryptosystem, the formula in \tboxlabel{\ref*{subsec:lattice-scheme}} is the same, but $\vec{s}, \vec{a}^{\langle i \rangle}, b^{\langle i \rangle}, m^{\langle i \rangle}, e^{\langle i \rangle}$ are replaced by polynomials $S, A^{\langle i \rangle}, B^{\langle i \rangle}, M^{\langle i \rangle}, E^{\langle i \rangle}$ as follows:

\begin{tcolorbox}[title={\textbf{\tboxlabel{\ref*{subsec:lattice-scheme2}} Lattice-based RLWE Cryptosystem}}]
\begin{itemize}
    \item \textbf{\underline{Encryption}: } $B^{\langle i \rangle} = S \cdot A^{\langle i \rangle} + \Delta \cdot M^{\langle i \rangle} + E^{\langle i \rangle}$, where $B^{\langle i \rangle}$ and $A^{\langle i \rangle}$ are publicly known also to the attacker, while $S, M^{\langle i \rangle}, E^{\langle i \rangle}$ are unknown (only known by the secret key owner). 

    $ $
    
    \item \textbf{\underline{Decryption}: } $ \dfrac{\lceil B^{\langle i \rangle} - S \cdot A^{\langle i \rangle} \rfloor_{\Delta}}{\Delta} = \dfrac{\lceil \Delta M^{\langle i \rangle} + E^{\langle i \rangle} \rfloor_{\Delta}}{\Delta} = M^{\langle i \rangle}$ \\$\Big($provided $||E^{\langle i \rangle}||_{\infty} < \dfrac{\Delta}{2}$, meaning each coefficient of $E^{\langle i \rangle}$ has a magnitude less than $\dfrac{\Delta}{2}\Big)$    

    $ $

    $\lceil  \rfloor_{\Delta}$ is equivalent to rounding each term's coefficient in the polynomial. 

\end{itemize}

\end{tcolorbox}

\clearpage

\section{LWE Cryptosystem}
\label{sec:lwe}

\textbf{- Reference:} 
\href{https://www.zama.ai/post/tfhe-deep-dive-part-1}{TFHE Deep Dive: Part I - Ciphertext types}~\cite{tfhe-1}

$ $

\subsection{Setup}
Let $[0, t-1]$ be the plaintext range, and $[0, q-1]$ the ciphertext range, where $t \ll q$ ($t$ is much smaller than $q$). Randomly pick a vector $\vec{s}$ of length $k$ comprising $k$ ternary numbers sampled from $\{-1, 0, 1\}$ as a secret key (denoted as $\vec{s} \xleftarrow{\$} \{-1, 0, 1\}^k$). Let $\Delta = \dfrac{q}{t}$, the scaling factor of plaintext.

\subsection{Encryption}
\label{subsec:lwe-enc}

\begin{enumerate}
\item Suppose we have a plaintext $m \in \mathbb{Z}_t$ to encrypt. \item Randomly pick a vector $\vec{a} \in \mathbb{Z}_q^k$ (of length $k$) as a one-time random public mask (denoted as $\vec{a} \xleftarrow{\$} \mathbb{Z}_q^k$).
\item Randomly pick a small one-time noise $e \in \mathbb{Z}_q$ sampled from the Gaussian distribution $\chi_\sigma$ (denoted as $e \xleftarrow{\chi_\sigma} \mathbb{Z}_q$). 
\item Scale $m$ by $\Delta$, which is to compute $\Delta \cdot m$. This converts $m \in \mathbb{Z}_t$ into $\Delta \cdot m \in \mathbb{Z}_q$.
\item Compute $b = \vec{a} \cdot \vec{s} + \Delta \cdot m + e \in \mathbb{Z}_q$. 
\end{enumerate}

$ $

The LWE encryption formula is summarized as follows:

$ $

\begin{tcolorbox}[title={\textbf{\tboxlabel{\ref*{subsec:lwe-enc}} LWE Encryption}}]
\textbf{\underline{Initial Setup}:} $\Delta = \dfrac{q}{t}$, $\vec{s} \xleftarrow{\$} \{-1, 0, 1\}^k$, where $t$ divides $q$

$ $

$ $

\textbf{\underline{Encryption Input}:} $m \in \mathbb{Z}_t$, $\vec{a} \xleftarrow{\$} \mathbb{Z}_q^k$, $e \xleftarrow{\chi_\sigma} \mathbb{Z}_q$

$ $

%\textbf{Encryption}

\begin{enumerate}
\item Scale up $m \longrightarrow \Delta \cdot m \text{ } \in \mathbb{Z}_q$

\item Compute $b = \vec{a} \cdot \vec{s} + \Delta  m + e \pmod q$
\item $\textsf{LWE}_{\vec{s},\sigma}(\Delta  m + e) = (\vec{a}, b) \text{ } \in \mathbb{Z}_q^{k + 1}$ 
\end{enumerate}

\end{tcolorbox}

\subsection{Decryption}
\label{subsec:lwe-dec} 

\begin{enumerate}
\item Given the ciphertext $(\vec{a}, b)$ where $b = \vec{a} \cdot \vec{s} + \Delta \cdot m + e \in \mathbb{Z}_q$, compute $b - \vec{a} \cdot \vec{s}$, which gives the same value as $\Delta \cdot m + e \in \mathbb{Z}_q$. 
\item Round $\Delta \cdot m + e \in \mathbb{Z}_q$ to the nearest multiple of $\Delta$ (i.e., round it as a base $\Delta$ number), which is denoted as $\lceil \Delta \cdot m + e \rfloor_{\Delta}$. This rounding operation successfully eliminates $e$ and gives $\Delta m$, provided $e$ is small enough to be $e < \dfrac{\Delta}{2}$. If $e \geq \dfrac{\Delta}{2}$, then some of the higher bits of the noise $e$ will overlap with the plaintext $m$, won't be blown away, and will corrupt some lower bits of the plaintext $m$.
\item Compute $\dfrac{\Delta m} {\Delta}$, which is equivalent to right-shifting $\lceil \Delta \cdot m + e \rfloor_{\Delta}$ by $\text{log}_2 \Delta$ bits. (Here we assume $\Delta$ is a power of 2; if $\Delta$ is not a power of 2, scaling up or down $m$ by $\Delta$ is equivalent to multiplying or dividing the value by $\Delta$.) 
\end{enumerate}

$ $

The LWE decryption formula is summarized as follows:

$ $

\begin{tcolorbox}[title={\textbf{\tboxlabel{\ref*{subsec:lwe-dec}} LWE Decryption}}]

\textbf{\underline{Decryption Input}:} $\textsf{ct} = (\vec{a}, b) \text{ } \in \mathbb{Z}_q^{k+1}$

$ $

%\textbf{Decryption}
\begin{enumerate}
\item $\textsf{LWE}^{-1}_{S,\sigma}(\textsf{ct}) = b - \vec{a}\cdot \vec{s} = \Delta  m + e \pmod q$

\item Scale down $\Bigg\lceil\dfrac{ \Delta  m + e } {\Delta}\Bigg\rfloor \bmod t = m \text{ } \in \mathbb{Z}_t$
\end{enumerate}

For correct decryption, the noise $e$ should be $e < \dfrac{\Delta}{2}$.

\end{tcolorbox}

During decryption, the secret key owner can subtract $\vec{a} \cdot \vec{s}$ from $b$ because he can directly compute $\vec{a} \cdot \vec{s}$ by using his secret key $\vec{s}$. 

The reason we scaled the plaintext $m$ by $\Delta$ is: (i) to left-shift $m$ by $\text{log}_2\Delta$ bits and separate it from the noise $e$ in the lower bits during encryption, whereas $e$ is essential to make it hard for the attacker to guess $m$ or $\vec{s}$; and (ii) to eliminate $e$ in the lower bits by right-shifting it by $\text{log}_2\Delta$ bits without compromising $m$ in the higher bits during decryption. The process of right-shifting (i.e., scaling) the plaintext $m$ by $\text{log}_2\Delta$ bits, followed by adding the noise $e$, is illustrated in \autoref{fig:scaling}.

\subsubsection{In the Case of $t$ not Dividing $q$}
\label{subsubsec:lwe-noise-bound}

In Summary~\ref*{subsec:lwe-enc} (\autoref{subsec:lwe-enc}), we assumed that $t$ divides $q$. In this case, there is no upper or lower limit on the size of plaintext $m$: its value is allowed to wrap around modulo $t$ indefinitely, yet the decryption works correctly. This is because any wrap-around of $m$
modulo $t$ is carried by the mod-$q$ reduction of the scaled plaintext during decryption.

On the other hand, suppose that $t$ does not divide $q$. In such a case, we set the scaling factor as $\Delta = \left\lfloor\dfrac{q}{t}\right\rfloor$. 
Then, provided $q \gg t$, the decryption works correctly even if $m$ is a large value that wraps around $t$. We will show why this is so. 

In this subsection's analysis, we choose $t$ to be an odd (prime) number, which is the general FHE practice for computational efficiency reasons (see \autoref{subsubsec:poly-vector-transformation-modulus}). 
We assume the use of the centered residue system, where the plaintext domain is $\left[-\dfrac{t-1}{2}, \dfrac{t-1}{2}\right]$ and the ciphertext domain is $\left[-\dfrac{q}{2}, \dfrac{q}{2} - 1\right]$.
We denote plaintext $m \bmod t$ as $m = m' + vt$, where $m' \in \mathbb{Z}_t$, and $v$ is an integer that represents the $t$-overflow portions of $m$. We set the plaintext scaling factor as $\Delta = \left\lfloor\dfrac{q}{t}\right\rfloor$. Then, the noise-added and $\Delta$-scaled plaintext can be expressed as follows:

$\left\lfloor\dfrac{q}{t}\right\rfloor\cdot m + e $

$= \left\lfloor\dfrac{q}{t}\right\rfloor\cdot m' + \left\lfloor\dfrac{q}{t}\right\rfloor\cdot vt + e$   \textcolor{red}{ $\rhd$ applying $m = m' + vt$}

$= \left\lfloor\dfrac{q}{t}\right\rfloor\cdot m' + \dfrac{q}{t}\cdot vt - \left(\dfrac{q}{t} - \left\lfloor\dfrac{q}{t}\right\rfloor\right)\cdot vt + e$  

$= \left\lfloor\dfrac{q}{t}\right\rfloor\cdot m' + qv - \left(\dfrac{q}{t} - \left\lfloor\dfrac{q}{t}\right\rfloor\right)\cdot vt + e$

$= \left\lfloor\dfrac{q}{t}\right\rfloor\cdot m' + qv - \epsilon \cdot vt + e$ \textcolor{red}{ $\rhd$ where $\epsilon = \dfrac{q}{t} - \left\lfloor\dfrac{q}{t}\right\rfloor$, a fractional value between $[0, 1)$}

$ $

Remember that the LWE decryption relation: $\dfrac{b - \vec{a}\cdot \vec{s} \bmod q}{\Delta} \bmod t = \dfrac{\Delta m + e \bmod q}{\Delta} \bmod t$. Therefore, from the above expression, we can decrypt the message by computing as follows:

$\left\lceil\dfrac{1}{\Delta} \cdot \left(b - \vec{a}\cdot\vec{s} \bmod q\right)\right\rfloor \bmod t$

$= \left\lceil\dfrac{1}{\Delta} \cdot \left(\Delta m + e \bmod q\right)\right\rfloor \bmod t$

$= \left\lceil\dfrac{1}{\left\lfloor\frac{q}{t}\right\rfloor} \cdot \left(\left\lfloor\dfrac{q}{t}\right\rfloor \cdot (m' + vt) + e \bmod q\right)\right\rfloor \bmod t$

$= \left\lceil\dfrac{1}{\left\lfloor\frac{q}{t}\right\rfloor} \cdot \left(\left\lfloor\dfrac{q}{t}\right\rfloor \cdot m' + \left\lfloor\dfrac{q}{t}\right\rfloor\cdot vt + e \bmod q\right)\right\rfloor \bmod t$

$= \left\lceil\dfrac{1}{\left\lfloor\frac{q}{t}\right\rfloor} \cdot \left(\left\lfloor\dfrac{q}{t}\right\rfloor \cdot m' + \left(\dfrac{q}{t} - \epsilon \right)\cdot vt + e \bmod q\right)\right\rfloor \bmod t$

$= \left\lceil\dfrac{1}{\left\lfloor\frac{q}{t}\right\rfloor} \cdot \left(\left\lfloor\dfrac{q}{t}\right\rfloor \cdot m' + vq - \epsilon vt + e \bmod q\right)\right\rfloor \bmod t$

$= \left\lceil\dfrac{1}{\left\lfloor\frac{q}{t}\right\rfloor} \cdot \left(\left\lfloor\dfrac{q}{t}\right\rfloor\cdot m'  - \epsilon vt + e \bmod q \right)\right\rfloor \bmod t$ 

$ $

In order for the decryption to work, we ideally want to eliminate the inner modulo $q$ reduction. That is, assuming the centered residue system $\left[-\dfrac{q}{2}, \dfrac{q}{2}-1\right]$, we want $-\dfrac{q}{2} \leq \left\lfloor\dfrac{q}{t}\right\rfloor\cdot m'  - \epsilon vt + e < \dfrac{q}{2}$, or more strictly, $\left|\left\lfloor\dfrac{q}{t}\right\rfloor\cdot m'  - \epsilon vt + e\right| < \dfrac{q}{2}$.

$ $

To ensure these conditions hold, we will make a special assumption: $|-\epsilon vt + e| < \dfrac{\Delta}{2}$. Applying this assumption to the expression above, we can derive the following relation:

$\left|\left\lfloor\dfrac{q}{t}\right\rfloor\cdot m'  - \epsilon vt + e\right| $

$ \leq \left|\left\lfloor\dfrac{q}{t}\right\rfloor\cdot m'\right| + \left|- \epsilon vt + e\right|$

$ \leq \left|\left\lfloor\dfrac{q}{t}\right\rfloor\cdot \dfrac{t-1}{2}\right| + \left|- \epsilon vt + e\right|$ \textcolor{red}{ $\rhd$ we assume $t$ is an odd prime, as that's the general FHE practice}

$ \leq \left|\dfrac{q}{t}\cdot \dfrac{t-1}{2}\right| + \left|- \epsilon vt + e\right|$

$ = \dfrac{q}{2} - \dfrac{q}{2t} + \left|- \epsilon vt + e\right|$

$ < \dfrac{q}{2} - \dfrac{q}{2t} + \dfrac{\Delta}{2}$ \textcolor{red}{ $\rhd$ applying our special assumption $|-\epsilon vt + e| < \dfrac{\Delta}{2}$}

$ = \dfrac{q}{2} + \dfrac{1}{2}\cdot \left(\Delta - \dfrac{q}{t}\right)$

$< \dfrac{q}{2}$ \textcolor{red}{ $\rhd$ since $\Delta < \dfrac{q}{t}$}

$ $

Therefore, if we assume $|-\epsilon vt + e| < \dfrac{\Delta}{2}$, then $\left\lfloor\dfrac{q}{t}\right\rfloor\cdot m'  - \epsilon vt + e \bmod q$ can be simplified to $\left\lfloor\dfrac{q}{t}\right\rfloor\cdot m'  - \epsilon vt + e$. We continue with the following derivation:

$\left\lceil\dfrac{1}{\left\lfloor\frac{q}{t}\right\rfloor} \cdot \left(\left\lfloor\dfrac{q}{t}\right\rfloor\cdot m'  - \epsilon vt + e \bmod q \right)\right\rfloor \bmod t$ 

$= \left\lceil\dfrac{1}{\left\lfloor\frac{q}{t}\right\rfloor} \cdot \left(\left\lfloor\dfrac{q}{t}\right\rfloor\cdot m'  - \epsilon vt + e \right)\right\rfloor \bmod t$

$= \left\lceil m' + \dfrac{-\epsilon  vt + e  }{\lfloor\frac{q}{t}\rfloor} \right\rfloor \bmod t$ 

$= m' + \left\lceil \dfrac{-\epsilon vt + e }{\lfloor\frac{q}{t}\rfloor} \right\rfloor \bmod t$ \textcolor{red}{ $\rhd$ since $\lceil m' \rfloor = m'$}

$= m' \bmod t$  \textcolor{red}{ $\rhd$ applying special assumption $|-\epsilon vt + e| < \dfrac{\Delta}{2} = \dfrac{\lfloor\frac{q}{t}\rfloor}{2}$}

To summarize, if we set the plaintext's scaling factor as $\Delta=\left\lfloor\dfrac{q}{t}\right\rfloor$ and $t$ is an odd (prime) number, the decryption works correctly as long as the following  error-bounding condition holds: $|-\epsilon vt + e| < \dfrac{\Delta}{2} = \dfrac{\lfloor\frac{q}{t}\rfloor}{2}$. This condition (i.e.,  decryption) breaks if: (1) the noise $e$ is too large relative to $q$; (2) the plaintext modulus $t$ is too large relative to $q$; or (3) the plaintext value wraps around $t$ too many times (i.e., $v$ is too large). A general solution to ensure all these error bound conditions is to set the ciphertext modulus $q$ to be sufficiently large. To put it differently, if $q \gg t$ and $q \gg e$, then the error bound holds. 

We can generalize the formula for the plaintext's scaling factor in Summary~\ref*{subsec:lwe-enc} (in \autoref{subsec:lwe-enc}) as $\left\lfloor\dfrac{q}{t}\right\rfloor$, where $t$ is an odd (prime) number.

\begin{tcolorbox}[title={\textbf{\tboxlabel{\ref*{subsubsec:lwe-noise-bound}} Noise Budget for an Odd Plaintext Modulus $\bm{t}$}}]

Given the plaintext's scaling factor $\Delta=\left\lfloor\dfrac{q}{t}\right\rfloor$ and $t$ is an odd (prime) number, the LWE decryption works correctly as long as the error-bounding condition holds: 

$|-\epsilon vt + e| < \dfrac{\Delta}{2}$

$ $

, where $\epsilon = \dfrac{q}{t} - \left\lfloor\dfrac{q}{t}\right\rfloor$ is a fractional value between $[0, 1)$, and $v$ accounts for the $t$-overflows of the plaintext $m$.

\end{tcolorbox}

\clearpage

\section{RLWE Cryptosystem}
\label{sec:rlwe}
The RLWE cryptosystem's ciphertext is a tuple $(A, B)$, where $B = S \cdot A + \Delta \cdot M + E$. The random public mask $A$ and the secret key $S$ are $(n-1)$-degree polynomials. The message $M$ and the noise $E$ are $(n-1)$-degree polynomials. Like in LWE, a new random public mask $A$ is created for each ciphertext, whereas the same secret key $S$ is used for all ciphertexts. In this section, we denote each ciphertext instance as $(A, B)$ instead of $(A^{\langle i \rangle}, b^{\langle i \rangle})$ for simplicity.

In RLWE, all polynomials are computed in the polynomial ring $\mathbb{Z}_q[x] / (x^n + 1)$, where $x^n + 1$ is a cyclotomic polynomial with $n = 2^f$ for some integer $f$ and the polynomial coefficients are in $\mathbb{Z}_q$. Thus, all polynomials in RLWE have the coefficient range $\mathbb{Z}_q$ and the maximum polynomial degree of $n -1$. For simplicity, we denote $\mathcal{R}_{\langle n,q \rangle} = \mathbb{Z}_q[x] / (x^n + 1)$.

\subsection{Setup}

Let $t$ be the size of plaintext, and $q$ the size of ciphertext, where $t < q$ ($t$ is much smaller than $q$) and $t | q$ (i.e., $t$ divides $q$). Randomly pick a $(n-1)$-degree polynomial $S \in \mathcal{R}_{\langle n, q \rangle}$ whose coefficients are either $\{-1, 0, 1\}$ as a secret key. Let $\Delta = \left\lfloor\dfrac{q}{t}\right\rfloor$ be the scaling factor of plaintext.

Notice that RLWE's setup parameters are similar to that of LWE. One difference is that $S$ is not a vector of length $k$ sampled from $\{-1, 0, 1\}$, but an $(n-1)$-degree polynomial encoding $n$ secret coefficients, where each coefficient is a randomly picked ternary number from $\{-1, 0, 1\}$ (denoted as $S \xleftarrow{\$} \mathcal{R}_{\langle n, \textit{tern} \rangle}$).

\subsection{Encryption}
\label{subsec:rlwe-enc}

\begin{enumerate}
\item Suppose we have an $(n-1)$-degree polynomial $M \in \mathcal{R}_{\langle n, t \rangle}$ whose coefficients represent the plaintext numbers to encrypt. 
\item Randomly pick an $(n-1)$-degree polynomial $A \in \mathcal{R}_{\langle n,q \rangle}$ as a one-time random public mask (denoted as $A \xleftarrow{\$} \mathcal{R}_{\langle n, q \rangle}$). 
\item Randomly pick a small polynomial $E \in \mathcal{R}_{\langle n,q \rangle}$ as a one-time noise, whose $n$ coefficients are small numbers in $\mathbb{Z}_q$ randomly sampled from the Gaussian distribution $\chi_\sigma$ (denoted as $E \xleftarrow{\chi_\sigma} \mathcal{R}_{\langle n, q \rangle}$). 
\item Scale $M$ by $\Delta$, which is to compute $\Delta \cdot M$. This converts $M \in \mathcal{R}_{\langle n,t \rangle}$ into $\Delta \cdot M \in \mathcal{R}_{\langle n,q \rangle}$.
\item Compute $B = A \cdot S + \Delta \cdot M + E \bmod \mathcal{R}_{\langle n,q \rangle}$ (i.e., reduce the degree by $n$ and the coefficient by modulo $q$). 
\item The final ciphertext is $(A, B)$.
\end{enumerate}

$ $

The RLWE encryption formula is summarized as follows:

$ $
\begin{tcolorbox}[title={\textbf{\tboxlabel{\ref*{subsec:rlwe-enc}} RLWE Encryption}}]
\textbf{\underline{Initial Setup}:} $\Delta = \left\lfloor\dfrac{q}{t}\right\rfloor$, $S \xleftarrow{\$} \mathcal{R}_{\langle n, \textit{tern} \rangle}$

$ $

$ $

\textbf{\underline{Encryption Input}:} $M \in \mathcal{R}_{\langle n, t \rangle}$, $A \xleftarrow{\$} \mathcal{R}_{\langle n, q \rangle}$, $E \xleftarrow{\chi_\sigma} \mathcal{R}_{\langle n, q \rangle}$

$ $

\begin{enumerate}
\item Scale up $M \longrightarrow \Delta  M \text{ } \in \mathcal{R}_{\langle n, q \rangle}$
\item Compute $B = A \cdot S + \Delta  M + E  \text{ } \bmod \mathcal{R}_{\langle n,q \rangle}$
\item $\textsf{RLWE}_{S,\sigma}(\Delta  M + E) = (A, B) \text{ } \in \mathcal{R}_{\langle n,q \rangle}^2$
\end{enumerate}

\end{tcolorbox}

\subsection{Decryption}
\label{subsec:rlwe-dec}

\begin{enumerate}
\item Given the ciphertext $(A, B)$ where $B = A \cdot S + \Delta \cdot M + E \in \mathcal{R}_{\langle n,q \rangle}$, compute $B - A \cdot S = \Delta \cdot M + E$. 
\item Round each coefficient of the polynomial $\Delta \cdot M + E \in \mathcal{R}_{\langle n,q \rangle}$ to the nearest multiple of $\Delta$ (i.e., round it as a base $\Delta$ number), which is denoted as $\lceil \Delta \cdot M + E \rfloor_{\Delta}$.  This rounding operation successfully eliminates $E$ and gives $\Delta \cdot M$. One caveat is that the noise $E$'s each coefficient $e_i$ should be small enough to be $|e_i| < \dfrac{\Delta}{2}$ in order to be eliminated during the rounding. Otherwise, some of $e_i$'s higher bits will overlap and corrupt the plaintext $m_i$ coefficient's lower bits and won't be blown away.
\item Compute $\dfrac{\Delta \cdot M}{\Delta}$, which is equivalent to scaling down each polynomial coefficient in $\Delta \cdot M$ by $\Delta$ (or right-shifting each coefficient by $\text{log}_2 \Delta$ bits if $\Delta$ is a power of 2).
\end{enumerate}

$ $

In summary, the RLWE decryption formula is summarized as follows:

\begin{tcolorbox}[title={\textbf{\tboxlabel{\ref*{subsec:rlwe-dec}} RLWE Decryption}}]

\textbf{\underline{Decryption Input}:} $\textsf{ct} = (A, B) \text{ } \in \mathcal{R}_{\langle n, q \rangle}^{2}$

$ $

\begin{enumerate}

\item $\textsf{RLWE}^{-1}_{S,\sigma}(\textsf{ct}) = B - A \cdot S = \Delta  M + E \text{ } \in \mathcal{R}_{\langle n,q \rangle}$ 
\item Scale down $\Bigg\lceil\dfrac{ \Delta M + E}{\Delta}\Bigg\rfloor \bmod t = M \text{ } \in \mathcal{R}_{\langle n,t \rangle}$

\end{enumerate}

For correct decryption, every noise coefficient $e_i$ of polynomial $E$ should be: $|e_i| < \dfrac{\Delta}{2}$. And in case $t$ does not divide $q$, $q$ should be sufficiently larger than $t$.

\end{tcolorbox}

\clearpage

\section{GLWE Cryptosystem} 
\label{sec:glwe}
The GLWE cryptosystem is a generalized form to encompass both the LWE and RLWE cryptosystems. The GLWE cryptosystem's ciphertext is a tuple $(\{A_i\}_{i=0}^{k-1}, B)$, where $B = \sum\limits_{i=0}^{k-1}{(A_i \cdot S_i)} + \Delta \cdot M + E$. The public key $\{A_i\}_{i=0}^{k-1}$ and the secret key $\{S_i\}_{i=0}^{k-1}$ are a list of $k$ $(n-1)$-degree polynomials, each. The message $M$ and the noise $E$ are $(n-1)$-degree polynomials, each. Like in LWE and RLWE, a random public mask $A$ is created for each ciphertext, whereas the same secret key $S$ is used for all ciphertexts. In this section, we denote each ciphertext instance as $(\{A_i\}_{i=0}^{k-1}, B)$ instead of $(\{A_i\}_{i=0}^{k-1, \langle j \rangle}, B^{\langle j \rangle})$ for simplicity.

\subsection{Setup}
Let $t$ be the size of plaintext, and $q$ the size of ciphertext, where $t < q$ ($t$ is much smaller than $q$) and $t | q$ (i.e., $t$ divides $q$). Randomly pick a list of $k$ $(n-1)$-degree polynomials as a secret key, where each polynomial coefficient is a randomly picked ternary number in $\{-1, 0, 1\}$ (i.e., $\{S_i\}_{i=0}^{k-1} \xleftarrow{\$} \mathcal{R}_{\langle n, \textit{tern} \rangle}^k$). Let $\Delta = \left\lfloor\dfrac{q}{t}\right\rfloor$ be the scaling factor of plaintext.

Notice that GLWE's setup parameters are similar to that of RLWE. One difference is that $S$ is not an $(n-1)$-degree polynomial encoding $n$ secret coefficients, but a list of $k$ such $(n-1)$-degree polynomials encoding total $n \cdot k$ secret coefficients.

\subsection{Encryption}
\label{subsec:glwe-enc}

Suppose we have an $(n-1)$-degree polynomial $M \in \mathcal{R}_{\langle n, t \rangle}$ whose coefficients represent the plaintext numbers to encrypt.

\begin{enumerate}
\item Randomly pick a list of $k$ $(n-1)$-degree polynomials $\{A_i\}_{i=0}^{k-1} \xleftarrow{\$} \mathcal{R}_{\langle n,q \rangle}^{k}$ as a one-time public key.
\item Randomly pick a small polynomial $E \xleftarrow{\chi_\sigma} \mathcal{R}_{\langle n,q \rangle}$ as a one-time noise, whose $n$ coefficients are small numbers in $\mathbb{Z}_q$ randomly sampled from the Gaussian distribution $\chi_\sigma$. 
\item Scale $M$ by $\Delta$, which is to compute $\Delta \cdot M$. This converts $M \in \mathcal{R}_{\langle n, t \rangle}$ into $\Delta \cdot M \in \mathcal{R}_{\langle n,q \rangle}$.
\item Compute $B = \sum\limits_{i=0}^{k-1}{(A_i \cdot S_i)} + \Delta \cdot M + E \in \mathcal{R}_{\langle n,q \rangle}$. 
\end{enumerate}

$ $

The GLWE encryption formula is summarized as follows:

$ $

\begin{tcolorbox}[title={\textbf{\tboxlabel{\ref*{subsec:glwe-enc}} GLWE Encryption}}]

\textbf{\underline{Initial Setup}:} $\Delta = \left\lfloor\dfrac{q}{t}\right\rfloor$, $\{S_i\}_{i=0}^{k-1} \xleftarrow{\$} \mathcal{R}_{\langle n, \textit{tern} \rangle}^k$

$ $

$ $

\textbf{\underline{Encryption Input}:} $M \in \mathcal{R}_{\langle n, t \rangle}$, $\{A_i\}_{i=0}^{k-1} \xleftarrow{\$} \mathcal{R}_{\langle n,q \rangle}^{k}$, $E \xleftarrow{\chi_\sigma} \mathcal{R}_{\langle n,q \rangle}$

$ $

\begin{enumerate}

\item Scale up $M \longrightarrow \Delta M \text { } \in \mathcal{R}_{\langle n, q\rangle}$

\item Compute $B = \sum\limits_{i=0}^{k-1}{(A_i \cdot S_i)} + \Delta  M + E \text{ } \in \mathcal{R}_{\langle n,q \rangle}$

\item $\textsf{GLWE}_{S,\sigma}(\Delta M + E) = (\{A_i\}_{i=0}^{k-1}, B) \text{ } \in \mathcal{R}_{\langle n,q \rangle}^{k + 1}$

\end{enumerate}

\end{tcolorbox}

\subsection{Decryption}
\label{subsec:glwe-dec}

\begin{enumerate}
\item Given the ciphertext $(\{A_i\}_{i=0}^{k-1}, B)$ where $B = \sum\limits_{i=0}^{k-1}{(A_i \cdot S_i)} + \Delta \cdot M + E \in \mathcal{R}_{\langle n,q \rangle}$, compute $B - \sum\limits_{i=0}^{k-1}{(A_i \cdot S_i)} = \Delta \cdot M + E$. 
\item Round each coefficient of the polynomial $\Delta \cdot M + E \in \mathcal{R}_{\langle n,q \rangle}$ to the nearest multiple of $\Delta$ (i.e., round it as a base $\Delta$ number), which is denoted as $\lceil \Delta \cdot M + E \rfloor_{\Delta}$. This operation successfully eliminates $E$ and gives $\Delta \cdot M$. One caveat is that $E$'s each coefficient $e_i$ has to be $|e_i| < \dfrac{\Delta}{2}$ to be eliminated during the rounding. Otherwise, some of $e_i$'s higher bits will overlap the plaintext $m_i$ coefficient's lower bit and won't be eliminated during decryption, corrupting the plaintext $m_1$.
\item Compute $\dfrac{\Delta \cdot M} {\Delta}$, which is equivalent to scaling down each polynomial coefficient in $\Delta \cdot M$ by $\Delta$. 
\end{enumerate}

$ $

In summary, the GLWE decryption formula is summarized as follows:

$ $

\begin{tcolorbox}[title={\textbf{\tboxlabel{\ref*{subsec:glwe-dec}} GLWE Decryption}}]
\textbf{\underline{Decryption Input}:} $\textsf{ct} = (\{A_i\}_{i=0}^{k-1}, B) \text{ } \in \mathcal{R}_{\langle n,q \rangle}^{k + 1}$

\begin{enumerate}
\item $\textsf{GLWE}^{-1}_{S,\sigma}(\textsf{ct}) = B - \sum\limits_{i=0}^{k-1}{(A_i \cdot S_i)} = \Delta  M + E \text{ } \in \mathcal{R}_{\langle n,q \rangle}$

\item Scale down 
$\Bigl\lceil \dfrac{ \Delta  M + E }{\Delta}\Bigr\rfloor \bmod t = M \text{ } \in \mathcal{R}_{\langle n, t \rangle}$

\end{enumerate}

For correct decryption, every noise coefficient $e_i$ of polynomial $E$ should be: $|e_i| < \dfrac{\Delta}{2}$.

\end{tcolorbox}

\subsubsection{Discussion}
\begin{enumerate}
\item \textbf{LWE} is a special case of GLWE where the polynomial ring's degree $n = 1$. That is, all polynomials in $\{A_i\}_{i=0}^{k-1}, \{S_i\}_{i=0}^{k-1}, E$, and $M$ are zero-degree polynomial constants. Instead, there are $k$ such constants for $A_i$ and $S_i$, so each of them forms a vector.
\item \textbf{RLWE} is a special case of GLWE where $k = 1$. That is, the secret key $S$ is a single polynomial $S_0$, and each encryption is processed by only a single polynomial $A_0$ as a public key.
\item \textbf{Size of $\bm{n}$:} A large polynomial degree $n$ increases the number of the secret key's coefficient terms (i.e., $S_i = s_{i,0} + s_{i,1}X + \cdots + s_{i, n-1}X^{n-1} $), which makes it more difficult to guess the complete secret key. The same applies to the noise polynomial $E$ and the public key polynomials $A_i$, thus making it harder to solve the search-hard problem (\autoref{subsec:lattice-overview}). Also, higher-degree polynomials can encode more plaintext terms in the same plaintext polynomial $M$, improving the throughput efficiency of processing ciphertexts.
%\item \textbf{Size of $\bm{q}$:} Increasing $q$ increases the ciphertext molulus space, making it hard to attack the cryptosystem. 
%Also, a large modulo space creates an enough separation between the plaintext bits and the noise bits in the ciphertext, which reduces the growth rate of the noise bits across ciphertext multiplications (will be discussed in \autoref{subsubsec:glwe-mult-plain-discussion}) and thus we can do a more number of ciphertext multiplications before the noise bits overflow to the plaintext bit area. 
\item \textbf{Size of $\bm{k}$:} A large $k$ increases the number of the secret key polynomials $(S_0, S_1, \ldots, S_k)$ and the number of the one-time public key polynomials $(A_0, A_1,\ldots, A_k)$, which makes it more difficult for the attacker to guess the complete secret keys. Meanwhile, there is only a single $M$ and $E$ polynomials per GLWE ciphertext, regardless of the size of $k$. 
\item \textbf{Reducing the Ciphertext Size:} The public key $\{A_i\}_{i=0}^{k-1}$ has to be created for each ciphertext, which is a big size. To reduce this size, each ciphertext can instead include the seed $d$ for the pseudo-random number generation hash function $H$. Then, the public key can be dynamically computed $k-1$ times upon each encryption \& decryption as $\{H(s), H(H(s)), H(H(H(s)), \ldots \}$. Note that $H$, by nature, generates the same sequence of numbers given the same random initial seed $d$. 
\end{enumerate}

\subsection{An Alternative Version of GLWE}
\label{subsec:glwe-alternative}

The following is an alternative version of \tboxlabel{\ref*{subsec:glwe-enc}}, where the sign of each $A_iS_i$ is flipped in the encryption and decryption formula as follows: 

\begin{tcolorbox}[title={\textbf{\tboxlabel{\ref*{subsec:glwe-alternative}} An Alternative GLWE Cryptosystem}}]

\textbf{\underline{Initial Setup}:} $\Delta = \left\lfloor\dfrac{q}{t}\right\rfloor$, $\{S_i\}_{i=0}^{k-1} \xleftarrow{\$} \mathcal{R}_{\langle n, \textit{tern} \rangle}^k$

$ $

\par\noindent\rule{\textwidth}{0.4pt}

\textbf{\underline{Encryption Input}:} $M \in \mathcal{R}_{\langle n, t \rangle}$, $\{A_i\}_{i=0}^{k-1} \xleftarrow{\$} \mathcal{R}_{\langle n,q \rangle}^{k}$, $E \xleftarrow{\chi_\sigma} \mathcal{R}_{\langle n,q \rangle}$

$ $

\begin{enumerate}
\item Scale up $M \longrightarrow \Delta M \text { } \in \mathcal{R}_{\langle n, q\rangle}$

\item Compute $B = \textcolor{red}{-\sum\limits_{i=0}^{k-1}{(A_i \cdot S_i)}} + \Delta  M + E \text{ } \in \mathcal{R}_{\langle n,q \rangle}$

\item $\textsf{GLWE}_{S,\sigma}(\Delta M + E) = (\{A_i\}_{i=0}^{k-1}, B) \text{ } \in \mathcal{R}_{\langle n,q \rangle}^{k + 1}$ 

\end{enumerate}

\par\noindent\rule{\textwidth}{0.4pt}

\textbf{\underline{Decryption Input}:} $\textsf{ct} = (\{A_i\}_{i=0}^{k-1}, B) \text{ } \in \mathcal{R}_{\langle n,q \rangle}^{k + 1}$

\begin{enumerate}
\item $\textsf{GLWE}^{-1}_{S,\sigma}(\textsf{ct}) = B \text{ } \textcolor{red}{ + \sum\limits_{i=0}^{k-1}{(A_i \cdot S_i)}} = \Delta  M + E \text{ } \in \mathcal{R}_{\langle n,q \rangle}$

\item Scale down 
$\Bigl\lceil \dfrac{ \Delta  M + E }{\Delta}\Bigr\rfloor = M  \text{ } \in \mathcal{R}_{\langle n, t \rangle}$

\end{enumerate}

For correct decryption, every noise coefficient $e_i$ of polynomial $E$ should be: $|e_i| < \dfrac{\Delta}{2}$.

\end{tcolorbox}

Even if the $A_iS_i$ terms flip their signs, the decryption stage cancels out those terms by adding their equivalent double-sign-flipped terms; thus, the same correctness of decryption is preserved as in the original version.

\subsection{Public Key Encryption}
\label{subsec:glwe-public-key-enc}

The encryption scheme in \autoref{subsec:glwe-enc} assumes that it is the secret key owner who encrypts each plaintext. In this section, we explain a public key encryption scheme in which we create a public key counterpart of the secret key. Anyone who knows the public key can encrypt the plaintext in such a way that only the secret key owner can decrypt it. The high-level idea is that a portion of the components to be used in the encryption stage is pre-computed at the setup stage and published as a public key. At the actual encryption stage, the public key is multiplied by an additional randomness ($U$) and added to additional noise ($E_1, \vec{E}_2$) to create unpredictable randomness for each encrypted ciphertext. The actual scheme is as follows: 

\begin{tcolorbox}[title={\textbf{\tboxlabel{\ref*{subsec:glwe-public-key-enc}} GLWE Public Key Encryption}}]

\textbf{\underline{Initial Setup}:} 
\begin{itemize}
\item The scaling factor $\Delta = \left\lfloor\dfrac{q}{t}\right\rfloor$
\item The secret key $\vec{S} = \{S_i\}_{i=0}^{k-1} \xleftarrow{\$} \mathcal{R}_{\langle n, \textit{tern} \rangle}^k$
\item The public key pair $(\mathit{PK}_1, \vv{\mathit{PK}}_2) \in \mathcal{R}_{\langle n, q \rangle}^{k+1}$ is generated as follows:

$\vec{A} = \{A_i\}_{i=0}^{k-1} \xleftarrow{\$} \mathcal{R}_{\langle n, q \rangle}^k$, \text{ } $E \xleftarrow{\chi_{\sigma}} \mathcal{R}_{\langle n, q \rangle}$

$\mathit{PK}_1 = \vec{A} \cdot \vec{S} + E \in \mathcal{R}_{\langle n, q \rangle}$

$\vv{\mathit{PK}}_2 = \vec{A} \in \mathcal{R}_{\langle n, q \rangle}^{k}$

\end{itemize}

\par\noindent\rule{\textwidth}{0.4pt}

\textbf{\underline{Encryption Input}:} $M \in \mathcal{R}_{\langle n, t \rangle}$, \text{ } $U \xleftarrow{\$} \mathcal{R}_{\langle n,\textit{tern} \rangle}, \text{ } E_1 \xleftarrow{\chi_{\sigma}} \mathcal{R}_{\langle n,q \rangle}, \text{ } \vec{E}_2 \xleftarrow{\chi_{\sigma}} \mathcal{R}_{\langle n,q \rangle}^k$

$ $

\begin{enumerate}

\item Scale up $M \longrightarrow \Delta M \text { } \in \mathcal{R}_{\langle n, q\rangle}$

\item Compute the following: 

$B = \mathit{PK}_1\cdot U + \Delta  M + E_1 \text{ } \in \mathcal{R}_{\langle n,q \rangle}$

$\vec{D} = \vv{{\mathit{PK}}}_2 \cdot U + \vec{E}_2 \in \mathcal{R}_{\langle n,q \rangle}^{k}$ \textcolor{red}{ $\rhd$ $\vv{\mathit{PK}}_2 \cdot U$ multiplies each element of $\vv{\mathit{PK}}_2$ by $U$}

\item $\textsf{GLWE}_{S,\sigma}(\Delta M + E_{\mathit{all}}) = (\vec{D}, B) \text{ } \in \mathcal{R}_{\langle n,q \rangle}^{k+1}$  \textcolor{red}{ $\rhd$ where $E_{\mathit{all}} = E \cdot U + E_1 - \vec{E}_2\cdot\vec{S}$}

\end{enumerate}

\par\noindent\rule{\textwidth}{0.4pt}

\textbf{\underline{Decryption Input}:} $\textsf{ct} = (\vec{D}, B) \text{ } \in \mathcal{R}_{\langle n,q \rangle}^{k + 1}$

\begin{enumerate}
\item $\textsf{GLWE}^{-1}_{S,\sigma}(\textsf{ct}) = B - \vec{D} \cdot \vec{S} = \Delta  M + E_{\mathit{all}} \text{ } \in \mathcal{R}_{\langle n,q \rangle}$

\item Scale down 
$\Bigl\lceil \dfrac{ \Delta  M + E_{\mathit{all}} }{\Delta}\Bigr\rfloor = M  \text{ } \in \mathcal{R}_{\langle n, t \rangle}$

\end{enumerate}

For correct decryption, every noise coefficient $e_i$ of polynomial $E_{\mathit{all}}$ should be: $|e_i| < \dfrac{\Delta}{2}$.

\end{tcolorbox}

The equation in the 1st step of the decryption process is derived as follows:

$\textsf{GLWE}^{-1}_{S,\sigma}\bm{(}\text{ } \textsf{ct} = (\vec{D}, B) \text{ }\bm{)} = B - \vec{D}\cdot\vec{S}$

$ = (\mathit{PK}_1\cdot U + \Delta  M + E_1) - (\vv{\mathit{PK}}_2 \cdot U + \vec{E}_2)\cdot \vec{S} $

$=  (\vec{A} \cdot \vec{S} + E)\cdot U + \Delta M + E_1 - (\vec{A}\cdot U)\cdot \vec{S} - \vec{E}_2\cdot\vec{S}$

$=  (U\cdot\vec{A}) \cdot \vec{S} + E \cdot U + \Delta M + E_1 - (U\cdot \vec{A}) \cdot \vec{S} - \vec{E}_2\cdot\vec{S}$

$= \Delta M + E \cdot U + E_1 - \vec{E}_2\cdot\vec{S}$

$= \Delta M + E_{\mathit{all}}$ \textcolor{red}{ $\rhd$ where $E_{\mathit{all}} = E \cdot U + E_1 - \vec{E}_2\cdot\vec{S}$}

$ $

\para{Security:} The GLWE encryption scheme's encryption formula (Summary~\ref{subsec:glwe-enc} in \autoref{subsec:glwe-enc}) is as follows: 

$\textsf{GLWE}_{S, \sigma}(\Delta M + E) = \bm{(} \text{ } \vec{A}, \text{ } B = \vec{A}\cdot\vec{S} + \Delta M + E \text{ }\bm{)}$

$ $

, where the hardness of the LWE and RLWE problems guarantees that guessing $\vec{S}$ is difficult given $\vec{A}$ and $E$ are randomly picked at each encryption. On the other hand, the public key encryption scheme is as follows: 

$\textsf{GLWE}_{S, \sigma}(\Delta M  + E_{\textit{all}}) = \bm{(} \text{ } \vec{D} = \vv{{\mathit{PK}}}_2 \cdot U + \vec{E}_2, \text{ } B = \mathit{PK}_1\cdot U + \Delta  M + E_1 \text{ } \bm{)}$

$ $

, where $\mathit{PK}_1$, $\vv{\mathit{PK}}_2$ are fixed and $U$, $E_1$, $\vec{E}_2$ are randomly picked at each encryption. Given the polynomial degree $n$ is large, both schemes provide the equivalent level of hardness to solve the problem. 

\clearpage

\section{GLev}
\label{sec:glev}
\begin{figure}[h!]
    \centering
  \includegraphics[width=1.0\linewidth]{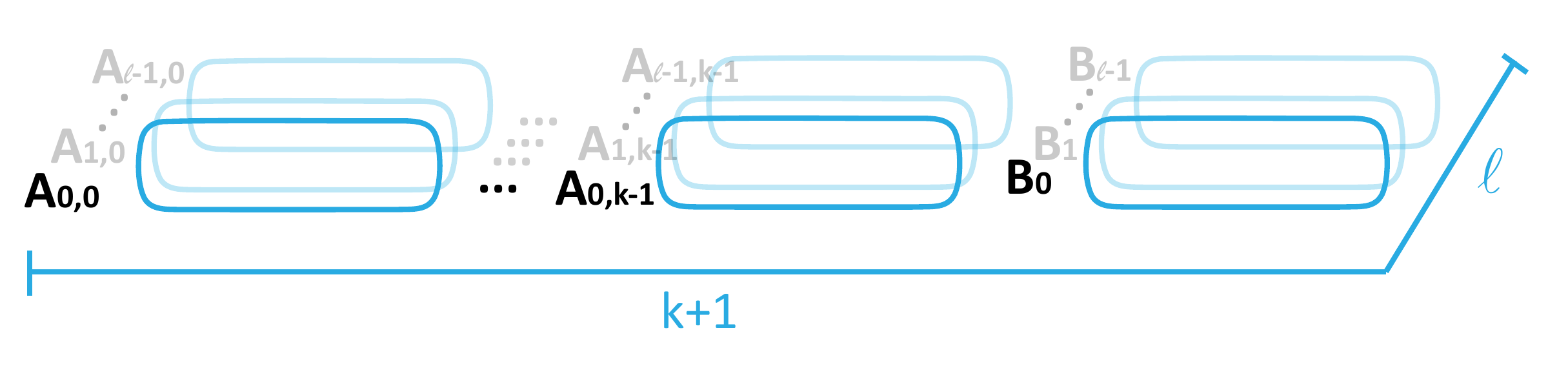}
  \caption{An illustration of a GLev ciphertext }
  \label{fig:glev}
\end{figure}

A GLev ciphertext is a list of GLWE ciphertexts that encrypt the list of plaintexts $\dfrac{q}{\beta^1}M, \dfrac{q}{\beta^2}M, \ldots, \dfrac{q}{\beta^l}M$, where $M$ is a plaintext encoded in a polynomial. Note that each $i$-th GLWE ciphertext of a GLev ciphertext uses a different plaintext scaling factor, which is: $\Delta_i = \dfrac{q}{\beta^i}$. The structure of GLev ciphertext is visually depicted in \autoref{fig:glev}.

Note that $\beta$ should be some value between $t$ and $q$. Specifically, $t$ should be smaller than or equal to $\beta$ because if $t$ is greater than $\beta$, then the higher bits of $M$ will overflow beyond $q$ when computing $\dfrac{q}{\beta^1}M$.

\subsection{Encryption}
\label{subsec:glev-enc}

\begin{tcolorbox}[title={\textbf{\tboxlabel{\ref*{subsec:glev-enc}} GLev Encryption}}]

$\textsf{GLev}_{S, \sigma}^{\beta, l}(M) = \Bigl \{\textsf{GLWE}_{S, \sigma}\left(\dfrac{q}{\beta^i} M + E_i\right)  \Bigr \}_{i=1}^{l} \in \mathcal{R}_{\langle n, q \rangle }^{(k+1) \cdot l}$
\end{tcolorbox}

\subsection{Decryption}

We decrypt the first GLWE ciphertext ($i=1$) using the secret $S$, with the scaling factor $\Delta_1 = \dfrac{q}{\beta}$. This is because while the ciphertext contains $l$ encryptions, the higher indices $i > 1$ have progressively smaller scaling factors $\Delta_i = q/\beta^i$. If $\Delta_i$ becomes smaller than the noise threshold, those specific components cannot be decrypted correctly.

\subsection{Lev and RLev}

Lev is GLev with $n=1$. RLev is GLev with $k=1$.

\clearpage

\section{GGSW}
\label{sec:ggsw}
\begin{figure}[h!]
    \centering
  \includegraphics[width=1.0\linewidth]{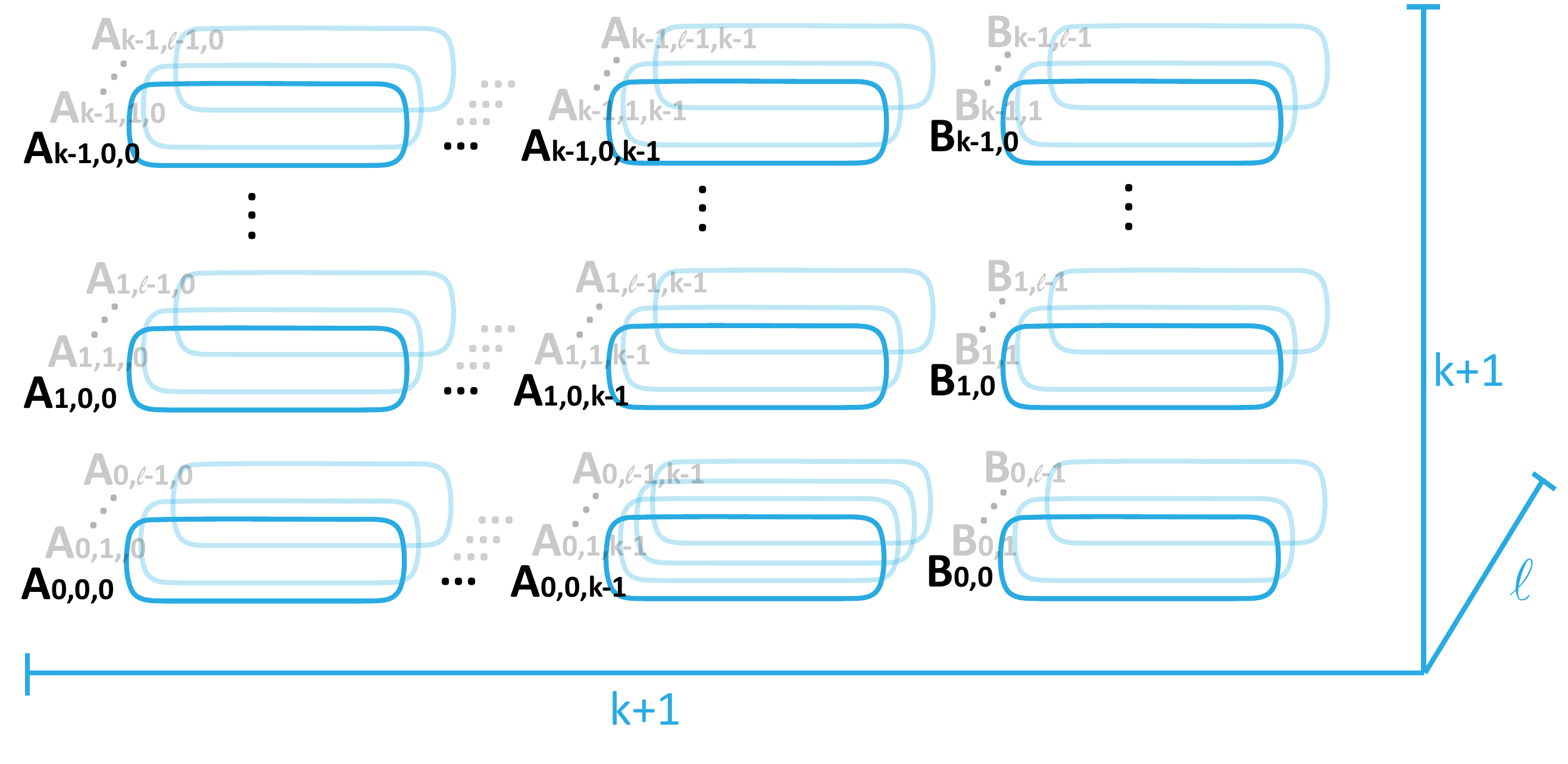}
  \caption{An illustration of a GGSW ciphertext}
  \label{fig:ggsw}
\end{figure}

The GGSW cryptosystem is a list of GLev ciphertexts. In the GGSW cryptosystem, the secret key $S$ is a list of $k$ polynomials (i.e., $S_0, S_1, ... \text{ } S_{k-1}$), and each $i$-th GLev ciphertext in the GGSW ciphertext encrypts the plaintext $-S_0 \cdot M, -S_1 \cdot M, \ldots, -S_{k-1} \cdot M$, and $M$. This is visually depicted in \autoref{fig:ggsw}.

\subsection{Encryption}
\label{subsec:ggsw-enc}

\begin{tcolorbox}[title={\textbf{\tboxlabel{\ref*{subsec:ggsw-enc}} GGSW Encryption}}]
$\textsf{GGSW}_{S, \sigma}^{\beta, l}(M) = \Bigl \{ \{ \textsf{GLev}_{S, \sigma}^{\beta, l}(-S_i \cdot M)  \}_{i=0}^{k-1}, \textsf{GLev}_{S, \sigma}^{\beta, l}(M) \Bigr \} \in \mathcal{R}_{\langle n, q \rangle }^{(k+1) \cdot l \cdot (k+1)}$
\end{tcolorbox}

\subsection{Decryption}

To recover the message $M$, it is sufficient to decrypt the last GLev ciphertext (the one encrypting $M$) using the secret $S$. Decrypting the other rows yields $-S_i \cdot M$, but recovering $M$ from these rows is only possible if $S_i$ is invertible (i.e., $S_i \neq 0$).

\subsection{GSW and RGSW}

GSW is GGSW with $n=1$. RGSW is GGSW with $k=1$.

\clearpage

\part{Generic Fully Homomorphic Encryption}
\label{part:generic-fhe}

\renewcommand{\thesection}{C-\arabic{section}}
\setcounter{section}{0}

This chapter explains the generic techniques of homomorphic computation adopted by various FHE schemes such as TFHE, CKKS, BGV, and BFV,

As we learned from \autoref{sec:glwe}, $\textsf{GLWE}_{S,\sigma}(\Delta M + E) = (A_0, A_1, \gap{$\cdots$} , A_{k-1}, B) \in \mathcal{R}_{\langle n, q \rangle }^{k + 1}$, where $\mathcal{R}_{\langle n,q \rangle} = \mathbb{Z}_q[x] / (x^n + 1)$, and $B$ is computed as $B = \sum\limits_{i=0}^{k-1}{(A_i \cdot S_i)} + \Delta \cdot M + E$. Each $A_i$ is an $(n-1)$-degree polynomial as a public key, whose each coefficient is uniformly randomly sampled from $\mathcal{R}_{\langle n, q \rangle }$. $E$ is an $(n-1)$-degree polynomial as a noise, whose each coefficient is sampled from $\mathcal{R}_{\langle n, q \rangle }$ based on the Gaussian distribution $\chi_\sigma$. $S$ is a list of $k$ $(n-1)$-degree polynomials as a secret key, such that $S = (S_0, S_1, \gap{$\cdots$} S_{k-1} ) \in \mathcal{R}_{\langle n, q \rangle }^k$, and each polynomial $S_i$'s each coefficient is a randomly sampled binary number in $\mathbb{Z}_2$ (i.e., $\{0, 1\}$). 

Based on this GLWE setup, this section will explain the following 5 homomorphic operations: ciphertext-to-ciphertext addition, ciphertext-to-plaintext addition, ciphertext-to-plaintext multiplication, ciphertext-to-ciphertext multiplication, and key switching.

\clearpage

\section{GLWE Ciphertext-to-Ciphertext Addition}
\label{sec:glwe-add-cipher}
\textbf{- Reference:} 
\href{https://www.zama.ai/post/tfhe-deep-dive-part-2}{TFHE Deep Dive - Part II - Encodings and linear leveled operations}~\cite{tfhe-2}

$ $

Suppose we have two GLWE ciphertexts encrypting two different plaintexts $M^{\langle 1 \rangle}, M^{\langle 2 \rangle}$:

$\textsf{GLWE}_{S, \sigma}(\Delta M^{\langle 1 \rangle} + E^{\langle 1 \rangle}) = \textsf{ct}^{\langle 1 \rangle} = ( A_0^{\langle 1 \rangle}, A_1^{\langle 1 \rangle}, \ldots, A_{k-1}^{\langle 1 \rangle}, B^{\langle 1 \rangle}) \in \mathcal{R}_{\langle n,q \rangle}^{k + 1}$

$\textsf{GLWE}_{S, \sigma}(\Delta M^{\langle 2 \rangle} + E^{\langle 2 \rangle}) = \textsf{ct}^{\langle 2 \rangle} = ( A_0^{\langle 2 \rangle}, A_1^{\langle 2 \rangle}, \ldots, A_{k-1}^{\langle 2 \rangle}, B^{\langle 2 \rangle}) \in \mathcal{R}_{\langle n,q \rangle}^{k + 1}$

$ $

\noindent Let's define the following ciphertext addition operation: 

$\textsf{ct}^{\langle 1 \rangle} + \textsf{ct}^{\langle 2 \rangle} = ( A_0^{\langle 1 \rangle} + A_0^{\langle 2 \rangle}, \text{ } A_1^{\langle 1 \rangle} + A_1^{\langle 2 \rangle}, \ldots, A_{k-1}^{\langle 1 \rangle} + A_{k-1}^{\langle 2 \rangle}, \text{ } B^{\langle 1 \rangle} + B^{\langle 2 \rangle} )$

$ $

\noindent Then, the following is true:

\begin{tcolorbox}[title={\textbf{\tboxlabel{\ref*{sec:glwe-add-cipher}} GLWE Homomorphic Addition}}]
$\textsf{GLWE}_{S, \sigma}(\Delta M^{\langle 1 \rangle} + E^{\langle 1 \rangle} ) + \textsf{GLWE}_{S, \sigma}(\Delta M^{\langle 2 \rangle} + E^{\langle 2 \rangle}) $

$ = ( \{A_i^{\langle 1 \rangle}\}_{i=0}^{k-1}, \text{ } B^{\langle 1 \rangle}) + (\{A_i^{\langle 2 \rangle}\}_{i=0}^{k-1}, \text{ } B^{\langle 2 \rangle}) $

$ = ( \{A_i^{\langle 1 \rangle} + A_i^{\langle 2 \rangle}\}_{i=0}^{k-1}, \text{ } B^{\langle 1 \rangle} + B^{\langle 2 \rangle} ) $

$= \textsf{GLWE}_{S, \sigma}(\Delta(M^{\langle 1 \rangle} + M^{\langle 2 \rangle}) + E^{\langle 3 \rangle})$ \textcolor{red}{ $\rhd$ where $E^{\langle 3 \rangle} = E^{\langle 1  \rangle} + E^{\langle 2 \rangle}$}
\end{tcolorbox}

This means that adding two GLWE ciphertexts (each of which encrypts $M^{\langle 1 \rangle}$ and $M^{\langle 2 \rangle}$) and decrypting the resulting ciphertext yields $M^{\langle 1 \rangle} + M^{\langle 2 \rangle}$. 

$ $

%\noindent \textbf{\underline{Proof}}
\begin{myproof}
\begin{enumerate}
\item Define the following notations: \\
$A_0^{\langle 3 \rangle} = A_0^{\langle 1 \rangle} + A_0^{\langle 2 \rangle}$ \\
$A_1^{\langle 3 \rangle} = A_1^{\langle 1 \rangle} + A_1^{\langle 2 \rangle}$ \\
$\vdots$ \\
$A_{k-1}^{\langle 3 \rangle} = A_{k-1}^{\langle 1 \rangle} + A_{k-1}^{\langle 2 \rangle}$ \\
$E^{\langle 3 \rangle} = E^{\langle 1 \rangle} + E^{\langle 2 \rangle}$ \\
$B^{\langle 3 \rangle} = B^{\langle 1 \rangle} + B^{\langle 2 \rangle}$
\item Derive the following: \\
$B^{\langle 3 \rangle} = B^{\langle 1 \rangle} + B^{\langle 2 \rangle}$ \\
$ = \sum\limits_{i=0}^{k-1}{(A_i^{\langle 1 \rangle} \cdot S_i)} + \Delta \cdot M^{\langle 1 \rangle} + E^{\langle 1 \rangle} + \sum\limits_{i=0}^{k-1}{(A_i^{\langle 2 \rangle} \cdot S_i)} + \Delta \cdot M^{\langle 2 \rangle} + E^{\langle 2 \rangle}$ \\ 
$= \sum\limits_{i=0}^{k-1}{((A_i^{\langle 1 \rangle} + A_i^{\langle 2 \rangle}) \cdot S_i)} + \Delta \cdot (M^{\langle 1 \rangle} + M^{\langle 2 \rangle}) + (E^{\langle 1 \rangle} + E^{\langle 2 \rangle})$ \textcolor{red}{ $\rhd$ commutative and distributive rules} \\
$= \sum\limits_{i=0}^{k-1}{(A_i^{\langle 3 \rangle} \cdot S_i)} + \Delta \cdot (M^{\langle 1 \rangle} + M^{\langle 2 \rangle}) + E^{\langle 3 \rangle}$ \\
\item Since $B^{\langle 3 \rangle} = \sum\limits_{i=0}^{k-1}{(A_i^{\langle 3 \rangle} \cdot S_i)} + \Delta \cdot (M^{\langle 1 \rangle} + M^{\langle 2 \rangle}) + E^{\langle 3 \rangle}$, 

this means that $(A_0^{\langle 3 \rangle}, A_1^{\langle 3 \rangle}, A_2^{\langle 3 \rangle}, \ldots, A_{k-1}^{\langle 3 \rangle}, B^{\langle 3 \rangle})$ form the ciphertext: $\textsf{GLWE}_{S, \sigma}(\Delta \cdot (M^{\langle 1 \rangle} + M^{\langle 2 \rangle}) + E^{\langle 3 \rangle})$. \\
\item Thus, \\
$\textsf{GLWE}_{S, \sigma}(\Delta M^{\langle 1 \rangle} + E^{\langle 1 \rangle}) + \textsf{GLWE}_{S, \sigma}(\Delta M^{\langle 2 \rangle} + E^{\langle 2 \rangle})$ \\
$ = ( A_0^{\langle 1 \rangle} + A_0^{\langle 2 \rangle}, \text{ } A_1^{\langle 1 \rangle} + A_1^{\langle 2 \rangle},\ldots, A_{k-1}^{\langle 1 \rangle} + A_{k-1}^{\langle 2 \rangle}, \text{ } B^{\langle 1 \rangle} + B^{\langle 2 \rangle} )$ \\
$ = ( A_0^{\langle 3 \rangle}, A_1^{\langle 3 \rangle}, A_2^{\langle 3 \rangle}, \ldots,  A_{k-1}^{\langle 3 \rangle}, B^{\langle 3 \rangle})$ \\
$ = ( \{A_i^{\langle 3 \rangle}\}_{i=0}^{k-1}, B^{\langle 3 \rangle})$ \\
$= \textsf{GLWE}_{S, \sigma}(\Delta (M^{\langle 1 \rangle} + M^{\langle 2 \rangle}) + E^{\langle 3 \rangle})$

%\begin{flushright}
%\qedsymbol{} 
%\end{flushright}

\end{enumerate}
\end{myproof}

\subsection{Discussion}
\label{subsubsec:glwe-add-cipher-discuss} 

\para{Noise Elimination:} If we decrypt $\textsf{GLWE}_{S, \sigma}(\Delta(M^{\langle 1 \rangle} + M^{\langle 2 \rangle}) + E^{\langle 3 \rangle})$ by using the secret key $S$, then we get the plaintext $M^{\langle 1 \rangle} + M^{\langle 2 \rangle}$. Meanwhile, $A_1^{\langle 3 \rangle}, A_2^{\langle 3 \rangle}, \ldots,  A_{k-1}^{\langle 3 \rangle}, E^{\langle 3 \rangle}$ gets eliminated by decryption (with rounding), regardless of whatever their randomly sampled values were during encryption.

$ $

\para{Noise Growth:} Note that after decryption, the original ciphertext $C$'s noise has increased from $E^{\langle 1 \rangle}$ and $E^{\langle 2 \rangle}$ to $E^{\langle 3 \rangle} = E^{\langle 1 \rangle} + E^{\langle 2 \rangle}$. However, if the noise is sampled from a Gaussian distribution with the mean $\mu = 0$, then the noise variance grows linearly with the number of additions and eventually consumes the budget $\Delta/2$. However, this growth rate is significantly lower than that of homomorphic multiplication, where noise typically grows multiplicatively.

$ $

\para{Hard Threshold on the Plaintext's Value Without Modulo Reduction $t$:} During homomorphic operations (e.g., addition or multiplication) and decryption, the $AS$ and $B$ terms in the $B = AS + \Delta M + E + vq$ relation are allowed to wrap around modulo $q$ indefinitely, because regardless of whatever their wrapping count is, the final decryption step will always subtract $B$ by $AS$, outputting $\Delta M + E + v'q = \Delta M + E \pmod q$, and the $v'q$ term is always exactly eliminated by modulo reduction by $q$. After that, we can correctly recover $M$ by computing $\left\lceil \dfrac{\Delta M + E \bmod q}{\Delta}\right\rfloor$, eliminating the noise $E$. However, as we learned in Summary~\ref*{subsubsec:lwe-noise-bound} (in \autoref{subsubsec:lwe-noise-bound}), if the error bound $|-\epsilon v t + e| < \dfrac{\Delta}{2}$ breaks (where $e$ can be any coefficient of $E$), then modulo reduction by $q$ starts to contaminate the scaled plaintext bits. This violation of the error bound occurs when the noise $e$ grows too much over homomorphic operations, or the ciphertext modulus $q$ is not sufficiently larger than the plaintext modulus $t$. If $q \gg t$, the scheme can take on a big $vt$ value (i.e., the plaintext value can wrap around the plaintext modulus $t$ many times across its homomorphic operations). The error bound constraint  $\dfrac{vt + e}{\lfloor\frac{q}{t}\rfloor} < \dfrac{1}{2}$ is used in the BFV scheme.  %This requirement on the valid range of the plaintext term is needed also to accomplish correct modulus switch of FHE ciphertexts to be later explained in \autoref{sec:modulus-rescaling}.  

\clearpage

\section{GLWE Ciphertext-to-Plaintext Addition}
\label{sec:glwe-add-plain}
Suppose we have a GLWE ciphertext \textsf{ct} and a new plaintext polynomial $\Lambda$ as follows:

$\textsf{ct} = \textsf{GLWE}_{S, \sigma}(\Delta M + E) = ( A_0, A_1, \ldots, A_{k-1}, B) \in \mathcal{R}_{\langle n,q \rangle}^{k + 1}$

$\Lambda$: a new plaintext polynomial

$\Delta \Lambda$: a $\Delta$-scaled new plaintext polynomial

$ $

\noindent Let's define the following ciphertext-to-plaintext addition operation: 

$\textsf{ct} + \Delta\Lambda = ( A_0, \text{ } A_1, \ldots, A_{k-1}, \text{ } B + \Delta \Lambda)$

$ $

\noindent Then, the following is true:

\begin{tcolorbox}[title={\textbf{\tboxlabel{\ref*{sec:glwe-add-plain}} GLWE Homomorphic Addition with a Plaintext}}]
$\textsf{GLWE}_{S, \sigma}(\Delta M + E) + \Delta\Lambda $

$=  (\{A_i\}_{i=0}^{k-1}, \text{ } B) + \Delta\Lambda$

$=  (\{A_i\}_{i=0}^{k-1}, \text{ } B + \Delta\Lambda)$

$= \textsf{GLWE}_{S, \sigma}(\Delta (M + \Lambda) + E)$
\end{tcolorbox}

This means that adding a ($\Delta$-scaled) plaintext polynomial $\Lambda$ to a GLWE ciphertext that encrypts $M$ and decrypting it yields $M + \Lambda$.

\begin{myproof}

\begin{enumerate}
\item Since $B = \sum\limits_{i=0}^{k-1}{(A_i \cdot S_i)} + \Delta \cdot M + E$,

$B + \Delta \cdot \Lambda = \sum\limits_{i=0}^{k-1}{(A_i \cdot S_i)} + \Delta \cdot M + E + \Delta \cdot \Lambda = \sum\limits_{i=0}^{k-1}{(A_i \cdot S_i)} + \Delta \cdot (M + \Lambda) + E$  \\

This means that $(A_0, A_1, ... \text{ } A_{k-1}
, B + \Delta \Lambda)$ form the ciphertext $\textsf{GLWE}_{S, \sigma}(\Delta (\Lambda + M) + E)$
\item Thus, \\
$\textsf{GLWE}_{S, \sigma}(\Delta M + E) + \Delta\Lambda$ \\
$ = (A_0, \text { } A_1, ... \text{ } A_{k-1}, \text { } B + \Delta \Lambda)$ \\
$ = (\{A_i\}_{i=0}^{k-1}, \text { } B + \Delta \Lambda)$ \\
$= \textsf{GLWE}_{S, \sigma}(\Delta (M + \Lambda) + E)$

%\begin{flushright}
%\qedsymbol{} 
%\end{flushright}
\end{enumerate}

\end{myproof}

\para{Noise Growth: } Note that after decryption, the original ciphertext $\textsf{ct} +\Delta\Lambda$'s noise $E$ stays the same as before. This means that ciphertext-to-plaintext addition does not increase the noise level.

\clearpage

\section{GLWE Ciphertext-to-Plaintext Multiplication}
\label{sec:glwe-mult-plain}
\textbf{- Reference:} 
\href{https://www.zama.ai/post/tfhe-deep-dive-part-2}{TFHE Deep Dive - Part II - Encodings and linear leveled operations}~\cite{tfhe-2}

$ $

Suppose we have a GLWE ciphertext \textsf{ct}:

$\textsf{ct} = \textsf{GLWE}_{S, \sigma}(\Delta M + E) = ( A_0, A_1, \ldots, A_{k-1}, B) \in \mathcal{R}_{\langle n,q \rangle}^{k + 1}$

$ $

\noindent and a new plaintext polynomial $\Lambda$ as follows: 

$\Lambda = \sum\limits_{i=0}^{n-1}(\Lambda_i \cdot X_i) \in \mathcal{R}_{\langle n, q \rangle}$

$ $

\noindent Let's define the following ciphertext-to-plaintext multiplication operation:

$\Lambda \cdot \textsf{ct} = (\Lambda \cdot A_0, \Lambda \cdot A_1, \ldots, \Lambda \cdot A_{k-1}, \Lambda \cdot B)$

$ $

\noindent We assume that we always do polynomial-to-polynomial multiplications efficiently in $O(n \log n)$ by using the NTT technique (\autoref{sec:ntt}). Then, the following is true: 

\begin{tcolorbox}[title={\textbf{\tboxlabel{\ref*{sec:glwe-mult-plain}} GLWE Ciphertext-to-Plaintext Multiplication}}]
$\Lambda \cdot \textsf{GLWE}_{S, \sigma}(\Delta M + E)$

$= \Lambda \cdot (\{A_i^{\langle 1 \rangle}\}_{i=0}^{k-1}, \text{ } B^{\langle 1 \rangle})$

$= (\{\Lambda\cdot A_i^{\langle 1 \rangle}\}_{i=0}^{k-1}, \text{ } \Lambda \cdot B^{\langle 1 \rangle})$

$= \textsf{GLWE}_{S, \sigma}(\Delta (M \cdot \Lambda) + \Lambda\cdot E )$
\end{tcolorbox}

This means that multiplying a plaintext polynomial $\Lambda$ by a GLWE ciphertext that encrypts $M$  and decrypting it yields $M \cdot \Lambda$.  

$ $

%\noindent \textbf{\underline{Proof}}
\begin{myproof}
\begin{enumerate}
\item Define the following notations: \\
$A_0' = \Lambda \cdot A_0$ \\
$A_1' = \Lambda \cdot A_1$ \\
$\vdots$ \\
$A_{k-1}' = \Lambda \cdot A_{k-1}$ \\
$E' = \Lambda \cdot E$ \\
$B' = \Lambda \cdot B$ \\
\item Derive the following: \\
$B' = \Lambda \cdot B$ \\
$= \Lambda \cdot (\sum\limits_{i=0}^{k-1}{(A_i \cdot S_i)} + \Delta \cdot M + E)$ 
$= \sum\limits_{i=0}^{k-1}{(\Lambda \cdot A_i \cdot S_i)} + \Delta \cdot \Lambda \cdot M + \Lambda \cdot E$   \\ \textcolor{red}{ $\rhd$ by the distributive property of a polynomial ring} \\
$= \sum\limits_{i=0}^{k-1}{((\Lambda \cdot A_i) \cdot S_i)} + \Delta \cdot (\Lambda \cdot M) + (\Lambda \cdot E)$  \\
$= \sum\limits_{i=0}^{k-1}{(A_i' \cdot S_i)} + \Delta \cdot (\Lambda \cdot M) + (E')$ \\
\item Since $B' = \sum\limits_{i=0}^{k-1}{(A_i' \cdot S_i)} + \Delta \cdot (\Lambda \cdot M) + (E')$, 

$(A_0', A_1', \ldots, A_{k-1}'
, B')$ form the ciphertext $\textsf{GLWE}_{S, \sigma}(\Delta \cdot \Lambda \cdot M)$.
\item Thus, \\
$\Lambda \cdot \textsf{GLWE}_{S, \sigma}(\Delta M + E)$ \\
$ = (\Lambda \cdot A_0, \text { } \Lambda \cdot A_1, \ldots, \Lambda \cdot A_{k-1}, \text { } \Lambda \cdot B)$ \\
$ = ( \{A'_{i}\}_{i=0}^{k-1}, \text { } \Lambda \cdot B)$ \\
$= \textsf{GLWE}_{S, \sigma}(\Delta (M \cdot \Lambda) + \Lambda \cdot E)$

%\begin{flushright}
%\qedsymbol{} 
%\end{flushright}
\end{enumerate}
\end{myproof}

If we decrypt $\textsf{GLWE}_{S, \sigma}(\Delta \cdot \Lambda \cdot M + \Lambda \cdot E)$ by using $S$, then we get the plaintext $\Lambda \cdot M$. Meanwhile, $A_0', A_1', \ldots, A_{k-1}', E'$ get eliminated by rounding during decryption, regardless of whatever their values were randomly sampled during encryption. 

The noise is a bigger problem now, because after decryption, the original ciphertext \textsf{ct}'s noise has increased from $E$ to $E' = \Lambda \cdot E$. This means that if we continue multiplication computations without decrypting the ciphertext to eliminate the noise $E'$, it will continue growing more and eventually the noise in the lower bit area in $B$ will overflow to the scaled plaintext bit area. If this happens, the noise $E'$ won't be eliminated during decryption, ending up corrupting the plaintext $M$. Therefore, if the constant $\Lambda$ is big, it is recommended to use gadget decomposition (\autoref{subsec:gadget-decomposition}), which we will explain in the next subsection. 

\subsection{Gadget Decomposition for Noise Suppression}
\label{subsubsec:gadget-decomposition-noise-suppression}

In the ciphertext-to-plaintext multiplication $\Lambda \cdot \textsf{GLWE}_{S, \sigma}(\Delta M)$, the noise $E$ grows to $E' = \Lambda \cdot E$. To limit this noise growth, we introduce a technique based on decomposing $\Lambda$ (\autoref{subsec:number-decomp}) and a GLev encryption (\autoref{subsec:glev-enc}) of $M$ as follows:

$\Lambda = \Lambda_1 \dfrac{q}{\beta^1} + \Lambda_2 \dfrac{q}{\beta^2} + \cdots + \Lambda_l \dfrac{q}{\beta^l} \longrightarrow \textsf{Decomp}^{\beta, l}(\Lambda) = (\Lambda_1, \Lambda_2, \cdots, \Lambda_l)$

$ $

$\textsf{GLev}_{S, \sigma}^{\beta, l}(\Delta M) = \Bigg\{ \textsf{GLWE}_{S, \sigma}\left(\Delta M \dfrac{q}{\beta^1} + E_1\right), \textsf{GLWE}_{S, \sigma}\left(\Delta M \dfrac{q}{\beta^2} + E_2\right), \cdots \textsf{GLWE}_{S, \sigma}\left(\Delta M \dfrac{q}{\beta^l} + E_l\right) \Bigg\}$

$ $

We will encrypt the plaintext $M$ as $\textsf{GLev}_{S, \sigma}^{\beta, l}(\Delta M)$ instead of $\textsf{GLWE}_{S, \sigma}(\Delta M)$, and compute $\textsf{Decomp}^{\beta, l}(\Lambda) \cdot \textsf{GLev}_{S, \sigma}^{\beta, l}(\Delta M)$ instead of $\Lambda \cdot \textsf{GLWE}_{S, \sigma}(\Delta M)$. Notice that the results of both computations are the same as follows:

$\textsf{Decomp}^{\beta, l}(\Lambda) \cdot \textsf{GLev}_{S, \sigma}^{\beta, l}(\Delta M)$

$= (\Lambda_1, \Lambda_2, \cdots, \Lambda_l) \cdot \left (\textsf{GLWE}_{S, \sigma}\left(\dfrac{q}{\beta} \Delta M + E_1\right), \text{ } \textsf{GLWE}_{S, \sigma}\left(\dfrac{q}{\beta^2} \Delta M + E_2\right), \text{ } \cdots, \text{ } \textsf{GLWE}_{S, \sigma}\left(\dfrac{q}{\beta^l} \Delta M + E_l\right) \right )$

$= \Lambda_1\cdot\textsf{GLWE}_{S, \sigma}\left(\dfrac{q}{\beta} \Delta M + E_1\right) +  \Lambda_2\cdot\textsf{GLWE}_{S, \sigma}\left(\dfrac{q}{\beta^2} \Delta M + E_2\right) + \cdots + \Lambda_l\cdot\textsf{GLWE}_{S, \sigma}\left(\dfrac{q}{\beta^l} \Delta M + E_l\right)$

$= \textsf{GLWE}_{S, \sigma}\left(\Lambda_1\cdot\dfrac{q}{\beta}\Delta M + \Lambda_1 E_1\right) +\textsf{GLWE}_{S, \sigma}\left(\Lambda_2\cdot\dfrac{q}{\beta^2}\Delta M + \Lambda_2 E_2\right)+ \cdots + \textsf{GLWE}_{S, \sigma}\left(\Lambda_l\cdot\dfrac{q}{\beta^l}\Delta M + \Lambda_l E_l\right)$

$= \textsf{GLWE}_{S, \sigma}\left(\Lambda_1\cdot\dfrac{q}{\beta} \Delta M + \Lambda_2\cdot\dfrac{q}{\beta^2} \Delta M + \cdots + \Lambda_l\cdot\dfrac{q}{\beta^l} \Delta M\right)$

$= \textsf{GLWE}_{S, \sigma}\left(\left(\Lambda_1\cdot\dfrac{q}{\beta} + \Lambda_2\cdot\dfrac{q}{\beta^2} + \cdots + \Lambda_l\cdot\dfrac{q}{\beta^l}\right)\cdot \Delta M  + E_{\textit{all}} \right)$ \textcolor{red}{ $\rhd$ where $E_{\textit{all}} = \sum\limits_{i=1}^l \Lambda_iE_i$}

$= \textsf{GLWE}_{S, \sigma}\left(\Lambda \cdot \Delta M + E_{\textit{all}}\right)$ \textcolor{red}{ $\rhd$ whose decryption is $\Lambda\cdot M$}

$ $

\begin{figure}[h!]
    \centering
  \includegraphics[width=0.8\linewidth]{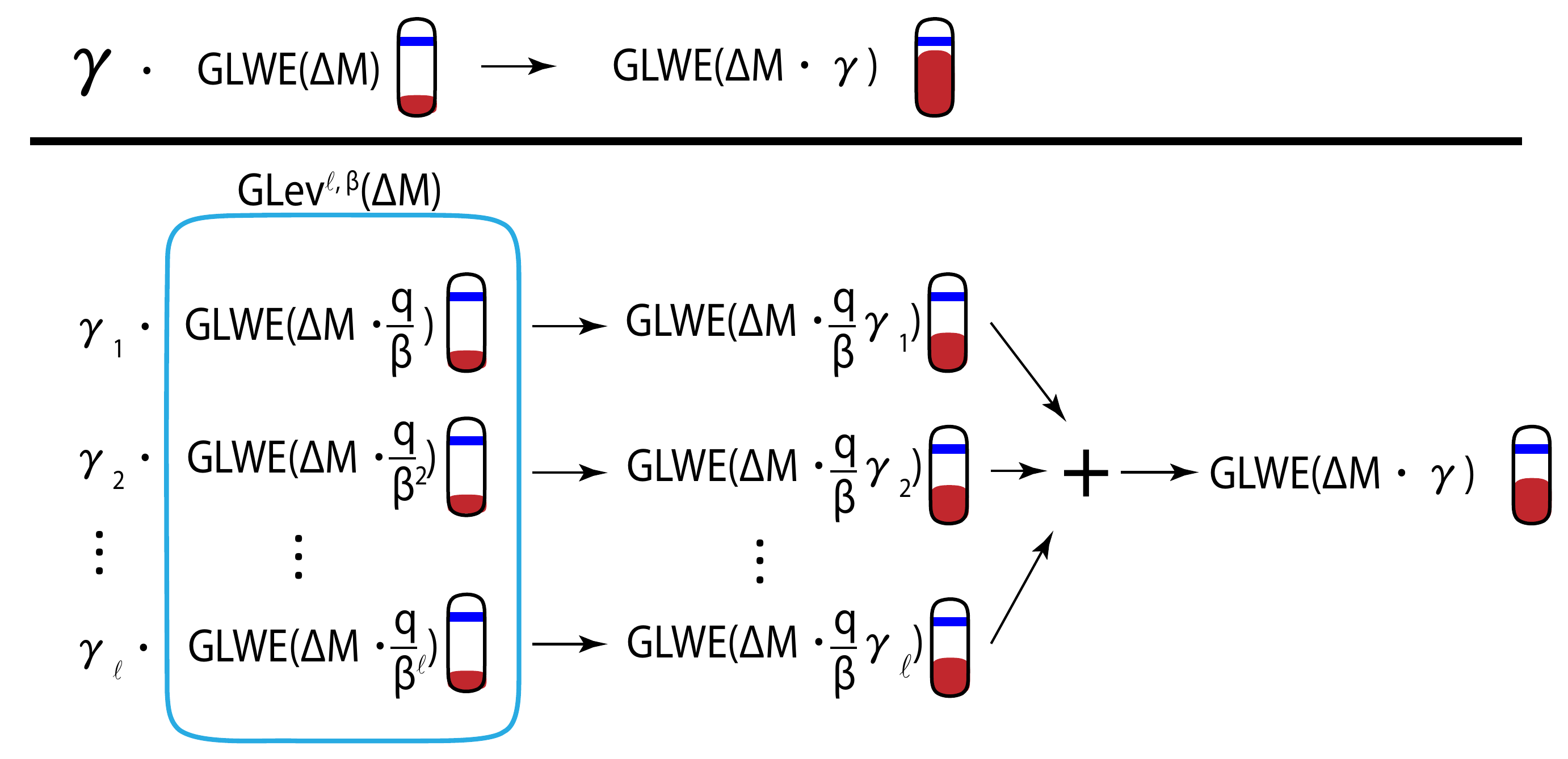}
  \caption{Noise reduction in ciphertext-to-plaintext multiplication by gadget decomposition.}
  \label{fig:decomp2}
\end{figure}

While the decrypted results are the same, as we decompose $\Lambda$ into smaller plaintext polynomials $\Lambda_1, \Lambda_2, \cdots, \Lambda_l$, the noise generated by each of $l$ plaintext-to-ciphertext multiplications becomes smaller.  Given the noise of each GLWE ciphertext in the GLev ciphertext is $E_i$, the final noise of the ciphertext-to-plaintext multiplication is $E_{\textit{all}} = \sum\limits_{i=1}^{l}\Lambda_i\cdot E_i$, which is much smaller than $\Lambda \cdot E$, because 
the coefficients of each decomposed polynomial $\Lambda_i$ are significantly smaller than those of $\Lambda$ (i.e., 
$\|\Lambda_i\|_\infty \le \beta/2$, whereas $\|\Lambda\|_\infty$ can be as large as $q/2$). This is visually depicted in~\autoref{fig:decomp2}.

\subsubsection{Discussion}
\label{subsubsec:glwe-mult-plain-discuss}

Nevertheless, the decomposition technique is still very useful: for GLWE key-switching (\autoref{sec:glwe-key-switching}), we will show how to key-switch by combining decomposed mask polynomials $\textsf{Decomp}^{\beta,l}(A_i)$
with a precomputed key-switching key
$\textsf{KSK}_i=\textsf{GLev}^{\beta,l}_{S',\sigma}(S_i)$,
so gadget decomposition can be repeatedly leveraged across key-switching calls even though each individual application outputs a standard GLWE ciphertext.

Meanwhile, for the technique to repeatedly re-initialize the noise $E$ of regular ciphertexts, we will describe TFHE's noise bootstrapping technique in \autoref{subsec:tfhe-noise-bootstrapping}.

\clearpage

\section{GLWE Modulus Switching}
\label{sec:modulus-switching}
In the GLWE cryptosystem, modulus switching is a process of changing a ciphertext's modulo domain to a smaller (or larger) one, while ensuring that the ciphertext still decrypts to the same plaintext. For example, suppose we have the ciphertext $\textsf{LWE}_{S, \sigma}(\Delta m) \in \mathbb{Z}_q^{k+1}$. If we switch the ciphertext's modulo from $q \rightarrow q'$, then the ciphertext is converted into $\textsf{LWE}_{S, \sigma}\left(\Delta\dfrac{q'}{q} m + e'\right) \in \mathbb{Z}_{q'}^{k+1}$. The ciphertext's all other components such as the noise ($e$) and public keys ($a_0, a_1, \cdots, a_{k-1}$, $b$) are scaled by $\dfrac{q'}{q}$, becoming $\left\lceil e\dfrac{q'}{q} \right\rfloor$, $\left(\left\lceil a_0\dfrac{q'}{q}\right\rfloor, \left\lceil a_1\dfrac{q'}{q}\right\rfloor, \cdots, \left\lceil a_{k-1}\dfrac{q'}{q}\right\rfloor, \left\lceil b\dfrac{q'}{q}\right\rfloor\right)$. To switch the modulo of an LWE ciphertext, we use the modulo rescaling technique learned from \autoref{sec:modulus-rescaling}. 
The same modulus switching technique can also be applied to RLWE ciphertexts. In this section, we will show how to switch (i.e., rescale) the modulo of LWE and RLWE ciphertexts and prove its correctness.  
%That is, after  LWE/RLWE ciphertexts, we will prove why they are still valid LWE/RLWE ciphertexts.

\subsection{LWE Modulus Switching}
\label{subsec:modulus-switch-lwe}

\textbf{- Reference:} 
\href{https://www.jeremykun.com/2022/07/16/modulus-switching-in-lwe/}{Modulus Switching in LWE}

$ $

Recall that the LWE cryptosystem (\autoref{subsec:lwe-enc}) comprises the following components:

\begin{itemize}
\item \textbf{\underline{Setup}:} $\Delta = \dfrac{q}{t}$, \text{ } $S = (s_0, s_1, \ldots, s_{k-1}) \xleftarrow{\$} \mathbb{Z}_2^{k}$

\item \textbf{\underline{Encryption Input}:} $m \in \mathbb{Z}_t$, \text{ } $\vv{a} = (a_0, a_1, \ldots, a_{k-1}) \xleftarrow{\$} \mathbb{Z}_q^{k}$, \text{ } $e \xleftarrow{\chi_\sigma} \mathbb{Z}_q$

$ $

\item \textbf{\underline{Encryption}:} $\textsf{LWE}_{\vv{s},\sigma}(\Delta  m + e) = (\vv{a}, b) \in \mathbb{Z}_q^{k + 1}$ \text{ } (where $b = \vv{a} \cdot \vv{s} + \Delta  m + e \in \mathbb{Z}_q$)

$ $

\item \textbf{\underline{Decryption}:} $\textsf{LWE}^{-1}_{S,\sigma}(\textsf{ct}) = b - \vv{a}\cdot \vv{s} = \Delta  m + e \text{ } \text{ } \in \mathbb{Z}_q$

\end{itemize}

$ $

In the LWE cryptosystem, modulus switching is a process of converting an original LWE ciphertext's modulo domain to a smaller modulo domain. This can be seen as scaling down all components, except for the plaintext $m$ and the secret key $S$, in the original LWE ciphertext to a smaller domain. This operation preserves the size and integrity of the original plaintext $m$, while the scaling factor $\Delta$ gets reduced to a smaller value $\hat{\Delta}$ and the noise $e$ to a smaller (reduced) noise $\hat{e}$ (note that noise alteration does not affect the original plaintext $m$, because the noise gets rounded away after decryption, anyway), and $\vv{a}$ also gets scaled down to a smaller $\vv{\hat{a}}$. Modulus switching is used for computational efficiency during TFHE's bootstrapping (which will be discussed in \autoref{subsec:tfhe-noise-bootstrapping}). Modulus switching is also used for implementing the ciphertext-to-ciphertext multiplication algorithm in BGV (\autoref{subsec:bfv-mult-cipher}) and CKKS (\autoref{subsec:ckks-mult-cipher}). 

The high-level idea of LWE modulus switch is to rescale the congruence relationship of the LWE scheme. LWE's homomorphic computation algorithms include the following: ciphertext-to-ciphertext addition, ciphertext-to-plaintext addition, ciphertext-to-plaintext multiplication, ciphertext-to-ciphertext multiplication. However, all congruence relationships used in these algorithms are essentially rewritten versions of the following single fundamental congruence relationship: $b = \vv{a}\cdot \vv{s} + \Delta m + e \bmod q$. Thus, modulus switch of an LWE ciphertext from $q \rightarrow q'$ is equivalent to rescaling the modulo of the above congruence relationship from $q \rightarrow q'$.

Based on this insight, the LWE cryptosystem's modulus switching from $q \rightarrow \hat{q}$ (where $q > \hat{q}$) is a process of converting the original LWE ciphertext $\textsf{LWE}_{S, \sigma}(\Delta m + e)$ as follows:

\begin{tcolorbox}[title={\textbf{\tboxlabel{\ref*{subsec:modulus-switch-lwe}} LWE Modulus Switching}}]

Given an LWE ciphertext $(\vv{a}, b)$ where $b = \vv{a}\cdot \vv{s} + \Delta m + e \bmod q$ and $m \in \mathbb{Z}_t$, modulus switch of the ciphertext from $q$ to $\hat q$ is equivalent to updating $(A, b)$ to $(\vv{\hat{a}}, \hat b)$ as follows:

$\vv{\hat{a}} = (\hat{a}_0, \hat{a}_1, \ldots, \hat{a}_{k-1})$, where each $\hat{a}_i = \Big\lceil a\dfrac{\hat{q}}{q} \Big\rfloor \in \mathbb{Z}_{\hat{q}}$ \textcolor{red}{ $\rhd$ $\lceil \rfloor$ means rounding to the nearest integer}

 $\hat{b} = \Big\lceil b \dfrac{\hat{q}}{q} \Big\rfloor \in \mathbb{Z}_{\hat{q}}$

$\textsf{LWE}_{{S},\sigma}(\hat{\Delta}  m + \hat{e} + \epsilon_{\textit{all}}) = (\hat{a}_0, \hat{a}_1, \ldots, \hat{b}) \in \mathbb{Z}_{\hat{q}}^{k + 1}$  \textcolor{red}{ $\rhd$  where $\epsilon_{\textit{all}}$ is a small rounding error}

$ $

The above update effectively changes $\hat e$ and $\hat \Delta$ as follows:

$\hat{e} = \Big\lceil e\dfrac{\hat{q}}{q} \Big\rfloor \in \mathbb{Z}_{\hat{q}}$, \text{ }

$\hat{\Delta} = \Delta\dfrac{\hat{q}}{q}$ \textcolor{red}{ $\rhd$ which should be an integer}

$ $

Meanwhile, $S$ and $m$ stay the same as before.

\end{tcolorbox}

$ $

 Note that in order for $(\hat{a}_0, \hat{a}_1, \ldots, \hat{b}) \in \mathbb{Z}_{\hat{q}}^{k + 1}$ to be a valid LWE ciphertext of $\hat{\Delta}m$, we need to prove that the following relationship holds:

$\hat{b} = \sum\limits_{i=0}^{k-1}\hat{a}_i \cdot s_i + \hat{\Delta} m + \hat{e} \in \mathbb{Z}_{\hat{q}}$

$ $

%\textbf{\underline{Proof}}
\begin{myproof}
\begin{enumerate}
\item Note the following: 

$\hat{b} = \Big\lceil b \dfrac{\hat{q}}{q} \Big\rfloor = b\dfrac{\hat{q}}{q} + \epsilon_b$ (where $-0.5 < \epsilon_b < 0.5$, a rounding drift error)

$\hat{a_i} = \Big\lceil a_i \dfrac{\hat{q}}{q} \Big\rfloor = a_i\dfrac{\hat{q}}{q} + \epsilon_{a_i}$ (where $-0.5 < \epsilon_{a_i} < 0.5$)

$\hat{e} = \Big\lceil e \dfrac{\hat{q}}{q} \Big\rfloor = e\dfrac{\hat{q}}{q} + \epsilon_e$ (where $-0.5 < \epsilon_e < 0.5$)
\item Note the following: 

$b = \vv{a} \cdot \vv{s} + \Delta  m + e = \sum\limits_{i=0}^{k-1}(a_is_i) + \Delta m + e  \in \mathbb{Z}_q$

$b = \sum\limits_{i=0}^{k-1}(a_is_i) + \Delta m + e + H \cdot q$ (where modulo $q$ is replaced by adding $H \cdot q$, some multiple of $q$)

\item According to step 1 and 2:

$\hat{b} = b\dfrac{\hat{q}}{q} + \epsilon_b \in \mathbb{Z}_{\hat{q}}$

$= \left(\sum\limits_{i=0}^{k-1}(a_is_i) + \Delta m + e + H \cdot q\right) \cdot \dfrac{\hat{q}}{q}  + \epsilon_b$

$= \dfrac{\hat{q}}{q} \cdot \sum\limits_{i=0}^{k-1}(a_is_i) + \dfrac{\hat{q}}{q} \cdot \Delta m + \dfrac{\hat{q}}{q} \cdot e + \dfrac{\hat{q}}{q} \cdot H \cdot q + \epsilon_b$

$= \sum\limits_{i=0}^{k-1}\left(\dfrac{\hat{q}}{q} \cdot a_is_i\right) + \hat{\Delta} m + (\hat{e} - \epsilon_e) + \hat{q}\cdot H + \epsilon_b$

$= \sum\limits_{i=0}^{k-1}\left((\hat{a}_i - \epsilon_{a_i}) \cdot s_i\right) + \hat{\Delta} m + \hat{e} - \epsilon_e + \hat{q}\cdot H + \epsilon_b$

$= \sum\limits_{i=0}^{k-1}(\hat{a}_is_i - \epsilon_{a_i}s_i) + \hat{\Delta} m + \hat{e} - \epsilon_e + \epsilon_b \in \mathbb{Z}_{\hat{q}}$

$= \sum\limits_{i=0}^{k-1}\hat{a}_is_i + \hat{\Delta} m + \left( \hat{e} - \epsilon_e + \epsilon_b - \sum\limits_{i=0}^{k-1}\epsilon_{a_i}s_i \right) \in \mathbb{Z}_{\hat{q}}$

$= \sum\limits_{i=0}^{k-1}\hat{a}_is_i + \hat{\Delta} m + \hat{e} + \epsilon_{\textit{all}} \in \mathbb{Z}_{\hat{q}}$ \textcolor{red}{ $\rhd$  where $\epsilon_{\textit{all}} = \left(- \epsilon_e + \epsilon_b - \sum\limits_{i=0}^{k-1}\epsilon_{a_i}s_i \right)$}

$ $

The biggest possible value for $\epsilon_{\textit{all}}$ is, 

$\epsilon_{\textit{all}} = |-0.5| + |0.5| + |-0.5 \cdot k| = 1 + 0.5k$ 

So, LWE modulus switching results in an approximate congruence relationship (\autoref{sec:modulus-rescaling}). However, if $\hat \Delta$ is large enough, $\epsilon_{\textit{all}} = 1 + 0.5k$ will be shifted to the right upon LWE decryption and get eliminated, and finally we can recover the original $m$. Also, in practice, the term $\sum\limits_{i=0}^{k-1}\epsilon_{a_i}s_i$ would 
remain small relative to the ciphertext modulus for a sufficiently large $k$, because each $a_i$ is uniformly sampled and $s_i$ is also uniformly sampled. 

\para{Caution:} If $\hat \Delta$ is not large enough, then $\epsilon_{\textit{all}}$ may not get eliminated during decryption and corrupt the plaintext $m$. Also, if $\Delta \rightarrow \hat\Delta$ shrinks too much, then the distance between $\hat\Delta m$ and $\hat e$ would become too narrow and the rounding process of $\hat e = \Big\lceil e \dfrac{\hat{q}}{q} \Big\rfloor$ may end up overlapping the least significant bit of $\hat \Delta m$, corrupting the plaintext. 

$ $

\item To summarize, $\hat{b}$ is approximately as follows:

$\hat{b} = \sum\limits_{i=0}^{k-1}\hat{a}_is_i + \hat{\Delta} m + \hat{e} + \epsilon_{\textit{all}}  \approx \sum\limits_{i=0}^{k-1}\hat{a}_i  s_i + \hat{\Delta} m + \hat{e} \in \mathbb{Z}_{\hat{q}}$

Thus, $(\hat{a}_0, \hat{a}_1, \ldots, \hat{b}) = \textsf{LWE}_{{S},\sigma}(\hat{\Delta} m + \hat{e} + \epsilon_{\textit{all}})$, decrypting which will give us $m$.

%\begin{flushright}
%\qedsymbol{} 
%\end{flushright}

\end{enumerate}
\end{myproof}

\subsection{Example}

Suppose we have the following LWE setup:

$ $

$t = 4$

$q = 64$

$n = 4$

$\Delta = \dfrac{q}{t} = 16$

$m = 1 \in \mathbb{Z}_t$

$S = (s_0, s_1, s_2, s_3) = (0, 1, 1, 0) \in \{-1, 0, 1\}^4$

$A = (a_0, a_1, a_2, a_3) = (-25, 12, -3, 7) \in \mathbb{Z}_q^4$

$e = 1 \in \mathbb{Z}_q$

$b = a_0s_0 + a_1s_1 + a_2s_2 + a_3s_3 + \Delta m + e = 26 \in \mathbb{Z}_q$

$\textsf{LWE}_{S, \sigma}(\Delta m + e) = \textsf{ct} = (a_0, a_1, a_2, a_3, b) = (-25, 12, -3, 7, 26) \in \mathbb{Z}_q^{n+1}$

$ $

Now, suppose we want modulus switching from $q=64$ to $\hat{q} = 32$, which gives:

$\hat{\Delta} = \Delta\cdot\dfrac{32}{64} = 8$

$\hat{e} = \Big\lceil 1 \cdot \dfrac{32}{64} \Big\rfloor = 1$

$\textsf{LWE}_{S, \sigma}(\hat{\Delta} m + \hat{e} + \epsilon_{\textit{all}}) = \hat{\textsf{ct}} = (\hat{a_0}, \hat{a_1}, \hat{a_2}, \hat{a_3}, \hat{b})$

$ = \left(\Big\lceil -25\cdot\dfrac{32}{64}\Big\rfloor, \Big\lceil12\cdot\dfrac{32}{64}\Big\rfloor, \Big\lceil-3\cdot\dfrac{32}{64}\Big\rfloor, \Big\lceil7\cdot\dfrac{32}{64}\Big\rfloor, \Big\lceil26\cdot\dfrac{32}{64}\Big\rfloor\right)$

$ = (-12, 6, -1, 4, 13) \in \mathbb{Z}_{\hat{q}}^{n+1}$

$ $

Now, verify if the following LWE constraint holds:

$\hat{b} = \hat{a}_0s_0 + \hat{a}_1s_1 + \hat{a}_2s_2 + \hat{a}_3s_3 + \hat{\Delta}m + \hat{e} \in \mathbb{Z}_{32}$

$13 = 0 + 6 - 1 + 0 + 8 \cdot 1 + 1 \in \mathbb{Z}_{32}$

$13 \approx 14 \in \mathbb{Z}_{32}$

We got this small difference of 1 due to the rounding drift error of:

$\hat {a_0} = \lceil -12.5 \rfloor = -12$, $\hat {a_2} = \lceil -1.5 \rfloor = -1$, $\hat{a_3} =  \lceil 3.5 \rfloor = 4$, and $\hat{e} = \lceil 0.5 \rfloor = 1 $

$ $

If we solve the LWE decryption formula:

$\hat b - (\hat{a}_0s_0 + \hat{a}_1s_1 + \hat{a}_2s_2 + \hat{a}_3s_3) = 13 - (0(-12) + 1(6) + 1(-1) + 0(4)) = 13 - (6 - 1) = 8 = \hat m + \hat e \in \mathbb{Z}_{32}$

$ $

$m = \left \lceil \dfrac{8}{\hat \Delta} \right \rfloor = \left \lceil \dfrac{8}{8} \right \rfloor = 1$, which is correct.

\subsection{Discussion}
\label{subsubsec:modulus-switch-lwe-discuss}

\begin{figure}[h!]
    \centering
  \includegraphics[width=0.7\linewidth]{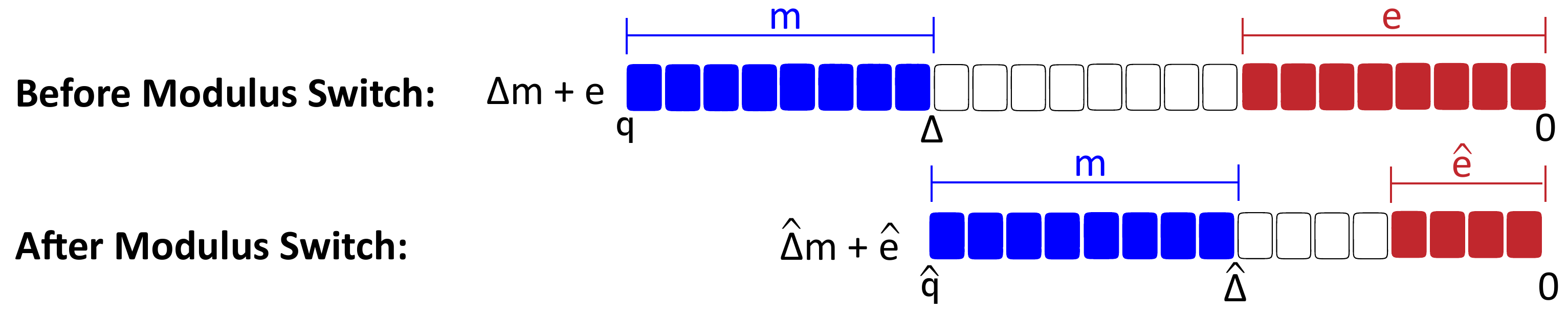}
  \caption{An illustration of scaled plaintext with a noise: $\Delta \cdot m + e \in \mathbb{Z}_q$}
  \label{fig:modulus-switch}
\end{figure}

\para{Reduced Distance between $m$ and $e$:} After modulus switching of an LWE ciphertext from $q \rightarrow \hat{q}$, the underlying plaintext (containing a noise) $\Delta m + e$ gets shrunk to $\hat{\Delta}m + \hat{e}$, as illustrated in \autoref{fig:modulus-switch}. Note that after the modulus switch from $q \rightarrow \hat{q}$, $\Delta m$ is down-scaled to $\hat{\Delta} m$ without losing its bit data. Notably, the plaintext value $m$ stays the same after the modulus switch, while its scaling factor $\Delta$ gets reduced to $\hat{\Delta}$ and the noise $e$ gets reduced to $\hat e$. However, after the modulus switch, the distance between $\hat e$'s MSB and $\hat \Delta m$'s LSB gets reduced compared to the distance between $e$'s MSB and $\Delta m$'s LSB.

%\para{Hard Threshold on the $\Delta m$ Term's Value:} In \autoref{subsubsec:glwe-add-cipher-discuss}, we learned that the value of $\Delta m$ should not overflow or underflow its MSB $q$ during homomorphic computation (i.e., $\Delta m$ should be within the range of modulo $q$ without wrapping around), because if it wraps around the MSB, then our derived congruence relation of modulus switch loses its correctness. This is necessary in order to get the correct functionality of modulus switch on LWE ciphertexts. In particular, if we need to handle the case of $\Delta m$'s overflow, then our congruence relation of modulus switch should include the additional term $H_2 \cdot \Delta$ to take into account the added multiples of $\Delta$ caused by the overflows of $\Delta m$. But problematically, this term does not get eliminated during the modulus switch as $H_2 \cdot \Delta$ is not divisible by $\hat{q}$, which breaks our derived congruence relation above. Therefore, the applications of GLWE homomorphic encryption should ensure to manage their scaled plaintext $\Delta m$ to be always within than its maximum upper and minimum lower boundaries of the ciphertext modulus domain (i.e., $[0, q - 1]$ or $[-\dfrac{q}{2}, \dfrac{q}{2} - 1]$) throughout all computations. Meanwhile, the masking terms $A_i\cdot S_i$ are allowed to overflow or underflow ciphertext indefinitely during homomorphic computations, because during the decryption procedure they will be exactly subtracted and become zeroed out modulo $q$.  

\subsection{RLWE Modulus Switching}
\label{subsec:modulus-switch-rlwe}

RLWE modulus switching is similar to LWE modulus switching. Recall that the RLWE cryptosystem (\autoref{subsec:rlwe-enc}) comprises the following components:

\begin{itemize}
\item \textbf{\underline{Setup}: } 
$\Delta = \dfrac{q}{t}$, \text{ } $S = s_0 + s_1X + s_2X^2 + \cdots + s_{n-1}X^{n-1} \xleftarrow{\$} \mathcal{R}_{\langle n, \textit{tern} \rangle}$

$ $

\item \textbf{\underline{Encryption Input}: } 

$M \in \mathcal{R}_{\langle n, t \rangle}$

$A = a_0 + a_1X + a_2X^2 + \cdots + a_{n-1}X^{n-1} \xleftarrow{\$} \mathcal{R}_{\langle n, q \rangle}$

$E = e_0 + e_1X + e_2X^2 + \cdots + e_{n-1}X^{n-1} \xleftarrow{\chi_\sigma} \mathcal{R}_{\langle n, q \rangle}$ 

$ $

\item \textbf{\underline{Encryption}: } 
$\textsf{RLWE}_{S,\sigma}(\Delta \cdot M + E) = (A, B) \in \mathcal{R}_{\langle n,q \rangle}^2$ 

, where $B = A \cdot S + \Delta \cdot M + E = b_0 + b_1X + b_2X^2 + \cdots + b_{n-1}X^{n-1}$

$ $

\item \textbf{\underline{Decryption}: } $\textsf{RLWE}^{-1}_{S,\sigma}(\textsf{ct}) = B - A \cdot S = \Delta  M + E \text{ } \in \mathcal{R}_{\langle n,q \rangle}$ 
\end{itemize}

$ $

RLWE modulus switching is done as follows:

\begin{tcolorbox}[title={\textbf{\tboxlabel{\ref*{subsec:modulus-switch-rlwe}} RLWE Modulus Switching}}]

For an RLWE ciphertext $(A, B)$ where $B = AS + \Delta M + E$ and $M \in \mathcal{R}_{\langle n, q \rangle}$, modulus switch of the ciphertext from $q$ to $\hat q$ is equivalent to updating $(A, B)$ to $(\hat A, \hat B)$ as follows:

$\hat{A} = \hat{a}_0 + \hat{a}_1X + \hat{a}_2X^2 + \cdots + \hat{a}_{n-1}X^{n-1}$, where each $\hat{a}_i = \Big\lceil a_i\dfrac{\hat{q}}{q} \Big\rfloor \in \mathbb{Z}_{\hat{q}}$ 

$\hat{B} = \hat{b}_0 + \hat{b}_1X + \hat{b}_2X^2 + \cdots + \hat{b}_{n-1}X^{n-1}$, where each $\hat{b}_i = \Big\lceil b_i\dfrac{\hat{q}}{q} \Big\rfloor \in \mathbb{Z}_{\hat{q}}$

$\textsf{RLWE}_{{S},\sigma}(\hat{\Delta}  M + \hat{E} + E^{\langle \epsilon_{\textit{all}} \rangle}) = (\hat{A}, \hat{B}) \in \mathcal{R}_{\langle n, \hat{q} \rangle}^{2}$ 

$ $

The above update effectively changes $\Delta$ and $E$ as follows:

$\hat{\Delta} = \Delta\dfrac{\hat{q}}{q}$ \textcolor{red}{ $\rhd$ which should be an integer}

$\hat{E} = \hat{e}_0 + \hat{e}_1X + \hat{e}_2X^2 + \cdots + \hat{e}_{n-1}X^{n-1}$, where each $\hat{e}_i = \Big\lceil e_i\dfrac{\hat{q}}{q} \Big\rfloor \in \mathbb{Z}_{\hat{q}}$

$ $

Meanwhile, $S$ and $M$ stay the same as before.

\end{tcolorbox}

The proof is similar to that of LWE modulus switching.

$ $

\textbf{\underline{Proof}}

\begin{enumerate}
\item Note the following: 

$\hat{b}_i = \Big\lceil b_i \dfrac{\hat{q}}{q} \Big\rfloor = b_i\dfrac{\hat{q}}{q} + \epsilon_{b_i}$ (where $-0.5 < \epsilon_{b_i} < 0.5$, a rounding drift error)

$\hat{a}_i = \Big\lceil a_i \dfrac{\hat{q}}{q} \Big\rfloor = a_i\dfrac{\hat{q}}{q} + \epsilon_{a_i}$ (where $-0.5 < \epsilon_{a_i} < 0.5$)

$\hat{e}_i = \Big\lceil e_i \dfrac{\hat{q}}{q} \Big\rfloor = e_i\dfrac{\hat{q}}{q} + \epsilon_{e_i}$ (where $-0.5 < \epsilon_{e_i} < 0.5$)

\item Note the following:

$B - A\cdot S$ 

$ = (b_0 + b_1X + \cdots + b_{n-1}X^{n-1} ) - (a_{0} + a_{1}X + \cdots + a_{n-1}X^{n-1})(s_{0} + s_{1}X + \cdots + s_{n-1}X^{n-1})$

$ = \left(b_0 - \left( \sum\limits_{i=0}^{0}(a_{0-i}s_{i}) - \sum\limits_{i=1}^{n-1}(a_{n-i}s_{i}) \right)\right)$

$ + \left(b_1 - \left( \sum\limits_{i=0}^{1}(a_{1-i}s_{i}) - \sum\limits_{i=2}^{n-1}(a_{n+1-i}s_{i})   \right) \right)\cdot X$

$ + \left(b_2 - \left( \sum\limits_{i=0}^{2}(a_{2-i}s_{i}) - \sum\limits_{i=3}^{n-1}(a_{n+2-i}s_{i})   \right) \right)\cdot X^2$

$\vdots$

$ + \left(b_{n-1} - \left(  \sum\limits_{i=0}^{n-1}(a_{n-1-i}s_{i}) -  \sum\limits_{i=n}^{n-1}(a_{n+n-1-i}s_{i})  \right) \right)\cdot X^{n-1}$ \textcolor{red}{ $\rhd$ Grouping the terms by same exponents}

$ $

$= \sum\limits_{h=0}^{n-1}  \left(b_h - \left( \sum\limits_{i=0}^{h}(a_{h-i}s_{i}) -  \sum\limits_{i=h+1}^{n-1}(a_{n+h-i}s_{i})  \right) \right)\cdot X^{h}  $

%$ $
$ $

Thus,

$ $

$B = \sum\limits_{h=0}^{n-1}  b_h  X^{h}  $

$A\cdot S = \sum\limits_{h=0}^{n-1}  \left(\sum\limits_{i=0}^{h}(a_{h-i}s_{i}) - \sum\limits_{i=h+1}^{n-1}(a_{n+h-i}s_{i}) \right)\cdot X^{h}  $

%$= \sum\limits_{h=0}^{n-1}  C_h \cdot X^{h}  $, where $C_h = b_h - \left(  \sum\limits_{i=0}^{h}(a_{h-i}s_{i}) - \sum\limits_{i=h+1}^{n-1}(a_{n+h-i}s_{i})  \right)$

$ $

\item Based on step 2,

$B = A \cdot S + \Delta  M + E$

$\sum\limits_{h=0}^{n-1}  b_h  X^{h}  = \sum\limits_{h=0}^{n-1}  \left(\sum\limits_{i=0}^{h}(a_{h-i}s_{i}) - \sum\limits_{i=h+1}^{n-1}(a_{n+h-i}s_{i}) \right)\cdot X^{h}  + \Delta \sum\limits_{h=0}^{n-1}  m_h  X^{h} + \sum\limits_{h=0}^{n-1}  e_h  X^{h}  \in \mathbb{Z}_q$

$\sum\limits_{h=0}^{n-1}  b_h  X^{h}  = \sum\limits_{h=0}^{n-1}  \left(\sum\limits_{i=0}^{h}(a_{h-i}s_{i}) - \sum\limits_{i=h+1}^{n-1}(a_{n+h-i}s_{i}) \right)\cdot X^{h}  + \Delta \sum\limits_{h=0}^{n-1}  m_h  X^{h} + \sum\limits_{h=0}^{n-1}  e_h  X^{h} + H \cdot q$

(where modulo $q$ is replaced by adding $H \cdot q$, an $(n-1)$-degree polynomial whose each coefficient $c_i$ is some multiple of $q$)

$ $

\item According to step 1 and 3, for each $j$ in $0 \leq j \leq n-1$:

$\hat{b}_j = b_j\dfrac{\hat{q}}{q} + \epsilon_{b_j} \in \mathbb{Z}_{\hat{q}}$

$= \left(\sum\limits_{i=0}^{j}(a_{j-i}s_{i}) - \sum\limits_{i=j+1}^{n-1}(a_{n+j-i}s_{i}) + \Delta m_j + e_j + c_j \cdot q\right) \cdot \dfrac{\hat{q}}{q} + \epsilon_{b_j}$

$= \dfrac{\hat{q}}{q} \sum\limits_{i=0}^{j}(a_{j-i}s_{i}) - \dfrac{\hat{q}}{q} \sum\limits_{i=j+1}^{n-1}(a_{n+j-i}s_{i}) + \dfrac{\hat{q}}{q} \cdot \Delta m_j + \dfrac{\hat{q}}{q} e_j + \dfrac{\hat{q}}{q} \cdot c_j \cdot q + \epsilon_{b_j}$

$=  \sum\limits_{i=0}^{j}\left(\dfrac{\hat{q}}{q}\cdot a_{j-i}s_{i}\right) - \sum\limits_{i=j+1}^{n-1}\left(\dfrac{\hat{q}}{q} \cdot a_{n+j-i}s_{i}\right) + \hat{\Delta} m_j + (\hat{e}_j - \epsilon_{e_j}) + \hat{q} \cdot c_j + \epsilon_{b_j}$

$=  \sum\limits_{i=0}^{j}((\hat{a}_{j-i}-\epsilon_{a_{j-i}})\cdot s_{i}) - \sum\limits_{i=j+1}^{n-1}((\hat{a}_{n+j-i}-\epsilon_{a_{n+j-i}})\cdot s_{i}) + \hat{\Delta} m_j + (\hat{e}_j - \epsilon_{e_j}) + \hat{q} \cdot c_j + \epsilon_{b_j}$

$=  \sum\limits_{i=0}^{j}(\hat{a}_{j-i}s_i) 
- \sum\limits_{i=j+1}^{n-1}(\hat{a}_{n+j-i}s_i)
- \sum\limits_{i=0}^{j}(\epsilon_{a_{j-i}} s_i)
+ \sum\limits_{i=j+1}^{n-1}(\epsilon_{a_{n+j-i}} s_i)
+ \hat{\Delta} m_j + (\hat{e}_j - \epsilon_{e_j}) + \hat{q} \cdot c_j + \epsilon_{b_j}$

$=  \left(\sum\limits_{i=0}^{j}(\hat{a}_{j-i}s_i) 
- \sum\limits_{i=j+1}^{n-1}(\hat{a}_{n+j-i}s_i)\right)
+ \hat{\Delta} m_j + \hat{e}_j 
+ \left(\epsilon_{b_j} - \epsilon_{e_j}
- \sum\limits_{i=0}^{j}(\epsilon_{a_{j-i}} s_i)
+ \sum\limits_{i=j+1}^{n-1}(\epsilon_{a_{n+j-i}} s_i)\right) + \hat{q} \cdot c_j
$

$=  \left(\sum\limits_{i=0}^{j}(\hat{a}_{j-i}s_i) 
- \sum\limits_{i=j+1}^{n-1}(\hat{a}_{n+j-i}s_i)\right)
+ \hat{\Delta} m_j + \hat{e}_j 
+ \epsilon_{\textit{all}} \in \mathbb{Z}_{\hat{q}}
$

\textcolor{red}{ $\rhd$ where $\epsilon_{\textit{all}} = \epsilon_{b_j} - \epsilon_{e_j}
- \sum\limits_{i=0}^{j}(\epsilon_{a_{j-i}} s_i)
+ \sum\limits_{i=j+1}^{n-1}(\epsilon_{a_{n+j-i}} s_i) \approx 0$}

$ $

\item To summarize, for each $0 \leq j \leq n-1 $, each polynomial degree coefficient $\hat{b_j}$ is approximately as follows:

$\hat{b}_j =  \left(\sum\limits_{i=0}^{j}(\hat{a}_{j-i}s_i) 
- \sum\limits_{i=j+1}^{n-1}(\hat{a}_{n+j-i}s_i)\right)
+ \hat{\Delta} m_j + \hat{e}_j 
+ \epsilon_{\textit{all}}$

$ \approx \left(\sum\limits_{i=0}^{j}(\hat{a}_{j-i}s_i) 
- \sum\limits_{i=j+1}^{n-1}(\hat{a}_{n+j-i}s_i)\right)
+ \hat{\Delta} m_j + \hat{e}_j$

Thus, $(\hat{A}, \hat{B}) = \textsf{RLWE}_{{S},\sigma}(\hat{\Delta} M + \hat{E} + E^{\langle \epsilon_{\textit{all}} \rangle})$, decrypting which will give us $M$.

\begin{flushright}
\qedsymbol{} 
\end{flushright}

\end{enumerate}

\subsection{GLWE Modulus Switching}
\label{subsec:modulus-switch-glwe}

GLWE modulus switching is an extension of RLWE modulus switching. The only difference is that while RLWE's $A$ and $S$ are a single polynomial each, GLWE's $A$ and $S$ are a list of $k$ polynomials each. Thus, the same modulus switching technique as RLWE can be applied to GLWE for its $k$ polynomials.  

Recall that the GLWE cryptosystem (\autoref{subsec:glwe-enc}) is comprised of the following components:

\begin{itemize}
\item \textbf{\underline{Initial Setup}: } $\Delta = \dfrac{q}{t}$, \text{ } $\{S_i\}_{i=0}^{k-1} \xleftarrow{\$} \mathcal{R}_{\langle n, \textit{tern} \rangle}^k$

$ $

\item \textbf{\underline{Encryption Input}: } $M \in \mathcal{R}_{\langle n, t \rangle}$, \text{ } $\{A_i\}_{i=0}^{k-1} \xleftarrow{\$} \mathcal{R}_{\langle n,q \rangle}^{k}$, \text{ } $E \xleftarrow{\chi_\sigma} \mathcal{R}_{\langle n,q \rangle}$

$ $

\item \textbf{\underline{Encryption}: } $\textsf{GLWE}_{S,\sigma}(\Delta M + E) = (\{A_i\}_{i=0}^{k-1}, B) \text{ } \in \mathcal{R}_{\langle n,q \rangle}^{k + 1}$ 

, where $B = \sum\limits_{i=0}^{k-1}{(A_i \cdot S_i)} + \Delta  M + E \text{ } \in \mathcal{R}_{\langle n,q \rangle}$

$ $

\item \textbf{\underline{Decryption}: } $\textsf{GLWE}^{-1}_{S,\sigma}(\textsf{ct}) = B - \sum\limits_{i=0}^{k-1}{(A_i \cdot S_i)} = \Delta  M + E \text{ } \in \mathcal{R}_{\langle n,q \rangle}$

\end{itemize}

$ $

GLWE modulus switching is done as follows:

\begin{tcolorbox}[title={\textbf{\tboxlabel{\ref*{subsec:modulus-switch-glwe}} GLWE Modulus Switching}}]

Given a GLWE ciphertext $(\{A_i\}_{i=0}^{k-1}, B)$ where $B = \vv A\cdot \vv S + \Delta M + E \bmod q$ and $M \in \mathcal{R}_{\langle n, q \rangle}$, the modulus switch of the ciphertext from $q$ to $\hat q$ is equivalent to updating $(\{A\}_{i=0}^{k-1}, B)$ to $(\{\hat{A}_i\}_{i=0}^{k-1}, \hat B)$ as follows:

$\hat{A_i} = \hat{a}_{i,0} + \hat{a}_{i,1}X + \hat{a}_{i,2}X^2 + \cdots + \hat{a}_{i, {n-1}}X^{n-1}$, where each $\hat{a}_{i,j} = \Big\lceil a_{i,j}\dfrac{\hat{q}}{q} \Big\rfloor \in \mathbb{Z}_{\hat{q}}$

$\hat{B} = \hat{b}_0 + \hat{b}_1X + \hat{b}_2X^2 + \cdots + \hat{b}_{n-1}X^{n-1}$, where each $\hat{b}_j = \Big\lceil b_j\dfrac{\hat{q}}{q} \Big\rfloor \in \mathbb{Z}_{\hat{q}}$

$\textsf{GLWE}_{{S},\sigma}(\hat{\Delta}  M + \hat{E} + E^{\langle \epsilon_{\textit{all}} \rangle}) = (\{\hat{A}_i\}_{i=0}^{k-1}, \hat{B}) \in \mathcal{R}_{\langle n, \hat{q} \rangle}^{k+1}$ 

$ $

The above update effectively changes $E$ and $\Delta$ as follows:

$\hat{E} = \hat{e}_0 + \hat{e}_1X + \hat{e}_2X^2 + \cdots + \hat{e}_{n-1}X^{n-1}$, where each $\hat{e}_j = \Big\lceil e_j\dfrac{\hat{q}}{q} \Big\rfloor \in \mathbb{Z}_{\hat{q}}$

$\hat{\Delta} = \Delta\dfrac{\hat{q}}{q}$ \textcolor{red}{ $\rhd$ which should be an integer}

$ $

Meanwhile, $\vv{S}$ and $M$ stay the same as before.

\end{tcolorbox}

The proof is similar to that of RLWE modulus switching. The modulus-switched GLWE ciphertext's culminating rounding drift error for each $j$-th polynomial coefficient in its congruence relationship (i.e., $B = \sum\limits_{i=0}^{k-1}A_i\cdot S_i + \Delta M + E$) is as follows:

$\epsilon_{j, all} = \epsilon_{b_j} - \epsilon_{e_j}
- \sum\limits_{l=0}^{k-1}\sum\limits_{i=0}^{j}(\epsilon_{a_{l,j-i}} \cdot s_{l,i})
+ \sum\limits_{l=0}^{k-1}\sum\limits_{i=j+1}^{n-1}(\epsilon_{a_{l,n+j-i}} \cdot s_{l,i})$

\textcolor{red}{ $\rhd$ derived from the proof step 4 of Summary~\ref*{subsec:modulus-switch-rlwe}: $\epsilon_{\textit{all}} = \epsilon_{b_j} - \epsilon_{e_j}
- \sum\limits_{i=0}^{j}(\epsilon_{a_{j-i}} s_i)
+ \sum\limits_{i=j+1}^{n-1}(\epsilon_{a_{n+j-i}} s_i)$}

$ $

Note that GLWE's modulus switching can have a bigger rounding drift error (about $k$ times) than that of RLWE's modulus switching. However, in the long run, the error remains relatively small to the ciphertext modulus, because the rounding errors are independent and uniform and their sum grows slowly (central limit theorem) relative to the modulus.

\clearpage

\section{GLWE Key Switching}
\label{sec:glwe-key-switching}
\textbf{- Reference:} 
\href{https://www.zama.ai/post/tfhe-deep-dive-part-3}{TFHE Deep Dive - Part III - Key switching and leveled multiplications}~\cite{tfhe-3}

$ $

Key switching is a process to change a GLWE ciphertext's secret key from $S$ to a new key $S'$ without decrypting the ciphertext. This is equivalent to converting a ciphertext $\textsf{GLWE}_{S,\sigma}(\Delta M + E)$ into a new ciphertext $\textsf{GLWE}_{S',\sigma}(\Delta M + E - E')$. 

$ $

\noindent Remember that $\textsf{GLWE}_{S,\sigma}(\Delta M + E) = (A_0, A_1, \ldots , A_{k-1}, B) \in \mathcal{R}_{\langle n, q \rangle }^{k + 1}$

, where $B = \sum\limits_{i=0}^{k-1}{(A_i \cdot S_i)} + \Delta \cdot M + E$

, and the secret key $S$ is a list of $k$ polynomials: $S = (S_0, S_1, \ldots, S_{k-1})$

$ $

\noindent Also, remember that GLev (\autoref{sec:glev}) is defined as follows: 

$\textsf{GLev}_{S, \sigma}^{\beta, l}(M) = \Bigl \{\text{GLWE}_{S, \sigma}\Bigl (\dfrac{q}{\beta^i} \cdot M + E_i \Bigr ) \Bigr \}_{i=1}^{l} \in \mathcal{R}_{\langle n, q \rangle }^{(k+1) \cdot l}$

$ $

\noindent Now, let us denote each of the $k$ key-switching keys as follows:

$\mathit{KSK}_i = \textsf{GLev}_{S', \sigma}^{\beta, l}(S_i)$ \\
$ = \left ( \textsf{GLWE}_{S', \sigma} \left ( 
 \dfrac{q}{\beta^1}S_i + E_{i, 1}\right ), \textsf{GLWE}_{S', \sigma} \left ( 
 \dfrac{q}{\beta^2}S_i + E_{i, 2}\right ), \ldots, \textsf{GLWE}_{S', \sigma} \left ( 
 \dfrac{q}{\beta^l}S_i  + E_{i, l} \right) \right ) \in \mathcal{R}_{\langle n, q \rangle }^{(k' + 1) \cdot l}$

$ $

\noindent , which is a list of GLWE encryptions of the secret key $S$ by $S'$, and the new secret key $S'$ is a list of $k'$ polynomials, $(S'_0, S'_1, \ldots, S'_{k'-1})$. Then, the GLWE ciphertext's key can be switched from $S \rightarrow S'$ as follows:

\begin{tcolorbox}[title={\textbf{\tboxlabel{\ref*{sec:glwe-key-switching}} GLWE Key Switching}}]

Given $\textsf{GLWE}_{S,\sigma}(\Delta M + E) = (A, B)$,

$\textsf{GLWE}_{S',\sigma}(\Delta M + E') = (\overbrace{0, 0, \ldots, 0}^{k'}, B) - \sum \limits_{i=0}^{k-1} A_i \cdot S_i$

\text{ } \text{ } $ = (0, 0, \ldots, 0, B) - \sum \limits_{i=0}^{k-1} \langle \textsf{Decomp}^{\beta, l}(A_i), \mathit{KSK}_i \rangle$
\end{tcolorbox}

\begin{myproof}

\begin{enumerate}
\item Given the principle of polynomial decomposition (\autoref{subsec:poly-decomp}) and the polynomial $A_i \in \mathbb{Z}_q[x] / (x^n + 1)$, its decomposition is as follows: 

$ $

$\textsf{Decomp}^{\beta, l}(A_i) = (A_{ i,1 }, A_{ i,2 }, \ldots, A_{ i,l })$, where

$ $

$A_i = A_{ i, 1 } \dfrac{q}{\beta^1} + A _{ i, 2 }\dfrac{q}{\beta^2} + \cdots + A_{ i, l } \dfrac{q}{\beta^l}  $

$ $

\item $\langle \textsf{Decomp}^{\beta, l}(A_i), \mathit{KSK}_i \rangle $ \\
$ =  A_{ i,1 } \cdot \textsf{GLWE}_{S', \sigma} \left ( 
 \dfrac{q}{\beta^1}S_i + E_{i, 1} \right ) + A_{ i,2 } \cdot \textsf{GLWE}_{S', \sigma} \left ( 
 \dfrac{q}{\beta^2}S_i  + E_{i, 2} \right ) + \cdots + A_{ i,l } \cdot \textsf{GLWE}_{S', \sigma} \left ( 
 \dfrac{q}{\beta^l}S_i  + E_{i, l} \right ) $ 

\textcolor{red}{ $\rhd$ where each \textsf{GLWE} ciphertext is an encryption of $S_i\dfrac{q}{\beta}, S_i\dfrac{q}{\beta^2}, \cdots, S_i\dfrac{q}{\beta^l}$ as plaintext with the plaintext scaling factor 1}

$ $
 
$= \textsf{GLWE}_{S', \sigma} \left ( 
 \left ( A_{ i,1 } \dfrac{q}{\beta^1} + A_{ i,2 } \dfrac{q}{\beta^2} + \cdots + A_{ i,l } \dfrac{q}{\beta^l} \right ) \cdot S_i + E_i \right )$ \textcolor{red}{ $\rhd$ where $E_i = \sum\limits_{j=1}^{l} A_{ i, j } E_{i, j}$}

$= \textsf{GLWE}_{S', \sigma} \left ( 
 \left ( A_i \right ) \cdot S_i + E_i \right ) = \textsf{GLWE}_{S', \sigma} ( 
 A_i  S_i  + E_i) $ 

$ $
 
 \item $\sum \limits_{i=0}^{k-1} \langle \textsf{Decomp}^{\beta, l}(A_i), \mathit{KSK}_i \rangle = \sum \limits_{i=0}^{k-1} (\textsf{GLWE}_{S', \sigma} ( 
 A_i  S_i  + E_i)) $ 
 
 $= \textsf{GLWE}_{S', \sigma} \left ( \sum \limits_{i=0}^{k-1}  
 A_i  S_i   + E_i \right ) $

$ $
 
 \item $B$ is equivalent to a trivial GLWE encryption with $S'$, where its every $A_0, A_1, \ldots, A_{k'-1}$ is 0. That is, $\textsf{GLWE}_{S', \sigma}(B) = (\overbrace{0, 0, \ldots}^{k'}, B)$. Note that $\textsf{GLWE}_{S', \sigma}(B)$ can be created without the knowledge of $S'$, because its all $A_iS_i$ terms are 0. 

$ $
 
 \item $\textsf{GLWE}_{S', \sigma}(B) - \textsf{GLWE}_{S', \sigma} \left( \sum \limits_{i=0}^{k-1} (A_i  S_i + E_i )\right) $
 
 $= \textsf{GLWE}_{S', \sigma}(B - \sum \limits_{i=0}^{k-1} (A_i  S_i + E_i)) $
 
 $= \textsf{GLWE}_{S', \sigma}(\Delta M + E')$ \textcolor{red}{ $\rhd$ where $E' = E - \sum\limits_{i=0}^{k-1}E_i$}

$ $

\item If we explicitly expand the above relation, 

$\textsf{GLWE}_{S', \sigma}(B) - \textsf{GLWE}_{S', \sigma} \left(\sum \limits_{i=0}^{k-1} (A_i  S_i + E_i)\right)$

$ = (\overbrace{0, 0, \ldots}^{k'}, B) - (\overbrace{A'_0, A'_1, \ldots, A'_{k'-1}}^{k'}, B')  $ \text { } \textcolor{red}{ $\rhd$ where $B' = \sum\limits_{i=0}^{k'-1} A'_iS'_i + \sum\limits_{i=0}^{k-1} A_iS_i + \sum\limits_{i=0}^{k-1}E_i$}

$(\overbrace{-A'_0, -A'_1, \ldots, -A'_{k'-1}}^{k'}, \text{ } B - B')$

$ $

The decryption of the above ciphertext gives us:

$ \textsf{GLWE}_{S', \sigma}^{-1} \bm{\Big(} \text{ } (\overbrace{-A'_0, -A'_1, \ldots, -A'_{k'-1}}^{k'}, \text{ } B - B') \text{ } \bm{\Big)}$

$ = B -  B' - \sum\limits_{i=0}^{k'-1} -A'_iS_i'$

$= B - \sum\limits_{i=0}^{k'-1} A'_iS'_i - \sum\limits_{i=0}^{k-1} A_iS_i - \sum\limits_{i=0}^{k-1}E_i + \sum\limits_{i=0}^{k'-1} A'_iS_i'$

$= B - \sum\limits_{i=0}^{k-1} A_iS_i - \sum\limits_{i=0}^{k-1}E_i$

$= \Delta M + E - \sum\limits_{i=0}^{k-1}E_i $

$= \Delta M + E' \approx \Delta M + E$ \textcolor{red}{ $\rhd$ where $E' = E - \sum\limits_{i=0}^{k-1}E_i$}

Strictly speaking, $B - \sum\limits_{i=0}^{k-1} A_iS_i = \Delta M + E + Kq$ where $K$ is a polynomial representing the wrap-around coefficient values as multiples of $q$. However, since all the above computations are done in modulo $q$, the $Kq$ term gets eliminated. 

$ $

\item Thus, $ (0, 0, \ldots , B) - \sum \limits_{i=0}^{k-1} \langle \textsf{Decomp}^{\beta, l}(A_i), \mathit{KSK}_i \rangle = \textsf{GLWE}_{S',\sigma}(\Delta M + E') \approx \textsf{GLWE}_{S',\sigma}(\Delta M + E) $ 

$\textsf{GLWE}_{S',\sigma}(\Delta M + E')$ is an encryption of plaintext $\Delta M$ with the plaintext scaling factor 1. However, we can re-interpret this ciphertext as an encryption of plaintext $M$ with the plaintext scaling factor $\Delta$. This way, we can recover the ciphertext's original scaling factor $\Delta$ without any additional computation. 
\end{enumerate}
\end{myproof}

\clearpage

%\section{GLWE Input Vector Rotation}
%\label{sec:glwe-input-rotation}
%\input{c06-glwe-input-rotation}
%\clearpage

\part{Fully Homomorphic Encryption Schemes}
\label{part:fhe-schemes}

\renewcommand{\thesection}{D-\arabic{section}}
\setcounter{section}{0}

This chapter explains the four most well-known FHE schemes: TFHE, CKKS, BGV, and BFV, as well as their RNS-variant versions. 

\newpage

\section{TFHE Scheme}
\label{sec:tfhe}

The TFHE scheme is designed for homomorphic addition and multiplication on integers (especially bit-wise computation, like logic circuits). Unlike BFV, GBV, or CKKS, TFHE is characterized by fast noise bootstrapping; therefore, it is efficient for processing deep multiplication depths. TFHE's noise bootstrapping technique can be further applied to functional encryption.

In TFHE, each plaintext is encrypted as an LWE ciphertext. Therefore, TFHE's ciphertext-to-ciphertext addition, ciphertext-to-plaintext addition, and ciphertext-to-plaintext multiplication are implemented based on GLWE's homomorphic addition and multiplication described in $\autoref{part:generic-fhe}$, with $n = 1$ to make GLWE an LWE.

This section will explain TFHE's novel components: key switching, ciphertext-to-ciphertext multiplication, coefficient extraction, and noise bootstrapping. 

$ $

\begin{tcolorbox}[
    title = \textbf{Required Background},    % box title
    colback = white,    % light background; tweak to taste
    colframe = black,  % frame colour
    boxrule = 0.8pt,     % line thickness
    left = 1mm, right = 1mm, top = 1mm, bottom = 1mm % inner padding
]
\begin{itemize}
  \item \autoref{sec:modulo}: \nameref{sec:modulo}
  \item \autoref{sec:group}: \nameref{sec:group}
  \item \autoref{sec:field}: \nameref{sec:field}
  \item \autoref{sec:order}: \nameref{sec:order}
  \item \autoref{sec:polynomial-ring}: \nameref{sec:polynomial-ring}
  \item \autoref{sec:decomp}: \nameref{sec:decomp}
  \item \autoref{sec:modulus-rescaling}: \nameref{sec:modulus-rescaling}
  \item \autoref{sec:lattice}: \nameref{sec:lattice}
  \item \autoref{sec:lwe}: \nameref{sec:lwe}
  \item \autoref{sec:rlwe}: \nameref{sec:rlwe}
  \item \autoref{sec:glwe}: \nameref{sec:glwe}
  \item \autoref{sec:glev}: \nameref{sec:glev}
  \item \autoref{sec:ggsw}: \nameref{sec:ggsw}
  \item \autoref{sec:glwe-add-cipher}: \nameref{sec:glwe-add-cipher}
  \item \autoref{sec:glwe-add-plain}: \nameref{sec:glwe-add-plain}
  \item \autoref{sec:glwe-mult-plain}: \nameref{sec:glwe-mult-plain}
  \item \autoref{subsec:modulus-switch-lwe}: \nameref{subsec:modulus-switch-lwe}
  \item \autoref{sec:glwe-key-switching}: \nameref{sec:glwe-key-switching}
\end{itemize}
\end{tcolorbox}

\clearpage

\subsection{Encryption and Decryption}
\label{subsec:tfhe-enc-dec}

TFHE encrypts and decrypts ciphertexts based on the LWE cryptosystem (\autoref{sec:lwe}), which is equivalent to the GLWE cryptosystem (\autoref{sec:glwe}) with $n = 1$. However, one distinction from the LWE cryptosystem is that TFHE samples the secret key elements from the binary set $\{0, 1\}$, not from the ternary set $\{-1, 0, 1\}$.

\begin{tcolorbox}[title={\textbf{\tboxlabel{\ref*{subsec:tfhe-enc-dec}} TFHE Encryption and Decryption}}]
\textbf{\underline{Initial Setup}:} $\Delta = \dfrac{q}{t}$, $\vec{s} \xleftarrow{\$} \mathbb{Z}_2^k$  \textcolor{red}{ $\rhd$ where $t$ divides $q$, and each element of $\vec{s}$ is a 0-degree polynomial}

$ $

\par\noindent\rule{\textwidth}{0.4pt}

\textbf{\underline{Encryption Input}:} $m \in \mathbb{Z}_t$, $\vec{a} \xleftarrow{\$} \mathbb{Z}_q^k$, $e \xleftarrow{\chi_\sigma} \mathbb{Z}_q$ \textcolor{red}{ $\rhd$ each element of $\vec{a}$ is a 0-degree polynomial}
\begin{enumerate}
\item Scale up $m \longrightarrow \Delta \cdot m \text{ } \in \mathbb{Z}_q$

\item Compute $b = \vec{a} \cdot \vec{s} + \Delta  m + e \text{ } \bmod q$
\item $\textsf{LWE}_{\vec{s},\sigma}(\Delta  m + e) = (\vec{a}, b) \text{ } \in \mathbb{Z}_q^{k + 1}$ 
\end{enumerate}

\par\noindent\rule{\textwidth}{0.4pt}

\textbf{\underline{Decryption Input}:} $\textsf{ct} = (\vec{a}, b) \text{ } \in \mathbb{Z}_q^{k+1}$
\begin{enumerate}
\item $\textsf{LWE}^{-1}_{\vec{s},\sigma}(\textsf{ct}) = b - \vec{a}\cdot \vec{s} = \Delta  m + e  \pmod q$

\item Scale down $\Bigg\lceil\dfrac{ \Delta  m + e } {\Delta}\Bigg\rfloor = m \text{ } \in \mathbb{Z}_t$ \textcolor{red}{ $\rhd$ i.e., modulus switch from $q \rightarrow t$}
\end{enumerate}

$ $

\textbf{{Condition for Correct Decryption}:}
\begin{itemize}
\item The noise $e$ grown over homomorphic operations should be: $e < \dfrac{\Delta}{2}$. 
\end{itemize}

\end{tcolorbox}

In this section, we will often write $\textsf{LWE}_{\vec{s},\sigma}(\Delta  m + e)$ as $\textsf{LWE}_{\vec{s},\sigma}(\Delta  m)$ for simplicity, because $\textsf{LWE}_{\vec{s},\sigma}(\Delta  m + e) \approx \textsf{LWE}_{\vec{s},\sigma}(\Delta  m)$ (i.e., they decrypt to the same message). Even in the case that we write $\textsf{LWE}_{\vec{s},\sigma}(\Delta  m)$ instead of $\textsf{LWE}_{\vec{s},\sigma}(\Delta  m + e)$, you should assume this as an encryption of $\Delta m + e$ (i.e., the noise is included inside the scaled message).

\subsection{Homomorphic Ciphertext-to-Ciphertext Addition}
\label{subsec:tfhe-add-cipher}

TFHE's ciphertext-to-ciphertext addition uses LWE's ciphertext-to-ciphertext addition scheme, which is equivalent to GLWE's ciphertext-to-ciphertext addition scheme (\autoref{sec:glwe-add-cipher}) with $n = 1$.

\begin{tcolorbox}[title={\textbf{\tboxlabel{\ref*{subsec:tfhe-add-cipher}} TFHE Ciphertext-to-Ciphertext Addition}}]
$\textsf{LWE}_{\vec{s}, \sigma}(\Delta m^{\langle 1 \rangle} + e^{\langle 1 \rangle} ) + \textsf{LWE}_{\vec{s}, \sigma}(\Delta m^{\langle 2 \rangle} + e^{\langle 2 \rangle}) $

$ = ( \vec{a}^{\langle 1 \rangle}, \text{ } b^{\langle 1 \rangle}) + (\vec{a}^{\langle 2 \rangle}, \text{ } b^{\langle 2 \rangle}) $

$ = ( \vec{a}^{\langle 1 \rangle} + \vec{a}^{\langle 2 \rangle}, \text{ } b^{\langle 1 \rangle} + b^{\langle 2 \rangle} ) $

$= \textsf{LWE}_{\vec{s}, \sigma}(\Delta(m^{\langle 1 \rangle} + m^{\langle 2 \rangle}) + e^{\langle 1 + 2 \rangle} )$
\label{Here}
\end{tcolorbox}

\subsection{Homomorphic Ciphertext-to-Plaintext Addition}
\label{subsec:tfhe-add-plain}

TFHE's ciphertext-to-plaintext addition (where $\lambda$ is a constant to add) uses LWE's ciphertext-to-plaintext addition scheme, which is equivalent to GLWE's ciphertext-to-plaintext addition scheme (\autoref{sec:glwe-add-plain}) with $n = 1$.  

\begin{tcolorbox}[title={\textbf{\tboxlabel{\ref*{subsec:tfhe-add-plain}} TFHE Ciphertext-to-Plaintext Addition}}]
$\textsf{LWE}_{\vec{s}, \sigma}(\Delta m + e) + \Delta \lambda $

$=  (\vec{a}, \text{ } b) + \Delta \lambda$

$=  (\vec{a}, \text{ } b + \Delta\lambda)$

$= \textsf{LWE}_{\vec{s}, \sigma}(\Delta (m + \lambda) + e )$
\end{tcolorbox}

\subsection{Homomorphic Ciphertext-to-Plaintext Multiplication}
\label{subsec:tfhe-mult-plain}

TFHE's ciphertext-to-plaintext multiplication uses LWE's ciphertext-to-plaintext multiplication scheme, which is equivalent to GLWE's ciphertext-to-plaintext multiplication scheme (\autoref{sec:glwe-mult-plain}) with $n = 1$.  

\begin{tcolorbox}[title={\textbf{\tboxlabel{\ref*{subsec:tfhe-mult-plain}} TFHE Ciphertext-to-Plaintext Multiplication}}]
$\textsf{LWE}_{\vec{s}, \sigma}(\Delta m + e) \cdot \lambda$

$= (\vec{a}, \text{ } b) \cdot \lambda$

$= (\lambda\cdot \vec{a}, \text{ } \lambda \cdot b)$

$= \textsf{LWE}_{\vec{s}, \sigma}(\Delta (m \cdot \lambda) + \lambda \cdot e )$
\end{tcolorbox}

\subsection{Homomorphic Key Switching}
\label{subsec:tfhe-key-switching}

\textbf{- Reference:} 
\href{https://www.zama.ai/post/tfhe-deep-dive-part-3}{TFHE Deep Dive - Part III - Key switching and leveled multiplications}~\cite{tfhe-3}

TFHE's key switching scheme changes an LWE ciphertext's secret key from $\vec{s}$ to $\vec{s}_{'}$, where the two key vectors may or may not have the same dimensions. This scheme is essentially LWE's key switching scheme. Specifically, this is equivalent to the alternative GLWE version's (\autoref{subsec:glwe-alternative}) key switching scheme (\autoref{sec:glwe-key-switching}) with $n = 1$ as follows:

\begin{tcolorbox}[title={\textbf{\tboxlabel{\ref*{subsec:tfhe-key-switching}} TFHE Key Switching}}]

Given $\textsf{LWE}_{\vec{s},\sigma}(\Delta m + e) = (\vec{a}, b)$,

$\textsf{LWE}_{\vec{s}_{'},\sigma}(\Delta m + e') = (0, b) - \bm{\langle} \textsf{Decomp}^{\beta, l}(\vec{a}), \text{ } \textsf{Lev}_{\vec{s}_{'}, \sigma}^{\beta, l}(\vec{s}) \bm{\rangle}$
\end{tcolorbox}

\subsection{Homomorphic Ciphertext-to-Ciphertext Multiplication}
\label{subsec:tfhe-mult-cipher}

\textbf{- Reference:} 
\href{https://www.zama.ai/post/tfhe-deep-dive-part-3}{TFHE Deep Dive - Part III - Key switching and leveled multiplications}~\cite{tfhe-3}

$ $

TFHE supports multiplication of two ciphertexts in the form: $\textsf{LWE}_{\vec{s}, \sigma}(\Delta m_1) \cdot \textsf{GSW}_{\vec{s}, \sigma}^{\beta, l}(m_2)$. 

$ $

\noindent The 1st term $\textsf{LWE}_{\vec{s}, \sigma}(\Delta m_1 + e_1)$ comes from one of the following: 
\begin{itemize}
\item A fresh LWE encryption (\autoref{subsec:glwe-enc}) of plaintext $m_1$. 
\item A homomorphically added result of two LWE ciphertexts (\autoref{sec:glwe-add-cipher}). 
\item A homomorphically multiplied result of a LWE ciphertext with a plaintext (\autoref{sec:glwe-mult-plain}). 
\end{itemize}

$ $

\noindent The 2nd term $\textsf{GSW}_{\vec{s}, \sigma}^{\beta, l}(m_2)$ comes from one of the following:
\begin{itemize}
\item A fresh GSW encryption (\autoref{subsec:ggsw-enc}) of plaintext $m_2$.
\item Converted from $\textsf{LWE}_{\vec{s}, \sigma}(\Delta m_2 + e_2)$ into $\textsf{GSW}_{\vec{s}, \sigma}^{\beta, l}(m_2)$ by \textit{circuit bootstrapping} (this will be covered in the future).
\end{itemize}

$ $

\noindent Remember the following: 

$\textsf{LWE}_{\vec{s}, \sigma}(\Delta m_1 + e_1) = (\vec{a}, b) \in \mathbb{Z}_{q}^{k + 1}$, where $b = \vec{a} \cdot \vec{s} + \Delta m_1 + e_1$ 

$ $

$\textsf{GSW}_{\vec{s}, \sigma}^{\beta, l}(m_2) = \Bigr \{ \{ \textsf{Lev}_{\vec{s}, \sigma}^{\beta, l} (-s_i \cdot m_2)  \}_{i=0}^{k-1}, \textsf{Lev}_{\vec{s}, \sigma}^{\beta, l}(m_2) \Bigl \} \in \mathbb{Z}_{q }^{(k+1) \cdot l \cdot (k'+1)}$ \textcolor{red}{ $\rhd$ from \autoref{subsec:ggsw-enc}}

$ $

\noindent Let's use the following notations:

$\textsf{GSW}_{\vec{s}, \sigma}^{\beta, l}(m_2) = {\bar{\textsf{ct}}} = (\bar{\textsf{ct}}_0,. \bar{\textsf{ct}}_1, \gap{$\cdots$} \bar{\textsf{ct}}_k)$ 

$\bar{\textsf{ct}}_i = \textsf{Lev}_{\vec{s}, \sigma}^{\beta, l}(-s_i \cdot m_2)$ for $0 \leq i \leq (k-1)$

$\bar{\textsf{ct}}_k = \textsf{Lev}_{\vec{s}, \sigma}^{\beta, l}(m_2)$

$\textsf{ct} = \textsf{LWE}_{\vec{s}, \sigma}(\Delta m_1 + e_1) = (\vec{a}, b) = (a_0, a_1, \gap{$\cdots$}, a_{k-1}, b) = (\textsf{ct}_0, \textsf{ct}_1, \cdots, \textsf{ct}_k)$

$ $

\noindent Let's define the following TFHE ciphertext multiplication operation: 

$\textsf{ct} \cdot {\bar{\textsf{ct}}} = \sum\limits_{i=0}^{k}\langle \textsf{Decomp}^{\beta, l}(\textsf{ct}_i), \bar{\textsf{ct}}_i \rangle$

$ $

\noindent Then, the following is true:

\begin{tcolorbox}[title={\textbf{\tboxlabel{\ref*{subsec:tfhe-mult-cipher}} TFHE Ciphertext-to-Ciphertext Multiplication}}]
$\textsf{ct} = \textsf{LWE}_{\vec{s}, \sigma}(\Delta m_1 + e_1) = (a_0, a_1, \cdots, a_{k-1}, b)$

$\bar{\textsf{ct}} = \textsf{GSW}_{\vec{s}, \sigma}^{\beta, l}(m_2) = \bm( \textsf{Lev}_{\vec{s}, \sigma}^{\beta, l}(-s_0\cdot m_2), \textsf{Lev}_{\vec{s}, \sigma}^{\beta, l}(-s_1\cdot m_2), \cdots, \textsf{Lev}_{\vec{s}, \sigma}^{\beta, l}(-s_{k-1}\cdot m_2), \textsf{Lev}_{\vec{s}, \sigma}^{\beta, l}(m_2)  \bm)$

$\textsf{LWE}_{\vec{s}, \sigma}(\Delta m_1 + e_1) \cdot \textsf{GSW}_{\vec{s}, \sigma}^{\beta, l}(m_2) $

$= \sum\limits_{i=0}^{k}\langle \textsf{Decomp}^{\beta, l}(\textsf{ct}_i), \bar{\textsf{ct}}_i \rangle $

$\approx \textsf{LWE}_{\vec{s}, \sigma}(\Delta m_1 m_2)$
\end{tcolorbox}

This means that multiplying two TFHE ciphertexts (one is in LWE and another in GSW) and decrypting the resulting LWE ciphertext gives the same result as multiplying their two original plaintexts.

%\textbf{\underline{Proof}}
\begin{myproof}
\begin{enumerate}
\item $\sum\limits_{i=0}^{k}\langle \textsf{Decomp}^{\beta, l}(\textsf{ct}_i), \bar{\textsf{ct}}_i \rangle$ \\
$= \langle \textsf{Decomp}^{\beta, l}(a_0), \bar{\textsf{ct}}_0 \rangle + \langle \textsf{Decomp}^{\beta, l}(a_1), \bar{\textsf{ct}}_1 \rangle + \gap{$\cdots$} + \langle \textsf{Decomp}^{\beta, l}(a_{k-1}), \bar{\textsf{ct}}_{k-1} \rangle + \langle \textsf{Decomp}^{\beta, l}(b), \bar{\textsf{ct}}_k \rangle$ \\
\textcolor{red}{ $\rhd$ expanding the dot product of two vectors}
\item For $i = k$: \\
$\textsf{Decomp}^{\beta, l}(b) = (b_1, b_2, \ldots , b_l)$, where $b = b_1\dfrac{q}{\beta^1} + b_2\dfrac{q}{\beta^2} + \cdots + b_l\dfrac{q}{\beta^l}$ \textcolor{red}{ $\rhd$ from \autoref{subsec:poly-decomp}}\\
$\bar{\textsf{ct}}_k = \textsf{Lev}_{\vec{s}, \sigma}^{\beta, l}(m_2) = \left(\textsf{LWE}_{\vec{s}, \sigma}\left(m_{2}\dfrac{q}{\beta^1} + e_{2, 1}\right), \textsf{LWE}_{\vec{s}, \sigma}\left(m_{2}\dfrac{q}{\beta^2} + e_{2, 2}\right), \ldots, \textsf{LWE}_{\vec{s}, \sigma}\left(m_{2}\dfrac{q}{\beta^l} + e_{2, l}\right) \right)$ \\
$ $

Therefore: \\
$\langle \textsf{Decomp}^{\beta, l}(b), \bar{\textsf{ct}}_k \rangle$\\
$= b_1 \cdot \textsf{LWE}_{\vec{s}, \sigma} \left (m_{2}\dfrac{q}{\beta^1}  + e_{2, 1}\right ) + b_2 \cdot \textsf{LWE}_{\vec{s}, \sigma} \left (m_{2}\dfrac{q}{\beta^2}  + e_{2, 2}\right ) + \gap{$\cdots$} + b_l \cdot \textsf{LWE}_{\vec{s}, \sigma} \left (m_{2}\dfrac{q}{\beta^l}  + e_{2, l}\right)$ \\
$= \textsf{LWE}_{\vec{s}, \sigma} \left (b_1m_{2}\dfrac{q}{\beta^1}  + b_1e_{2, 1}\right ) + \textsf{LWE}_{\vec{s}, \sigma} \left (b_2m_{2}\dfrac{q}{\beta^2}  + b_2e_{2,2}\right ) + \gap{$\cdots$} + \textsf{LWE}_{\vec{s}, \sigma} \left (b_lm_{2}\dfrac{q}{\beta^l} + b_le_{2, l}\right)$ \textcolor{red}{ $\rhd$ from \autoref{sec:glwe-mult-plain}} \\
$= \textsf{LWE}_{\vec{s}, \sigma} \left (b_1m_{2}\dfrac{q}{\beta^1} + b_2m_{2}\dfrac{q}{\beta^2} + \gap{$\cdots$} + b_lm_{2}\dfrac{q}{\beta^l} + b_1e_{2,1} + \cdots + b_le_{2,l}\right)$ \textcolor{red}{ $\rhd$ from \autoref{sec:glwe-add-cipher}}\\
$= \textsf{LWE}_{\vec{s}, \sigma} \left (m_{2} \cdot \left ( b_1\dfrac{q}{\beta^1} + b_2\dfrac{q}{\beta^2} + \gap{$\cdots$} + b_l\dfrac{q}{\beta^l} + e^{\langle k \rangle} \right)\right)$ \textcolor{red}{ $\rhd$ where $e^{\langle k \rangle} = \sum\limits_{i=1}^{l}b_ie_{2, i}$}\\
$= \textsf{LWE}_{\vec{s}, \sigma} (m_{2}b + e^{\langle k \rangle})$ \textcolor{red}{ $\rhd$ from \autoref{subsec:poly-decomp}}

\item For $0 \leq i \leq (k - 1)$: \\
$\textsf{Decomp}^{\beta, l}(a_i) = (a_{\langle i, 1 \rangle}, a_{\langle i, 2 \rangle}, \gap{$\cdots$}, a_{\langle i, l \rangle})$, where $a_i = a_{\langle i, 1 \rangle}\dfrac{q}{\beta^1} + a_{\langle i, 2 \rangle}\dfrac{q}{\beta^2} + \gap{$\cdots$} + a_{\langle i, l \rangle}\dfrac{q}{\beta^l}$ \\
$\bar{\textsf{ct}}_i =  \textsf{Lev}_{\vec{s}, \sigma}^{\beta, l}(-s_im_2) $

$= \left(\textsf{LWE}_{\vec{s}, \sigma}\left(-s_im_2\dfrac{q}{\beta^1} + e_{i, 1}\right), \textsf{LWE}_{\vec{s}, \sigma}\left(-s_im_2\dfrac{q}{\beta^2} + e_{i, 2}\right), \gap{$\cdots$}, \textsf{LWE}_{\vec{s}, \sigma}\left(-s_im_2\dfrac{q}{\beta^l} + e_{i, l}\right) \right)$ \\

$ $
Therefore: \\
$\langle \textsf{Decomp}^{\beta, l}(a_0), \bar{\textsf{ct}}_0 \rangle + \langle \textsf{Decomp}^{\beta, l}(a_1), \bar{\textsf{ct}}_1 \rangle + \gap{$\cdots$} + \langle \textsf{Decomp}^{\beta, l}(a_{k-1}), \bar{\textsf{ct}}_{k-1} \rangle$ \\
$= \sum\limits_{i=0}^{k-1}\langle \textsf{Decomp}^{\beta, l}(a_i), \bar{\textsf{ct}}_i \rangle$ \\
$= \sum\limits_{i=0}^{k-1}\Bigg(a_{\langle i, 1\rangle} \cdot \textsf{LWE}_{\vec{s}, \sigma}\left(-s_im_2\dfrac{q}{\beta^1} + e_{i,1}\right) + a_{\langle i, 2\rangle} \cdot \textsf{LWE}_{\vec{s}, \sigma}\left(-s_im_2\dfrac{q}{\beta^2} + e_{i,2}\right) + $

$ \cdots +  a_{\langle i, l\rangle} \cdot \textsf{LWE}_{\vec{s}, \sigma}\left(-s_im_2\dfrac{q}{\beta^l} + e_{i,l}\right)\Bigg)$ \\
$= \sum\limits_{i=0}^{k-1}\Bigg(\textsf{LWE}_{\vec{s}, \sigma}\left(-a_{\langle i, 1\rangle}s_im_2\dfrac{q}{\beta^1} + a_{\langle i, 1 \rangle}e_{i,1}\right) + \textsf{LWE}_{\vec{s}, \sigma}\left(-a_{\langle i, 2\rangle}s_im_2\dfrac{q}{\beta^2} + a_{\langle i, 2 \rangle}e_{i,2}\right) + $

$\gap{$\cdots$} + \textsf{LWE}_{\vec{s}, \sigma}\left(-a_{\langle i, l\rangle}s_im_2\dfrac{q}{\beta^l} + a_{\langle i, l \rangle}e_{i,l}\right)\Bigg)$ \\
$= \sum\limits_{i=0}^{k-1}\textsf{LWE}_{\vec{s}, \sigma}\left(-a_{\langle i, 1\rangle}s_im_2\dfrac{q}{\beta^1} -a_{\langle i, 2\rangle}s_im_2\dfrac{q}{\beta^2} + \gap{$\cdots$} -a_{\langle i, l\rangle}s_im_2\dfrac{q}{\beta^l} + a_{\langle i, 1 \rangle}e_{i,1} + \cdots + a_{\langle i, l \rangle}e_{i,l}\right)$ \\
$= \sum\limits_{i=0}^{k-1}\textsf{LWE}_{\vec{s}, \sigma}\left(-s_im_2 \cdot \left(a_{\langle i, 1\rangle}\dfrac{q}{\beta^1} + a_{\langle i, 2\rangle}\dfrac{q}{\beta^2} + \gap{$\cdots$} + a_{\langle i, l\rangle}\dfrac{q}{\beta^l}\right) + a_{\langle i, 1 \rangle}e_{i,1} + \cdots + a_{\langle i, l \rangle}e_{i,l}\right)$ \\
$= \sum\limits_{i=0}^{k-1}\textsf{LWE}_{\vec{s}, \sigma}(-s_im_2a_i + e^{\langle i \rangle})$ \textcolor{red}{ $\rhd$ where $e^{\langle i \rangle} = a_{\langle i, 1 \rangle}e_{i,1} + \cdots + a_{\langle i, l \rangle}e_{i,l}$}
\item According to step 2 and 3, \\
$\sum\limits_{i=0}^{k}\langle \textsf{Decomp}^{\beta, l}(\textsf{ct}_i), \bar{\textsf{ct}}_i \rangle$ \\
$= \sum\limits_{i=0}^{k-1}\textsf{LWE}_{\vec{s}, \sigma}(-s_im_2a_i + e^{\langle i \rangle}) + \textsf{LWE}_{\vec{s}, \sigma} (m_{2}b + e^{\langle k \rangle})$ \\ 
$= \textsf{LWE}_{\vec{s}, \sigma}\left(\sum\limits_{i=0}^{k-1}(-s_im_2a_i) + m_{2}b + \sum\limits_{i=0}^ke^{\langle i \rangle}\right)$ \textcolor{red}{ $\rhd$ addition of two LWE ciphertexts} \\ 
$= \textsf{LWE}_{\vec{s}, \sigma}\left(m_2b - \sum\limits_{i=0}^{k-1}m_2a_is_i + \sum\limits_{i=0}^ke^{\langle i \rangle}\right)$  \\ 
$= \textsf{LWE}_{\vec{s}, \sigma}\left(m_2(b - \sum\limits_{i=0}^{k-1}a_is_i) + \sum\limits_{i=0}^ke^{\langle i \rangle}\right)$\\
$= \textsf{LWE}_{\vec{s}, \sigma}\left(m_2(\Delta m_1 + e_1) + \sum\limits_{i=0}^ke^{\langle i \rangle}\right)$ \\
$= \textsf{LWE}_{\vec{s}, \sigma}\left(\Delta m_1m_2 + m_2e_1 + \sum\limits_{i=0}^ke^{\langle i \rangle}\right)$ \\
$\approx \textsf{LWE}_{\vec{s}, \sigma}(\Delta m_1m_2)$ \textcolor{red}{ $\rhd$ given $m_2e_1 + \sum\limits_{i=0}^ke^{\langle i \rangle}$ is small and thus $m_2e_1$ is also small}
\end{enumerate}

\end{myproof}

%This issue of noise growth is similar to that in ciphertext-to-plaintext multiplication (\autoref{subsubsec:glwe-mult-plain-discussion}). 
To reduce the noise growth, noise bootstrapping is needed (will be discussed in \autoref{subsec:tfhe-noise-bootstrapping}).

\subsubsection{Generalization to GLWE-to-GGSW Multiplication}
\label{subsubsec:tfhe-glwe-to-ggsw-multiplication}

We can further generalize TFHE's LWE-to-GSW multiplication to GLWE-to-GGSW multiplication between the following two ciphertexts: $\textsf{GLWE}_{\vec{S}, \sigma}(\Delta M_1) \cdot \textsf{GGSW}_{\vec{S}, \sigma}^{\beta, l}(M_2)$, where $M_1$, $M_2$, and $S$ are $(n-1)$-degree polynomials.

$ $

\noindent The 1st term $\textsf{GLWE}_{\vec{S}, \sigma}(\Delta M_1)$ comes from one of the following: 
\begin{itemize}
\item A fresh GLWE encryption (\autoref{subsec:glwe-enc}) of plaintext $M_1$. 
\item A homomorphically added result of two GLWE ciphertexts (\autoref{sec:glwe-add-cipher}). 
\item A homomorphically multiplied result of a GLWE ciphertext with a plaintext (\autoref{sec:glwe-mult-plain}). 
\end{itemize}

$ $

\noindent The 2nd term $\textsf{GGSW}_{\vec{S}, \sigma}^{\beta, l}(M_2)$ comes from one of the following:
\begin{itemize}
\item A fresh GGSW encryption (\autoref{subsec:ggsw-enc}) of plaintext $M_2$.
\item Converted from $\textsf{GLWE}_{\vec{S}, \sigma}(\Delta M_2)$ into $\textsf{GGSW}_{\vec{S}, \sigma}^{\beta, l}(M_2)$ by \textit{circuit bootstrapping} (this will be covered in the future).
\end{itemize}

$ $

\noindent Remember the following: 

$\textsf{GLWE}_{\vec{S}, \sigma}(\Delta M_1) = (A_0, A_1, \gap{$\cdots$}, A_{k-1}, B) \in \mathcal{R}_{n, q}^{k + 1}$, where $B = \sum\limits_{i=0}^{k-1}(A_i \cdot S_i) + \Delta M_1 + E$ \\ \textcolor{red}{ $\rhd$ from \autoref{subsec:glwe-enc}} 

$ $

$\textsf{GGSW}_{\vec{S}, \sigma}^{\beta, l}(M_2) = \Bigr \{ \{ \textsf{GLev}_{\vec{S}, \sigma}^{\beta, l} (-S_i \cdot M_2)  \}_{i=0}^{k-1}, \textsf{GLev}_{\vec{S}, \sigma}^{\beta, l}(M_2) \Bigl \} \in \mathcal{R}_{\langle n, q \rangle }^{(k+1) \cdot l \cdot (k'+1)}$ \textcolor{red}{ $\rhd$ from \autoref{subsec:ggsw-enc}}

$ $

\noindent Let's use the following notations:

$\textsf{GGSW}_{\vec{S}, \sigma}^{\beta, l}(M_2) = {\bar{C}} = (\bar{C_0},. \bar{C_1}, \gap{$\cdots$} \bar{C_k})$ 

$\bar{C_i} = \textsf{GLev}_{\vec{S}, \sigma}^{\beta, l}(-S_i \cdot M_2)$ for $0 \leq i \leq (k-1)$

$\bar{C_k} = \textsf{GLev}_{\vec{S}, \sigma}^{\beta, l}(M_2)$

$\textsf{ct} = \textsf{GLWE}_{\vec{S}, \sigma}(\Delta M_1) = (C_0, C_1, \gap{$\cdots$}, C_k) = (A_0, A_1, \gap{$\cdots$}, A_{k-1}, B)$

$ $

\noindent Let's define the following TFHE ciphertext multiplication operation: 

$\textsf{ct} \cdot {\bar{C}} = \sum\limits_{i=0}^{k}\langle \textsf{Decomp}^{\beta, l}(C_i), \bar{C_i} \rangle$

$ $

\noindent Then, the following is true:

\begin{tcolorbox}[title={\textbf{\tboxlabel{\ref*{subsubsec:tfhe-glwe-to-ggsw-multiplication}} Generalization to GLWE-to-GGSW Multiplication}}]
$\textsf{ct} = \textsf{GLWE}_{\vec{S}, \sigma}(\Delta M_1) = (A_0, A_1, \cdots, A_{k-1}, B)$

$\bar{C} = \textsf{GGSW}_{\vec{S}, \sigma}^{\beta, l}(M_2)$

$ = \bm(\textsf{GLev}_{\vec{S}, \sigma}^{\beta, l}(-S_0\cdot M_2), \textsf{GLev}_{\vec{S}, \sigma}^{\beta, l}(-S_1\cdot M_2), \cdots, \textsf{GLev}_{\vec{S}, \sigma}^{\beta, l}(-S_{k-1}\cdot M_2), \textsf{GLev}_{\vec{S}, \sigma}^{\beta, l}(M_2)\bm)$

$ $

$\textsf{GLWE}_{\vec{S}, \sigma}(\Delta M_1) \cdot \textsf{GGSW}_{\vec{S}, \sigma}^{\beta, l}(M_2) = \sum\limits_{i=0}^{k}\langle \textsf{Decomp}^{\beta, l}(C_i), \bar{C_i} \rangle \approx \textsf{GLWE}_{\vec{S}, \sigma}(\Delta M_1 M_2)$
\end{tcolorbox}

This means that multiplying two TFHE ciphertexts (one is in GLWE and another in GGSW) and decrypting the resulting GLWE ciphertext gives the same result as multiplying their two original plaintexts.

%\textbf{\underline{Proof}}
\begin{myproof}
\begin{enumerate}
\item $\sum\limits_{i=0}^{k}\langle \textsf{Decomp}^{\beta, l}(C_i), \bar{C_i} \rangle$ \\
$= \langle \textsf{Decomp}^{\beta, l}(A_0), \bar{C_0} \rangle + \langle \textsf{Decomp}^{\beta, l}(A_1), \bar{C_1} \rangle + \gap{$\cdots$} + \langle \textsf{Decomp}^{\beta, l}(A_{k-1}), \bar{C}_{k-1} \rangle + \langle \textsf{Decomp}^{\beta, l}(B), \bar{C_k} \rangle$ \\
\textcolor{red}{ $\rhd$ expanding the dot product of two vectors}
\item For $i = k$: \\
$\textsf{Decomp}^{\beta, l}(B) = (B_1, B_2, \gap{$\cdots$}, B_l)$, where $B = B_1\dfrac{q}{\beta^1} + B_2\dfrac{q}{\beta^2} + \gap{$\cdots$} + B_l\dfrac{q}{\beta^l}$ \textcolor{red}{ $\rhd$ from \autoref{subsec:poly-decomp}}\\
$\bar{C}_k = \textsf{GLev}_{\vec{S}, \sigma}^{\beta, l}(M_2) = \left(\textsf{GLWE}_{\vec{S}, \sigma}\left(M_{2}\dfrac{q}{\beta^1}\right), \textsf{GLWE}_{\vec{S}, \sigma}\left(M_{2}\dfrac{q}{\beta^2}\right), \gap{$\cdots$}, \textsf{GLWE}_{\vec{S}, \sigma}\left(M_{2}\dfrac{q}{\beta^l}\right) \right)$ 

\textcolor{red}{ $\rhd$ we omit the noise terms $E_{2,1}, \ldots, E_{2,l}$ in each GLWE ciphertext for simplicity} \\
$ $

Therefore: \\
$\langle \textsf{Decomp}^{\beta, l}(B), \bar{C_k} \rangle$\\
$= B_1 \cdot \textsf{GLWE}_{\vec{S}, \sigma} \left (M_{2}\dfrac{q}{\beta^1} \right ) + B_2 \cdot \textsf{GLWE}_{\vec{S}, \sigma} \left (M_{2}\dfrac{q}{\beta^2} \right ) + \gap{$\cdots$} + B_l \cdot \textsf{GLWE}_{\vec{S}, \sigma} \left (M_{2}\dfrac{q}{\beta^l}\right)$ \\
$= \textsf{GLWE}_{\vec{S}, \sigma} \left (B_1M_{2}\dfrac{q}{\beta^1} \right ) + \textsf{GLWE}_{\vec{S}, \sigma} \left (B_2M_{2}\dfrac{q}{\beta^2} \right ) + \gap{$\cdots$} + \textsf{GLWE}_{\vec{S}, \sigma} \left (B_lM_{2}\dfrac{q}{\beta^l}\right)$ \textcolor{red}{ $\rhd$ from \autoref{sec:glwe-mult-plain}} \\
$= \textsf{GLWE}_{\vec{S}, \sigma} \left (B_1M_{2}\dfrac{q}{\beta^1} + B_2M_{2}\dfrac{q}{\beta^2} + \gap{$\cdots$} + B_lM_{2}\dfrac{q}{\beta^l}\right)$ \textcolor{red}{ $\rhd$ from \autoref{sec:glwe-add-cipher}}\\
$= \textsf{GLWE}_{\vec{S}, \sigma} \left (M_{2} \cdot \left ( B_1\dfrac{q}{\beta^1} + B_2\dfrac{q}{\beta^2} + \gap{$\cdots$} + B_l\dfrac{q}{\beta^l} \right)\right)$ \\
$= \textsf{GLWE}_{\vec{S}, \sigma} (M_{2}B)$ \textcolor{red}{ $\rhd$ from \autoref{subsec:poly-decomp}}
\item For $0 \leq i \leq (k - 1)$: \\
$\textsf{Decomp}^{\beta, l}(A_i) = (A_{\langle i, 1 \rangle}, A_{\langle i, 2 \rangle}, \gap{$\cdots$}, A_{\langle i, l \rangle})$, where $A_i = A_{\langle i, 1 \rangle}\dfrac{q}{\beta^1} + A_{\langle i, 2 \rangle}\dfrac{q}{\beta^2} + \gap{$\cdots$} + A_{\langle i, l \rangle}\dfrac{q}{\beta^l}$ \\
$\bar{C_i} =  \textsf{GLev}_{\vec{S}, \sigma}^{\beta, l}(-S_iM_2) = \left(\textsf{GLWE}_{\vec{S}, \sigma}\left(-S_iM_2\dfrac{q}{\beta^1}\right), \textsf{GLWE}_{\vec{S}, \sigma}\left(-S_iM_2\dfrac{q}{\beta^2}\right), \gap{$\cdots$}, \textsf{GLWE}_{\vec{S}, \sigma}\left(-S_iM_2\dfrac{q}{\beta^l}\right) \right)$ \\

$ $
Therefore: \\
$\langle \textsf{Decomp}^{\beta, l}(A_0), \bar{C_0} \rangle + \langle \textsf{Decomp}^{\beta, l}(A_1), \bar{C_1} \rangle + \gap{$\cdots$} + \langle \textsf{Decomp}^{\beta, l}(A_{k-1}), \bar{C}_{k-1} \rangle$ \\
$= \sum\limits_{i=0}^{k-1}\langle \textsf{Decomp}^{\beta, l}(A_i), \bar{C_i} \rangle$ \\
$= \sum\limits_{i=0}^{k-1}\left(A_{\langle i, 1\rangle} \cdot \textsf{GLWE}_{\vec{S}, \sigma}\left(-S_iM_2\dfrac{q}{\beta^1}\right) + A_{\langle i, 2\rangle} \cdot \textsf{GLWE}_{\vec{S}, \sigma}\left(-S_iM_2\dfrac{q}{\beta^2}\right) + \gap{$\cdots$} + A_{\langle i, l\rangle} \cdot \textsf{GLWE}_{\vec{S}, \sigma}\left(-S_iM_2\dfrac{q}{\beta^l}\right)\right)$ \\
$= \sum\limits_{i=0}^{k-1}\left(\textsf{GLWE}_{\vec{S}, \sigma}\left(-A_{\langle i, 1\rangle}S_iM_2\dfrac{q}{\beta^1}\right) + \textsf{GLWE}_{\vec{S}, \sigma}\left(-A_{\langle i, 2\rangle}S_iM_2\dfrac{q}{\beta^2}\right) + \gap{$\cdots$} + \textsf{GLWE}_{\vec{S}, \sigma}\left(-A_{\langle i, l\rangle}S_iM_2\dfrac{q}{\beta^l}\right)\right)$ \\
$= \sum\limits_{i=0}^{k-1}\textsf{GLWE}_{\vec{S}, \sigma}\left(-A_{\langle i, 1\rangle}S_iM_2\dfrac{q}{\beta^1} + -A_{\langle i, 2\rangle}S_iM_2\dfrac{q}{\beta^2} + \gap{$\cdots$} + -A_{\langle i, l\rangle}S_iM_2\dfrac{q}{\beta^l}\right)$ \\
$= \sum\limits_{i=0}^{k-1}\textsf{GLWE}_{\vec{S}, \sigma}\left(-S_iM_2 \cdot \left(A_{\langle i, 1\rangle}\dfrac{q}{\beta^1} + A_{\langle i, 2\rangle}\dfrac{q}{\beta^2} + \gap{$\cdots$} + A_{\langle i, l\rangle}\dfrac{q}{\beta^l}\right)\right)$ \\
$= \sum\limits_{i=0}^{k-1}\textsf{GLWE}_{\vec{S}, \sigma}(-S_iM_2A_i)$
\item According to step 2 and 3, \\
$\sum\limits_{i=0}^{k}\langle \textsf{Decomp}^{\beta, l}(C_i), \bar{C_i} \rangle$ \\
$= \sum\limits_{i=0}^{k-1}\textsf{GLWE}_{\vec{S}, \sigma}(-S_iM_2A_i) + \textsf{GLWE}_{\vec{S}, \sigma} (M_{2}B)$ \\ 
$= \textsf{GLWE}_{\vec{S}, \sigma}\Big(\sum\limits_{i=0}^{k-1}(-S_iM_2A_i) + M_{2}B\Big)$ \textcolor{red}{ $\rhd$ addition of two GLWE ciphertexts} \\ 
$= \textsf{GLWE}_{\vec{S}, \sigma}\Big(BM_2 - \sum\limits_{i=0}^{k-1}M_2A_iS_i\Big)$  \\ 
$= \textsf{GLWE}_{\vec{S}, \sigma}\Big(M_2(B - \sum\limits_{i=0}^{k-1}A_iS_i)\Big)$\\
$= \textsf{GLWE}_{\vec{S}, \sigma}(M_2(\Delta M_1 + E))$ \\
$= \textsf{GLWE}_{\vec{S}, \sigma}(\Delta M_1M_2 + M_2E)$ \\
$\approx \textsf{GLWE}_{\vec{S}, \sigma}(\Delta M_1M_2)$ \textcolor{red}{ $\rhd$ given $E$ is small and thus $M_2E$ is also small}
\end{enumerate}

\end{myproof}

\subsection{Coefficient Extraction}
\label{subsec:tfhe-extraction}

\textbf{- Reference:} 
\href{https://www.zama.ai/post/tfhe-deep-dive-part-4}{TFHE Deep Dive - Part IV - Programmable Bootstrapping}~\cite{tfhe-4}

$ $

In TFHE, coefficient extraction is a process of extracting a coefficient of a polynomial that is encrypted as GLWE ciphertext. The extracted coefficient is in the form of LWE ciphertext (\autoref{sec:lwe}). %We will explain how to extract the coefficient of a plaintext polynomial $M$ from a GLWE ciphertext and RLWE ciphertext, respectively. 

%\subsubsection{Coefficient Extraction from a GLWE Ciphertext}
%\label{subsubsec:tfhe-extraction-glwe}

Note that in the GLWE cryptosystem, plaintext $M$ is encoded as a polynomial, where each coefficient encodes the plaintext value $m_0, m_1, \cdots, m_{n-1}$.

%\subsubsection{Overview}
%\label{subsec:tfhe-extraction-overview}

Suppose we have a GLWE ciphertext setup as the following: \\ 
$M = \sum\limits_{j=0}^{n-1}m_jX^j \in \mathcal{R}_{\langle n, q \rangle}$ 

$S = \left(S_0 = \sum\limits_{j=0}^{n-1}s_{0,j}X^j, S_1 = \sum\limits_{j=0}^{n-1}s_{1,j}X^j, \gap{$\cdots$}, S_{k-1} = \sum\limits_{j=0}^{n-1}s_{k-1,j}X^j \right)$ 

$\textsf{GLWE}_{\vec{S}, \sigma}(\Delta M) = \left(A_0 = \sum\limits_{j=0}^{n-1}a_{0,j}X^j, A_1 = \sum\limits_{j=0}^{n-1}a_{1,j}X^j, \gap{$\cdots$}, A_{k-1} = \sum\limits_{j=0}^{n-1}a_{k-1,j}X^j, B = \sum\limits_{j=0}^{n-1}b_{j}X^j\right)$ 

$B = \sum\limits_{i=0}^{k-1}A_iS_i + \Delta M + E$ 

$E = \sum\limits_{i=0}^{n-1}e_iX^i$ 

$ $

\noindent Note that:

$\Delta M + E = B - \sum\limits_{i=0}^{k-1}A_iS_i$ 

$ = (\Delta m_0 + \Delta m_1X + \gap{$\cdots$} + \Delta m_{n-1}X^{n-1}) + (e_0 + e_1X + \gap{$\cdots$} + e_{n-1}X^{n-1})$

$= (\Delta m_0 + e_0) + (\Delta m_1 + e_1)X + \gap{$\cdots$} + (\Delta m_{n-1} + e_{n-1})X^{n-1}$

$ $

\noindent Another way to write the formula is:

$B - \sum\limits_{i=0}^{k-1}A_iS_i$ 

$ = (b_0 + b_1X + \gap{$\cdots$} + b_{n-1}X^{n-1} )$ 

$ - (a_{0,0} + a_{0,1}X + \gap{$\cdots$} + a_{0, n-1}X^{n-1})(s_{0,0} + s_{0,1}X + \gap{$\cdots$} + s_{0, n-1}X^{n-1})$ 

$ - (a_{1,0} + a_{1,1}X + \gap{$\cdots$} + a_{1, n-1}X^{n-1})(s_{1,0} + s_{1,1}X + \gap{$\cdots$} + s_{1, n-1}X^{n-1})$

$ - \gap{$\cdots$} $ 

$ - (a_{k-1,0} + a_{k-1,1}X + \gap{$\cdots$} + a_{k-1, n-1}X^{n-1})(s_{k-1,0} + s_{k-1,1}X + \gap{$\cdots$} + s_{k-1, n-1}X^{n-1})$ 

$ $

$ = \left(b_0 - \left( \sum\limits_{i=0}^{k-1} \sum\limits_{j=0}^{0}(a_{i,0-j}s_{i,j}) - \sum\limits_{i=0}^{k-1} \sum\limits_{j=1}^{n-1}(a_{i,n-j}s_{i,j}) \right)\right)$

$ + \left(b_1 - \left( \sum\limits_{i=0}^{k-1} \sum\limits_{j=0}^{1}(a_{i,1-j}s_{i,j}) - \sum\limits_{i=0}^{k-1} \sum\limits_{j=2}^{n-1}(a_{i,n+1-j}s_{i,j})   \right) \right)\cdot X$

$ + \left(b_2 - \left( \sum\limits_{i=0}^{k-1} \sum\limits_{j=0}^{2}(a_{i,2-j}s_{i,j}) - \sum\limits_{i=0}^{k-1} \sum\limits_{j=3}^{n-1}(a_{i,n+2-j}s_{i,j})   \right) \right)\cdot X^2$ 

$ $

$\gap{$\cdots$}$ 

$ $

$ + \left(b_{n-1} - \left(  \sum\limits_{i=0}^{k-1} \sum\limits_{j=0}^{n-1}(a_{i,n-1-j}s_{i,j}) - \sum\limits_{i=0}^{k-1} \sum\limits_{j=n}^{n-1}(a_{i,n+(n-1)-j}s_{i,j})  \right) \right)\cdot X^{n-1}$ 

\textcolor{red}{ $\rhd$ Grouping the terms by same exponents}

$ $

$ $

$= \sum\limits_{h=0}^{n-1}  \left(b_h - \left(  \sum\limits_{i=0}^{k-1} \sum\limits_{j=0}^{h}(a_{i,h-j}s_{i,j}) - \sum\limits_{i=0}^{k-1} \sum\limits_{j=h+1}^{n-1}(a_{i,n+h-j}s_{i,j})  \right) \right)\cdot X^{h}  $

$ $

$= \sum\limits_{h=0}^{n-1}  C_h \cdot X^{h}  $, where $C_h = b_h - \left(  \sum\limits_{i=0}^{k-1} \sum\limits_{j=0}^{h}(a_{i,h-j}s_{i,j}) - \sum\limits_{i=0}^{k-1} \sum\limits_{j=h+1}^{n-1}(a_{i,n+h-j}s_{i,j})  \right)$

$ $

\noindent In the above $(n-1)$-degree polynomial, notice that each $X^h$ term's coefficient, $C_h$, can be expressed as an LWE ciphertext $\textsf{ct}_h$ as follows:

$S' = (s_{0,0}, s_{0,1}, \gap{$\cdots$}, s_{0,n-1}, s_{1,0}, s_{1,1}, \gap{$\cdots$}, s_{1, n-1}, \gap{$\cdots$}, s_{k-1, n-1}) = (s'_0, s'_1, \gap{$\cdots$}, s'_{nk-1} ) \in \mathbb{Z}_q^{nk}$

$\textsf{ct}_h = (a'_0, a'_1, \gap{$\cdots$}, a'_{nk-1}, b_h) \in \mathbb{Z}_q^{nk + 1}$

\[
    \text{, where } a'_{n \cdot i + j} =   
\begin{cases}
    a_{i,h - j} \text{ (if } 0 \leq j \leq h\text{)}\\
    -a_{i,n + h - j} \text{ (if } h+1 \leq j \leq n-1\text{)}\\
\end{cases}
\centering , b_h \text{ is directly obtained from the polynomial } B
\]

\noindent Note that $b_h - \sum\limits_{i=0}^{nk-1}s'_ia'_i = \Delta m_h + e_h$. This means that $C_h$ can be replaced by its encrypted version, $\textsf{LWE}_{\vec{s}_{'}, \sigma}(\Delta m_h)$, an LWE ciphertext $\textsf{ct}_h$ encrypting the $h$-th coefficient of $M$. Therefore, we just extracted $\textsf{LWE}_{\vec{s}_{'}, \sigma}(\Delta m_h)$ from $\textsf{GLWE}_{\vec{S}, \sigma}(\Delta M)$. This operation is called coefficient extraction, which does not add any noise because it simply extracts an LWE ciphertext by reordering the polynomial of the GLWE ciphertext. 

Once we have $\textsf{LWE}_{\vec{s}_{'}, \sigma}(\Delta m_h)$, we can key-switch it from $\vec{s}_{'} \rightarrow \vec{s}$ (\autoref{subsec:tfhe-key-switching}). 

\begin{tcolorbox}[title={\textbf{\tboxlabel{\ref*{subsec:tfhe-extraction}} GLWE Ciphertext's Coefficient Extraction}}]
Given the following GLWE ciphertext: 

$M = \sum\limits_{j=0}^{n-1}m_jX^j \in \mathcal{R}_{\langle n, t \rangle}$

$\vec{S} = \left(S_0 = \sum\limits_{j=0}^{n-1}s_{0,j}X^j, S_1 = \sum\limits_{j=0}^{n-1}s_{1,j}X^j, \gap{$\cdots$}, S_{k-1} = \sum\limits_{j=0}^{n-1}s_{k-1,j}X^j \right)$

$\textsf{GLWE}_{\vec{S}, \sigma}(\Delta M) = \left(A_0 = \sum\limits_{j=0}^{n-1}a_{0,j}X^j, A_1 = \sum\limits_{j=0}^{n-1}a_{1,j}X^j, \gap{$\cdots$}, A_{k-1} = \sum\limits_{j=0}^{n-1}a_{k-1,j}X^j, B = \sum\limits_{j=0}^{n-1}b_{j}X^j\right)$

$B = \sum\limits_{i=0}^{k-1}A_iS_i + \Delta M + E \bmod q$, \text{ } $E = \sum\limits_{i=0}^{n-1}e_iX^i$

$ $

$\textsf{LWE}_{\vec{s}_{'}, \sigma}(\Delta m_h)$ is an LWE ciphertext that encrypts $\Delta M$'s $h$-th coefficient (i.e., $\Delta m_h$). $\textsf{LWE}_{\vec{s}_{'}, \sigma}(\Delta m_h)$ can be extracted from $\textsf{GLWE}_{\vec{S}, \sigma}(\Delta M)$ as follows: 

$ $

$\vec{s}_{'} = (s_{0,0}, s_{0,1}, \gap{$\cdots$}, s_{0,n-1}, s_{1,0}, s_{1,1}, \gap{$\cdots$}, s_{1, n-1}, \gap{$\cdots$}, s_{k-1, n-1}) = (s'_0, s'_1, \gap{$\cdots$}, s'_{nk-1} ) \in \mathbb{Z}_q^{nk}$

$\textsf{LWE}_{\vec{s}_{'}, \sigma}(\Delta m_h) = (a_0', a_1', \gap{$\cdots$} , a_{nk-1}', b_h) \in \mathbb{Z}_q^{nk + 1}$

\[
    \text{, where } a'_{n \cdot i + j} =   
\begin{cases}
    a_{i,h - j} \text{ (if } 0 \leq j \leq h\text{)}\\
    -a_{i,n + h - j} \text{ (if } h+1 \leq j \leq n-1\text{)}\\
\end{cases}
, b_h \text{ is obtained from the polynomial } B
\]

Once we have $\textsf{LWE}_{\vec{s}_{'}, \sigma}(\Delta m_h)$, key-switch it from $\vec{s}_{'} \rightarrow \vec{s}$ (\autoref{subsec:tfhe-key-switching}).

\end{tcolorbox}

\subsection{Noise Bootstrapping}
\label{subsec:tfhe-noise-bootstrapping}

\textbf{- Reference:} 
\href{https://www.zama.ai/post/tfhe-deep-dive-part-4}{TFHE Deep Dive - Part IV - Programmable Bootstrapping}~\cite{tfhe-4}

$ $

Continuing homomorphic additions of TFHE ciphertexts does not necessarily increase the noise $e$, because $e$ is randomly generated over the Gaussian distribution, thus adding up many noises would give the mean value of 0. On the other hand, continuing homomorphic multiplications increases the noise, because the noise terms get multiplied, growing its magnitude. Thus, we need to somehow \textit{reset} the noise before it trespasses on the higher bits where plaintext $m$ resides (i.e., preventing the red noise bits from overflowing to the blue plaintext bits as shown in \autoref{fig:scaling}). The process of re-initializing the noise to a smaller value is called noise bootstrapping.

As explained in the beginning of this section, TFHE uses LWE (which is GLWE with polynomial degree 0) to encrypt \& decrypt a plaintext. That is, each plaintext is $m$ (a single number), encoded as a zero-degree polynomial. Further, the secret key S that encrypts each $m$ is a vector $ \{s_0, s_1, \text{ } \cdots \text{ }, s_{k-1} \}$ instead of a polynomial. On the other hand, TFHE's noise bootstrapping uses homomorphic addition between GLWE ciphertexts and homomorphic multiplication between GLWE and GGSW ciphertexts.  

%Nonetheless, the reason why we learned TFHE ciphertext addition (\autoref{subsec:glwe-add} and \autoref{subsec:glwe-add-plain}), multiplication (\autoref{subsec:tfhe-mult-cipher} and \autoref{subsec:glwe-mult-plain}), and polynomial rotation (\autoref{subsec:coeff-rotation}) in terms of GLWE is because the bootstrapping procedure we will explain in this subsection requires homomorphic addition and multiplication of GLWE ciphertexts.

Suppose we have a TFHE ciphertext as follows: 

$ $

$\textsf{LWE}_{\vec{s}, \sigma}(\Delta m) = (a_0, a_1, \gap{$\cdots$} a_{k-1}, b)$

$b = \sum\limits_{i=0}^{k-1} a_is_i + \Delta m + e_b$

$\vec{s} = (s_0, s_1, \gap{$\cdots$} s_{k-1})$

$ $

, where $e_b$ is a big noise accumulated over a series of many ciphertext (or plaintext) multiplications. The goal of noise bootstrapping is to convert $(a_0, a_1, \gap{$\cdots$} a_{k-1}, b)$ into $(a_0', a_1', \gap{$\cdots$} a_{k-1}', b')$ such that: 

$ $

$b' = \sum\limits_{i=0}^{k-1} a_i's_i + \Delta m + e_s$

$ $

, where $e_s$ is a re-initialized noise. 

$ $
\subsubsection{Overview}
\label{subsec:bootstrapping-overview}

To implement noise bootstrapping, we create a specially designed $(n-1)$-degree polynomial $V(X)$ called a Lookup Table (LUT). Before explaining $V(X)$, we will first motivate the idea based on a preliminary LUT polynomial $V_q(X)$. Imagine that the polynomial $V_q(X)$'s each degree term $X^{j}$ has its exponent $j = \Delta m_i + e_*$, a plaintext $m_i$ with some noise $e_* \in \mathbb{Z}_{\Delta}$, and its corresponding coefficient $v_{j} = m_i$, which is a noise-free plaintext. Therefore, the $(q-1)$-degree polynomial $V_q(X)$ is defined as follows:

$V_q(X) = v_0 + v_1X^1 + v_2X^2 + \gap{$\cdots$} + v_{q-1}X^{q-1}$

\text{ } $= m_0X^{\Delta m_0 + e_0} + m_0X^{\Delta m_0 + e_1} + m_0X^{\Delta m_0 + e_2} + \gap{$\cdots$} + m_0X^{\Delta m_0 + e_{\Delta - 1}}$ \textcolor{red}{ $\rhd$ total $\Delta$ terms}

\text{ } $ + \text{ } m_1X^{\Delta m_1 + e_0} + m_1X^{\Delta m_1 + e_1} + m_1X^{\Delta m_1 + e_2} + \gap{$\cdots$} + m_1X^{\Delta m_1 + e_{\Delta - 1}}$ \textcolor{red}{ $\rhd$ total $\Delta$ terms}

\text{ } $ + \gap{$\cdots$} $

\text{ } $ + \text{ } m_{t - 1}X^{\Delta m_{t - 1} + e_0} + m_{t - 1}X^{\Delta m_{t - 1} + e_1} + m_{t - 1}X^{\Delta m_{t - 1} + e_2} + \gap{$\cdots$} + m_{t - 1}X^{\Delta m_{t - 1} + e_{\Delta - 1}}$ \textcolor{red}{ $\rhd$ total $\Delta$ terms}

$ $

In the above formula, each $m_i$ and $e_k$ represents every possible plaintext message and error values (where $m_i \in \mathbb{Z}_t$ and $e_k \in \mathbb{Z}_{\Delta}$). %Note that $\mathbb{Z}_{q} = \{0, 1, \cdots, q-1\} = \{\Delta m_0 + e_0, \Delta m_0 + e_1, \cdots, \Delta m_{t-1} + e_{\Delta - 1}\}$. 

We design $V_q(X)$ to have the special property that each $v_{j}X^{j}$ term represents the special mapping (exponent, coefficient) $= (\Delta m_i + e_{*}, m_i)$, where $e_*$ can be any value in $\mathbb{Z}_{\Delta}$. During the TFHE setup stage, we GLWE-encrypt $V_q(X)$ by using our newly defined GLWE key $\vec{S}_{bk}$, a \textit{bootstrapping key}, which is different from the LWE secret key $\vec{s}$. $\vec{S}_{bk}$ is a list of $(n-1)$-degree polynomials with binary coefficients. Later, during the noise bootstrapping stage, we will rotate the coefficients of $V$ by $\Delta m + e$ positions to the left by computing $V \cdot X^{-(\Delta m + e)} = V'$, using the polynomial coefficient rotation method 1 technique (Summary~\ref*{subsec:coeff-rotation}.1 in \autoref{subsec:coeff-rotation}). Then, we will extract the polynomial's constant term's coefficient (i.e., the left-most 0-degree term's coefficient in the rotated polynomial $V'$) by using the coefficient extraction technique (\autoref{subsec:tfhe-extraction}). Further, we will encrypt $V_q(X)$ as a GLWE ciphertext at the TFHE setup stage, and thus the rotated $V'_q(X)$'s extracted constant term's coefficient is an LWE encryption of $m$ (i.e., $\textsf{LWE}_{\vec{s}, \sigma}(\Delta m)$) with a re-initialized (i.e., completely reduced) noise. 

To summarize, the noise bootstrapping procedure can be conceptually understood (at least for now) as follows: 

$ $

\begin{enumerate}
\item \textbf{\underline{Input}:} $\textsf{LWE}_{\vec{s}, \sigma}(\Delta m + e)$ as a noisy ciphertext encrypting $m$
\item Convert the input into the form of $X^{-(\Delta m + e)}$ as a rotator of $V_q(X)$ (Lookup Table).
\item Rotate $V_q$ to the left by $\Delta m + e$ positions by computing $V_q \cdot X^{-(\Delta m + e)} = V'_q$.
\item Extract the rotated $V'_q(X)$'s constant term's coefficient $m$ as an LWE encryption, which is $\textsf{LWE}_{\vec{s}, \sigma}(\Delta m)$.
\item \textbf{\underline{Output}:} $\textsf{LWE}_{\vec{s}, \sigma}(\Delta m)$ as an LWE encryption of the plaintext $m$ with a re-initialized noise
\end{enumerate}

$ $

The output $\textsf{LWE}_{\vec{s}, \sigma}(\Delta m)$ encrypts the same plaintext message as the input ciphertext, but with completely reduced noise. Therefore, the output $\textsf{LWE}_{\vec{s}, \sigma}(\Delta m)$ can be used for subsequent TFHE homomorphic operations (e.g., addition or multiplication). During this noise bootstrapping process, the polynomial $V_q$ is used as a \textit{dictionary} that contains the mappings from the noisy plaintext $\Delta m + e$ (i.e., as $\Delta m + e = j$ where $v_jX^j$) to the noise-free plaintext $m$ (i.e., as $m = v_j$ where $v_jX^j$). Therefore, $V_q$ is called the Lookup Table (LUT).  

Then, what should be the degree of $V_q(X)$? In order for $V_q(X)$ to encode all possible mappings from $\Delta m + e \in \mathbb{Z}_q$ to $m \in \mathbb{Z}_t$, $V_q(X)$ should be a $(q-1)$-degree polynomial. However, $q$ is a very big number, and it is computationally infeasible to manage a $(q-1)$-degree polynomial. Thus, in practice, we instead use a much smaller polynomial $V(X)$ whose degree is only $n-1$. Remember that according to our TFHE setup, $n \ll q$. Therefore, we need a way to \textit{compress} the big ciphertext space $\Delta m + e \in \mathbb{Z}_q$ into a much smaller space $\mathbb{Z}_n$ and encode the compressed values as the exponents of $X^j$ in a \textit{proportionally} correct way. For this proportional compression of $\mathbb{Z}_q \rightarrow \mathbb{Z}_n$, we will use the LWE modulus switching technique learned from \autoref{subsec:modulus-switch-lwe}. 

\subsubsection{Modulus Switch for Noise Bootstrapping}
\label{subsec:bootstrapping-modulus-switch}

To avoid using the giant $(q-1)$-degree polynomial $V_q$, we will compress $q$ possible ciphertext elements $\Delta m + e \in \mathbb{Z}_q$ into $n$ distinct exponents of the $(n-1)$-degree polynomial $V$, where each $v_jX^j$ term in $V$ represents a mapping from $j \rightarrow v_j$ (i.e., noisy plaintext to noise-free plaintext). However, notice that when we rotate the coefficients of the $(n-1)$-degree polynomial $V$ to the left, as $v_jX^j$ rotates across the boundary between $X^0$ and $X^{n-1}$ degree terms, $v_j$'s sign flips to $-v_j$ (as shown in the example of \autoref{subsec:coeff-rotation-ex}). Due to this coefficient sign flip, the $(n-1)$-degree polynomial $V$ can theoretically encode total $2n$ distinct coefficient states as follows: $(v_0, v_1, v_2, \gap{$\cdots$}, v_{n-1}, -v_0, -v_1, \gap{$\cdots$}, -v_{n-1})$. To move each of these $2n$ distinct coefficients to the constant term's coefficient position in $V$ (i.e., shifting the coefficient $v_j$ to the leftmost term in $V$), the rotating computation of $V \cdot X^{-j}$ can use $2n$ distinct $j$ values, which are $\{0, 1, 2, \gap{$\cdots$}, n-1, n, \gap{$\cdots$}, 2n-1\}$, to move each of $(v_0, v_1, v_2, \gap{$\cdots$}, v_{n-1}, -v_0, -v_1, \gap{$\cdots$}, -v_{n-1})$ coefficients to the constant term's position. This implies that the exponent $j$ in $V\cdot X^{-j}$ can use any of the $2n$ distinct values to cover all possible $2n$ (sign-flipped) coefficient states of $V$. Also, remember that $j = \Delta m + e$.  Therefore, we will switch the modulo of $\Delta m + e$ from $q \rightarrow 2n$. Using the LWE modulus switching technique (\autoref{subsec:modulus-switch-lwe}), our original LWE ciphertext $\textsf{LWE}_{\vec{s}, \sigma}(\Delta m + e) = ({a}_0, {a}_1, \gap{$\cdots$} {a}_{k-1}, {b}) \in \mathbb{Z}_q^{k+1}$ (i.e., the initial input to the noise bootstrapping procedure) will be converted into the following: 

$ $

$\textsf{LWE}_{\vec{s}, \sigma}(\hat{\Delta} m) = (\hat{a}_0, \hat{a}_1, \gap{$\cdots$} \hat{a}_{k-1}, \hat{b}) \in \mathbb{Z}_{2n}^{k+1}$

$\vec{s} = (s_0, s_1, \gap{$\cdots$} s_{k-1}) \in \mathbb{Z}_2^{k}$  \textcolor{red}{ $\rhd$ the secret key stays the same, as each $s_i$ is binary}

$\hat{\Delta} = \Delta\dfrac{2n}{q} = \dfrac{2n}{t} \in \mathbb{Z}_{2n}$

$\hat{a}_i = \left \lceil a_i\dfrac{2n}{q} \right \rfloor \in \mathbb{Z}_{2n}$

$\hat{e} = \left \lceil e\dfrac{2n}{q} \right \rfloor \in \mathbb{Z}_{2n}$

$\hat{b} = \left \lceil b\dfrac{2n}{q} \right \rfloor \approx \sum\limits_{i=0}^{k-1} \hat{a}_is_i + \hat{\Delta} m + \hat{e}  \in \mathbb{Z}_{2n}$

$ $

\para{The degree of $V(X)$ and Security}: If our goal were to design the minimal-degree polynomial $V$ whose coefficients map all possible values of the plaintext $m$, then it would be sufficient to design a ${t}$-degree polynomial $V$. Nonetheless, the reason why we choose the degree of $V$ to be $2n$ instead of ${t}$ is to guarantee an enough security level-- the higher the polynomial degree $n$ is, the safer our scheme is against attacks.

\subsubsection{Halving the Plaintext Space To be Used}
\label{subsec:tfhe-zero-padding} 

Problematically, the LUT polynomial $V(X)$  rotates \textit{negacyclically}, that is, $V(X)\cdot X^{n} = -V(X)$ (i.e., coefficients flip their signs with the rotation period of $n$). More generally:

$V(X)\cdot X^{-j}= V(X)\cdot X^{2n - j} = V(X)\cdot X^{4n - j} = \cdots$

$
= V(X)\cdot X^{-(j \bmod 2n)} =  
\begin{cases}
    v_j  + v_{j+1}X + \cdots, \text{for } 0\leq i<n\\
    -v_{j} - v_{j-1}X - \cdots, \text{for } n\leq j<2n
\end{cases}
$

$ $

\noindent , where $v_j$ denotes the constant term's coefficient after rotating the polynomial $V(X)$ by $j$ positions to the left. Problematically, $v_j$ flips its sign whenever its rotation crosses the boundary between $X^0$ and $X^{n-1}$. Given the modulus-switched values $\hat{a}_j$, $\hat{e}$, and $\hat{b}$, we design the following LUT polynomial $V(X)$:

$ $

$V(X) = v_0 + v_1X^1 + v_2X^2 + \gap{$\cdots$} + v_{n-1}X^{n-1}$

\text{ } $= m_0X^{\hat{\Delta} m_0 + \hat{e}_0} + m_0X^{\hat{\Delta} m_0 + \hat{e}_1} + m_0X^{\hat{\Delta} m_0 + \hat{e}_2} + \gap{$\cdots$} +  m_0X^{\hat{\Delta} m_0 + \hat{e}_{\hat{\Delta} - 1}}$ \textcolor{red}{ $\rhd$ total $\hat{\Delta}$ terms}

\text{ } $ + \text{ } m_1X^{\hat{\Delta} m_1 + \hat{e}_0} + m_1X^{\hat{\Delta} m_1 + \hat{e}_1} + m_1X^{\hat{\Delta} m_1 + \hat{e}_2} + \gap{$\cdots$} + m_1X^{\hat{\Delta} m_1 + \hat{e}_{\hat{\Delta} - 1}}$ \textcolor{red}{ $\rhd$ total $\hat{\Delta}$ terms}

\text{ } $ + \gap{$\cdots$} $

\text{ } $ + \text{ } m_{t/2 - 1}X^{\hat{\Delta} m_{t/2 - 1} + \hat{e}_0} + m_{t/2 - 1}X^{\hat{\Delta} m_{t/2 - 1} + \hat{e}_1} + m_{t/2 - 1}X^{\hat{\Delta} m_{t/2 - 1} + \hat{e}_2} + \gap{$\cdots$} + m_{t/2 - 1}X^{\hat{\Delta} m_{t/2 - 1} + \hat{e}_{\hat{\Delta} - 1}}$ \textcolor{red}{ $\rhd$ total $\hat{\Delta}$ terms}

$ $

Remember that by computing $V(X)\cdot X^{-j}$ for  $j= \{0, 1, \cdots , n-1\}$, we can rotate $V(X)$'s coefficients to the left by $\{0, 1, \cdots n-1 \}$ positions. For each $j$-slot rotation of $V(X)$ where $j= \{0, 1, \cdots , n-1\}$, the rotated polynomial $V'(X)$ gets the following values as the constant-term's coefficient: 

$\overbrace{\overbrace{\underbrace{\Delta m_0, \Delta m_0, \gap{$\cdots$}}_{\text{coeff. of } X^{\hat{\Delta}m_0 + \hat{e}_*}}}^{\hat{\Delta} \text{ repetitions}} \overbrace{\underbrace{\Delta m_1, \Delta m_1, \gap{$\cdots$}}_{\text{coeff. of } X^{\hat{\Delta}m_1 + \hat{e}_*}}}^{\hat{\Delta} \text{ repetitions}} \gap{$\cdots$} \overbrace{\underbrace{\Delta m_{{t}/{2}-1}, \Delta m_{{t}/{2}-1}, \gap{$\cdots$}}_{\text{coeff. of } X^{\hat{\Delta}m_{{t}/{2}-1}+ \hat{e}_*} }}^{\hat{\Delta} \text{ repetitions}}}^{\text{$V'(X)$'s constant term's coefficient for $j = \{0, 1, \cdots n-1\}$ rotations}}$

$ $

In the above expression, $\hat{e}_*$ is a noise that can range from $[0, \hat{\Delta})$. Note that all of $\hat\Delta m_i + \hat{e}_0, \hat\Delta m_i + \hat{e}_1, \cdots,\hat\Delta m_i + \hat{e}_{\hat\Delta - 1}$ exponents are designed to be mapped to the same coefficient value, $m_i$, which aligns with the fact that their underlying plaintext $m_i$ is the same when decrypted (once their associated noise $\hat{e}_*$ gets eliminated). This is why each $m_i$ is redundantly used $\hat\Delta$ times in a row as coefficients in $V(X)$. We can view this sequential repetition of coefficients as having a robustness of mapping each $\hat{\Delta}m_i + \hat{e}_*$ to $m_i$ against any noise $\hat{e}_* \in \mathbb{Z}_{\hat{\Delta}}$. 

So far, the above sequence of $m_0, m_1, \cdots, m_{{t}/{2} - 1}$ coefficients is what we expect $V'(X)$ (i.e., the rotated polynomial) to return as its constant term's coefficient for each of $0, 1, \cdots, n-1$ rotations (where each $m_i + 1 = m_{i+1}$). However, the correctness of the coefficient mappings breaks when the rotation count is between $[n, 2n -1)$, because their coefficients flip their signs when they cross the term's boundary from $X^0$ to $X^{n-1}$, due to the polynomial ring's negacyclic nature. Specifically, the constant term's coefficient values of the rotated polynomial $V'(X)$ are as follows for each rotation of $n, n+1, \cdots, 2n-1$ positions:

$ $

$\overbrace{\overbrace{\underbrace{-m_{0}, -m_{0}, \gap{$\cdots$}}_{-\text{coeff. of } X^{\hat{\Delta}m_{0} + \hat{e}_*}}}^{\hat{\Delta} \text{ repetitions}} \overbrace{\underbrace{-m_{1}, -m_{1}, \gap{$\cdots$}}_{-\text{coeff. of } X^{\hat{\Delta}m_{1} + \hat{e}_*}}}^{\hat{\Delta} \text{ repetitions}} \gap{$\cdots$} \overbrace{\underbrace{-m_{t/2 - 1}, -m_{t/2 - 1}, \gap{$\cdots$}}_{-\text{coeff. of } X^{\hat{\Delta}m_{t/2 - 1} + \hat{e}_*} }}^{\hat{\Delta} \text{ repetitions}}}^{\text{$V'(X)$'s constant term's coefficient after each of $n, n+1, \cdots 2n-1$ rotations}}$ 

$ $

As we can see above, the rotated $V'(X)$'s constant term's coefficient shows a negacyclic pattern with the rotation period of $n$, where the second $n$-rotation group's coefficients are exactly the negated values of the first $n$-rotation group's values. Let's understand why this negacyclic behavior breaks the (exponent, coefficient) = $(\hat{\Delta} m + \hat{e}, m)$ mappings. Since TFHE's plaintext and ciphertext values are defined in rings, as we rotate the LUT polynomial $V(X)$, we ideally want the rotated polynomial $V'(X)$'s constant term's value (i.e., mapped plaintext value) to wrap around in a circular manner, representing a ring pattern (with sequential $\hat{\Delta}$ repetitions of each value to be resistant against up to a $\hat{e}_* \in Z_{\hat{\Delta}}$ noise). However, the negacyclic nature of a polynomial ring makes the constant term's value of the second-half rotation group problematic, because they are exact negations of those of the first-half rotation group, breaking the circular wrapping-around ring pattern between the first-group values and the second-group values. 
%Specifically, as we gradually rotate $V$ with the rotation count $[0, 2n)$, the constant term's coefficient (i.e., $m_i$) should be ideally a value that is either the same as before or greater than the previous value ($m_{i-1}$), all the way to $2n$ rotations (to represent a ring pattern). This constraint holds for the first-half $i = [0, n)$, as the rotated $V'(X)$'s constant term's coefficient is $m_0, \cdots, m_1, \cdots, m_{t/2 - 1}$. However, for the second-half $i = [n, 2n)$, the rotated $V'(X)$'s constant term's coefficient is $-m_0, \cdots, -m_1, \cdots, -m_{t/2 - 1}$, which violates our constraint that $m_i \leq m_{i+1}$. Therefore, the combination of the first-half and the second-half domains of $i$ cannot mathematically implement a continuous modulo value range $\mathbb{Z}_t$. 

To summarize the problem, $V(X)$ has a limitation in becoming a perfect LUT, because it preserves the correct mappings of (exponent, coefficient) = $(\hat{\Delta} m + \hat{e}, m)$ only for one half of $i \in \mathbb{Z}_t$, not for the other half. 

\para{Solution:} 
Good news is that we have observed that $V(X)$'s mappings of (exponent, coefficient) = $(\hat{\Delta} m + \hat e, \Delta m)$ preserve their ring-pattern consistency if $V(X)$ is rotated no more than $n-1$ positions (i.e., the first-half rotation group). Therefore, the easiest solution to avoid the negacyclic problem of the LUT polynomial rotation is that the application of TFHE restricts $V(X)$ to be rotated no more than $n-1$ positions \textit{by design} during the noise bootstrapping. To enforce this, when the TFHE application's computation pipeline processes plaintext values (in its original plaintext computation logic before considering any homomorphic operations), the application should ensure to involve only some pre-defined contiguous $\dfrac{t}{2}$ modulo values within $\mathbb{Z}_t$ as the possible inputs and outputs of each computation step. This constraint effectively ensures the possible values of $\hat\Delta m + \hat e$ to be contiguous $n$ values within $\mathbb{Z}_{2n}$. Since the LUT polynomial $V(X)$ gets rotated by computing $V(X)\cdot X^{-(\hat\Delta m + \hat e)}$, as the application restricts $\hat\Delta m + \hat e$ to be at most $n-1$ (out of $2n - 1$) by its application design, $V(X)$ will be rotated at most $n-1$ positions during the noise bootstrapping. Thus, we can prevent the occurrences of the problematic $\{n, n+1, \gap{$\cdots$}, 2n -1 \}$ rotations that flip the signs of coefficients.  

To summarize, at the cost of halving the application's usable plaintext values to some contiguous $\dfrac{t}{2}$ values within $\mathbb{Z}_t$, we can prevent $V(X)$'s negacyclic rotation problem, and thereby preserve $V(X)$'s correct mappings of (exponent, coefficient) = $(\hat{\Delta} m + \hat e, m)$.

\begin{comment}
Meanwhile, the ciphertext space $q$ is unmodified, because If we take a simple example of $t = 8$, our TFHE application can design the following encoding scheme:

%Note that during the blind rotation (\autoref{subsec:bootstrapping-blind-rotation}), each step's cumulative rotation positions of the homomorphic MUX gate may end up rotating $V(X)$ more than $n$ positions. However, enforcing $0 \leq \hat\Delta m + e < n$ guarantees the rotation-finished $V'(X)$ to have the cumulative rotation count to be some integer within $[0, n) \bmod 2n$.

\begin{multicols}{2}
\begin{itemize}
 \item $0000_2 \rightarrow 000_2 (= 0)$
 \item $0001_2 \rightarrow 001_2 (= 1)$
 \item $0010_2 \rightarrow 010_2 (= 2)$
 \item $0011_2 \rightarrow 011_2 (= 3)$
 \item $0100_2 \rightarrow 100_2 (= -4)$
 \item $0101_2 \rightarrow 101_2 (= -3)$
 \item $0110_2 \rightarrow 110_2 (= 2)$
 \item $0111_2 \rightarrow 111_2 (= 1)$
\end{itemize}
\end{multicols}

By designing such binary encoding, we can encode signed integers in $\dfrac{t}{2}$ out of $t$ plaintext states (where the MSB is always 0). 
\end{comment}

Considering all these, our final LUT polynomial $V(X)$ is as follows:

\begin{tcolorbox}[title={\textbf{\tboxlabel{\ref*{subsec:tfhe-zero-padding}} Structure of Lookup Table Polynomial $\bm{V(X)}$}}]

$ $

$V(X) = v_0 + v_1X^1 + v_2X^2 + \gap{$\cdots$} + v_{n-1}X^{n-1}$

\text{ } $= m_0X^{\hat{\Delta} m_0 + \hat{e}_0} + m_0X^{\hat{\Delta} m_0 + \hat{e}_1} + m_0X^{\hat{\Delta} m_0 + \hat{e}_2} + \gap{$\cdots$} + m_0X^{\hat{\Delta} m_0 + \hat{e}_{\hat{\Delta} - 1}}$

\text{ } $ + \text{ } m_1X^{\hat{\Delta} m_1 + \hat{e}_0} + m_1X^{\hat{\Delta} m_1 + \hat{e}_1} + m_1X^{\hat{\Delta} m_1 + \hat{e}_2} + \gap{$\cdots$} + m_1X^{\hat{\Delta} m_1 + \hat{e}_{\hat{\Delta} - 1}}$

\text{ } $ + \gap{$\cdots$} $

\text{ } $ + m_{{t}/{2} - 1}X^{\hat{\Delta} m_{{t}/{2} - 1} + \hat{e}_0} + m_{{t}/{2} - 1}X^{\hat{\Delta} m_{{t}/{2} - 1} + \hat{e}_1} + m_{{t}/{2} - 1}X^{\hat{\Delta} m_{{t}/{2} - 1} + \hat{e}_2}$

\text{ } \text{ } $ + \gap{$\cdots$} + m_{{t}/{2} - 1}X^{\hat{\Delta} m_{{t}/{2} - 1} + \hat{e}_{\hat{\Delta} - 1}}$

$ $

\noindent , where $\hat\Delta = \Delta \cdot \dfrac{2n}{q} = \dfrac{2n}{t}$. The application should ensure that $m_0, m_1, \cdots, m_{t/2 - 1}$ are some contiguous modulo-$\dfrac{t}{2}$ values in $\mathbb{Z}_t$. This constraint ensures $V(X)$'s rotation positions $\hat{\Delta} m_i + \hat {e}_{*}$ (where $e_* \in \mathbb{Z}_{\hat{\Delta}}$) to be most $n$ contiguous possibilities, preventing $V(X)$ from making more than 1 full-cycle rotation that triggers a negacyclic problem.

\end{tcolorbox}

Another name for the LUT polynomial $V(X)$ is an accumulator.

\subsubsection{Blind Rotation}
\label{subsec:bootstrapping-blind-rotation} 

Blind rotation refers to rotating the coefficients of an \textit{encrypted} polynomial so that it is not possible to know how many positions the polynomial's coefficients have been rotated. After the rotation, it is not possible to see which coefficient has moved to which degree term. Blind rotation uses the basic polynomial rotation method 1 technique (Summary~\ref*{subsec:coeff-rotation} in \autoref{subsec:coeff-rotation}), with the difference that the $V(X)\cdot X^{-i}$ computation is done homomorphically. 

Note that $X^{\hat{\Delta} m + \hat{e}_b} = X^{\hat{b} - \sum_{i=0}^{k-1}{\hat{a}_is_i}}$, so we can rotate $V$ by computing $V \cdot X^{-(\hat{b} - \sum_{i=0}^{k-1}{\hat{a}_is_i})} = V \cdot X^{-\hat{b} + \sum_{i=0}^{k-1}{\hat{a}_is_i}}$. In fact, we cannot directly compute the LWE decryption formula $-\hat{b} + \sum_{i=0}^{k-1}{\hat{a}_is_i}$ (or $\hat{b} - \sum_{i=0}^{k-1}{\hat{a}_is_i}$) without the knowledge of the LWE secret key $S$. Nevertheless, there is a mathematical work-around to compute $V \cdot X^{-\hat{b} + \sum_{i=0}^{k-1}{\hat{a}_is_i}}$ without the knowledge of the secret key $S$, provided we are given $\{GGSW_{\vec{S}_{bk}, \sigma}^{\beta, l}(s_i)\}_{i=0}^{k-1}$ at the TFHE setup stage. Note that $\{GGSW_{\vec{S}_{bk}, \sigma}^{\beta, l}(s_i)\}_{i=0}^{k-1}$ is a GGSW encryption of the LWE secret key $S$, encrypted by the GLWE secret key $\vec{S}_{bk}$ (i.e., a \textit{bootstrapping} key). We use $\vec{S}_{bk}$ to homomorphically compute $V \cdot X^{-\hat{b} + \sum_{i=0}^{k-1}{\hat{a}_is_i}}$ (i.e., blindly rotate the coefficients of $V$ to the left by $\hat{b} + \sum_{i=0}^{k-1}{\hat{a}_is_i}$ positions), according to the following procedure:

\begin{enumerate}
\item GLWE-encrypt the polynomial $V$ with the bootstrapping key $\vec{S}_{bk}$ at the TFHE setup stage, so that each coefficient of $V$ gets encrypted.
\item Compute $V_0 = V \cdot X^{-\hat{b}}$, which is basically rotating $V$'s polynomials by $\hat{b}$ positions to the left. Since $\hat{b}$ is a known value visible in the LWE ciphertext, we can directly compute the rotation of $V_0 = V \cdot X^{-\hat{b}}$. 
\item Compute $V_1 = V_0 \cdot X^{\hat{a}_0s_0} = s_0 \cdot (V_0 \cdot X^{\hat{a}_0} - V_0) + V_0$.  
This formula works for the special case where $s_0 \in \{0, 1\}$: if $s_0 = 0$, then $V_1 = V_0$; else if $s_0 = 1$, then $V_1 = V_0 \cdot X^{\hat{a}_0}$. Computing $s_0 \cdot (V_0 \cdot X^{\hat{a}_0} - V_0) + V_0$ is done as a TFHE homomorphic addition and multiplication. We call this blind rotation of $V_0$, where the selection bit $s_0$ (i.e., the 1st element of the secret key vector $S$) is encrypted as a GGSW ciphertext by using $\vec{S}_{bk}$ (i.e. the bootstrapping key) and $V_0$ is an encrypted polynomial as a GLWE ciphertext. Multiplying GLWE-encrypted $V_0$ with $x^{\hat{a}_0}$ is done by GLWE ciphertext-to-plaintext multiplication (\autoref{sec:glwe-mult-plain}), and subtracting the result by GLWE-encrypted $V_0$ is done by GLWE ciphertext-to-ciphertext addition/subtraction (\autoref{sec:glwe-add-cipher}), and multiplying the result by GGSW-encrypted $s_0$ is done by GLWE-to-GGSW multiplication (\autoref{subsubsec:tfhe-glwe-to-ggsw-multiplication}), and adding the result with GLWE-encrypted $V_0$ is done by GLWE ciphertext-to-ciphertext addition. If $s_0 = 1$, then the formula's $X^{\hat{a}_0}$ term gets multiplied to $V_0$, which effectively rotates $V_0$'s coefficients by $\hat{a}_0$ positions to the right. Else if $s_0 = 0$, then $V_0$ does not get rotated and stays the same. In both cases, the resulting $V_1$ is encrypted as a new GLWE ciphertext. During this blind rotation, unless we have the knowledge of $s_0$ and $\vec{S}_{bk}$, it is impossible to know whether $V_0$ has been rotated or not, and also how many positions have been rotated.
\item By using the same blind rotation method as in the previous step, compute the GLWE encryption of $V_2 = V_1 \cdot X^{\hat{a}_1s_1} = s_1 \cdot (V_1 \cdot X^{\hat{a}}_1 - V_1) + V_1$. Note that we have the following publicly known components: $\hat{a}_1$, a GLWE encryption of $V_1$, and a GGSW ciphertext of $s_1$ encrypted by using $\vec{S}_{bk}$. 
\item Continue to compute the GLWE encryption of $V_3, V_4, \gap{$\cdots$}, V_k$ in the same manner, and we finally get a GLWE encryption of $V_k$, whose computed value is equivalent to:

$V' = V_k$

$\text{ } = V_{k-1} \cdot X^{\hat{a}_{k-1}s_{k-1}}$

$\text{ } = V_0 \cdot X^{\hat{a}_0s_0}X^{\hat{a}_1s_1}\cdots X^{\hat{a}_{k-1}s_{k-1}}$

$\text{ } = V \cdot X^{- \hat{b} + \sum_{i=0}^{k-1}{\hat{a}_is_i}} $

$\text{ } = V \cdot X^{-(\hat{\Delta} m + \hat{e}_b)}$

\end{enumerate}

This means that the GLWE encryption of the final polynomial $V_k$ will have the coefficient $m$ in its constant term, as $V(X)$ is designed to have the mapping ($\hat\Delta m + \hat e_{*}, m$).

Note that while we restrict the application's plaintext space usage to some contiguous $\dfrac{t}{2}$ modulo values within $\mathbb{Z}_t$ (\autoref{subsec:tfhe-zero-padding}), this restriction does not exist in the ciphertext space. That is, it is acceptable for blind rotations to rotate $V(X)$ more than $n$ positions during the intermediate steps because their invalid positions can be brought back to valid ones by subsequent steps. Therefore, what matters for the noise bootstrapping correctness is only the completed state $V_k(X)$. The rotation-completed $V_k(X)$ must have been rotated $\hat\Delta m + \hat e$ positions to the left. Therefore, we only need to ensure that  $\hat\Delta m + \hat e$ falls within our pre-defined contiguous $\dfrac{t}{2}$ modulo range within the $\mathbb{Z}_t$ domain, which is equivalent to ensuring that the aggregate rotation count is at most $n-1$ positions to avoid the extraction of any double-signed contradicting coefficients. 

\begin{figure}[h!]
    \centering
  \includegraphics[width=0.2\linewidth]{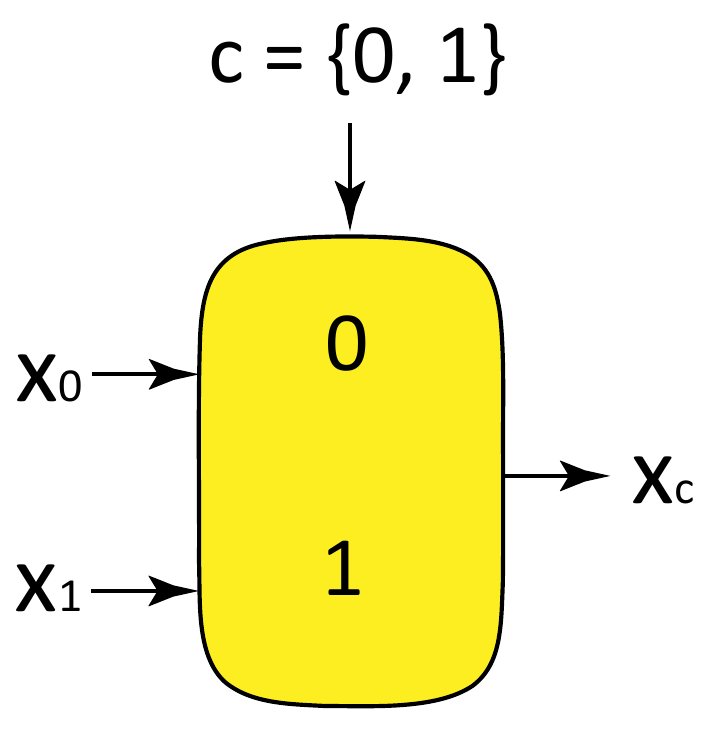}
  \caption{An illustration of the MUX logic gate }
  \label{fig:mux}
\end{figure}

\para{Homomorphic MUX Logic Gate:} In step 3, the formula $s_0 \cdot (V_0 \cdot X^{\hat{a}_0} - V_0) + V_0$ implements the MUX logic gate as shown in \autoref{fig:mux}, where in our case $s_0$ is the selection bit that chooses between the two inputs: $V_0$ and $V_0 \cdot X^{\hat{a}_0}$. If $s_0$ is 1, then the output is $V_0 \cdot X^{\hat{a}_0}$; otherwise, the output is $V_0$. In our design, the homomorphic computation of $s_0 \cdot (V_0 \cdot X^{\hat{a}_0} - V_0) + V_0$ effectively implements a homomorphic MUX gate, where the two inputs are LWE-encrypted and the selection bit is GGSW-encrypted.

\para{Dimensions of the GLWE and GGSW Ciphertexts:} The GLWE ciphertext that encrypts the LUT polynomial $V(X)$ has a $k' \times n$ dimension, where $k'$ is the length of $\vec{A}$ (i.e., the number of public mask polynomials) and $n$ is the polynomial degree of $A, B,$ and $V$. The dimension of the GGSW ciphertext that encrypts each element of $\vec{S}_{\textit{bk}}$ (i.e., each coefficient of the bootstrapping key polynomials) is $(k' + 1) \times l \times (k' + 1)$. In practice, we let $k' = 1$, which simplifies these ciphertexts as RLWE and RGSW ciphertexts. The reason we set $k' = 1$ is for computational efficiency. 

\subsubsection{Coefficient Extraction}

Next, we use the coefficient extraction technique (\autoref{subsec:tfhe-extraction}) to extract the constant term's coefficient of the rotated polynomial $V'(X)$' as an encryption of $m$: $\textsf{LWE}_{\vec{s}', \sigma}(\Delta m)$, where $\vec{s}'$ is a vector of length $k'\cdot n$. At this point, the original $\textsf{LWE}$ ciphertext's old noise $\hat{e}_b$ has disappeared, and the bootstrapped new ciphertext $\textsf{LWE}_{\vec{s}', \sigma}(\Delta m)$ has newly generated small noise $e_s$. 

In fact, the homomorphic MUX logic in the blind rotation procedure (\autoref{subsec:bootstrapping-blind-rotation}) involves numerous ciphertext multiplications and additions, which can accumulate additional noise until we reach the point of coefficient extraction. To limit the accumulating noise during the series of MUX logic operations, we can carefully adjust the security parameters. For example, we can design a narrower Gaussian distribution for sampling the noise $e$, while designing a sufficiently large $n$ to compensate for the reduced noise. Meanwhile, increasing $q$ can make the system less secure because it becomes more vulnerable to lattice reduction attacks.

\subsubsection{Key Switching}
\label{subsec:bootstrapping-key-switch} 

The output of the coefficient extraction is $\textsf{LWE}_{\vec{s}', \sigma}(\Delta m)$, which is encrypted by a secret key $\vec{s}'$ whose length is $k'\cdot n$. For consistency, we need to switch its key from $\vec{s}' \rightarrow \vec{s}$, where $\vec{s}$ is our original $k$-length key. This key switching can be done by using the technique explained in Summary~\ref*{subsec:tfhe-key-switching} (in \autoref{subsec:tfhe-key-switching}).

\subsubsection{Noise Bootstrapping Summary}
\label{subsec:tfhe-summary}

TFHE's noise bootstrapping procedure is summarized as follows:

\begin{tcolorbox}[title={\textbf{\tboxlabel{\ref*{subsec:tfhe-summary}} TFHE Noise Bootstrapping Procedure}}]

\textbf{\underline{Lookup Table Encryption}:} Encrypt the LUT polynomial $V(X)$ as a GLWE ciphertext by using the bootstrapping key $\vec{S}_{bk}$ whose length is $k'$.

\begin{enumerate}
\item \textbf{\underline{Modulus Switch}:} Change the modulus of the TFHE ciphertext $\textsf{LWE}_{\vec{s}, \sigma}(\Delta m + e_b)$ from $q \rightarrow 2n$ to get $\textsf{LWE}_{\vec{s}, \sigma}(\hat{\Delta} m + \hat e_b)$, where $\hat\Delta = \Delta \cdot \dfrac{2n}{q} = \dfrac{2n}{t}$.

$ $

\item \textbf{\underline{Blind Rotation}:} Rotate the GLWE-encrypted polynomial $V$ by $(\hat{b} - \sum\limits_{i=0}^{k-1}{\hat{a}_is_i}) = (\hat{\Delta}m + \hat{e}_b)$ positions to the left, by recursively computing: 

$V_0 = V \cdot X^{-\hat{b}}$

$V_1 = V_0 \cdot X^{\hat{a}_0s_0} = s_0 \cdot (V_0 \cdot X^{\hat{a}_0} - V_0) + V_0$

\text{ } $\vdots$

$V_k = V_{k-1} \cdot X^{\hat{a}_{k-1}s_{k-1}} = s_{k-1} \cdot (V_{k-1} \cdot X^{\hat{a}_{k-1}} - V_{k-1}) + V_{k-1}$

$= V_0 \cdot X^{\hat{a}_0s_0}X^{\hat{a}_1s_1}\cdots X^{\hat{a}_{k-1}s_{k-1}}$

$= V \cdot X^{- \hat{b} + \sum_{i=0}^{k-1}{\hat{a}_is_i}} $

$ = V \cdot X^{-(\hat{\Delta}m + \hat{e}_b)}$

$ $

\noindent Each step of the actual blind rotation above is computed as the following TFHE ciphertext-to-ciphertext multiplication and addition:

$\textsf{GLWE}_{\vec{S}_{bk}, \sigma}(V_{i+1}) = \textsf{GGSW}_{\vec{S}_{bk}, \sigma}^{\beta, l}(s_i) \cdot (\textsf{GLWE}_{\vec{S}_{bk}, \sigma}(V_i) \cdot X^{\hat{a}_i} - \textsf{GLWE}_{\vec{S}_{bk}, \sigma}(V_i)) + \textsf{GLWE}_{\vec{S}_{bk}, \sigma}(V_i)$

$ $

\item \textbf{\underline{Coefficient Extraction}:} Homomorphically extract the constant term's coefficient $m$ from the rotated polynomial $V_k$, which is $\textsf{LWE}_{\vec{s}', \sigma}(\Delta m)$, where $\vec{s}'$ is a vector of length $k' \cdot n$. 

$ $

\item \textbf{\underline{Key Switching}:} Homomorphically switch the key of the LWE ciphertext from $\vec{s}' \rightarrow \vec{s}$. 

%\item \textbf{Rescaling:} Multiply $\textsf{LWE}_{\vec{s}, \sigma}(\hat{\Delta}m)$ by $\dfrac{q}{2n}$ to get:  

%$\dfrac{q}{2n} \cdot LWE_{S, \sigma}(\hat{\Delta}m)$

%$\textsf{LWE}_{\vec{s}, \sigma}(\dfrac{q}{2n} \cdot \hat{\Delta}m)$

%$ = LWE_{S, \sigma}(\Delta m)$

%, whose noise get reinitialized to $e_s$. 
\end{enumerate}

$ $

\para{\underline{Halving the Usable Plaintext Range}: } Problematically, the LUT polynomial V rotates negacyclically. To avoid this problem, we require the application to ensure that the plaintext $m$ uses only contiguous $\dfrac{t}{2}$ modulo values within $\mathbb{Z}_t$. This way, we avoid rotating $V(X)$ more than $n-1$ positions that cause coefficient extraction of double-signed contradicting coefficients. 

$ $

\para{\underline{Choice of $k'$}:} For computational efficiency, $k'$ is set to be 1, which simplifies the GLWE and GGSW ciphertexts as RLWE and RGSW ciphertexts.  
%\item Piping the LUT output to a \textbf{negacyclic function}.
%\item Multiplying the LUT output with \textbf{another negacyclically rotating polynomial}. 
%\end{itemize}
\end{tcolorbox}

\subsubsection{Example: Noise Bootstrapping}
\label{subsec:tfhe-noise-bootstrapping-ex}

Suppose the GLWE security setup: $n = 16$, $t = 8$, $q = 64$, $k = 8$

$\mathbb{Z}_{t=8} = \{-4, -3, -2, -1, 0 , 1, 2, 3\}$

$\mathbb{Z}_{q=64} = \{ -32, -31, -30, \gap{$\cdots$}, 29, 30, 31 \}$

$\Delta = \dfrac{q}{t} = \dfrac{64}{8} = 8$

$ $

And suppose we have the following LWE ciphertext:

$\vec{s} = (1, 0, 0, 1, 1, 1, 0, 1) = \mathbb{Z}_2^{k=8}$

$m = 1  \in \mathbb{Z}_{t=8}$

$\Delta m = 1 \cdot 8 = 8 \in \mathbb{Z}_{q=64}$

$\textsf{LWE}_{\vec{s}, \sigma}(\Delta m) = (a_0, a_1, a_2, a_3, a_4, a_5, a_6, a_7, b) 
= (8, -28, 4, -32, 0, 31, -6, 7, 24) \in \mathbb{Z}_{q=64}^{k+1=9}$

$e = 2 \in \mathbb{Z}_{q=64}$ (should be the case that $|e| < \dfrac{\Delta}{2} = 4$ for correct decryption)

$b = \sum\limits_{i=0}^{7}a_is_i + \Delta m + e = (8 -32 + 31 +7) + 8 + 2 = 24 \in \mathbb{Z}_{q=64}$

$ $

Now then, the TFHE noise bootstrapping procedure is as follows:

$ $

\begin{enumerate}
\item \textbf{Modulus Switch:} Switch the modulus of $\textsf{LWE}_{\vec{s}, \sigma}(\Delta m)$ From $q \rightarrow 2n$, which is from $64 \rightarrow 32$. After the modulus switch, the original LWE ciphertext is converted as follows:

$\mathbb{Z}_{2n=32} = \{-16, -15, -14, \gap{$\cdots$}, 13, 14, 15\}$

$\vec{s} = (1, 0, 0, 1, 1, 1, 0, 1) = \mathbb{Z}_2^{k=8}$

$\hat{\Delta} = \Delta\dfrac{2n}{64} = 8\dfrac{32}{64} = 4$

$\hat{\Delta}m = 4 \cdot 1 = 4 \in \mathbb{Z}_{2n=32}$

$\hat{e} =  \left\lceil e \dfrac{2n}{q} \right\rfloor = \left\lceil 2\dfrac{32}{64} \right\rfloor = 1 \in \mathbb{Z}_{2n=32}$

$ $

$\textsf{LWE}_{\vec{s}, \sigma}(\hat{\Delta}m) = (\hat{a}_0, \hat{a}_1, \hat{a}_2, \hat{a}_3, \hat{a}_4, \hat{a}_5, \hat{a}_6, \hat{a}_7, \hat{b}) \in \mathbb{Z}_{2n=32}^{k+1=9}$

$ = (\Big\lceil 8\dfrac{32}{64} \Big\rfloor, \Big\lceil -28\dfrac{32}{64}\Big\rfloor, \Big\lceil 4\dfrac{32}{64}\Big\rfloor, \Big\lceil -32\dfrac{32}{64}\Big\rfloor, \Big\lceil 0\dfrac{32}{64}\Big\rfloor, \Big\lceil 31\dfrac{32}{64}\Big\rfloor, \Big\lceil -6\dfrac{32}{64}\Big\rfloor, \Big\lceil 7\dfrac{32}{64}\Big\rfloor, \Big\lceil 24\dfrac{32}{64}\Big\rfloor)$

$ $

$= (4, {-14}, 2, -16, 0, 16, {-3}, 4, 12)$

$ $

Note that $\sum\limits_{i=0}^{7}(\hat{a}_is_i) + \hat{\Delta}m + \hat{e} = (4 - 16 + 16 + 4) + 4 + 1 = 13 \in \mathbb{Z}_{2n=32}$

$ $

$\hat b = 12 \approx 13 = \sum\limits_{i=0}^{7}(\hat{a}_is_i) + \hat{\Delta}m + \hat{e}$

This small difference in $\hat b$ comes from the aggregated noises of rounding $\hat a_0, \hat a_1, \cdots , \hat e$ during the modulus switch. 

$ $

\item \textbf{Blind Rotation:} We assume that the application avoids the problem of negacyclic polynomial rotation by ensuring that the usable plaintext values are the following contiguous $\dfrac{8}{2}$ modulo values within $\mathbb{Z}_8 = \{-4, -3, -2, -1, 0, 1, 2, 3\}$, which are $\{-2, -1, 0, 1\}$. This implies that the only possible values of $i=\hat\Delta m + \hat e$ in $V(X)\cdot X^{i}$ will be: $i = \{-8, -7, \cdots, 6, 7\}$. Based on these requirements, \autoref{tab:lut} is the Lookup Table polynomial $V(X)$ that maps $\hat{\Delta} m + \hat{e}$ to $\Delta m$.

\begin{table}[h]
\centering
\footnotesize
%\noindent\adjustbox{max width=\columnwidth}{
\begin{tabular}{|c||c|c|c|c|c|c|c|c|c|c|} % left align
\hline
% remove line in a table, \usepackage{adjustbox}
\multicolumn{10}{|l|}{$V(X) = v_0 + v_1X + v_2X^2 + v_3X^3 + v_4X^4 + v_5X^5 + v_6X^6 + v_7X^7$}  \\ 
\multicolumn{10}{|l|}{\textcolor{white}{......} $ + v_8X^8 + v_9X^9 + v_{10}X^{10} + v_{11}X^{11} + v_{12}X^{12} + v_{13}X^{13} + v_{14}X^{14} + v_{15}X^{15}$}  \\ 
\multicolumn{10}{|l|}{\textcolor{white}{....}$= 0 + 0X +0X^2 +0X^3 +1X^4 +1X^5 +1X^6 +1X^7$}  \\ 
\multicolumn{10}{|l|}{\textcolor{white}{......}$+ 2X^8  + 2X^9 + 2X^{10} + 2X^{11} + 1X^{12} + 1X^{13} + 1X^{14} + 1X^{15}$} \\ 
\hline
\hline
\textbf{\boldmath$i = \hat{\Delta}m + \hat{e}$} & $-8$ & $-7$ & $-6$ & $-5$ & $-4$ & $-3$ & $-2$ & $-1$ \\
(in $V \cdot X^{-i}$) & ($\textcolor{orange}{110}\textcolor{green}{00}_2$) & ($\textcolor{orange}{110}\textcolor{green}{01}_2$) & ($\textcolor{orange}{110}\textcolor{green}{10}_2$) & ($\textcolor{orange}{110}\textcolor{green}{11}_2$) & ($\textcolor{orange}{111}\textcolor{green}{00}_2$) & ($\textcolor{orange}{111}\textcolor{green}{01}_2$) & ($\textcolor{orange}{111}\textcolor{green}{10}_2$) & ($\textcolor{orange}{111}\textcolor{green}{11}_2$)\\
\hline
\textbf{constant term's} & $-2$ & $-2$ & $-2$ & $-2$ & $-1$ & $-1$ & $-1$ & $-1$ \\
 \textbf{coeff. of $V\cdot X^{-i}$}& $\textcolor{orange}{110}_2$ & $\textcolor{orange}{110}_2$ & $\textcolor{orange}{110}_2$ & $\textcolor{orange}{110}_2$ & $\textcolor{orange}{111}_2$ & $\textcolor{orange}{111}_2$ & $\textcolor{orange}{111}_2$ & $\textcolor{orange}{111}\textcolor{green}{00}_2$ \\
\hline
\textbf{$\bm{m}$ (plaintext)} & $-2$ & $-2$ & $-2$ & $-2$ & $-1$ & $-1$ & $-1$ & $-1$ \\
\hline
\hline
\textbf{\boldmath$i = \hat{\Delta}m + \hat{e}$} & $0$ & $1$ & $2$ & $3$ & $4$ & $5$ & $6$ & $7$ \\
(in $V \cdot X^{-i}$) & ($\textcolor{orange}{000}_2$)& ($\textcolor{orange}{000}_2$)& ($\textcolor{orange}{000}_2$)& ($\textcolor{orange}{000}_2$)& ($\textcolor{orange}{001}_2$)& ($\textcolor{orange}{001}_2$)& ($\textcolor{orange}{001}_2$)&($\textcolor{orange}{001}_2$)\\
\hline
\textbf{constant term's} & $0$ & $0$ & $0$ & $0$ & $1$ & $1$ & $1$ & $1$ \\
\textbf{coeff. of $V\cdot X^{-i}$}& $\textcolor{orange}{000}\textcolor{green}{00}_2$ & $\textcolor{orange}{000}\textcolor{green}{00}_2$ & $\textcolor{orange}{000}\textcolor{green}{00}_2$ & $\textcolor{orange}{000}\textcolor{green}{00}_2$ & $\textcolor{orange}{001}\textcolor{green}{00}_2$ & $\textcolor{orange}{001}\textcolor{green}{00}_2$ & $\textcolor{orange}{001}\textcolor{green}{00}_2$ & $\textcolor{orange}{001}\textcolor{green}{00}_2$ \\
\hline
\textbf{$\bm{m}$ (plaintext)} & $0$ & $0$ & $0$ & $0$ & $1$ & $1$ & $1$ & $1$ \\
\hline
\end{tabular}
\centering
\caption{The Lookup Table for $n=16, q=64, t=8$ LWE setup.
\textcolor{orange}{Orange} is the plaintext $m$'s bits. \textcolor{green}{Green} is the noise $e$'s bits. %\textcolor{cyan}{Cyan} is the 0-padding bit at the MSB of $\hat\Delta m + e$ (and thus the MSB of $m$), ensured by the application's usage of plaintext numbers.
}
\label{tab:lut}
\end{table}

Note that $V(X)$'s coefficients for the $X^8 \sim X^{15}$ terms are $\{2, 1\}$ instead of $\{-2, -1\}$, so that if $V$ gets rotated by $\{-8,-7,-6,-5,4,5,6,7\}$ slots to the left, the constant term's coefficient flips its sign to $\{-2, -1\}$ due to wrapping around the boundary of the $n$ exponent. 

During the actual bootstrapping, we will do a blind rotation of \autoref{tab:lut}'s $V(X)$ (which is GLWE-encrypted) by $\hat{b} - \sum\limits_{i=0}^{7}\hat{a}_is_i = 4$ positions to the left, which is computed as follows:

$\hat\Delta m + \hat e = \hat{b} - \sum\limits_{i=0}^{7}\hat{a}_is_i = 12 - (4 - 16 + 16 + 4) = 4 \text{ mod 32} \in \mathbb{Z}_{2n=32}$

In \autoref{tab:lut}, if the rotation count $i = 4$, the corresponding constant term's coefficient is $v_4 = 1 = m$. As $\Delta = 4$, we finally get $\textsf{LWE}_{\vec{s},\sigma}(\Delta m) = 1$.

$ $

The actual blind rotation is computed as follows:

$\vec{s} = (1, 0, 0, 1, 1, 1, 0, 1)$

$\textsf{LWE}_{\vec{s}, \sigma}(\hat{\Delta}m) = (\hat{a}_0, \hat{a}_1, \hat{a}_2, \hat{a}_3, \hat{a}_4, \hat{a}_5, \hat{a}_6, \hat{a}_7, \hat{b}) =  (4, -14, 2, -16, 0, 16, -3, 4, 12)$

$V_0 = V \cdot X^{-\hat{b}} = V \cdot X^{-12} = v_{12} + v_{13}X + v_{14}X^2 + \cdots $

$V_1 = V_0 \cdot X^{\hat{a}_0s_0} = s_0 \cdot (V_0 \cdot X^{\hat{a}_0} - V_0) + V_0 = V_0 \cdot X^{4} = v_{8} + v_{9}X + v_{10}X^2 + \cdots$ 

$V_2 = V_1 \cdot X^{\hat{a}_1s_1} = s_1 \cdot (V_1 \cdot X^{\hat{a}_1} - V_1) + V_1 = V_1 = v_{8} + v_{9}X + v_{10}X^2 + \cdots$

$V_3 = V_2 \cdot X^{\hat{a}_2s_2} = s_2 \cdot (V_2 \cdot X^{\hat{a}_2} - V_2) + V_2 = V_2 = v_{8} + v_{9}X + v_{10}X^2 + \cdots$

$V_4 = V_3 \cdot X^{\hat{a}_3s_3} = s_3 \cdot (V_3 \cdot X^{\hat{a}_3} - V_3) + V_3 = V_3 \cdot X^{-16} = -v_{8} - v_{9}X - v_{10}X^2 - \cdots$

$V_5 = V_4 \cdot X^{\hat{a}_4s_4} = s_4 \cdot (V_4 \cdot X^{\hat{a}_4} - V_4) + V_4 = V_4 \cdot X^{0} = -v_{8} - v_{9}X - v_{10}X^2 - \cdots$

$V_6 = V_5 \cdot X^{\hat{a}_5s_5} = s_5 \cdot (V_5 \cdot X^{\hat{a}_5} - V_5) + V_5 = V_5 \cdot X^{16} = v_{8} + v_{9}X + v_{10}X^2 + \cdots$

$V_7 = V_6 \cdot X^{\hat{a}_6s_6} = s_6 \cdot (V_6 \cdot X^{\hat{a}_6} - V_6) + V_6 = V_6 = v_{8} + v_{9}X + v_{10}X^2 + \cdots$

$V_8 = V_7 \cdot X^{\hat{a}_7s_7} = s_7 \cdot (V_7 \cdot X^{\hat{a}_7} - V_7) + V_7 = V_7 \cdot X^{4} = v_{4} + v_5X + v_6X^2 + \cdots $

$ $

The final output of blind rotation is the GLWE ciphertext of $V_8$, $\textsf{GLWE}_{\vec{S}, \sigma}(V_8)$, whose constant term's coefficient is $v_4 = m = 1$. 

$ $

Each step of the actual blind rotation above is computed as the following TFHE ciphertext-to-ciphertext multiplication:

$\textsf{GLWE}_{\vec{S}, \sigma}(V_{i+1}) = \textsf{GGSW}_{\vec{S}, \sigma}^{\beta, l}(s_i) \cdot (\textsf{GLWE}_{\vec{S}, \sigma}(V_i) \cdot X^{\hat{a}_i} - \textsf{GLWE}_{\vec{S}, \sigma}(V_i)) + \textsf{GLWE}_{\vec{S}, \sigma}(V_i)$

$ $

We will leave this computation for the reader's exercise.

$ $

\item \textbf{Coefficient Extraction:} At the end of blind rotation, we finally get the following GLWE ciphertext:

$\textsf{GLWE}_{\vec{S}, \sigma}(V_8)$

$ = \textsf{GLWE}_{\vec{S}, \sigma}\bm(\hat{\Delta} \cdot (v_4 + v_5X + v_6X^2 + v_7X^3 + v_8X^{4} + v_9X^{5} + v_{10}X^{6} + v_{11}X^{7} + v_{12}X^{8} + v_{13}X^{9} + v_{14}X^{10} + v_{15}X^{11} - v_{0}X^{12} - v_{1}X^{13} - v_{2}X^{14} - v_{3}X^{15})\bm)$

$ = \textsf{GLWE}_{\vec{S}, \sigma}\bm(\hat{\Delta}\cdot(1 + 1X + 1X^2 + 1X^3 + 2X^{4} + 2X^{5} + 2X^{6} + 2X^{7} + 1X^{8} + 1X^{9} + 1X^{10} + 1X^{11} - 0X^{12} - 0X^{13} - 0X^{14} - 0X^{15})\bm)$

$ = \left(A_0 = \sum\limits_{j=0}^{15}(a_{0,0} + a_{0,1}X + \cdots), A_1 = \cdots, A_{k-1} = \cdots, B = \sum\limits_{j=0}^{15}b_{j}X^j\right)$

Now, we extract the constant term's coefficient of the encrypted polynomial $\textsf{GLWE}_{\vec{S}, \sigma}(\hat{\Delta} \cdot (1 + 1X + 1X^2 + \cdots))$ by using the coefficient extraction formula (Summary~\ref{subsec:tfhe-extraction}). Specifically, we will extract the constant term's coefficient, which corresponds to $\textsf{LWE}_{\vec{s}, \sigma}(\Delta m_0)$. We extract $\textsf{LWE}_{\vec{s}, \sigma}(\Delta m_0)$ by computing the following:

$\textsf{LWE}_{\vec{s}, \sigma}(\Delta m_0) = (a_0', a_1', \gap{$\cdots$} , a_{nk-1}', b_h)$ \textcolor{red}{ $\rhd$ where $h = 0$}

\[
    \text{, where } a'_{n \cdot i + j} =   
    \begin{cases}
      a_{i,0 - j} \text{ (if } 0 \leq j \leq 0\text{)}\\
      -a_{i,n + 0 - j} \text{ (if } 0+1 \leq j \leq n-1\text{)}\\
  \end{cases}
, b_0 \text{ is obtained from the polynomial } B
\]

\end{enumerate}

\subsubsection{Discussion}
\label{subsec:tfhe-noise-bootstrapping-discussion}

\begin{itemize}
\item \textbf{Programmable Bootstrapping}: While the bootstrapping (\autoref{subsec:tfhe-noise-bootstrapping}) uses a simple Lookup Table $V(X)$ which maps $\Delta m + e$ to $\Delta m$, we can edit the coefficients of $V(X)$ to make $\Delta m + e$ map to different values. For example, an altered mappings between the inputs and outputs to LUT can implement logic gates such as AND, OR, XOR, CMUX, etc, which will be explained in \autoref{subsec:tfhe-noise-bootstrapping-gate}. Such edited mappings between the exponents and coefficients in $V(X)$ are called programmable bootstrapping. If we encrypt $V(X)$ as a GLWE ciphertext, we can hide the mappings as well as each input instance, which effectively implements \textit{functional encryption}. Note that both the vanilla bootstrapping (\autoref{subsec:tfhe-noise-bootstrapping}) and programmable bootstrapping (\autoref{subsec:tfhe-noise-bootstrapping-gate}) generate the same amount of noise. 
\item \textbf{Bootstrapping Noise}: During the bootstrapping's LUT polynomial $V(X)$ rotation, we perform many TFHE multiplications in the homomorphic MUX gates to derive $V_0 \cdots V_k$, which inevitably creates additional noises before the noise gets re-initialized at the end. However, a careful parameter choice can limit the growth of this additional noise during modulus switch and blind rotation. 
\end{itemize}

\subsubsection{Application: Gate Bootstrapping}
\label{subsec:tfhe-noise-bootstrapping-gate}

Besides implementing the homomorphic MUX logic gate used during blind rotation (\autoref{subsec:bootstrapping-blind-rotation}), it is possible to leverage the LUT polynomial $V(X)$ to implement other homomorphic logic gates such as AND, NAND, OR, XOR, etc. When implementing these gates, each ciphertext is an encryption of a single-bit plaintext (or several bits can be bundled up in a linear combination formula and be processed simultaneously by using LUT). Suppose $q = 32$, $t = 8$, $m \in \mathbb{Z}_8 = \{-4,-3, -2,-1,0,1,2,3\}$, $\Delta = \dfrac{q}{t} = 4$, $\hat\Delta = \dfrac{\Delta\cdot 2n}{q} = 2$, and we encode the gate input into LWE plaintext as $0 \rightarrow -1$, and $1 \rightarrow 1$, and the maximum (accumulated) noise $e = [-1, 1]$.

\begin{table}[ht]
  \centering
  \resizebox{\columnwidth}{!}{%
    \begin{tabular}{c c | r r | c | c | c | c}
      \hline
        \multicolumn{8}{c}{Lookup Table Polynomial $V(X) = 1 + 1X + 1X^2 + 1X^3 + 1X^4 + 1X^5 + 1X^6 + 1X^7$}\\
      \hline
      \multicolumn{2}{c|}{\textbf{Decoded}} & 
      \multicolumn{2}{c|}{\textbf{Encoded}} & 
      \multicolumn{1}{c|}{\begin{tabular}{c}
        \textbf{Linear} \\ \textbf{combination} \\
        \textbf{of encodings}
      \end{tabular}} & 
      \multicolumn{1}{c|}{\begin{tabular}{c}
        \textbf{Scaled} \textbf{Encoded} \\
         \textbf{Combination} 
      \end{tabular}} &
      \textbf{Bootstrapping} & 
      \begin{tabular}{c}
        \textbf{Decoded} \\ \textbf{result} \\
      \end{tabular} \\
        \hline
      $d_1$ & $d_2$ & $m_1$ & $m_2$ & $m_1 + m_2 - 1$ & $\hat\Delta\cdot(m_1+m_2-1)$ &
      $V(X)\cdot X^{-\hat\Delta\cdot(m_1 + m_2 - 1)+e}$ &   $d_1 \land d_2$ \\
      \hline
      0 & 0 &
      $-1$ & $-1$ &
      $-3$ &
      $-6$ &
      constant term's coeff. is $-1$ &
      0 \\
      0 & 1 & 
      $-1$ & $1$ &
      $-1$ &
      $-2$ &
      constant term's coeff. is $-1$ &
      0 \\
      1 & 0 & 
      $1$ & $-1$ &
      $-1$ &
      $-2$ &
      constant term's coeff. is $-1$ &
      0 \\
      1 & 1 & 
      $1$ & $1$ &
      $1$ &
      $2$ &
      constant term's coeff. is $1$ &
      1 \\
      \hline
    \end{tabular}
  }
  \caption{An example truth table for an AND operation with an additional encoding.}
  \label{tab:gate-and}
\end{table}

%\begin{figure}[h!]
%    \centering
%  \includegraphics[width=\linewidth]{figures/torus-gate.png}
%  \caption{An illustration of an AND gate in Torus \href{https://assets-global.website-files.com/622ef9de9152c97467eac748/6298a348a0c75850f0688c83_NAND_bootstrap.png}{(Source)}}
%  \label{fig:torus-gate}
%\end{figure}

\autoref{tab:gate-and} is a programmable bootstrapping design for an AND logic gate. For this application, we define the LUT polynomial $V$ as $V(X) = \sum\limits_{i=0}^{7}X^i$. The LUT polynomial $V(X)$ maps one half of the plaintext domain to $1$, while the other half to $-1$ (as the terms wrap around the boundary of $X^7$). In this design setup, each bit is separately encrypted as independent TFHE ciphertext. Gate inputs 0 and 1 are encoded as $-1$ and $1$, respectively. The linear combination (i.e., homomorphic computation formula) for an AND gate is $\textsf{LWE}_{\vec{s}, \sigma}(\Delta m_1) + \textsf{LWE}_{\vec{s}, \sigma}(\Delta m_2) - 1$. Its output is positive if both inputs are positive (i.e. $1$, in which case the blind rotation will rotate $V$ to the left by $\hat\Delta\cdot 1 + e$ positions and the constant term's coefficient will be $1$. Thus, the output of blind rotation and coefficient extraction will be $\textsf{LWE}_{\vec{s}, \sigma}(\Delta \cdot 1)$ with a reduced noise, which is an encoding of $1$. This design can tolerate the maximum noise of $|e| = 1$. To endure bigger noises, we should increase $q$ and $n$. 

Note that the AND gate's LUT layout is negacyclic, which is a special case, thus we could use the entire $2n=16$ coefficient states in $V(X)$ for the AND gate mapping function's outputs, by leveraging $V(X)$'s innate property of negacyclic rotation. However, in many use cases, the LUT layout is not necessarily negacyclic like this AND gate example. Even our noise bootstrapping's LUT layout (\autoref{subsec:bootstrapping-overview}) was not negacyclic, but a unity function (as it simply removes the noise). Thus, for most use cases, we need to use only $\dfrac{t}{2}$ out of $t$ plaintext space to avoid more than $n-1$ rotations of $V(X)$ (\autoref{subsec:tfhe-zero-padding}). 

Besides the AND gate, other logic gates can be built in a similar manner, each of which is based on a different linear combination formula and LUT layout. 

\para{Division:} TFHE does not support direct division of plaintext numbers of any size. This is because TFHE's LWE vector elements are in the $Z_q$ ring, where each element $g$ does not necessarily have a multiplicative inverse $g^{-1}$, which makes it hard to multiply $g^{-1}$ to the target number to divide. Instead, division can be implemented as binary division based on the gates implemented by gate bootstrapping. To support binary division, each plaintext has to be a single bit and encrypted as an independent ciphertext. Or multiple bits can be bundled up and processed concurrently by designing a linear combination formula, similar to the linear combination that we designed for processing 2 input bits of an AND gate.

\subsubsection{Application: Neural Networks Bootstrapping}
\label{subsec:tfhe-neural-network}

\begin{figure}[h!]
    \centering
  \includegraphics[width=0.8\linewidth]{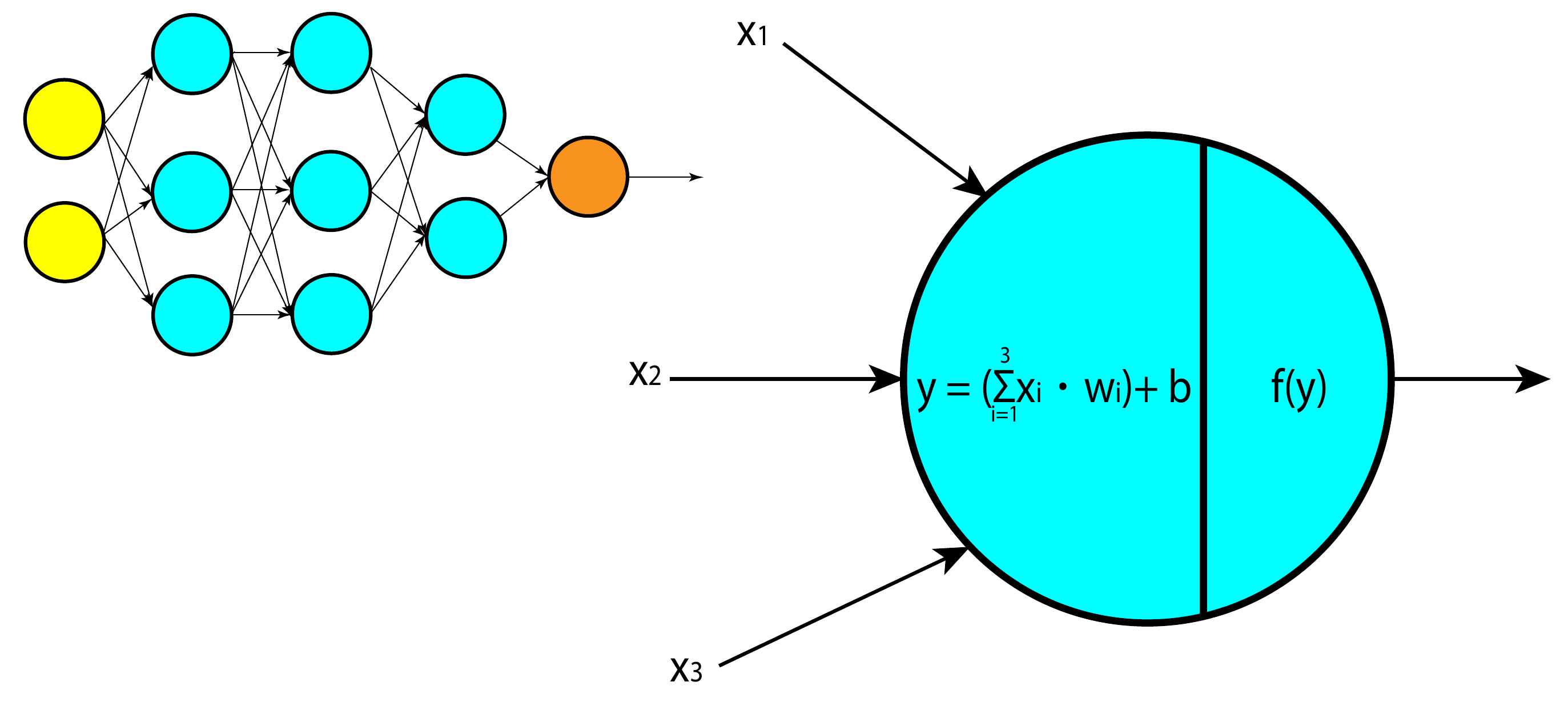}
  \caption{An illustration of neural networks }
  \label{fig:neural-network}
\end{figure}

\begin{figure}[h!]
    \centering
  \includegraphics[width=0.8\linewidth]{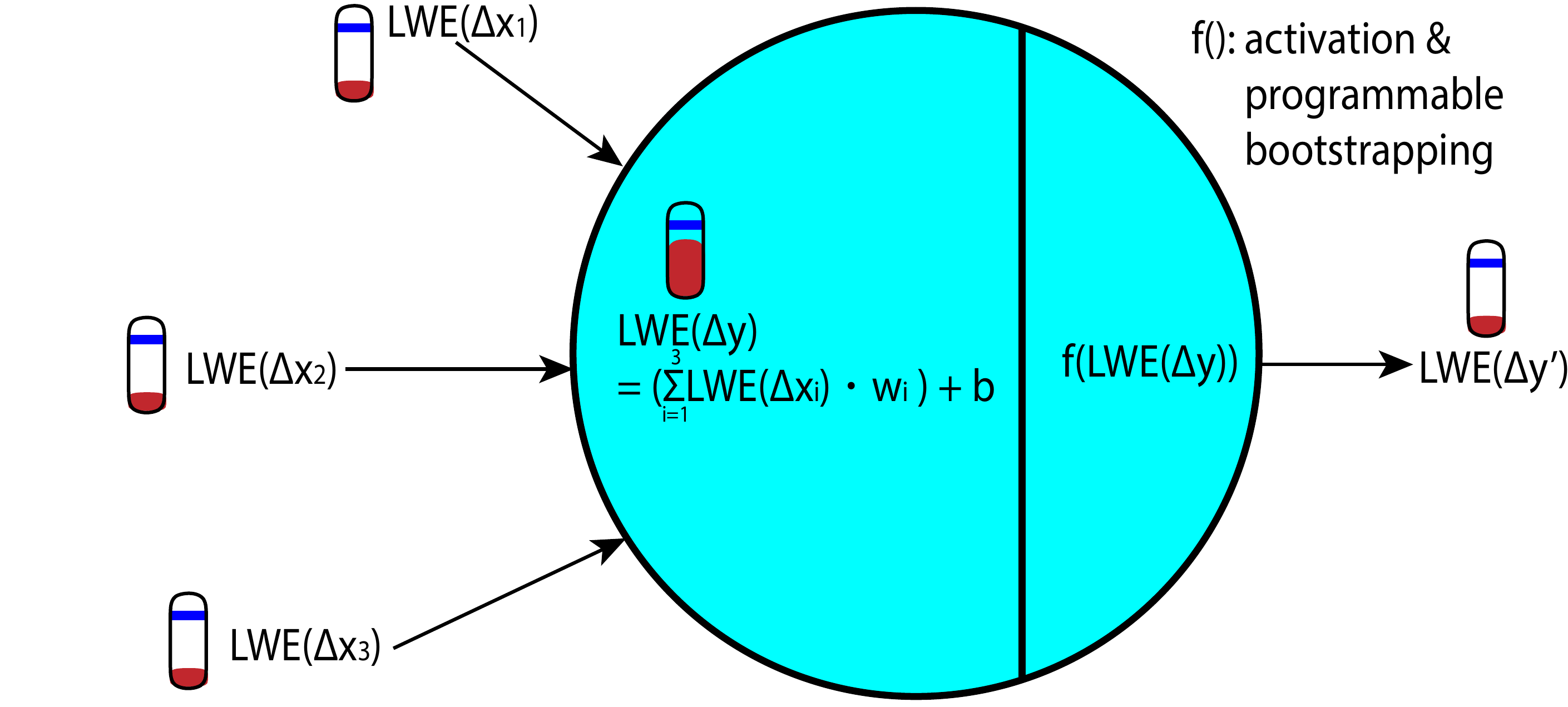}
  \caption{An illustration of neural networks's programmable bootstrapping}
  \label{fig:neural-network2}
\end{figure}

Homomorphic encryption can be applied to the neurons of deep neural networks, in which each neuron is generally comprised of two steps of computation: 

\begin{enumerate}
  \item \textbf{Linear Combination of Input Values:} An input feature value (or intermediate value) set $(x_1, x_2, \gap{$\cdots$}, x_n)$, a weight set $(w_1, w_2, \gap{$\cdots$}, w_n)$, and a bias $b$ are computed as: $y = \sum\limits_{i=1}^{n}x_iw_i + b$.
\item \textbf{Activation Function:} $f(y)$ is computed, where $f$ is a non-linear activation function such as the $\sin$ function, ReLU, sigmoid, hyperbolic tangent, etc. 
\end{enumerate}

TFHE can homomorphically compute the 1st step's linear combination formula: $y = \sum\limits_{i=1}^{n}x_iw_i + b$ as $\sum\limits_{i=1}^{n}\textsf{LWE}_{\vec{s}, \sigma}(x_i) \cdot w_i + b$, which can be implemented as ciphertext addition (\autoref{sec:glwe-add-cipher}) and ciphertext-to-plaintext multiplication (\autoref{sec:glwe-mult-plain}). 

However, the 2nd step's non-linear functions cannot be expressed as addition and multiplication of ciphertexts. To address this issue, the activation function can be evaluated as a programmable bootstrapping, such that the output of the bootstrapping matches or (or is similar) to the output of the activation function. If we use bootstrapping at the 2nd step, noises can be refreshed at the end of every neuron, thus we can potentially handle neural networks of any depth without worrying about the noise growth.

\subsection{TFHE on a Discrete Torus}
\label{subsec:torus}

\textbf{- Reference:} 
\href{https://eprint.iacr.org/2021/1402.pdf}{Guide to
Fully Homomorphic Encryption
over the [Discretized] Torus}~\cite{torus}

$ $

Torus $\mathbb{T}$ is a continuous real number domain between 0 and 1 that wraps around, that is $[0, 1)$. 

A discrete torus $\mathbb{T}_t$ is a finite real number set: $\Big(0, \dfrac{1}{t}, \dfrac{2}{t}, \gap{$\cdots$}, \dfrac{t - 1}{t}\Big)$

In the previous subsections, we learned the TFHE scheme based on the following setup:

$ $

$m \in \mathbb{Z}_t$ 
$\vec{s} = \{0, 1\}^{k}$
$e \in \mathbb{Z}_q$

$\textsf{LWE}_{\vec{s}, \sigma}(\Delta m) = (a_0, a_1, \gap{$\cdots$}, a_k, b) \in \mathbb{Z}_t^{k + 1}$

$ $

However, the original TFHE scheme is designed based on a discrete torus:

$ $

$m \in \mathbb{T}_{t}, \text{ } \vec{s} \in \{0, 1\}^{k}, \text{ } e \in \mathbb{T}_{q}$

$ $

$\textsf{LWE}_{\vec{s}, \sigma}(m) = \textsf{ct} = (a_0, a_1, \gap{$\cdots$}, a_k, b) \in \mathbb{T}_{q}^{k+1}$

$b = \sum\limits_{i=0}^{k}(a_is_i) + m + e \in \mathbb{T}_q$

$\textsf{LWE}_{\vec{s}, \sigma}^{-1}(\textsf{ct}) =  \Big\lceil b - \sum\limits_{i=0}^{k-1}(a_is_i) \Big\rfloor_{\frac{1}{t}} = \Big\lceil m + e \Big\rfloor_{\frac{1}{t}} = m$, given $e < \dfrac{1}{2t}$  

\textcolor{red}{ $\rhd$ where $\lceil x \rfloor_{\frac{1}{t}}$ means rounding $x$ to the nearest multiple of $\dfrac{1}{t}$}

$ $

The original TFHE's difference is that all values (either polynomial coefficients or vector elements) are computed in a floating point modulo 1 (i.e., $[0, 1)$) instead of a big integer (i.e., $[0, q)$). This means the plaintext also has to be encoded as values within $[0, 1)$ instead of integers within $[0, q)$. Note that in the original TFHE scheme, there is no need for the scaling factor $\Delta$, because the continuous domain of torus $[0, 1)$ provides a floating-point precision up to $q$ discrete fractional values, and its decryption process can successfully blow away the noise $E$ as far as each coefficient (or vector element) $e_i$ in $E$ is smaller than $\dfrac{1}{2t}$. 

Both the torus-based and integer-ring-based TFHE schemes are built based on the same fundamental principles.

\newpage

\section{BFV Scheme}
\label{sec:bfv}
The BFV scheme is designed for homomorphic addition and multiplication of integers. BFV's encoding scheme does not require such approximation issues because BFV is designed to encode only integers. Therefore, BFV guarantees exact encryption and decryption. BFV is suitable for use cases where the encrypted and decrypted values should exactly match (e.g., voting, financial computation), whereas CKKS is suitable for the use cases that tolerate tiny errors (e.g., data analytics, machine learning).  

In BFV, each plaintext is encrypted as an RLWE ciphertext. Therefore, BFV's ciphertext-to-ciphertext addition, ciphertext-to-plaintext addition, and ciphertext-to-plaintext multiplication are implemented based on GLWE's homomorphic addition and multiplication (as we learned in $\autoref{part:generic-fhe}$), with $k = 1$ to make GLWE an RLWE.

\begin{tcolorbox}[
    title = \textbf{Required Background},    % box title
    colback = white,    % light background; tweak to taste
    colframe = black,  % frame colour
    boxrule = 0.8pt,     % line thickness
    left = 1mm, right = 1mm, top = 1mm, bottom = 1mm % inner padding
]

\begin{itemize}
\item \autoref{sec:modulo}: \nameref{sec:modulo}
\item \autoref{sec:group}: \nameref{sec:group}
\item \autoref{sec:field}: \nameref{sec:field}
\item \autoref{sec:order}: \nameref{sec:order}
\item \autoref{sec:polynomial-ring}: \nameref{sec:polynomial-ring}
\item \autoref{sec:decomp}: \nameref{sec:decomp}
\item \autoref{sec:roots}: \nameref{sec:roots}
\item \autoref{sec:cyclotomic}: \nameref{sec:cyclotomic}
\item \autoref{sec:cyclotomic-polynomial-integer-ring}: \nameref{sec:cyclotomic-polynomial-integer-ring}
\item \autoref{sec:matrix}: \nameref{sec:matrix}
\item \autoref{sec:euler}: \nameref{sec:euler}
\item \autoref{sec:modulus-rescaling}: \nameref{sec:modulus-rescaling}
\item \autoref{sec:chinese-remainder}: \nameref{sec:chinese-remainder}
\item \autoref{sec:polynomial-interpolation}: \nameref{sec:polynomial-interpolation}
\item \autoref{sec:ntt}: \nameref{sec:ntt}
\item \autoref{sec:lattice}: \nameref{sec:lattice}
\item \autoref{sec:rlwe}: \nameref{sec:rlwe}
\item \autoref{sec:glwe}: \nameref{sec:glwe}
\item \autoref{sec:glwe-add-cipher}: \nameref{sec:glwe-add-cipher}
\item \autoref{sec:glwe-add-plain}: \nameref{sec:glwe-add-plain}
\item \autoref{sec:glwe-mult-plain}: \nameref{sec:glwe-mult-plain}
\item \autoref{subsec:modulus-switch-rlwe}: \nameref{subsec:modulus-switch-rlwe}
\item \autoref{sec:glwe-key-switching}: \nameref{sec:glwe-key-switching}
\end{itemize}
\end{tcolorbox}

\clearpage

\subsection{Single Value Encoding}
\label{subsec:bfv-single-encoding}

BFV supports two encoding schemes: single value encoding and batch encoding. In this subsection, we will explain the single value encoding scheme.

\begin{tcolorbox}[title={\textbf{\tboxlabel{\ref*{subsec:bfv-single-encoding}} BFV Encoding}}]

\textbf{\underline{Input Integer}:} Decompose the input integer $m \in \mathbb{Z}_{\geq 0}$ (i.e., 0 or any positive integer) as follows:

$m = b_{n-1}\cdot 2^{n-1} + b_{n-2}\cdot 2^{n-2} + \cdots + b_1\cdot 2^1 + b_0\cdot 2^0$  \text { } , where each $b_i \in \{0, 1\}$

$ $

\textbf{\underline{Encoded Polynomial}:} $M(X) = b_0 + b_1X + b_2X^2 + \cdots + b_{n-1}X^{n-1} \in \mathcal{R}_{\langle n, t\rangle}$

$ $

\textbf{\underline{Decoding}:} $M(X=2) = m$

\end{tcolorbox}

Let's analyze whether the encoding scheme in Summary~\ref*{subsec:bfv-single-encoding} ensures correct decoding after addition and multiplication. This is equivalent to showing that:

$\text{Decode}(\sigma(m_1) + \sigma(m_2)) = m_1 + m_2$ \textcolor{red}{$\rhd$ where $\sigma$ is the encoding function}

$\text{Decode}(\sigma(m_1) \cdot \sigma(m_2)) = m_1 \cdot m_2$

Let $\sigma(m_1) = M_1(X)$ and $\sigma(m_2) = M_2(X)$. Then, 

$\text{Decode}(\sigma(m_1) + \sigma(m_2)) = \text{Decode}(M_1(X) + M_2(X))$

$= \text{Decode}(M_{1+2}(X))$ \textcolor{red}{$\rhd$ where $M_{1+2}(X) = M_1(X) + M_2(X)$}

$= M_{1+2}(2)$ \textcolor{red}{$\rhd$ since decoding is evaluating the polynomial at $X = 2$}

$= M_{1}(2) + M_{2}(2)$ \textcolor{red}{$\rhd$ since evaluating $M_{1+2}(X)$ at $X=2$ is computationally the same as splitting $M_{1+2}(X)$ into $M_{1}(X)$ and $M_{2}(X)$, evaluating $M_1(2)$ and $M_2(2)$ and summing them}

$ $

Similarly, the decoding preserves correctness over multiplication as well:

$\text{Decode}(\sigma(m_1) \cdot \sigma(m_2)) = \text{Decode}(M_1(X) \cdot M_2(X))$

$= \text{Decode}(M_{1\cdot2}(X))$ \textcolor{red}{$\rhd$ where $M_{1\cdot2}(X) = M_1(X) \cdot M_2(X)$}

$= M_{1\cdot2}(2)$ \textcolor{red}{$\rhd$ since decoding is evaluating the polynomial at $X = 2$}

$= M_{1}(2) \cdot M_{2}(2)$ \textcolor{red}{$\rhd$ since evaluating $M_{1\cdot2}(X)$ at $X=2$ is computationally the same as splitting $M_{1\cdot2}(X)$ into $M_{1}(X)$ and $M_{2}(X)$, evaluating $M_1(2)$ and $M_2(2)$ and multiplying them}

$ $

Therefore, the single value encoding scheme preserves the correctness of decoding after addition and multiplication. 

$ $

The encoding scheme in Summary~\ref*{subsec:bfv-single-encoding} can be validly used for fully homomorphic encryption only if the multiplication of the encoded polynomials do not exceed the polynomial ring's degree $n$, because once the degree gets reduced due to an overflow, the evaluated values of polynomials lose consistency. Also, the coefficients of polynomials should not wrap modulo $t$ after additions or multiplications. 
Due to these constraints, the single value encoding is not a good choice for fully homomorphic encryption. Also, the single value encoding is computationally inefficient, because each polynomial can encode only a single value even if it holds $n$ coefficients.

\subsection{Batch Encoding}
\label{subsec:bfv-batch-encoding}

While the single-value encoding scheme (\autoref{subsec:bfv-single-encoding}) encodes \& decodes each individual value one at a time, the batch encoding scheme does the same for a huge list of values simultaneously using a large dimensional vector. Therefore, batch encoding is more efficient than single-value encoding. Furthermore, batch-encoded values can be homomorphically added or multiplied simultaneously element-wise by vector-to-vector addition and Hadamard product. Therefore, the homomorphic operation of batch-encoded values can be processed more efficiently in a SIMD (single-instruction-multiple-data) manner than single-value encoded ones. 

BFV's encoding converts an $n$-slot integer input vector $\vec{v} = (v_0, v_1, v_2, \cdots v_{n-1})$ modulo $t$ into another $n$-slot vector $\vec{m} = (m_0, m_1, m_2, \cdots m_{n-1})$ modulo $t$, which are the coefficients of the encoded $(n-1)$-degree (or lesser-degree) polynomial $M(X) \in \mathbb{Z}_t[X] / (X^n + 1)$. 

$ $

\subsubsection{\textsf{Encoding\textsubscript{1}}}  
\label{subsubsec:bfv-encoding-1} 

In \autoref{subsec:poly-vector-transformation}, we learned that an $(n-1)$-degree (or lesser degree) polynomial can be isomorphically mapped to an $n$-slot vector based on the mapping $\sigma$ (we notate $\sigma_c$ in \autoref{subsec:poly-vector-transformation} as $\sigma$ for simplicity): 

$\sigma: M(X) \in \mathbb{Z}_t[X]/(X^n + 1) \longrightarrow  (M(\omega),M(\omega^3),M(\omega^5), \cdots, M(\omega^{2n-1})) \in \mathbb{Z}_t^n$

$ $

, which evaluates the polynomial $M(X)$ at $n$ distinct $(\mu=2n)$-th primitive roots of unity: $\omega, \omega^3, \omega^5, \cdots, \omega^{2n-1}$. Let $\vec{m}$ be a vector that contains $n$ coefficients of the polynomial $M(X)$. Then, we can express the mapping $\sigma$ as follows: 

$\vec{v} = W^T \cdot \vec{m}$

$ $

, where $W^T$ is as follows: 

$W^T = \begin{bmatrix}
1 & (\omega) & (\omega)^2 & \cdots & (\omega)^{n-1}\\
1 & (\omega^3) & (\omega^3)^2 & \cdots & (\omega^3)^{n-1}\\
1 & (\omega^5) & (\omega^5)^2 & \cdots & (\omega^5)^{n-1}\\
\vdots & \vdots & \vdots & \ddots & \vdots \\
1 & (\omega^{2n-1}) & (\omega^{2n-1})^2 & \cdots & (\omega^{2n-1})^{n-1}\\
\end{bmatrix}$  \textcolor{red}{ \text{ } \# $W^T$ is a transpose of $W$ described in \autoref{subsec:poly-vector-transformation}}

$ $

Note that the dot product between each row of $W^T$ and $\vec{m}$ computes the evaluation of $M(X)$ at each $X = \{w, w^3, w^5, \cdots, w^{2n-1}\}$. In the BFV encoding scheme, the \textsf{Encoding\textsubscript{1}} process encodes an $n$-slot input vector $\vec{v}$ $\in \mathbb{Z}_t$ into a plaintext polynomial $M(X) \in \mathbb{Z}_t[X] / (X^n + 1)$, and the \textsf{Decoding\textsubscript{2}} process decodes $M(X)$ back to $\vec{v}$. Since $W^T \cdot \vec{m}$ gives us $\vec{v}$ which is a decoding of $M(X)$, we call $W^T$ a decoding matrix. Meanwhile, the goal of \textsf{Encoding\textsubscript{1}} is to encode $\vec{v}$ into $M(X)$ so that we can do homomorphic computations based on $M(X)$. Given the relation $\vec{v} = W^T \cdot \vec{m}$, the encoding formula can be derived as follows:

$(W^T)^{-1} \cdot \vec{v} = (W^T)^{-1} W^T \cdot \vec{m}$

$\vec{m} = (W^T)^{-1} \cdot \vec{v}$

$ $

Therefore, we need to find out what $(W^T)^{-1}$ is, the inverse of $W^T$ as the encoding matrix. But we already learned from Theorem~\ref*{subsec:vandermonde-euler} (in \autoref{subsec:vandermonde-euler}) that $V^{-1} = \dfrac{V^T \cdot I_n^R}{n}$, where $V = W^T$ and $V^T = W$. In other words, $(W^T)^{-1} = \dfrac{W \cdot I_n^R}{n}$. Therefore, we can express the BFV encoding formula as: 

$\vec{m} = (W^T)^{-1} \cdot \vec{v} = \dfrac{W \cdot I_n^R \cdot \vec{v}}{n}$, \text{ } where

$ $

$W =  \begin{bmatrix}
1 & 1 & 1 & \cdots & 1\\
(\omega) & (\omega^3) & (\omega^5) & \cdots & (\omega^{2n-1})\\
(\omega)^2 & (\omega^3)^2 & (\omega^5)^2 & \cdots & (\omega^{2n-1})^2\\
\vdots & \vdots & \vdots & \ddots & \vdots \\
(\omega)^{n-1} & (\omega^3)^{n-1} & (\omega^5)^{n-1} & \cdots & (\omega^{2n-1})^{n-1}\\
\end{bmatrix}$

$= \begin{bmatrix}
1 & 1 & \cdots & 1 & 1 & \cdots & 1 & 1\\
(\omega) & (\omega^3) & \cdots & (\omega^{\frac{n}{2} - 1}) & (\omega^{-(\frac{n}{2} - 1)}) & \cdots & (\omega^{-3}) & (\omega^{-1})\\
(\omega)^2 & (\omega^3)^2 & \cdots & (\omega^{\frac{n}{2} - 1})^2 & (\omega^{-(\frac{n}{2} - 1)})^2 & \cdots & (\omega^{-3})^2 & (\omega^{-1})^2\\
\vdots & \vdots & \vdots & \ddots & \vdots \\
(\omega)^{n-1} & (\omega^3)^{n-1} & \cdots & (\omega^{\frac{n}{2} - 1})^{n-1} & (\omega^{-(\frac{n}{2} - 1)})^{n-1} & \cdots & (\omega^{-3})^{n-1} & (\omega^{-1})^{n-1}\\
\end{bmatrix}$

$ $

, where $\omega$ is a primitive $2n$-th root of unity modulo $t$ (which implies $t \equiv 1 \bmod 2n$). This implies that $\omega = g^{\frac{t - 1}{2n}} \bmod t$ ($g$ is a generator of $\mathbb{Z}_t^{\times}$ (see \autoref{subsubsec:poly-vector-transformation-modulus}).

In \autoref{subsec:poly-vector-transformation}, we learned that $W$ is a valid basis of the $n$-dimensional vector space. Therefore, $\dfrac{W \cdot \vec{v}}{n} = \vec{m}$ is guaranteed to be a unique vector corresponding to each $\vec{v}$ in the $n$-dimensional vector space $\mathbb{Z}_t^{n}$ (refer to Theorem~\ref*{subsec:projection} in \autoref{subsec:projection}), and thereby the polynomial $M(X)$ comprising the $n$ elements of $\vec{m}$ as coefficients is a unique polynomial bi-jective to $\vec{v}$. 

Note that by computing $\dfrac{W \cdot I_n^R \cdot \vec{v}}{n}$
, we transform the input slot vector $\vec{v}$ into another vector $\vec{m}$ in the same vector space $\mathbb{Z}_t^{n}$, while preserving isomorphism between these two vectors (i.e., bi-jective one-to-one mappings and homomorphism on the $(+, \cdot)$ operations). 

\subsubsection{\textsf{Encoding\textsubscript{2}}}  

Once we have the $n$-slot vector $\vec{m}$, we scale (i.e., multiply) it by some scaling factor $\Delta = \left\lfloor \dfrac{q}{t} \right\rfloor$, where $q$ is the ciphertext modulus. We scale $\vec{m}$ by $\Delta$ and make it $\Delta \vec{m}$. 
The $n$ integers in $\Delta\vec{m}$ will be used as $n$ coefficients of the plaintext polynomial for RLWE encryption. The finally encoded plaintext polynomial $\Delta M = \sum\limits_{i=0}^{n-1}\Delta m_iX^i$.

\subsubsection{\textsf{Decoding\textsubscript{1}}}  
\label{subsubsec:bfv-enc-dec-decoding1} 

Once an RLWE ciphertext is (first-half) decrypted to $\Delta M = \sum\limits_{i=0}^{n-1}\Delta m_iX^i$, we compute $\dfrac{\Delta\vec{m}}{\Delta} = \vec{m}$.

\subsubsection{\textsf{Decoding\textsubscript{2}}}

In \autoref{subsubsec:bfv-encoding-1}, we already derived the decoding formula that transforms an $(n-1)$-degree polynomial having integer modulo $t$ coefficients into an $n$-slot input vector as follows: 

$\vec{v} = W^T \cdot \vec{m}$

\subsubsection{Summary}
\label{subsubsec:bfv-encoding-summary}

\begin{tcolorbox}[title={\textbf{\tboxlabel{\ref*{subsubsec:bfv-encoding-summary}} BFV's Encoding and Decoding}}]

\textbf{\underline{Input}:} An $n$-slot integer modulo $t$ vector $\vec{v} = (v_0, v_1, \cdots, v_{n-1}) \in \mathbb{Z}_t^n$

\par\noindent\rule{\textwidth}{0.4pt}

\textbf{\underline{Encoding}} 
\begin{enumerate}

\item Convert $\vec{v} \in \mathbb{Z}_t^n$ into $\vec{m} \in \mathbb{Z}_t^n$ by applying the transformation $\vec{m} = n^{-1}\cdot W \cdot I_n^R \cdot \vec{v}$

, where $W$ is a basis of the $n$-dimensional vector space crafted as follows: 

$W =  \begin{bmatrix}
1 & 1 & 1 & \cdots & 1\\
(\omega) & (\omega^3) & (\omega^5) & \cdots & (\omega^{2n-1})\\
(\omega)^2 & (\omega^3)^2 & (\omega^5)^2 & \cdots & (\omega^{2n-1})^2\\
\vdots & \vdots & \vdots & \ddots & \vdots \\
(\omega)^{n-1} & (\omega^3)^{n-1} & (\omega^5)^{n-1} & \cdots & (\omega^{2n-1})^{n-1}\\
\end{bmatrix}$

$= \begin{bmatrix}
1 & 1 & \cdots & 1 & 1 & \cdots & 1 & 1\\
(\omega) & (\omega^3) & \cdots & (\omega^{\frac{n}{2} - 1}) & (\omega^{-(\frac{n}{2} - 1)}) & \cdots & (\omega^{-3}) & (\omega^{-1})\\
(\omega)^2 & (\omega^3)^2 & \cdots & (\omega^{\frac{n}{2} - 1})^2 & (\omega^{-(\frac{n}{2} - 1)})^2 & \cdots & (\omega^{-3})^2 & (\omega^{-1})^2\\
\vdots & \vdots & \vdots & \ddots & \vdots \\
(\omega)^{n-1} & (\omega^3)^{n-1} & \cdots & (\omega^{\frac{n}{2} - 1})^{n-1} & (\omega^{-(\frac{n}{2} - 1)})^{n-1} & \cdots & (\omega^{-3})^{n-1} & (\omega^{-1})^{n-1}\\
\end{bmatrix}$

$ $

, where $\omega$ is a primitive $2n$-th root of unity modulo $t$. This implies that $\omega = g^{\frac{t - 1}{2n}} \bmod t$ ($g$ is a generator of $\mathbb{Z}_t^{\times}$ (see \autoref{subsubsec:poly-vector-transformation-modulus}).

$ $

\item Convert $\vec{m}$ into a scaled integer vector $\Delta\vec{m}$, where $1 \leq \Delta \leq \lfloor \dfrac{q}{t}\rfloor$ is a scaling factor. If $\Delta$ is too close to 1 (i.e., $t$ is too big), the \textbf{noise budget} will become too small (making decryption fail easily). If $\Delta$ is too close to $q$ (i.e., $t$ is too small), the \textbf{message capacity} will become too small (i.e., the plaintext modulus $t$ limits the range of values that can be encoded). The finally encoded plaintext polynomial $\Delta M = \sum\limits_{i=0}^{n-1} \Delta m_i X^i \text{ } \in \mathbb{Z}_q[X] / (X^n + 1)$. %The rounding of $\lceil \Delta \vec{m} \rfloor$ during the encoding process makes CKKS an approximate FHE scheme. 

\end{enumerate}

\par\noindent\rule{\textwidth}{0.4pt}

\textbf{\underline{Decoding}:} From the plaintext polynomial $\Delta M = \sum\limits_{i=0}^{n-1}\Delta m_iX^i$, recover $\vec{m} = \dfrac{\Delta \vec{m}}{\Delta}$. Then,
compute $\vec{v} = W^T \cdot \vec{m}$.

\end{tcolorbox}

%\para{BFV's Encoding Range:} \autoref{tab:polynomial-encoding-capacity-integer} depicts the range of input values that can be encoded by the polynomials $M(X) \in \mathcal{R}_{\langle n, t \rangle}$ over $X \in \mathbb{Z}_t$. As illustrated in \autoref{tab:polynomial-encoding-capacity-integer}, the encoding range of integer inputs is determined by the size of the $n$ and $t$ parameters. Therefore, these two parameters should be chosen sufficiently large enough to encode all values that can be used by an application. 

However, Summary~\ref*{subsubsec:bfv-encoding-summary} is not the final version of BFV's batch encoding. In \autoref{subsec:bfv-rotation}, we will explain how to homomorphically rotate the input vector slots without decrypting the ciphertext that encapsulates it. To support such homomorphic rotation, we will need to slightly update the encoding scheme explained in Summary~\ref*{subsubsec:bfv-encoding-summary}. We will explain how to do this in \autoref{subsec:bfv-rotation}, and BFV's final encoding scheme is summarized in Summary~\ref*{subsubsec:bfv-rotation-summary} in \autoref{subsubsec:bfv-rotation-summary}.

\subsection{Encryption and Decryption}
\label{subsec:bfv-enc-dec}

BFV encrypts and decrypts ciphertexts based on the RLWE cryptosystem (\autoref{sec:rlwe}) with the sign of each $A\cdot S$ term flipped in the encryption and decryption formula. Specifically, this is equivalent to the alternative version of the GLWE cryptosystem (\autoref{subsec:glwe-alternative}) with $k = 1$. Thus, BFV's encryption and decryption formulas are as follows:

\begin{tcolorbox}[title={\textbf{\tboxlabel{\ref*{subsec:bfv-enc-dec}} BFV Encryption and Decryption}}]

\textbf{\underline{Initial Setup}:} $\Delta=\left\lfloor\dfrac{q}{t}\right\rfloor \text{ is a plaintext scaling factor for polynomial encoding}, \text{ } S \xleftarrow{\$} \mathcal{R}_{\langle n, \textit{tern} \rangle}$

, where plaintext modulus $t$ is either a prime ($p$) or a power of prime ($p^r$), and ciphertext modulus $q \gg t$. As for the coefficients of polynomial $S$, they are ternary (i.e., $\{-1, 0, 1\}$).

\par\noindent\rule{\textwidth}{0.4pt}

\textbf{\underline{Encryption Input}:} $\Delta M \in \mathcal{R}_{\langle n, q \rangle}$, $A_i \xleftarrow{\$} \mathcal{R}_{\langle n, q \rangle}$, $E \xleftarrow{\chi_\sigma} \mathcal{R}_{\langle n, q \rangle}$

$ $

\begin{enumerate}

\item Compute $B = -A \cdot S + \Delta M + E \text{ } \in \mathcal{R}_{\langle n,q \rangle}$

\item $\textsf{RLWE}_{S,\sigma}(\Delta M + E) = (A, B) \text{ } \in \mathcal{R}_{\langle n,q \rangle}^2$ 

\end{enumerate}

\par\noindent\rule{\textwidth}{0.4pt}

\textbf{\underline{Decryption Input}:} $\textsf{ct} = (A, B) \text{ } \in \mathcal{R}_{\langle n,q \rangle}^2$ 

$\textsf{RLWE}^{-1}_{S,\sigma}(\textsf{ct}) = \left\lceil\dfrac{B + A \cdot S \bmod q}{\Delta}\right\rfloor \bmod t = \left\lceil\dfrac{\Delta M + E}{\Delta}\right\rfloor \bmod t = M \bmod t$

(The noise $E = \sum\limits_{i=0}^{n-1}e_iX^i$ gets eliminated by the rounding process)

$ $

\textbf{\underline{Conditions for Correct Decryption}:}

As explained in Summary~\ref*{subsubsec:lwe-noise-bound} (in \autoref{subsubsec:lwe-noise-bound}), the noise bound is $|-\epsilon k_i t + e_i| < \dfrac{\Delta}{2}$, where $k_i$ is each coefficient of the polynomial $K$ that accounts for the $t$-multiple overflows of the coefficients of the plaintext polynomial updated across homomorphic operations. 

\end{tcolorbox}

In this section, we will often write $\textsf{RLWE}_{S,\sigma}(\Delta  M + E)$ as $\textsf{RLWE}_{S,\sigma}(\Delta  M)$ for simplicity, because $\textsf{RLWE}_{S,\sigma}(\Delta M + E) \approx \textsf{RLWE}_{S,\sigma}(\Delta M)$ (i.e., they decrypt to the same message). Even in the case that we write $\textsf{RLWE}_{S,\sigma}(\Delta  M)$ instead of $\textsf{RLWE}_{S,\sigma}(\Delta  M + E)$, you should assume this as an encryption of $\Delta  M + E$ (i.e., the noise is included inside the scaled message).

We will explain the conditions for BFV's correct decryption in more detail in \autoref{subsubsec:bfv-noise-analysis}.

\subsection{Ciphertext-to-Ciphertext Addition}
\label{subsec:bfv-add-cipher}

BFV's ciphertext-to-ciphertext addition uses RLWE's ciphertext-to-ciphertext addition scheme with the sign of the $A\cdot S$ term flipped in the encryption and decryption formula. Specifically, this is equivalent to the alternative GLWE version's (\autoref{subsec:glwe-alternative}) ciphertext-to-ciphertext addition scheme with $k = 1$.

\begin{tcolorbox}[title={\textbf{\tboxlabel{\ref*{subsec:bfv-add-cipher}} BFV Ciphertext-to-Ciphertext Addition}}]
$\textsf{RLWE}_{S, \sigma}(\Delta M^{\langle 1 \rangle} + E^{\langle 1 \rangle} ) + \textsf{RLWE}_{S, \sigma}(\Delta M^{\langle 2 \rangle} + E^{\langle 2 \rangle}) $

$ = ( A^{\langle 1 \rangle}, \text{ } B^{\langle 1 \rangle}) + (A^{\langle 2 \rangle}, \text{ } B^{\langle 2 \rangle}) $

$ = ( A^{\langle 1 \rangle} + A^{\langle 2 \rangle}, \text{ } B^{\langle 1 \rangle} + B^{\langle 2 \rangle} ) $

$= \textsf{RLWE}_{S, \sigma}(\Delta(M^{\langle 1 \rangle} + M^{\langle 2 \rangle}) + E^{\langle 1 \rangle} + E^{\langle 2 \rangle})$
\end{tcolorbox}

\subsubsection{Noise Bound Analysis}
\label{subsubsec:bfv-noise-analysis}

In the last part of Summary~\ref*{subsec:bfv-enc-dec} (in \autoref{subsec:bfv-enc-dec}), we learned the noise bound conditions for BFV's correct decryption. In this subsection, we will explain how this condition holds in more detail by walking through BFV's ciphertext-to-ciphertext addition. 

Let's denote the homomorphically added ciphertext as follows: 

$ ( A^{\langle 3 \rangle}, \text{ } B^{\langle 3 \rangle}) = ( A^{\langle 1 \rangle} + A^{\langle 2 \rangle}, \text{ } B^{\langle 1 \rangle} + B^{\langle 2 \rangle} ) \bmod q$

$ $

Applying the first step of decryption to it yields the following intermediate result:

$B^{\langle 3 \rangle} + A^{\langle 3 \rangle}\cdot S \bmod q$

$B^{\langle 1 \rangle} + B^{\langle 2 \rangle} + A^{\langle 1 \rangle}\cdot S + A^{\langle 2 \rangle}\cdot S \bmod q$

$= (-A^{\langle 1 \rangle}\cdot S + \Delta M^{\langle 1\rangle} + E^{\langle 1 \rangle}) + (- A^{\langle 2 \rangle}\cdot S  + \Delta M^{\langle 2\rangle}  + E^{\langle 2 \rangle}) + A^{\langle 1 \rangle}\cdot S + A^{\langle 2 \rangle}\cdot S \bmod q$

$= \Delta M^{\langle 1\rangle} + E^{\langle 1 \rangle} + \Delta M^{\langle 2\rangle} + E^{\langle 2 \rangle} \bmod q $

$ $

The second step of decryption is to divide each coefficient of the above intermediate polynomial by $\Delta$, round it, and reduce it modulo $t$ as follows:

$ $

$\left\lceil\dfrac{\Delta (M^{\langle 1\rangle} + M^{\langle 2\rangle}) + E^{\langle 1 \rangle} + E^{\langle 2 \rangle} \bmod q}{\Delta}\right\rfloor \bmod t$

$ $

Correct decryption requires the above result to match the value $M^{\langle 1 + 2 \rangle} = M^{\langle 1 \rangle} + M^{\langle 2 \rangle} \bmod t$, where $M^{\langle 1 + 2 \rangle}$ is the modulo $t$-reduced final polynomial. Let's define $\epsilon = \dfrac{q}{t} - \left\lfloor\dfrac{q}{t}\right\rfloor = \dfrac{q}{t} - \Delta$. Given $q \gg t$, $\epsilon$ is a fractional value between $[0, 1)$. Now, we can re-write the above decryption term as follows:

$ $

$\left\lceil\dfrac{\Delta (M^{\langle 1\rangle} + M^{\langle 2\rangle}) + E^{\langle 1 \rangle} + E^{\langle 2 \rangle} \bmod q}{\Delta}\right\rfloor \bmod t$ 

$ $

$\left\lceil\dfrac{(\frac{q}{t} - \epsilon)\cdot (M^{\langle 1\rangle} + M^{\langle 2\rangle}) + E^{\langle 1 \rangle} + E^{\langle 2 \rangle} \bmod q}{\Delta}\right\rfloor \bmod t$ \textcolor{red}{ $\rhd$ applying $\Delta = \left\lfloor\dfrac{q}{t}\right\rfloor = \dfrac{q}{t} - \epsilon$}

$ $

$ = \left\lceil\dfrac{(\frac{q}{t} - \epsilon)\cdot (M^{\langle 1 + 2\rangle} + t\cdot K) + E^{\langle 1 \rangle} + E^{\langle 2 \rangle} \bmod q}{\Delta}\right\rfloor \bmod t$ 

\textcolor{red}{ $\rhd$ where $t \cdot K$ represents the $t$-multiple overflows generated by the modulo addition of $M^{\langle 1\rangle} + M^{\langle 2\rangle}$ }

$ $

$ $

$ = \left\lceil\dfrac{\frac{q}{t} \cdot M^{\langle 1 + 2\rangle}  - \epsilon \cdot M^{\langle 1 + 2\rangle}  + \frac{q}{t} \cdot t\cdot K  - \epsilon \cdot t\cdot K + E^{\langle 1 \rangle} + E^{\langle 2 \rangle} \bmod q}{\Delta}\right\rfloor \bmod t$ 

$ $

$ = \left\lceil\dfrac{\frac{q}{t} \cdot M^{\langle 1 + 2\rangle}  - \epsilon \cdot M^{\langle 1 + 2\rangle}  - \epsilon \cdot t\cdot K + E^{\langle 1 \rangle} + E^{\langle 2 \rangle} \bmod q}{\Delta}\right\rfloor \bmod t$ 

\textcolor{red}{ $\rhd$ since $\frac{q}{t}\cdot t = q$, and $q\cdot K \bmod q = 0$}

$ $

$ $

$ = \left\lceil\dfrac{\left\lfloor\frac{q}{t}\right\rfloor \cdot M^{\langle 1 + 2\rangle} + \epsilon \cdot M^{\langle 1 + 2\rangle} - \epsilon \cdot M^{\langle 1 + 2\rangle}  - \epsilon \cdot t\cdot K + E^{\langle 1 \rangle} + E^{\langle 2 \rangle} \bmod q}{\Delta}\right\rfloor \bmod t$  

\textcolor{red}{ $\rhd$ applying  $\dfrac{q}{t} = \left\lfloor\dfrac{q}{t}\right\rfloor + \epsilon$}

$ $

$ $

$ = \left\lceil\dfrac{\left\lfloor\frac{q}{t}\right\rfloor \cdot M^{\langle 1 + 2\rangle} - \epsilon \cdot t\cdot K + E^{\langle 1 \rangle} + E^{\langle 2 \rangle} \bmod q}{\Delta}\right\rfloor \bmod t$  

$ $

$ $

$ = \left\lceil\dfrac{\left\lfloor\frac{q}{t}\right\rfloor \cdot M^{\langle 1 + 2\rangle} - \epsilon \cdot t\cdot K + E^{\langle 1 \rangle} + E^{\langle 2 \rangle}}{\Delta}\right\rfloor \bmod t$  

\textcolor{red}{ $\rhd$ applying the special assumption $|\epsilon \cdot t \cdot K + E^{\langle 1 \rangle} + E^{\langle 2 \rangle}| < \dfrac{\Delta}{2}$ to all $n$ coefficients (see \autoref{subsubsec:lwe-noise-bound})}

$ $

$ $

$ = M^{\langle 1 + 2\rangle} + \left\lceil\dfrac{E^{\langle 1 \rangle} + E^{\langle 2 \rangle} - \epsilon \cdot t\cdot K}{\Delta}\right\rfloor \bmod t$  \textcolor{red}{ $\rhd$ since  $\Delta = \left\lfloor\dfrac{q}{t}\right\rfloor$, and $\lceil M^{\langle 1 + 2\rangle} \rfloor = M^{\langle 1 + 2\rangle}$}

$ $

$ $

$ = M^{\langle 1 + 2\rangle} \bmod t$
\textcolor{red}{ $\rhd$ applying the special assumption $\epsilon \cdot t \cdot K + E^{\langle 1 \rangle} + E^{\langle 2 \rangle} < \dfrac{\Delta}{2}$ to all $n$ coefficients}

$ $

The above final expression implies that correct decryption (i.e., $M^{\langle 1 + 2\rangle}$) is preserved if 
the special assumption $\epsilon \cdot t \cdot K + E^{\langle 1 \rangle} + E^{\langle 2 \rangle} < \dfrac{\Delta}{2}$ holds (for all $n$ coefficients of the polynomial). At a high level, the greater the ciphertext modulus $q$ becomes compared to the plaintext modulus $t$, the greater the scaling factor $\Delta$ becomes, which can sustain a greater noise budget ($E^{\langle 1 \rangle} + E^{\langle 2 \rangle}$) and greater wrapping around $t$-multiple overflows of the plaintext ($\epsilon \cdot t\cdot K$). 

This noise bound principle not only applies to homomorphic addition but also to homomorphic multiplication and rotation, which will be explained in later subsections. The term $E^{\langle 1 \rangle} + E^{\langle 2 \rangle}$ can be generalized as the cumulative noise across all homomorphic operations (e.g., additions, multiplications, rotations), and the term $\epsilon \cdot t\cdot K$ can be generalized as the amount of $t$-multiple overflows of each coefficient of the plaintext polynomial computed across homomorphic operations.

\subsection{Ciphertext-to-Plaintext Addition}
\label{subsec:bfv-add-plain}

BFV's ciphertext-to-plaintext addition uses RLWE's ciphertext-to-plaintext addition scheme with the sign of the $A\cdot S$ term flipped in the encryption and decryption formulas. Specifically, this is equivalent to the alternative GLWE version's (\autoref{subsec:glwe-alternative}) ciphertext-to-plaintext addition scheme (\autoref{sec:glwe-add-plain}) with $k = 1$.

\begin{tcolorbox}[title={\textbf{\tboxlabel{\ref*{subsec:bfv-add-plain}} BFV Ciphertext-to-Plaintext Addition}}]
$\textsf{RLWE}_{S, \sigma}(\Delta M + E) + \Delta\Lambda $

$=  (A, \text{ } B) + \Delta\Lambda$

$=  (A, \text{ } B + \Delta\cdot\Lambda)$

$= \textsf{RLWE}_{S, \sigma}(\Delta (M + \Lambda) + E)$
\end{tcolorbox}

\subsection{Ciphertext-to-Plaintext Multiplication}
\label{subsec:bfv-mult-plain}

BFV's ciphertext-to-plaintext multiplication uses RLWE's ciphertext-to-plaintext multiplication scheme with the sign of the $A\cdot S$ term flipped in the encryption and decryption formula. Specifically, this is equivalent to the alternative GLWE version's (\autoref{subsec:glwe-alternative}) ciphertext-to-plaintext multiplication scheme (\autoref{sec:glwe-mult-plain}) with $k = 1$.

\begin{tcolorbox}[title={\textbf{\tboxlabel{\ref*{subsec:bfv-mult-plain}} BFV Ciphertext-to-Plaintext Multiplication}}]
$\textsf{RLWE}_{S, \sigma}(\Delta M + E) \cdot \Lambda$

$= (A, \text{ } B) \cdot \Lambda$

$= (A \cdot \Lambda, \text{ }  B \cdot \Lambda )$

$= \textsf{RLWE}_{S, \sigma}(\Delta (M \cdot \Lambda) + \Lambda E)$
\end{tcolorbox}

\subsection{Ciphertext-to-Ciphertext Multiplication}
\label{subsec:bfv-mult-cipher}

\noindent \textbf{- Reference 1:} 
\href{https://www.inferati.com/blog/fhe-schemes-bfv}{Introduction to the BFV encryption scheme}~\cite{inferati-bfv}

\noindent \textbf{- Reference 2:} 
\href{https://eprint.iacr.org/2012/144.pdf}{Somewhat Partially Fully Homomorphic Encryption}~\cite{cryptoeprint:2012/144}

%\noindent \textbf{- Reference 3:} 
%\href{https://eprint.iacr.org/2016/510.pdf}{A Full RNS Variant $of FV like Somewhat Homomorphic Encryption Schemes}~\cite{cryptoeprint:2016/510}

Given two ciphertexts $\textsf{RLWE}_{S, \sigma}(\Delta M^{\langle 1\rangle })$ and $\textsf{RLWE}_{S, \sigma}(\Delta M^{\langle 2\rangle})$, the goal of ciphertext-to-ciphertext multiplication is to derive a new ciphertext whose decryption is $\Delta M^{\langle 1 \rangle} M^{\langle 2 \rangle}$. Ciphertext-to-ciphertext multiplication is more complex than ciphertext-to-plaintext multiplication. It comprises four steps: (1) \textsf{ModRaise}; (2) polynomial multiplication; (3) relinearization; and (4) rescaling. 

For better understanding, we will explain BFV's ciphertext-to-ciphertext multiplication based on the alternate version of RLWE (Theorem~\ref*{subsec:glwe-alternative} in \autoref{subsec:glwe-alternative}), where the sign of the $AS$ term is flipped in the encryption and decryption formulas.

\subsubsection{\textsf{ModRaise}}
\label{subsubsec:bfv-mult-cipher-modraise}

We learned from Summary~\ref*{subsec:bfv-enc-dec} (in \autoref{subsec:bfv-enc-dec}) that a BFV ciphertext whose ciphertext modulus is $q$ has the (decryption) relation: $\Delta M + E = A\cdot S + B - K \cdot q$, where $K \cdot q$ stands for modulo reduction by $q$. \textsf{ModRaise} is a process of forcibly raising the modulus of a ciphertext from $q \rightarrow Q$, where $q \ll Q$. Suppose we modify the modulus of ciphertext $(A, B)$ from $q$ to $Q$, where $Q = q \cdot \Delta$ (remember $\Delta = \left\lfloor\dfrac{q}{t}\right\rfloor$). Then, the decryption of the \textit{mod-raised} ciphertext will output $A\cdot S + B \bmod Q$. However, since each polynomial coefficient of $A$ and $B$ is less than $q$ and each polynomial coefficient of $S$ is either $\{-1, 0, 1\}$, the resulting polynomial of $A\cdot S + B$ is guaranteed to have each coefficient strictly less than $Q$ even without modulo reduction by $Q$-- this is because $(q-1)\cdot n + (q-1) < Q$, where $(q-1)\cdot n$ is the maximum possible coefficient of $A \cdot S$ and $(q-1)$ is the maximum possible coefficient of $B$. And as mentioned before, we know the relation: $A\cdot S + B = \Delta M + E + Kq$. Therefore, the decryption of the \textit{mod-raised} ciphertext $(A, B) \bmod Q$ is as follows:

$\Delta M + E + Kq \bmod Q = \Delta M + E + Kq$ \textcolor{red}{ $\rhd$  since $\Delta M + E + Kq < Q$} 

$ $

The first step of BFV's ciphertext-to-ciphertext multiplication is to \textit{mod-raise} the two input ciphertexts $(A^{\langle 1 \rangle}, B^{\langle 1 \rangle}) \bmod q$ and $(A^{\langle 2 \rangle}, B^{\langle 2 \rangle}) \bmod q$  
from $q \rightarrow Q$ (where $Q = q\cdot \Delta$) as follows: 

$(A^{\langle 1 \rangle}, B^{\langle 1 \rangle}) \bmod Q$

$(A^{\langle 2 \rangle}, B^{\langle 2 \rangle}) \bmod Q$

$ $

After \textsf{ModRaise}, the decryption of these two ciphertexts would be the following: 

$A^{\langle 1 \rangle} \cdot S + B^{\langle 1 \rangle} = \Delta M^{\langle 1 \rangle} + E^{\langle 1 \rangle} + K_1q < Q$

$A^{\langle 2 \rangle} \cdot S + B^{\langle 2 \rangle} = \Delta M^{\langle 2 \rangle} + E^{\langle 2 \rangle} + K_2q < Q$

$ $

Therefore, the \textit{mod-raised} ciphertexts have the following form: 

$\textsf{RLWE}_{S, \sigma}(\Delta M^{\langle 1 \rangle} + K_1q) = (A^{\langle 1 \rangle}, B^{\langle 1 \rangle}) \bmod Q$

$\textsf{RLWE}_{S, \sigma}(\Delta M^{\langle 2 \rangle} + K_2q) = (A^{\langle 2 \rangle}, B^{\langle 2 \rangle}) \bmod Q$

\subsubsection{Polynomial Multiplication}
\label{subsubsec:bfv-mult-cipher-multiplication}

Our next goal is to derive a new ciphertext which encrypts 
$(\Delta M^{\langle 1 \rangle} + E^{\langle 1 \rangle} + K_1q) \cdot (\Delta M^{\langle 2 \rangle} + E^{\langle 2 \rangle} + K_2q)$. 

First, we can derive the following relation: 

$(\Delta M^{\langle 1 \rangle} + E^{\langle 1 \rangle} + K_1q) \cdot (\Delta M^{\langle 2 \rangle} + E^{\langle 2 \rangle} + K_2q)$

$ = (A^{\langle 1 \rangle} \cdot S + B^{\langle 1 \rangle}) \cdot (A^{\langle 2 \rangle} \cdot S + B^{\langle 2 \rangle})$

$ = \underbrace{B^{\langle 1 \rangle}B^{\langle 2 \rangle}}_{D_0}  + \underbrace{(B^{\langle 2 \rangle}A^{\langle 1 \rangle} + B^{\langle 1 \rangle}A^{\langle 2 \rangle})}_{D_1} \cdot S + \underbrace{(A^{\langle 1 \rangle} \cdot A^{\langle 2 \rangle})}_{D_2} \cdot S \cdot S $

$= D_0 + D_1\cdot S + D_2\cdot S^2$

$ $

Meanwhile, we also have the following relations:

$\textsf{RLWE}_{S, \sigma}^{-1}(\Delta M^{\langle 1 \rangle} + K_1q) = \Delta M^{\langle 1 \rangle} + E^{\langle 1 \rangle} + K_1q$

$\textsf{RLWE}_{S, \sigma}^{-1}(\Delta M^{\langle 2 \rangle} + K_2q) = \Delta M^{\langle 2 \rangle} + E^{\langle 2 \rangle} + K_2q$

$ $

Combining all these, we reach the following relation: 

$\textsf{RLWE}_{S, \sigma}^{-1}(\Delta M^{\langle 1 \rangle} + K_1q) \cdot \textsf{RLWE}_{S, \sigma}^{-1}(\Delta M^{\langle 2 \rangle} + K_2q) = D_0 + D_1\cdot S + D_2\cdot S^2$

$ $

Notice that $D_0, D_1,$ and $D_2$ are known values as ciphertext components, whereas $S$ is only known to the private key owner. Therefore, we can view $D_0 + D_1\cdot S + D_2\cdot S^2$ as a decryption formula such that given the ciphertext components $D_0, D_1, D_2$ and the private key $S$, one can derive $(\Delta M^{\langle 1 \rangle} + E^{\langle 1 \rangle} + K_1q) \cdot (\Delta M^{\langle 2 \rangle} + E^{\langle 2 \rangle} + K_2q)$. In other words, we can let $(D_0, D_1, D_2)$ be a new form of ciphertext which can be decrypted by $S$ into $(\Delta M^{\langle 1 \rangle} + E^{\langle 1 \rangle} + K_1q) \cdot (\Delta M^{\langle 2 \rangle} + E^{\langle 2 \rangle} + K_2q)$. 

However, $(D_0, D_1, D_2)$ is not in the RLWE ciphertext format, because it has 3 components instead of 2. Having 3 ciphertext components is computationally inefficient, as its decryption involves a square root of $S$ (i.e., $ S^2$). Over consequent ciphertext-to-ciphertext multiplications, this $S$ term will double its exponents as $S^4, S^8, \cdots$ as well as the number of ciphertext components, which would exponentially increase the computational overhead of decryption. Therefore, we want to convert the intermediate ciphertext format $(D_0, D_1, D_2)$ into a regular BFV ciphertext format that has two polynomials as ciphertext components. This conversion process is called a relinearization process (which will be explained in the next subsection).

\subsubsection{Relinearization}
\label{subsubsec:bfv-mult-cipher-relinearization}

Relinearization is a process of converting the polynomial triplet $(D_0, D_1, D_2) \in \mathcal{R}_{\langle n, Q \rangle}^{3}$ into two RLWE ciphertexts $\textsf{ct}_\alpha$ and $\textsf{ct}_\beta$ which hold the relation: $D_0 + D_1 S + D_2 S^2 = \textsf{RLWE}^{-1}_{S, \sigma}(\textsf{ct}_\alpha + \textsf{ct}_\beta)$. 

In the formula $D_0 + D_1 S + D_2 S^2$, we can re-write $D_0 + D_1 S$ as a \textit{synthetic} RLWE ciphertext $\textsf{ct}_\alpha = (D_1, D_0)$, which can be decrypted by $S$ into $D_1 S + D_0$. Similarly, our next task is to derive a synthetic RLWE ciphertext $\textsf{ct}_\beta$ whose decryption is $D_2 \cdot S^2$ (i.e., $\textsf{RLWE}_{S, \sigma}^{-1}(\textsf{ct}_\beta) = D_2\cdot S^2$). 

A naive way of creating a ciphertext that encrypts $D_2 \cdot S^2$ is as follows: we encrypt $S^2$ into an RLWE ciphertext as $\textsf{RLWE}_{S, \sigma}(S^2) = (A^{\langle s \rangle}, B^{\langle s \rangle})$ such that $A^{\langle s \rangle}\cdot S + B^{\langle s \rangle} = S^2 + E^{\langle s \rangle} \bmod Q$ (where the ciphertext modulus is $Q$ and the plaintext scaling factor $\Delta = 1$). Then, we perform a ciphertext-to-plaintext multiplication (\autoref{sec:glwe-mult-plain}) with $D_2$, treating $D_2$ as a plaintext polynomial in modulo $Q$. However, this approach does not work in practice, because computing $D_2 \cdot \textsf{RLWE}_{S, \sigma}(S^2)$ generates a huge noise as follows:

$D_2 \cdot (A^{\langle s \rangle}, B^{\langle s \rangle}) = (D_2\cdot A^{\langle s \rangle}, D_2 \cdot B^{\langle s \rangle})$

$ $

, whose decryption is:

$D_2\cdot A^{\langle s \rangle} \cdot S + D_2\cdot B^{\langle s \rangle} = D_2 \cdot S^2 + D_2\cdot E^{\langle s \rangle} \pmod Q$ 

$ $

. In the above decrypted expression $D_2S^2 + D_2E^{\langle s \rangle} \bmod Q$, the term $D_2S^2$ is okay to be reduced modulo $Q$, because this term is originally allowed to be reduced modulo $Q$ in the final decryption formula $D_0 + D_1S + D_2S^2 \bmod Q$ as well. However, the problematic term is the noise $D_2\cdot E^{\langle s \rangle}$, because its coefficients can be any value in $[0, Q - 1]$ (since each coefficient of polynomial $D_2 = A^{\langle 1 \rangle} A^{\langle 2 \rangle}$ can be any value in $[0, Q - 1]$). Such a huge noise is not allowed for correct final decryption. 

To avoid this noise issue, an improved solution is to express the RLWE ciphertext that encrypts $D_2 S^2$ as additions of multiple RLWE ciphertexts with small noises by using the gadget decomposition technique (\autoref{subsec:gadget-decomposition}). For this, we use an RLev ciphertext (\autoref{sec:glev}) that encrypts $S^2$. Suppose our gadget vector is $\vec{g} = \Bigg(\dfrac{Q}{\beta}, \dfrac{Q}{\beta^2}, \dfrac{Q}{\beta^3}, \cdots, \dfrac{Q}{\beta^l}\Bigg)$. Remember that our goal is to find $\textsf{ct}_\beta = \textsf{RLWE}_{S, \sigma}( S^2 \cdot D_2)$ given known $D_2$, unknown $S$, and known $\textsf{RLev}_{S, \sigma}^{\beta, l}(S^2) = \left\{\textsf{RLWE}_{S, \sigma}\left(\dfrac{Q}{\beta^i}\cdot S\right)\right\}_{i=1}^{l}$. Then, we can derive $\textsf{ct}_\beta$ as follows:

$\textsf{ct}_\beta = \textsf{RLWE}_{S, \sigma}(S^2 \cdot D_2)$

$= \textsf{RLWE}_{S, \sigma}\left( S^2 \cdot \left(D_{2,1}\dfrac{Q}{\beta} + D_{2,2}\dfrac{Q}{\beta^2} + \cdots D_{2,l}\dfrac{Q}{\beta^l}\right)\right)$ \textcolor{red}{  $\rhd$ by decomposing $D_2$}

$= \textsf{RLWE}_{S, \sigma}\left( S^2 \cdot D_{2,1} \cdot \dfrac{Q}{\beta}\right) + \textsf{RLWE}_{S, \sigma}\left( S^2 \cdot D_{2,2} \cdot \dfrac{Q}{\beta^2}\right) + \cdots + \textsf{RLWE}_{S, \sigma}\left( S^2 \cdot D_{2,l} \cdot \dfrac{Q}{\beta^l}\right)$ 

$= D_{2,1}\cdot\textsf{RLWE}_{S, \sigma}\left( S^2 \cdot \dfrac{Q}{\beta}\right) + D_{2,2}\cdot\textsf{RLWE}_{S, \sigma}\left( S^2 \cdot \dfrac{Q}{\beta^2}\right) + \cdots + D_{2,l}\cdot\textsf{RLWE}_{S, \sigma}\left( S^2 \cdot \dfrac{Q}{\beta^l}\right)$ 
\textcolor{red}{ $\rhd$ where each \textsf{RLWE} ciphertext is an encryption of $S^2\dfrac{Q}{\beta}, S^2\dfrac{Q}{\beta^2}, \cdots, S^2\dfrac{Q}{\beta^l}$ as plaintext with the plaintext scaling factor $\Delta = 1$}

$ $

$= \bm{\langle} \textsf{Decomp}^{\beta, l}(D_2), \text{ } \textsf{RLev}_{S, \sigma}^{\beta, l}( S^2) \bm{\rangle}$ \textcolor{red}{  $\rhd$ inner product of \textsf{Decomp} and \textsf{RLev} treating them as vectors}

$ $

If we decrypt the above, we get the following:

$\textsf{RLWE}_{S, \sigma}^{-1}(\textsf{ct}_\beta = \bm{\langle} \textsf{Decomp}^{\beta, l}(D_2), \text{ } \textsf{RLev}_{S, \sigma}^{\beta, l}( S^2) \bm{\rangle} \bm{)}$ \textcolor{red}{ $\rhd$ the scaling factors of $\textsf{RLev}_{S, \sigma}^{\beta, l}( S^2)$ are all 1}

$= D_{2,1}\cdot\left(E_1' +  S^2\dfrac{Q}{\beta}\right) + D_{2,2}\cdot\left(E_2' +  S^2\dfrac{Q}{\beta^2}\right) + \cdots + D_{2,l}\cdot\left(E_l' +  S^2\dfrac{Q}{\beta^l}\right)$ \textcolor{red}{ \text{ } \# where each $E_i'$ is a noise embedded in $\textsf{RLWE}_{S, \sigma}\left(S^2\cdot\dfrac{Q}{\beta^i}\right)$}

$ $

$= \sum\limits_{i=1}^{l} (E_i'\cdot D_{2,i}) +  S^2\cdot\left(D_{2,1}\dfrac{Q}{\beta} + D_{2,2}\dfrac{Q}{\beta^2} + \cdots + D_{2, l}\dfrac{Q}{\beta^l}\right)$

$= \sum\limits_{i=1}^{l} \epsilon_{i} + D_2\cdot S^2$ \textcolor{red}{ \text{ } \# where each $\epsilon_i = E_i'\cdot D_{2,i}$}

$\approx D_2\cdot S^2$ \textcolor{red}{ \text{ } \# $\sum\limits_{i=1}^{l} \epsilon_{i} \ll D_2\cdot E''$, where $E''$ is the noise that could've been embedded in $\textsf{RLWE}_{S, \sigma}\bm(S^2\bm)$}

$ $

Therefore, we get the following comprehensive relation:

$\textsf{RLWE}_{S, \sigma}^{-1}(\Delta M^{\langle 1 \rangle} + K_1q) \cdot \textsf{RLWE}_{S, \sigma}^{-1}(\Delta M^{\langle 2 \rangle} + K_2q) \bmod Q$

$= (\Delta M^{\langle 1 \rangle} + E^{\langle 1 \rangle} + K_1q) \cdot (\Delta M^{\langle 2 \rangle} + E^{\langle 2 \rangle} + K_2q) \bmod Q$

$ = (A^{\langle 1 \rangle} \cdot S + B^{\langle 1 \rangle}) \cdot (A^{\langle 2 \rangle} \cdot S + B^{\langle 2 \rangle})  \bmod Q$

$ = D_0 + D_1\cdot S + D_2\cdot S^2  \bmod Q$ \textcolor{red}{ $\rhd$ $D_0 = B^{\langle 1 \rangle} B^{\langle 2 \rangle}$, \text{ } $D_1 = A^{\langle 1 \rangle} B^{\langle 2 \rangle} + A^{\langle 2 \rangle} B^{\langle 1 \rangle}$, \text{ } $D_2 = A^{\langle 1 \rangle} A^{\langle 2 \rangle}$}

$ = \textsf{RLWE}_{S, \sigma}^{-1}(\textsf{ct}_\alpha) + \textsf{RLWE}_{S, \sigma}^{-1}(\textsf{ct}_\beta) - \sum\limits_{i=1}^{l} (E_i'\cdot D_{2,i})  \bmod Q$ 

\textcolor{red}{ $\rhd$ $\textsf{ct}_\alpha = (D_1, D_0) = (A_\alpha, B_\beta)$, \text{ } $\textsf{ct}_\beta = \langle \textsf{Decomp}^{\beta, l}(D_2), \textsf{RLev}_{S, \sigma}^{\beta, l}(S^2) \rangle = (A_\beta, B_\beta)$}

$ $

$ = \textsf{RLWE}_{S, \sigma}^{-1}(\textsf{ct}_\alpha + \textsf{ct}_\beta) - \sum\limits_{i=1}^{l} (E_i'\cdot D_{2,i})  \bmod Q$

$ = \textsf{RLWE}_{S, \sigma}^{-1}\bm((A_{\alpha+\beta}, B_{\alpha+\beta})\bm) - \sum\limits_{i=1}^{l} (E_i'\cdot D_{2,i}) \bmod Q$  \textcolor{red}{ $\rhd$ $A_{\alpha+\beta} = A_{\alpha} + A_\beta$, \text{ } $B_{\alpha + \beta} = B_\alpha + B_\beta$}

$ $

From the above, we extract the following relation: 

$(\Delta M^{\langle 1 \rangle} + E^{\langle 1 \rangle} + K_1q) \cdot (\Delta M^{\langle 2 \rangle} + E^{\langle 2 \rangle} + K_2q) \bmod Q$

$=  \Delta^2M^{\langle 1 \rangle}M^{\langle 2 \rangle} + \Delta\cdot (M^{\langle 1 \rangle}E^{\langle 2 \rangle} + M^{\langle 2 \rangle}E^{\langle 1 \rangle}) + q\cdot(\Delta M^{\langle 1 \rangle}K_2 + \Delta M^{\langle 2 \rangle}K_1 + E^{\langle 1 \rangle}K_2 + E^{\langle 2 \rangle}K_1) + K_1 K_2 q^2 + E^{\langle 1 \rangle} E^{\langle 2 \rangle} \bmod Q$

$= \textsf{RLWE}_{S, \sigma}^{-1}\bm((A_{\alpha+\beta}, B_{\alpha+\beta})\bm) - \sum\limits_{i=1}^{l} (E_i'\cdot D_{2,i}) \bmod Q$  

$ $

We can re-write the above relation as follows:

$\textsf{RLWE}_{S, \sigma}^{-1}\bm((A_{\alpha+\beta}, B_{\alpha+\beta})\bm) = A_{\alpha+\beta}\cdot S + B_{\alpha+\beta} \bmod Q$

$ = \Delta^2M^{\langle 1 \rangle}M^{\langle 2 \rangle} + \Delta\cdot (M^{\langle 1 \rangle}E^{\langle 2 \rangle} + M^{\langle 2 \rangle}E^{\langle 1 \rangle}) + q\cdot(\Delta M^{\langle 1 \rangle}K_2 + \Delta M^{\langle 2 \rangle}K_1 + E^{\langle 1 \rangle}K_2 + E^{\langle 2 \rangle}K_1) + K_1 K_2 q^2 + E^{\langle 1 \rangle} E^{\langle 2 \rangle} + \sum\limits_{i=1}^{l} (E_i'\cdot D_{2,i}) \bmod Q$ 

$ $

To verbally interpret the above relation, decrypting the synthetically generated ciphertext $(A_{\alpha+\beta}, B_{\alpha+\beta})$ and applying a reduction modulo $Q$ to it gives us $\Delta^2 M^{\langle 1 \rangle}M^{\langle 2 \rangle}$ with a lot of noise terms. Meanwhile, as explained in the beginning of this subsection, our goal is to derive a ciphertext whose decryption is $\Delta M^{\langle 1 \rangle}M^{\langle 2 \rangle}$, also ensuring that the decrypted ciphertext's noise is small enough to be fully eliminated by scaling down the plaintext by $\Delta$ at the end. This goal is accomplished by the final rescaling step to be explained in the next subsection.

\subsubsection{Rescaling}
\label{subsubsec:bfv-mult-cipher-rescaling}

The rescaling step is equivalent to converting the ciphertext $(A_{\alpha+\beta}, B_{\alpha+\beta}) \bmod Q$ into $\left(\left\lceil\dfrac{A_{\alpha+\beta}}{\Delta}\right\rfloor, \left\lceil\dfrac{B_{\alpha+\beta}}{\Delta}\right\rfloor\right) \bmod q$, where $\Delta = \left\lfloor\dfrac{q}{t}\right\rfloor \approx \dfrac{q}{t}$. The decryption of this rescaled ciphertext (and finally scaling down by $\Delta$) is $\Delta M^{\langle 1 \rangle}M^{\langle 2 \rangle}$. This is demonstrated below: 

$ $

$ \left\lceil\dfrac{A_{\alpha+\beta}}{\Delta}\right\rfloor \cdot S + \left\lceil\dfrac{B_{\alpha+\beta}}{\Delta}\right\rfloor \bmod q$ \textcolor{red}{ $\rhd$ decryption of ciphertext $\left(\left\lceil\dfrac{A_{\alpha+\beta}}{\Delta}\right\rfloor, \left\lceil\dfrac{B_{\alpha+\beta}}{\Delta}\right\rfloor\right) \bmod q$}

$ $

$ = \left\lceil\dfrac{A_{\alpha+\beta}}{\Delta}\right\rfloor \cdot S + \left\lceil\dfrac{B_{\alpha+\beta}}{\Delta}\right\rfloor + K_3q$ \textcolor{red}{ $\rhd$ where $K_3q$ stands for modulo reduction by $q$}

$ $

$ = \left\lceil\dfrac{A_{\alpha+\beta}}{\Delta}\right\rfloor \cdot S + \left\lceil\dfrac{B_{\alpha+\beta}}{\Delta}\right\rfloor + \dfrac{K_3Q}{\Delta}$ \textcolor{red}{ $\rhd$ since $Q = \Delta \cdot q$}

$ $

$ = \left\lceil\dfrac{1}{\Delta}\cdot (A_{\alpha+\beta}\cdot S + B_{\alpha+\beta} + K_3Q)\right\rfloor + E_r$ \textcolor{red}{ $\rhd$ $E_r$ is a rounding error}

$ $

$ = \left\lceil\dfrac{1}{\Delta}\cdot (A_{\alpha+\beta}\cdot S + B_{\alpha+\beta} \bmod Q)\right\rfloor + E_r$ 

$ $

$ = \Bigg\lceil\dfrac{1}{\Delta}\cdot ( \Delta^2M^{\langle 1 \rangle}M^{\langle 2 \rangle} + \Delta\cdot (M^{\langle 1 \rangle}E^{\langle 2 \rangle} + M^{\langle 2 \rangle}E^{\langle 1 \rangle}) + $

\text{ } \text{ } $ q\cdot(\Delta M^{\langle 1 \rangle}K_2 + \Delta M^{\langle 2 \rangle}K_1 + E^{\langle 1 \rangle}K_2 + E^{\langle 2 \rangle}K_1) + K_1 K_2 q^2 + E^{\langle 1 \rangle} E^{\langle 2 \rangle} + \sum\limits_{i=1}^{l} (E_i'\cdot D_{2,i}) \bmod Q
)\Bigg\rfloor + E_r$

\textcolor{red}{ $\rhd$ as we derived at the end of \autoref{subsubsec:bfv-mult-cipher-relinearization}}

$ $

$ = \Bigg\lceil\dfrac{1}{\Delta}\cdot ( \Delta^2M^{\langle 1 \rangle}M^{\langle 2 \rangle} + \Delta\cdot (M^{\langle 1 \rangle}E^{\langle 2 \rangle} + M^{\langle 2 \rangle}E^{\langle 1 \rangle}) + $

\text{ } \text{ } $ q\cdot(\Delta M^{\langle 1 \rangle}K_2 + \Delta M^{\langle 2 \rangle}K_1 + E^{\langle 1 \rangle}K_2 + E^{\langle 2 \rangle}K_1) + K_1 K_2 q^2 + E^{\langle 1 \rangle} E^{\langle 2 \rangle} + \sum\limits_{i=1}^{l} (E_i'\cdot D_{2,i}) + K_4Q
)\Bigg\rfloor + E_r$

\textcolor{red}{ $\rhd$ where $K_4Q$ stands for modulo reduction by $Q$}

$ $

$ $

$ = \Bigg\lceil\Delta M^{\langle 1 \rangle}M^{\langle 2 \rangle} + (M^{\langle 1 \rangle}E^{\langle 2 \rangle} + M^{\langle 2 \rangle}E^{\langle 1 \rangle}) + q\cdot (M^{\langle 1 \rangle}K_2 +  M^{\langle 2 \rangle}K_1) $

\text{ } \text{ } $+ \dfrac{1}{\Delta}\cdot q\cdot(E^{\langle 1 \rangle}K_2 + E^{\langle 2 \rangle}K_1) + \dfrac{1}{\Delta}\cdot (K_1K_2q^2 
 + E^{\langle 1 \rangle} E^{\langle 2 \rangle} + \sum\limits_{i=1}^{l} (E_i'\cdot D_{2,i})) + K_5q\Bigg\rfloor $ 

\text{ } \text{ }  \textcolor{red}{ $\rhd$ where $K_5q = K_4q + M^{\langle 1 \rangle}q K_2 + M^{\langle 2 \rangle} K_1q$}

$ $

$ $

$ = \Bigg\lceil\Delta M^{\langle 1 \rangle}M^{\langle 2 \rangle} + (M^{\langle 1 \rangle}E^{\langle 2 \rangle} + M^{\langle 2 \rangle}E^{\langle 1 \rangle}) + q\cdot (M^{\langle 1 \rangle}K_2 +  M^{\langle 2 \rangle}K_1) + (t + \epsilon)\cdot(E^{\langle 1 \rangle}K_2 + E^{\langle 2 \rangle}K_1) $

\text{ } \text{ } $+  \dfrac{1}{\Delta}\cdot (K_1K_2q^2 
 + E^{\langle 1 \rangle} E^{\langle 2 \rangle} + \sum\limits_{i=1}^{l} (E_i'\cdot D_{2,i})) + K_5q\Bigg\rfloor $ 

\text{ } \text{ }  \textcolor{red}{ $\rhd$ where $\epsilon = \dfrac{q}{\Delta} - \dfrac{q}{\dfrac{q}{t}} = \dfrac{q}{\left\lfloor\dfrac{q}{t}\right\rfloor} - \dfrac{q}{\dfrac{q}{t}} \approx 0$, thus we substituted $\dfrac{q}{\Delta} = \epsilon + t$}

 $ $

 $ $

$ = \Bigg\lceil\Delta M^{\langle 1 \rangle}M^{\langle 2 \rangle} + (M^{\langle 1 \rangle}E^{\langle 2 \rangle} + M^{\langle 2 \rangle}E^{\langle 1 \rangle}) + q\cdot (M^{\langle 1 \rangle}K_2 +  M^{\langle 2 \rangle}K_1) + (t + \epsilon)\cdot(E^{\langle 1 \rangle}K_2 + E^{\langle 2 \rangle}K_1)$

\text{ } \text{ } $ + \dfrac{K_1K_2q^2}{\Delta}   
 + \dfrac{E^{\langle 1 \rangle} E^{\langle 2 \rangle} + \sum\limits_{i=1}^{l} (E_i'\cdot D_{2,i})}{\Delta} + K_5q\Bigg\rfloor $

\text{ } \text{ }  \textcolor{red}{ $\rhd$ Now, let $\epsilon' = \dfrac{K_1K_2q^2}{\Delta} - \dfrac{K_1K_2q^2}{\dfrac{q}{t}} = \dfrac{K_1K_2q^2}{\left\lfloor\dfrac{q}{t}\right\rfloor} - \dfrac{K_1K_2q^2}{\dfrac{q}{t}} \approx 0$.}

\text{ } \text{ }  \textcolor{red}{ Thus, we will substitute $\dfrac{K_1K_2q^2}{\Delta} = \dfrac{K_1K_2q^2}{\dfrac{q}{t}} + \epsilon' = K_1K_2qt + \epsilon'$}
 
 $ $

 $ $

$ = \Delta M^{\langle 1 \rangle}M^{\langle 2 \rangle} + (M^{\langle 1 \rangle}E^{\langle 2 \rangle} + M^{\langle 2 \rangle}E^{\langle 1 \rangle}) + q\cdot (M^{\langle 1 \rangle}K_2 +  M^{\langle 2 \rangle}K_1) $

\text{ } \text{ } $+ (t + \epsilon)\cdot(E^{\langle 1 \rangle}K_2 + E^{\langle 2 \rangle}K_1) + K_1K_2qt + \epsilon'   
 + K_5q + \Bigg\lceil\dfrac{E^{\langle 1 \rangle} E^{\langle 2\rangle} + \sum\limits_{i=1}^{l} (E_i'\cdot D_{2,i}) }{\Delta}\Bigg\rfloor $

 $ $

 $ $

$ = \Delta M^{\langle 1 \rangle}M^{\langle 2 \rangle} + \epsilon'' + K_6q$

\textcolor{red}{ $\rhd$ where $K_6q = K_5q + q\cdot(M^{\langle 1 \rangle}K_2 +  M^{\langle 2 \rangle}K_1) + K_1K_2qt, $}

\textcolor{red}{$ \epsilon'' = M^{\langle 1 \rangle}E^{\langle 2 \rangle} + M^{\langle 2 \rangle}E^{\langle 1 \rangle} + (t + \epsilon)\cdot(E^{\langle 1 \rangle}K_2 + E^{\langle 2 \rangle}K_1) + \Bigg\lceil\dfrac{E^{\langle 1 \rangle} E^{\langle 2 \rangle} + \sum\limits_{i=1}^{l} (E_i'\cdot D_{2,i})}{\Delta}\Bigg\rfloor$}

$ $

$ $

$ = \Delta M^{\langle 1 \rangle}M^{\langle 2 \rangle} + \epsilon'' \bmod q$

$ $

In conclusion, the ciphertext $\left(\left\lceil\dfrac{A_{\alpha+\beta}}{\Delta}\right\rfloor, \left\lceil\dfrac{B_{\alpha+\beta}}{\Delta}\right\rfloor\right) \bmod q$ successfully decrypts to $\Delta M^{\langle 1 \rangle}M^{\langle 2 \rangle}$ if $\epsilon'' < \dfrac{\Delta}{2} \approx \dfrac{q}{2t}$. 

$ $

\para{Noise Analysis:} Among the terms of $\epsilon''$, let's analyze the noise growth of the $(t + \epsilon)\cdot(E^{\langle 1 \rangle}K_2 + E^{\langle 2 \rangle}K_1)$ term after ciphertext-to-ciphertext multiplication. Each coefficient of $K_1$ is at most $n$, because $A^{\langle 1 \rangle}\cdot S + B^{\langle 1 \rangle} = \Delta M + E^{\langle 1 \rangle} + K_1q$, where the maximum possible coefficient value of $A^{\langle 1 \rangle}\cdot S + B^{\langle 1 \rangle}$ is $q\cdot n$. And the same is true for the coefficients of $K_2$. Therefore, after scaling down $(t + \epsilon)\cdot(E^{\langle 1 \rangle}K_2 + E^{\langle 2 \rangle}K_1)$ by $\Delta$ upon the final decryption stage, this term's down-scaled noise gets bound by: 

$\dfrac{1}{\Delta}\cdot(t + \epsilon)\cdot(E^{\langle 1 \rangle}K_2 + E^{\langle 2 \rangle}K_1) \approx \dfrac{t}{q}\cdot(t + \epsilon)\cdot(E^{\langle 1 \rangle}K_2 + E^{\langle 2 \rangle}K_1) < \dfrac{nt \cdot (t + \epsilon)}{q} \cdot (E^{\langle 1 \rangle} + E^{\langle 2 \rangle})$

$ $

This implies that for correct decryption, $\dfrac{nt \cdot (t + \epsilon)}{q} \cdot (E^{\langle 1 \rangle} + E^{\langle 2 \rangle})$ has to be smaller than $0.5$. In other words, $E^{\langle 1 \rangle} + E^{\langle 2 \rangle}$ has to be smaller than $\dfrac{q}{2nt \cdot (t + \epsilon)}$. We can do noise analysis for all other terms for $\epsilon''$ in a similar manner. Importantly, upon decryption, the aggregation of all these noise terms' down-scaled values has to be smaller than $0.5$ for correct decryption. 

$ $

\para{Modulus Switch v.s. Rescaling: } 
Notice in the rescaling process, multiplying $\dfrac{1}{\Delta}$ to $A_{\alpha + \beta}$ and $B_{\alpha + \beta}$ results in two effects: (1) converts $\Delta^2 M^{\langle 1 \rangle} M^{\langle 2 \rangle}$ into $\Delta M^{\langle 1 \rangle}  M^{\langle 2 \rangle}$; (2) switches the modulus of the \textit{mod-raised} ciphertexts from $Q \rightarrow q$. 
In fact, modulus switch and rescaling are closely equivalent to each other. Modulus switch is a process of changing a ciphertext's modulus (e.g., $q \rightarrow q'$), while preserving the property that the decryption of both ciphertexts results in the same plaintext. On the other hand, rescaling refers to the process of changing the scaling factor of a plaintext within a ciphertext (e.g., $\Delta \rightarrow \Delta'$). Modulus switch inevitably changes the scaling factor of the plaintext within the target ciphertext, and rescaling also inevitably changes the modulus of the ciphertext that contains the plaintext (as shown in \autoref{subsubsec:bfv-mult-cipher-rescaling}). Therefore, these two terms can be used interchangeably.

\subsubsection{Summary}
\label{subsubsec:bfv-mult-cipher-summary}

To put all things together, BFV's ciphertext-to-ciphertext multiplication is summarized as follows:

\begin{tcolorbox}[title={\textbf{\tboxlabel{\ref*{subsubsec:bfv-mult-cipher-summary}} BFV's Ciphertext-to-Ciphertext Multiplication}}]

Suppose we have the following two RLWE ciphertexts:

$\textsf{RLWE}_{S, \sigma}(\Delta M^{\langle 1 \rangle}) = (A^{\langle 1 \rangle}, B^{\langle 1 \rangle}) \bmod q$, \text{ } where $B^{\langle 1 \rangle} = -A^{\langle 1 \rangle} \cdot S + \Delta M^{\langle 1 \rangle} + E^{\langle 1 \rangle} \bmod q$

$\textsf{RLWE}_{S, \sigma}(\Delta M^{\langle 2 \rangle}) = (A^{\langle 2 \rangle}, B^{\langle 2 \rangle})  \bmod q$, \text{ } where $B^{\langle 2 \rangle} = -A^{\langle 2 \rangle} \cdot S + \Delta M^{\langle 2 \rangle} + E^{\langle 2 \rangle} \bmod q$

$ $

Multiplication between these two ciphertexts is performed as follows:

$ $

\begin{enumerate}

\item \textbf{\underline{\textsf{ModRaise}}}

Forcibly raise the modulus of the ciphertexts $(A^{\langle 1 \rangle}, B^{\langle 1 \rangle}) \bmod q$ and $(A^{\langle 2 \rangle}, B^{\langle 2 \rangle}) \bmod q$ to $Q$ (where $Q = q \cdot \Delta$) as follows:

$(A^{\langle 1 \rangle}, B^{\langle 1 \rangle}) \bmod Q$

$(A^{\langle 2 \rangle}, B^{\langle 2 \rangle}) \bmod Q$

$ $

\item \textbf{\underline{Multiplication}}

Compute the following polynomial multiplications in modulo $Q$: 

$D_0 = B^{\langle 1 \rangle}B^{\langle 2 \rangle} \bmod Q$

$D_1 = B^{\langle 2 \rangle}A^{\langle 1 \rangle} +  B^{\langle 1 \rangle}A^{\langle 2 \rangle} \bmod Q$

$D_2 = A^{\langle 1 \rangle} \cdot A^{\langle 2 \rangle} \bmod Q$

$ $

\item \textbf{\underline{Relinearization}}

Compute the following:

$ \textsf{ct}_\alpha = (D
_1, D_0)$

$ \textsf{ct}_\beta = \bm{\langle} \textsf{Decomp}^{\beta, l}(D_2), \text{ } \textsf{RLev}_{S, \sigma}^{\beta, l}( S^2) \bm{\rangle}$.

$ \textsf{ct}_{\alpha + \beta} = \textsf{ct}_{\alpha} + \textsf{ct}_{\beta}$

$ $

Then, the following property holds: 

$\textsf{RLWE}^{-1}_{S, \sigma}\bm(\textsf{RLWE}_{S, \sigma}(\Delta^2 \cdot M^{\langle 1 \rangle} \cdot M^{\langle 2 \rangle})\bm) \approx \textsf{RLWE}^{-1}_{S, \sigma}\bm{(}\textsf{ct}_{\alpha + \beta}\bm)$

$ $

\item \textbf{\underline{Rescaling}}

Update $\textsf{ct}_{\alpha + \beta} = (A_{\alpha + \beta}, B_{\alpha + \beta}) \bmod Q$ to $\textsf{ct'}_{\alpha + \beta} = \left(\left\lceil\dfrac{A_{\alpha + \beta}}{\Delta}\right\rfloor, \left\lceil\dfrac{B_{\alpha + \beta}}{\Delta}\right\rfloor\right) \bmod q$. 

$ $

This plaintext rescaling process can be also viewed as a modulus switch of the ciphertext $\textsf{ct}_{\alpha + \beta}$ from $Q \rightarrow q$.

$ $

\end{enumerate}

Note that after the ciphertext-to-ciphertext multiplication, the plaintext scaling factor $\Delta=\left\lfloor\dfrac{q}{t}\right\rfloor$, the ciphertext modulus $q$, and the private key $S$ stay the same as before.

\end{tcolorbox}

\para{The Purpose of \textsf{ModRaise}: } When we multiply polynomials at the second step of ciphertext-to-ciphertext multiplication (\autoref{subsubsec:bfv-mult-cipher-multiplication}), the underlying plaintext within the ciphertext temporarily grows to $\Delta^2 M^{\langle 1 \rangle} M^{\langle 2 \rangle}$, which exceeds the allowed maximum boundary $q$ for the plaintext (Summary~\ref*{subsec:bfv-enc-dec} in \autoref{subsec:bfv-enc-dec}). After this point, applying modulo-$q$ reduction to the intermediate result will irrevocably corrupt the plaintext. To avoid the corruption of the plaintext when it grows to $\Delta^2 M^{\langle 1 \rangle} M^{\langle 2 \rangle}$, we temporarily increase the ciphertext modulus from $q \rightarrow Q$, which is sufficiently large to hold $\Delta^2 M^{\langle 1 \rangle} M^{\langle 2 \rangle}$ without wrapping around the boundary of the ciphertext modulus. 

$ $

\para{Swapping the Order of \textsf{Relinearization} and \textsf{Rescaling}: } The order of relinearization and rescaling is interchangeable. Running rescaling before relinearization reduces the size of the ciphertext modulus, and therefore the subsequent relinearization can be executed faster.

\subsubsection{Application to TFHE's Ciphertext-to-Ciphertext Multiplication}
\label{subsubsec:bfv-mult-cipher-tfhe}

In \autoref{subsec:tfhe-mult-cipher}, we learned how TFHE performs ciphertext-to-ciphertext multiplication between LWE and GSW ciphertexts. However, this method is computationally expensive because it requires circuit bootstrapping to convert an LWE ciphertext into a GSW ciphertext. Such an overhead can be avoided if we directly apply BFV's ciphertext-to-ciphertext multiplication strategy to TFHE's LWE ciphertexts. Given two LWE ciphertexts to multiply, they hold the following relations:

$\textsf{LWE}_{\vec{s}, \sigma}(\Delta m^{\langle 1 \rangle} + e^{\langle 1 \rangle}) = (\vec{a}^{\langle 1 \rangle}, b^{\langle 1 \rangle }) \Rightarrow \,\,\, \vec{a}^{\langle 1 \rangle} \cdot \vec{s} + b^{\langle 1 \rangle} = \Delta m^{\langle 1 \rangle} + e^{\langle 1 \rangle} + k_1q$

$\textsf{LWE}_{\vec{s}, \sigma}(\Delta m^{\langle 2 \rangle} + e^{\langle 2 \rangle})  = (\vec{a}^{\langle 2 \rangle}, b^{\langle 2 \rangle}) \Rightarrow \,\,\, \vec{a}^{\langle 2 \rangle} \cdot \vec{s} + b^{\langle 2 \rangle} = \Delta m^{\langle 2 \rangle} + e^{\langle 2 \rangle} + k_2q$

$ $

Multiplying these two equations yields the following relation:

$(\Delta m^{\langle 1 \rangle} + e^{\langle 1 \rangle} + k_1q) \cdot (\Delta m^{\langle 2 \rangle} + e^{\langle 2 \rangle} + k_2q)$ \textcolor{red}{ $\rhd \approx \Delta^2 m^{\langle 1 \rangle} m^{\langle 2 \rangle} \bmod q$ }

$ = (\vec{a}^{\langle 1 \rangle} \cdot \vec{s} + b^{\langle 1 \rangle}) \cdot (\vec{a}^{\langle 2 \rangle} \cdot \vec{s} + b^{\langle 2 \rangle})$ 

$ = b^{\langle 1 \rangle}b^{\langle 2 \rangle}  + (b^{\langle 2 \rangle}\vec{a}^{\langle 1 \rangle} + b^{\langle 1 \rangle}\vec{a}^{\langle 2 \rangle}) \cdot \vec{s} + (\vec{a}^{\langle 1 \rangle} \cdot \vec{s}) \cdot (\vec{a}^{\langle 2 \rangle} \cdot \vec{s}) $

$ = \underbrace{b^{\langle 1 \rangle}b^{\langle 2 \rangle}}_{d_0}  + \underbrace{(b^{\langle 2 \rangle}\vec{a}^{\langle 1 \rangle} + b^{\langle 1 \rangle}\vec{a}^{\langle 2 \rangle})}_{d_1} \cdot \vec{s} + \underbrace{(\vec{a}^{\langle 1 \rangle} \otimes \vec{a}^{\langle 2 \rangle})}_{\vec{d}_2} \cdot (\vec{s} \otimes \vec{s}) $

$= d_0 + d_1\cdot \vec{s} + \vec{d}_2\cdot \vec{s}^{\otimes}$ \textcolor{red}{ $\rhd$ where $\vec{s}^{\otimes} = \vec{s} \otimes \vec{s}$}

$ $

, where $\otimes$ denotes an outer product of two vectors. For example, given two $n$-length vectors $\vec{v}$ and $\vec{u}$, $\vec{v} \otimes \vec{u}$ is equivalent to an $n^2$-length vector that concatenates the following $n$ distinct $n$-length vectors: $v_0\cdot\vec{u}, \,\, v_1\cdot\vec{u}, \,\, \cdots v_{n-1}\cdot\vec{u}$. Notice that the above relation is similar to that we derived in BFV's ciphertext-to-ciphertext multiplication. Therefore, similar to the remaining steps in BFV, we can relinearize and rescale this LWE term as follows:

\begin{tcolorbox}[title={\textbf{\tboxlabel{\ref*{subsubsec:bfv-mult-cipher-summary}} TFHE's Ciphertext-to-Ciphertext Multiplication - Method 2}}]

Suppose we have the following two LWE ciphertexts:

$\textsf{LWE}_{\vec{s}, \sigma}(\Delta M^{\langle 1 \rangle}) = (\vec{a}^{\langle 1 \rangle}, b^{\langle 1 \rangle}) \bmod q$, \text{ } where $b^{\langle 1 \rangle} = -\vec{a}^{\langle 1 \rangle} \cdot \vec{s} + \Delta m^{\langle 1 \rangle} + e^{\langle 1 \rangle} \bmod q$

$\textsf{LWE}_{\vec{s}, \sigma}(\Delta M^{\langle 2 \rangle}) = (\vec{a}^{\langle 2 \rangle}, b^{\langle 2 \rangle}) \bmod q$, \text{ } where $b^{\langle 2 \rangle} = -\vec{a}^{\langle 2 \rangle} \cdot \vec{s} + \Delta m^{\langle 2 \rangle} + e^{\langle 2 \rangle} \bmod q$

$ $

Multiplication between these two LWE ciphertexts is performed as follows:

$ $

\begin{enumerate}

\item \textbf{\underline{\textsf{ModRaise}}}

Forcibly raise the modulus of the ciphertexts $(\vec{a}^{\langle 1 \rangle}, b^{\langle 1 \rangle}) \bmod q$ and $(\vec{a}^{\langle 2 \rangle}, b^{\langle 2 \rangle}) \bmod q$ to $Q$ (where $Q = q \cdot \Delta$) as follows:

$(\vec{a}^{\langle 1 \rangle}, b^{\langle 1 \rangle}) \bmod Q$

$(\vec{a}^{\langle 2 \rangle}, b^{\langle 2 \rangle}) \bmod Q$

$ $

\item \textbf{\underline{Multiplication}}

Compute the following polynomial multiplications in modulo $Q$: 

$d_0 = b^{\langle 1 \rangle}b^{\langle 2 \rangle} \bmod Q$

$d_1 = b^{\langle 2 \rangle}\vec{a}^{\langle 1 \rangle} +  b^{\langle 1 \rangle}\vec{a}^{\langle 2 \rangle} \bmod Q$

$\vec{d}_2 = \vec{a}^{\langle 1 \rangle} \otimes \vec{a}^{\langle 2 \rangle} \bmod Q$

$ $

\item \textbf{\underline{Relinearization}}

Compute the following:

$ \textsf{ct}_\alpha = (d
_1, d_0)$

$ \textsf{ct}_\beta = \bm{\langle} \textsf{Decomp}^{\beta, l}(\vec{d}_2), \text{ } \textsf{Lev}_{\vec{s}, \sigma}^{\beta, l}(\vec{s}^{\otimes}) \bm{\rangle}$.

$ \textsf{ct}_{\alpha + \beta} = \textsf{ct}_{\alpha} + \textsf{ct}_{\beta}$

$ $

Then, the following property holds:

$\textsf{LWE}^{-1}_{\vec{s}, \sigma}\bm(\textsf{LWE}_{\vec{s}, \sigma}(\Delta^2 \cdot m^{\langle 1 \rangle} \cdot m^{\langle 2 \rangle})\bm) \approx \textsf{LWE}^{-1}_{\vec{s}, \sigma}\bm{(}\textsf{ct}_{\alpha + \beta}\bm)$

$ $

\item \textbf{\underline{Rescaling}}

Update $\textsf{ct}_{\alpha + \beta} = (\vec{a}_{\alpha + \beta}, b_{\alpha + \beta}) \bmod Q$ to $\textsf{ct'}_{\alpha + \beta} = \left(\left\lceil\dfrac{\vec{a}_{\alpha + \beta}}{\Delta}\right\rfloor, \left\lceil\dfrac{b_{\alpha + \beta}}{\Delta}\right\rfloor\right) \bmod q$. 

$ $

This plaintext rescaling process can be also viewed as a modulus switch of the ciphertext $\textsf{ct}_{\alpha + \beta}$ from $Q \rightarrow q$.

$ $

\end{enumerate}

Note that after the ciphertext-to-ciphertext multiplication, the plaintext scaling factor $\Delta=\dfrac{q}{t}$, the ciphertext modulus $q$, and the private key $\vec{s}$ stay the same as before.

\end{tcolorbox}

\subsection{Homomorphic Key Switching}
\label{subsec:bfv-key-switching}

BFV's key switching scheme changes an RLWE ciphertext's secret key from $S$ to $S'$. This scheme is essentially RLWE's key switching scheme with the sign of the $A\cdot S$ term flipped in the encryption and decryption formula. Specifically, this is equivalent to the alternative GLWE version's (\autoref{subsec:glwe-alternative}) key switching scheme (\autoref{sec:glwe-key-switching}) with $k = 1$ as follows:

\begin{tcolorbox}[title={\textbf{\tboxlabel{\ref*{subsec:bfv-key-switching}} BFV's Key Switching}}]
$\textsf{RLWE}_{S',\sigma}(\Delta M) = (0, B) + \bm{\langle} \textsf{Decomp}^{\beta, l}(A), \text{ } \textsf{RLev}_{S', \sigma}^{\beta, l}(S) \bm{\rangle}$
\end{tcolorbox}

\subsection{Homomorphic Rotation of Input Vector Slots}
\label{subsec:bfv-rotation}

In this section, we will explain how to homomorphically rotate the elements of an input vector $\vec{v}$ after it is already encoded as a polynomial and encrypted as an RLWE ciphertext. In \autoref{subsec:coeff-rotation}, we learned how to rotate the coefficients of a polynomial. However, rotating the plaintext polynomial $M(X)$ or RLWE ciphertext polynomials $(A(X), B(X))$ does not necessarily rotate the input vector, which is the source of them. 

The key requirement of homomorphic rotation of input vector slots (i.e., input vector) is that this operation should be performed on the RLWE ciphertext such that after this operation, if we decrypt the RLWE ciphertext and decode it, the recovered input vector will be in a rotated state as we expect. We will divide this task into the following two sub-problems:

\begin{enumerate}[leftmargin=3\parindent]
\item How to indirectly rotate the input vector by updating the plaintext polynomial $M$ to $M'$?

\item How to indirectly update the plaintext polynomial $M$ to $M'$ by updating the RLWE ciphertext polynomials $(A, B)$ to $(A', B')$?
\end{enumerate}

\subsubsection{Rotating Input Vector Slots by Updating the Plaintext Polynomial}

In this task, our goal is to modify the plaintext polynomial $M(X)$ such that the first-half elements of the input vector $\vec{v}$ are shifted to the left by $h$ positions in a wrapping manner among them, and the second-half elements of $\vec{v}$ are also shifted to the left by $h$ positions in a wrapping manner among them. Specifically, if $\vec{v}$ is defined as follows:

$\vec{v} = (v_0, v_1, \cdots, v_{n-1})$

$ $

Then, we will denote the $h$-shifted vector $\vec{v}^{\langle h \rangle}$ as follows:  

$\vec{v}^{\langle h \rangle} = (\underbrace{v_h, v_{h+1}, \cdots, v_0, v_1, \cdots, v_{h-2}, v_{h-1},}_{\text{The first-half $\dfrac{n}{2}$ elements $h$-rotated to the left}} \text{ } \underbrace{v_{\frac{n}{2}+h}, v_{\frac{n}{2}+h+1}, \cdots, v_{\frac{n}{2}+h-2}, v_{\frac{n}{2}+h-1}}_{\text{The second-half $\dfrac{n}{2}$ elements $h$-rotated to the left}})$

$ $

Remember from \autoref{subsec:bfv-batch-encoding} that the BFV encoding scheme's components are as follows:

$\vec{v} = (v_0, v_1, v_2, \cdots, v_{n-1})$ \textcolor{red}{  $\rhd$ $n$-slot input vector}

$ $

$W =  \begin{bmatrix}
1 & 1 & 1 & \cdots & 1\\
(\omega) & (\omega^3) & (\omega^5) & \cdots & (\omega^{2n-1})\\
(\omega)^2 & (\omega^3)^2 & (\omega^5)^2 & \cdots & (\omega^{2n-1})^2\\
\vdots & \vdots & \vdots & \ddots & \vdots \\
(\omega)^{n-1} & (\omega^3)^{n-1} & (\omega^5)^{n-1} & \cdots & (\omega^{2n-1})^{n-1}\\
\end{bmatrix}$

$= \begin{bmatrix}
1 & 1 & \cdots & 1 & 1 & \cdots & 1 & 1\\
(\omega) & (\omega^3) & \cdots & (\omega^{\frac{n}{2} - 1}) & (\omega^{-(\frac{n}{2} - 1)}) & \cdots & (\omega^{-3}) & (\omega^{-1})\\
(\omega)^2 & (\omega^3)^2 & \cdots & (\omega^{\frac{n}{2} - 1})^2 & (\omega^{-(\frac{n}{2} - 1)})^2 & \cdots & (\omega^{-3})^2 & (\omega^{-1})^2\\
\vdots & \vdots & \vdots & \ddots & \vdots \\
(\omega)^{n-1} & (\omega^3)^{n-1} & \cdots & (\omega^{\frac{n}{2} - 1})^{n-1} & (\omega^{-(\frac{n}{2} - 1)})^{n-1} & \cdots & (\omega^{-3})^{n-1} & (\omega^{-1})^{n-1}\\
\end{bmatrix}$

, where $\omega = g^{\frac{t - 1}{2n}} \bmod t$ ($g$ is a generator of $\mathbb{Z}_t^{\times}$)

\textcolor{red}{ $\rhd$ The encoding matrix that converts $\vec{v}$ into $\vec{m}$ ( i.e., $n$ coefficients of the plaintext polynomial $M(X)$ )}

$ $

$\Delta\vec{m}  = n^{-1}\cdot\Delta W\cdot I_n^R\cdot\vec{v}$  

\textcolor{red}{  $\rhd$ A vector containing the  scaled $n$ integer coefficients of the plaintext polynomial}

$ $

$\Delta M = \sum\limits_{i=0}^{n-1} ( \Delta m_i  X^i)$ 

\textcolor{red}{ $\rhd$ The integer polynomial that isomorphically encodes the input vector $\vec{v}$}

$ $

We learned from \autoref{subsec:poly-vector-transformation} that decoding the polynomial $M(X)$ $\vec{v}$ is equivalent to evaluating $M(X)$ at the following $n$ distinct primitive $(\mu=2n)$-th root of unity: $\{\omega, \omega^3, \omega^5, \cdots, \omega^{2n-3}, \omega^{2n-1} \}$. Thus, the above decoding process is equivalent to the following:

$\vec{v} = \dfrac{W^T \Delta \vec{m}}{\Delta} = \bm{(}W^T_0\cdot \vec{m}, \text{ } W^T_1\cdot \vec{m}, \text{ } W^T_2\cdot \vec{m}, \text{ } \cdots, \text{ } W^T_{n-1}\cdot \vec{m}\bm{)}$  

$= \bm{(} \text{ } M(\omega), \text{ } M(\omega^3), \text{ } M(\omega^5), \cdots, M(\omega^{2n-3}), \text{ } M(\omega^{2n-1}) \text{ } \bm{)}$

$= \bm{(} \text{ } M(\omega), \text{ } M(\omega^3), \text{ } M(\omega^5), \cdots, M(\omega^{n-3}), \text{ } M(\omega^{n-1}), \text{ } M(\omega^{-(n-1)}), \text{ } M(\omega^{-(n-3)}), \cdots, M(\omega^{-3}), \text{ } M(\omega{-1}) \text{ } \bm{)}$

%%$\textsf{LWE}_{S, \sigma}(\Delta M) = (A, B) \in \mathcal{R}_{\langle n, q \rangle }^{2}$ \textcolor{red}{  $\rhd$ An LWE ciphertext that encrypts $\Delta M$}

%$A\cdot S + B = \Delta M + E$

$ $

Now, our next task is to modify $M(X)$ to $M'(X)$ such that decoding $M'(X)$ will give us a modified input vector $\vec{v}^{\langle h \rangle}$ that is a rotation of the first half elements of $\vec{v}$ by $h$ positions to the left (in a wrapping manner among them), and the second half elements of it also rotated by $h$ positions to the left (in a wrapping manner among them). To accomplish this rotation, we will take a 2-step solution: 

\begin{enumerate}[leftmargin=3\parindent]
\item To convert $M(X)$ into $M'(X)$, we will define the new mapping $\sigma_M$ as follows: 

$\sigma_M : (M(X), h) \in (\mathcal{R}_{\langle n, t \rangle}, \mathbb{Z}_{n}) \longrightarrow M'(X) \in \mathcal{R}_{\langle n, t \rangle}$

, where $h$ is the number of rotation positions to be applied to $\vec{v}$.
\item To decode $M'(X)$ into the rotated input vector $\vec{v}^{\langle h \rangle}$, we need to re-design our decoding scheme by modifying \textsf{Encoding\textsubscript{1}}'s (\autoref{subsubsec:bfv-encoding-1}) isomorphic mapping $\sigma : M(X) \in \mathcal{R}_{\langle n, t \rangle} \longrightarrow \vec{v} \in \mathbb{Z}^n$ 
\end{enumerate}

$ $

\para{Converting $\bm{M(X)}$ into $\bm{M'(X)}$:} Our first task is to convert $M(X)$ into $M'(X)$, which is equivalent to applying our new mapping $\sigma_M : (M(X), h) \in (\mathcal{R}_{\langle n, t \rangle}, \mathbb{Z}_{n}) \longrightarrow M'(X) \in \mathcal{R}_{\langle n, t \rangle}$, such that decoding $M'(X)$ gives a rotated input vector $\vec{v}^{\langle h \rangle}$ whose first half of the elements in $\vec{v}$ are rotated by $h$ positions to the left (in a wrapping manner among them), and the second half of the elements are also rotated by $h$ positions to the left (in a wrapping manner among them). To design $\sigma_M$ that satisfies this requirement, we will use the number $5^j$ which has the following two special properties (based on number theory):
\begin{itemize}
\item $(5^j \bmod 2n)$ and $(-5^j \bmod 2n)$ generate all odd numbers between $[0, 2n)$ for the integer $j$ where $0 \leq j < \dfrac{n}{2}$. 
\item For each integer $j$ where $0 \leq j < \dfrac{n}{2}$, $(5^j \bmod 2n) + (-5^j \bmod 2n) \equiv 0 \bmod 2n$. 
\end{itemize}
For example, suppose the modulus $2n = 16$. Then, 

\begin{multicols}{2}
\noindent $5^0 \bmod 16 = 1$
\newline $5^1 \bmod 16 = 5$
\newline $5^2 \bmod 16 = 9$
\newline $5^3 \bmod 16 = 13$
\newline $-(5)^0 \bmod 16 = 15$
\newline $-(5)^1 \bmod 16 = 11$
\newline $-(5)^2 \bmod 16 = 7$
\newline $-(5)^3 \bmod 16 = 3$
\end{multicols}

As shown above, $0 \leq j < 4$ generate all odd numbers between $[0, 16)$. Also, for each $j$ in $0 \leq j < 4$, $(5^j \bmod 16) + (-5^j \bmod 16) = 16$. 

$ $

Let's define $J(h) = 5^h \bmod 2n$, \text{ } and \text{ } $J_*(h) = -5^h \bmod 2n$. Based on $J(h)$ and $J_*(h)$, we will define the mapping $\sigma_M : M(X) \rightarrow M'(X)$ as follows:

$\sigma_{M} : M(X) \rightarrow M(X^{J(h)})$

$ $

Given a plaintext polynomial $M(X)$, in order to give its decoded version of input vector $\vec{v}$ the effect of the first half of the elements being rotated by $h$ positions to the left (in a wrapping manner) and the second half of the elements also being rotated by $h$ positions to the left (in a wrapping manner), we update the current plaintext polynomial $M(X)$ to a new polynomial $M'(X) = M(X^{J(h)}) = M(X^{5^h})$ by applying the $\sigma_M$ mapping, where $h$ is the number of positions for left rotations for the first half and second half of the elements of $\vec{v}$.

$ $

\para{Decoding $\bm{M'(X)}$ into \boldmath$\vec{v}^{\langle h \rangle}$}: Our second task is to modify our original decoding scheme in order to successfully decode $M'(X)$ into the rotated input vector $\vec{v}^{\langle h \rangle}$. For this, we will modify our original isomorphic mapping $\sigma : M(X) \longrightarrow \vec{v}$, from:

$\sigma: M(X) \in \mathcal{R}_{\langle n, q \rangle} \longrightarrow  \bm{(}M(\omega), M(\omega^3),M(\omega^5), \cdots, M(\omega^{2n-1})\bm{)} \in \mathbb{Z}^n$ \textcolor{red}{ \text{ } \# designed in \autoref{subsec:poly-vector-transformation}}

$ $

, to the following: 

$\sigma_J: M(X) \in \mathcal{R}_{\langle n, t \rangle} \longrightarrow  \bm{(}M(\omega^{J(0)}),M(\omega^{J(1)}),M(\omega^{J(2)}), \cdots, M(\omega^{J(\frac{n}{2} - 1)}), $

\textcolor{white}{$\sigma_J: M(X) \in \mathcal{R}_{\langle n, q \rangle} \longrightarrow  \bm{(}$} $M(\omega^{J_*(0)}), \text{ } M(\omega^{J_*(1)}), \text{ } M(\omega^{J_*(2)}), \cdots, M(\omega^{J_*(\frac{n}{2} - 1}) \text{ } \bm{)} \in \mathbb{Z}^{n}$

$ $

The common aspect between $\sigma$ and $\sigma_J$ is that they both evaluate the polynomial $M(X)$ at $n$ distinct primitive $(\mu=2n)$-th roots of unity (i.e., $w^i$ for all odd $i$ between $[0,2n]$ ). In the case of the $\sigma_J$ mapping, note that $J(j) = 5^j \bmod 2n$ \text{ } and \text{ } $J_*(j) = -5^j \bmod 2n$ \text{ } for each $j$ in $0 \leq j < \dfrac{n}{2} $ \text{ }  cover all odd numbers between $[0, 2n]$. Therefore, $\omega^{J(j)}$ and $\omega^{J_*(j)}$ between $0 \leq j < \frac{n}{2}$ cover all $n$ distinct primitive $(\mu=2n)$-th roots of unity.

Meanwhile, the difference between $\sigma$ and $\sigma_J$ is the order of the output vector elements. In the $\sigma$ mapping, the order of evaluated coordinates for $M(X)$ is $\omega, \omega^3, \cdots, \omega^{2n - 1}$, whereas in the $\sigma_J$ mapping, the order of evaluated coordinates is $\omega^{J(0)}, \omega^{J(1)}, \cdots, \omega^{J(\frac{n}{2} - 1)}, \omega^{J_*(0)}, \omega^{J_*(1)}, \cdots, \omega^{J_*(\frac{n}{2} - 1)}$. We will later explain why we modified the ordering like this. 

In the original \textsf{Decoding\textsubscript{2}} process (\autoref{subsec:bfv-batch-encoding}), applying the $\sigma$ mapping to a plaintext polynomial $M(X)$ was equivalent to computing the following: 

$\vec{v}_{'} = \bm{(} \text{ } M(\omega), \text{ } M(\omega^3), \text{ } M(\omega^5), \cdots, M(\omega^{2n-3}), \text{ } M(\omega^{2n-1}) \bm{)}$ 

$ $

$ = \bm{(} \text{ } M(\omega), \text{ } M(\omega^3), \text{ } M(\omega^5), \cdots, M(\omega^{n-1}), \text{ } M(\omega^{-(n-1)}), \cdots, \text{ } M(\omega^{-3}), \text{ } M(\omega^{-1}) \text{ } \bm{)}$ 

$ $

$= \bm{(} \text{ } W^T_0\cdot \vec{m}, \text{ } W^T_1\cdot \vec{m}, \text{ } W^T_2\cdot \vec{m}, \text{ } \cdots, \text{ } W^T_{n-1}\cdot \vec{m} \text{ } \bm{)} $ 
$ $ \textcolor{red}{ $\rhd$ where $W_i^T$ is the $(i+1)$-th row of $W^T$}

$ = W^T \vec{m} $
%$= \dfrac{I_n^R}{n}\cdot \bm{(} \text{ } M(\omega), \text{ } M(\omega^3), \text{ } M(\omega^5), \cdots, M(\omega^{2n-3}), \text{ } M(\omega^{2n-1}) \text{ } )$

$ $

Similarly, the modified $\sigma_J$ mapping to the plaintext polynomial $M(X)$ is equivalent to computing the following:

$\vec{v}_{'} =  \bm{(} \text{ } 
M(\omega^{J(0)}), \text{ } M(\omega^{J(1)}), \text{ } M(\omega^{J(2)}), \cdots,  M(\omega^{J(\frac{n}{2}-1)}), \text{ } M(\omega^{J_*(0)}), \text{ } M(\omega^{J_*(1)}), \cdots,  M(\omega^{J_*(\frac{n}{2}-1)}) \text{ } \bm{)}$

$ $

$= \bm{(} \text{ } \hathat{W}^*_0\cdot \vec{m}, \text{ } \hathat{W}^*_1\cdot \vec{m}, \text{ } \hathat{W}^*_2\cdot \vec{m}, \text{ } \cdots, \text{ } \hathat{W}^*_{n-1}\cdot \vec{m} \text{ } \bm{)} $ 

$ $

$ = \hathat{W}^* \vec{m} $, where 

$ $

\noindent $\hathat{W}^* = \begin{bmatrix}
1 & (\omega^{J(0)}) & (\omega^{J(0)})^2 & \cdots & (\omega^{J(0)})^{n-1}\\
1 & (\omega^{J(1)}) & (\omega^{J(1)})^2 & \cdots & (\omega^{J(1)})^{n-1}\\
1 & (\omega^{J(2)}) & (\omega^{J(2)})^2 & \cdots & (\omega^{J(2)})^{n-1}\\
\vdots & \vdots & \vdots & \ddots & \vdots \\
1 & (\omega^{J(\frac{n}{2}-1)}) & (\omega^{J(\frac{n}{2}-1)})^2 & \cdots & (\omega^{J(\frac{n}{2}-1)})^{n-1}\\
1 & (\omega^{J_*(0)}) & (\omega^{J_*(0)})^2 & \cdots & (\omega^{J_*(0)})^{n-1}\\
1 & (\omega^{J_*(1)}) & (\omega^{J_*(1)})^2 & \cdots & (\omega^{J_*(1)})^{n-1}\\
1 & (\omega^{J_*(2)}) & (\omega^{J_*(2)})^2 & \cdots & (\omega^{J_*(2)})^{n-1}\\
\vdots & \vdots & \vdots & \ddots & \vdots \\
1 & (\omega^{J_*(\frac{n}{2}-1)}) & (\omega^{J_*(\frac{n}{2}-1)})^2 & \cdots & (\omega^{J_*(\frac{n}{2}-1)})^{n-1}\\
\end{bmatrix}$

$ $

$ $

\noindent $\hathat{W} = \begin{bmatrix}
1 & 1 & \cdots & 1 & 1 & 1 & \cdots & 1\\
(\omega^{J(\frac{n}{2} - 1)}) & (\omega^{J(\frac{n}{2} - 2)}) & \cdots & (\omega^{J(0)}) & (\omega^{J_*(\frac{n}{2} - 1)}) & (\omega^{J_*(\frac{n}{2} - 2)}) & \cdots & (\omega^{J_*(0)})\\
(\omega^{J(\frac{n}{2} - 1)})^2 & (\omega^{J(\frac{n}{2} - 2)})^2 & \cdots & (\omega^{J(0)})^2 & (\omega^{J_*(\frac{n}{2} - 1)})^2 & (\omega^{J_*(\frac{n}{2} - 2)})^2 & \cdots & (\omega^{J_*(0)})^2 \\
\vdots & \vdots & \ddots & \vdots & \vdots & \vdots & \vdots & \vdots \\
(\omega^{J(\frac{n}{2} - 1)})^{n-1} & (\omega^{J(\frac{n}{2} - 2)})^{n-1} & \cdots & (\omega^{J(0)})^{n-1} & (\omega^{J_*(\frac{n}{2} - 1)})^{n-1} & (\omega^{J_*(\frac{n}{2} - 2)})^{n-1} & \cdots  & (\omega^{J_*(0)})^{n-1}
\end{bmatrix}$

$ $

$ $

From this point, we will replace $W$ in the \textsf{Encoding\textsubscript{1}} process (\autoref{subsec:bfv-batch-encoding}) by $\hathat{W}$, and $W^T$ in the \textsf{Decoding\textsubscript{2}} process by $\hathat{W}^*$. 

$ $

To demonstrate that $\hathat{W}$ is a valid encoding matrix like $W$ and $\hathat{W}^*$ is a valid decoding matrix like $W^T$, we need to prove the following 2 aspects:
\begin{itemize}[leftmargin=3\parindent]

\item \textbf{${\hathat{\bm W}}$ is a basis of the $\bm{n}$-dimensional vector space:} This is true, because $\hathat{W}$ is simply a row-wise re-ordering of $W$, which is still a basis of the $\bm{n}$-dimensional vector space.

\item \boldmath{$\hathat{W}^{*} \cdot \hathat{W} = n \cdot I_n^R$ (to satisfy Theorem~\ref*{subsec:vandermonde-euler} in \autoref{subsec:vandermonde-euler})}: This proof is split into 2 sub-proofs: 
\begin{enumerate}[leftmargin=2\parindent]
\item \boldmath{${\hathat{ W}^{ *} \cdot \hathat{ W}}$ has value $\bm{n}$ along the anti-diagonal line:} Each element along the anti-diagonal line of \boldmath$\hathat{ W}^{ *} \cdot \hathat{ W}$ is computed as $\sum\limits_{i=0}^{n-1} \omega^{2nik} = n$ where $k$ is some integer.
\item \boldmath{${\hathat{ W}^{ *} \cdot \hathat{ W}}$ has value 0 at all other coordinates:} All other elements except for the ones along the anti-diagonal lines are $\sum\limits_{i=0}^{n-1}\omega^{2i} \dfrac{\omega^n - 1}{\omega - 1} = 0$ (by Geometric Sum). 
\end{enumerate}

We provide \href{https://github.com/fhetextbook/fhe-textbook/blob/main/source%20code/bfv_j_matrix_inverse_proof.py}{\underline{the Python script}} that empirically demonstrates this. 

\end{itemize}

Therefore, $\hathat{W}^*$ and $\hathat{W}$ are valid encoding \& decoding matrices that transform $\vec{v}$ into $M(X)$. 

$ $

Now, let's think about what will be the structure of $\vec{v}^{\langle h \rangle}$ (i.e., the first-half elements of $\vec{v}$ being rotated $h$ positions to the left in a wrapping manner among them and the second-half elements of it also being rotated $h$ positions to the left in a wrapping manner among them). Remember that $\vec{v}$ is as follows:

$\vec{v} = \bm{(} \text{ } 
M(\omega^{J(0)}), \text{ } M(\omega^{J(1)}), \cdots,  M(\omega^{J(\frac{n}{2}-1)}), \text{ } 
M(\omega^{J_*(0)}), \text{ } M(\omega^{J_*(1)}),  \cdots,  M(\omega^{J_*(\frac{n}{2}-1)}) \bm{)}$

\text{ } \text{ } $= \bm{(} \text{ } \hathat{W}^*_0\cdot \vec{m}, \text{ } \hathat{W}^*_1\cdot \vec{m}, \text{ } \hathat{W}^*_2\cdot \vec{m}, \text{ } \cdots, \text{ } \hathat{W}^*_{n - 1}\cdot \vec{m} \text{ } \bm{)} $ 

$ $

Thus, the state of $\vec{v}^{\langle h \rangle}$ which is equivalent to rotating $\vec{v}$ by $h$ positions to the left for the first-half and second-half element groups will be the following: 

$\vec{v}^{\langle h \rangle} = \bm{(} \text{ } M(\omega^{J(h)}), \text{ } M(\omega^{J(h+1)}), \cdots, M(\omega^{J(\frac{n}{2}-1)}), \text{ } M(\omega^{J(0)}), \text{ }M(\omega^{J(1)}), \cdots, \text{ } M(\omega^{J(h-2)}), \text{ } M(\omega^{J(h-1)}),$

\text{ }\text{ }\text{ } $ M(\omega^{J_*(h)}), \text{ } M(\omega^{J_*(h+1)}), \cdots, M(\omega^{J_*(\frac{n}{2}-1)}), \text{ } M(\omega^{J_*(0)}), \text{ }M(\omega^{J_*(1)}), \cdots, \text{ } M(\omega^{J_*(h-2)}), \text{ } M(\omega^{J_*(h-1)}) \text{ } \bm{)}$

$ $

Notice that the above computation of $\vec{v}^{\langle h \rangle}$ is equivalent to vertically rotating the upper $\frac{n}{2}$ rows of $\hathat{W}^*$ by $h$ positions upward (in a wrapping manner among them), rotating the lower $\frac{n}{2}$ rows of $\hathat{W}^{*}$ by $h$ positions upward (in a wrapping manner among them), and multiplying the resulting matrix with $\vec{m}$. However, it is not desirable to directly modify the decoding matrix $\hathat{W}^*$ like this in practice, because then the decoding matrix loses its consistency. Therefore, instead of directly modifying $\hathat{W}^*$, we will modify $\vec{m}$ to $\vec{m}_{'}$ (i.e., modify $M(X)$ to $M'(X)$) such that the relation $\vec{v}^{\langle h \rangle} = \hathat{W}^* \cdot \vec{m}_{'}$ holds. Let's extract the upper-half rows of $\hathat{W}^*$ and denote this $\dfrac{n}{2} \times n$ matrix as $\hathat{H}_1^*$. Then, $\hathat{H}_1^*$ is equivalent to a Vandermonde matrix (Definition~\ref*{subsec:vandermonde} in \autoref{subsec:vandermonde}) in the form of $V(\omega^{J(0)}, \omega^{J(1)}, \cdots, \omega^{J(\frac{n}{2}-1)})$. Similarly, let's extract the lower-half rows of $\hathat{W}^*$ and denote this $\dfrac{n}{2} \times n$ matrix as $\hathat{H}_2^*$. Then, $\hathat{H}_2^*$ is equivalent to a Vandermonde matrix (Definition~\ref*{subsec:vandermonde} in \autoref{subsec:vandermonde}) in the form of $V(\omega^{J_*(0)}, \omega^{J_*(1)}, \cdots, \omega^{J_*(\frac{n}{2}-1)})$. 

Now, let's vertically rotate the rows of $\hathat{H}_1^*$ by $h$ positions upward and denote it as $\hathat{H}_1^{*\langle h \rangle}$; and vertically rotate the rows of $\hathat{H}_2^*$ by $h$ positions upward and denote it as $\hathat{H}_2^{*\langle h \rangle}$. And let's denote the $\dfrac{n}{2}$-dimensional vector comprising the first-half elements of $\vec{v}^{\langle h \rangle}$ as $\vec{v}_1^{\langle h \rangle}$, and the $\dfrac{n}{2}$-dimensional vector comprising the second-half elements of $\vec{v}^{\langle h \rangle}$ as $\vec{v}_2^{\langle h \rangle}$. 
Then, computing (i.e., decoding) $\vec{v}_1^{\langle h \rangle} = \hathat{H}_1^{*\langle h \rangle}\cdot \vec{m}$ is equivalent to modifying $M(X)$ to $M'(X) = M(X^{J(h)})$ (whose coefficient vector is $\vec{m}^{\langle h \rangle}$) and then computing (i.e., decoding) $\vec{v}_1^{\langle h \rangle} = \hathat{H}_1^{*}\cdot \vec{m}^{\langle h \rangle}$. This is because:

$\vec{v}_1^{\langle h \rangle} =\hathat{H}_1^{*\langle h \rangle}\cdot \vec{m}$

$ = \bm{(} \text{ } \hathat{W}^*_{h}\cdot\vec{m}, \text{ } \hathat{W}^*_{h+1}\cdot\vec{m}, \text{ } \hathat{W}^*_{h+2}\cdot\vec{m}, \cdots, \hathat{W}^*_{\frac{n}{2}-1}\cdot\vec{m}, \text{ } \hathat{W}^*_{0}\cdot\vec{m}, \text{ }\hathat{W}^*_{1}\cdot\vec{m}, \cdots, \text{ } \hathat{W}^*_{h-2}\cdot\vec{m}, \text{ } \hathat{W}^*_{h-1}\cdot\vec{m}  \bm{)}$

$ $

$= \bm{(} \text{ } M((\omega^{J(h)})^{J(0)}), \text{ } M((\omega^{J(h)})^{J(1)}), \text{ } M((\omega^{J(h)})^{J(2)}), \cdots, M((\omega^{J(h)})^{J(\frac{n}{2}-1)}) \text{ } \bm{)}$

\textcolor{red}{ $\rhd$ This is equivalent to evaluating the polynomial $M(X^{J(h)})$ at the following $\dfrac{n}{2}$ distinct $(\mu=2n)$-th roots of unity: $\omega^{J(0)}, \omega^{J(1)},  \omega^{J(2)}, \cdots, \omega^{J(\frac{n}{2} - 1)}$}

$ $

$= \bm{(} \text{ } M(\omega^{J(h)\cdot J(0)}), \text{ } M(\omega^{J(h)\cdot J(1)}), \text{ } M(\omega^{J(h)\cdot J(2)}), \cdots, M(\omega^{J(h)\cdot J(\frac{n}{2}-1)}) \text{ } \bm{)}$

$= \bm{(} \text{ } M(\omega^{5^h\cdot 5^0}), \text{ } M(\omega^{5^h\cdot 5^1}), \text{ } M(\omega^{5^h\cdot 5^2}), \cdots, M(\omega^{5^h\cdot 5^{n/2-1}}) \text{ } \bm{)}$

$= \bm{(} \text{ } M(\omega^{5^{h}}), \text{ } M(\omega^{5^{h+1}}), \text{ } M(\omega^{5^{h+2}}), \cdots, M(\omega^{5^{h+n/2-1}}) \text{ } \bm{)}$ \textcolor{red}{ $\rhd$ note that $5^{\frac{n}{2}} \bmod 2n = 1$}

$ $

$= \bm{(} \text{ } M(\omega^{J(h)}), \text{ } M(\omega^{J(h+1)}), \text{ } M(\omega^{J(h+2)}), \cdots, M(\omega^{J(\frac{n}{2}-1)}), \text{ } M(\omega^{J(0)}), \text{ }M(\omega^{J(1)}),$ 
                                    
$\cdots, \text{ } M(\omega^{J(h-2)}), \text{ } M(\omega^{J(h-1)})  \bm{)}$ 

$ $

$= \tilde{H}_1^{*}\cdot \vec{m}^{\langle h \rangle}$ \textcolor{red}{ \text{ } \# where $\vec{m}^{\langle h \rangle}$ contains the $n$ coefficients of the polynomial $M(X^{J(h)})$ }

$ $

Similarly, computing (i.e., decoding) $\vec{v}_2^{\langle h \rangle} = \hathat{H}_2^{*\langle h \rangle}\cdot \vec{m}$ is equivalent to modifying $M(X)$ to $M'(X) = M(X^{J(h)})$ (whose coefficient vector is $\vec{m}^{\langle h \rangle}$) and then computing (i.e., decoding) $\vec{v}_2^{\langle h \rangle} = \hathat{H}_2^{*}\cdot \vec{m}^{\langle h \rangle}$. This is because,

$\vec{v}_2^{\langle h \rangle} =\hathat{H}_2^{*\langle h \rangle}\cdot \vec{m}$

$ $

$ = \bm{(} \text{ } \hathat{W}^*_{\frac{n}{2} +h}\cdot\vec{m}, \text{ } \hathat{W}^*_{\frac{n}{2} + h+1}\cdot\vec{m}, \text{ } \hathat{W}^*_{\frac{n}{2} +h+2}\cdot\vec{m}, \cdots, \hathat{W}^*_{n -1}\cdot\vec{m}, \text{ } \hathat{W}^*_{\frac{n}{2}}\cdot\vec{m}, \text{ }\hathat{W}^*_{\frac{n}{2} + 1}\cdot\vec{m}, \cdots,$

\text{ }\text{ }\text{ }\text{ }$ \text{ } \hathat{W}^*_{\frac{n}{2} + h-2}\cdot\vec{m}, \text{ } \hathat{W}^*_{\frac{n}{2} + h-1}\cdot\vec{m}  \bm{)}$

$ $

$= \bm{(} \text{ } M((\omega^{J(h)})^{J_*(0)}), \text{ } M((\omega^{J(h)})^{J_*(1)}), \text{ } M((\omega^{J(h)})^{J_*(2)}), \cdots, M((\omega^{J(h)})^{J_*(\frac{n}{2}-1)}) \text{ } \bm{)}$

\textcolor{red}{ $\rhd$ This is equivalent to evaluating the polynomial $M(X^{J(h)})$ at the following $\dfrac{n}{2}$ distinct $(\mu=2n)$-th roots of unity: $\omega^{J_*(0)}, \omega^{J_*(1)},  \omega^{J_*(2)}, \cdots, \omega^{J_*(\frac{n}{2} - 1)}$}

$ $

$= \bm{(} \text{ } M(\omega^{J(h)\cdot J_*(0)}), \text{ } M(\omega^{J(h)\cdot J_*(1)}), \text{ } M(\omega^{J(h)\cdot J_*(2)}), \cdots, M(\omega^{J(h)\cdot J_*(\frac{n}{2}-1)}) \text{ } \bm{)}$

$= \bm{(} \text{ } M(\omega^{5^h\cdot -5^0}), \text{ } M(\omega^{5^h\cdot -5^1}), \text{ } M(\omega^{5^h\cdot -5^2}), \cdots, M(\omega^{5^h\cdot -5^{n/2-1}}) \text{ } \bm{)}$

$= \bm{(} \text{ } M(\omega^{-5^{h}}), \text{ } M(\omega^{-5^{h+1}}), \text{ } M(\omega^{-5^{h+2}}), \cdots, M(\omega^{-5^{h+n/2-1}}) \text{ } \bm{)}$ \textcolor{red}{ $\rhd$ note that $-(5^{\frac{n}{2}} \bmod 2n) = -1$}

$ $

$= \bm{(} \text{ } M(\omega^{J_*(h)}), \text{ } M(\omega^{J_*(h+1)}), \text{ } M(\omega^{J_*(h+2)}), \cdots, M(\omega^{J_*(\frac{n}{2}-1)}), \text{ } M(\omega^{J_*(0)}), \text{ }M(\omega^{J_*(1)}),$ 
                                    
$\cdots, \text{ } M(\omega^{J_*(h-2)}), \text{ } M(\omega^{J_*(h-1)})  \bm{)}$ 

$ $

$= \hathat{H}_2^{*}\cdot \vec{m}^{\langle h \rangle}$

$ $

The above derivations demonstrate that $\vec{v}_1^{\langle h \rangle} = \hathat{H}_1^{*}\cdot \vec{m}^{\langle h \rangle}$, and $\vec{v}_2^{\langle h \rangle} = \hathat{H}_2^{*}\cdot \vec{m}^{\langle h \rangle}$. Combining these two findings, we reach the conclusion that $\vec{v}^{\langle h \rangle} = \hathat{W}^* \cdot \vec{m}^{\langle h \rangle}$: rotating the first-half elements of the input vector $\vec{v}$ by $h$ positions to the left and the second-half elements of it by $h$ positions also to the left is equivalent to updating the plaintext polynomial $M(X)$ to $M(X^{J(h)})$ and then decoding it with the decoding matrix $\hathat{W}^*$.

However, now a new problem is that we cannot directly update the plaintext $M(X)$ to $M(X^{J(h)})$, because $M(X)$ is encrypted as an RLWE ciphertext. Therefore, we need to instead update the RLWE ciphertext components $(A, B)$ to \textit{indirectly} by updating $M(X)$ to $M(X^{J(X)})$. We will explain this in the next subsection.

\subsubsection{Updating the Plaintext Polynomial by Updating the Ciphertext Polynomials}

Given an RLWE ciphertext $\textsf{ct} = (A, B)$, our goal is to update $\textsf{ct} = (A, B)$ to $C^{\langle h \rangle} = (A^{\langle h \rangle}, B^{\langle h \rangle})$ such that decrypting it gives the input vector $\vec{v}^{\langle h \rangle}$. That is, the following relation should hold: 

$\textsf{RLWE}^{-1}_{S, \sigma}\bm{(} \text{ } C^{\langle h \rangle} = (A^{\langle h \rangle}, B^{\langle h \rangle}) \text{ }\bm{)} = \Delta M(X^{J(h)}) + E'$

$ $

Remember that in the RLWE cryptosystem (\autoref{sec:rlwe})'s alternative version (\autoref{subsec:glwe-alternative}), the plaintext and ciphertext pair have the following relation:

$\Delta M(X) + E(X) = A(X)\cdot S(X) + B(X) \approx \Delta M(X)$

$ $

If we apply $X = X^{J(h)}$ in the above relation, we can derive the following relation: 

$\Delta M(X^{J(h)}) + E(X^{J(h)}) = A(X^{J(h)}) \cdot S(X^{J(h)}) + B(X^{J(h)}) \approx \Delta M(X^{J(h)})$

$ $

This relation implies that if we decrypt the ciphertext $C^{\langle h \rangle} = (A(X^{J(h)}), B(X^{J(h)}))$ with $S(X^{J(h)})$ as the secret key, then we get $\Delta M(X^{J(h)})$. Therefore, $C^{\langle h \rangle} = (A(X^{J(h)}), B(X^{J(h)}))$ is the RLWE ciphertext we are looking for, because decrypting it and then decoding its plaintext $M(X^{J(h)})$ will give us the input vector $\vec{v}^{\langle h \rangle}$. 

We can easily convert $\textsf{ct} = (A(X), B(X))$ into $C^{\langle h \rangle} = (A(X^{J(h)}), B(X^{J(h)}))$ by applying $X^{J(h)}$ to $X$ for each of the $A(X)$ and $B(X)$ polynomials. However, after that, notice that the decryption key of the RLWE ciphertext $C^{\langle h \rangle} = (A(X^{J(h)}), B(X^{J(h)}))$ has been changed from $S(X)$ to $S(X^{J(h)})$. Thus, we need to additionally switch the ciphertext $C^{\langle h \rangle}$'s key from $S(X^{J(h)}) \rightarrow S(X)$, which is equivalent to converting $\textsf{RLWE}_{S(X^{J(h)}), \sigma}\bm{(} C^{\langle h \rangle} = (A(X^{J(h)}), B(X^{J(h)})) \bm{)}$ into $\textsf{RLWE}_{S, \sigma}\bm{(} C^{\langle h \rangle} = (A(X^{J(h)}), B(X^{J(h)})) \bm{)}$. For this, we will apply the BFV key switching technique (Summary~\ref*{subsec:bfv-key-switching}) learned in \autoref{subsec:bfv-key-switching} as follows: 

$ $

\noindent $\underbrace{\textsf{RLWE}_{S(X),\sigma}\bm{(}\Delta M(X^{J(h)})\bm{)}}_{\substack{\text{the result of} \\ \text{plaintext-to-ciphertext addition}}} = \underbrace{\bm{(} \text{ } 0, B(X^{J(h)}) \text{ } \bm{)}}_{\substack{\text{the plaintext } B(X^{J(h)}) \\ \text{(trivial ciphertext)}}} + \bm{\langle} \text{ } \underbrace{\textsf{Decomp}^{\beta, l}\bm{(}A(X^{J(h)})\bm{)}, \text{ } \textsf{RLev}_{S(X), \sigma}^{\beta, l}\bm{(}S(X^{J(h)})\bm{)} \text{ } \bm{\rangle}}_{\substack{\substack{\text{an RLWE ciphertext encrypting $A(X^{J(h)}) \cdot S(X^{J(h)})$}\\ \text{which is key-switched from $S(X^{J(h)})\rightarrow S(X)$}}}}$

\subsubsection{Summary}
\label{subsubsec:bfv-rotation-summary}

We summarize the procedure of rotating the BFV input vectors as follows: 

\begin{tcolorbox}[title={\textbf{\tboxlabel{\ref*{subsec:bfv-rotation}} BFV's Homomorphic Rotation of Input Vector Slots}}]

To support input vector slot rotation, we update the original encoding matrix in \textsf{Encoding\textsubscript{1}} as follows: 

$ $

{\footnotesize{\noindent $\hathat{W} = \begin{bmatrix}
1 & 1 & \cdots & 1 & 1 & 1 & \cdots & 1\\
(\omega^{J(\frac{n}{2} - 1)}) & (\omega^{J(\frac{n}{2} - 2)}) & \cdots & (\omega^{J(0)}) & (\omega^{J_*(\frac{n}{2} - 1)}) & (\omega^{J_*(\frac{n}{2} - 2)}) & \cdots & (\omega^{J_*(0)})\\
(\omega^{J(\frac{n}{2} - 1)})^2 & (\omega^{J(\frac{n}{2} - 2)})^2 & \cdots & (\omega^{J(0)})^2 & (\omega^{J_*(\frac{n}{2} - 1)})^2 & (\omega^{J_*(\frac{n}{2} - 2)})^2 & \cdots & (\omega^{J_*(0)})^2 \\
\vdots & \vdots & \ddots & \vdots & \vdots & \vdots & \ddots & \vdots \\
(\omega^{J(\frac{n}{2} - 1)})^{n-1} & (\omega^{J(\frac{n}{2} - 2)})^{n-1} & \cdots & (\omega^{J(0)})^{n-1} & (\omega^{J_*(\frac{n}{2} - 1)})^{n-1} & (\omega^{J_*(\frac{n}{2} - 2)})^{n-1} & \cdots  & (\omega^{J_*(0)})^{n-1}
\end{bmatrix}$}}

$ $

$ $

, and update the original decoding matrix in \textsf{Decoding\textsubscript{2}} as follows:

$\hathat{W}^* = \begin{bmatrix}
1 & (\omega^{J(0)}) & (\omega^{J(0)})^2 & \cdots & (\omega^{J(0)})^{n-1}\\
1 & (\omega^{J(1)}) & (\omega^{J(1)})^2 & \cdots & (\omega^{J(1)})^{n-1}\\
1 & (\omega^{J(2)}) & (\omega^{J(2)})^2 & \cdots & (\omega^{J(2)})^{n-1}\\
\vdots & \vdots & \vdots & \ddots & \vdots \\
1 & (\omega^{J(\frac{n}{2}-1)}) & (\omega^{J(\frac{n}{2}-1)})^2 & \cdots & (\omega^{J(\frac{n}{2}-1)})^{n-1}\\
1 & (\omega^{J_*(0)}) & (\omega^{J_*(0)})^2 & \cdots & (\omega^{J_*(0)})^{n-1}\\
1 & (\omega^{J_*(1)}) & (\omega^{J_*(1)})^2 & \cdots & (\omega^{J_*(1)})^{n-1}\\
1 & (\omega^{J_*(2)}) & (\omega^{J_*(2)})^2 & \cdots & (\omega^{J_*(2)})^{n-1}\\
\vdots & \vdots & \vdots & \ddots & \vdots \\
1 & (\omega^{J_*(\frac{n}{2}-1)}) & (\omega^{J_*(\frac{n}{2}-1)})^2 & \cdots & (\omega^{J_*(\frac{n}{2}-1)})^{n-1}\\
\end{bmatrix}$

$ $

$ $

, where $J(h)$ is the \textit{rotation helper formula}: $J(h) = 5^h \bmod 2n$, 
\text{ } $J_*(h) = -5^h \bmod 2n$. 

$ $

Using $\hathat{W}$, the encoding is perform as: $\vec{m} = n^{-1}\cdot \hathat{W} \cdot I_n^R \cdot \vec{v}$. Using $\hathat{W}^*$, the decoding is performed as $\vec{v} = \hathat{W}^* \cdot \vec{m}$

$ $

Suppose we have an RLWE ciphertext and a key-switching key as follows:

$\textsf{RLWE}_{S, \sigma}(\Delta M) = (A, B)$, \text{ } $\textsf{RLev}_{S, \sigma}^{\beta, l}\bm(S(X^{J(h)})\bm)$

$ $

Then, the procedure of rotating the first-half elements of the ciphertext's original input vector $\vec{v}$ by $h$ positions to the left (in a wrapping manner among them) and the second-half elements of $\vec{v}$ by $h$ positions to the left (in a wrapping manner among them) is as follows: 

\begin{enumerate}
\item Update $A(X)$, $B(X)$ to $A(X^{J(h)})$, $B(X^{J(h)})$. 
\item Perform the following key switching (\autoref{subsec:ckks-key-switching}) from $S(X^{J(h)})$ to $S(X)$:

$\textsf{RLWE}_{S(X),\sigma}\bm{(}\Delta M(X^{J(h)})\bm{)} = \bm{(} 0, B(X^{J(h)}) \bm{)} \text{ } + \text{ } \bm{\langle}  \textsf{Decomp}^{\beta, l}\bm{(}A(X^{J(h)})\bm{)}, \text{ } \textsf{RLev}_{S(X), \sigma}^{\beta, l}\bm{(}S(X^{J(h)})\bm{)} \bm{\rangle}$
\end{enumerate}

\end{tcolorbox}

\subsubsection{Encoding Example}
\label{subsubsec:bfv-batch-encoding-ex}

%$bfv_vector = [10, 0, 5, 13]
%bfv_vector2 = [2, 4, 8, 6]

In the above example, we use the unsigned modulo representation (e.g., $[0, t-1], [0, q]$) when computing $\bmod t$ and $\bmod q$. However, the correctness of the result holds even if we use the centered modulo representation (e.g., $[-\dfrac{t}{2}, \dfrac{t}{2})$, $[-\dfrac{q}{2}, \dfrac{q}{2})$). In many actual FHE libraries, centered modulo arithmetic is often used to manage noise growth more effectively.

Suppose we have the following setup: 

$\mu = 8, \, n = \dfrac{\mu}{2} = 4, \, t = 17, \, q = 2^8 = 256, \, n^{-1} = 13, \, \Delta = \left\lfloor\dfrac{q}{t}\right\rfloor = 15$

$ $

$\mathcal{R}_{\langle 4, 17 \rangle} = \mathbb{Z}_{17}[X] / (X^4 + 1)$

$ $

The roots of $X^4 + 1 \pmod{17}$ are $X = \{2, 8, 15, 9\}$, as demonstrated as follows:

$2^4 \equiv 8^4 \equiv 15^4 \equiv 9^4 \equiv 16 \equiv -1 \bmod{17}$

$ $

Definition~\ref*{subsec:cyclotomic-def} (in \autoref{subsec:cyclotomic-def}) states that the roots of the $\mu$-th cyclotomic polynomial are the primitive $\mu$-th roots of unity. And Definition~\ref*{subsec:roots-def} (in \autoref{subsec:roots-def}) states that the order of the primitive $\mu$-th roots of unity is $\mu$. These definitions apply to both the cyclotomic polynomials over $X\in\mathbb{C}$ (complex numbers) and the cyclotomic polynomials over $X \in \mathbb{Z}_t$ (ring). 

Since $\{2, 8, 15, 9\}$ are the roots of the $(\mu=8)$-th cyclotomic polynomial $X^4 + 1$ over the ring $\mathbb{Z}_{17}$, they are also the $(\mu=8)$-th primitive roots of unity of $\mathbb{Z}_{17}$. Therefore, their order (\autoref{subsec:order-def}) is $\mu=8$ as demonstrated as follows: 

$2^8 \equiv 8^8 \equiv 15^8 \equiv 9^8 \equiv 1 \bmod 17$

$2^4 \equiv 8^4 \equiv 15^4 \equiv 9^4 \equiv 16 \not\equiv 1 \bmod 17$

$ $

Definition~\ref*{subsec:cyclotomic-def} (in \autoref{subsec:cyclotomic-def}) and Theorem~\ref*{subsec:cyclotomic-theorem} (\autoref{subsec:cyclotomic-theorem}) also state that for each primitive $\mu$-th root of unity $\omega$, $\{\omega^k\}_{\textsf{gcd}(k, \mu) = 1}$ generates all roots of the $\mu$-th cyclotomic polynomial. Notice that in the case of the $(\mu=8)$-th cyclotomic polynomial $X^4 + 1$, its roots $\{2, 8, 15, 9\}$ generate all $(\mu=8)$-th roots of unity as follows:

$\{2^1, 2^3, 2^5, 2^7\} \equiv \{8^1, 8^3, 8^5, 8^7\} \equiv \{15^1, 15^3, 15^5, 15^7\} \equiv \{9^1, 9^3, 9^5, 9^7\} \equiv \{2, 8, 15, 9\} \bmod 17$

$ $

Among $\{2, 8, 15, 9\}$ as the roots of $X^4 + 1$, let's choose $\omega = 9$ as the base root to construct the encoding matrix $\hathat{W}$ and the decoding matrix $\hathat{W}^*$ as follows: 

$\hathat{W} = \begin{bmatrix}
1 & 1 & 1 & 1\\
\omega^{J(1)} & \omega^{J(0)} & \omega^{J_*(1)} & \omega^{J_*(0)}\\
(\omega^{J(1)})^2 & (\omega^{J(0)})^2 & (\omega^{J_*(1)})^2 & (\omega^{J_*(0)})^2\\
(\omega^{J(1)})^3 & (\omega^{J(0)})^3 & (\omega^{J_*(1)})^3 & (\omega^{J_*(0)})^3\\
\end{bmatrix}$ \textcolor{red}{  $\rhd$ where $J(h) = 5^h \bmod 8$}

$ = \begin{bmatrix}
1 & 1 & 1 & 1\\
9^{5} & 9^{1} & 9^{3} & 9^{7}\\
(9^{5})^2 & (9^{1})^2 & (9^{3})^2 & (9^{7})^2\\
(9^{5})^3 & (9^{1})^3 & (9^{3})^3 & (9^{7})^3\\
\end{bmatrix} \equiv \begin{bmatrix}
1 & 1 & 1 & 1\\
8 & 9 & 15 & 2\\
13 & 13 & 4 & 4\\
2 & 15 & 9 & 8\\
\end{bmatrix} \bmod{17}$ 

$ $ 

$\hathat{W}^* = \begin{bmatrix}
1 & \omega^{J(0)} & (\omega^{J(0)})^2 & (\omega^{J(0)})^3\\
1 & \omega^{J(1)} & (\omega^{J(1)})^2 & (\omega^{J(1)})^3\\
1 & \omega^{J_*(0)} & (\omega^{J_*(0)})^2 & (\omega^{J_*(0)})^3\\
1 & \omega^{J_*(1)} & (\omega^{J_*(1)})^2 & (\omega^{J_*(1)})^3\\
\end{bmatrix} \equiv \begin{bmatrix}
1 & 9 & 13 & 15\\
1 & 8 & 13 & 2\\
1 & 2 & 4 & 8\\
1 & 15 & 4 & 9\\
\end{bmatrix}  \bmod{17}$

$ $

Notice that Theorem~\ref*{subsec:vandermonde-euler-integer-ring} (in \autoref{subsec:vandermonde-euler-integer-ring}) is demonstrated as follows:

$\hathat{W}^* \cdot \hathat{W} = \begin{bmatrix}
1 & 9 & 13 & 15\\
1 & 8 & 13 & 2\\
1 & 2 & 4 & 8\\
1 & 15 & 4 & 9\\
\end{bmatrix} \cdot \begin{bmatrix}
1 & 1 & 1 & 1\\
8 & 9 & 15 & 2\\
13 & 13 & 4 & 4\\
2 & 15 & 9 & 8\\
\end{bmatrix} = \begin{bmatrix}
0 & 0 & 0 & 4\\
0 & 0 & 4 & 0\\
0 & 4 & 0 & 0\\
4 & 0 & 0 & 0\\
\end{bmatrix} = n I_n^{R} \pmod{17}$

$ $

Now suppose that we encode the following two input vectors (i.e., input vector slots) in $\mathbb{Z}_{17}$:

$\vec{v}_1 = (10, 3, 5, 13)$

$\vec{v}_2 = (2, 4, 3, 6)$

$\vec{v}_1 + \vec{v}_2 = (10, 3, 5, 13) + (2, 4, 3, 6) \equiv (12, 7, 8, 2) \bmod 17$

$ $

These two vectors are encoded as follows:

$ $

$\vec{m}_1 = n^{-1} \hathat{W} \cdot I_n^R \cdot \vec{v} = 13 \cdot \begin{bmatrix}
1 & 1 & 1 & 1\\
8 & 9 & 15 & 2\\
13 & 13 & 4 & 4\\
2 & 15 & 9 & 8\\
\end{bmatrix} \cdot \begin{bmatrix}
0 & 0 & 0 & 1\\
0 & 0 & 1 & 0\\
0 & 1 & 0 & 0\\
1 & 0 & 0 & 0\\
\end{bmatrix} \cdot \begin{bmatrix}
10\\
3\\
5\\
13\\
\end{bmatrix} \equiv (12, 11, 12, 1)  \bmod 17$

$ $

$\vec{m}_2 = n^{-1} \hathat{W} \cdot I_n^R \cdot \vec{v} = 13 \cdot \begin{bmatrix}
1 & 1 & 1 & 1\\
8 & 9 & 15 & 2\\
13 & 13 & 4 & 4\\
2 & 15 & 9 & 8\\
\end{bmatrix} \cdot \begin{bmatrix}
0 & 0 & 0 & 1\\
0 & 0 & 1 & 0\\
0 & 1 & 0 & 0\\
1 & 0 & 0 & 0\\
\end{bmatrix} \cdot \begin{bmatrix}
2\\
4\\
3\\
6\\
\end{bmatrix} \equiv (8, 5, 14, 6) \bmod 17$

$ $

$ $

$\Delta M_1(X) = 15\cdot(12 + 11X + 12X^2 + 1X^3) = 180 + 165X + 180X^2 + 15X^3 \pmod{q}$ \textcolor{red}{  $\rhd$ where $q = 256$}

$\Delta M_2(X) = 15\cdot(8 + 5X + 14X^2 + 6X^3) = 120 + 75X + 210X^2 + 90X^3 \pmod{q}$

$\Delta M_{1+2}(X) = \Delta M_1(X) + \Delta M_2(X) = \Delta (M_1(X) + M_2(X)) = 44 + 240X + 134X^2 + 105X^3 \pmod{q}$

$M_{1+2}(X) =\left\lceil\dfrac{44 + 240X + 134X^2 + 105X^3}{15}\right\rfloor \bmod 17$

$= \left\lceil2.933 + 16X + 8.933X^2 + 7X^3\right\rfloor \bmod 17$

$3 + 16X + 9X^2 + 7X^3 \bmod 17 $

$ $

This polynomial matches the value of $\vec{m}_{1+2}$ as follows:

$\vec{m}_1 + \vec{m}_2 = (12, 11, 12, 1) + (8, 5, 14, 6) = (20, 16, 26, 7) \equiv (3, 16, 9, 7) \bmod 17$

$ $

Finally, we decode $\vec{m}_{1+2}$ as follows:

$\vec{v}_{1+2} = \hathat{W}^* \cdot \vec{m}_{1+2} = \begin{bmatrix}
1 & 9 & 13 & 15\\
1 & 8 & 13 & 2\\
1 & 2 & 4 & 8\\
1 & 15 & 4 & 9\\
\end{bmatrix} \cdot \begin{bmatrix}3\\16\\9\\7\end{bmatrix} = (12, 7, 8, 2) \bmod{17}$

$ $

This result matches the expected vector sum $\vec{v}_1 + \vec{v}_2$.

\subsubsection{Rotation Example}
\label{subsubsec:bfv-rotation-ex}

Also in the above example, we use the unsigned modulo representation (e.g., $[0, t-1], [0, q]$) when computing $\bmod t$ and $\bmod q$. However, the correctness of the result holds even if we use the centered modulo representation (e.g., $[-\dfrac{t}{2}, \dfrac{t}{2})$, $[-\dfrac{q}{2}, \dfrac{q}{2})$).

Suppose we have the following setup: 

$\mu = 16, n = \dfrac{\mu}{2} = 8, \text{ } t = 17, \text{ } q = 2^8 = 256, \text{ } n^{-1} = 15, \text{ } \Delta = 15$

$ $

$\mathcal{R}_{\langle 8, 17 \rangle} = \mathbb{Z}_{17}[X] / (X^8 + 1)$

$ $

The roots of $X^8 + 1 \pmod{17}$ are $X = \{3, 5, 6, 7, 10, 11, 12, 14\}$, as demonstrated as follows:

$3^8 \equiv 5^8 \equiv 6^8 \equiv 7^8 \equiv 10^8 \equiv 11^8 \equiv 12^8 \equiv 14^8 \equiv 16 \equiv -1 \bmod{17}$

$ $

Among $\{3, 5, 6, 7, 10, 11, 12, 14\}$ as the roots of $X^8 + 1$, let's choose $\omega = 3$ as the base root to construct the encoding matrix $\hathat{W}$ and the decoding matrix $\hathat{W}^*$ as follows: 

$\hathat{W} = \begin{bmatrix}
1 & 1 & 1 & 1 & 1 & 1 & 1 & 1\\
\omega^{J(3)} & \omega^{J(2)} & \omega^{J(1)} & \omega^{J(0)} & \omega^{J_*(3)} & \omega^{J_*(2)} & \omega^{J_*(1)} & \omega^{J_*(0)}\\
(\omega^{J(3)})^2 & (\omega^{J(2)})^2 & (\omega^{J(1)})^2 & (\omega^{J(0)})^2 & (\omega^{J_*(3)})^2 & (\omega^{J_*(2)})^2 & (\omega^{J_*(1)})^2 & (\omega^{J_*(0)})^2\\
(\omega^{J(3)})^3 & (\omega^{J(2)})^3 & (\omega^{J(1)})^3 & (\omega^{J(0)})^3 & (\omega^{J_*(3)})^3 & (\omega^{J_*(2)})^3 & (\omega^{J_*(1)})^3 & (\omega^{J_*(0)})^3\\
(\omega^{J(3)})^4 & (\omega^{J(2)})^4 & (\omega^{J(1)})^4 & (\omega^{J(0)})^4 & (\omega^{J_*(3)})^4 & (\omega^{J_*(2)})^4 & (\omega^{J_*(1)})^4 & (\omega^{J_*(0)})^4\\
(\omega^{J(3)})^5 & (\omega^{J(2)})^5 & (\omega^{J(1)})^5 & (\omega^{J(0)})^5 & (\omega^{J_*(3)})^5 & (\omega^{J_*(2)})^5 & (\omega^{J_*(1)})^5 & (\omega^{J_*(0)})^5\\
(\omega^{J(3)})^6 & (\omega^{J(2)})^6 & (\omega^{J(1)})^6 & (\omega^{J(0)})^6 & (\omega^{J_*(3)})^6 & (\omega^{J_*(2)})^6 & (\omega^{J_*(1)})^6 & (\omega^{J_*(0)})^6\\
(\omega^{J(3)})^7 & (\omega^{J(2)})^7 & (\omega^{J(1)})^7 & (\omega^{J(0)})^7 & (\omega^{J_*(3)})^7 & (\omega^{J_*(2)})^7 & (\omega^{J_*(1)})^7 & (\omega^{J_*(0)})^7\\
\end{bmatrix}$ 

$ \equiv \begin{bmatrix}
1 & 1 & 1 & 1 & 1 & 1 & 1 & 1\\
12&14&5&3&10&11&7&6\\
8&9&8&9&15&2&15&2\\
11&7&6&10&14&5&3&12\\
13&13&13&13&4&4&4&4\\
3&12&14&5&6&10&11&7\\
2&15&2&15&9&8&9&8\\
7&6&10&11&5&3&12&14\\
\end{bmatrix} \bmod{17}$ 

$ $ 

$\hathat{W}^* = \begin{bmatrix}
1 & \omega^{J(0)} & (\omega^{J(0)})^2 & (\omega^{J(0)})^3 & (\omega^{J(0)})^4 & (\omega^{J(0)})^5 & (\omega^{J(0)})^6 & (\omega^{J(0)})^7\\
1 & \omega^{J(1)} & (\omega^{J(1)})^2 & (\omega^{J(1)})^3 & (\omega^{J(1)})^4& (\omega^{J(1)})^5& (\omega^{J(1)})^6& (\omega^{J(1)})^7\\
1 & \omega^{J(2)} & (\omega^{J(2)})^2 & (\omega^{J(2)})^3 & (\omega^{J(2)})^4 & (\omega^{J(2)})^5 & (\omega^{J(2)})^6 & (\omega^{J(2)})^7\\
1 & \omega^{J(3)} & (\omega^{J(3)})^2 & (\omega^{J(3)})^3 & (\omega^{J(3)})^4& (\omega^{J(3)})^5& (\omega^{J(3)})^6& (\omega^{J(3)})^7\\
1 & \omega^{J_*(0)} & (\omega^{J_*(0)})^2 & (\omega^{J_*(0)})^3 & (\omega^{J_*(0)})^4 & (\omega^{J_*(0)})^5 & (\omega^{J_*(0)})^6 & (\omega^{J_*(0)})^7\\
1 & \omega^{J_*(1)} & (\omega^{J_*(1)})^2 & (\omega^{J_*(1)})^3 & (\omega^{J_*(1)})^4& (\omega^{J_*(1)})^5& (\omega^{J_*(1)})^6& (\omega^{J_*(1)})^7\\
1 & \omega^{J_*(2)} & (\omega^{J_*(2)})^2 & (\omega^{J_*(2)})^3 & (\omega^{J_*(2)})^4 & (\omega^{J_*(2)})^5 & (\omega^{J_*(2)})^6 & (\omega^{J_*(2)})^7\\
1 & \omega^{J_*(3)} & (\omega^{J_*(3)})^2 & (\omega^{J_*(3)})^3 & (\omega^{J_*(3)})^4& (\omega^{J_*(3)})^5& (\omega^{J_*(3)})^6& (\omega^{J_*(3)})^7\\
\end{bmatrix}$

$ \equiv \begin{bmatrix}
1&3&9&10&13&5&15&11\\
1&5&8&6&13&14&2&10\\
1&14&9&7&13&12&15&6\\
1&12&8&11&13&3&2&7\\
1&6&2&12&4&7&8&14\\
1&7&15&3&4&11&9&12\\
1&11&2&5&4&10&8&3\\
1&10&15&14&4&6&9&5\\
\end{bmatrix}  \bmod{17}$

$ $

$ $

Notice that Theorem~\ref*{subsec:vandermonde-euler-integer-ring} (in \autoref{subsec:vandermonde-euler-integer-ring}) is demonstrated as follows:

$\hathat{W}^* \cdot \hathat{W} = \begin{bmatrix}
1&3&9&10&13&5&15&11\\
1&5&8&6&13&14&2&10\\
1&14&9&7&13&12&15&6\\
1&12&8&11&13&3&2&7\\
1&6&2&12&4&7&8&14\\
1&7&15&3&4&11&9&12\\
1&11&2&5&4&10&8&3\\
1&10&15&14&4&6&9&5\\
\end{bmatrix} \cdot \begin{bmatrix}
1 & 1 & 1 & 1 & 1 & 1 & 1 & 1\\
12&14&5&3&10&11&7&6\\
8&9&8&9&15&2&15&2\\
11&7&6&10&14&5&3&12\\
13&13&13&13&4&4&4&4\\
3&12&14&5&6&10&11&7\\
2&15&2&15&9&8&9&8\\
7&6&10&11&5&3&12&14\\
\end{bmatrix}$

$ = \begin{bmatrix}
0 & 0 & 0 & 0 & 0 & 0 & 0 & 8\\
0 & 0 & 0 & 0 & 0 & 0 & 8 & 0\\
0 & 0 & 0 & 0 & 0 & 8 & 0 & 0\\
0 & 0 & 0 & 0 & 8 & 0 & 0 & 0\\
0 & 0 & 0 & 8 & 0 & 0 & 0 & 0\\
0 & 0 & 8 & 0 & 0 & 0 & 0 & 0\\
0 & 8 & 0 & 0 & 0 & 0 & 0 & 0\\
8 & 0 & 0 & 0 & 0 & 0 & 0 & 0\\
\end{bmatrix} = n I_n^{R} \pmod{17}$

$ $

Now suppose that we encode the following input vector (i.e., input vector slots) in $\mathbb{Z}_{17}$:

$\vec{v} = (1, 2, 3, 4, 5, 6, 7, 8)$

By rotating this vector by 3 positions to the left (i.e., the first-half slots and the second-half slots separately wrapping around within their own group), we get a new vector:

$\vec{v}_r = (4, 1, 2, 3, 8, 5, 6, 7) \bmod 17$

$ $

$\vec{v}$ is encoded as follows:

$ $

$\vec{m} = n^{-1} \hathat{W} \cdot I_n^R \cdot \vec{v} $

$= 15 \cdot  \begin{bmatrix}
1 & 1 & 1 & 1 & 1 & 1 & 1 & 1\\
12&14&5&3&10&11&7&6\\
8&9&8&9&15&2&15&2\\
11&7&6&10&14&5&3&12\\
13&13&13&13&4&4&4&4\\
3&12&14&5&6&10&11&7\\
2&15&2&15&9&8&9&8\\
7&6&10&11&5&3&12&14\\
\end{bmatrix} \cdot \begin{bmatrix}
0 & 0 & 0 & 0 & 0 & 0 & 0 & 1\\
0 & 0 & 0 & 0 & 0 & 0 & 1 & 0\\
0 & 0 & 0 & 0 & 0 & 1 & 0 & 0\\
0 & 0 & 0 & 0 & 1 & 0 & 0 & 0\\
0 & 0 & 0 & 1 & 0 & 0 & 0 & 0\\
0 & 0 & 1 & 0 & 0 & 0 & 0 & 0\\
0 & 1 & 0 & 0 & 0 & 0 & 0 & 0\\
1 & 0 & 0 & 0 & 0 & 0 & 0 & 0\\
\end{bmatrix} \cdot \begin{bmatrix}
1\\
2\\
3\\
4\\
5\\
6\\
7\\
8\\
\end{bmatrix}$

$ \equiv (13, 16, 10, 5, 9, 12, 7, 1)  \bmod 17$

$ $

$\Delta M(X) = 15\cdot(13 + 16X + 10X^2 + 5X^3 + 9X^4 + 12X^5 + 7X^6 + X^7)$

$ = 195 + 240X + 150X^2 + 75X^3 + 135X^4 + 180X^5 + 105X^6 + 15X^7 \pmod{q}$ \textcolor{red}{  $\rhd$ where $q = 256$}

$ $

This polynomial matches the value of $\Delta M(X^{J(3)})$ as follows:

$\Delta M(X^{J(3)}) = \Delta M(X^{13}) $

$= 195 + 180X - 150X^2 - 15X^3 + 135X^4 - 240X^5 - 105X^6 + 75X^7 \pmod{q}$

$ $

%Note that the coefficients of the scaled polynomials $\Delta M(X^{J(3)})$ are still within the range of the ciphertext domain $q=64$ (which must hold throughout all homomorphic computations to preserve correctness). 

Now, we decode $M(X^{J(3)})$ as follows: 

$\vec{m}_{J(3)} = \dfrac{\Delta \vec{m}_{J(3)}}{\Delta} = \dfrac{(195, 180, -150, -15, 135, -240, -105, 75)}{15} = (13, 12, -10, -1, 9, -16, -7, 5) \pmod{17}$

$ $

$\vec{v}_{J(3)} = \hathat{W}^* \cdot \vec{m}_{J(3)} = \begin{bmatrix}
1&3&9&10&13&5&15&11\\
1&5&8&6&13&14&2&10\\
1&14&9&7&13&12&15&6\\
1&12&8&11&13&3&2&7\\
1&6&2&12&4&7&8&14\\
1&7&15&3&4&11&9&12\\
1&11&2&5&4&10&8&3\\
1&10&15&14&4&6&9&5\\
\end{bmatrix} \cdot \begin{bmatrix}
13\\
12\\
-10\\
-1\\
9\\
-16\\
-7\\
5\end{bmatrix}$

$= (4, 1, 2, 3, 8, 5, 6, 7) \pmod{17}$

$= \vec{v}_r$

$ $

The decoded $\vec{v}_{J(3)}$ matches the expected rotated input vector $\vec{v}_r$.

$ $

In practice, we do not directly update $\Delta M(X)$ to $\Delta M(X^{J(3)})$, because we would not have access to the plaintext polynomial $M(X)$ unless we have the secret key $S(X)$. Therefore, we instead update $\textsf{ct}=\bm(A(X), B(X)\bm)$ to $\textsf{ct}^{\langle h=3 \rangle}=\bm(A(X^{J(3)}), B(X^{J(3)})\bm)$, which is equivalent to homomorphically rotating the encrypted input vector slots. Then, decrypting $\textsf{ct}^{\langle h=3 \rangle}$ and decoding it would output $\vec{v}_r$.

$ $

\para{Source Code:} Examples of BFV's batch encoding and homomorphic input vector rotation can be executed by running \href{https://github.com/fhetextbook/fhe-textbook/blob/main/source%20code/bfv.py}{\underline{this Python script}} (BFV addition) and \href{https://github.com/fhetextbook/fhetextbook.github.io/blob/main/source%20code/bfv_rotation_only.py}{\underline{this one}} (BFV rotation only).

\subsection{Application: Matrix Multiplication}
\label{subsec:bfv-matrix-multiplication}

BFV has no clean way to do a homomorphic dot product between two vectors (i.e., $\vec{v}_1 \cdot \vec{v}_2$), because the last step of a vector dot product requires summation of all slot-wise intermediate values (i.e., $v_{1,1}v_{2,1} + v_{1,2}v_{2,2} + \cdots + v_{1,n}v_{2,n}$). However, each slot in BFV's batch encoding is independent from each other, which cannot be simply added up across slots (i.e., input vector elements). Instead, we need $n$ copies of the multiplied ciphertexts and properly align their slots by many rotation operations before adding them up. Meanwhile, the homomorphic input vector slot rotation scheme can be effectively used when we homomorphically multiply a plaintext matrix with an encrypted vector. Remember that given a matrix $A$ and vector $\vec{x}$ (Definition~\ref*{subsec:matrix-arithmetic} in \autoref{subsec:matrix-arithmetic}):

$A = \begin{bmatrix}
 a_{\langle 0, 0\rangle} & a_{\langle 0, 1\rangle} & a_{\langle 0, 2\rangle} & \cdots & a_{\langle 0, n-1\rangle}\\
 a_{\langle 1, 0\rangle} & a_{\langle 1, 1\rangle} & a_{\langle 1, 2\rangle} & \cdots & a_{\langle 1, n-1\rangle} \\
 a_{\langle 2, 0\rangle} & a_{\langle 2, 1\rangle} & a_{\langle 2, 2\rangle} & \cdots & a_{\langle 2, n-1\rangle} \\
\vdots & \vdots & \vdots & \ddots & \vdots \\
 a_{\langle m-1, 0\rangle} & a_{\langle m-1, 1\rangle} & a_{\langle m-1, 2\rangle} & \cdots & a_{\langle m-1, n-1\rangle} \\
\end{bmatrix} = \begin{bmatrix} 
\vec{a}_{\langle 0, * \rangle} \\ 
\vec{a}_{\langle 1, * \rangle} \\ 
\vec{a}_{\langle 2, * \rangle} \\ 
\vdots\\
\vec{a}_{\langle m-1, * \rangle} \\ 
\end{bmatrix}, \text{ } \vec{x} = (x_0, x_1, \cdots, x_{n-1})$

$ $

The result of $A \cdot \vec{x}$ is an $m$-dimensional vector computed as:

$A \cdot \vec{x} = \Big(\vec{a}_{\langle 0, * \rangle} \cdot \vec{x}, \text{ } \vec{a}_{\langle 1, * \rangle} \cdot \vec{x}, \text{ }\cdots,\text{ } \vec{a}_{\langle m-1, * \rangle} \cdot \vec{x} \Big) = \left(\sum\limits_{i=0}^{n-1} a_{0, i} \cdot  x_i, \text{ } \sum\limits_{i=0}^{n-1}  a_{1, i} \cdot x_i, \cdots, \text{ } \sum\limits_{i=0}^{n-1} a_{m-1, i} \cdot x_i \right)$

$ $

Let's define $\rho(\vec{v}, h)$ as the rotation of $\vec{v}$ by $h$ positions to the left. And remember that the Hadamard dot product (Definition~\ref*{subsec:vector-arithmetic} in \autoref{subsec:vector-arithmetic}) is defined as slot-wise multiplication of two vectors: 

$\vec{a} \odot \vec{b} = (a_0 b_0, \text{ } a_1 b_1, \text{ } \cdots, \text{ } a_{n-1} b_{n-1})$

$ $

Let's define $n$ distinct diagonal vector $\vec{u}_i$ extracted from matrix $A$ as follows: 

$\vec{u}_i = \{a_{\langle j \bmod m, \text{ } (i + j)\bmod n \rangle}\}_{j = 0}^{n-1}$

$ $

Then, the original matrix-to-vector multiplication formula can be equivalently constructed as follows:

$A \cdot \vec{x} = \vec{u}_0 \odot \rho(\vec{x}, 0) \text{ } + \text{ } \vec{u}_1 \odot \rho(\vec{x}, 1) \text{ } + \text{ } \cdots \text{ } + \text{ } \vec{u}_{n-1} \odot \rho(\vec{x}, n-1)$

, whose computation result is equivalent to $A \cdot \vec{x}$. 
The above formula is compatible with homomorphic computation, because BFV supports Hadamard dot product between input vectors as a ciphertext-to-plaintext multiplication between their polynomial-encoded forms (\autoref{subsec:bfv-mult-plain}), and BFV also supports $\rho(\vec{v}, h)$ as homomorphic input vector slot rotation (\autoref{subsec:bfv-rotation}). After homomorphically computing the above formula, we can consider only the first $m$ (out of $n$) resulting input vector slots to store the result of $A \cdot \vec{x}$.

\subsection{Noise Bootstrapping}
\label{subsec:bfv-bootstrapping}

\noindent \textbf{- Reference 1:} 
\href{https://eprint.iacr.org/2022/1363.pdf}{Bootstrapping for BGV and BFV Revisited}~\cite{cryptoeprint:2022/1363}

\noindent \textbf{- Reference 2:} 
\href{https://eprint.iacr.org/2014/873.pdf}{Bootstrapping for HELib}~\cite{10.1007/s00145-020-09368-7}

\noindent \textbf{- Reference 3:} 
\href{https://arxiv.org/pdf/1906.02867}{A Note on Lower Digits Extraction Polynomial for Bootstrapping}~\cite{huo2019notelowerdigitsextraction}

\noindent \textbf{- Reference 4:} 
\href{https://eprint.iacr.org/2024/1587.pdf}{Fully Homomorphic Encryption for Cyclotomic Prime Moduli}~\cite{cryptoeprint:2024/1587}

%\noindent \textbf{- Reference 4:} 
%\href{https://s-space.snu.ac.kr/bitstream/10371/152893/1/000000155273.pdf}{Bootstrapping Methods for Homomorphic Encryption}

%\noindent \textbf{- Reference 5:} 
%\href{https://www.esat.kuleuven.be/cosic/publications/thesis-418.pdf}{Bootstrapping Algorithms for BGV and FV}

$ $

In BFV, continuous ciphertext-to-ciphertext multiplication increases the noise in a multiplicative manner, and once the noise overflows the message bits, then the message gets corrupted. Bootstrapping is a process of resetting the grown noise. 

\subsubsection{High-level Idea}
\label{subsubsec:bfv-bootstrapping-high-level}

In this subsection, we will assume the plaintext modulus $t = p$, a prime number. Although $t$ can be generalized as $t = p^r$ where $r \in \mathbb{Z}$ and $r \geq 1$ (Summary~\ref*{subsec:bfv-enc-dec} in \autoref{subsec:bfv-enc-dec}), we will explain BFV's bootstrapping assuming $t=p$ for simplicity, and then generalize $t$ as $t = p^r$ in the end. 

The core idea of the BFV bootstrapping is to homomorphically evaluate a special polynomial $G_\varepsilon(x)$, a digit extraction polynomial modulo $p^\varepsilon$ (for some positive integer $\varepsilon$), where the input to $G_\varepsilon(x)$ is a noisy plaintext value modulo $p^\varepsilon$ and the output is a noise-free plaintext value modulo $p^\varepsilon$, shifted to the right by 1 base-$p$ digit. Here, where the noise located at the least significant digits in a base-$p$ (prime) representation is zeroed out and shifted right. For example, $G_\varepsilon(3p^3 + 4p^2 + 6p + 2) = 3p^2 + 4p^1 + 6 \bmod p^\varepsilon$. Given that the noise resides in the least significant $\varepsilon-1$ digits in base-$p$ representation, we can homomorphically and recursively evaluate $G_\varepsilon(x)$ total $\varepsilon-1$ times in a row, which zeroes out and removes the least significant (base-$p$) $\varepsilon-1$ digits of input $x$. To homomorphically evaluate the noisy plaintext through $G_\varepsilon(x) \bmod p^\varepsilon$, we need to first switch the plaintext modulus from $t$ to $p^\varepsilon$, where $q \gg p^{\varepsilon} > p = t$. The larger $\varepsilon$ is, the more likely it is that the noise gets successfully zeroed out; however, the computation overhead becomes larger. If $\varepsilon$ is small, the computation gets faster, but the digit-wise distance between the noise and the plaintext decreases, potentially corrupting the plaintext during bootstrapping, because removing the most significant noise digit may also remove the least significant plaintext digit. Therefore, $\varepsilon$ should be chosen carefully. 

$ $

The technical details of the BFV bootstrapping procedure are as follows. Suppose we have an RLWE ciphertext $(A, B)  = \textsf{RLWE}_{S, \sigma}\bm(\Delta M\bm) \bmod q$, where $A\cdot S + B = \Delta M + E$, \text{ } $\Delta = \left\lfloor\dfrac{q}{t}\right\rfloor$, and $t = p$ (i.e., the plaintext modulus is a prime).

\begin{enumerate}
\item \textbf{Modulus Switch (\boldmath$q \rightarrow p^\varepsilon$):} Scale down the ciphertext modulus from $(A, B) \bmod q$ to $\left(\left\lceil \dfrac{p^\varepsilon}{q}\cdot A\right\rfloor, \left\lceil \dfrac{p^\varepsilon}{q}\cdot B\right\rfloor\right) = (A', B') \bmod p^\varepsilon$, where $p^\varepsilon \ll q$. The purpose of this modulus switch is to change the plaintext modulus to $p^\varepsilon$, which is required to use the digit extraction polynomial $G_\varepsilon(x)$ (because we need to represent the input to $G_\varepsilon(x)$ as a base-$p$ number in order to interpret it as a $\bmod p^\varepsilon$ value). Notice that $\textsf{RLWE}_{S, \sigma}^{-1}\bm(\textsf{ct} = (A', B') \bm) = p^{\varepsilon-1}M + E'$, where $E' \approx \dfrac{p^\varepsilon}{q}\cdot E  + \left(\left\lfloor\dfrac{q}{p}\right\rfloor\cdot\dfrac{p^\varepsilon}{q} - p^{\varepsilon-1}\right)\cdot M$ \textcolor{red}{ $\rhd$ the modulus switch error of $E \rightarrow E'$ plus the modulus switch error of the plaintext's scaling factor $\left\lfloor\dfrac{q}{t}\right\rfloor \rightarrow p^{\varepsilon-1}$}

$ $

\item \textbf{Homomorphic Decryption:} Suppose we have the \textit{bootstrapping key} $\textsf{RLWE}_{\Delta' S, \sigma}(S) \bmod q$, which is the secret key $S$ encrypted by $S$ (itself) modulo $q$ with the plaintext scaling factor $\Delta'$. Using this \textit{encrypted} secret key, we \textit{homomorphically} decrypt $(A', B')$ as follows:

$A' \cdot \textsf{RLWE}_{S, \sigma}\bm(\Delta' S\bm) + B'$ \textcolor{red}{ $\rhd$ where $\textsf{RLWE}_{S, \sigma}(\Delta' S)$ is a modulo-$q$ ciphertext that encrypts a modulo-$p^\varepsilon$ plaintext $S$ whose scaling factor $\Delta' = \dfrac{q}{p^\varepsilon}$ }

$ $

$= \textsf{RLWE}_{S, \sigma}\bm(\Delta' (A' \cdot S)\bm) + B'  \bmod q$

$= \textsf{RLWE}_{S, \sigma}(\Delta' \cdot \bm(A'\cdot S + B')\bm) \bmod q$

$ = \textsf{RLWE}_{S, \sigma}\bm(\Delta' \cdot (p^{\varepsilon-1} M  + E' + Kp^\varepsilon)\bm) \bmod q$ \textcolor{red}{ $\rhd$ $K$ is some integer polynomials to represent the coefficient values that wrap around $p^\varepsilon$}

%$ $

%$ = \textsf{RLWE}_{S, \sigma}\bm\left(\dfrac{q}{p} M  + \dfrac{q}{p^\varepsilon} E' + Kq\bm\right) \bmod q$

%$ = \textsf{RLWE}_{S, \sigma}\bm\left(\dfrac{q}{p} M  + E'' + Kq\bm\right) \bmod q$ \textcolor{red}{ $\rhd$ where $E'' = \dfrac{q}{p^\varepsilon} E'$}

$ $

Let's see how we derived the above relation. Suppose we compute $A'\cdot S + B' \bmod p^\varepsilon$, whose output will be $p^{\varepsilon-1}M + E'$. Now, instead of using the plaintext secret key $S$, we use an encrypted secret key $\textsf{RLWE}_{S, \sigma}(\Delta' S)$, where $S$ is a plaintext modulo $p^\varepsilon$, its scaling factor $\Delta' = \left\lfloor\dfrac{q}{p^\varepsilon}\right\rfloor$, and the ciphertext encrypting $S$ is in modulo $q$. Then, the result of homomorphically computing $A'\cdot \textsf{RLWE}_{S, \sigma}(\Delta' S) + B'$ will be an encryption of $p^{\varepsilon-1}M + E' + Kp^\varepsilon$, where $Kp^\varepsilon$ stands for the wrapping-around coefficient values of the multiples of $p^\varepsilon$. %Finally, we replace $A'\cdot \textsf{RLWE}_{S, \sigma}(S)$ with $\langle \textsf{Decomp}^{\beta, l}(A'), \textsf{RLev}_{S, \sigma}^{\beta, l}(S)\rangle$, which is computationally equivalent to $A' \cdot \textsf{RLWE}_{S, \sigma}(S)$ but the noise becomes much smaller (we will explain why later in \autoref{subsubsec:bfv-bootstrapping-homomorphic-decryption}).

Notice that we did not reduce $Kp^\varepsilon$ by modulo $p^\varepsilon$ during homomorphic decryption (without modulo-$q$ reduction), because such a homomorphic modulo reduction is not directly doable. Instead, we will handle $Kp^\varepsilon$ in the later digit extraction step.

For simplicity, we will denote $Z = p^{\varepsilon-1} M + E' \bmod p^\varepsilon$.

$ $

\item \textbf{\textsf{CoeffToSlot}:} Move the (encrypted) polynomial $Z$'s coefficients $z_0, z_i, \cdots, z_{n-1}$ to the input vector slots of an RLWE ciphertext. This is done by computing: 

$\textsf{RLWE}_{S, \sigma}(Z) \cdot n^{-1}\cdot \hathat{W}\cdot I_R^n$

, where $n^{-1}\cdot \hathat{W}\cdot I_R^n$ is the batch encoding matrix that converts input vector slot values into polynomial coefficients (Summary~\ref*{subsubsec:bfv-rotation-summary} in \autoref{subsubsec:bfv-rotation-summary}). 

$ $

\item \textbf{Digit Extraction:} We design a polynomial $G_\varepsilon(x)$ (i.e., a digit extraction polynomial) that zeros out the least significant base-$p$  digit(s) modulo $p^{\varepsilon}$. The digit extraction procedure recursively applies $G_{2} \circ G_{3} \circ \cdots \circ G_{\varepsilon-1} \circ G_\varepsilon(x)$ to the input $x$ to eliminate the least significant (base-$p$) $\varepsilon-2$ digits of input $x$ and shifts them to the right. Regarding the input $x$, we assume the noise resides in the least significant (base-$p$) $\varepsilon-2$ digits, and the message $m$ resides in the higher base-$p$ digits modulo $p^{\varepsilon}$. Therefore, digit extraction has the effect of zeroing out and deleting (i.e., shifting to the right) the least significant $\varepsilon-1$ digits of the base-$p$ representation of $p^{\varepsilon-1}m + e$. This digit extraction is performed homomorphically. Throughout the digit extraction, the scaled plaintext message with noise stored at the input vector slots of the ciphertext gets updated from $p^{\varepsilon-1} M  + E' + Kp^\varepsilon$ to $M  + K'p$. During digit extraction, we also use a special method called scaling factor re-interpretation, which conceptually increases the scaling factor $\Delta'=\left\lfloor\dfrac{q}{p^\varepsilon}\right\rfloor$ over each round of digit extraction to $\left\lfloor\dfrac{q}{p^\varepsilon-1}\right\rfloor, \left\lfloor\dfrac{q}{p^\varepsilon-2}\right\rfloor, \ldots, \left\lfloor\dfrac{q}{p}\right\rfloor$, which equivalently has the conceptual effect of dividing the scaled message and noise stored in the plaintext slots by $p$ (i.e., shift to the right by 1 base-$p$ digit) as follows: 

\textbf{Input:} $\left\lfloor\dfrac{q}{p^{\varepsilon}}\right\rfloor \cdot \left (p^{\varepsilon-1}M + E' + K p^{\varepsilon}\right )$

\textbf{Round 1:} $\left\lfloor\dfrac{q}{p^{\varepsilon - 1}}\right\rfloor \cdot \left (p^{\varepsilon-2}M + \left\lfloor\dfrac{E'}{p}\right\rfloor + K p^{\varepsilon-1}\right )$

\textbf{Round 2:} $\left\lfloor\dfrac{q}{p^{\varepsilon - 2}}\right\rfloor \cdot \left (p^{\varepsilon-3}M + \left\lfloor\dfrac{E'}{p^2}\right\rfloor + K p^{\varepsilon-2}\right )$

$\vdots$

\textbf{Round $\varepsilon-1$:} $\left\lfloor\dfrac{q}{p}\right\rfloor \cdot (M + K p)$

$ $

Importantly, the scaling factor re-interpretation does not require any actual computation; we only change the way of interpreting the ciphertext. We will cover this in more detail. later.

$ $

\item \textbf{\textsf{SlotToCoeff}:} Homomorphically move each input vector slot's value back to the (encrypted) polynomial coefficient position. This is done by homomorphically multiplying the decoding matrix $\hathat{W}^*$ to the ciphertext (Summary~\ref*{subsubsec:bfv-rotation-summary} in \autoref{subsubsec:bfv-rotation-summary}). The final output of this step is $ = \textsf{RLWE}_{S, \sigma}\left(\left\lfloor\dfrac{q}{p}\right\rfloor M  + E^{\langle b \rangle}\right)$, where $E^{\langle \text{b} \rangle}$ is a new small noise term generated during the homomorphic operation of \textsf{CoeffToSlot}, digit extraction, and \textsf{SlotToCoeff}. The size of $E^{\langle \text{b} \rangle}$ is fixed and smaller than $E$ and $E'$. 

\end{enumerate}

Next, we will explain each step in more detail. 

\subsubsection{Modulus Switch}
\label{subsubsec:bfv-bootstrapping-modulus-switch}

The first step of the BFV bootstrapping is to do a modulus switch from $q$ to some prime power modulus $p^\varepsilon$ where $p^\varepsilon \ll q$. Before the bootstrapping, suppose the encrypted plaintext with noise is $\Delta M + E \bmod q$. Then, after the modulus switch from $q \rightarrow p^\varepsilon$, the plaintext would scale down to $p^{\varepsilon-1}M + E' \bmod p^\varepsilon$, where $E'$ roughly contains $\left\lceil\dfrac{p^\varepsilon}{q} E\right\rfloor$ plus the modulus switching noise of the plaintext's scaling factor $\Delta \rightarrow p^{\varepsilon-1}$. The goal of the BFV bootstrapping is to zero out this noise $E'$.  

\subsubsection{Homomorphic Decryption}
\label{subsubsec:bfv-bootstrapping-homomorphic-decryption}

Let's denote the modulus-switched noisy plaintext as $Z = p^{\varepsilon-1}M + E' \bmod p^\varepsilon$. We further denote polynomial $Z$'s each degree term's coefficient $z_i$ as base-$p$ number as follows: 

$z_i = z_{i, \varepsilon-1}p^{\varepsilon-1} + z_{i, \varepsilon-2}p^{\varepsilon-2} + \cdots + z_{i,1}p + z_{i,0} \bmod p^\varepsilon$

$ $

Then, $z_i \bmod p^\varepsilon$ is a base-$p$ number comprising $\varepsilon$ digits: $\{z_{i,\varepsilon-1}, z_{i,\varepsilon-2}, \cdots, z_{i,0}\}$

We assume that the highest base-$p$ digit index for the noise is $\varepsilon-2$, which is equivalent to the noise budget, and the pure plaintext portion solely resides at the base-$p$ digit index $\varepsilon-1$ (i.e., the most significant base-$p$ digit in modulo $p^\varepsilon$). Given this assumption, we extract the noise-free plaintext by computing the following:

$\left\lceil \dfrac{z_i}{p^{\varepsilon-1}} \right\rfloor \bmod p = z_{i,\varepsilon-1}$ \textcolor{red}{ $\rhd$ where the noise is assumed to be smaller than $\dfrac{p^{\varepsilon-1}}{2}$}

The above formula is equivalent to shifting the base-$p$ number $z_i$ by $\varepsilon-1$ digits to the right (and rounding the decimal value). However, remember that we don't have direct access to polynomial $Z = p^{\varepsilon-1}M + E' \bmod p^\varepsilon$ unless we have the secret key $S$ to decrypt the ciphertext storing the plaintext. Instead, we can only derive $Z$ as an encrypted form. Specifically, we can \textit{homomorphically} decrypt the modulus-switched ciphertext $(A', B')$ by using the \textit{encrypted} secret key $\textsf{RLWE}_{S, \sigma}(\Delta' S)$ as a \textit{bootstrapping key}. For this ciphertext $\textsf{RLWE}_{S, \sigma}(\Delta' S)$, the plaintext modulus is $p^\varepsilon$, the plaintext scaling factor is $\Delta' = \dfrac{q}{p^\varepsilon}$, and the ciphertext modulus is $q$. With this encrypted secret key $S$, we homomorphically decrypt the encrypted $Z$ as follows:

$\left(\left\lceil \dfrac{p^\varepsilon}{q}\cdot A\right\rfloor, \left\lceil \dfrac{p^\varepsilon}{q}\cdot B\right\rfloor\right) = (A', B') \bmod p^\varepsilon$ 

$ $

$A' \cdot \textsf{RLWE}_{S, \sigma}\bm(\Delta' S\bm) + B' \bmod q$

$= \textsf{RLWE}_{S, \sigma}\bm(\Delta' (A' \cdot S)\bm) + B' \bmod q$

$= \textsf{RLWE}_{S, \sigma}(\Delta' \cdot \bm(A'\cdot S + B')\bm) \bmod q$

$ = \textsf{RLWE}_{S, \sigma}\bm(\Delta' \cdot (p^{\varepsilon-1} M  + E' + Kp^\varepsilon)\bm) \bmod q$ \textcolor{red}{ $\rhd$ $K$ is some integer polynomials to represent the coefficient values that wrap around modulo $p^\varepsilon$ as multiples of $p^\varepsilon$}

$ $

$ = \textsf{RLWE}_{S, \sigma}\bm(\Delta' Z\bm) \bmod q$

$ $

During this homomorphic decryption, we did not reduce the plaintext result by modulo $p^\varepsilon$, because the homomorphic decryption is a ciphertext-to-plaintext multiplication and addition done in the ciphertext modulus $q$ (not $p^{\varepsilon}$) by using $A'$ and $B'$ as plaintexts (with the plaintext modulus $p^e$) and $\textsf{RLWE}_{S, \sigma}(\Delta' S)$ as a ciphertext (with the ciphertext modulus $q$). This is why the wrapping term $Kp^\varepsilon$ is preserved in the plaintext after the homomorphic decryption-- we will handle this term at the later stage of bootstrapping. Also, notice that the computation of $A'\cdot \textsf{RLWE}_{S, \sigma}(\Delta' S)$ would not generate much noise. This is because $A'$ is a plaintext modulo $p^\varepsilon$, and thus the new noise generated by ciphertext-to-plaintext multiplication is $A' \cdot E_s$ (where $E_s$ is the encryption noise of $\textsf{RLWE}_{S, \sigma}(\Delta' S)$). Since the ciphertext modulus $q \gg p^\varepsilon$, $q \gg A' \cdot E_s$. 

Once we have derived $\textsf{RLWE}_{S, \sigma}\bm(\Delta' Z\bm)$, our next step is to remove the noise in the lower $\varepsilon-1$ digits (in terms of base-$p$ representation) of each $z_i$ for $0 \leq i \leq n - 1$. This is equivalent to transforming noisy $\textsf{RLWE}_{S, \sigma}(\Delta' Z) = \textsf{RLWE}_{S, \sigma}\bm(\Delta' \cdot (p^{\varepsilon-1} M  + E' + Kp^\varepsilon)\bm)$ into noise-free $\textsf{RLWE}_{S, \sigma}\bm (\Delta M)\bm )$ where $\Delta = \dfrac{q}{p}$. BFV's solution to do this is to design a $p$-degree polynomial function which computes the same logical result as $\left\lceil \dfrac{z_i}{p^{\varepsilon-1}} \right\rfloor \bmod p$. We will later explain how to design this polynomial by using the digit extraction polynomial $G_\varepsilon(x)$ (\autoref{subsubsec:bfv-bootstrapping-digit-extraction}).

However, in order to \textit{homomorphically} evaluate this polynomial at each coefficient $z_i$ given the ciphertext $\textsf{RLWE}_{S, \sigma}(\Delta' Z)$, we need to move polynomial $Z$'s each coefficient $z_i$ to the input vector slots of an RLWE ciphertext. This is because BFV supports homomorphic batched $(+, \cdot)$ operations based on input vector slots of ciphertexts as operands. Therefore, we need to evaluate the noise-removing polynomial $G_\varepsilon(x)$ based on the values stored in the input vector slots of a ciphertext. 

In the next sub-section, we will explain the \textsf{CoeffToSlot} procedure, a process of moving polynomial coefficients into input vector slots of a ciphertext \textit{homomorphically}. 

\subsubsection{\textsf{CoeffToSlot} and \textsf{SlotToCoeff}}
\label{subsubsec:bfv-bootstrapping-coefftoslot}

The goal of the \textsf{CoeffToSlot} step is to homomorphically move polynomial $Z$'s coefficients $z_i$ to input vector slots.

In Summary~\ref*{subsubsec:bfv-rotation-summary} (in \autoref{subsubsec:bfv-rotation-summary}), we learned that the encoding formula for converting a vector of input slots $\vec{v}$ into a vector of polynomial coefficients $\vec{m}$ is: $\vec{m} = n^{-1}\cdot\hathat{W} \cdot I_n^R \cdot \vec{v}$, where $\hathat{W}$ is a basis of the $n$-dimensional vector space crafted as follows: 

$\hathat{W} = \begin{bmatrix}
1 & 1 & \cdots & 1 & 1 & 1 & \cdots & 1\\
(\omega^{J(\frac{n}{2} - 1)}) & (\omega^{J(\frac{n}{2} - 2)}) & \cdots & (\omega^{J(0)}) & (\omega^{J_*(\frac{n}{2} - 1)}) & (\omega^{J_*(\frac{n}{2} - 2)}) & \cdots & (\omega^{J_*(0)})\\
(\omega^{J(\frac{n}{2} - 1)})^2 & (\omega^{J(\frac{n}{2} - 2)})^2 & \cdots & (\omega^{J(0)})^2 & (\omega^{J_*(\frac{n}{2} - 1)})^2 & (\omega^{J_*(\frac{n}{2} - 2)})^2 & \cdots & (\omega^{J_*(0)})^2 \\
\vdots & \vdots & \ddots & \vdots & \vdots & \vdots & \ddots & \vdots \\
(\omega^{J(\frac{n}{2} - 1)})^{n-1} & (\omega^{J(\frac{n}{2} - 2)})^{n-1} & \cdots & (\omega^{J(0)})^{n-1} & (\omega^{J_*(\frac{n}{2} - 1)})^{n-1} & (\omega^{J_*(\frac{n}{2} - 2)})^{n-1} & \cdots  & (\omega^{J_*(0)})^{n-1}
\end{bmatrix}$

$ $

\textcolor{red}{ $\rhd$ where the rotation helper function $J(h) = 5^h \bmod 2n$}

$ $

Therefore, given the input ciphertext $\textsf{ct} = \textsf{RLWE}_{S, \sigma}\bm(\Delta' Z\bm) \bmod q$, we can understand its input vector slots as storing some values such that multiplying $n^{-1}\cdot\hathat{W} \cdot I_n^R$ to each of them turns them into a coefficient $z_i$ of polynomial $Z$. This implies that if we \textit{homomorphically} multiply $n^{-1}\cdot\hathat{W} \cdot I_n^R$ to the input vector slots of $\textsf{RLWE}_{S, \sigma}\bm(\Delta' Z\bm)$, then the resulting ciphertext's $n$-slot input vector slots will contain the $n$ coefficients of $Z$, which is equivalent to moving the coefficients of $Z$ to the input vector slots. Therefore, the \textsf{CoeffToSlot} step is equivalent to homomorphically computing $n^{-1}\cdot\hathat{W} \cdot I_n^R \cdot \textsf{RLWE}_{S, \sigma}\bm(\Delta' Z\bm)$. We can homomorphically compute matrix-vector multiplication by using the technique explained in \autoref{subsec:bfv-matrix-multiplication}.

After the \textsf{CoeffToSlot} step, we can homomorphically eliminate the noise in the lower (base-$p$) $\varepsilon-1$ digits of each $z_i$ by homomorphically evaluating the noise-removing polynomial (to be explained in the next subsection).  

After we get noise-free coefficients of $Z$, we need to move them back from the input vector slots to their original coefficient positions. This step is called \textsf{SlotToCoeff}, which is an exact inverse procedure of \textsf{CoeffToSlot}. We also learned in Summary~\ref*{subsubsec:bfv-rotation-summary} (in \autoref{subsubsec:bfv-rotation-summary}) that the inverse matrix of $n^{-1}\cdot\hathat{W} \cdot I_n^R$ is $\hathat{W}^*$, where: 

$\hathat{W}^* = \begin{bmatrix}
1 & (\omega^{J(0)}) & (\omega^{J(0)})^2 & \cdots & (\omega^{J(0)})^{n-1}\\
1 & (\omega^{J(1)}) & (\omega^{J(1)})^2 & \cdots & (\omega^{J(1)})^{n-1}\\
1 & (\omega^{J(2)}) & (\omega^{J(2)})^2 & \cdots & (\omega^{J(2)})^{n-1}\\
\vdots & \vdots & \vdots & \ddots & \vdots \\
1 & (\omega^{J(\frac{n}{2}-1)}) & (\omega^{J(\frac{n}{2}-1)})^2 & \cdots & (\omega^{J(\frac{n}{2}-1)})^{n-1}\\
1 & (\omega^{J_*(0)}) & (\omega^{J_*(0)})^2 & \cdots & (\omega^{J_*(0)})^{n-1}\\
1 & (\omega^{J_*(1)}) & (\omega^{J_*(1)})^2 & \cdots & (\omega^{J_*(1)})^{n-1}\\
1 & (\omega^{J_*(2)}) & (\omega^{J_*(2)})^2 & \cdots & (\omega^{J_*(2)})^{n-1}\\
\vdots & \vdots & \vdots & \ddots & \vdots \\
1 & (\omega^{J_*(\frac{n}{2}-1)}) & (\omega^{J_*(\frac{n}{2}-1)})^2 & \cdots & (\omega^{J_*(\frac{n}{2}-1)})^{n-1}\\
\end{bmatrix}$

Therefore, the \textsf{SlotToCoeff} step is equivalent to homomorphically multiplying $\hathat{W}^*$ to the output of the noise-eliminating polynomial evaluation.  

In the next subsection, we will learn how to design the core algorithm of BFV, the noise elimination polynomial, based on the digit extraction polynomial $G_\varepsilon(x)$.

\subsubsection{Digit Extraction}
\label{subsubsec:bfv-bootstrapping-digit-extraction}

Remember that we defined polynomial $Z$ as the scaled noisy plaintext: $Z = p^{\varepsilon-1}M + E' + Kp^\varepsilon \equiv p^{\varepsilon-1}M + E' \bmod p^\varepsilon$, and each $z_i$ is the $i$-th coefficient of $Z$ (where $0 \leq i \leq n -1$). The goal of the digit extraction step is to homomorphically zero out and delete (i.e., shift to the right) the lower (base-$p$) $\varepsilon-1$  digits of each $z_i$, where the noise resides.%, and shift the resulting number by 1 digit to the right.  

First, we always think of $z_i$ as a base-$p$ representation (since this is a modulo-$p^\varepsilon$ value) as follows:

$z_i = z_{i, \varepsilon-1}p^{\varepsilon-1} + z_{i, \varepsilon-2}p^{\varepsilon-2} + \cdots + z_{i, 2}p^2 + z_{i, 1}p + z_{i, 0} \bmod p^\varepsilon$

$ $

Next, we define a new notation that denotes $z_i$ in a different way as follows: 

$z_i = d_0 + d_*p^{\varepsilon'}$

, where $d_0 \in \mathbb{Z}_p$, and $d_* \in \mathbb{Z}$, and $\varepsilon'$ is $z_i$'s least significant base-$p$ digit index whose value is non-zero after digit index 0. Therefore, each $z_i \in \mathbb{Z}_{p^\varepsilon}$ is mapped to a unique set of $(d_0, d_*, \varepsilon')$.

Next, we define a \textit{lifting} polynomial $F_{\varepsilon'}$ in terms of $z_i$ and its associated $(d_0, d_*, \varepsilon')$ as follows:

$F_{\varepsilon'}(z_i) \equiv d_0 \bmod p^{\varepsilon'+1}$

%, where $F_{\varepsilon'}(z_i)$ satisfies the above relation for all $\varepsilon'$ values such that $1 \leq \varepsilon' \leq e - 1$. 
Verbally speaking, $F_{\varepsilon'}(z_i)$ processes $z_i$ in such a way that it keeps $z_{i,0}$ (i.e., $z_{i}$'s value at the base-$p$ digit index 0) the same as before, then finds the next least significant base-$p$ digit whose value is non-zero (whose digit index is denoted as $\varepsilon'$) and zeros it, during which the subsequent higher significant base-$p$ digits may be updated to arbitrary values (i.e., the function doesn't care about those values whose base-$p$ digit index is higher than $\varepsilon'$ because they fall outside the modulo $p^{\varepsilon' + 1}$ range). 

We will show an example of how $z_i$ is updated if it is evaluated by the $F_{\varepsilon'}$ function recursively a total of $\varepsilon-1$ times in a row as follows:

$\underbrace{F_{\varepsilon-1} \cdots F_{3} \circ F_{2} \circ  F_{1}}_{\varepsilon-1 \text{ times}}(z_i)$

$ $

\para{1st Recursion:} $F_{1}(z_i) = c_{i,\varepsilon-1}p^{\varepsilon-1} + c_{i,\varepsilon-2}p^{\varepsilon-2} + \cdots + c_{i, 2}p^2 + 0p + z_{i, 0} \bmod p^\varepsilon$ 

\textcolor{red}{ $\rhd$ $F_{1}(z_i) \equiv z_{i,0} \bmod p^2$}

$ $

\para{2nd Recursion:} $F_{2} \circ F_{1}(z_i) = c'_{i,\varepsilon-1}p^{\varepsilon-1} + c'_{i,\varepsilon-2}p^{\varepsilon-2} + \cdots + 0p^2 + 0p + z_{i, 0} \bmod p^\varepsilon$  

\textcolor{red}{ $\rhd$ $F_{1} \circ F_{2}(z_i) \equiv z_{i,0} \bmod p^3$} 

$ $

\para{3rd Recursion:} $F_{3} \circ F_{2} \circ F_{1}(z_i) = c''_{i,\varepsilon-1}p^{\varepsilon-1} + c''_{i,\varepsilon-2}p^{\varepsilon-2} + \cdots + 0p^3 + 0p^2 + 0p + z_{i, 0} \bmod p^\varepsilon$  

\textcolor{red}{ $\rhd$ $F_{3} \circ F_{2} \circ F_{1}(z_i) \equiv z_{i,0} \bmod p^4$} 

$\vdots$

\para{$\bm{(\varepsilon-1)}$-th Recursion:} $\underbrace{F_{\varepsilon-1}  \circ \cdots \circ F_{3} \circ F_{2} \circ F_{1}}_{\varepsilon-1 \text{ times}}(z_i) = 0p^{\varepsilon-1} + 0p^{\varepsilon-2} + \cdots + 0p^2 + 0p + z_{i, 0} \bmod p^\varepsilon$ 

\textcolor{red}{ $\rhd$ $F_{\varepsilon-1} \cdots F_{3} \circ F_{2} \circ F_{1}(z_i) \equiv z_{i,0} \bmod p^\varepsilon$} 

$ $

In the above recursive computation, notice that the order of using function $F_{\varepsilon'}$ is specifically $F_{1} \rightarrow F_{2} \rightarrow F_{3} \rightarrow \cdots \rightarrow F_{\varepsilon -1}$. We choose this specific order because we assume that for the initial input $z_i$, we do not know its associated $\varepsilon'$ value (i.e., the least significant base-$p$ digit index whose value is non-zero after digit index 0). If we choose the order $F_{1} \rightarrow F_{2} \rightarrow F_{3} \rightarrow \cdots \rightarrow F_{\varepsilon -1}$, then regardless of the value of $z_i$, we obtain the universal guaranty that the final output will be $z_{i,0} \bmod p^\varepsilon$ (i.e., the value of the base-$p$ digit index 0).

$ $

Now, we define the digit extraction function $G_{\varepsilon, v}(z_i)$ as follows:

$G_{\varepsilon}(z_i) = \left\lfloor\dfrac{z_i}{p}\right\rfloor_p = (z_i - (\underbrace{F_{\varepsilon-1}  \circ F_{\varepsilon-2}  \circ F_{\varepsilon-3}  \cdots \circ F_{3} \cdots \circ F_{2}  \cdots \circ F_{1}}_{\varepsilon - 1 \text{ times}}(z_i)) \cdot |p^{-1}|_{q} $

, where $\left\lfloor\right\rfloor_p$ denotes division by $p$ and rounding down to the nearest multiple of $p$. Verbally speaking, $G_{\varepsilon, v}(z_i)$ is equivalent to zeroing out the least significant base-$p$ digit of $z_i$ and then shifting to the right by 1 base-$p$ digit. The shifting is done by inverse $p$ multiplication (i.e., $|p^{-1}|_{q}$). Notice that the last base-$p$ digit of $z_i - (F_{\varepsilon-1}  \circ F_{\varepsilon-2}  \circ F_{\varepsilon-3}  \cdots \circ F_{1}(z_i))$ is 0, which is exactly divisible by $p$. Thus, multiplying by the inverse $p$ has the effect of exact division (i.e., shifting the whole base-$p$ representation by 1 digit to the right). The reason why the inverse $p$ is in modulo $q$ is that we are currently under the BFV ciphertext relation: $\textsf{CoeffToSlot}\bm((A', B')\bm) = (A^{\langle c\rightarrow s \rangle}, B^{\langle c\rightarrow s \rangle}) \bmod q$, whose plaintext slots store the coefficients of $\Delta' \cdot (p^{\varepsilon - 1}M + E' + K^{\langle 1 \rangle}p^{\varepsilon}) \bmod q$. Upon homomorphically computing $(z_i - (F_{\varepsilon-1}  \circ F_{\varepsilon-2}  \circ F_{\varepsilon-3}  \cdots \circ F_{3} \cdots \circ F_{2}  \cdots \circ F_{1})$ as part of the 1st round of digit extraction, the ciphertext is updated to $(A^{\langle g_1 \rangle}, B^{\langle g_1 \rangle}) \bmod q$, whose plaintext slots store the coefficients of $\Delta' \cdot (p^{\varepsilon - 1}M + \lfloor E' \rfloor_p + K^{\langle 1 \rangle}p^{\varepsilon}) \bmod q$, where $\lfloor E' \rfloor_p$ is equivalent to rounding $E'$ down to the nearest multiple of $p$. At this point (i.e., just before inverse-$p$ multiplication in the first round), the ciphertext holds the following relation:

$A^{\langle g_1 \rangle}\cdot S  + B^{\langle g_1 \rangle} = \Delta' \cdot (p^{\varepsilon - 1}M + \left\lfloor{E'}\right\rfloor_p + K^{\langle 1 \rangle}p^{\varepsilon}) + E^{\langle g_1 \rangle} \bmod q$

$ $

, where $E^{\langle g_1 \rangle}$ is the combined noise of the \textsf{SlotToCoeff} step and the first part of the 1st round of digit extraction. We could consider multiplying $|p^{-1}|_q$ to both sides of the relation to scale down $p^{\varepsilon - 1}M, \left\lfloor{E'}\right\rfloor_p,$ and $K^{\langle 1 \rangle}p^{\varepsilon}$ by $p$. However, this causes a noise explosion problem, because this scaling will be also applied to the $E^{\langle g_1 \rangle}$ term, and $|p^{-1}|_qE^{\langle g_1 \rangle}$ is a huge value. To selectively apply the inverse-$p$ multiplication only to those terms of interest, we use the scaling factor re-interpretation technique. 

$ $

\para{Scaling Factor Re-interpretation:} Although we previously explained that  $G_{\varepsilon}$ involves the multiplication by $|p^{-1}|_{q}$, technically, we skip this multiplication and instead conceptually re-interpret the scaling factor $\Delta'$ as $\left\lfloor\dfrac{q}{p^\varepsilon}\right\rfloor \rightarrow \left\lfloor\dfrac{q}{p^\varepsilon-1}\right\rfloor \rightarrow \left\lfloor\dfrac{q}{p^\varepsilon-2}\right\rfloor \rightarrow \cdots, \left\lfloor\dfrac{q}{p}\right\rfloor$ across $\varepsilon-1$ rounds of digit extraction. During this re-interpretation, each step's $p$ being decreased in the denominator of the scaling factor is conceptually placed back into the bracket. Applying the re-interpretation of the scaling factor, the digit extraction procedure (without explicit multiplication by $|p^{-1}|_q$ at each round) is performed as follows:

\textbf{Input:} $\left\lfloor\dfrac{q}{p^{\varepsilon}}\right\rfloor \cdot (p^{\varepsilon - 1}M + E' + Kp^{\varepsilon}) \bmod q$

\textbf{1st Round:} $G_{\varepsilon}(z_i) \xRightarrow{\text{effect}} \left\lfloor\dfrac{q}{p^{\varepsilon-1}}\right\rfloor \cdot (p^{\varepsilon - 2}M + \left\lfloor\dfrac{E'}{p}\right\rfloor + K^{\langle 1 \rangle}p^{\varepsilon-1}) \bmod q$

\textbf{2nd Round:} $G_{\varepsilon-1} \circ G_{\varepsilon}(z_i) \xRightarrow{\text{effect}} \left\lfloor\dfrac{q}{p^{\varepsilon-2}}\right\rfloor \cdot (p^{\varepsilon - 3}M + \left\lfloor\dfrac{E'}{p^2}\right\rfloor + K^{\langle 2 \rangle}p^{\varepsilon-2}) \bmod q$

\textbf{3rd Round:} $G_{\varepsilon-2} \circ G_{\varepsilon-1} \circ G_{\varepsilon}(z_i)  \xRightarrow{\text{effect}} \left\lfloor\dfrac{q}{p^{\varepsilon-3}}\right\rfloor \cdot (p^{\varepsilon - 4}M + \left\lfloor\dfrac{E'}{p^3}\right\rfloor + K^{\langle 3 \rangle}p^{\varepsilon-3}) \bmod q$

$\vdots$

\textbf{$\bm{\varepsilon - 1}$-th Round:} $G_{2}\circ \cdots \circ G_{\varepsilon} (z_i) \xRightarrow{\text{effect}} \left\lfloor\dfrac{q}{p}\right\rfloor \cdot (M + K^{\langle \varepsilon - 1 \rangle}p)  \bmod q$

$ $

$ $

Note that the scaling factor re-interpretation does not involve any actual computation, but only changes our way of interpreting the scaling factor. Notice that at the end of the final round, the plaintext scaling factor becomes $\left\lfloor\dfrac{q}{p}\right\rfloor$, which is the desired plaintext scaling factor for standard BFV ciphertexts. Therefore, the scaling factor re-interpretation has three benefits: (1) we skip explicit multiplication by $|p^{-1}|_q$ at each round; (2) we prevent noise explosion; (3) at the end of all rounds, the plaintext scaling factor becomes the desired value for standard BFV ciphertexts, gracefully completing the bootstrapping procedure. 

Meanwhile, there are two important points to be aware of. First, each $i$-th round of scaling factor re-interpretation creates a small drift error in the scaling factor due to the difference $\epsilon_i = p^{-1}\cdot \left\lfloor\dfrac{q}{p^{\varepsilon - i}}\right\rfloor - \left\lfloor\dfrac{q}{p^{\varepsilon - i + 1}}\right\rfloor$, where $0 \leq \epsilon_{i} < p$. Therefore, an additional drift error $E_i^{\langle \text{re} \rangle}$ is created at each $i$-th round, which is small enough to be bounded by: 

$E_i^{\langle \text{re} \rangle} \leq \epsilon_i\cdot (p^{\varepsilon - i-1}M + \left\lfloor\dfrac{E'}{p^{i}}\right\rfloor + K^{\langle i \rangle} p^{\varepsilon - i}) < p^{\varepsilon-1}M + \left\lfloor\dfrac{E'}{p}\right\rfloor + K^{\langle i \rangle}p^{\varepsilon + 1} \ll \left\lfloor\dfrac{q}{p^{\varepsilon}}\right\rfloor$

$ $

Second, at each $i$-th round of digit extraction, the plaintext operands used for ciphertext-to-plaintext homomorphic operations should encode their values with the specific scaling factor interpreted at that round: $\Delta_i' = \left\lfloor\dfrac{q}{p^{\varepsilon-i+1}}\right\rfloor$. 

$ $

Now, our final remaining task is to design the actual \textit{lifting} polynomial $F_{\varepsilon'}(z_i)$ that implements $G_{\varepsilon}$.  

$ $

\para{Designing \boldmath$F_{\varepsilon'}(z_i)$:} We will derive $F_{\varepsilon'}(z_i)$ based on the following steps. 

\begin{enumerate}
\item \textbf{\textit{Claim:}} $z_i^{p} \equiv z_{i,0} \bmod p$ 

\begin{proof}
It's true that $z_i \equiv z_{i,0} \bmod p$. Fermat's Little Theorem states $a^{p} \equiv a \bmod p$ for all $a \in \mathbb{Z}_p$ and prime $p$. Therefore, $z_i \equiv z_{i,0} \equiv z_{i,0}^{p} \equiv z^{p} \bmod p$.
\end{proof}

\item \textbf{\textit{Claim:}} $z_i^p \equiv z_{i, 0}^p \bmod p^{\varepsilon'+1}$ 

\begin{myproof}
$(z_{i,0} + kp^{\varepsilon'})^p \bmod p^{\varepsilon'+1}= \sum\limits_{j=0}^{p}  \binom{p}{j}\cdot  z_{i,0}^{j} \cdot (kp^{\varepsilon'})^{p-j} \bmod p^{\varepsilon'+1}$ \textcolor{red}{ $\rhd$ binomial expansion formula}

$ \equiv z_{i,0}^{p} \bmod p^{\varepsilon' + 1}$ \textcolor{red}{ $\rhd$ all terms where $j < p$ are $0 \bmod p^{\varepsilon' + 1}$}
\end{myproof}

$ $

\item \textbf{\textit{Claim:}} Given $p$ and $\varepsilon'$ are fixed, there exists $\varepsilon'+1$ polynomials $f_0, f_1, f_2, \cdots, f_{\varepsilon'}$ (where each polynomial is at most $p-1$ degrees) such that any $z_i$ (i.e., any number whose base-$p$ representation has 0s between the base-$p$ digit index greater than 0 and smaller than $\varepsilon'$) can be expressed as the following formula:

$z_i^p \equiv \sum\limits_{j=0}^{\varepsilon'}f_j(z_{i,0})\cdot p^j \bmod p^{\varepsilon'+1}$

\begin{myproof}
$z_i^p \bmod p^{\varepsilon'+1}$ can be expressed as a base-$p$ number as follows:

$z_i^p \bmod p^{\varepsilon'+1} = c_0 + c_1p + c_2p^2 + \cdots + c_{\varepsilon'}p^{\varepsilon'}$

$ $

Based on step 3's claim ($z_i^p \equiv z_{i, 0}^p \bmod p^{\varepsilon'+1}$), we know that the value of $z_i^p \bmod p^{\varepsilon' + 1}$ depends only on $z_{i,0}$ (given $p$ and $\varepsilon'$ are fixed). Therefore, we can imagine that there exists some function $f(z_{i,0})$ whose input is $z_{i,0} \in [0, p-1]$ and the output is $z_i^p \in [0, p^{\varepsilon'+1} - 1]$. Alternatively, we can imagine that there exist $\varepsilon'+1$ different functions $f_0, f_1, \cdots, f_{\varepsilon'}$ such that each $f_j$ is a polynomial whose input is $z_{i,0} \in [0, p-1]$ and the output is $c_i \in [0, p-1]$, and $z_i^p \equiv \sum\limits_{j=0}^{\varepsilon'}f_j(z_{i,0})\cdot p^j \bmod p^{\varepsilon'+1}$. In this case, the input and output domain of each polynomial $f_j$ is $[0, p - 1]$. Therefore, we can design each $f_j$ as a $(p-1)$-degree polynomial and derive each $f_j$ based on polynomial interpolation (\autoref{sec:polynomial-interpolation}) by using $p$ coordinate values. 

Note that whenever we increase $\varepsilon'$ to $\varepsilon' + 1$, we add a new polynomial $f_{\varepsilon' + 1}$. However, the previous polynomials $f_0, f_1, \cdots, f_{\varepsilon'}$ stay the same as before, because increasing $\varepsilon'$ by 1 only adds a new base-$p$ constant $c_{\varepsilon' + 1}$ for the highest base-$p$ digit, while keeping the lower-digit constants $c_0, c_1, \cdots, c_{\varepsilon'}$ the same as before. Therefore, the polynomials $f_0, f_1, \cdots, f_{\varepsilon'}$, each of which computes  $c_0, c_1, \cdots, c_{\varepsilon'}$, also stay the same as before. 

\end{myproof}

\item \textbf{\textit{Claim:}} The formula in step 4's claim can be further concretized as follows:

$z_i^p \equiv z_{i,0} + \sum\limits_{j=1}^{\varepsilon'}f_j(z_{i,0})\cdot p^j \bmod p^{\varepsilon'+1}$

\begin{myproof}
According to step 2's claim ($z_i^{p} \equiv z_{i,0} \bmod p$), we know that the following base-$p$ representation of $z_i^p$:

$z_i^p \equiv c_0 + c_1p + c_2p^2 + \cdots + c_{\varepsilon'}p^{\varepsilon'} \bmod p^{\varepsilon'+1} $

$ $

will be the following:

$z_i^p \equiv z_{i,0} + c_1p + c_2p^2 + \cdots + c_{\varepsilon'}p^{\varepsilon'} \bmod p^{\varepsilon'+1} $

$ $

, since step 2's claim implies that the least significant base-$p$ digit of $z_i^p$ in the base-$p$ representation is always $z_{i,0}$. Thus, the formula in step 4's claim:

$z_i^p \equiv \sum\limits_{j=0}^{\varepsilon'}f_j(z_{i,0})\cdot p^j \bmod p^{\varepsilon'+1}$

$ $

can be further concretized as follows:

$ $

$z_i^p \equiv z_{i,0} + \sum\limits_{j=1}^{\varepsilon'}f_j(z_{i,0})\cdot p^j \bmod p^{\varepsilon'+1}$

\end{myproof}

\item \textbf{\textit{Claim:}} $z_i^p - \sum\limits_{j=1}^{\varepsilon'}f_j(z_i)\cdot p^j \equiv z_{i,0} \bmod p^{\varepsilon'+1}$

\begin{myproof}

\item Remember that in step 1, we defined $z_i$ as: $z_i = z_{i,0} + \sum\limits_{j=\varepsilon'}^{\varepsilon-1} z_{i,j}p^j$. Therefore, $z_i \bmod p^{\varepsilon'} = z_{i,0}$. This implies that for each polynomial $f_j$, the following is true:

$f_j(z_{i,0}) \equiv f_j(z_i) \bmod p^{\varepsilon'}$

$ $

Next, we derive the following:

$f_j(z_{i,0})\cdot p^j \equiv f_j(z_i)\cdot p^j \bmod p^{\varepsilon' + 1}$ (where $j \geq 1$)

$ $

The above is true because: 

$f_j(z_{i,0}) \equiv f_j(z_i) \bmod p^{\varepsilon'}$

$f_j(z_{i,0}) = f_j(z_i) + q\cdot p^{\varepsilon'}$ (for some integer $q$)

$f_j(z_{i,0})\cdot p^j = f_j(z_i)\cdot p^j + q\cdot p^{\varepsilon'}\cdot p^j$ \textcolor{red}{ $\rhd$ multiplying $p^j$ to both sides}

$f_j(z_{i,0})\cdot p^j = f_j(z_i)\cdot p^j + q\cdot p^{j-1} \cdot p^{\varepsilon'+1}$ \textcolor{red}{ $\rhd$ $f_j(z_{i,0})\cdot p^j$ and $f_j(z_i)\cdot p^j$ differ by some multiple of $p^{\varepsilon' + 1}$}

$ $

Therefore, $f_j(z_{i,0})\cdot p^j \equiv f_j(z_i)\cdot p^j \bmod p^{\varepsilon' + 1}$.

$ $

Now, given step 5's claim ($z_i^p \equiv z_{i,0} + \sum\limits_{j=1}^{\varepsilon'}f_j(z_{i,0})\cdot p^j \bmod p^{\varepsilon'+1}$), we can derive the following:

$z_i^p - \sum\limits_{j=1}^{\varepsilon'}f_j(z_{i})\cdot p^j \bmod p^{\varepsilon' + 1}$

$\equiv (z_{i,0} + \sum\limits_{j=1}^{\varepsilon'}f_j(z_{i,0})\cdot p^j) - \sum\limits_{j=1}^{\varepsilon'}f_j(z_{i})\cdot p^j \bmod p^{\varepsilon' + 1}$ \textcolor{red}{ $\rhd$ applying step 5's claim}

$\equiv (z_{i,0} + \sum\limits_{j=1}^{\varepsilon'}f_j(z_i)\cdot p^j) - \sum\limits_{j=1}^{\varepsilon'}f_j(z_i)\cdot p^j \bmod p^{\varepsilon' + 1}$ \textcolor{red}{ $\rhd$ applying $f_j(z_{i,0})\cdot p^j \equiv f_j(z_i)\cdot p^j \bmod p^{\varepsilon' + 1}$}

$\equiv z_{i,0} \bmod p^{\varepsilon' + 1}$

\end{myproof}

\item Finally, we define the lifting polynomial $F_{\varepsilon'}(z_i)$ as follows:

$F_{\varepsilon'}(z_i) = z_i^p - \sum\limits_{j=1}^{\varepsilon'}f_j(z_i)\cdot p^j$

$\textcolor{white}{F_{\varepsilon'}(z_i) } \equiv z_{i,0} \bmod p^{\varepsilon' + 1}$

The above relation implies that $F_{\varepsilon'}(x) \bmod p^{\varepsilon'+1}$ is equivalent to the least significant base-$p$ digit of $x$ (according to step 6's claim). Therefore, if we plug in $z_i$ into $F_{\varepsilon'}(x)$ and regard $\varepsilon' = 1$, then the output is some number whose least significant base-$p$ digit is $z_{i,0} \bmod p^2$ and the 2nd least significant base-$p$ digit is $0 \bmod p^2$. As we recursively apply the output back to $F_{\varepsilon'}(x)$ and increment $\varepsilon'$ by 1, we iteratively zero out the 2nd least significant base-$p$ digit, the 3rd least significant base-$p$ digit, and so on. We repeat this process for $\varepsilon-1$ times to zero out the upper $\varepsilon-1$ base-$p$ digits, keeping only the least significant digit as it is (i.e., $z_{i,0}$). Therefore, $F_{\varepsilon'}(x)$ is a valid lifting polynomial that can be iteratively used to extract the least significant digit of $z_i \bmod p^\varepsilon$.

\end{enumerate}

We use $F_{\varepsilon'}(x)$ as the internal helper function within the digit extraction function $G_{\varepsilon}(z_i)$ that calls $F_{\varepsilon'}(x)$ a total of $v-1$ times.

\subsubsection{Summary}
\label{subsubsec:bfv-bootstrapping-summary}

We summarize the BFV bootstrapping procedure (with the generalization of $t = p^r$) as follows. 

\begin{tcolorbox}[title={\textbf{\tboxlabel{\ref*{subsubsec:bfv-bootstrapping-summary}} BFV Bootstrapping}}]

Suppose we have an RLWE ciphertext $(A, B)  = \textsf{RLWE}_{S, \sigma}\bm(\Delta M + E\bm) \bmod q$, where $\Delta = \left\lfloor\dfrac{q}{t}\right\rfloor$ and $t = p^r$ (i.e., the plaintext modulus is a power of some prime), $r \in \mathbb{Z}$, and $r \geq 1$. 

$ $

\begin{enumerate}
\item \textbf{\underline{Modulus Switch} (from \boldmath$q \rightarrow p^\varepsilon$):} Scale down the ciphertext from $(A, B)$ to $\left(\left\lceil \dfrac{p^\varepsilon}{q}\cdot A\right\rfloor, \left\lceil \dfrac{p^\varepsilon}{q}\cdot B\right\rfloor\right) = (A', B')$ \textcolor{red}{ $\rhd$ where $p^\varepsilon \ll q$} 

$ $

$A'S + B' = p^{\varepsilon-r}M + E' \bmod p^\varepsilon$ \textcolor{red}{ $\rhd$ where $E' \approx \dfrac{p^\varepsilon}{q}\cdot E  + \left(\left\lfloor\dfrac{q}{p^r}\right\rfloor\cdot\dfrac{p^\varepsilon}{q} - p^{\varepsilon-r}\right)\cdot M$, which is a modulus switch noise plus the rounding noise caused by treating $\Delta=\left\lfloor\dfrac{q}{p^r}\right\rfloor \approx \dfrac{q}{p^r}$.}

$ $

\item \textbf{\underline{Homomorphic Decryption}:} With the bootstrapping key $\textsf{RLWE}_{S, \sigma}(\Delta' S) \bmod q$, homomorphically decrypt $(A', B') \bmod p^\varepsilon$ as follows:

$A' \cdot \textsf{RLWE}_{S, \sigma}(\Delta' S)  + B' = \textsf{RLWE}_{S, \sigma}\bm(\Delta' \cdot (p^{\varepsilon-r} M + E' + Kp^\varepsilon)\bm) \bmod q$ \textcolor{red}{ $\rhd$ where $\Delta' = \left\lfloor\dfrac{q}{p^\varepsilon}\right\rfloor$}

$ $

Now, we denote the modulus-switched noisy plaintext polynomial as $Z = p^{\varepsilon-r} M + E' + Kp^\varepsilon$.

$ $

\item \textbf{\textsf{\underline{CoeffToSlot}}:} Move the (encrypted) polynomial $Z$'s coefficients $z_0, z_i, \cdots, z_{n-1}$ to the input vector slots. This is done by computing: 

$\textsf{RLWE}_{S, \sigma}(\Delta' Z) \cdot n^{-1}\cdot \hathat{W}\cdot I_R^n$

$= \textsf{RLWE}_{S, \sigma}(\Delta' Z^{\langle1\rangle})$

, where $n^{-1}\cdot \hathat{W}\cdot I_R^n$ is the batch encoding matrix (Summary~\ref*{subsubsec:bfv-rotation-summary} in \autoref{subsubsec:bfv-rotation-summary}). 

$ $

\item \textbf{\underline{Digit Extraction}:} We design a polynomial $G_{\varepsilon}(z_i)$ (a digit extraction polynomial) as follows:

$z_i = d_0 + \left(\sum\limits_{j=\varepsilon'}^{\varepsilon-r} d_* p^j\right)$ \textcolor{red}{ $\rhd$ where $d_0 \in \mathbb{Z}_{p}$, and $\varepsilon'$ is $z_i$'s first least significant base-$p$ digit index whose value is non-zero (after digit index 0) currently being processed by the $F_{\varepsilon'}(Z_i)$ function}

$F_{\varepsilon'}(z_i) \equiv d_0 \bmod p^{\varepsilon'+1}$ \textcolor{red}{ $\rhd$ a $(p-1)$-degree polynomial recursively used to finally extract the value $d_0 \bmod p^{\varepsilon}$}

$G_{\varepsilon}(z_i) \equiv (z_i - \underbrace{F_{\varepsilon-1} \circ F_{\varepsilon-2} \circ F_{\varepsilon-3} \cdots F_{r}}_{\varepsilon - r \text{ times}} (z_i)) \cdot |p^{-1}|_q \bmod p^\varepsilon$

$ $

We homomorphically evaluate the digit extraction polynomial $G_{\varepsilon}$ recursively total $\varepsilon-r$ times at each coefficient $z_i$ of $Z$ stored at input vector slots, which zeros out and right-shifts the least significant (base-$p$) $\varepsilon-r$ digits of $z_i$ as follows:

$G_{r + 1} \circ G_{r + 2} \circ \cdots \circ G_{\varepsilon-1} \circ G_{\varepsilon} (z_i)$

$= m_i + k_i'p$

$ $

, provided $E'$'s each coefficient is smaller than $\left\|\dfrac{p^{\varepsilon-r}}{2}\right\|$. At this point, each input vector slot contains the noise-removed coefficient $m_i + k_i^{\langle \varepsilon - r - 1\rangle}p$. 

\para{\underline{Scaling Factor Re-interpretation}:} At each $i$-th round of digit extraction, we do not explicitly multiply $|p^{-1}|_q$ to perform division by $p$, but instead conceptually borrow this term from the denominator of the plaintext scaling factor $\Delta'$, conceptually updating the scaling factor to $\left\lfloor\dfrac{q}{p^{\varepsilon-i}}\right\rfloor$. This implies that at $i$-th round of digit extraction, the plaintext operations used for ciphertext-to-plaintext homomorphic operations should encode their values by using the scaling factor $\left\lfloor\dfrac{q}{p^{\varepsilon-i}}\right\rfloor$. At the end of digit extraction, the plaintext scaling factor becomes $\Delta' = \Delta = \left\lfloor\dfrac{q}{p^r}\right\rfloor$.

$ $

\item \textbf{\textsf{\underline{SlotToCoeff}}:} Homomorphically move each input vector slot's value $m_i + k_i^{\langle \varepsilon - r - 1\rangle}p^\varepsilon$ back to the (encrypted) polynomial coefficient positions. This is done by multiplying $\hathat{W}^*$ to the output ciphertext of the digit extraction step, where $\hathat{W}^*$ is the decoding matrix (Summary~\ref*{subsubsec:bfv-rotation-summary} in \autoref{subsubsec:bfv-rotation-summary}). The output of this computation is $\textsf{RLWE}_{S, \sigma}\bm(\Delta\cdot (M + K^{\langle \varepsilon - r - 1\rangle}p) + E^{\langle \text{b} \rangle}) \bm) = \textsf{RLWE}_{S, \sigma}\bm(\Delta\cdot (M + E^{\langle \text{final} \rangle}) \bm) \bmod q$, where $E^{\langle \text{b} \rangle}$ is the new noise generated during bootstrapping (step $3\sim 5$).

\end{enumerate}

\end{tcolorbox}

\para{The purpose of Homomorphic Decryption:} Its purpose is to temporarily adjust the scaling factor of the ciphertext to preserve the correctness of bootstrapping. Before homomorphic decryption, the ciphertext encrypts $p^{\varepsilon-r} M$, a message with the scaling factor $p^{\varepsilon-r}$. After homomorphic decryption, the ciphertext encrypts $\Delta' p^{\varepsilon-r} M$, the message $p^{\varepsilon-r} M$ with the scaling factor $\Delta' = \left\lfloor\dfrac{q}{p^{\varepsilon}}\right\rfloor$. This specific adjustment is needed for the scaling factor re-interpretation during each round of digit extraction to conceptually divide by $p$ (i.e., multiply by $|p^{-1}|_q$) and eventually adjust the scaling factor back to $\left\lfloor\dfrac{q}{p^r}\right\rfloor$ as the original form for standard BFV ciphertexts.

\newpage

\section{CKKS Scheme}
\label{sec:ckks}
The CKKS scheme is designed for homomorphic addition and multiplication of complex numbers that contain imaginary components. Therefore, unlike BFV, BGV, or TFHE, which can only compute over integers, CKKS can compute real-world floating point arithmetic, such as in machine learning.

The CKKS scheme's goal is to homomorphically compute the addition and multiplication of complex numbers. However, while our targeted inputs are complex numbers, CKKS's plaintext space is defined as a $(n-1)$-degree polynomial ring with real-number coefficients having limited precision; that is, $\mathcal{R}_{\langle n \rangle} = \mathbb{R}[x] / (x^n + 1)$. Therefore, CKKS designs its unique encoding scheme, which converts the input complex numbers into integers that can be used as coefficients of a polynomial in $\mathcal{R}_{\langle n \rangle}$.

$ $

\noindent Overall, CKKS's encryption procedure is as follows:

\begin{enumerate}
\item \underline{\textsf{Encoding\textsubscript{1}}:} Encode the targeted input complex number as a real number
\item \underline{\textsf{Encoding\textsubscript{2}}:} Encode the real number as an integer
\item \underline{\textsf{Encryption}:} Encrypt the integer using RLWE 
\end{enumerate}
The encrypted RLWE ciphertext supports homomorphic addition and multiplication. 

$ $

\noindent At the end of all homomorphic operations, CKKS's decryption procedure is as follows: 

\begin{enumerate}
\item \underline{\textsf{Decryption}:} Decrypt the RLWE ciphertext into a plaintext integer
\item \underline{\textsf{Decoding\textsubscript{1}}:} Decode the integer to a real number
\item \underline{\textsf{Decoding\textsubscript{2}}:} Decode the real number to a complex number
\end{enumerate}

$ $

Remember that BFV is an exact encryption scheme based on rings. On the other hand, CKKS introduces a drifting error while its encoding process of rounding square-root values (included in the Euler's formula) to the nearest integer. Therefore, its decryption is not exactly the same as before encryption. Such a small error occurring during encryption and decryption makes CKKS an \textit{approximate} encryption scheme. 

CKKS internally uses the same schemes as BFV for encryption, decryption, ciphertext-to-plaintext addition, ciphertext-to-ciphertext addition, and ciphertext-to-plaintext multiplication. Meanwhile, CKKS uses slightly different schemes than BFV for encoding the input vector (i.e., input vector slots) rotation (if BFV uses the batch encoding scheme), ciphertext-to-ciphertext multiplication, and bootstrapping. This difference comes from the fact that CKKS handles homomorphic operations over complex numbers as inputs, whereas BFV handles homomorphic operations over rings.

\begin{tcolorbox}[
    title = \textbf{Required Background},    % box title
    colback = white,    % light background; tweak to taste
    colframe = black,  % frame colour
    boxrule = 0.8pt,     % line thickness
    left = 1mm, right = 1mm, top = 1mm, bottom = 1mm % inner padding
]

\begin{itemize}
\item \autoref{sec:modulo}: \nameref{sec:modulo}
\item \autoref{sec:group}: \nameref{sec:group}
\item \autoref{sec:field}: \nameref{sec:field}
\item \autoref{sec:order}: \nameref{sec:order}
\item \autoref{sec:polynomial-ring}: \nameref{sec:polynomial-ring}
\item \autoref{sec:decomp}: \nameref{sec:decomp}
\item \autoref{sec:roots}: \nameref{sec:roots}
\item \autoref{sec:cyclotomic}: \nameref{sec:cyclotomic}
\item \autoref{sec:matrix}: \nameref{sec:matrix}
\item \autoref{sec:euler}: \nameref{sec:euler}
\item \autoref{sec:modulus-rescaling}: \nameref{sec:modulus-rescaling}
\item \autoref{sec:chinese-remainder}: \nameref{sec:chinese-remainder}
\item \autoref{sec:taylor-series}: \nameref{sec:taylor-series}
\item \autoref{sec:polynomial-interpolation}: \nameref{sec:polynomial-interpolation}
\item \autoref{sec:ntt}: \nameref{sec:ntt}
\item \autoref{sec:lattice}: \nameref{sec:lattice}
\item \autoref{sec:rlwe}: \nameref{sec:rlwe}
\item \autoref{sec:glwe}: \nameref{sec:glwe}
\item \autoref{sec:glev}: \nameref{sec:glev}
\item \autoref{sec:glwe-add-cipher}: \nameref{sec:glwe-add-cipher}
\item \autoref{sec:glwe-add-plain}: \nameref{sec:glwe-add-plain}
\item \autoref{sec:glwe-mult-plain}: \nameref{sec:glwe-mult-plain}
\item \autoref{subsec:modulus-switch-rlwe}: \nameref{subsec:modulus-switch-rlwe}
\item \autoref{sec:glwe-key-switching}: \nameref{sec:glwe-key-switching}
\item \autoref{sec:bfv}: \nameref{sec:bfv}
\end{itemize}
\end{tcolorbox}

\clearpage

\subsection{Encoding and Decoding}
\label{subsec:ckks-encoding-decoding}

CKKS's encoding and decoding is fundamentally very similar to BFV's batch encoding scheme. BFV designs its batch encoding scheme (Summary~\ref*{subsec:bfv-enc-dec} in \autoref{subsec:bfv-enc-dec}) based on the updated $\hathat W$ and $\hathat W^*$ matrices (Summary~\ref*{subsubsec:bfv-rotation-summary} in \autoref{subsec:bfv-rotation}). That is, BFV decodes a polynomial into an input slot vector by evaluating the polynomial at each root of $X^n+1$, which is the primitive $(\mu=2n)$-th root of unity (i.e., $\vec{v} = \hathat W^* \cdot \vec{m}$), and encodes an input slot vector into a polynomial by inversing this operation (i.e., $\vec{m} = n^{-1} \cdot \hathat W \cdot I_n^R \cdot \vec{v}$). This encoding and decoding scheme is designed based on Summary~\ref*{subsec:poly-vector-transformation-complex} (\autoref{subsec:poly-vector-transformation-complex}) which designs the isomorphic mapping between $n$-slot vectors in a ring (finite field) and $(n-1)$-degree (or lesser degree) polynomials as follows: 

$\sigma: f(x) \in \mathbb{Z}_t[X] / F(X) \text{ } \longrightarrow \text{ } (f(\omega^1), f(\omega^3)), \cdots, f(\omega^{2n-1})) \in \mathbb{Z}_t^n$

, where $\omega = g^{\frac{t - 1}{2n}}$ is a root of (i.e., primitive $(\mu=2n)$-th root of unity) of the $(\mu=2n)$-th cyclotomic polynomial $X^n + 1$ defined over a prime modulo $t$ ring. 

$ $

CKKS's batch encoding scheme uses exactly the same formula for encoding and decoding (i.e., $\vec{v} = W^T \cdot \vec{m}$ and $\vec{m} = \dfrac{W \cdot I_n^R \cdot \vec{v}}{n}$), but the $n$ input slot vector comprises not in a ring (i.e., $\mathbb{Z}^n_p$), but complex numbers (i.e., $\mathbb{\hat C}^n$). In Summary~\ref*{subsec:poly-vector-transformation-complex} (\autoref{subsec:poly-vector-transformation-complex}), we also designed the mapping $\sigma_c$ between polynomials and vectors over complex numbers as follows:

$\sigma_c: f(X) \in \mathbb{R}[X]/(X^n + 1) \longrightarrow (f(\omega),f(\omega^3),f(\omega^5), \cdots, f(\omega^{2n-1})) \in \mathbb{\hat{C}}^{n} \text{ } (\longrightarrow \mathbb{C}^{\frac{n}{2}})$

, where $\omega = e^{i\pi/n}$ is a root (i.e., the primitive $(\mu=2n)$-th root) of the $(\mu=2n)$-th cyclotomic polynomial $X^n + 1$ defined over complex numbers, and $\mathbb{\hat{C}}^{n}$ is $n$-dimensional complex special vector space whose second-half elements are reverse-ordered conjugates of the first-half elements. And $\mathbb{\hat{C}}^{n}$ is isomorphic to $\mathbb{{C}}^{\frac{n}{2}}$, because the second-half elements of $\mathbb{\hat{C}}^{n}$ are automatically determined by its first-half elements. Therefore, the $\sigma_c$ mapping is essentially an isomorphism between $\dfrac{n}{2}$-slot complex vectors $\vec{v} \in \mathbb{C}^{\frac{n}{2}}$ and $(n-1)$-degree (or lesser degree) real-number polynomials $\mathbb{R}[X] / (X^n + 1)$. Therefore, CKKS' batch encoding scheme encodes an $\dfrac{n}{2}$-slot complex input vector into an $(n-1)$-degree (or lesser degree) real-number polynomial, and the decoding process is a reverse of this. 

In addition, remember that in BFV, we updated $W$ and $W^T$ to $\hathat W$ and $\hathat W^*$ (Summary~\ref*{subsubsec:bfv-rotation-summary} in \autoref{subsubsec:bfv-rotation-summary}) to support homomorphic rotation of input vector slots. Likewise, the CKKS batch encoding scheme uses $\hathat W$ and $\hathat W^*$ instead of $W$ and $W^T$ in order to support homomorphic rotation. Therefore, the CKKS batch encoding scheme's isomorphic mapping is updated as follows:

$\sigma_c: f(X) \in \mathbb{R}[X]/(X^n + 1) \longrightarrow \bm ( f(\omega^{J(0)}),f(\omega^{J(1)}),f(\omega^{J(2)}), \cdots, f(\omega^{J(\frac{n}{2} - 1)}), $

\textcolor{white}{$ f(X) \in \mathbb{R}[X]/(X^n + 1) \longrightarrow \bm ( $} $\cdots, f(\omega^{J_*(0)}),f(\omega^{J_*(1)}),f(\omega^{J_*(2)}), \cdots, f(\omega^{J_*(\frac{n}{2} - 1)}) \bm) \in \mathbb{\hat{C}}^{n} \text{ } (\longrightarrow \mathbb{C}^{\frac{n}{2}})$

, where $J(h) = 5^h \bmod 2n$, a rotation helper formula.

$ $

The encoding schemes of BFV and CKKS have the following differences: 

\begin{itemize}
\item \textbf{Type of Input Slot Values:} BFV's input slot values are ${n}$ integers modulo $t$, which are encoded into $n$ polynomial coefficients (i.e., modulo-$t$ integers). On the other hand, CKKS's input slot values are $\dfrac{n}{2}$ complex numbers, which are encoded into $n$ polynomial coefficients (i.e., real numbers). 

\item \textbf{Type of Polynomial Coefficients:} BFV's encoded polynomial coefficients are integer moduli, whereas CKKS's encoded polynomial coefficients are real numbers. 

\item \textbf{Scaling Factor:} Both BFV and CKKS scales their encoded polynomial coefficients $\vec{m}$ by $\Delta$ to $\lceil \Delta\cdot \vec{m} \rfloor$. BFV's suggested scaling factor is $\Delta = \lfloor\dfrac{q_0}{t}\rfloor$, but CKKS's scaling factor $\Delta$ has no suggested formula because its polynomial coefficients are real numbers not bound by modulus, and thus it can be any value provided that the scaled coefficients do not overflow or underflow the range $[1, q_0 - 1]$ (or $\left[-\dfrac{q_0}{2}, \dfrac{q_0}{2}\right)$). 

\item \textbf{Encoding Precision:} In the case of BFV, during its decoding process, BFV's down-scaled polynomial coefficients $\dfrac{ \Delta\cdot \vec{m} }{\Delta}$ preserve the precision of input values 
%as far as $\Delta$ is sufficiently large enough to round off the scaling error (as explained in \autoref{subsubsec:bfv-enc-dec-decoding1})
. On the other hand, CKKS's down-scaled polynomial coefficients may lose their precision if their original input values have too many decimal digits so that the scaling factor cannot left-shift all of them to make them part of the integer domain, which means that some lower decimal digits of the input value may be rounded off, which loses precision of the original input. For example, suppose the polynomial coefficient $m_i = \dfrac{1}{3} = 0.33333\cdots$, and the scaling factor $\Delta = 100$. Then, the scaled coefficient $\lceil \Delta m_i \rfloor = 33$, and down-scaling it gives $\dfrac{33}{100} = 0.33$. Since $0.33 \neq 0.33333\cdots$, CKKS's encoding and decoding process does not always guarantee exact precision. Due to this encoding error, CKKS is called an \textit{approximate} encryption scheme. The impact of this encoding error can grow over homomorphic operations which increases the magnitude of error and the decoded result would gradually become more deviated from the expected exact value. One way to reduce CKKS's encoding error is to increase $\Delta$, and thereby left-shift more decimal digits to make them part of the scaled integer digits.  

\end{itemize}

$ $

\para{Structure of \boldmath$\vec{v}_{'} \in \mathbb{\hat C}^n$:} Note that the original decoding scheme for $\vec{v}_{'}$ described in Summary~\ref*{subsec:poly-vector-transformation-complex} (\autoref{subsec:poly-vector-transformation-complex}) was: 

$\vec{v}_{'} = \bm{(} \text{ } M(\omega), \text{ } M(\omega^3), \text{ } M(\omega^5), \cdots, M(\omega^{2n-3}), \text{ } M(\omega^{2n-1}) \bm{)}$

, which decodes to a Hermitian vector:

$\vec{v}_{'} = (v_0, v_1, \cdots, v_{\frac{n}{2} - 1}, \overline v_{\frac{n}{2} - 1}, \cdots, \overline v_1, \overline v_0 )$

, whose second-half elements are reverse-ordered conjugates of the first-half elements. 

$ $

However, by replacing $W$ and $W^T$ with $\hathat W$ and $\hathat W^*$, we changed the above decoding scheme to the following that supports homomorphic rotation:

$\vec{v}_{'} =  \bm{(} \text{ } 
M(\omega^{J(0)}), \text{ } M(\omega^{J(1)}), \text{ } M(\omega^{J(2)}), \cdots,  M(\omega^{J(\frac{n}{2}-1)}), \text{ } M(\omega^{J_*(0)}), \text{ } M(\omega^{J_*(1)}), \cdots,  M(\omega^{J_*(\frac{n}{2}-1)}) \text{ } \bm{)}$

$\textcolor{white}{\vec{v}_{'}} =  \bm{(} \text{ } 
M(\omega^{J(0)}), \text{ } M(\omega^{J(1)}), \text{ } M(\omega^{J(2)}), \cdots,  M(\omega^{J(\frac{n}{2}-1)}), \text{ } M(\overline\omega^{J(0)}), \text{ } M(\overline\omega^{J(1)}), \cdots,  M(\overline\omega^{J(\frac{n}{2}-1)}) \text{ } \bm{)}$ 

\textcolor{red}{ $\rhd$ because $\omega^{-1} = (e^{\frac{i\pi}{n}})^{-1} = e^{\frac{-i\pi}{n}} = \overline\omega$, given  $J(h) = 5^h \bmod 2n$, and $J_*(h) = -5^h \bmod 2n$}

, which decodes to a \textit{forward-ordered} (not reverse-ordered) Hermitian vector as follows:

$\vec{v}_{'} = (v_0, v_1, \cdots, v_{\frac{n}{2} - 1}, \overline v_0, \overline v_1, \cdots, \overline v_{\frac{n}{2} - 1})$

, whose second-half elements are conjugates of the first-half elements with the same order. Upon homomorphic rotation (which will be explained in \autoref{subsec:ckks-rotation}), just like in BFV's homomorphic rotation, the first-half elements and the second-half elements of $\vec{v}_{'}$ rotate within their own group in a wrapping manner.

We summarize CKKS's encoding and decoding procedure as follows, which is similar to BFV's encoding and decoding procedure (described in Summary~\ref*{subsubsec:bfv-encoding-summary} in \autoref{subsubsec:bfv-encoding-summary}):

\begin{tcolorbox}[title={\textbf{\tboxlabel{\ref*{subsec:ckks-encoding-decoding}} CKKS's Encoding and Decoding}}]

\textbf{\underline{Input}:} An $\dfrac{n}{2}$-slot complex vector $\vec{v} = (v_0, v_1, \cdots, v_{\frac{n}{2}-1}) \in \mathbb{C}^{\frac{n}{2}}$

\par\noindent\rule{\textwidth}{0.4pt}

\textbf{\underline{Encoding}:}

$ $

\begin{enumerate}
\item Convert (i.e., isomorphically transform) $\vec{v}$ into an $n$-slot \textit{forward-ordered} Hermitian vector $\vec{v}_{'}$ as follows:

$\vec{v}_{'} = (v_0, v_1, \cdots, v_{\frac{n}{2}-1}, \overline v_0, \overline v_1, \cdots, \overline v_{\frac{n}{2}-1}) \in \mathbb{\hat C}^{n}$

\item Convert $\vec{v}_{'}$ into a real number vector $\vec{m}$ by applying the transformation $\vec{m} = \dfrac{\hathat W \cdot I_n^R \cdot \vec{v}_{'}}{n}$

, where $\hathat{W}$ is a basis of the $n$-dimensional vector space crafted as follows: 

$ $

\noindent{\footnotesize{\noindent $\noindent\hathat W = \begin{bmatrix}
1 & 1 & \cdots & 1 & 1 & 1 & \cdots & 1\\
(\omega^{J(\frac{n}{2} - 1)}) & (\omega^{J(\frac{n}{2} - 2)}) & \cdots & (\omega^{J(0)}) & (\omega^{J_*(\frac{n}{2} - 1)}) & (\omega^{J_*(\frac{n}{2} - 2)}) & \cdots & (\omega^{J_*(0)})\\
(\omega^{J(\frac{n}{2} - 1)})^2 & (\omega^{J(\frac{n}{2} - 2)})^2 & \cdots & (\omega^{J(0)})^2 & (\omega^{J_*(\frac{n}{2} - 1)})^2 & (\omega^{J_*(\frac{n}{2} - 2)})^2 & \cdots & (\omega^{J_*(0)})^2 \\
\vdots & \vdots & \ddots & \vdots & \vdots & \ddots & \vdots & \vdots \\
(\omega^{J(\frac{n}{2} - 1)})^{n-1} & (\omega^{J(\frac{n}{2} - 2)})^{n-1} & \cdots & (\omega^{J(0)})^{n-1} & (\omega^{J_*(\frac{n}{2} - 1)})^{n-1} & (\omega^{J_*(\frac{n}{2} - 2)})^{n-1} & \vdots  & (\omega^{J_*(0)})^{n-1}
\end{bmatrix}$}}

\textcolor{red}{ $\rhd$ where $\omega = e^{i\pi/n} = \cos \left(\dfrac{\pi}{n}\right) + i\sin \left(\dfrac{\pi}{n}\right)$,  $J(h) = 5^h \bmod 2n$,  and  $J_*(h) = -5^h \bmod 2n$}

\small{\noindent $ = \begin{bmatrix}
1 & 1 & \cdots & 1 & 1 & 1 & \cdots & 1\\
(\omega^{J(\frac{n}{2} - 1)}) & (\omega^{J(\frac{n}{2} - 2)}) & \cdots & (\omega^{J(0)}) & (\overline\omega^{J(\frac{n}{2} - 1)}) & (\overline\omega^{J(\frac{n}{2} - 2)}) & \cdots & (\overline\omega^{J(0)})\\
(\omega^{J(\frac{n}{2} - 1)})^2 & (\omega^{J(\frac{n}{2} - 2)})^2 & \cdots & (\omega^{J(0)})^2 & (\overline\omega^{J(\frac{n}{2} - 1)})^2 & (\overline\omega^{J(\frac{n}{2} - 2)})^2 & \cdots & (\overline\omega^{J(0)})^2 \\
\vdots & \vdots & \ddots & \vdots & \vdots & \ddots & \vdots & \vdots \\
(\omega^{J(\frac{n}{2} - 1)})^{n-1} & (\omega^{J(\frac{n}{2} - 2)})^{n-1} & \cdots & (\omega^{J(0)})^{n-1} & (\overline\omega^{J(\frac{n}{2} - 1)})^{n-1} & (\overline\omega^{J(\frac{n}{2} - 2)})^{n-1} & \vdots  & (\overline\omega^{J(0)})^{n-1}
\end{bmatrix}$}

\textcolor{red}{ $\rhd$ because $\omega^{-1} = e^{\frac{-i\pi}{n}} = \overline{e^{\frac{i\pi}{n}}} = \overline\omega$}

$ $

\item Convert $\vec{m}$ into a scaled integer vector $\lceil \Delta\vec{m} \rfloor \approx \Delta \vec{m}$, where $\Delta$ is a scaling factor bigger than 1 such that $\Delta m_i$ never overflows or underflows $q_0$ (i.e., $0 \leq \Delta m_i < q_0$ or $-\dfrac{q_0}{2} \leq \Delta m_i < \dfrac{q_0}{2}$) in all cases, even across all homomorphic operations. The finally encoded plaintext polynomial is $\Delta M = \sum\limits_{i=0}^{n-1} \lceil \Delta m_i \rfloor X^i \text{ } \in \mathbb{Z}_q[X] / (X^n + 1)$. The rounding process of $\lceil \Delta \vec{m} \rfloor$ during the encoding process causes an encoding error, which makes CKKS an approximate encryption scheme. 

\end{enumerate}

\par\noindent\rule{\textwidth}{0.4pt}

\textbf{\underline{Decoding}: } From the plaintext polynomial $\Delta M = \sum\limits_{i=0}^{n-1}\Delta m_iX^i$, recover $\vec{m} = \dfrac{\Delta \vec{m}}{\Delta}$. Then,
compute $\vec{v}_{'} = \hathat W^* \cdot \vec{m}$, where:

\noindent $\hathat{W}^* = \begin{bmatrix}
1 & (\omega^{J(0)}) & (\omega^{J(0)})^2 & \cdots & (\omega^{J(0)})^{n-1}\\
1 & (\omega^{J(1)}) & (\omega^{J(1)})^2 & \cdots & (\omega^{J(1)})^{n-1}\\
1 & (\omega^{J(2)}) & (\omega^{J(2)})^2 & \cdots & (\omega^{J(2)})^{n-1}\\
\vdots & \vdots & \vdots & \ddots & \vdots \\
1 & (\omega^{J(\frac{n}{2}-1)}) & (\omega^{J(\frac{n}{2}-1)})^2 & \cdots & (\omega^{J(\frac{n}{2}-1)})^{n-1}\\
1 & (\omega^{J_*(0)}) & (\omega^{J_*(0)})^2 & \cdots & (\omega^{J_*(0)})^{n-1}\\
1 & (\omega^{J_*(1)}) & (\omega^{J_*(1)})^2 & \cdots & (\omega^{J_*(1)})^{n-1}\\
1 & (\omega^{J_*(2)}) & (\omega^{J_*(2)})^2 & \cdots & (\omega^{J_*(2)})^{n-1}\\
\vdots & \vdots & \vdots & \ddots & \vdots \\
1 & (\omega^{J_*(\frac{n}{2}-1)}) & (\omega^{J_*(\frac{n}{2}-1)})^2 & \cdots & (\omega^{J_*(\frac{n}{2}-1)})^{n-1}\\
\end{bmatrix}$

$ $

\noindent $ = \begin{bmatrix}
1 & (\omega^{J(0)}) & (\omega^{J(0)})^2 & \cdots & (\omega^{J(0)})^{n-1}\\
1 & (\omega^{J(1)}) & (\omega^{J(1)})^2 & \cdots & (\omega^{J(1)})^{n-1}\\
1 & (\omega^{J(2)}) & (\omega^{J(2)})^2 & \cdots & (\omega^{J(2)})^{n-1}\\
\vdots & \vdots & \vdots & \ddots & \vdots \\
1 & (\omega^{J(\frac{n}{2}-1)}) & (\omega^{J(\frac{n}{2}-1)})^2 & \cdots & (\omega^{J(\frac{n}{2}-1)})^{n-1}\\
1 & (\overline\omega^{J(0)}) & (\overline\omega^{J(0)})^2 & \cdots & (\overline\omega^{J(0)})^{n-1}\\
1 & (\overline\omega^{J(1)}) & (\overline\omega^{J(1)})^2 & \cdots & (\overline\omega^{J(1)})^{n-1}\\
1 & (\overline\omega^{J(2)}) & (\overline\omega^{J(2)})^2 & \cdots & (\overline\omega^{J(2)})^{n-1}\\
\vdots & \vdots & \vdots & \ddots & \vdots \\
1 & (\overline\omega^{J(\frac{n}{2}-1)}) & (\overline\omega^{J(\frac{n}{2}-1)})^2 & \cdots & (\overline\omega^{J(\frac{n}{2}-1)})^{n-1}\\
\end{bmatrix}$

$ $

$ $

, and extract only the first $\dfrac{n}{2}$ elements in the \textit{forward-ordered} Hermitian vector $\vec{v}_{'}$ to recover the input vector $\vec{v}$. 

\end{tcolorbox}

\para{CKKS's Approximation Property:} In the encoding process, when we convert $\vec{v}_{'} \rightarrow \vec{m} \rightarrow \Delta\vec{m}$, we multiply $\vec{v}_{'}$ by $\hathat{W}$ which contains complex numbers with infinite decimals (e.g., $\sqrt{2}$) coming from Euler's formula, which we should round to the nearest integer by computing $\lceil \Delta m \rfloor$ (which we will denote as $\Delta m$ throughout this section for simplicity) and thus we lose some precision. This implies that if we later decode $\Delta\vec{m}$ into $\vec{v}_{'d}$, this value would be slightly different from the original input vector $\vec{v}_{'}$. As CKKS's encoding scheme is subject to such a small rounding error, the decryption does not perfectly match the original input vector. Such errors also propagate across homomorphic computations, because those computations are done based on approximately encoded plaintext $\lceil \Delta \vec{m} \rfloor$. As these errors are caused by throwing away the infinitely long decimal digits, they can be corrected during the decoding process only if we use an infinitely big scaling factor $\Delta$, which is impossible because $\Delta m_i$ should not overflow the ciphertext modulus $q_0$ of the lowest multiplicative level. Due to this limitation, CKKS is considered an \textit{approximate} homomorphic encryption.

\subsubsection{Example}
\label{subsubsec:ckks-encoding-ex}

Suppose our input complex vector's dimension $\dfrac{n}{2} = 2$, the bounding polynomial degree $n$ = 4, and the scaling factor $\Delta = 1024$. 

Our basis of the $n$-dimensional vector space 

$\hathat{W}= \begin{bmatrix}
1 & 1 & 1 & 1\\
\omega^{J(1)} & \omega^{J(0)} & \overline{\omega}^{J(1)} & \overline{\omega}^{J(0)}\\
(\omega^{J(1)})^2 & ({\omega^{J(0)}})^2 & (\overline{\omega}^{J(1)})^2 & (\overline{\omega}^{J(0)})^2\\
(\omega^{J(1)})^3 & (\omega^{J(0)})^3 & (\overline{\omega}^{J(1)})^3 & (\overline{\omega}^{J(0)})^3\\
\end{bmatrix}$
$= \begin{bmatrix}
1 & 1 & 1 & 1\\
\omega^5 & \omega & \overline{\omega}^5 & \overline{\omega}\\
\omega^2 & \omega^2 & \overline{\omega^2} & \overline{\omega}^2\\
\omega^7 & \omega^3 & \overline{\omega}^7 & \overline{\omega}^3\\
\end{bmatrix}$

, where $\omega = e^{i\pi/n} = \cos \left(\dfrac{\pi}{n}\right) + i\sin \left(\dfrac{\pi}{n}\right)$

$ $

Given this setup, suppose we have the input complex vector $\vec{v} = (1.1 + 4.3i, \text{ } 3.5 - 1.4i)$ to encode. 

$ $

First, construct the forward-ordered Hermitian vector $\vec{v}_{'} = (1.1 + 4.3i, \text{ } 3.5 - 1.4i, \text{ } 1.1 - 4.3i, \text{ } 3.5 + 1.4i)$. 

$ $

Next, convert the complex vector $\vec{v}_{'}$ into a real number vector $\vec{m}$ by applying the transformation:

$\vec{m} = \dfrac{\hathat{W} \cdot I_n^R \cdot \vec{v}_{'}}{n} = \dfrac{1}{4} \cdot \begin{bmatrix}
1 & 1 & 1 & 1\\
\omega^5 & \omega & \overline{\omega}^5 & \overline{\omega}\\
\omega^2 & \omega^2 & \overline{\omega}^2 & \overline{\omega}^2\\
\omega^7 & \omega^3 & \overline{\omega}^7 & \overline{\omega}^3\\
\end{bmatrix} \cdot 
\begin{bmatrix}
0 & 0 & 0 & 1 \\
0 & 0 & 1 & 0 \\
0 & 1 & 0 & 0 \\
1 & 0 & 0 & 0 
\end{bmatrix}
\cdot
\begin{bmatrix} 1.1 + 4.3i\\3.5 - 1.4i\\1.1 - 4.3i\\3.5 + 1.4i \end{bmatrix}$

$ $

$= \dfrac{W \cdot I_n^R \cdot \vec{v}_{'}}{n} = \dfrac{1}{4} \cdot\begin{bmatrix}
1 & 1 & 1 & 1\\
\overline{\omega} & \overline{\omega}^5 & \omega  & \omega^5\\
\overline{\omega}^2 & \overline{\omega}^2 & \omega^2 & \omega^2\\
\overline{\omega}^3 & \overline{\omega}^7 & \omega^3 & \omega^7\\
\end{bmatrix} 
\cdot
\begin{bmatrix} 1.1 + 4.3i\\3.5 - 1.4i\\1.1 - 4.3i\\3.5 + 1.4i \end{bmatrix}$

$ $

$= \dfrac{1}{4} \cdot
\begin{bmatrix} 
(1.1 + 4.3i) + (3.5 - 1.4i) + (1.1 - 4.3i) + (3.5 + 1.4i)\\
(1.1 + 4.3i)\overline{\omega} + (3.5 - 1.4i)\overline{\omega}^5 + (1.1 - 4.3i)\omega+ (3.5 + 1.4i)\omega^5 \\
(1.1 + 4.3i)\overline{\omega^2} + (3.5 - 1.4i)\overline{\omega^2} + (1.1 - 4.3i)\omega^2+ (3.5 + 1.4i)\omega^2 \\
(1.1 + 4.3i)\overline{\omega}^3 + (3.5 - 1.4i)\overline{\omega}^7 + (1.1 - 4.3i)\omega^3 + (3.5 + 1.4i)\omega^7 \\
\end{bmatrix}$

$ $

$= \dfrac{1}{4} \cdot
\begin{bmatrix} 
(1.1 + 4.3i) + (3.5 - 1.4i) + (1.1 - 4.3i) + (3.5 + 1.4i)\\
(1.1 + 4.3i) + (3.5 - 1.4i)\overline{\omega}^5 + (1.1 - 4.3i)\omega+ (3.5 + 1.4i)\omega^5 \\
(1.1 + 4.3i)\overline{\omega^2} + (3.5 - 1.4i)\overline{\omega^2} + (1.1 - 4.3i)\omega^2+ (3.5 + 1.4i)\omega^2 \\
(1.1 + 4.3i)\overline{\omega}^3 + (3.5 - 1.4i)\overline{\omega}^7 + (1.1 - 4.3i)\omega^3 + (3.5 + 1.4i)\omega^7 \\
\end{bmatrix}$

$ $

$= \dfrac{1}{4} \cdot
\begin{bmatrix} 
9.2\\
1.1(\overline{\omega} + \omega) + 4.3i(\overline{\omega} - \omega) +
3.5(\overline{\omega}^5 + \omega^5) - 1.4i(\overline{\omega}^5 - \omega^5)\\
1.1(\overline{\omega}^2 + {\omega}^2) + 4.3i(\overline{\omega}^2 - {\omega}^2) +
3.5(\overline{\omega}^2 + {\omega}^2) - 1.4i(\overline{\omega}^2 - {\omega}^2)\\
1.1(\overline{\omega}^3 + {\omega}^3) + 4.3i(\overline{\omega}^3 - {\omega}^3) +
3.5(\overline{\omega}^7 + {\omega}^7) - 1.4i(\overline{\omega}^7 - {\omega}^7)\\
\end{bmatrix}$

$ $

$= \dfrac{1}{4} \cdot
\begin{bmatrix} 
9.2\\
1.1\left(2\cos\dfrac{\pi}{4}\right) - 4.3i\left(2i\sin\dfrac{\pi}{4}\right) + 3.5\left(2\cos\dfrac{5\pi}{4}\right) + 1.4i\left(2i\sin\dfrac{5\pi}{4}\right)\\
1.1\left(2\cos\dfrac{\pi}{2}\right) - 4.3i\left(2i\sin\dfrac{\pi}{2}\right) + 3.5\left(2\cos\dfrac{\pi}{2}\right) + 1.4i\left(2i\sin\dfrac{\pi}{2}\right)\\
1.1\left(2\cos\dfrac{3\pi}{4}\right) - 4.3i\left(2i\sin\dfrac{3\pi}{4}\right) + 3.5 \left(2\cos\dfrac{7\pi}{4}\right) + 1.4i\left(2i\sin\dfrac{7\pi}{4}\right)\\
\end{bmatrix}$

$ $

$= \dfrac{1}{4} \cdot
\begin{bmatrix} 
9.2\\
1.1\left(2\dfrac{\sqrt{2}}{2}\right) + 4.3\left(2\dfrac{\sqrt{2}}{2}\right) + 3.5\left(-2\dfrac{\sqrt{2}}{2}\right) - 1.4\left(-2\dfrac{\sqrt{2}}{2}\right)\\
1.1(2\cdot 0) + 4.3(2\cdot1) + 3.5(2\cdot 0) - 1.4(2\cdot 1)\\
1.1\left(2-\dfrac{\sqrt{2}}{2}\right) + 4.3\left(2\dfrac{\sqrt{2}}{2}\right) + 3.5 \left(2\dfrac{\sqrt{2}}{2}\right) - 1.4\left(-2\dfrac{\sqrt{2}}{2}\right)\\
\end{bmatrix}$

$ $

$= 0.25 \cdot
\begin{bmatrix} 
9.2\\
1.1\sqrt{2} + 4.3{\sqrt{2}} - 3.5\sqrt{2} + 1.4\sqrt{2}\\
1.1(0) + 4.3(2) - 3.5(0) - 1.4(2)\\
-1.1\sqrt{2} + 4.3\sqrt{2} + 3.5\sqrt{2} + 1.4\sqrt{2}\\
\end{bmatrix} = 
\begin{bmatrix} 
2.3\\
0.825\sqrt{2}\\
1.45\\
2.025\sqrt{2}\\
\end{bmatrix} \approx (2.3, \text{ } 1.1657, \text{ } 1.45, \text{ } 2.8638)$

$ $

Convert the real number vector $\vec{m}$ into a scaled integer vector $\Delta\vec{m}$ by $\Delta$-scaling and rounding as follows:

$\Delta\vec{m} \approx \lceil \Delta \vec{m} \rfloor = \lceil 1024 \cdot (2.3, \text{ } 1.1657, \text{ } 1.45, \text{ } 2.8638) \rfloor = (2355, \text{ } 1195, \text{ } 1485, \text{ } 2933)$

$ $

Finally, $\vec{v} = (1.1 + 4.3i, \text{ } 3.5 - 1.4i)$ has been encoded into the plaintext polynomial $M(X)$ as follows: 

$\Delta M(X) = 2355 +  1195X + 1485X^2 + 2933X^3 \in \mathcal{R}_{\langle 4 \rangle}$

$ $

To decode $\vec{m}$, we compute:

$\vec{v}_{'} = \dfrac{W^T\cdot\Delta\vec{m}}{\Delta} = \begin{bmatrix}
1,\omega,\omega^2, \omega^3\\
1,\omega^3, \omega^6,\omega\\
1,\overline{\omega}, \overline{\omega^2}, \overline{\omega^3}\\
1,\overline{\omega^3}, \overline{\omega^6}, \overline{\omega}
\end{bmatrix}$
$\cdot \begin{bmatrix}
2355\\1195\\1485\\2933
\end{bmatrix}\cdot \dfrac{1}{1024}$

$= 
\begin{bmatrix}
1,\dfrac{\sqrt{2}}{2} + \dfrac{i\sqrt{2}}{2},i, -\dfrac{\sqrt{2}}{2} + \dfrac{i\sqrt{2}}{2}\\
1,-\dfrac{\sqrt{2}}{2} + \dfrac{i\sqrt{2}}{2}, -i,\dfrac{\sqrt{2}}{2} + \dfrac{i\sqrt{2}}{2}\\
1,\dfrac{\sqrt{2}}{2} - \dfrac{i\sqrt{2}}{2}, -i, -\dfrac{\sqrt{2}}{2} - \dfrac{i\sqrt{2}}{2}\\
1,-\dfrac{\sqrt{2}}{2} - \dfrac{i\sqrt{2}}{2}, i, \dfrac{\sqrt{2}}{2} - \dfrac{i\sqrt{2}}{2}
\end{bmatrix}$
$\cdot \begin{bmatrix}
2.2998046875\\1.1669921875\\1.4501953125\\2.8642578125
\end{bmatrix}$

$ $

$ \approx (1.0997+4.3007i, \text{ } 3.5000-1.4003i, \text{ } 1.0997-4.3007i, \text{ } 3.5000+1.4003i)$

$ $

Extract the first $\dfrac{n}{2} = 2$ elements in the Hermitian vector $\vec{v}_{'}$ to recover the input vector:

$(1.0997+4.3007i, \text{ } 3.5000-1.4003i)$

$\approx (1.1 + 4.3i, \text{ } 3.5 - 1.4i) = \vec{v}$ \textcolor{red}{  $\rhd$ The original input vector}

$ $

Because of the rounding drifts for converting square roots into integers, the decoded value is slightly different from the original input complex values. This is why CKKS is called an approximate homomorphic encryption.

$ $

\para{Source Code:} Examples of CKKS encoding can be executed by running \href{https://github.com/fhetextbook/fhe-textbook/blob/main/source%20code/ckks.py}{\underline{this Python script}}.

\subsection{Encryption and Decryption}
\label{subsec:ckks-enc-dec}

CKKS's encryption and decryption schemes are similar to BFV's encryption and decryption schemes (Summary~\ref*{subsec:bfv-enc-dec} in \autoref{subsec:bfv-enc-dec}).

\begin{tcolorbox}[title={\textbf{\tboxlabel{\ref*{subsec:ckks-enc-dec}} CKKS Encryption and Decryption}}]

\textbf{\underline{Initial Setup}:} 

$\Delta \text{ is a plaintext scaling factor for polynomial encoding}, \text{ } S \xleftarrow{\$} \mathcal{R}_{\langle n, \textit{tern} \rangle}$. The coefficients of the polynomial $S$ are ternary (i.e., $\{-1, 0, 1\}$).

\par\noindent\rule{\textwidth}{0.4pt}

\textbf{\underline{Encryption Input}:} $\Delta M \in \mathcal{R}_{\langle n, q \rangle}$, $A_i \xleftarrow{\$} \mathcal{R}_{\langle n, q \rangle}$, $E \xleftarrow{\xi_\sigma} \mathcal{R}_{\langle n, q \rangle}$

\begin{enumerate}
%\item Scale up $M \rightarrow \Delta M \text { } \in \mathcal{R}_{\langle n, q\rangle}$

\item Compute $B = -A \cdot S + \Delta M + E \text{ } \in \mathcal{R}_{\langle n,q \rangle}$

\item $\textsf{RLWE}_{S,\sigma}(\Delta M + E) = (A, B) \text{ } \in \mathcal{R}_{\langle n,q \rangle}^2$ 

\end{enumerate}

\par\noindent\rule{\textwidth}{0.4pt}

\textbf{\underline{Decryption Input}:} $\textsf{ct} = (A, B) \text{ } \in \mathcal{R}_{\langle n,q \rangle}^2$

$\textsf{RLWE}^{-1}_{S,\sigma}(\textsf{ct}) = \left\lceil\dfrac{B + A \cdot S \bmod q}{\Delta}\right\rfloor_{\frac{1}{\Delta}} = \left\lceil\dfrac{\Delta M + E}{\Delta}\right\rfloor_{\frac{1}{\Delta}} \approx M$

\textcolor{red}{ $\rhd$ $\lceil x\rfloor_{k}$ means rounding $x$ to the nearest multiple of $k$}
%\item Scale down $\Bigl\lceil \dfrac{ \Delta  M + E }{\Delta}\Bigr\rfloor = M  \text{ } \in \mathcal{R}_{\langle n, t \rangle}$

$ $

\textbf{\underline{Property of Approximate Decryption}:}
\begin{itemize}
\item Unlike BFV, CKKS's each plaintext value $m_i$ is originally not in a modulus ring, but a real number with infinite decimal digits. Therefore, it's not possible to exactly decrypt the ciphertext to the same original value.
\item If each coefficient of the noise $E$ is smaller than $\dfrac{\Delta}{2}$, then the decryption ensures the precision level with the multiple of $\dfrac{1}{\Delta}$.  
\end{itemize}

\end{tcolorbox}

In this section, we will often write $\textsf{RLWE}_{S,\sigma}(\Delta  M + E)$ as $\textsf{RLWE}_{S,\sigma}(\Delta  M)$ for simplicity, because $\textsf{RLWE}_{S,\sigma}(\Delta M + E) \approx \textsf{RLWE}_{S,\sigma}(\Delta M)$ (i.e., they decrypt to approximately the same message). Even in the case that we write $\textsf{RLWE}_{S,\sigma}(\Delta  M)$ instead of $\textsf{RLWE}_{S,\sigma}(\Delta  M + E)$, you should assume this as an encryption of $\Delta  M + E$ (i.e., the noise is included inside the scaled message).

\subsection{Ciphertext-to-Ciphertext Addition}
\label{subsec:ckks-add-cipher}

CKKS's ciphertext-to-ciphertext addition scheme is exactly the same as BFV's ciphertext-to-ciphertext addition scheme (Summary~\ref*{subsec:bfv-add-cipher} in \autoref{subsec:bfv-add-cipher}).

\begin{tcolorbox}[title={\textbf{\tboxlabel{\ref*{subsec:ckks-add-cipher}} CKKS Ciphertext-to-Ciphertext Addition}}]
$\textsf{RLWE}_{S, \sigma}(\Delta M^{\langle 1 \rangle} ) + \textsf{RLWE}_{S, \sigma}(\Delta M^{\langle 2 \rangle} ) $

$ = ( A^{\langle 1 \rangle}, \text{ } B^{\langle 1 \rangle}) + (A^{\langle 2 \rangle}, \text{ } B^{\langle 2 \rangle}) $

$ = ( A^{\langle 1 \rangle} + A^{\langle 2 \rangle}, \text{ } B^{\langle 1 \rangle} + B^{\langle 2 \rangle} ) $

$= \textsf{RLWE}_{S, \sigma}(\Delta(M^{\langle 1 \rangle} + M^{\langle 2 \rangle}) )$
\end{tcolorbox}

\subsection{Ciphertext-to-Plaintext Addition}
\label{subsec:ckks-add-plain}

CKKS's ciphertext-to-plaintext addition scheme is exactly the same as BFV's ciphertext-to-plaintext addition scheme (Summary~\ref*{subsec:bfv-add-plain} in \autoref{subsec:bfv-add-plain}).

\begin{tcolorbox}[title={\textbf{\tboxlabel{\ref*{subsec:ckks-add-plain}} CKKS Ciphertext-to-Plaintext Addition}}]
$\textsf{RLWE}_{S, \sigma}(\Delta M) + \Delta\Lambda $

$=  (A, \text{ } B) + \Delta\Lambda$

$=  (A, \text{ } B + \Delta\cdot\Lambda)$

$= \textsf{RLWE}_{S, \sigma}(\Delta (M + \Lambda) )$
\end{tcolorbox}

\subsection{Ciphertext-to-Ciphertext Multiplication}
\label{subsec:ckks-mult-cipher}

CKKS's ciphertext-to-ciphertext multiplication is partially different from that of BFV. In the case of BFV, its ciphertext modulus remains the same after each multiplication. On the other hand, CKKS reduces its ciphertext modulus size by 1 after each multiplication (which is equivalent to reducing its multiplicative level by 1). When the level reaches 0, no further multiplication can be performed (unless we bootstrap the modulus). This difference arises because the two schemes use different strategies in handling their plaintext scaling factors-- BFV's $\Delta = \left\lfloor\dfrac{q}{t}\right\rfloor$, whereas CKKS's $\Delta$ can be any value such that $\Delta \ll q_0$, where $q_0$ is the lowest multiplicative level's ciphertext modulus. However, both schemes use a similar relinearization technique. 

To make it easy to understand, we will explain CKKS's ciphertext-to-ciphertext multiplication based on this alternate version of RLWE (Theorem~\ref*{subsec:glwe-alternative} in \autoref{subsec:glwe-alternative}),
where the sign of the $AS$ term is flipped in the encryption and decryption formulas.

Suppose we have the following two (CKKS) RLWE ciphertexts:

$\textsf{RLWE}_{S, \sigma}(\Delta M^{\langle 1 \rangle}) = (A^{\langle 1 \rangle}, B^{\langle 1 \rangle})$, \text{ } where $B^{\langle 1 \rangle} = -A^{\langle 1 \rangle} \cdot S + \Delta M^{\langle 1 \rangle} + E^{\langle 1 \rangle}$

$\textsf{RLWE}_{S, \sigma}(\Delta M^{\langle 2 \rangle}) = (A^{\langle 2 \rangle}, B^{\langle 2 \rangle})$, \text{ } where $B^{\langle 2 \rangle} = -A^{\langle 2 \rangle} \cdot S + \Delta M^{\langle 2 \rangle} + E^{\langle 2 \rangle}$

$ $

\noindent RLWE ciphertext-to-ciphertext multiplication comprises the following 2 steps:

$ $

\begin{enumerate}
\item Find a formula for the \textit{synthetic} ciphertext that is equivalent to $\textsf{RLWE}_{S, \sigma}(\Delta^2 \cdot M^{\langle 1 \rangle} \cdot M^{\langle 2 \rangle})$ by leveraging the following congruence relation: 

$\textsf{RLWE}_{S, \sigma}(\Delta^2 \cdot M^{\langle 1 \rangle} \cdot M^{\langle 2 \rangle}) = \textsf{RLWE}_{S, \sigma}(\Delta \cdot M^{\langle 1 \rangle}) \cdot \textsf{RLWE}_{S, \sigma}(\Delta \cdot M^{\langle 2 \rangle})$ 

$ $

\item Rescale $\textsf{RLWE}_{S, \sigma}(\Delta^2 \cdot M^{\langle 1 \rangle} \cdot M^{\langle 2 \rangle})$ to $\textsf{RLWE}_{S, \sigma}(\Delta \cdot M^{\langle 1 \rangle} \cdot M^{\langle 2 \rangle})$.
\end{enumerate}

$ $

\noindent We will explain each of these steps.

\subsubsection{Synthetic Ciphertext Derivation}
\label{subsubsec:ckks-mult-cipher-relation}

The 1st step of RLWE ciphertext-ciphertext multiplication is to find a way to express the following congruence relation: 

$\textsf{RLWE}_{S, \sigma}(\Delta^2 \cdot M^{\langle 1 \rangle} \cdot M^{\langle 2 \rangle}) = \textsf{RLWE}_{S, \sigma}(\Delta \cdot M^{\langle 1 \rangle}) \cdot \textsf{RLWE}_{S, \sigma}(\Delta \cdot M^{\langle 2 \rangle})$

$ $

in terms of our following known values: $A^{\langle 1 \rangle}, \text{ } B^{\langle 1 \rangle}, \text{ } A^{\langle 2 \rangle}, \text{ } B^{\langle 2 \rangle}, \text{ } S$. First, notice that the following is true:

$ $

\hspace{-5mm}\noindent $\textsf{RLWE}^{-1}_{S, \sigma}(\text{ } \textsf{RLWE}_{S, \sigma}(\Delta^2 \cdot M^{\langle 1 \rangle} \cdot M^{\langle 2 \rangle}) \textsf{ } )= \textsf{RLWE}^{-1}_{S, \sigma}(\text{ }  \textsf{RLWE}_{S, \sigma}(\Delta \cdot M^{\langle 1 \rangle}) \text{ } ) \cdot \textsf{RLWE}^{-1}_{S, \sigma}(\text{ } \textsf{RLWE}_{S, \sigma}(\Delta \cdot M^{\langle 2 \rangle})  \text{ } )$

$ $

\noindent , because encrypting and decrypting the multiplication of two plaintexts should give the same result as decrypting two encrypted plaintexts and then multiplying them. As the encryption and decryption functions cancel out, we get the following:

\noindent $\Delta^2 \cdot M^{\langle 1 \rangle} \cdot M^{\langle 2 \rangle} \approx (\Delta \cdot M^{\langle 1 \rangle} + E^{\langle 1 \rangle} ) \cdot ( \Delta \cdot M^{\langle 2 \rangle} + E^{\langle 2 \rangle} ) $

$ = \textsf{RLWE}^{-1}_{S, \sigma}(\text{ }  \textsf{RLWE}_{S, \sigma}(\Delta \cdot M^{\langle 1 \rangle}) \text{ } ) \cdot \textsf{RLWE}^{-1}_{S, \sigma}(\text{ } \textsf{RLWE}_{S, \sigma}(\Delta \cdot M^{\langle 2 \rangle})  \text{ } )$

\textcolor{red}{ $\rhd$ where $(\Delta \cdot M^{\langle 1 \rangle} + E^{\langle 1 \rangle} ) \cdot ( \Delta \cdot M^{\langle 2 \rangle} + E^{\langle 2 \rangle} ) = \Delta^2\cdot M^{\langle 1 \rangle} \cdot M^{\langle 2 \rangle} + \Delta \cdot M^{\langle 1 \rangle} \cdot E^{\langle 2 \rangle} + \Delta \cdot M^{\langle 2 \rangle} \cdot E^{\langle 1 \rangle} + E^{\langle 1 \rangle} \cdot E^{\langle 2 \rangle}$, where $E^{\langle 1 \rangle} \cdot E^{\langle 2 \rangle}$ is small enough to be eliminated upon decryption, and $\Delta \cdot M^{\langle 1 \rangle} \cdot E^{\langle 2 \rangle}$ and $\Delta \cdot M^{\langle 2 \rangle} \cdot E^{\langle 1 \rangle}$ will be scaled down to $M^{\langle 1 \rangle} \cdot E^{\langle 2 \rangle}$ and $M^{\langle 2 \rangle} \cdot E^{\langle 1 \rangle}$ upon modulus switch later, becoming sufficiently small to be eliminated during decryption}

$ $

Remember from \autoref{subsec:glwe-alternative} the following: 

$\textsf{RLWE}^{-1}_{S,\sigma}\bm{(} \text{ } C = (A, B) \text{ } \bm{)} = \Delta  M + E = B + A\cdot S$ 

$ $

Thus, the above congruence relation can be rewritten as follows:

$ $

$\Delta^2 \cdot M^{\langle 1 \rangle} \cdot M^{\langle 2 \rangle}$ \textcolor{red}{$ \approx (\Delta \cdot M^{\langle 1 \rangle} + E^{\langle 1 \rangle} ) \cdot ( \Delta \cdot M^{\langle 2 \rangle} + E^{\langle 2 \rangle} )$}

$ = (B^{\langle 1 \rangle}  + A^{\langle 1 \rangle} \cdot S - E^{\langle 1 \rangle}) \cdot (B^{\langle 2 \rangle}  + A^{\langle 2 \rangle} \cdot S - E^{\langle 2 \rangle})$

$ \approx (B^{\langle 1 \rangle}  + A^{\langle 1 \rangle} \cdot S) \cdot (B^{\langle 2 \rangle}  + A^{\langle 2 \rangle} \cdot S)$

$ = B^{\langle 1 \rangle}B^{\langle 2 \rangle}  + (B^{\langle 2 \rangle}A^{\langle 1 \rangle} + B^{\langle 1 \rangle}A^{\langle 2 \rangle}) \cdot S  + (A^{\langle 1 \rangle}S)\cdot(A^{\langle 2 \rangle}S)$

$ $

$ = \underbrace{B^{\langle 1 \rangle}B^{\langle 2 \rangle}}_{D_0}  + \underbrace{(B^{\langle 2 \rangle}A^{\langle 1 \rangle} + B^{\langle 1 \rangle}A^{\langle 2 \rangle})}_{D_1} \cdot S + \underbrace{(A^{\langle 1 \rangle} \cdot A^{\langle 2 \rangle})}_{D_2} \cdot \underbrace{(S \cdot S)}_{ S^2}$

%\textcolor{red}{ $\rhd$ where $\otimes$ is a vector outer-product such that for $\vec{u} = (u_0, u_1, \cdots, u_{k-1})$ and $\vec{v} = (v_0, v_1, \cdots, v_{k-1})$,}

%\textcolor{red}{ \text{ } $\vec{u} \otimes \vec{v} = (u_0v_0, u_0v_1, \cdots, \text{ } u_1v_0, u_1v_1, \cdots, \text{ } u_{k-1}v_0, u_{k-1}v_1,  \cdots, u_{k-1}v_{k-1})$}

$= {D_0 + D_1\cdot S} + D_2\cdot S^2$

$ $

$= \textsf{RLWE}_{S, \sigma}^{-1}\bm{(}\text{ } C_\alpha = (D_1, D_0) \text{ }\bm{)} + D_2\cdot S^2$ \textcolor{red}{ \text{ } \# since $D_0 + D_1\cdot S = \textsf{RLWE}_{S, \sigma}^{-1}\bm{(}\text{ } C_\alpha=(D_1, D_0) \text{ } \bm{)}$ }

$ $

In the final step above, we converted $D_0 + D_1\cdot S$ into $\textsf{RLWE}_{S, \sigma}^{-1}\bm{(}\text{ } C_\alpha=(D_1, D_0) \text{ } \bm{)}$, where $C_\alpha$ is the synthetic RLWE ciphertext $(D_1, D_0)$ encrypted by $S$. Similarly, our next task is to derive a synthetic RLWE ciphertext $C_\beta$ such that $D_2\cdot S^2 = \textsf{RLWE}_{S, \sigma}^{-1}(C_\beta)$. The reason why we want this synthetic ciphertext is that we do not want the square of $S$ (i.e., $ S^2$), because if we continue to keep $ S^2$, then over more consequent ciphertext-to-ciphertext multiplications, this term will aggregate exponentially growing bigger exponents such as $S^4, S^8, \cdots...$, which would exponentially increase the computational overhead of decryption. In the next subsection, we will explain how to derive the synthetic RLWE ciphertext $C_\beta$ such that $D_2\cdot S^2 = \textsf{RLWE}_{S, \sigma}^{-1}(C_\beta)$.

\subsubsection{Relinearization  Method 1 -- Ciphertext Decomposition}
\label{subsubsec:relinearization-gadget-decomposition}

As explained in BFV's ciphertext-to-ciphertext multiplication (\autoref{subsubsec:bfv-mult-cipher-relinearization}), relinearization is a process of converting the polynomial triplet $(D_0, D_1, D_2) \in \mathcal{R}_{\langle n, q \rangle}^{3}$, which can be decrypted into $\Delta M$ using $S$ and $S^2$ as keys, into the polynomial pairs $(C_\alpha, C_\beta)\in \mathcal{R}_{\langle n, q \rangle}^{2}$, which can be decrypted into the same $\Delta M$ by using $S$ as key. In the previous subsection, we learned that we can convert $D_0$ and $D_1$ into $C_\alpha$ simply by viewing $D_0$ and $D_1$ as  $C_\alpha = (D_1, D_0)$. The process of converting $D_2$ into $C_\beta$ is exactly the same as the technique explained in \autoref{subsubsec:bfv-mult-cipher-relinearization}, which applies the gadget decomposition (\autoref{subsec:gadget-decomposition}) on $D_2$ and computes an inner product with the RLev encryption (\autoref{sec:glev}) of $S^2$. Specifically, we compute the following:

$\textsf{RLWE}_{S, \sigma}^{-1}(C_\beta = \bm{\langle} \textsf{Decomp}^{\beta, l}(D_2), \text{ } \textsf{RLev}_{S, \sigma}^{\beta, l}( S^2) \bm{\rangle} \bm{)}$ \textcolor{red}{ $\rhd$ the scaling factors of $\textsf{RLev}_{S, \sigma}^{\beta, l}( S^2)$ are all 1}

$= D_{2,1}(E_1' +  S^2\dfrac{q}{\beta}) + D_{2,2}(E_2' +  S^2\dfrac{q}{\beta^2}) + \cdots + D_{2,l}(E_l' +  S^2\dfrac{q}{\beta^l})$

$= \sum\limits_{i=1}^{l} (E_i'\cdot D_{2,i}) +  S^2\cdot(D_{2,1}\dfrac{q}{\beta} + D_{2,2}\dfrac{q}{\beta^2} + \cdots + D_{2, l}\dfrac{q}{\beta^l})$

=============

$= D_{2,1}(E_1' +  S^2 \beta^0) + D_{2,2}(E_2' +  S^2 \beta^1) + \cdots + D_{2,l}(E_l' +  S^2 \beta^{l-1})$

$= \sum\limits_{i=1}^{l} (E_i'\cdot D_{2,i}) +  S^2\cdot(D_{2,1}\beta^0 + D_{2,2}\beta^1 + \cdots + D_{2, l}\beta^{l-1})$

=============

$= \sum\limits_{i=1}^{l} \epsilon_{i} + D_2\cdot S^2$ \textcolor{red}{ \text{ } \# where $\epsilon_i = E_i'\cdot D_{2,i}$}

$\approx D_2\cdot S^2$ \textcolor{red}{ \text{ } \# because $\sum\limits_{i=1}^{l} \epsilon_{i} \ll D_2\cdot E''$ (where $E''$ is the noise embedded in $\textsf{RLWE}_{S, \sigma}\bm(S^2\bm)$}

$ $

Finally, we get the following relation:

$\textsf{RLWE}_{S, \sigma}(\Delta^2 \cdot M^{\langle 1 \rangle} \cdot M^{\langle 2 \rangle}) \approx C_\alpha + C_\beta$ \text{ } , where \text{ } $C_\alpha = (D_1, D_0), \text{ } C_\beta = \bm{\langle}  \textsf{Decomp}^{\beta, l}(D_2), \text{ } \textsf{RLev}_{S, \sigma}^{\beta, l}( S^2) \bm{\rangle}$

$ $

In the next subsection, we introduce another (older) relinearization technique. 

\subsubsection{Relinearization Method 2 -- Ciphertext Modulus Switch}
\label{subsubsec:relinearization-modulus-switch}

At the setup stage of the RLWE scheme, we craft a special pair of polynomials modulo $q$ as follows: 

$A'  \xleftarrow{\$} \mathcal{R}_{\langle n,q\rangle }^{k}$

$E' \xleftarrow{\sigma} \mathcal{R}_{\langle n,q\rangle }$

$\mathit{evk} = (A', \text{ } -A'\cdot S + E' +  S^2) \in \mathcal{R}_{\langle n, q\rangle}^2$

$\mathit{evk}$ is called an evaluation key, which is essentially a RLWE ciphertext of $ S^2$ encrypted by the secret key $S$ without any scaling factor $\Delta$. Remember that our goal is to find a synthetic RLWE ciphertext $C_\beta$ such that decrypting it gives us $D_2\cdot S^2$, that is: $\textsf{RLWE}_{S, \sigma}^{-1}(C_\beta) = D_2\cdot S^2$. Let's suppose  that $C_\beta = D_2 \cdot \mathit{evk}$. Then, decrypting $C_\beta$ gives us the following:

$\textsf{RLWE}_{S, \sigma}^{-1}\bm{(}C_\beta = (D_2 \cdot \mathit{evk})\bm{)} = \textsf{RLWE}_{S, \sigma}^{-1}\bm{(} \text{ } C = (D_2A', \text{ } -D_2A'\cdot S + D_2E' + D_2\cdot S^2) \text{ } \bm{)}$

$= D_2A'\cdot S - D_2A'\cdot S + D_2E' + D_2\cdot S^2$

$= D_2E' + D_2\cdot S^2$

$ $

But unfortunately, $D_2E' + D_2\cdot S^2 \not\approx D_2\cdot S^2$, because $D_2E' \not\approx 0$ (as $D_2$ is not necessarily a small number).This is because $D_2 = A^{\langle 1 \rangle} \cdot A^{\langle 2 \rangle}$, $D_2E'$ can be any arbitrary value between $[0, q)$.

To solve the above problem, we modify the evaluation key as a set of polynomials in big modulo $g$ as follows:

$A'  \xleftarrow{\$} \mathcal{R}_{\langle n,q\rangle }^{k}$

$E' \xleftarrow{\sigma} \mathcal{R}_{\langle n,q\rangle }$

$g \xleftarrow{\$} \mathbb{Z}_{q_L^2}$ \textcolor{red}{  $\rhd$ where $g$ is some large integer power of 2, $q_L$ is the largest modulo before any relinearization}

$\mathit{evk_g} = (A', -A'\cdot S + E' + g S^2) \in \mathcal{R}_{\langle n, gq\rangle}^2$

$ $

$\mathit{evk_g}$ is essentially an RLWE ciphertext of $g S^2$ encrypted by $S$. We can derive the following:

$\mathit{evk_g} = (A', -A'\cdot S + E' + g S^2) \in \mathcal{R}_{\langle n, gq\rangle}^2$

$= (A'  \text{ mod } gq, \text{ } -A'\cdot S + E' + g S^2 \text{ mod } gq)$

$= (A' + k_2gq, \text{ } -A'\cdot S + E' + g S^2 + k_1gq)$  \text{ } (for some integers $k_1, k_2$)

$ $

Note that $D_2 = A^{\langle 1 \rangle} \cdot A^{\langle 2 \rangle} \in \mathcal{R}_{\langle n, q \rangle}$

$= D_2 \text{ mod } q$

$= D_2 + k_3q$ \text{ } (for some integer $k_3$)

$ $

Now, let's multiply $D_2$ to each component of $\mathit{evk_g}$ as follows: 

$(A' + k_2gq, \text{ } -A'\cdot S + E' + g S^2 + k_1gq) \cdot (D_2 + k_3q)$

$= (\text{ } D_2A' + D_2k_2gq + k_3qA' + k_3qk_2gq,$ 

\text{ } $ -D_2A'\cdot S + D_2E' + gD_2\cdot S^2 + D_2k_1gq
-k_3qA'\cdot S + k_3qE' + k_3qg S^2 + k_3qk_1gq)$

$ $

Now, we switch the modulus of this RLWE ciphertext from $gq \rightarrow q$ based on the technique in \autoref{subsec:modulus-switch-glwe}:

%$(-D_2A'\cdot S + D_2E' + D_2g S^2 + D_2k_1gq -k_3qA'\cdot S + k_3qE' + k_3qg S^2 + k_3qk_1gq,$ 

%\text{ } $D_2A' + D_2k_2gq + k_3qA' + k_3qk_2gq)$

%$ $

$ \Bigg(\Bigg\lceil\dfrac{D_2A'}{g}\Bigg\rfloor + \Bigg\lceil\dfrac{D_2k_2gq}{g}\Bigg\rfloor + \Bigg\lceil\dfrac{k_3qA'}{g}\Bigg\rfloor + \Bigg\lceil\dfrac{k_3qk_2gq}{g}\Bigg\rfloor, \text{ } \text{ }-\Bigg\lceil\dfrac{D_2A'\cdot S}{g}\Bigg\rfloor + \Bigg\lceil\dfrac{D_2E'}{g}\Bigg\rfloor + \Bigg\lceil\dfrac{gD_2\cdot S^2}{g}\Bigg\rfloor + \Bigg\lceil\dfrac{D_2k_1gq}{g}\Bigg\rfloor
-\Bigg\lceil\dfrac{k_3qA'\cdot S}{g}\Bigg\rfloor + \Bigg\lceil\dfrac{k_3qE'}{g}\Bigg\rfloor + \Bigg\lceil\dfrac{k_3qg S^2}{g}\Bigg\rfloor + \Bigg\lceil\dfrac{k_3qk_1gq}{g}\Bigg\rfloor\Bigg)$

$ $

$=\Bigg(\Bigg\lceil\dfrac{D_2A'}{g}\Bigg\rfloor + D_2k_2q +\Bigg\lceil\dfrac{k_3qA'}{g}\Bigg\rfloor + k_3qk_2q,$

$  \text{ } -\Bigg\lceil\dfrac{D_2A'\cdot S}{g}\Bigg\rfloor + \Bigg\lceil\dfrac{D_2E'}{g}\Bigg\rfloor + D_2\cdot S^2 + D_2k_1q
-\Bigg\lceil\dfrac{k_3qA'\cdot S}{g}\Bigg\rfloor + \Bigg\lceil\dfrac{k_3qE'}{g}\Bigg\rfloor + k_3q S^2 + k_3qk_1q\Bigg)$

$ $

$= \Bigg(\Bigg\lceil\dfrac{D_2A'}{g}\Bigg\rfloor +\Bigg\lceil\dfrac{k_3qA'}{g}\Bigg\rfloor \text{ mod } q,  \text{ } -\Bigg\lceil\dfrac{D_2A'\cdot S}{g}\Bigg\rfloor + \Bigg\lceil\dfrac{D_2E'}{g}\Bigg\rfloor + D_2\cdot S^2
-\Bigg\lceil\dfrac{k_3qA'\cdot S}{g}\Bigg\rfloor + \Bigg\lceil\dfrac{k_3qE'}{g}\Bigg\rfloor \text{ mod } q\Bigg)$

$ $

$= C_\beta \in \mathcal{R}_{n, q}^2$

Now, we finally got $C_\beta$ which is in the form of RLWE ciphertext modulo $q$. Remember that our goal is to express $D_2\cdot S^2$ as a decryption of RLWE ciphertext. If we treat $C_\beta$ as a synthetic RLWE ciphertext and decrypt it, we get the following:

\noindent $\textsf{RLWE}^{-1}_{S, \sigma}(C_\beta)$ \textcolor{red}{  $\rhd$ where $C_\beta$ is treated as a synthetic RLWE ciphertext}

\noindent $=  \textsf{RLWE}^{-1}_{S, \sigma}\Bigg( \Bigg(-\Bigg\lceil\dfrac{D_2A'\cdot S}{g}\Bigg\rfloor + \Bigg\lceil\dfrac{D_2E'}{g}\Bigg\rfloor + D_2\cdot S^2
-\Bigg\lceil\dfrac{k_3qA'\cdot S}{g}\Bigg\rfloor + \Bigg\lceil\dfrac{k_3qE'}{g}\Bigg\rfloor, \text{ } \Bigg\lceil\dfrac{D_2A'}{g}\Bigg\rfloor +\Bigg\lceil\dfrac{k_3qA'}{g}\Bigg\rfloor\Bigg)\Bigg)$

$ $

\noindent $=  -\Bigg\lceil\dfrac{D_2A'\cdot S}{g}\Bigg\rfloor + \Bigg\lceil\dfrac{D_2E'}{g}\Bigg\rfloor + D_2\cdot S^2
-\Bigg\lceil\dfrac{k_3qA'\cdot S}{g}\Bigg\rfloor + \Bigg\lceil\dfrac{k_3qE'}{g}\Bigg\rfloor + \Bigg\lceil\dfrac{D_2A'}{g}\Bigg\rfloor\cdot S + \Bigg\lceil\dfrac{k_3qA'}{g}\Bigg\rfloor\cdot S$

$ $

\noindent $\approx \Bigg\lceil\dfrac{D_2E'}{g}\Bigg\rfloor + D_2\cdot S^2
 + \Bigg\lceil\dfrac{k_3qE'}{g}\Bigg\rfloor$ \textcolor{red}{  $\rhd$ $-\Bigg\lceil\dfrac{D_2A'\cdot S}{g}\Bigg\rfloor + \Bigg\lceil\dfrac{D_2A'}{g}\Bigg\rfloor\cdot S = -\Bigg\lceil\dfrac{k_3qA'\cdot S}{g}\Bigg\rfloor + \Bigg\lceil\dfrac{k_3qA'}{g}\Bigg\rfloor\cdot S \approx 0$}

$ $

\noindent $\approx D_2\cdot S^2$ \textcolor{red}{  $\rhd$ 
  $\Bigg\lceil\dfrac{D_2E'}{g}\Bigg\rfloor \approx 0$, \text{ } $\Bigg\lceil\dfrac{k_3qE'}{g}\Bigg\rfloor \approx 0$}

$ $

As shown in the above, decrypting $C_\beta$ gives us $D_2\cdot S^2$. Therefore, we reach the following conclusion: 

$\Delta^2 \cdot M^{\langle 1 \rangle} \cdot M^{\langle 2 \rangle} \approx \textsf{RLWE}^{-1}_{S, \sigma}(C_\alpha) + \textsf{RLWE}^{-1}_{S, \sigma}(C_\beta)$ \text{ } , where $C_\alpha = (D_1, D_0), \text{ } C_\beta = \Bigg\lceil\dfrac{D_2 \cdot \mathit{evk_g}}{g}\Bigg\rfloor$

$ $

Therefore, we finally get the following congruence relation:

$\textsf{RLWE}_{S, \sigma}(\Delta^2 \cdot M^{\langle 1 \rangle} \cdot M^{\langle 2 \rangle}) \approx C_\alpha + C_\beta$ \text{ } , where $C_\alpha = (D_1, D_0), \text{ } C_\beta = \Bigg\lceil\dfrac{D_2 \cdot \mathit{evk_g}}{g}\Bigg\rfloor$

$ $

$ $

Our last step of ciphertext-to-ciphertext multiplication is to convert $\textsf{RLWE}_{S, \sigma}(\Delta^2 \cdot M^{\langle 1 \rangle} \cdot M^{\langle 2 \rangle})$ into $\textsf{RLWE}_{S, \sigma}(\Delta \cdot M^{\langle 1 \rangle} \cdot M^{\langle 2 \rangle})$, because if the result of ciphertext-to-ciphertext multiplication is $M^{\langle 1 \rangle} \cdot M^{\langle 2 \rangle} = M^{\langle 3 \rangle}$, then for consistency purposes, the resulting RLWE ciphertext is supposed to be: 

$\textsf{RLWE}_{S, \sigma}(\Delta \cdot M^{\langle 1 \rangle} \cdot M^{\langle 2 \rangle}) = \textsf{RLWE}_{S, \sigma}(\Delta \cdot M^{\langle 3 \rangle})$

We will explain this process in the next subsection.

\subsubsection{Rescaling}
\label{subsubsec:ckks-mult-cipher-rescale}

To convert $\textsf{RLWE}_{S, \sigma}(\Delta^2 \cdot M^{\langle 1 \rangle} \cdot M^{\langle 2 \rangle})$ into $\textsf{RLWE}_{S, \sigma}(\Delta \cdot M^{\langle 1 \rangle} \cdot M^{\langle 2 \rangle})$, we cannot simply divide the ciphertext $\textsf{RLWE}_{S, \sigma}(\Delta^2 \cdot M^{\langle 1 \rangle} \cdot M^{\langle 2 \rangle})$ by $\Delta$, because as explained in \autoref{subsec:modulo-division}, modulo arithmetic does not support direct division. Multiplying the RLWE ciphertext by $\Delta^{-1}$ (i.e., an inverse of $\Delta$) does not work either, because the only useful property we can use for inverse multiplication is: $a \cdot a^{-1} \equiv 1$. If an inverse is multiplied to any other values other than its counterpart, the result is an arbitrary value. For example, if $\Delta^{-1}$ is multiplied to a noise (i.e., $\Delta^{-1}E$), then the result can be a very huge value. Thus, multiplying the RLWE ciphertext by $\Delta^{-1}$ does not help due to the unpredictable result of the noise term. 

The safest way to convert $\textsf{RLWE}_{S, \sigma}(\Delta^2 \cdot M^{\langle 1 \rangle} \cdot M^{\langle 2 \rangle})$ into $\textsf{RLWE}_{S, \sigma}(\Delta \cdot M^{\langle 1 \rangle} \cdot M^{\langle 2 \rangle})$ is modulus switch (\autoref{subsec:modulus-switch-glwe}), which is essentially modulo rescaling (\autoref{sec:modulus-rescaling}). For this to work, the RLWE setup stage should design the ciphertext domain $q$ as $q_0\cdot\Delta^{L}$, where $L$ is denoted as the level of multiplication, and $q_0 \gg \Delta$ (which is important for the accuracy of homomorphic modulo reduction during bootstrapping in \autoref{subsec:ckks-bootstrapping}). Upon each ciphertext-to-ciphertext multiplication, we switch the modulus of the RLWE ciphertext from $q_0\cdot\Delta^{i} \rightarrow q_0\cdot\Delta^{i-1}$, which effectively converts the plaintext's squared scaling factor $\Delta^2$ (in $\textsf{RLWE}_{S, \sigma}(\Delta^2 \cdot M^{\langle 1 \rangle} \cdot M^{\langle 2 \rangle})$) into $\Delta$ (in $\textsf{RLWE}_{S, \sigma}(\Delta \cdot M^{\langle 1 \rangle} \cdot M^{\langle 2 \rangle})$). Once the RLWE ciphertext's level reaches 0 (i.e., ciphertext modulus $q_0$), we cannot do any more ciphertext-to-ciphertext multiplication, in which case we need a special process called bootstrapping to re-initialize the modulus level to $L$. 

However, one problem with this setup is that $\Delta^L$ will be a huge number. Performing homomorphic addition or multiplication over modulo $\Delta^L$ is computationally expensive. To reduce the overhead of ciphertext size, we use the Chinese remainder theorem (\autoref{sec:chinese-remainder}): given an integer $x \text{ mod } W$ where $W$ is a multiplication of $L+1$ co-primes such that $W = w_0w_1w_2w_3\cdots w_L$, the following congruence relationships hold:

$x \equiv d_0 \text{ mod } w_0$ 

$x \equiv d_1 \text{ mod } w_1$ 

$x \equiv d_2 \text{ mod } w_2$ 

$\vdots$

$x \equiv d_L \text{ mod } w_L$ 

$ $

, where $x = \sum\limits_{m=0}^{L} d_my_mz_m \text{ mod } W, \text{ } \text{ } y_m = \dfrac{W}{w_m}, \text{ } \text{ } z_m = y_m^{-1} \text{ mod } w_m$, and $w_0 = q_0$ 

$ $

In other words, $x \bmod W$ can be isomorphically mapped to a vector of smaller numbers $(d_0, d_1, \cdots, d_l)$ each in modulo $w_0, w_1, \cdots, w_l$, addition/multiplication with big elements in modulo $W$ can be done by using their encoded smaller-magnitude CRT vectors element-wise, and later decode the intended big-number result. By leveraging this property, we design the CKKS scheme's maximal ciphertext modulus as $W= \prod\limits_{m=0}^{L}w_m$, where $L$ is the maximum multiplicative level, $w_0 = q_0 \gg \Delta$, and all other $w_i \approx \Delta$. Then, whenever reaching from the $l$-th to the next lower $l-1$-th multiplicative level, we switch its modulus from $q=\prod\limits_{m=0}^{l}w_m$ to $\hat q =\prod\limits_{m=0}^{l - 1}w_m$ as follows:

$(\text{ }C = (A, B)\text{ }) \in \mathcal{R}_{\langle n, q \rangle} \rightarrow \textsf{RLWE}_{S, \sigma}(\text{ }\hat{C} = (\hat{A}, \hat{B})\text{ }) \in \mathcal{R}_{\langle n, \hat q \rangle}$

$q = \prod\limits_{m=0}^{l}w_m$, \textcolor{red}{ \text{ } \# where all $w_m$ are prime numbers,  $w_0 = q_0 \gg \Delta\cdot p$ to ensure the scaled plaintext $\Delta M$ during homomorphic operations never overflows the ciphertext modulus even at the lowest multiplicative level, and all other $w_i \approx \Delta$}

$\hat{q} = \dfrac{q}{w_l}$

$\hat{A_i} = \left\lceil\dfrac{\hat q}{q}\cdot A_i\right\rfloor = \hat{a}_{i,0} + \hat{a}_{i,1}X + \hat{a}_{i,2}X^2 + \cdots + \hat{a}_{i, {n-1}}X^{n-1}$, where each $\hat{a}_{i,j} = \Big\lceil a_{i,j}\dfrac{\hat{q}}{q} \Big\rfloor = \Big\lceil \dfrac{a_{i,j}}{w_l} \Big\rfloor \in \mathbb{Z}_{\hat{q}}$

$\hat{B} = \left\lceil\dfrac{\hat q}{q}\cdot B\right\rfloor = \hat{b}_0 + \hat{b}_1X + \hat{b}_2X^2 + \cdots + \hat{b}_{n-1}X^{n-1}$, where each $\hat{b}_j = \Big\lceil b_j\dfrac{\hat{q}}{q} \Big\rfloor = \Big\lceil \dfrac{b_j}{w_l} \Big\rfloor \in \mathbb{Z}_{\hat{q}}$

$\textsf{RLWE}_{{S},\sigma}(\Delta  M) = (\hat{A}, \hat{B}) \in \mathcal{R}_{\langle n, \hat{q} \rangle}$

$ $

The above update of $(\{A_i\}_{i=0}^{k-1}, B)$ to $(\{\hat{A}_i\}_{i=0}^{k-1}, \hat B)$ effectively changes $\Delta, E$ to $\hat\Delta, \hat E$ as follows:

$\hat{E} = \hat{e}_0 + \hat{e}_1X + \hat{e}_2X^2 + \cdots + \hat{e}_{n-1}X^{n-1}$, where each $\hat{e}_j = \Big\lceil e_j\dfrac{\hat{q}}{q} \Big\rfloor = \Big\lceil \dfrac{e_j}{w_l} \Big\rfloor \in \mathbb{Z}_{\hat{q}}$

$\hat{\Delta} = \left\lceil\Delta^2\dfrac{\hat{q}}{q}\right\rfloor = \left\lceil\dfrac{\Delta^2}{w_l}\right\rfloor \approx \Delta$ \textcolor{red}{ $\rhd$ If we treat $\hat\Delta$ as $\Delta$, the rounding error slightly increases the noise $\hat E$ to $\hat E + E_\Delta$, while the decryption of $(\hat A, \hat B)$ outputs the same $M$}

$ $

Note that after the rescaling, the plaintext scaling factor of $(\hat A, \hat B)$ is also updated to $\hat{\Delta}$. Meanwhile, $M$ and $S$ stay the same as before.

$ $

After we switch the modulus of the ciphertext $C$ from $q \rightarrow \hat{q}$ by multiplying $\dfrac{\hat{q}}{q}$ to $A$ and $B$, the encrypted original plaintext term $\Delta^2 M^{\langle 1 \rangle} M^{\langle 2 \rangle}$ will become $\Delta^2 M^{\langle 1 \rangle} M^{\langle 2 \rangle} \cdot \dfrac{\hat{q}}{q} = \dfrac{\Delta^2 M^{\langle 1 \rangle} M^{\langle 2 \rangle}}{w_l} = (\Delta + \epsilon_\Delta)\cdot M^{\langle 1 \rangle} M^{\langle 2 \rangle}$, where $\epsilon_\Delta \approx 0$, because as explained before, we chose $\{w_i\}_{i=1}^{L}$ such that $w_i \approx \Delta$. Therefore, $(\Delta + \epsilon_\Delta)\cdot M^{\langle 1 \rangle} M^{\langle 2 \rangle} = \Delta M^{\langle 1 \rangle} M^{\langle 2 \rangle} + \epsilon_\Delta M^{\langle 1 \rangle} M^{\langle 2 \rangle}$, where $\epsilon_\Delta M^{\langle 1 \rangle} M^{\langle 2 \rangle} \approx 0$, which becomes part of the noise term of the modulus-switched (i.e., rescaled) new ciphertext $\hat{C}$. 

$ $

The benefit of this design of the CRT (Chinese remainder problem)-based ciphertext modulus and rescaling is that we can isomorphically decompose the huge coefficients (bigger than 64 bits) of polynomials in ciphertexts into $l$-dimensional Chinese remainder vectors (Theorem~\ref{sec:chinese-remainder}.2 in \autoref{sec:chinese-remainder}) and perform element-wise addition or multiplication for computing coefficients over the small vector elements. This promotes computational efficiency for homomorphic addition and multiplication over a large ciphertext modulus (although the number of addition/multiplication operations increases). This technique is called Residue Number System (RNS). When CRT is used in ciphertexts, the security regarding the ciphertext modulus depends on the smallest and the largest CRT elements. 

\para{Initial Scaling Factor $\bm\Delta$ Upon Encryption:} To support multi-level multiplicative levels (using CRT), we need to modify the generic scaling factor setup presented in Summary~\ref*{subsec:glwe-enc} (\autoref{subsec:glwe-enc}) from $\Delta = \dfrac{q}{t}$ to $\Delta = w_L$. 

\para{Noise Growth:} Upon each step of rescaling during ciphertext-ciphertext multiplication, the noise also gets scaled down by $\dfrac{1}{\Delta}$ (or by $\dfrac{1}{w_l}$ at multiplicative level $l$ in the case of using CRT). 
Therefore, rescaling reduces the absolute magnitude of the noise by a factor of $\Delta$ (or $w_l$). However, during each ciphertext-to-ciphertext multiplication, the encrypted (noisy) plaintext is $(\Delta M_1 + E_1)\cdot(\Delta M_2 + E_2) = \Delta^2 M_1 M_2 + \Delta\cdot (M_1E_2 + M_2E_1) + E_1E_2$, and rescaling roughly has the effect of dividing this by $\Delta$, which approximately gives us $\Delta M_1 M_2 + M_1E_2 + M_2E_1 + \dfrac{E_1E_2}{\Delta}$. Because of the $(M_1E_2 + M_2E_1)$ term, the noise actually grows compared to before ciphertext-to-ciphertext multiplication. Therefore, ciphertext-to-ciphertext multiplication inevitably increases the noise. 

\subsubsection{Comparing BFV and CKKS Bootstrapping}
\label{subsubsec:bfv-bootstrapping-ckks-comparison}

CKKS bootstrapping shares several common aspects with BFV bootstrapping. CKKS's \textsf{ModRaise} and \textsf{Homomorphic Decryption} steps are equivalent to BFV's \textsf{Homomorphic Decryption (without modulo-$q$ reduction)} step. BFV homomorphically multiplies polynomial $A$ and $B$ whose coefficients are in $\mathbb{Z}_{p^e}$ with the encrypted secret key whose ciphertext modulus is $q$, which generates the modulo wrap-around coefficient values $p^eK$. Similarly, CKKS coefficients are in $\mathbb{Z}_{q_0}$ with the encrypted secret key whose ciphertext modulus is $q_L$, which generates the modulo wrap-around coefficient values $q_0K$. However, they use different strategies to handle their modulo wrap-around values. CKKS uses evaluation of the sine function having a period of $q_0$ to approximately eliminate $q_0K$ (i.e., \textsf{EvalExp}). On the other hand, BFV uses digit extraction to scale down $p^eK$ by $p^{e-1}$ and then treats the remaining small $pK$ as part of the modulo wrap-around value of the plaintext. The requirement of the digit extraction algorithm is that the plaintext inputs should be represented as base-$p$ values, and because of this, BFV bootstrapping includes the initial step of modulus switch from $q \rightarrow p^e$, where $p^e$ is used as the plaintext modulus after homomorphic decryption. 

Both BFV and CKKS use the same strategy for their \textsf{CoeffToSlot}, \textsf{SlotToCoeff}, and \textsf{Scaling Factor Re-interpretation} steps. 

\para{Multiplicative Level:} A critical difference between BFV and CKKS is that in BFV, the ciphertext modulus $q$ stays the same after ciphertext-ciphertext multiplication, and there is no restriction on the number of ciphertext-ciphertext multiplications. On the other hand, in CKKS, the ciphertext modulus changes from $q_l \rightarrow q_{l-1}$ after each multiplication, and when it reaches $q_0$, no more multiplication can be done, unless we reset the ciphertext modulus to $q_L$ by using the modulus bootstrapping technique (\autoref{subsec:ckks-bootstrapping}). 

\para{Limitation of Noise Handling:} Although CKKS's rescaling during ciphertext-to-ciphertext multiplication reduces the magnitude of noise $E$ by $\Delta$, it also reduces the ciphertext modulus by the same amount, and thus the relative noise-to-ciphertext-modulus ratio does not get decreased by rescaling. In order to reduce (or reset) the noise-to-modulus ratio, we need to do bootstrapping (\autoref{subsec:ckks-bootstrapping}), which will be explained at the end of this section.

\subsubsection{Summary}
\label{subsubsec:ckks-mult-cipher-summary}

To put all things together, CKKS's ciphertext-to-ciphertext multiplication is summarized as follows:

\begin{tcolorbox}[title={\textbf{\tboxlabel{\ref*{subsubsec:ckks-mult-cipher-summary}} CKKS's Ciphertext-to-Ciphertext Multiplication}}]

Suppose we have the following two RLWE ciphertexts:

$\textsf{RLWE}_{S, \sigma}(\Delta M^{\langle 1 \rangle}) = (A^{\langle 1 \rangle}, B^{\langle 1 \rangle})$, \text{ } where $B^{\langle 1 \rangle} = -A^{\langle 1 \rangle} \cdot S + \Delta M^{\langle 1 \rangle} + E^{\langle 1 \rangle}$

$\textsf{RLWE}_{S, \sigma}(\Delta M^{\langle 2 \rangle}) = (A^{\langle 2 \rangle}, B^{\langle 2 \rangle})$, \text{ } where $B^{\langle 2 \rangle} = -A^{\langle 2 \rangle} \cdot S + \Delta M^{\langle 2 \rangle} + E^{\langle 2 \rangle}$

$ $

Multiplication between these two ciphertexts is performed as follows:

$ $

\setlist[itemize]{leftmargin=*}
\setlist[enumerate]{leftmargin=*}
\begin{enumerate}
\item \textbf{\underline{Basic Multiplication}}

Compute the following:

$ $

$D_0 = B^{\langle 1 \rangle}B^{\langle 2 \rangle}$

$D_1 = B^{\langle 2 \rangle}A^{\langle 1 \rangle} + B^{\langle 1 \rangle}A^{\langle 2 \rangle}$

$D_2 = A^{\langle 1 \rangle} \cdot A^{\langle 2 \rangle}$

$ $

, where $\Delta^2M^{\langle 1 \rangle}M^{\langle 2 \rangle} \approx \underbrace{B^{\langle 1 \rangle}B^{\langle 2 \rangle}}_{D_0}  + \underbrace{(B^{\langle 2 \rangle}A^{\langle 1 \rangle} + B^{\langle 1 \rangle}A^{\langle 2 \rangle})}_{D_1} \cdot S + \underbrace{(A^{\langle 1 \rangle} \cdot A^{\langle 2 \rangle})}_{D_2} \cdot \underbrace{S \cdot S}_{ S^2}$

$= D_0 + D_1\cdot S + D_2\cdot S^2$

$ $

\item \textbf{\underline{Relinearization}} 

$\textsf{RLWE}_{S, \sigma}(\Delta^2 \cdot M^{\langle 1 \rangle} \cdot M^{\langle 2 \rangle}) \approx \textsf{RLWE}_{S, \sigma}\bm{(}\text{ }(D_0 + D_1\cdot S + D_2\cdot S^2)\text{ }\bm{)} \approx C_\alpha + C_\beta$

$ $

$, \text{ where } \text{ } C_\alpha = (D
_1, D_0),$

\text{ }\text{ }\text{ }\text{ }\text{ }\text{ }\text{ }\text{ }\text{ } $C_\beta = \bm{\langle} \textsf{Decomp}^{\beta, l}(D_2), \text{ } \textsf{RLev}_{S, \sigma}^{\beta, l}( S^2) \bm{\rangle} \text{ or }   \Bigg\lceil\dfrac{D_2 \cdot \mathit{evk_g}}{g}\Bigg\rfloor$,

\text{ }\text{ }\text{ }\text{ }\text{ }\text{ }\text{ }\text{ }\text{ }  $\mathit{evk_g} = (A', -A'\cdot S + E' + g S^2) \in \mathcal{R}_{\langle n, gq\rangle}^2, \text{ } \text{ } g  = q_L^2, \text{ } L: \text{ the maximum level}$

$ $

\item \textbf{\underline{Rescaling}}

Switch the relinearlized ciphertext's modulus from $q \rightarrow \hat q$ 
by updating $(A, B)$ to $(\hat{A}, \hat B)$ as follows:

$ $

$(\text{ }C = (A, B)\text{ }) \in \mathcal{R}_{\langle n, q \rangle} \rightarrow (\text{ }\hat{C} = (\hat{A}, \hat{B})\text{ }) \in \mathcal{R}_{\langle n, \hat q \rangle}$ 

$q = \prod\limits_{m=0}^{l}w_m$, \textcolor{red}{ \text{ } \# where all $w_m$ are prime numbers,  $w_0 = q_0 \gg \Delta\cdot p$ to ensure the plaintext $\Delta M$ during homomorphic operations never overflows the ciphertext modulus even at the lowest multiplicative level, and all other $w_i \approx \Delta$}

$\hat{q} = \dfrac{q}{w_l}$

$\hat{A} = \left\lceil\dfrac{\hat q}{q}\cdot A\right\rfloor = \hat{a}_{0} + \hat{a}_{1}X + \hat{a}_{2}X^2 + \cdots + \hat{a}_{{n-1}}X^{n-1}$, where each $\hat{a}_{i} = \Big\lceil a_{i}\dfrac{\hat{q}}{q} \Big\rfloor = \Big\lceil \dfrac{a_{i}}{w_l} \Big\rfloor \in \mathbb{Z}_{\hat{q}}$

$\hat{B} = \left\lceil\dfrac{\hat q}{q}\cdot B\right\rfloor = \hat{b}_0 + \hat{b}_1X + \hat{b}_2X^2 + \cdots + \hat{b}_{n-1}X^{n-1}$, where each $\hat{b}_i = \Big\lceil b_i\dfrac{\hat{q}}{q} \Big\rfloor = \Big\lceil \dfrac{b_i}{w_l} \Big\rfloor \in \mathbb{Z}_{\hat{q}}$

$ $

The above update of $(A, B)$ to $(\hat{A}, \hat B)$ effectively changes $\Delta, E$ to $\hat\Delta, \hat E$ as follows:

$\hat{E} = \hat{e}_0 + \hat{e}_1X + \hat{e}_2X^2 + \cdots + \hat{e}_{n-1}X^{n-1}$, where each $\hat{e}_i = \Big\lceil e_i\dfrac{\hat{q}}{q} \Big\rfloor = \Big\lceil \dfrac{e_i}{w_l} \Big\rfloor \in \mathbb{Z}_{\hat{q}}$

$\hat{\Delta} = \left\lceil\Delta^2\dfrac{\hat{q}}{q}\right\rfloor = \left\lceil\dfrac{\Delta^2}{w_l}\right\rfloor \approx \Delta$ \textcolor{red}{ $\rhd$ This rounding error slightly increases the noise $\hat E$ to $\hat E + E_\Delta$, while the decryption of $(\hat A, \hat B)$ outputs the same plaintext $M$}

$ $

Note that after the rescaling, the ciphertext modulus changes from $q \rightarrow \hat q$, and the plaintext scaling factor of $(\hat A, \hat B)$ is also updated to $\hat{\Delta}$. Meanwhile, the plaintext $M$ and the secret key $S$ stay the same as before.

$ $

\end{enumerate}
\setlist[itemize]{leftmargin=\leftmargini}
\setlist[enumerate]{leftmargin=\leftmargini}

\para{Swapping the Order of \textsf{Relinearization} and \textsf{Rescaling}: } The order of relinearization and rescaling is interchangeable. Running rescaling before relinearization reduces the size of the ciphertext modulus, and therefore the subsequent relinearization can be executed faster. 

\end{tcolorbox}

\subsection{Ciphertext-to-Plaintext Multiplication}
\label{subsec:ckks-mult-plain}

Remember that BFV's ciphertext-to-plaintext multiplication (\autoref{subsec:bfv-mult-plain}) is performed as follows:

$\textsf{RLWE}_{S, \sigma}(\Delta M) \cdot \Lambda$

$= (A, \text{ } B) \cdot \Lambda$

$= (A \cdot \Lambda, \text{ }  B \cdot \Lambda )$

$= \textsf{RLWE}_{S, \sigma}(\Delta (M \cdot \Lambda) )$

, where the plaintext polynomial $\Lambda$ is not scaled by $\Delta$. However, the above relation cannot be used in CKKS's ciphertext-to-plaintext multiplication because when CKKS encodes the input vector slots into polynomial coefficients, the encoding is computed as $\vec{m} = \dfrac{\hathat W \cdot I_n^R \cdot \vec{v}_{'}}{n}$ (Summary~\ref*{subsec:ckks-encoding-decoding} in \autoref{subsec:ckks-encoding-decoding}), where the $n$-th root-of-unity base $\omega = e^{i\pi/n} = \cos \left(\dfrac{\pi}{n}\right) + i\sin \left(\dfrac{\pi}{n}\right)$. Since $\omega$ is usually not an integer, the encoded polynomial $\Lambda$'s coefficients are usually not integers (will usually have infinite decimal digits). Therefore, we need to follow the CKKS encoding procedure's last step (Summary~\ref*{subsec:ckks-encoding-decoding} in \autoref{subsec:ckks-encoding-decoding}), which scales $\Lambda$ by $\Delta$ to shift an enough number of its decimal values to the integer digits, which effectively approximates the decimal coefficients to integers with high precision. Then, the resulting encrypted plaintext becomes $\Delta M \cdot \Delta \Lambda = \Delta^2 M \Lambda$. To convert $\Delta^2 M \Lambda$ into $\Delta M \Lambda$, we need to do a rescaling operation as we did in CKKS's ciphertext-to-ciphertext multiplication's (Summary~\ref*{subsubsec:ckks-mult-cipher-summary} in \autoref{subsubsec:ckks-mult-cipher-summary}) last step. Therefore, CKKS's ciphertext-to-plaintext multiplication consumes one multiplicative level (whereas BFV's ciphertext-to-plaintext multiplication does not consume any multiplicative level). CKKS's ciphertext-to-plaintext multiplication is summarized as follows:

\begin{tcolorbox}[title={\textbf{\tboxlabel{\ref*{subsec:ckks-mult-plain}} CKKS Ciphertext-to-Plaintext Multiplication}}]

\setlist[itemize]{leftmargin=*}
\setlist[enumerate]{leftmargin=*}
\begin{enumerate}
\item \textbf{\underline{Basic Multiplication}}

$\textsf{RLWE}_{S, \sigma}(\Delta M) \cdot \Delta\Lambda$

$= (A, \text{ } B) \cdot \Delta\Lambda$

$= (A \cdot \Delta\Lambda, \text{ }  B \cdot \Delta\Lambda )$

$= \textsf{RLWE}_{S, \sigma}(\Delta^2 (M \cdot \Lambda) )$

$ $

\item \textbf{\underline{Rescaling}}

Switch the relinearlized ciphertext's modulus from $q \rightarrow \hat q$ as done in CKKS's ciphertext-to-ciphertext multiplication's (Summary~\ref*{subsubsec:ckks-mult-cipher-summary} in \autoref{subsubsec:ckks-mult-cipher-summary}) last step. 

\end{enumerate}
\setlist[itemize]{leftmargin=\leftmargini}
\setlist[enumerate]{leftmargin=\leftmargini}

\end{tcolorbox}

\subsection{\textsf{ModDrop}}
\label{subsec:ckks-moddrop}

Remember that CKKS's ciphertext decryption relation is as follows:

$\Delta M + E = A \cdot S + B \bmod q_l$

$\Delta M + E = A \cdot S + B - K\cdot q_l$ \textcolor{red}{ $\rhd$ where $K\cdot q_l$ represents a modulo reduction by $q_l$}

$ $

\textsf{ModDrop} is an operation of lowering the multiplicative level of a ciphertext by sequentially throwing away its modulus's one or more prime elements $\left(\text{i.e.,} \left\{\dfrac{q_i}{q_{i-1}}\right\}_{i=0}^{L}\right)$ except for the last one $q_0$, while ensuring that the plaintext's scaling factor $\Delta$ stays the same as before. Specifically, a \textsf{ModDrop} operation that decreases its modulus from $q_l \rightarrow q_{l-1}$ is performed by updating the ciphertext $(A, B)$ to a new one: $\bm(A' = A \bmod q_{l-1}$, $B' = B \bmod q_{l-1})$. After the \textsf{ModDrop}, the ciphertext's modulus decreases from $q_l \rightarrow q_{l-1}$, yet its decryption relation still holds the same as follows:

$A' \cdot S + B' - K\cdot q_l $

$= (A \bmod q_{l-1}) \cdot S + (B \bmod q_{l-1}) - K\cdot q_l$

$= (A - K_A\cdot q_{l-1}) \cdot S + (B - K_B\cdot q_{l-1}) - K\cdot q_l$ 

$= A\cdot S + B - (K_A + K_B + K\dfrac{q}{q_{l-1}})\cdot q_{l-1}$
\textcolor{red}{ $\rhd$ where $\dfrac{q}{q_{l-1}}$ is an integer (the $l$-the prime element of $q_L$)}

$= A\cdot S + B - K'\cdot q_{l-1}$ \textcolor{red}{ $\rhd$ where $K' = K_A + K_B + K\dfrac{q}{q_{l-1}}$ is an integer}

$= A\cdot S + B \bmod q_{l-1}$

$= \Delta M + E$ \textcolor{red}{ $\rhd$ since $\Delta M + E < q_0 < q_{l-1}$}

$ $

As shown above, $(A', B') \bmod q_{l-1}$ decrypts to the same $\Delta M + E$, a scaled plaintext with an error. 

$ $

CKKS's \textsf{ModDrop} is summarized as follows:

\begin{tcolorbox}[title={\textbf{\tboxlabel{\ref*{subsec:ckks-moddrop}} CKKS's \textsf{ModDrop}}}]

Given a CKKS ciphertext with the $l$-th multiplicative level $\textsf{RLWE}_{S, \sigma}(\Delta M) = (A, B) \bmod q_l$, a \textsf{ModDrop} operation is as follows: 

$(A', B') \bmod q_{l-1} = (A \bmod q_{l-1}, B \bmod q_{l-1})$

$ $

, after which the ciphertext's multiplicative level decreases by 1, while the plaintext's scaling factor $\Delta$ and the noise are unaffected. 

\end{tcolorbox}

\subsubsection{Difference between Modulus Switch and \textsf{ModDrop}}
\label{subsubsec:ckks-moddrop-vs-modswitch}

In CKKS, both modulus switch (i.e., rescaling explained in \autoref{subsubsec:ckks-mult-cipher-rescale}) and \textsf{ModDrop} lower a ciphertext's modulus from $q_l \rightarrow q_{l-1}$. However, the key difference is that rescaling also decreases the plaintext's scaling factor by the $\dfrac{q_l}{q_{l-1}} \approx \Delta$, whereas \textsf{ModDrop} does not affect the plaintext's scaling factor and the noise. Therefore, rescaling is used only during ciphertext-to-ciphertext multiplication when scaling down the plaintext's scaling factor in the intermediate ciphertext from $\Delta^2 \rightarrow \Delta$. Meanwhile, \textsf{ModDrop} is used to reduce the modulo computation time during an application's routine when it becomes certain that the ciphertext will not undergo any additional ciphertext-to-ciphertext multiplication (i.e., no need to further decrease the ciphertext's modulus).

\subsection{Homomorphic Key Switching}
\label{subsec:ckks-key-switching}

CKKS's homomorphic key switching scheme changes an RLWE ciphertext's secret key from $S$ to $S'$. This scheme is exactly the same as BFV's key switching scheme (Summary~\ref*{subsec:bfv-key-switching} in \autoref{subsec:bfv-key-switching}).

\begin{tcolorbox}[title={\textbf{\tboxlabel{\ref*{subsec:ckks-key-switching}} CKKS's Key Switching}}]
$\textsf{RLWE}_{S',\sigma}(\Delta M) = (0, B) + \bm{\langle} \textsf{Decomp}^{\beta, l}(A), \text{ } \textsf{RLev}_{S', \sigma}^{\beta, l}(S) \bm{\rangle}$
\end{tcolorbox}

\subsection{Homomorphic Rotation of Input Vector Slots}
\label{subsec:ckks-rotation}

CKKS's batch encoding scheme (Summary~\ref*{subsec:ckks-encoding-decoding} in \autoref{subsec:ckks-encoding-decoding}) implicitly supports homomorphic rotation of input slot vectors like that of BFV's homomorphic rotation (Summary~\ref*{subsubsec:bfv-rotation-summary} in \autoref{subsubsec:bfv-rotation-summary}). This is because CKKS uses the same encoding and decoding matrices ($\hathat W$ and $\hathat{W}^*$) designed for the BFV encoding and decoding scheme that supports homomorphic rotation of input vector slots. Although the roots of the $(\mu=2n)$-th cyclotomic polynomial $X^n + 1$ are different for the BFV and CKKS schemes (as one is designed over $X \in \mathbb{Z}_t$ and the other is over $X \in \mathbb{R}$), CKKS still can use the same $\hathat W$ (and $\hathat{W}^*$) matrices as BFV, because the $(\mu=2n)$-th cyclotomic polynomial over $X \in \mathbb{Z}_t$ exhibits the same essential properties as the $(\mu=2n)$-th cyclotomic polynomial over $X \in \mathbb{R}$ (as explained in \autoref{sec:cyclotomic-polynomial-integer-ring}). Especially, the roots of both $(\mu=2n)$ cyclotomic polynomials are the primitive $(\mu=2n)$-th roots of unity having the order $2n$, and those $n$ distinct roots are defined as $\omega^1, \omega^3, \cdots, \omega^{2n-1}$, where $\omega$ can be any root. Therefore, substituting CKKS's $(\mu=2n)$-th roots of unity into the $\omega$ terms in BFV's encoding matrix $\hathat{W}$ (and decoding matrix $\hathat{W}^*$) preserves the same computational correctness for the encoding and decoding schemes, as well as for input vector slot rotation. 

Importantly, the $\hathat{W}$ and $\hathat{W}^*$ matrices in both the BFV and CKKS schemes satisfy the exact requirement for supporting input vector slot rotation. That is, given the following relations:

\begin{itemize}
    \item The $\vec{v} \rightarrow \vec{m}$ encoding formula: $\vec{m} = n^{-1}\cdot I_n^R\cdot \hathat{W}\cdot \vec{v}$
    \item The $\vec{m} \rightarrow \vec{v}$ decoding formula: $\vec{v} = \hathat{W}^* \cdot \vec{m}$
    \item The encoded polynomial $M(X) = \sum\limits_{i=0}^{n-1}m_iX^i$
\end{itemize}    

, updating the polynomial $M(X)$ to $M(X^{J(h)})$ results in the effect of rotating the first half of the $n$ input vector slots ($\vec{v} \in \mathbb{Z}^n_p$ in the case of BFV, and the forward-ordered Hermitian vector $\vec{v}_{'} \in \mathbb{\hat C}^{n} \rightarrow \mathbb{C}^{\frac{n}{2}}$ in the case of CKKS) by $h$ positions to the left (in a wrapping manner among them) and the second half of the slots also by $h$ positions to the right (in a wrapping manner among them). 

BFV uses CKKS's same rotation scheme described in Summary~\ref*{subsubsec:bfv-rotation-summary} (in \autoref{subsubsec:bfv-rotation-summary}) as follows:

\begin{tcolorbox}[title={\textbf{\tboxlabel{\ref*{subsec:ckks-rotation}} CKKS's Homomorphic Rotation of Input Vector Slots}}]

Suppose we have an RLWE ciphertext and a key-switching key as follows:

$\textsf{RLWE}_{S, \sigma}(\Delta M) = (A, B)$, \text{ } $\textsf{RLev}_{S, \sigma}^{\beta, l}(S^{J(h)})$

$ $

Then, the procedure of rotating all $\dfrac{n}{2}$ elements of the ciphertext's original input vector $\vec{v}$ by $h$ positions to the left is as follows: 

\setlist[itemize]{leftmargin=*}
\setlist[enumerate]{leftmargin=*}
\begin{enumerate}
\item Update $A(X)$, $B(X)$ to $A(X^{J(h)})$, $B(X^{J(h)})$. 
\item Perform the following key switching (\autoref{subsec:ckks-key-switching}) from $S(X^{J(h)})$ to $S(X)$:

$\textsf{RLWE}_{S(X),\sigma}\bm{(}\Delta M(X^{J(h)})\bm{)} = \bm{(} 0, B(X^{J(h)}) \bm{)} \text{ } + \text{ } \bm{\langle}  \textsf{Decomp}^{\beta, l}\bm{(}A(X^{J(h)})\bm{)}, \text{ } \textsf{RLev}_{S(X), \sigma}^{\beta, l}\bm{(}S(X^{J(h)})\bm{)} \bm{\rangle}$
\end{enumerate}
\setlist[itemize]{leftmargin=\leftmargini}
\setlist[enumerate]{leftmargin=\leftmargini}

\end{tcolorbox}

\para{Rotation within Half Slots:} Like BFV, CKKS rotates the first half of the forward-ordered Hermitian input vector slots $\vec{v}_{'} \in \mathbb{\hat{C}}^n$ and the second half of its slots separately in a partitioned manner. This is because the first half rows of $\hathat{W}^*$ comprise the terms $\omega^{J(h)}$ for $h = \{0, 1, \cdots, \dfrac{n}{2} - 1\}$ (i.e., evaluates $M(X)$ at $X=\{\omega^{J(0)}, \omega^{J(1)}, \cdots, \omega^{J(\frac{n}{2}-1)}\}$), whereas the second half rows of $\hathat{W}^*$ comprise the terms $\omega^{J_*(h)}$ (i.e., evaluates $M(X)$ at $X=\{\omega^{J_*(0)}, \omega^{J_*(1)}, \cdots, \omega^{J_*(\frac{n}{2}-1)}\}$), and the computed values of $J(h)$ and $J_*(h)$ repeat (i.e., rotate) within their own rotation group across $h=\{0, 1, \cdots, \dfrac{n}{2} - 1\}$. Because of this structure of $\hathat W$ and $\hathat W^*$, BFV and CKKS cannot design a wrapping rotation scheme across all $n$ slots of the input vector homogeneously, but can instead design a wrapping rotation scheme across each group of the first-half and the second-half $\dfrac{n}{2}$ slots of the input vector in a partitioned manner. That being said, CKKS can meaningfully only use the first $\dfrac{n}{2}$ slots for homomorphic computations anyway, because the latter $\dfrac{n}{2}$ slots are conjugates of the first $\dfrac{n}{2}$ slots which cannot be chosen by the user but are deterministically configured based on the first $\dfrac{n}{2}$. On the other hand, in BFV, the user can choose the entire $\dfrac{n}{2}$ according to his/her needs, so BFV's utility of slots is full $n$. Therefore, the user can use BFV's first-half slots and second-half slots together to perform parallel computations.

\subsubsection{Example}
\label{subsubsec:ckks-rotation-ex}

In this subsection, we will show the following 2 examples:
\begin{enumerate}
\item Encode an input vector $\vec{v}$ into a plaintext polynomial $M(X)$ based on our updated updated encoding \& decoding matrices $\hathat W$ and $\hathat W^*$
\item Rotate all elements of the input vector $\vec{v}$ $h$ positions to the left by updating the encoded plaintext $M(X)$ to $M(X^{J(h)})$
\end{enumerate}

$ $

We will use the same example of the input vector $\vec{v}$ used in \autoref{subsubsec:ckks-encoding-ex}: $\vec{v}^{\langle h=1 \rangle} = (1.1 + 4.3i, 3.5 - 1.4i)$.

Remember that the encoded plaintext polynomial of $\vec{v}$ is as follows: 

$\Delta M(X) = 2355 + 1195X + 1485X^2 + 2933X^3 \in \mathcal{R}_{\langle 4 \rangle} \in \mathbb{R}[X] / X^4 + 1$

Suppose we want to rotate the input vector $\vec{v}$ by 1 position to the left as follows: 

$\vec{v}^{\langle h=1 \rangle} = (3.5 - 1.4i, 1.1 + 4.3i)$

$ $

Therefore, we update $\Delta M(X)$ to $\Delta M(X^{J(1)})$ as follows:

$\Delta M(X^{J(1)}) = \Delta M(X^{5}) = 2355 + 1195(X^5) + 1485(X^5)^2 + 2933(X^5)^3$ 

$= 2355 + 1195X^5 + 1485X^{10} + 2933X^{15}$

$= 2355 + 1195X\cdot(-1) + 1485X^{2}\cdot(-1)\cdot(-1) + 2933X^{3}\cdot(-1)\cdot(-1)\cdot(-1)$

$= 2355 - 1195X + 1485X^2 - 2933X^3$

$ $

The rotated \textit{forward-ordered} Hermitian input vector is computed as follows:

$\dfrac{\hathat W^* \cdot \Delta \vec{m}}{\Delta} = \begin{bmatrix}
1,\omega,\omega^2, \omega^3\\
1,\omega^5, \omega^{10},\omega^{15}\\
1,\overline{\omega}, \overline{\omega^2}, \overline{\omega^3}\\
1,\overline{\omega^5}, \overline{\omega^{10}}, \overline{\omega^{15}}
\end{bmatrix} \cdot \begin{bmatrix}
2355\\- 1195\\1485\\-2933
\end{bmatrix}\cdot \dfrac{1}{1024}$

$= \dfrac{\hathat W^* \cdot \Delta \vec{m}}{\Delta} = \begin{bmatrix}
1,\omega,\omega^2, \omega^3\\
1,\omega^5, \omega^{2},\omega^{7}\\
1,\overline{\omega}, \overline{\omega^2}, \overline{\omega^3}\\
1,\overline{\omega^5}, \overline{\omega^{2}}, \overline{\omega^{7}}
\end{bmatrix} \cdot \begin{bmatrix}
2355\\- 1195\\1485\\-2933
\end{bmatrix}\cdot \dfrac{1}{1024}$

$= \dfrac{\hathat W^* \cdot \Delta \vec{m}}{\Delta} = \begin{bmatrix}
1,\omega,\omega^2, \omega^3\\
1,\omega^5, \omega^{2},\omega^{7}\\
1,\omega^7, \omega^6, \omega^5\\
1,\omega^3, \omega^{6}, \omega
\end{bmatrix} \cdot \begin{bmatrix}
2355\\- 1195\\1485\\-2933
\end{bmatrix}\cdot \dfrac{1}{1024}$

$= 
\begin{bmatrix}
1,\dfrac{\sqrt{2}}{2} + \dfrac{i\sqrt{2}}{2},i, -\dfrac{\sqrt{2}}{2} + \dfrac{i\sqrt{2}}{2}\\
1,-\dfrac{\sqrt{2}}{2} - \dfrac{i\sqrt{2}}{2}, i,\dfrac{\sqrt{2}}{2} - \dfrac{i\sqrt{2}}{2}\\
1,\dfrac{\sqrt{2}}{2} - \dfrac{i\sqrt{2}}{2}, -i, -\dfrac{\sqrt{2}}{2} - \dfrac{i\sqrt{2}}{2}\\
1,-\dfrac{\sqrt{2}}{2} + \dfrac{i\sqrt{2}}{2}, -i, \dfrac{\sqrt{2}}{2} + \dfrac{i\sqrt{2}}{2}
\end{bmatrix}$
$\cdot \begin{bmatrix}
2.2998046875\\-1.1669921875\\1.4501953125\\-2.8642578125
\end{bmatrix}$

$ $

$ \approx (3.500 - 1.4003i, \text{ } 1.0997 + 4.3007i,\text{ } 3.500 + 1.4003i, \text{ } 1.0997 - 4.3007i)$

$ $

Extract the first $\dfrac{n}{2} = 2$ elements in the above Hermitian vector to recover the input vector:

$(3.500 - 1.4003i, \text{ } 1.0997 + 4.3007i)$

$ \approx (3.5 - 1.4i, \text{ } 1.1 + 4.3i)$
$ = \vec{v}^{\langle h=1 \rangle}$ \textcolor{red}{  $\rhd$ The original input vector $\vec{v}$ rotated by 1 position to the left}

$ $

In practice, we do not directly update $\Delta M(X)$ to $\Delta M(X^{J(1)})$, because we would not have access to the plaintext polynomial $M(X)$ unless we have the secret key $S(X)$. Therefore, we instead update $\textsf{ct}=\bm(A(X), B(X)\bm)$ to $\textsf{ct}^{\langle h=1 \rangle}=\bm(A(X^{J(1)}), B(X^{J(1)})\bm)$, which is equivalent to homomorphically rotating the encrypted input vector slots. Then, decrypting $\textsf{ct}^{\langle h=1 \rangle}$ and decoding it would output $\vec{v}^{\langle h=1 \rangle}$.

$ $

\para{Source Code:} Examples of CKKS's homomorphic input vector slot rotation can be executed by running \href{https://github.com/fhetextbook/fhe-textbook/blob/main/source%20code/ckks.py}{\underline{this Python script}}.

\subsection{Contemplation on CKKS Encoding}
\label{subsec:ckks-encoding-contemplate}

\para{Why CKKS's Encoding Uses the $\bm{(\mu=2n)}$-th Cyclotomic Polynomial:} At this point, it becomes clear why the CKKS encoding and decoding scheme uses the (power-of-2)-th cyclotomic polynomial (i.e., $X^n + 1$) over $X \in \mathbb{C}$ (complex numbers) (\autoref{subsec:ckks-encoding-decoding}). The first reason is that CKKS's first requirement for designing a valid encoding and decoding formula for an input complex vector is to isomorphically convert it into a unique real number vector (and we scale this real number vector as an integer vector and use it as a list of coefficients for polynomial encoding, because CKKS's homomorphic encryption and decryption are supported only based on polynomials with integer coefficients). As for the decoding formula of an input complex vector, our high-level idea was to treat the encoded real number vector as coefficients of an $(n-1)$-degree polynomial and evaluate this polynomial at $n$ distinct $X$ coordinates, whose resulting set of $n$ distinct $Y$ values is guaranteed to be unique within the $n$-th degree polynomial ring. Based on this insight, we designed a decoding matrix (\autoref{subsec:ckks-encoding-decoding}) in the form of a Vandermonde matrix (\autoref{subsec:vandermonde}). Then, the encoding formula is equivalent to multiplying the input complex vector by the inverse of this decoding matrix. However, in linear algebra, not all matrices are guaranteed to have a counterpart inverse matrix. Therefore, for the guarantee of the existence of a valid encoding matrix (i.e., an inverse of the decoding matrix), we leveraged the following arithmetic property: if a Vandermonde matrix $V = \mathit{Vander}(x_0, x_1, \cdots, x_{n-1})$ is made of $n$ distinct primitive $(\mu=2n)$-th roots of unity (where $n$ is a power of 2), then such a Vandermonde matrix is guaranteed to have an inverse (\autoref{subsec:vandermonde-euler}) counterpart. In fact, the $(\mu=2n)$-th roots of unity are $n$ distinct roots of the $(\mu=2n)$-th cyclotomic polynomial: $X^n + 1$ (\autoref{subsec:cyclotomic-def}). Therefore, CKKS uses $X^n + 1$ as the polynomial ring of its subsequent encryption and decryption scheme (\autoref{subsec:ckks-enc-dec}) as well. 

The CKKS encoding's second reason for using the $(\mu=2n)$-th cyclotomic polynomial is to design a valid input vector slot rotation scheme (\autoref{subsec:ckks-rotation}). In this rotation scheme, updating the encoded polynomial $M(X)$ to $M(X^{J(h)})$ (where $J(h) = 5^h \bmod 2n$) is equivalent to updating the CKKS decoding process's each evaluation coordinate of $M(X)$ from $x_i$ to $x_i^{J(h)}$ (where each $x_i$ is the primitive $(\mu=2n)$-th roots of unity), which gives the same effect as vertically rotating the encoding matrix (i.e., the inverse of the Vandermonde matrix whose roots are the primitive $(\mu=2n)$-th roots of unity) upward by $h$ positions. And this vertical rotation of the encoding matrix (while the input vector is fixed) gives the same effect of rotating the input vector $\vec{v}$ by $h$ positions to the left (without modifying the encoding matrix). Therefore, the $(\mu=2n)$-th cyclotomic polynomial $X^n + 1$ is an ideal tool to design input vector slot rotation.

\subsection{Homomorphic Conjugation}
\label{subsec:ckks-conjugation}

As explained in Summary~\ref*{subsec:ckks-encoding-decoding} (\autoref{subsec:ckks-encoding-decoding}), given the $\dfrac{n}{2}$-slot input vector $\vec{v} = (v_0, v_1, \cdots, v_{\frac{n}{2}-1})$, its corresponding $n$-slot Hermitian vector is $\vec{v}_{'} = (v_0, v_1, \cdots, v_{\frac{n}{2}-1}, \overline{v}_0, \overline{v}_1, \cdots, \overline{v}_{\frac{n}{2}-1})$. To compute the conjugation of $\vec{v}$, which is essentially conjugating $\vec{v}_{'}$, we can conjugate $M(X)$ as follows:

$\overline{\vec{v}}_{'} = (\overline{v}_0, \overline{v}_1, \cdots, \overline{v}_{\frac{n}{2}-1}, v_0, v_1, \cdots, v_{\frac{n}{2}-1})$

$= \bm{(}M(\overline{\omega}), M(\overline{\omega^{3}}), \cdots, M(\overline{\omega^{n-1}}), M(\omega), M(\omega^{3}), \cdots, M(\omega^{n-1})\bm{)}$ \textcolor{red}{ $\rhd$ where $\omega = e^{\frac{i\pi}{n}}$}

$= \bm{(}M((\omega)^{-1}), M((\omega^{3})^{-1}), \cdots, M((\omega^{n-1})^{-1}), M(\omega), M(\omega^{3}), \cdots, M(\omega^{n-1})\bm{)}$ 

\textcolor{red}{ $\rhd$ since $\overline{\omega^k} = e^{\overline{\frac{ki\pi}{n}}} = e^{-\frac{ki\pi}{n}} = \omega^{-k}$ and $\omega^{k} = (\overline{\omega^k})^{-1}$ for $k = \{1, 3, \cdots, n-1\}$}

$ $

$=\{M(X^{-1})\}$ 

\textcolor{red}{ $\rhd$ where $X = \{\omega, \omega^{3}, \cdots, \omega^{n - 1}, \omega^{-1}, \omega^{-3}, \cdots, \omega^{-(n-1)}\}  = \{\omega, \omega^{3}, \cdots, \omega^{n - 1}, \omega^{2n-1}, \omega^{2n-3}, \cdots, \omega^{n+1} \}$}

$ $

Therefore, homomorphic conjugation of the input vector is equivalent to updating the ciphertext $\bm{(}A(X), B(X)\bm{)}$ to $\bm{(}A(X^{-1}), B(X^{-1})\bm{)}$ and then key-switching it from $S(X^{-1}) \rightarrow S(X)$.

\begin{tcolorbox}[title={\textbf{\tboxlabel{\ref*{subsec:ckks-conjugation}} CKKS's Homomorphic Conjugation}}]

Homomorphic conjugation of the input vector of a ciphertext is equivalent to the following:

\begin{enumerate}
\item Update the ciphertext $\bm{(}A(X), B(X)\bm{)}$ to $\bm{(}A(X^{-1}), B(X^{-1})\bm{)}$.
\item Key-switch $\bm{(}A(X^{-1}), B(X^{-1})\bm{)}$ from $S(X^{-1})$ to $S(X)$.

\end{enumerate}

\end{tcolorbox}

\subsection{Sparsely Packing Ciphertexts}
\label{subsec:ckks-sparse-packing}

In \autoref{subsec:ckks-encoding-decoding}, we learned the CKKS encoding scheme, which encodes an input vector with $\dfrac{n}{2}$ slots (i.e., $\dfrac{n}{2}$-slot input vector) into an $(n-1)$-degree polynomial. While the polynomial ring's degree $n$ is fixed at the setup stage of CKKS as a security parameter, some applications may only need to use fewer than $\dfrac{n}{2}$ input vector slots. Suppose we only need to use $\frac{n'}{2}$ slots out of $\frac{n}{2}$ slots, where $n'$ is some number that divides $n$. Then, the corresponding input vector and encoded polynomial acquire a special property as described below:

\begin{tcolorbox}[title={\textbf{\tboxlabel{\ref*{subsec:ckks-sparse-packing}} CKKS's Sparsely Packing Polynomial and Ciphertext}}]

Suppose that an $\frac{n}{2}$-slot input vector gets encoded into a polynomial in $\mathbb{R}[X] / (X^n + 1)$. And suppose that $n'$ is some number that divides $n$. We define polynomial $M(X) \in \mathbb{R}[X] / (X^n + 1)$ as the one which has non-zero constants at the terms whose power is a multiple of $\frac{n}{n'}$ and all other terms have zero constants (i.e., $M(X) = c_0 + c_{\frac{n}{n'}}X^{\frac{n}{n'}} + c_{\frac{2n}{n'}}X^{\frac{2n}{n'}} + \cdots + c_{n - \frac{n}{n'}}X^{n - \frac{n}{n'}}$). We can express $M(X)$ as some $M_Y(Y) \in \mathbb{R}[Y] / (Y^{n'} + 1)$ where $Y = X^{\frac{n}{n'}}$ and thus $M(X) = M_Y(X^{\frac{n}{n'}})$. 

$ $

Then, the following are true:

\begin{enumerate}
    \item Every $M_Y(Y) \in \mathbb{R}[Y]/(Y^{n'} + 1)$ is isomorphically mapped to (i.e., decoded into) some $\dfrac{n}{2}$-slot input vector which comprises $\dfrac{n}{n'}$ repetitions of $\dfrac{n'}{2}$ consecutive slot values.
    \item Conversely, if an $\dfrac{n}{2}$-slot input vector comprises $\dfrac{n}{n'}$ repetitions of the first $\dfrac{n'}{2}$ consecutive slot values, then the vector gets encoded into some $M_Y(Y) \in \mathbb{R}[Y] / (Y^{n'} + 1)$ (i.e., some polynomial in $\mathbb{R}[X] / (X^n + 1)$ that has zero constants at the terms whose degree is not a multiple of $\dfrac{n}{n'}$). 
\end{enumerate}

\end{tcolorbox}

We could show both directions of proof: (1) the forward (decoding-direction) proof; and (2) the backward (encoding-direction) proof. However, it is sufficient to prove only either direction because the encoding ($\sigma^{-1}$) and decoding ($\sigma$) processes are isomorphic. Among these two, we will show only the forward proof for simplicity. 

\subsubsection{Forward Proof: Decoding of Sparsely Packed Ciphertext}

We will prove that for each $M_Y(Y) \in \mathbb{R}[Y] / (Y^{n'} + 1)$ (i.e., a polynomial in $\mathbb{R}[X] / (X^n + 1)$ that has non-zero constants only at those terms with a power that is a multiple of $\dfrac{n}{n'}$ and zero constants in all other terms), the polynomial is decoded into some $\dfrac{n}{2}$-slot input vector which comprises $\dfrac{n}{n'}$ repetitions of the first $\dfrac{n'}{2}$ consecutive slot values. 

To decode $M(X)$ into an input vector, we need to evaluate $M(X)$ at $\frac{n}{2}$ distinct roots of $X^n + 1$ (i.e., $n$ distinct primitive $(\mu=2n)$-th roots of unity), which are: 

$\bm{(} \text{ } 
M(\omega^{J(0)}), \text{ } M(\omega^{J(1)}), \cdots,  M(\omega^{J(\frac{n}{2}-1)})\bm{)}$ 

\textcolor{red}{ $\rhd$ where $\omega=e^{\frac{i\pi}{n}}$, the base and generator of the primitive ($\mu=2n$)-th roots of unity}

$ $

But since $M(X) = M_Y(X^{\frac{n}{n'}})$, the above evaluation is equivalent to evaluating: 

$\bm{(} \text{ } 
M_Y((\omega^{J(0)})^{\frac{n}{n'}}), \text{ } M_Y((\omega^{J(1)})^{\frac{n}{n'}}), \cdots,  M_Y((\omega^{J(\frac{n}{2}-1)})^{\frac{n}{n'}}) \bm{)}$

$ $

$= \bm{(} \text{ } 
M_Y((\omega^{\frac{n}{n'}})^{J(0)}), \text{ } M_Y((\omega^{\frac{n}{n'}})^{J(1)}), \cdots,  M_Y((\omega^{\frac{n}{n'}})^{J(\frac{n}{2}-1)})\bm{)}$

$ $

$= \bm{(} \text{ } 
M_Y(\xi^{J(0)}), \text{ } M_Y(\xi^{J(1)}), \cdots,  M_Y(\xi^{J(\frac{n}{2}-1)})\bm{)}$ 

\textcolor{red}{ $\rhd$ where $\xi = e^{\frac{i\pi}{n'}}$, the base and generator of the primitive $(\mu=2n')$-th roots of unity}

$ $

Notice that $\xi = \omega^{\frac{n}{n'}}$. Therefore, the above evaluation of $M_Y(Y)$ outputs $\dfrac{n}{n'}$ repeated values of $M_Y(Y)$ evaluated at $\frac{n'}{2}$ distinct primitive $(\mu=2n')$-th roots of unity.

\subsection{Modulus Bootstrapping}
\label{subsec:ckks-bootstrapping}

\textbf{- Reference:} 
\href{https://eprint.iacr.org/2018/153.pdf}{Bootstrapping for Approximate Homomorphic Encryption}~\cite{ckks}

During CKKS's ciphertext-to-ciphertext multiplication, each ciphertext is associated with a particular multiplicative level and it decreases by 1 upon each ciphertext-to-ciphertext multiplication (by its internal modulus rescaling operation). Reaching multiplicative level 0 is equivalent to reaching the end of a ciphertext's modulus chain and no more ciphertext-to-ciphertext multiplication can be performed. To continue with further ciphertext-to-ciphertext multiplication, CKKS provides a special operation called \textit{bootstrapping}, which is a process of resetting the ciphertext's end-of-chain modulus $q_0$ to the initial maximum modulus $q_L$ (which is either $q_0\cdot\Delta^{L}$ in the vanilla rescaling scheme, or $\prod\limits_{m=0}^{L}w_m$ in the case of using CRT, as explained in \autoref{subsubsec:ckks-mult-cipher-rescale}).

Suppose we have a ciphertext $(A, B)$ with multiplicative depth 0. If we decrypt a ciphertext whose multiplicative level is 0 (i.e., the ciphertext's modulus is $q_0$), then decrypting it \textit{without} reduction modulo $q_0$ would output:

$\textsf{RLWE}^{-1}\textbf{(}\textsf{ct} = (A, B)\textbf{)}$

$ = B + A\cdot S = \Delta M + E + q_0\cdot K$ \textcolor{red}{ $\rhd$ since $B + A\cdot S \bmod q_0 = \Delta M + E$}

, where $q_0 \cdot K$ accounts for wrap-around modulo $q_0$ values-- each coefficient of polynomial $q_0K$ is some multiple of $q_0$. CKKS's bootstrapping procedure is equivalent to \textit{safely} transforming a ciphertext's modulus from $q_0$ to $q_L$ (where $q_L \gg q_0$).

\subsubsection{High-level Idea}
\label{subsubsec:ckks-bootstrapping-high-level}

As the first step of bootstrapping, we forcibly change the modulus of the ciphertext $(A, B)$ from $q_0$ to $q_L$. Then, its decryption with reduction modulo $q_L$ would output:

$\textsf{RLWE}^{-1}\textbf{(}\textsf{ct} = (A, B)\textbf{)} \bmod q_L$

$ = B + A\cdot S \bmod q_L$

$= \Delta M + E + q_0 K \bmod q_L$

Here, we assume that $q_L$ is large enough such that $\Delta M + E + q_0K \ll q_L$. This is true given $S$ has small coefficients which are either $\{-1, 0, 1\}$, and thus the coefficients of $B + A\cdot S$ would not grow much.

In the $\Delta M + E + q_0 K \bmod q_L$ term, notice that because of the $q_0K$ term which is not modulo-reduced by $q_0$ anymore, the ciphertext's decrypted plaintext polynomial's each $i$-th term would get a corrupted coefficient $\Delta m_i + e_i + q_0\cdot k_i \bmod q_L$ instead of $\Delta m_i + e_i \bmod q_L$. So, we now need to eliminate the garbage term $q_0\cdot k_i \bmod q_L$ in each coefficient and distill the pure plaintext coefficient $\Delta m_i + e_i$.

\begin{figure}[h!]
    \centering
  \includegraphics[width=1.0\linewidth]{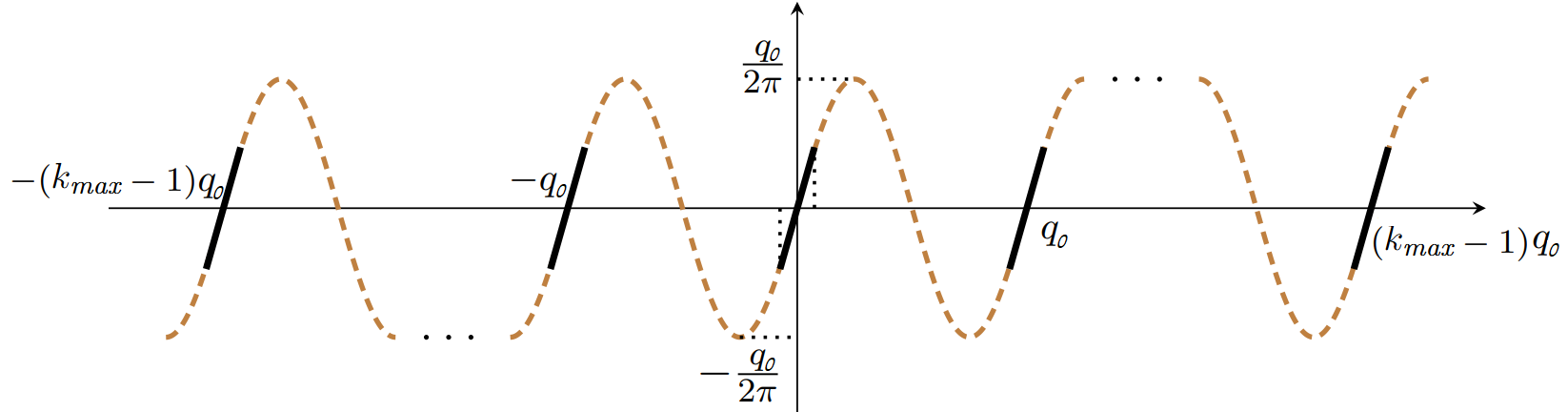}
  \caption{Sine function $f(x) = \dfrac{q_0}{2\pi}\cdot \sin \left(\dfrac{2\pi x}{q_0}\right)$ such that $f(\Delta m_i + e_i + q_0k_i) \approx \Delta m_i + e_i$ (provided $\Delta m_i + e_i \ll q_0$) \href{https://eprint.iacr.org/2018/153.pdf}{(Source)}}
  \label{fig:modulo-reduction-sine}
\end{figure}

 To do so, we will take an approximated approach by using a sine function described in \autoref{fig:modulo-reduction-sine}, which has a period of $q_0$ with the amplitude $\dfrac{q_0}{2\pi}$. This sine function has the following two useful properties:
 
 \begin{enumerate}
\item When $f(x)$ is evaluated at $x$ values near the multiple of $q_0$, the result approximates to that of a line function $y = x$. This is because the derivative (slope) of $\sin x$ is $y' = \cos x$, and if $x$ is a multiple of $2\pi$, the slope is: $y' = \cos 2\pi = 1$. 
\item The evaluation of $f(x)$ eliminates the multiples of $q_0$ from the input (i.e., modulo reduction $q_0$)
\end{enumerate}

$ $

Combining these two properties, given input $x = \Delta m_i + e_i + q_0k_i$, 

$f(\Delta m_i + e_i + q_0k_i) = \dfrac{q_0}{2\pi}\cdot \sin \left(\dfrac{2\pi \cdot (\Delta m_i + e_i + q_0k_i))}{q_0}\right) = \dfrac{q_0}{2\pi}\cdot \sin \left(\dfrac{2\pi \cdot (\Delta m_i + e_i))}{q_0}\right) \approx \Delta m_i + e_i$

$ $

, provided $\Delta m_i + e_i$ is very close to 0 relative to $q_0$ (i.e., $\Delta m_i + e_i \ll q_0$). This is true, because by construction of the CKKS scheme, the plaintext modulus (even with scaling it up by $\Delta$), is significantly smaller than the ciphertext modulus. Therefore, to remove $q_0K$ from $\Delta M + E + q_0K$, we can update each coefficient of the polynomial $\Delta M + E + q_0K$ by evaluating it with the $f(x)$ sine function. However, we cannot directly update the coefficients of the polynomial, because the CKKS scheme (the RLWE scheme in general) only supports the input vector's slot-wise $(+, \cdot)$ operations. Therefore, to update the polynomial coefficients, we need to express the update logic in terms of slot-wise input vector arithmetic $(+, \cdot)$. Considering all these, CKKS's overall bootstrapping procedure is described in \autoref{tab:ckks-bootstrapping-procedure}.

\begin{table}
\begin{tabular}{|ll|}
\hline
1& \textsf{\textbf{\underline{ModRaise:}}} Given ciphertext $(A, B) \bmod q_0$, we forcibly raise its modulus from $q_0$ to $q_L$. \\
& Then, it ends up encrypting $\Delta M + E + q_0k$ instead of $\Delta M + E$.\\
2& \textsf{\textbf{\underline{CoeffToSlot:}}} Based on step 1's ciphertext $(A, B) \bmod q_L$, we generate a new ciphertext\\
& that encrypts an input vector whose its each $i$-th slot stores $\Delta m_i + e_i + q_0k_i$.\\
& This is equivalent to moving the coefficients of polynomial $\Delta M + E + q_0K$ to\\
& the input vector slots.\\
3& \textsf{\textbf{\underline{EvalExp:}}} We convert the sine function into an approximated polynomial by using \\
& the Taylor series, as well as other optimizations such as\\ 
& Euler's formula ($e^{i\theta} = \cos(\theta) + i\cdot\sin(\theta)$). Then, we generate a CKKS plaintext that encodes \\
& this approximated sine function, and then use this to homomorphically evaluate \\
& step 2's encrypted vector elements (to homomorphically remove every coefficient's $q_0k_i$).\\
4& \textsf{\textbf{\underline{SlotToCoeff:}}} Based on the resulting ciphertext from step 3, we generate a new ciphertext \\
& whose encrypted polynomial's each $i$-th coefficient is (approximately) $\Delta m_i + e_i$. \\
& This is equivalent to moving the $q_0k_i$-eliminated values stored in the input vector slots in \\
&step 3 back to the positions of the polynomial coefficients. The final ciphertext is \\
& our goal ciphertext that (approximately) encrypts $\Delta M + E$ under modulus $q_L$.\\
\hline
\end{tabular}
\caption{High-level Description of CKKS's Bootstrapping Procedure}
\label{tab:ckks-bootstrapping-procedure}
\end{table}
%CKKS's bootstrapping procedure described above breaks down into the following 5 steps: \textsf{ModRaise}, \textsf{CoeffToSlot}, \textsf{EvalExp}, \textsf{ImgExt}, and \textsf{SlotToCoeff}.

\subsubsection{Mathematical Expansion of the High-level Idea}
\label{subsubsec:ckks-bootstrapping-high-level-correctness}

We will mathematically walk through how the bootstrapping procedure (\autoref{tab:ckks-bootstrapping-procedure}) correctly updates the modulus of the input ciphertext from $q_0$ to $q_L$. 

For ease of understanding, we will first explain how we would do modulus bootstrapping for a ciphertext with multiplicative level 0 (i.e., its modulus is $q_0$) in case we have access to the secret key $S(X)$. Using this key, we can decrypt the ciphertext as follows:

$\textsf{RLWE}^{-1}\textbf{(}\textsf{ct} = (A, B)\textbf{)}$ \textcolor{red}{ $\rhd$ where $\textsf{ct}=(A, B) = \textsf{RLWE}_{S, \sigma}\bm(\Delta M \bm)$}

$ = B + A\cdot S = \Delta M + E \bmod q_0$ 

$ = \Delta M + E + q_0K$ \textcolor{red}{ $\rhd$ where $q_0K$ accounts for any potential wrap-around modulo $q_0$ values}

Our initial goal is to bootstrap the modulus of the ciphertext from $q_0$ to $q_L$ by using only the following three tools:

\begin{itemize}
\item Secret key $S$
\item Batch-encoding ($\sigma^{-1}$) and decoding ($\sigma$) formulas
\item Batched slot-wise $(+, \cdot)$ operation of input vectors based on their batch-encoded polynomials
\end{itemize}

$ $

After explaining the above, we will then explain how to achieve the same bootstrapping without having access to the secret key $S$.

$ $

\para{\textsf{\textbf{\underline{ModRaise}:}}} This step forcibly changes the ciphertext's modulus from $q_0$ to $q_L$ and then decrypts the ciphertext as follows:

$\textsf{RLWE}^{-1}\textbf{(}\textsf{ct} = (A, B)\textbf{)} = B + A\cdot S = \Delta M + E + q_0 K \bmod q_L$

Notice that the ciphertext's decrypted plaintext polynomial's each $i$-th coefficient gets corrupted to $\Delta m_i + e_i + q_0\cdot k_i \bmod q_L$. So, we now need to eliminate the garbage term $q_0k_i \bmod q_L$ in each coefficient and distill the pure plaintext coefficient $\Delta m_i + e_i$. 

$ $

\para{\textsf{\textbf{\underline{CoeffToSlot}:}}} This step generates a new plaintext polynomial whose each $i$-th input vector slot stores the corrupted coefficient $(m_i + e_i + q_0k_i)$. The trick of doing this is to apply CKKS's batch-encoding mapping $\sigma^{-1}$ (which represents the transformation $\vec{m} = \dfrac{\hathat W \cdot I_n^R \cdot \vec{v}_{'}}{n}$ as explained in \autoref{subsec:ckks-encoding-decoding}) to the input vector slots that encode the polynomial $\Delta M + E + q_0 K \bmod q_L$. Let $\vec{v}_c$ be the input vector that corresponds to polynomial $\Delta M + E + q_0 K$. Then, $\vec{v}_c$ and $\Delta M + E + q_0 K$ satisfy the following relation over the encoding mapping $\sigma^{-1}$: 

$\sigma^{-1}\bm(\vec{v}_c\bm) = M_c = \sum\limits_{i=0}^{n-1}(\Delta m_i + e_i + q_0k_i)\cdot X^{i}$ \textcolor{red}{ $\rhd$ i.e., polynomial $\Delta M + E + q_0K$} 

This implies that if we \textit{homomorphically} apply the $\sigma^{-1}$ transformation to the elements of the input vector $\vec{v}_c$, then the resulting input vector $\vec{v}_s$ will store $\vec{v}_c$'s encoded polynomial coefficient values as follows:

$\sigma^{-1} \circ \vec{v}_c = \vec{v}_s = (\Delta m_0 + e_0 + q_0k_0, \text{ } \Delta m_1 + e_1 + q_0k_1, \text{ } \cdots, \text{ } \Delta m_{n-1} + e_{n-1} + q_0k_{n-1})$  

\textcolor{red}{ $\rhd$ where $\circ$ represents a linear transformation operation comprising $(+, \cdot)$}

$ $

However, remember that at the end of the \textsf{ModRaise} step, we get the decrypted (but corrupted by $q_0k$) polynomial $M_c = \textsf{RLWE}^{-1}\textbf{(}\textsf{ct} = (A, B)\bm) = \Delta M + E + q_0K$ and we are not allowed to decode it into $\vec{v}_c$. Therefore, we will instead \textit{encode} the matrix $\hathat W \cdot I_n^R$ in the encoding transformation $\sigma^{-1}$ ($\vec{m} = \dfrac{\hathat W \cdot I_n^R \cdot \vec{v}_{'}}{n}$) into its equivalent polynomials (treating a matrix as a combination of vectors) and then perform batched slot-wise $(+, \cdot)$ operation between $M_c$ and the polynomial version of $\hathat W \cdot I_n^R$. We express this polynomial-based computation as follows:

$M_s = \sigma^{-1}_{\sigma^{-1}} \circ M_c \bmod q_L$
\textcolor{red}{ $\rhd$ $\sigma^{-1}_{\sigma^{-1}}$ is the polynomial-encoded version of the $\sigma^{-1}$ transformation}

Then, the resulting polynomial $M_s$'s corresponding input vector slots (i.e., the decoded version of $M_s$) will store $\vec{v}_s = (\Delta m_0 + e_0 + q_0k_0, \text{ } \Delta m_1 + e_1 + q_0k_1, \text{ } \cdots, \text{ } \Delta m_{n-1} + e_{n-1} + q_0k_{n-1})$. In other words, the above computation effectively \textit{moves} the coefficients of $M_c$ to the input vector slots of a new plaintext polynomial. 

However, remember that in CKKS, an input vector can store only up to $\dfrac{n}{2}$ slots, whereas we need to store a total of $n$ coefficients of $M_c$ in the input vector slots. Therefore, we technically need to create 2 pieces of $M_s$ as $M_{s1}$ and $M_{s2}$, where the input vector of $M_{s1}$ stores $(\Delta m_0 + e_0 + q_0k_0, \text{ } \Delta m_1 + e_1 + q_0k_1, \text{ } \cdots, \text{ } \Delta m_{\frac{n}{2}-1} + e_{\frac{n}{2}-1} + q_0k_{\frac{n}{2}-1})$, and the input vector of $M_{s2}$ stores $(\Delta m_{\frac{n}{2}} + e_{\frac{n}{2}} + q_0k_{\frac{n}{2}}, \text{ } \cdots, \text{ } \Delta m_{n-1} + e_{n-1} + q_0k_{n-1})$.

$ $

\para{\textsf{\textbf{\underline{EvalExp}:}}}
Our next step is to update $\vec{v}_s$'s each element $m_i + e_i + q_0k_i$ to $m_i + e_i$ by evaluating it with the sine function $f(x)$. Since the output of the \textsf{CoeffToSlot} step is polynomial $M_s$ (technically $M_{s1}$ and $M_{s2}$), we need to apply the evaluation transformation in an encoded form. First, we approximate $f(x)$ as a linear combination comprising only $(+, \cdot)$ operations by using the Taylor series and Euler's formula (will be explained later). Then, we encode (i.e., $\sigma$) the approximated formula into a polynomial form, and we denote it as $\sigma_f$. Finally, we apply the $\sigma_f$ transformation to $M_s$ as follows: 

$\sigma^{-1}_f \circ M_s \bmod q_L$  \textcolor{red}{ $\rhd$ Applying the sine function's linear transformation to $\vec{v}_s$'s each slot storing $\Delta m_i + e_i + q_0k_i$}

$= M_t = \sigma\bm{(}\vec{v}_t\bm{)} = \sigma\bm{(}(\Delta m_i + e_i)_{i=0}^{n-1}\bm{)} \bmod q_L$

$ $

After the linear transformation by the sine function, notice that each $q_0 k_i$ term gets eliminated from $\vec{v}_s$'s slots (i.e. modulo reduction by $q$) and the resulting vector $\vec{v}_t$ stores only the $\Delta m_i + e_i$ terms.

$ $

\para{\textsf{\textbf{\underline{SlotToCoeff}:}}}
Now that we have a polynomial $M_t$ whose corresponding input vector $\vec{v}_t$'s slots store garbage-removed coefficients of (i.e., $\Delta m_i + e_i$) our initial plaintext polynomial, our next step is to put these coefficients stored in $\vec{v}_t$ back to the polynomial. This is an exact reverse operation of \textsf{CoeffToSlot} as follows:

$\sigma^{-1}_{\sigma} \circ M^t = M_b$ 
\textcolor{red}{ $\rhd$ $\sigma^{-1}_{\sigma}$ is a polynomial-encoded form of the batch-decoding formula $\vec{v}_{'} = \hathat W^* \cdot \vec{m}$ (\autoref{subsec:ckks-encoding-decoding})}

The result is polynomial $M_b$ whose coefficients are garbage-eliminated (i.e., $q_0k_i$-free) versions of $M_c$. Finally, we re-encrypt $M_b$ as $\textsf{RLWE}_{S, \sigma}(M_b)$ as the final modulus-bootstrapped ciphertext.

$ $

\para{\textsf{\textbf{\underline{Bootstrapping Without a Secret Key}:}}} So far, we have assumed that we have access to the secret key $S$. With decryption and re-encryption enabled, the above bootstrapping steps described are mathematically equivalent to computing the following:

\begin{enumerate}
\item \textsf{\textbf{\underline{INPUT}:}} $\textsf{ct} = (A, B) \bmod q_0$ \textcolor{red}{ $\rhd$ where $\textsf{ct}=(A, B) = \textsf{RLWE}_{S, \sigma}\bm(\Delta M \bm)$}

\item \textsf{\textbf{\underline{ModRaise}:}} $\textsf{ct} = (A, B) \bmod q_L$

\item \textsf{\textbf{\underline{Decryption}:}} 
$\textsf{RLWE}^{-1}\textbf{(}\textsf{ct} = (A, B)\textbf{)} \textbf{)} \bmod q_L$

\item \textsf{\textbf{\underline{CoeffToSlot}:}} $\sigma^{-1}_{\sigma^{-1}} \circ \textsf{RLWE}^{-1}\textbf{(}\textsf{ct} = (A, B)\textbf{)} \textbf{)} \bmod q_L$

\item \textsf{\textbf{\underline{EvalExp}:}} $\sigma^{-1}_f \circ (\sigma^{-1}_{\sigma^{-1}} \circ \textsf{RLWE}^{-1}\textbf{(}\textsf{ct} = (A, B)\textbf{)} \textbf{)}) \bmod q_L$

\item \textsf{\textbf{\underline{SlotToCoeff}:}} $\sigma^{-1}_{\sigma} \circ (\sigma^{-1}_f \circ (\sigma^{-1}_{\sigma^{-1}} \circ \textsf{RLWE}^{-1}\textbf{(}\textsf{ct} = (A, B)\textbf{)} \textbf{)})) \bmod q_L$

\item \textsf{\textbf{\underline{Re-encryption}:}} $\textsf{RLWE}_{S, \sigma}\bm(\sigma^{-1}_{\sigma} \circ (\sigma^{-1}_f \circ (\sigma^{-1}_{\sigma^{-1}} \circ \textsf{RLWE}^{-1}\textbf{(}\textsf{ct} = (A, B)\textbf{)} \textbf{)}))\bm) \bmod q_L$

\end{enumerate}

$ $

However, the ultimate goal of CKKS bootstrapping is to reset the modulus of a ciphertext from $q_0$ to $q_L$ without having access to $S$.

Meanwhile, one important insight is that CKKS's \textsf{ModRaise} procedure on the ciphertext $(A, B) \bmod q_0$ from $q_0 \rightarrow q_L$ effectively transforms the ciphertext into a new one which is an encryption of $\Delta M + q_0K$. Before \textsf{ModRaise}, ciphertext $(A, B) \bmod q_0$'s decryption relation is as follows:

$A\cdot A + B = \Delta M + E + Kq_0 \bmod q_0 = \Delta M + E$ 

$ $

After \textsf{ModRaise} to $(A, B) \bmod q_L$, its decryption relation is as follows:

$A\cdot S + B = \Delta M + E + Kq_0 \bmod q_L = \Delta M + E + Kq_0$ \textcolor{red}{ $\rhd$ because $\Delta M + E + Kq_0 \ll q_L$}

$ $

Therefore, the \textit{mod-raised} ciphertext $(A, B) \bmod q_L = \textsf{RLWE}_{S, \sigma}(\Delta M + Kq_0)$ with noise $E$. Thus, CKKS's \textit{homomorphic} bootstrapping strategy is to run the subsequent \textsf{CoeffToSlot}, \textsf{EvalExp}, and \textsf{SlotToCoeff} steps \textsf{homomorphically} based on the ciphertext $(A, B) \bmod q_L$. Running these 3 steps consumes a few multiplicative levels due to the ciphertext-to-ciphertext multiplication operations when homomorphically multiplying the coefficient-to-slot and slot-to-coefficient transformation matrices and homomorphically computing powers of $X$ (i.e., $X^k$) during sine approximation. Therefore, upon completion of these 3 steps, the ciphertext modulus reduces from $q_L \rightarrow q_l$ (where $l$ is some integer such that $l < L$).

Note that the result of homomorphic bootstrapping is equal to the explicit bootstrapping based on decryption \& re-encryption (if we ignore the small differences in the final ciphertext modulus and the noise). In the following subsections, we will explain the algebraic details of \textsf{CoeffToSlot}, \textsf{EvalExp} and \textsf{SlotToCoeff} steps. 

\subsubsection{Details: \textsf{CoeffToSlot}}
\label{subsubsec:ckks-bootstrapping-coefftoslot-details}

Homomorphically moving the coefficients of $M_c$ (i.e., $\Delta m_i + e_i + q_0k_i$ for $0 \leq i \leq n - 1$) to a new ciphertext's input vector slots is mathematically equivalent to homomorphically computing $\sigma^{-1}_{\sigma^{-1}} \circ (\textsf{RLWE}_{S, \sigma}\bm{(}\textsf{ct} = (A, B)\bm))$, which is equivalent to applying the encoding formula to the input vector slot values of $\textsf{RLWE}_{S, \sigma}\bm{(}\textsf{ct} = (A, B)\bm)$.

As explained in Summary~\ref*{subsec:ckks-rotation} (in \autoref{subsec:ckks-rotation}), the encoding formula for converting an input vector into a list of polynomial coefficients is $\vec{m} = \dfrac{\hathat W \cdot I_n^R \cdot \vec{v}_{'}}{n}$, where $\hathat W$ is a basis of the $n$-dimensional vector space crafted as follows: 

$ $

\noindent{\small{$\noindent\hathat W = \begin{bmatrix}
1 & 1 & \cdots & 1 & 1 & 1 & \cdots & 1\\
(\omega^{J(\frac{n}{2} - 1)}) & (\omega^{J(\frac{n}{2} - 2)}) & \cdots & (\omega^{J(0)}) & (\omega^{J_*(\frac{n}{2} - 1)}) & (\omega^{J_*(\frac{n}{2} - 2)}) & \cdots & (\omega^{J_*(0)})\\
(\omega^{J(\frac{n}{2} - 1)})^2 & (\omega^{J(\frac{n}{2} - 2)})^2 & \cdots & (\omega^{J(0)})^2 & (\omega^{J_*(\frac{n}{2} - 1)})^2 & (\omega^{J_*(\frac{n}{2} - 2)})^2 & \cdots & (\omega^{J_*(0)})^2 \\
\vdots & \vdots & \ddots & \vdots & \vdots & \vdots & \ddots & \vdots \\
(\omega^{J(\frac{n}{2} - 1)})^{n-1} & (\omega^{J(\frac{n}{2} - 2)})^{n-1} & \cdots & (\omega^{J(0)})^{n-1} & (\omega^{J_*(\frac{n}{2} - 1)})^{n-1} & (\omega^{J_*(\frac{n}{2} - 2)})^{n-1} & \cdots  & (\omega^{J_*(0)})^{n-1}
\end{bmatrix}$}}

$ $

\textcolor{red}{ $\rhd$ where the rotation helper function $J(h) = 5^h \bmod 2n$}

$ $

Therefore, given the input ciphertext $\textsf{ct}_{\textsf{c}} = \textsf{RLWE}_{S, \sigma}\bm{(}\textsf{ct} = (A, B)\bm) \bmod q_L$ whose plaintext polynomial $M_c$ contains corrupted coefficients, computing $\sigma^{-1}_{\sigma^{-1}} \text{ } \circ \text{ } \textsf{ct}_{\textsf{c}}$ is equivalent to computing $\dfrac{\hathat W \cdot I_n^R \cdot \textsf{ct\textsubscript{c}}}{n}$. However, one problem here is that each CKKS ciphertext encodes only $\dfrac{n}{2}$ input vector slots, whereas our goal is to move $n$ (corrupted) coefficients of the plaintext polynomial $M_c$ encrypted in $\textsf{ct\textsubscript{c}}$. Therefore, we will instead generate 2 ciphertexts, $\textsf{ct}_{\textsf{s1}}$ and $\textsf{ct}_{\textsf{s2}}$, such that each $\textsf{ct}_{\textsf{s1}}$'s input vector slots store $(\Delta m_i + e_i + q_0k_i)_{0 \leq i < \frac{n}{2}}$ and $\textsf{ct}_{\textsf{s2}}$'s input vector slots store $(\Delta m_i + e_i + q_0k_i)_{\frac{n}{2} \leq i < n}$. 

We will leverage the following matrix split technique. When we split an $n \times n$ matrix $\hathat{W}$ into four $\frac{n}{2} \times \frac{n}{2}$ blocks to multiply it by the vector $\vec{v}'$, the standard matrix multiplication is as follows:

$$\begin{bmatrix} \hathat{W}_{11} & \hathat{W}_{12} \\ \hathat{W}_{21} & \hathat{W}_{22} \end{bmatrix} \begin{bmatrix} \vec{a} \\ \vec{b} \end{bmatrix} = \begin{bmatrix} \hathat{W}_{11}\vec{a} + \hathat{W}_{12}\vec{b} \\ \hathat{W}_{21}\vec{a} + \hathat{W}_{22}\vec{b} \end{bmatrix}$$

In our case, $\vec{a} = (v_0, \dots, v_{\frac{n}{2}-1})$ (the first half of $\vec{v}'$), and $\vec{b} = (\overline{v}_0, \dots, \overline{v}_{\frac{n}{2}-1})$ (the second half of $\vec{v}'$).

$ $

Therefore, We split the $n \times n$ matrix $\hathat W$ into four $\dfrac{n}{2} \times \dfrac{n}{2}$ matrices as follows:

\begin{itemize}
\item $[\hathat W]_{11}$: a matrix comprising the upper left-half section of $\hathat W$
\item $[\hathat W]_{12}$: a matrix comprising the upper right-half section of $\hathat W$
\item $[\hathat W]_{21}$: a matrix comprising the lower left-half section of $\hathat W$
\item $[\hathat W]_{22}$: a matrix comprising the lower right-half section of $\hathat W$
\end{itemize}

$ $

Then, we can compute $\textsf{ct}_{\textsf{s1}}$ and $\textsf{ct}_{\textsf{s2}}$ as follows:

$\textsf{ct}_{\textsf{s1}} = \dfrac{[\hathat W]_{11} \cdot I_{\frac{n}{2}}^R \cdot \mathit{\overline{\textsf{ct\textsubscript{c}}}} + [\hathat W]_{12} \cdot I_{\frac{n}{2}}^R \cdot \textsf{ct\textsubscript{c}}}{n}$

$\textsf{ct}_{\textsf{s2}} = \dfrac{[\hathat W]_{21} \cdot I_{\frac{n}{2}}^R \cdot \mathit{\overline{\textsf{ct\textsubscript{c}}}} + [\hathat W]_{22} \cdot I_{\frac{n}{2}}^R \cdot \textsf{ct\textsubscript{c}}}{n}$

, where $\mathit{\overline{\textsf{ct\textsubscript{c}}}}$ can be computed by applying homomorphic conjugation to \textsf{ct\textsubscript{c}} (\autoref{subsec:ckks-conjugation}). Each homomorphic matrix-vector multiplication (e.g., $[\hathat W]_{21} \cdot I^R_{\frac{n}{2}}\overline{\textsf{ct\textsubscript{c}}}$) can be done in an efficient manner that reduces the number of homomorphic rotations (\autoref{subsec:bfv-matrix-multiplication}).

\subsubsection{Details: \textsf{EvalExp}}
\label{subsubsec:ckks-bootstrapping-evalexp-details}

We will use the sine function $f(x) = \dfrac{q_0}{2\pi}\cdot \sin\left(\dfrac{2\pi x}{q_0}\right)$ to approximately eliminate $q_0k_i$ from $\Delta m_i + e_i + q_0k_i$ by computing $f(\Delta m_i + e_i + q_0k_i) \approx \Delta m_i + e_i$. This approximation works if $\Delta m_i + e_i$ is very close to $x=0$ relative to $q_0$ (i.e., $\Delta m_i + e_i \ll q_0$). Still, the elimination of $q_0k_i$ is approximate (i.e., $\approx \Delta m_i + e_i$), because $f(x)$ is $y \approx x$ nearby $x=0$, not exactly $y = x$. 

One issue is that we need to evaluate $f(x)$ homomorphically based on \textsf{ct\textsubscript{s1}} and \textsf{ct\textsubscript{s2}} as inputs (i.e., $f(\textsf{ct\textsubscript{s1}})$ and $f(\textsf{ct\textsubscript{s2}})$), but FHE supports only $(+, \cdot)$ operations, whereas the sine graph cannot be formulated by only $(+, \cdot)$. Therefore, we will approximate the sine function $f(x)$ by using the Taylor series (\autoref{sec:taylor-series}): 

$f(x) = f(a) + \dfrac{f'(a)}{1!}(x-a) + \dfrac{f''(a)}{2!}(x-a)^2 + \dfrac{f'''(a)}{3!}(x-a)^3 + \cdots = \sum\limits_{d=0}^{\infty}\dfrac{f^{(d)}(a)}{d!}(x-a)^d$ 

$ $

If we approximate $f(x)$ around $x = 0$, then the approximated polynomial is as follows: 

$f(x) = \dfrac{q_0}{2\pi}\cdot \sin\left(\dfrac{2\pi}{q_0}\cdot 0\right) + \dfrac{q_0}{2\pi}\cdot\dfrac{2\pi}{q_0}\cdot\dfrac{\cos\left(\dfrac{2\pi}{q_0} \cdot 0\right)}{1!}\cdot x + \dfrac{q_0}{2\pi}\cdot \left(\dfrac{2\pi}{q_0}\right)^{2}\cdot\dfrac{-\sin\left(\dfrac{2\pi}{q_0} \cdot 0\right)}{2!}\cdot x^2 + \cdots$

$ = \dfrac{q_0}{2\pi}\cdot \sum\limits_{j=0}^{\infty}\left(\dfrac{(-1)^j}{(2j + 1)!}\cdot \left(\dfrac{2\pi x}{q_0}\right)^{2j + 1}\right)$

$ \approx \dfrac{q_0}{2\pi}\cdot \sum\limits_{j=0}^{h}\left(\dfrac{(-1)^j}{(2j + 1)!}\cdot \left(\dfrac{2\pi x}{q_0}\right)^{2j + 1}\right) = \hat f(x)$

$ $

, where $\hat{f}(x)$ is a $(2h + 1)$-degree polynomial. 

Remember that in the RLWE cryptosystem, $B + AS \bmod q_0 = \Delta M + E$, or $B + AS = \Delta M + E + q_0K$ with some polynomial $K$ representing the wrapping around values of modulo $q_0$. Since the secret key $S$ is an $(n-1)$-degree polynomial whose coefficients are small (i.e., $s_i \in \{-1, 0, 1\}$), the coefficients of $K$ will have some \textit{reasonably small} upper bound, which decreases with the sparsity of $S$ (i.e., the frequency of 0 coefficients in $S$). Therefore, the degree of our approximated $\hat f(x)$ only needs to be high enough to accurately evaluate $y$ values between $-q_0\cdot \mathit{k_{max}} \leq x \leq q_0\cdot \mathit{k_{max}}$.
The required minimum degree of our approximated Taylor polynomial $\hat f(x)$ increases with $q_0\mathit{k_{max}}$ (i.e., the upper bound of $x$). Our one issue is that the computation overhead for homomorphic evaluation of a polynomial generally increases exponentially with the degree of the polynomial, which will slow down bootstrapping. To reduce this computation cost, we will leverage Euler's formula (\autoref{sec:euler}) and its square arithmetic:

\[
\begin{cases}
e^{i\cdot \theta} = \cos \theta + i \cdot \sin \theta\\
(e^{i \cdot \theta})^2 = e^{i\cdot 2\theta}\\
\end{cases}
\]

By substituting $\theta = \dfrac{2\pi x}{q_0}$, we will use Euler's formula. We will also approximate $e^{i\theta}$ with the Taylor series, but instead of directly approximating $e^{i\theta}$, we will first approximate $e^{\frac{i\theta}{2^r}}$ for some large $2^r$. After that, we will iteratively square $e^{\frac{i\theta}{2^r}}$ a total $r$ times. Then, we get an approximation of $(e^{\frac{i\theta}{2^r}})^{2^r} = e^{i\theta}$. The reason why we start with the approximation of $e^{\frac{i\theta}{2^r}}$ instead of $e^{i\theta}$ is that its approximation requires a small degree of polynomial, as $\dfrac{\theta}{2^r}$ (i.e., the input to the complex exponential function) is small provided $2^r$ is sufficiently large. Specifically, we learned that $x$ ($= \Delta m_i + e_i + q_0k_i$) is upper-bounded by $q_0\mathit{k_{max}}$, thus $\theta = \dfrac{2\pi x}{q_0}$ is upper-bounded by $\dfrac{2\pi\mathit{k_{max}}}{2^r}$. As the targeted range of $x$ for approximation in $f(x)$ is small, we need a small degree of Taylor series polynomial. 

Using the Taylor series with degree $d_0$ around $x = 0$, we can approximate $e^{\frac{2\pi i x}{2^r q_0}}$ as: 

$f_e(x) = e^{\frac{2\pi i x}{2^r q_0}} \approx \sum\limits_{d=0}^{d_0}\dfrac{1}{d!}\left(\dfrac{2\pi i x}{2^rq_0}\right)^d = \hat{f}_e(x)$

$ $

Then, we iteratively square $\hat{f}_e$ total $r$ times to get:

$(\hat{f}_e(x))^{2^r} \approx (f_e(x))^{2^r} = e^{i\frac{2\pi x}{q_0}} = e^{i\theta}$

$ $

Then, based on Euler's formula $e^{i\cdot \theta} = \cos \theta + i \cdot \sin \theta$, we can derive the following relations:

$ $

$\overline{e^{i\cdot \theta}} = \cos \theta + \overline{i \cdot \sin \theta} $

$ e^{-i\cdot \theta} = \cos \theta - i \cdot \sin \theta$

$ e^{i\cdot \theta} - e^{-i\cdot \theta} = (\cos \theta + i \cdot \sin \theta) - (\cos \theta - i \cdot \sin \theta) = 2i\sin \theta$

$\sin \theta = \dfrac{-i}{2} \cdot (e^{i\cdot \theta} - e^{-i\cdot \theta})$

$\dfrac{q_0}{2\pi}\cdot \sin \theta = \dfrac{q_0}{2\pi}\cdot \dfrac{-i}{2} \cdot (e^{i\cdot \theta} - e^{-i\cdot \theta})$

$ $

Substituting $\theta = \dfrac{2\pi x}{q_0}$, we finally get: 

$\dfrac{q_0}{2\pi}\cdot \sin \left(\dfrac{2\pi x}{q_0}\right) = \dfrac{q_0}{2\pi}\cdot \dfrac{-i}{2} \cdot (e^{i\cdot \frac{2\pi x}{q_0}} - e^{-i\cdot \frac{2\pi x}{q_0}})$

$ $

Using the final relation above, the \textsf{EvalExp} step homomorphically evaluates the approximation of $\dfrac{q_0}{2\pi}\cdot \sin \left(\dfrac{2\pi x}{q_0}\right)$ where $x = \Delta m_i + e_i + q_0k_i$ as follows: 

\begin{enumerate}
\item Homomorphically approximately compute $\hat{f}(x) = e^{i\cdot \frac{2\pi x}{q_0}}$.
\item Homomorphically approximately compute $\overline{\hat{f}(x)} = e^{-i\cdot \frac{2\pi x}{q_0}}$ by applying homomorphic conjugation. (\autoref{subsec:ckks-conjugation}) to $\hat{f}(x)$
\item Homomorphically compute $\hat{f}(x) - \overline{\hat{f}(x)} = e^{i\cdot \frac{2\pi x}{q_0}} -  e^{-i\cdot \frac{2\pi x}{q_0}}$, and then multiply the result by $\dfrac{-i}{2}$ encoded as CKKS plaintext.
\end{enumerate}

$ $

The result of \textsf{EvalExp} is two ciphertexts whose input vector slots store the bootstrapped coefficients of $M_c$, which are modulo-reduced $q_0$ from $\Delta m_i + e_i + q_0k_i$ to $\Delta m_i + e_i + e_{bi}$. Note that $e_{bi}$ is a bootstrapping error introduced by the following three factors: (1) the intrinsic homomorphic $(+, \cdot)$ computation noises of the \textsf{CoeffToSlot}, \textsf{EvalExp}, and \textsf{SlotToCoeff} steps; (2) the \textsf{EvalExp} step's Taylor polynomial approximation error of the exponential function $e^{i\theta}$; (3) the \textsf{EvalExp} step's sine graph error, since the graph is not exactly $y=x$ around $x=0$, but only $y \approx x$.

Note that since the output of the \textsf{CoeffToSlot} step was split into 2 ciphertexts (\textsf{ct\textsubscript{s1}} and \textsf{ct\textsubscript{s2}}), the output of the \textsf{EvalExp} step is also in 2 ciphertexts: (\textsf{ct\textsubscript{b1}} and \textsf{ct\textsubscript{b2}}). The input vector slots of \textsf{ct\textsubscript{b1}} store $(\Delta m_i + e_i + e_{bi})_{i=0}^{\frac{n}{2} - 1}$, whereas the input vector slots of \textsf{ct\textsubscript{b2}} store $(\Delta m_i + e_i + e_{bi})_{i=\frac{n}{2}}^{n - 1}$.

\subsubsection{Details: \textsf{SlotToCoeff}}
\label{subsubsec:ckks-bootstrapping-slottocoeff-details}

This step is the exact inverse of the \textsf{CoeffToSlot} step, which is moving the bootstrapped (i.e. modulo-reduced $q_0$) coefficients of $M_v$ stored in the input vector slots back to the final plaintext polynomial $M_f$. Remember that the decoding formula from a polynomial to an input vector (\autoref{subsec:ckks-encoding-decoding}) is $\vec{v}_{'} = \hathat W^* \cdot \vec{m}$, where:

$\hathat{W}^* = \begin{bmatrix}
1 & (\omega^{J(0)}) & (\omega^{J(0)})^2 & \cdots & (\omega^{J(0)})^{n-1}\\
1 & (\omega^{J(1)}) & (\omega^{J(1)})^2 & \cdots & (\omega^{J(1)})^{n-1}\\
1 & (\omega^{J(2)}) & (\omega^{J(2)})^2 & \cdots & (\omega^{J(2)})^{n-1}\\
\vdots & \vdots & \vdots & \ddots & \vdots \\
1 & (\omega^{J(\frac{n}{2}-1)}) & (\omega^{J(\frac{n}{2}-1)})^2 & \cdots & (\omega^{J(\frac{n}{2}-1)})^{n-1}\\
1 & (\omega^{J_*(\frac{n}{2}-1)}) & (\omega^{J_*(\frac{n}{2}-1)})^2 & \cdots & (\omega^{J_*(\frac{n}{2}-1)})^{n-1}\\
\vdots & \vdots & \vdots & \ddots & \vdots \\
1 & (\omega^{J_*(1)}) & (\omega^{J_*(1)})^2 & \cdots & (\omega^{J_*(1)})^{n-1}\\
1 & (\omega^{J_*(0)}) & (\omega^{J_*(0)})^2 & \cdots & (\omega^{J_*(0)})^{n-1}\\
\end{bmatrix}$

$ $

$ $

We denote $[\hathat W^*]_{11}$, $[\hathat W^*]_{12}$, $[\hathat W^*]_{21}$, and $[\hathat W^*]_{22}$ as $\dfrac{n}{2} \times \dfrac{n}{2}$ matrices corresponding to the upper-left, upper-right, lower-left, and lower-right sections of $\hathat W^*$. Then, homomorphically applying the decoding formula results in the final bootstrapped ciphertext \textsf{ct\textsubscript{final}} modulo $q_L$ whose plaintext polynomial is garbage-eliminated from $\Delta M + E + q_0K \bmod q_L$ to $\Delta M + E + E_b \bmod q_l$ (where $E_b$ is the bootstrapping error polynomial). Note that the ciphertext modulus changed from $q_L \rightarrow q_l$ (for some $l < L$) because we consumed some multiplicative levels for computing ciphertext-to-ciphertext multiplications during polynomial evaluation (i.e., $X^k$).  

$$\begin{bmatrix} \hathat{W}^*_{11} & \hathat{W}^*_{12} \\ \hathat{W}^*_{21} & \hathat{W}^*_{22} \end{bmatrix} \begin{bmatrix} \vec{m}_1 \\ \vec{m}_2 \end{bmatrix} = \begin{bmatrix} \hathat{W}^*_{11}\vec{m}_1 + \hathat{W}^*_{12}\vec{m}_2 \\ \hathat{W}^*_{21}\vec{m}_1 + \hathat{W}^*_{22}\vec{m}_2 \end{bmatrix}$$

$ $

In our case, $\vec{m}_1 = \textsf{ct\textsubscript{b1}}$, and $\vec{m}_2 = \textsf{ct\textsubscript{b2}}$. We can derive  \textsf{ct\textsubscript{final}} by homomorphically applying the decoding transformation $\hathat{W}^*$ to the results of the \textsf{EvalExp} step (\textsf{ct\textsubscript{b1}} and \textsf{ct\textsubscript{b2}}) as follows:

$\textsf{ct\textsubscript{final}} = \textsf{RLWE}_{S, \sigma}\bm(\Delta M + E + E_{b}\bm) = [\hathat W^*]_{11} \cdot \textsf{ct\textsubscript{b1}} + [\hathat W^*]_{12} \cdot \textsf{ct\textsubscript{b2}}$

$\overline{\textsf{ct\textsubscript{final}}} = \textsf{RLWE}_{S, \sigma}\bm(\Delta M + E + E_{b}\bm) = [\hathat W^*]_{21} \cdot \textsf{ct\textsubscript{b1}} + [\hathat W^*]_{22} \cdot \textsf{ct\textsubscript{b2}}$

$ $

Note that we do not need to compute $\overline{\textsf{ct\textsubscript{final}}}$ (i.e., a homomorphic conjugation of $\textsf{ct\textsubscript{final}}$), because we only need to derive the $\dfrac{n}{2}$ input vector slots whose decoding would result in the $n$ coefficients $(\Delta m_i + e_i + e_{bi})_{i=0}^{n-1}$ of the final $(n-1)$-degree polynomial. Once we generate a new ciphertext \textsf{ct\textsubscript{final}} whose $\dfrac{n}{2}$ input vector slots store $(\Delta m_i + e_i + e_{bi})_{i=0}^{n-1}$, then its latter $\dfrac{n}{2}$ conjugate slots get automatically filled with the conjugates of the first $\dfrac{n}{2}$ slot values.

$\textsf{ct\textsubscript{final}} = \textsf{RLWE}_{S, \sigma}(\Delta M + E + E_b)$ 
$= \textsf{SlotToCoeff}(\textsf{ct\textsubscript{b1}}, \textsf{ct\textsubscript{b2}})$

\subsubsection{Reducing the Bootstrapping Overhead by Sparsely Packing Ciphertext}
\label{subsubsec:ckks-bootstrapping-time-reduction}

In many cases, the application of CKKS may use only a small number of input vector slots (e.g., $\dfrac{n'}{2}$) out of $\dfrac{n}{2}$ slots. Suppose that such $n'$ is some number that divides $n$. Then, we can do a series of homomorphic rotations and multiplications to make the input vector slots store $\dfrac{n}{n'}$ repetitions of the $\dfrac{n'}{2}$-slot values. Specifically, we can do this in total $\dfrac{n}{n'}$ rounds of rotations and additions: initially, we zero-mask between the $\frac{n'}{2}$-th slot and the $\frac{n}{2} - 1$-th slots and save as \textsf{ct}, and then in each $i$-th round we compute $\textsf{ct} = \textsf{ct} +  \textsf{Rotate}\bm(\textsf{ct}, -n' \cdot 2^i\bm)$. 

Then, we apply the optimization of sparsely packing ciphertext in Summary~\ref{subsec:ckks-sparse-packing} (\autoref{subsec:ckks-sparse-packing}): if an $\dfrac{n}{2}$-slot input vector is structured as $\dfrac{n}{n'}$ consecutive repetitions of the first $\dfrac{n'}{2}$ slot values, then its encoded polynomial $M(X) \in \mathbb{Z}[X] / (X^{n} + 1)$ has the structure such that all its coefficients whose degree term is not a multiple of $\dfrac{n}{n'}$ are zero as follows:

$M(X) = c_0 + c_{\frac{n}{n'}}X^{\frac{n}{n'}} + c_{\frac{2n}{n'}}X^{\frac{2n}{n'}} + \cdots + c_{n - \frac{n}{n'}}X^{n - \frac{n}{n'}}$. 

Remember that in the \textsf{CoeffToSlot} step (\autoref{subsubsec:ckks-bootstrapping-coefftoslot-details}), we use the formula $\vec{m} = \dfrac{\hathat W \cdot I_n^R \cdot \vec{v}_{'}}{n}$ to move the $q_0k$-contaminated polynomial's coefficients to the input vector slots. But by the principle of sparsely packed ciphertext, we know that all the slots of $\vec{m}$ which are not a multiple of $\dfrac{n}{n'}$ slots would store a zero coefficient. This means that we will get the same computation result even if we only compute the $\vec{m} = \dfrac{\hathat W \cdot I_n^R \cdot \vec{v}_{'}}{n}$ formula with the rows of $\hathat W$ whose row index is a multiple of $\dfrac{n}{n'}$. Mathematically, we can update the encoding formula to $\vec{m_s} = \dfrac{ \text{\rotatecharone{E}} \cdot I_{n'}^R \cdot \vec{v}_{'}}{n'}$ where the $n \times \frac{n}{n'}$ matrix \rotatecharone{E} is an elimination of all those columns from $\hathat W$ whose column index is not a multiple of $\dfrac{n}{n'}$:

{\footnotesize{\hspace{-1cm}\noindent $\text{\rotatecharone{E}} = \begin{bmatrix}
1 & 1 & \cdots & 1 & 1 & \cdots & 1 & 1\\
(\xi^{J(\frac{0\cdot n}{n'}-n')}) & (\xi^{J(\frac{1\cdot n}{n'})}) & \cdots & (\xi^{J(n-\frac{n}{n'})}) & (\xi^{J_*(n-\frac{n}{n'})}) & \cdots & (\xi^{J_*(\frac{1\cdot n}{n'})}) & (\xi^{J_*(\frac{0\cdot n}{n'}-n')})\\
(\xi^{J(\frac{0\cdot n}{n'}-n')})^2 & (\xi^{J(\frac{1\cdot n}{n'}-n')})^2 & \cdots & (\xi^{J(n-\frac{n}{n'})})^2 & (\xi^{J_*(n-\frac{n}{n'})})^2 & \cdots & (\xi^{J_*(\frac{1\cdot n}{n'}-n')})^2 & (\xi^{J_*(\frac{0\cdot n}{n'}-n')})^2 \\
\vdots & \vdots & \ddots & \vdots & \vdots & \ddots & \vdots & \vdots \\
(\xi^{J(\frac{0\cdot n}{n'}-n')})^{n-1} & (\xi^{J(\frac{1\cdot n}{n'}-n')})^{n-1} & \cdots & (\xi^{J(n-\frac{n}{n'})})^{n-1} & (\xi^{J_*(n-\frac{n}{n'})})^{n-1} & \vdots & (\xi^{J_*(\frac{1\cdot n}{n'}-n')})^{n-1} & (\xi^{J_*(\frac{0\cdot n}{n'}-n')})^{n-1} 
\end{bmatrix}$}}

$ $

Remember that in the original \textsf{CoeffToSlot} step (\autoref{subsubsec:ckks-bootstrapping-coefftoslot-details}), we had to split \textsf{ct\textsubscript{s}} into \textsf{ct\textsubscript{s1}} and \textsf{ct\textsubscript{s2}} because in CKKS each input vector can store a maximum of $\frac{n}{2}$ slots but we need to move a total of $n$ coefficient values to the input vector slots for bootstrapping. On the other hand, the computation result of the above updated encoding formula (using a sparsely packed ciphertext) is $\vec{m}_s$, having only $\frac{n}{n'}$ coefficient slots instead of $n$ coefficient slots, and each slot index $i$ in $\vec{m}_s$ corresponds to the encoded polynomial's coefficient with degree term $i \cdot \dfrac{n}{n'}$ (we do not compute any other coefficient terms, because we know that they are 0 anyway, so no need to bootstrap them). And notice that $\frac{n}{n'} \leq \frac{n}{2}$, because $n'$ divides $n$. Therefore, without computing two ciphertexts $\textsf{ct}_{\textsf{s1}} = [\hathat WI_n^R]_{11} \cdot \textsf{ct\textsubscript{c}} + [\hathat WI_n^R]_{12} \cdot I_{\frac{n}{2}}^R \cdot \mathit{\overline{\textsf{ct\textsubscript{c}}}}$ and $\textsf{ct}_{\textsf{s2}} = [\hathat WI_n^R]_{21} \cdot \textsf{ct\textsubscript{c}} + [\hathat WI_n^R]_{22} \cdot I_{\frac{n}{2}}^R \cdot \mathit{\overline{\textsf{ct\textsubscript{c}}}}$ separately, we can directly compute $\textsf{ct\textsubscript{c}} = \dfrac{ \text{\rotatecharone{E}} \cdot I_{n'}^R \cdot \textsf{ct\textsubscript{c}}}{n'}$, because all coefficients for bootstrapping fit in $\dfrac{n}{2}$ slots. Therefore, the number of homomorphic computations and memory requirement for the \textsf{CoeffToSlot} step can be reduced by half. And the same is true for the \textsf{EvalExp} step (\autoref{subsubsec:ckks-bootstrapping-evalexp-details}).

Similarly, as for the \textsf{SlotToCoeff} step (\autoref{subsubsec:ckks-bootstrapping-slottocoeff-details}), we update the decoding formula $\vec{v}_{'} = \hathat W^* \cdot \vec{m}$ to $\vec{v}_{'} = \text{\rotatecharone{E}}^{T} \cdot \vec{m}_c$. This again reduces the number of homomorphic computations and memory requirements for the \textsf{SlotToCoeff} step by half. Notice that $\text{\rotatecharone{E}}^{T}$ is a matrix where those columns whose column index is not a multiple of $\dfrac{n}{n'}$ are zero. This zero-enforcement to the columns of $\text{\rotatecharone{E}}^{T}$ still outputs the same computation result, because $\vec{m}_c$ is a vector such that those slots whose slot index is not a multiple of $\dfrac{n}{n'}$ are zero, which makes the computation result with their corresponding columns of $\text{\rotatecharone{E}}^{T}$ (i.e., the columns whose index is not a multiple of $\dfrac{n}{n'}$) zero, anyway.

\subsubsection{Summary}
\label{subsubsec:ckks-bootstrapping-summary}

We summarize the CKKS bootstrapping procedure as follows. 

\begin{tcolorbox}[title={\textbf{\tboxlabel{\ref*{subsubsec:ckks-bootstrapping-summary}} CKKS Bootstrapping}}]

\setlist[itemize]{leftmargin=*}
\setlist[enumerate]{leftmargin=*}
\begin{enumerate}
\item \textsf{\textbf{\underline{INPUT}:}} $\textsf{ct} = (A, B) \bmod q_0$ \textcolor{red}{ $\rhd$ where $\textsf{ct}=(A, B) = \textsf{RLWE}_{S, \sigma}\bm(\Delta M \bm)$}

$ $

, which satisfies the decryption relation: $A\cdot S + B = \Delta M + E + Kq_0$ 

$ $

\item \textsf{\textbf{\underline{ModRaise}:}} View the polynomials $A$ and $B$ as plaintext polynomials whose each coefficient is in $\mathbb{Z}_{q_L}$ (i.e., $(A, B) \bmod q_L$). This change of viewpoint automatically changes the ciphertext as $\textsf{RLWE}_{S, \sigma}(\Delta M + Kq_0)$. The \textsf{ModRaise} step does not require any actual computation. 

$ $

\item \textsf{\textbf{\underline{CoeffToSlot}:}} 

Move the coefficients of the encrypted plaintext $\Delta M + E + q_0K$ to the input vector slots by homomorphically multiplying $n^{-1}\cdot \hathat W \cdot I_n^R$ to it follows: 

$\textsf{RLWE}_{S, \sigma}(Z_1) = n^{-1}\cdot \hathat W \cdot I_n^R \cdot \textsf{RLWE}_{S, \sigma}(\Delta M + E + q_0K) \bmod q_L$

$ $

\item \textsf{\textbf{\underline{EvalExp}:}} 

Remove the wrap-around garbage value $q_0K$ in $\Delta M + E + q_0K$ by homomorphically evaluating the polynomial $\sigma_f$ which approximates a sine function with period $q_0$ as follows: 

$\textsf{RLWE}_{S, \sigma}(Z_2) = \sigma_f \circ \textsf{RLWE}_{S, \sigma}(Z_1)  \bmod q_l$

$ $

This step is equivalent to \textit{homomorphically} performing modulo reduction by $q_0$ to the input value. This step reduces the ciphertext modulus from $q_L \rightarrow q_l$ as it consumes multiplicative levels when homomorphically evaluating the polynomial approximation of the sine function.

$ $

\item \textsf{\textbf{\underline{SlotToCoeff}:}} 

Move the modulo-$q_0$-reduced plaintext value $\Delta M + E$ stored in the input vector slots back to the plaintext coefficient positions by homomorphically multiplying the encoding matrix $\hathat W^*$ as follows:

$\textsf{RLWE}_{S, \sigma}(\Delta M + E) = \hathat W^* \cdot \textsf{RLWE}_{S, \sigma}(Z_2)  \bmod q_l$

$ $

\end{enumerate}
\setlist[itemize]{leftmargin=\leftmargini}
\setlist[enumerate]{leftmargin=\leftmargini}

$ $

\textbf{Limitation:} The noise slowly grows over each bootstrapping due to the bootstrapping error and will eventually overflow the message and the ciphertext modulus. 

\end{tcolorbox}

\para{Comparison between BFV and CKKS Bootstrapping:} In the case of CKKS's bootstrapping, it does not reduce the magnitude of the old noise $E$ and keeps it the same as before, because the sine approximation function converts $\Delta M + E + Kq_0$ into $\Delta M + E$. However, as the ciphertext modulus gets increased from $q_0 \rightarrow q_L$, the noise-to-ciphertext-modulus ratio decreases, since $\dfrac{E}{q_L} \ll \dfrac{E}{q_0}$. On the other hand, the bootstrapping procedure introduces a new bootstrapping noise, which can be viewed as a fixed amount. However, this fixed amount of new noise accumulates over each bootstrapping. Therefore, after a very large number of bootstrappings, the noise will eventually overflow the message and even the ciphertext modulus. 

In the case of BFV's bootstrapping, it reduces the noise, but does not change the ciphertext modulus. However, there is no need to reset the ciphertext modulus, because BFV does not have a leveled ciphertext modulus chain, and BFV's ciphertext-to-ciphertext multiplication does not consume ciphertext modulus. Furthermore, since BFV's bootstrapping directly removes the noise, the noise is guaranteed to be kept under a certain threshold even after an infinite number of bootstrappings. 

Another important difference is that CKKS's bootstrapping does not require homomorphic decryption, primarily because it maintains the plaintext's scaling factor to be the same across the entire bootstrapping procedure. On the other hand, BFV's bootstrapping needs to change the plaintext's scaling factor to run the digit extraction algorithm. Therefore, homomorphic decryption is required to change the plaintext scaling factor ($p^{\varepsilon}$) while preserving the same ciphertext modulus ($q$).

\subsubsection{Reducing the Bootstrapping Noise}
\label{subsubsec:ckks-bootstrapping-noise-reduction}

As explained in \autoref{subsubsec:ckks-bootstrapping-evalexp-details}, the bootstrapping procedure generates three types of noises:

\begin{itemize}
\item \underline{Type-1 Noise}: the intrinsic homomorphic $(+, \cdot)$ computation noises of the \textsf{CoeffToSlot}, \textsf{EvalExp}, and \textsf{SlotToCoeff} steps
\item \underline{Type-2 Noise}: the \textsf{EvalExp} step's approximation error of the exponential function $e^{i\theta}$
\item \underline{Type-3 Noise}: the \textsf{EvalExp} step's sine graph error (i.e., not exactly $y=x$ around $x=0$, but only $y \approx x$)
\end{itemize}

$ $

The Type-1 noise is inevitable by the design of FHE. The Type-2 noise can be either avoided or unavoidable depending on the tradeoff setup between the bootstrapping accuracy and efficiency. Unlike these two types of noises, the Type-3 noise can be effectively reduced by newer bootstrapping techniques.  

\begin{figure}[h!]
    \centering
  \includegraphics[width=1.0\linewidth]{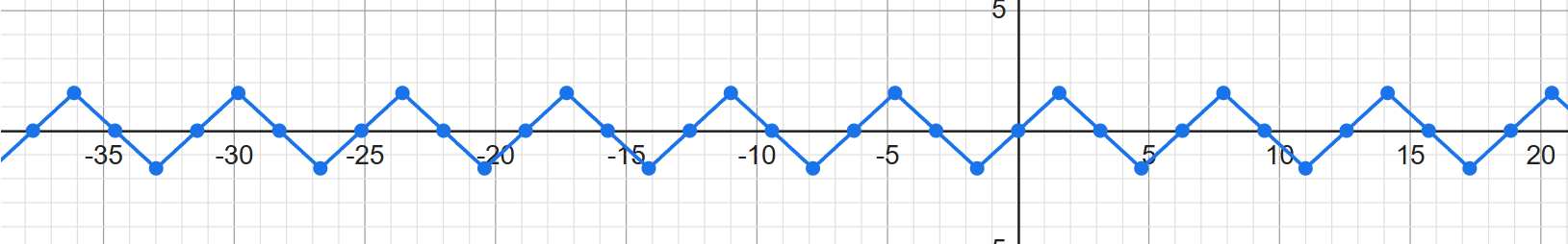}
  \caption{Arc-sine graph for smaller approximation error \href{https://www.google.com/search?sca_esv=c744cb070de47b7e&sxsrf=ADLYWIKKaF93fcAFvJC4ehJso1E5qFTaow:1732546684173&q=arcsin(sin(x))&source=lnms&fbs=AEQNm0DmKhoYsBCHazhZSCWuALW8l8eUs1i3TeMYPF4tXSfZ95GzcfXnm5XYTvJV_9Qreh2py964ICpZJthXkELijctC8pFBYULoa3-fvQmwK0VJF0ntzsbN_W2CCJL9N57SWFNwWI58jCKaBJSdPgkprHQVK8H1PYOYWXMHTCCV-rDbC44rR6ANM870jZCZRtTKwFWtMIe2&sa=X&ved=2ahUKEwjZy6qt3_eJAxXbhlYBHXD2IvEQ0pQJegQIDxAB&cshid=1732546705369143&biw=1280&bih=635&dpr=1.5}{(Source)}}
  \label{fig:arc-sine}
\end{figure}

$ $

\para{$\bm{\arcsin(\sin(x))}$ Approximation (\href{https://dl.acm.org/doi/10.1007/978-3-030-77870-5_22}{EUROCRYPT 2021}~\cite{cryptoeprint:2020/552}):} Using the $\arcsin(\sin(x))$ function instead of the $\sin x$ function can reduce the Type-3 noise, because its line is not curved but straight, as shown in \autoref{fig:arc-sine} (comprising a series of $y = x$ and $y = -x$ segments). This technique also uses the Remez algorithm that evenly distributes the approximation error over a specified region. However, one downside of this technique is that it consumes 3 multiplicative levels. 

$ $

\para{Meta-BTS (\href{https://dl.acm.org/doi/10.1145/3548606.3560696}{CCS 2022}~\cite{cryptoeprint:2022/1167}):} This is thus far the most computationally efficient and accurate bootstrapping technique, whose procedure is as follows:
\begin{enumerate}
\item Perform the regular bootstrapping based on the sine graph to the input ciphertext.
\item Rescale step 1's bootstrapped ciphertext to modulus $q_0$.
\item Subtract step 2's ciphertext from the initial un-bootstrapped ciphertext (where both ciphertexts are modulo $q_0$), whose result is a modulo $q_0$ ciphertext storing the bootstrapping error.
\item Bootstrap the output ciphertext of step 3 (storing the bootstrapping error) to modulus $q_l$.
\item Subtract step 4's ciphertext from step 1's ciphertext (where both ciphertexts are modulo $q_l$), which gives a new modulo $q_l$ ciphertext with a reduced bootstrapping error.
\end{enumerate}

$ $

\para{Limitation in Noise Handling:} The above bootstrapping techniques can reduce the Type-3 noise, because the bootstrapping error is smaller than the plaintext message and a smaller input value $x$ to the approximating sine function outputs a value closer to $y=x$. Running this algorithm multiple times, the Type-3 noise becomes exponentially smaller, because the size of the target plaintext (i.e., the extracted bootstrapping error as the output of step 3 above) is much smaller than before. Meanwhile, Type-1 and Type-2 noises do not decrease over multiple bootstrapping rounds, relatively keeping  their same level, because each round generates new Type-1 and Type-2 noises.

\clearpage

\section{BGV Scheme}
\label{sec:bgv}

Similar to BFV, the BGV scheme is designed for homomorphic addition and multiplication of integers. Unlike CKKS, BGV guarantees exact encryption and decryption. From this view, BGV is similar to BFV. However, the major difference between these two schemes is that BFV stores the plaintext value in the MSBs (most significant bits) and the noise in the low-digit area (least significant bits), while BGV stores them the other way around: the plaintext value in the low-digit area and the noise in the MSBs. Technically, while BFV scales the plaintext polynomial by $\Delta$, BGV scales the noise polynomial by $\Delta$. Therefore, these two schemes use slightly different strategies to store and manage the plaintext and noise within a ciphertext. 

BGV internally uses almost the same strategy as BFV for plaintext encoding, ciphertext-to-plaintext addition, ciphertext-to-ciphertext addition, ciphertext-to-plaintext multiplication, and input vector rotation. On the other hand, BGV's encryption and decryption are slightly different from BFV's scheme, because its scaling target is not the plaintext, but the noise. Also, unlike BFV where ciphertext-to-ciphertext multiplication has no limit on the number, BGV's ciphertext-to-ciphertext multiplication is leveled, switching the modulus to a lower level like CKKS, and thus it is limited. Furthermore, BGV's modulus switch and bootstrapping are partially different from BFV's.

\begin{tcolorbox}[
    title = \textbf{Required Background},    % box title
    colback = white,    % light background; tweak to taste
    colframe = black,  % frame colour
    boxrule = 0.8pt,     % line thickness
    left = 1mm, right = 1mm, top = 1mm, bottom = 1mm % inner padding
]

\begin{itemize}
\item \autoref{sec:modulo}: \nameref{sec:modulo}
\item \autoref{sec:group}: \nameref{sec:group}
\item \autoref{sec:field}: \nameref{sec:field}
\item \autoref{sec:order}: \nameref{sec:order}
\item \autoref{sec:polynomial-ring}: \nameref{sec:polynomial-ring}
\item \autoref{sec:decomp}: \nameref{sec:decomp}
\item \autoref{sec:roots}: \nameref{sec:roots}
\item \autoref{sec:cyclotomic}: \nameref{sec:cyclotomic}
\item \autoref{sec:cyclotomic-polynomial-integer-ring}: \nameref{sec:cyclotomic-polynomial-integer-ring}
\item \autoref{sec:matrix}: \nameref{sec:matrix}
\item \autoref{sec:euler}: \nameref{sec:euler}
\item \autoref{sec:modulus-rescaling}: \nameref{sec:modulus-rescaling}
\item \autoref{sec:chinese-remainder}: \nameref{sec:chinese-remainder}
\item \autoref{sec:polynomial-interpolation}: \nameref{sec:polynomial-interpolation}
\item \autoref{sec:ntt}: \nameref{sec:ntt}
\item \autoref{sec:lattice}: \nameref{sec:lattice}
\item \autoref{sec:rlwe}: \nameref{sec:rlwe}
\item \autoref{sec:glwe}: \nameref{sec:glwe}
\item \autoref{sec:glwe-add-cipher}: \nameref{sec:glwe-add-cipher}
\item \autoref{sec:glwe-add-plain}: \nameref{sec:glwe-add-plain}
\item \autoref{sec:glwe-mult-plain}: \nameref{sec:glwe-mult-plain}
\item \autoref{subsec:modulus-switch-rlwe}: \nameref{subsec:modulus-switch-rlwe}
\item \autoref{sec:glwe-key-switching}: \nameref{sec:glwe-key-switching}
\item \autoref{sec:bfv}: \nameref{sec:bfv}
\item \autoref{sec:ckks}: \nameref{sec:ckks}
\end{itemize}
\end{tcolorbox}

\clearpage

\subsection{Encoding and Decoding}
\label{subsec:bgv-encoding-decoding}

BGV uses almost the same plaintext encoding scheme as BFV as described in Summary~\ref{subsubsec:bfv-encoding-summary} in \autoref{subsubsec:bfv-encoding-summary}, with the only difference that the scaling factor $\Delta = \left\lfloor\dfrac{q}{t}\right\rfloor$ is not applied to the plaintext polynomial $M(X)$ like BFV does. Instead, BGV applies its own scaling factor $\Delta = t$ to the noise polynomial $E(X)$ whenever it encrypts a new ciphertext (will be explained in \autoref{subsec:bgv-enc-dec}). 

The following is BGV's encoding and decoding scheme. 

\begin{tcolorbox}[title={\textbf{\tboxlabel{\ref*{subsec:bgv-encoding-decoding}} BGV's Encoding and Decoding}}]

\textbf{\underline{Input}:} An $n$-slot integer modulo $t$ vector $\vec{v} = (v_0, v_1, \cdots, v_{n-1}) \in \mathbb{Z}_t^n$

$ $

\textbf{\underline{Encoding}:} 

 Convert $\vec{v} \in \mathbb{Z}_t^n$ into $\vec{m} \in \mathbb{Z}_t^n$ by applying the transformation $\vec{m} = \dfrac{\hathat W \cdot I_n^R \cdot \vec{v}}{n}$

, where $\hathat W$ is a basis of the $n$-dimensional vector space crafted as follows: 

{\footnotesize{$\hathat W = \begin{bmatrix}
1 & 1 & \cdots & 1 & 1 & 1 & \cdots & 1\\
(\omega^{J(\frac{n}{2} - 1)}) & (\omega^{J(\frac{n}{2} - 2)}) & \cdots & (\omega^{J(0)}) & (\omega^{J_*(\frac{n}{2} - 1)}) & (\omega^{J_*(\frac{n}{2} - 2)}) & \cdots & (\omega^{J_*(0)})\\
(\omega^{J(\frac{n}{2} - 1)})^2 & (\omega^{J(\frac{n}{2} - 2)})^2 & \cdots & (\omega^{J(0)})^2 & (\omega^{J_*(\frac{n}{2} - 1)})^2 & (\omega^{J_*(\frac{n}{2} - 2)})^2 & \cdots & (\omega^{J_*(0)})^2 \\
\vdots & \vdots & \ddots & \vdots & \vdots & \ddots & \vdots & \vdots \\
(\omega^{J(\frac{n}{2} - 1)})^{n-1} & (\omega^{J(\frac{n}{2} - 2)})^{n-1} & \cdots & (\omega^{J(0)})^{n-1} & (\omega^{J_*(\frac{n}{2} - 1)})^{n-1} & (\omega^{J_*(\frac{n}{2} - 2)})^{n-1} & \vdots  & (\omega^{J_*(0)})^{n-1}
\end{bmatrix}$}}

$ $

, where $\omega$ is a primitive $2n$-th root of unity modulo $t$ (which implies $t \equiv 1 \bmod 2n$). This implies that $\omega = g^{\frac{t - 1}{2n}} \bmod t$ ($g$ is a generator of $\mathbb{Z}_t^{\times}$ (see \autoref{subsubsec:poly-vector-transformation-modulus}). The final output is $M = \sum\limits_{i=0}^{n-1} m_iX^i \text{ } \in \mathbb{Z}_t[X] / (X^n + 1)$, 
which we can also treat as 

$M = \sum\limits_{i=0}^{n-1} m_iX^i \text{ } \in \mathbb{Z}_q[X] / (X^n + 1)$ during encryption/decryption later, because the initial fresh coefficients $m_i$ are guaranteed to be smaller than any $q$ where $q = \{q_0, q_1, \cdots, q_L\}$.

$ $

\textbf{\underline{Decoding}:} For the plaintext polynomial $M = \sum\limits_{i=0}^{n-1} m_iX^i$, 
compute $\vec{v} = \hathat W^* \cdot \vec{m}$, where 

$\hathat{W}^* = \begin{bmatrix}
1 & (\omega^{J(0)}) & (\omega^{J(0)})^2 & \cdots & (\omega^{J(0)})^{n-1}\\
1 & (\omega^{J(1)}) & (\omega^{J(1)})^2 & \cdots & (\omega^{J(1)})^{n-1}\\
1 & (\omega^{J(2)}) & (\omega^{J(2)})^2 & \cdots & (\omega^{J(2)})^{n-1}\\
\vdots & \vdots & \vdots & \ddots & \vdots \\
1 & (\omega^{J(\frac{n}{2}-1)}) & (\omega^{J(\frac{n}{2}-1)})^2 & \cdots & (\omega^{J(\frac{n}{2}-1)})^{n-1}\\
1 & (\omega^{J_*(0)}) & (\omega^{J_*(0)})^2 & \cdots & (\omega^{J_*(0)})^{n-1}\\
1 & (\omega^{J_*(1)}) & (\omega^{J_*(1)})^2 & \cdots & (\omega^{J_*(1)})^{n-1}\\
1 & (\omega^{J_*(2)}) & (\omega^{J_*(2)})^2 & \cdots & (\omega^{J_*(2)})^{n-1}\\
\vdots & \vdots & \vdots & \ddots & \vdots \\
1 & (\omega^{J_*(\frac{n}{2}-1)}) & (\omega^{J_*(\frac{n}{2}-1)})^2 & \cdots & (\omega^{J_*(\frac{n}{2}-1)})^{n-1}\\
\end{bmatrix}$

\end{tcolorbox}

\subsection{Encryption and Decryption}
\label{subsec:bgv-enc-dec}

BGV's encryption and decryption scheme is very similar to BFV's scheme (Summary~\ref*{subsec:bfv-enc-dec} in \autoref{subsec:bfv-enc-dec}) with a small difference: while BFV scales the plaintext polynomial $M(X)$ by $\Delta$, BGV scales the noise polynomial $E(X)$ by $\Delta$. In BFV, each encoded plaintext polynomial $M(X)$ is scaled by $ \Delta = \left\lfloor \dfrac{q}{t} \right\rfloor$. This strategy effectively shifts each plaintext coefficient value to the most significant bits while keeping the noise in the least significant bits. On the other hand, BGV does not scale the plaintext polynomial $M(X)$, but instead it scales each new noise $E(X)$ by $\Delta = t$, making the noise $\Delta E(X)$, which is newly generated upon each new ciphertext creation. This different scaling strategy effectively shifts the noise (i.e., $e_i$) to the most significant bits by scaling it by $\Delta = t$ while keeping the plaintext value (i.e., $m_i$) $M(X)$'s each coefficient in the least significant bits. 

Also, in BGV, the ciphertext modulus $q$ is leveled like CKKS's one: $q \in \{q_0, q_1, \cdots, q_L\}$, where each $q_l = \prod\limits_{i=0}^l w_i$ (where each $w_i$ is a CRT modulus). 

BGV's encryption decryption process is described as follows:

\begin{tcolorbox}[title={\textbf{\tboxlabel{\ref*{subsec:bgv-enc-dec}} BGV Encryption and Decryption}}]

\textbf{\underline{Initial Setup}:} 

\begin{itemize}
\item The plaintext modulus $t = p$ (a prime)

\item The ciphertext modulus $q$ is leveled like in CKKS: $q \in \{q_0, q_1, \cdots, q_L\}$, where each $q_l = \prod\limits_{i=0}^l w_i$ (each $w_i$ is a CRT modulus), and each $q_l \equiv 1 \bmod t$ (will be explained in \autoref{subsec:bgv-modulus-switch})

\item The noise scaling factor $\Delta=t$ 
\item The secret key $S \xleftarrow{\$} \mathcal{R}_{\langle n, \textit{tern} \rangle}$. The coefficients of the polynomial $S$ ternary (i.e., $\{-1, 0, 1\}$).

\end{itemize}

\par\noindent\rule{\textwidth}{0.4pt}

\textbf{\underline{Encryption Input}:} $M \in \mathcal{R}_{\langle n, q \rangle}$, $A \xleftarrow{\$} \mathcal{R}_{\langle n, q \rangle}$, $E \xleftarrow{\chi_\sigma} \mathcal{R}_{\langle n, q \rangle}$

\begin{enumerate}
%\item Scale up $M \rightarrow \Delta M \text { } \in \mathcal{R}_{\langle n, q\rangle}$

\item Compute $B = -A \cdot S + M + \Delta E \text{ } \in \mathcal{R}_{\langle n,q \rangle}$

\item $\textsf{RLWE}_{S,\sigma}(M + \Delta E) = (A, B) \text{ } \in \mathcal{R}_{\langle n,q \rangle}^2$ 

\end{enumerate}

\par\noindent\rule{\textwidth}{0.4pt}

\textbf{\underline{Decryption Input}:} $\textsf{ct} = (A, B) \text{ } \in \mathcal{R}_{\langle n,q \rangle}^2$

\begin{enumerate}
\item $\textsf{RLWE}^{-1}_{S,\sigma}(\textsf{ct}) = B + A \cdot S  = M + \Delta E \pmod q$

\item Compute $M + \Delta E \bmod t$ to get $M$. \textcolor{red}{ $\rhd$ modulo reduction of $M + \Delta E$ by $t$}

\end{enumerate}

$ $

The final output is $M(X) \in \mathbb{Z}_t[X] / (X^n + 1)$.

$ $

\textbf{\underline{Conditions for Correct Decryption}:}

%\begin{enumerate}
%\item 
Each coefficient $\Delta e_i + m_i$ that contains the scaled noise and the plaintext should not overflow or underflow its ciphertext's any current moment's multiplicative level $l$'s ciphertext modulus $q_l$ (i.e., $\Delta e_i + m_i < q_l$, or $|\Delta e_i + m_i| < \frac{q_l}{2}$ in the signed modulo representation). 

%\end{enumerate}

\end{tcolorbox}

When restoring the plaintext at the end of the decryption process, while BFV shifts down the plaintext and the noise to the lower bit area (which effectively rounds off the noise), BGV computes $\text{ mod } p$, which effectively modulo-reduces the accumulated noise because every coefficient of $E$ is a multiple of $t$ (i.e., $\Delta$). Finally, only each coefficient of the plaintext polynomial $m_i$ remains in the low-digit area without any noise $e_i$.

\subsection{Ciphertext-to-Ciphertext Addition}
\label{subsec:bgv-add-cipher}

BGV's ciphertext-to-ciphertext addition scheme is exactly the same as BFV's scheme (Summary~\ref*{subsec:bfv-add-cipher} in \autoref{subsec:bfv-add-cipher}).

\begin{tcolorbox}[title={\textbf{\tboxlabel{\ref*{subsec:bgv-add-cipher}} BGV Ciphertext-to-Ciphertext Addition}}]
$\textsf{RLWE}_{S, \sigma}(M^{\langle 1 \rangle} + \Delta E^{\langle 1 \rangle} ) + \textsf{RLWE}_{S, \sigma}(M^{\langle 2 \rangle} + \Delta E^{\langle 2 \rangle}) $

$ = ( A^{\langle 1 \rangle}, \text{ } B^{\langle 1 \rangle}) + (A^{\langle 2 \rangle}, \text{ } B^{\langle 2 \rangle}) $

$ = ( A^{\langle 1 \rangle} + A^{\langle 2 \rangle}, \text{ } B^{\langle 1 \rangle} + B^{\langle 2 \rangle} ) $

$= \textsf{RLWE}_{S, \sigma}\bm((M^{\langle 1 \rangle} + M^{\langle 2 \rangle})  + \Delta E^{\langle 1 \rangle} + \Delta E^{\langle 2 \rangle} \bm)$
\end{tcolorbox}

\subsection{Ciphertext-to-Plaintext Addition}
\label{subsec:bgv-add-plain}

BGV's ciphertext-to-plaintext addition scheme is almost the same as BFV's scheme (Summary~\ref*{subsec:bfv-add-plain} in \autoref{subsec:bfv-add-plain}). However, one difference is that it's not the case that the plaintext polynomial $\Lambda(X)$ to be added is scaled up by $\Delta$, but it remains as $\Lambda(X)$.

\begin{tcolorbox}[title={\textbf{\tboxlabel{\ref*{subsec:bgv-add-plain}} BGV Ciphertext-to-Plaintext Addition}}]
$\textsf{RLWE}_{S, \sigma}(M + \Delta E) + \Lambda $

$=  (A, \text{ } B) + \Lambda$

$=  (A, \text{ } B + \Lambda)$

$= \textsf{RLWE}_{S, \sigma}\bm((M + \Lambda) + \Delta E \bm)$
\end{tcolorbox}

\subsection{Ciphertext-to-Plaintext Multiplication}
\label{subsec:bgv-mult-plain}

BGV's ciphertext-to-plaintext multiplication scheme is exactly the same as BFV's scheme (Summary~\ref*{subsec:bfv-mult-plain} in \autoref{subsec:bfv-mult-plain}).

\begin{tcolorbox}[title={\textbf{\tboxlabel{\ref*{subsec:bgv-mult-plain}} BGV Ciphertext-to-Plaintext Multiplication}}]
$\textsf{RLWE}_{S, \sigma}(M + \Delta E) \cdot \Lambda$

$= (A, \text{ } B) \cdot \Lambda$

$= (A \cdot \Lambda, \text{ }  B \cdot \Lambda )$

$= \textsf{RLWE}_{S, \sigma}((M \cdot \Lambda) + \Delta E\cdot \Lambda )$
\end{tcolorbox}

Notice that BGV's ciphertext-to-plaintext multiplication does not consume any multiplicative level. 

\subsection{\textsf{ModDrop}}
\label{subsec:bgv-moddrop}

BGV's \textsf{ModDrop} works similarly to that of CKKS's \textsf{ModDrop} (Summary~\ref*{subsec:ckks-moddrop} in \autoref{subsec:ckks-moddrop}). Remember that CKKS's ciphertext decryption relation is as follows:

$M + \Delta E = A \cdot S + B \bmod q_l$

$M + \Delta E = A \cdot S + B - K\cdot q_l$ \textcolor{red}{ $\rhd$ where $K\cdot q_l$ represents a modulo reduction by $q_l$}

BGV's \textsf{ModDrop} operation decreases its modulus from $q_l \rightarrow q_{l-1}$ is performed by updating the ciphertext $(A, B)$ to a new one: $\bm(A' = A \bmod q_{l-1}$, $B' = B \bmod q_{l-1})$. After the \textsf{ModDrop}, the ciphertext's modulus decreases from $q_l \rightarrow q_{l-1}$, yet its decryption relation still holds the same as follows:

$A' \cdot S + B' - K\cdot q_l $

$= (A \bmod q_{l-1}) \cdot S + (B \bmod q_{l-1}) - K\cdot q_l$

$= (A - K_A\cdot q_{l-1}) \cdot S + (B - K_B\cdot q_{l-1}) - K\cdot q_l$ 

$= A\cdot S + B - (K_A \cdot S + K_B + K\dfrac{q}{q_{l-1}})\cdot q_{l-1}$
\textcolor{red}{ $\rhd$ where $\dfrac{q}{q_{l-1}} = w_l$ (i.e., the $l$-th prime element of $q_L$)}

$= A\cdot S + B - K'\cdot q_{l-1}$ \textcolor{red}{ $\rhd$ where $K' = K_A\cdot S + K_B + K\dfrac{q}{q_{l-1}}$, $\dfrac{q}{q_{l-1}} = w_l$ }

$= A\cdot S + B \bmod q_{l-1}$

$= M + \Delta E$ \textcolor{red}{ $\rhd$ since $M + \Delta E < q_0 < q_{l-1}$}

$ $

As shown above, $(A', B') \bmod q_{l-1}$ decrypts to the same $M + \Delta E$, a plaintext with a scaled error. However, the noise budget (i.e., allowed threshold of the noise) decreases because the ciphertext modulus-to-noise ratio decreases. 

$ $

BGV's \textsf{ModDrop} is summarized as follows:

\begin{tcolorbox}[title={\textbf{\tboxlabel{\ref*{subsec:bgv-moddrop}} BGV's \textsf{ModDrop}}}]

Given a BGV ciphertext with the $l$-th multiplicative level $\textsf{RLWE}_{S, \sigma}(M + \Delta E) = (A, B) \bmod q_l$, a \textsf{ModDrop} operation is as follows: 

$(A', B') \bmod q_{l-1} = (A \bmod q_{l-1}, B \bmod q_{l-1})$

$ $

After this, the ciphertext's multiplicative level decreases by 1, the noise's scaling factor $\Delta$ and the plaintext are unaffected, and the noise budget (i.e., allowed noise threshold) decreases. 

\end{tcolorbox}

\subsection{Modulus Switch}
\label{subsec:bgv-modulus-switch}

\noindent \textbf{- Reference 1:} 
\href{https://eprint.iacr.org/2020/1481}{Design and implementation of HElib: a homomorphic encryption library}~\cite{bgv-modswitch1}

\noindent \textbf{- Reference 2:} 
\href{https://eprint.iacr.org/2011/277.pdf}{Fully Homomorphic Encryption without Bootstrapping}~\cite{bgv-modswitch2}

\noindent \textbf{- Reference 3:} 
\href{https://eprint.iacr.org/2012/099.pdf}{Homomorphic Evaluation of the AES Circuit}~\cite{bgv-modswitch3}

Remember that the requirement of modulus switch is that while we change the ciphertext modulus from $q$ to $\hat q$, it should decrypt to the same plaintext $M$. BGV's modulus switch is similar to that of the RLWE modulus switch (\autoref{subsec:modulus-switch-rlwe}), but there is an additional requirement, because BGV applies the scaling factor $\Delta$ not to plaintext $M$, but to noise $E$. In the case of BFV or CKKS, their decryption process only needs to round off the noise in the low-digit area. However, in the case of BGV, the plaintext is in the low-digit area and its decryption process has to remove the noise in the higher-bit area by modulo-$t$ reduction (i.e., the plaintext modulus). More concretely, BGV's modulus switch from $(A, B) \bmod q_{l}$ $\rightarrow$  $(\hat{A}, \hat{B}) \bmod \hat{q}$ should satisfy the decryption relation such that $((\hat{A} \cdot S + \hat{B}) \bmod \hat{q}) \bmod t = M$. In BGV's modulus switch, $\hat{q}$ does not have to be one of the multiplicative levels of the ciphertext, and $\hat{q}$ only needs to satisfy the relationship: $\hat{q} < q_l$ and $\hat{q} \equiv 1 \bmod t$. BGV's modulus switch procedure is as follows:

$ $

\begin{enumerate}

\item The input ciphertext is $\textsf{ct} = (A, B) \bmod q_l$. We compute new polynomials $A'$ and $B'$ as follows:

$(A', B') = \left(\left\lceil\dfrac{\hat{q}}{q_l}\cdot A\right\rfloor, \left\lceil\dfrac{\hat{q}}{q_l}\cdot B\right\rfloor\right) \pmod{\hat{q}}$

$ $

And we compute the rounding error $\epsilon_A, \epsilon_B$ as follows: 

$\epsilon_A = \dfrac{\hat{q}}{q_l}\cdot A - A'$

$\epsilon_B = \dfrac{\hat{q}}{q_l}\cdot B - B'$

$ $

, which we rewrite as follows:

$\hat{q} A = q_{l} A' + q_{l}\epsilon_A = q_{l} A' + \epsilon'_A$ \textcolor{red}{ $\rhd$ we denote $\epsilon'_A = q_{l}\epsilon_A$, where $\epsilon'_A \in \mathbb{Z}_{q_l}$}

$\hat{q} B = q_{l} B' + q_{l}\epsilon_B = q_{l} B' + \epsilon'_B$ \textcolor{red}{ $\rhd$ we denote $\epsilon'_B = q_{l}\epsilon_B$, where $\epsilon'_B \in \mathbb{Z}_{q_l}$}

$ $

\item We compute new polynomials $H_A$ and $H_B$ as follows:

$H_A = q_l^{-1}\cdot\epsilon'_A \bmod t$

$H_B = q_l^{-1}\cdot\epsilon'_B \bmod t$

$ $

\item We propose the final mod-switched ciphertext $\hat{\textsf{ct}}$ as follows:

$\hat{\textsf{ct}} = (\hat{A}, \hat{B}) = (A' + H_A, \text{ } B' + H_B) \bmod \hat{q}$

$ $

Note that the computation result of $A' + H_A$ and $B' + H_B$ alone can exceed the range $\mathbb{Z}_{\hat{q}}$, because $A', B' \in \mathbb{Z}_{\hat{q}}$ and $H_A, H_B \in \mathbb{Z}_t$. Therefore, our goal is reduce $A' + H_A$ and $B' + H_B$ modulo $\hat{q}$  to derive $\hat{A} \in \mathbb{Z}_{\hat{q}}$ and $\hat{B} \in \mathbb{Z}_{\hat{q}}$. 

$ $

\item From now on, we will verify that $\hat{\textsf{ct}}$ is a valid ciphertext satisfying BGV's required decryption relation. First, we can derive the relationship among $\textsf{ct} = (A, B)$, $A' + H_A$, and $B' + H_B$ as follows:

$\hat{q}\cdot \textsf{ct} \bmod t $

$= (\hat{q}A, \hat{q}B) \bmod t$

$= (q_{l} A' + \epsilon'_A, \text{ } q_{l} B' + \epsilon'_B) \bmod t$ \textcolor{red}{ $\rhd$ applying step 1's result: $\hat{q} A = q_{l}A' + \epsilon'_A$, \text{ }  $\hat{q} B = q_{l}B' + \epsilon'_B$}

$ = (q_{l} A' + q_lH_A, \text{ } q_{l} B' + q_lH_B) \bmod t$ 
\textcolor{red}{ $\rhd$ applying step 2's result: $H_A = q_l^{-1}\cdot\epsilon'_A \bmod t$, \text{ } $H_B = q_l^{-1}\cdot\epsilon'_B \bmod t$}

$ $

$ = q_l\cdot(A' + H_A, \text{ } B' + H_B) \bmod t$

%$ = q_l \cdot \hat{\textsf{ct}} \bmod t$

$ $

So, $\hat{q}\cdot\textsf{ct} = q_l\cdot(A' + H_A, \text{ } B' + H_B) \bmod t$. But in BGV, $q_l \equiv q_2 \equiv \cdots \equiv \hat{q} \equiv \cdots q_L \equiv 1 \bmod t$. Thus, the following holds:

$A \equiv A' + H_A \bmod t$

$B \equiv B' + H_B \bmod t$

$ $

\item We can derive the decryption relation of $\hat{\textsf{ct}}$ from the decryption relation of $\textsf{ct}$ as follows:

$M = (A\cdot S + B \bmod q_l) \bmod t$ \textcolor{red}{ $\rhd$ The BGV decryption relation of $\textsf{ct} = (A, B) \bmod q_l$}

$= (A\cdot S + B - K\cdot q_l) \bmod t$ \textcolor{red}{ $\rhd$ where $K\cdot q_l$ represents the modulo-$q_l$ reduction}

$= ((A' + H_A)\cdot S + B' + H_B - K\cdot q_{l}) \bmod t$ \textcolor{red}{ $\rhd$ applying step 4's result: $A \equiv A' + H_A \bmod t$, \text{ } $B \equiv B' + H_B \bmod t$}

$= ((A' + H_A)\cdot S + B' + H_B - K\cdot \hat{q}) \bmod t$ \textcolor{red}{ $\rhd$ since in BGV, $q_0\equiv q_1\equiv \cdots q_L \equiv 1 \bmod t$, and we chose $\hat{q}$ such that $\hat{q} \equiv 1 \bmod t$}

$ $

\item Now, if we can prove that $(A' + H_A)\cdot S + B' + H_B - K\cdot \hat{q} = (A' + H_A)\cdot S + B' + H_B \bmod \hat{q}$ (i.e., $K\cdot\hat{q}$ reduces $(A' + H_A)\cdot S + B' + H_B$ modulo $\hat{q}$), then this sufficiently leads to the following conclusion: 

$(\hat{A}\cdot S + \hat{B} \bmod \hat{q}) \bmod t$

$= ((A' + H_A)\cdot S + B' + H_B \bmod \hat{q}) \bmod t$

$= M$ \textcolor{red}{ $\rhd$ i.e., $(\hat{A}, \hat{B}) = (A' + H_A, B' + H_B) \bmod \hat{q}$ is a valid ciphertext that decrypts to $M$}

$ $
\begin{comment}
\item The following is also true: 

$(A' + H_A)\cdot S + B' + H_B \bmod \hat{q}$

$ = |A' + H_A|_{\hat{q}}\cdot S + |B' + H_B|_{\hat{q}} \bmod \hat{q}$ 

\textcolor{red}{ $\rhd$ where $|A' + H_A|_{\hat{q}} = A' + H_A \bmod \hat{q}$, \text{ } $|B' + H_B|_{\hat{q}} = B' + H_B \bmod \hat{q}$} 

$ $

$ = \hat{A}\cdot S + \hat{B} \bmod \hat{q}$ \textcolor{red}{ $\rhd$ applying step 3: $\hat{A} = A' + H_A \bmod \hat{q}$, \text{ } $\hat{B} = B' + H_B \bmod \hat{q}$}

$ $

Therefore, proving $(A' + H_A)\cdot S + B' + H_B - K\cdot \hat{q} = (A' + H_A)\cdot S + B' + H_B \bmod \hat{q}$ is equivalent to proving $(A' + H_A)\cdot S + B' + H_B - K\cdot \hat{q} = \hat{A}\cdot S + \hat{B} \bmod \hat{q}$. 

$ $
\end{comment}

\item We will prove that $(A' + H_A)\cdot S + B' + H_B - K\cdot \hat{q} = (A' + H_A)\cdot S + B' + H_B \bmod \hat{q}$ as follows: 

$(A' + H_A)\cdot S + B' + H_B - K\cdot \hat{q}$

$ = (\dfrac{\hat{q}}{q_{l}}\cdot A - \dfrac{\epsilon'_A}{q_{l}} + H_A)\cdot S + (\dfrac{\hat{q}}{q_{l}}\cdot B - \dfrac{\epsilon'_B}{q_{l}} + H_B) - K\cdot \hat{q}$ 

\textcolor{red}{ $\rhd$ applying step 1's result: $A' = \dfrac{\hat{q}}{q_{l}}\cdot A - \dfrac{\epsilon'_A}{q_{l}}, \text{ } B' = \dfrac{\hat{q}}{q_{l}}\cdot B - \dfrac{\epsilon'_B}{q_{l}}$}

$ $

$ $

$ = \left(\dfrac{\hat{q}}{q_{l}}\cdot A\cdot S + \dfrac{\hat{q}}{q_{l}}\cdot B  - K\cdot \hat{q}\right) + H_A\cdot S + H_B - \dfrac{\epsilon'_A}{q_{l}}\cdot S - \dfrac{\epsilon'_B}{q_{l}}$ \textcolor{red}{ $\rhd$ rearranging the terms}

$ = \dfrac{\hat{q}}{q_{l}}\cdot(A\cdot S + B  - K\cdot q_{l}) + H_A\cdot S + H_B - \dfrac{\epsilon'_A}{q_{l}}\cdot S - \dfrac{\epsilon'_B}{q_{l}}$ \textcolor{red}{ $\rhd$ taking out the common factor $\dfrac{\hat{q}}{q_{l}}$}

$ = \dfrac{\hat{q}}{q_{l}}\cdot(A\cdot S + B \bmod q_l) + H_A\cdot S + H_B - \dfrac{\epsilon'_A\cdot S + \epsilon'_B}{q_{l}}$ \textcolor{red}{ $\rhd$ since $A\cdot S + B - K\cdot q_{l} = A\cdot S + B \bmod q_l$}

$ $

For successful decryption, every coefficient of the resulting polynomial of the above expression has to be within the range $\mathbb{Z}_{\hat{q}}$ (which means that $K\cdot\hat{q}$ has successfully reduced $(A' + H_A)\cdot S + B' + H_B$ modulo $\hat{q}$). The first term $\dfrac{\hat{q}}{q_{l}}\cdot(A\cdot S + B \bmod q_l)$ can be viewed as the original ciphertext \textsf{ct}'s noise (with the plaintext message) scaled down by $\dfrac{\hat{q}}{q_l}$, which is guaranteed to be within the $\mathbb{Z}_{\hat{q}}$ range. The coefficients of the second term $H_A \cdot S$ are also small, because $H_A \in \mathbb{Z}_t^n$ and $S \in \{-1, 0, 1\}^n$. The coefficients of the third term $H_B$ are also small, because $H_B \in \mathbb{Z}_t^n$. The coefficients of the last term $- \dfrac{\epsilon'_A\cdot S + \epsilon'_B}{q_{l}}$ are also small, because $\dfrac{\epsilon'_A}{q_l}$ and $\dfrac{\epsilon'_B}{q_l}$ are $\in \mathbb{Z}_{\frac{q_l}{\hat{q}}}$. 

$ $

Therefore, $(A' + H_A)\cdot S + B' + H_B - K\cdot \hat{q} = (A' + H_A)\cdot S + B' + H_B \bmod \hat{q}$ (provided the above error thresholds hold). 

$ $

\item Finally, we combine the results of step 6 and 7 as follows: 

$\bm((\hat{A}\cdot S + \hat{B}) \bmod \hat{q} \bm) \bmod t$

$= \bm((A' + H_A)\cdot S + B' + H_B \bmod \hat{q}\bm) \bmod t$ \textcolor{red}{ $\rhd$ since $\hat{A} \equiv A' + H_A \bmod \hat{q}$, and $\hat{B} \equiv B' + H_B \bmod \hat{q}$}

$ = ((A' + H_A)\cdot S + B' + H_B - K\cdot \hat{q}) \bmod t$ \textcolor{red}{ $\rhd$ by applying step 7}

$ = M$ \textcolor{red}{ $\rhd$ by applying step 5}

$ $

Hence, decrypting $(\hat{A}, \hat{B}) \bmod \hat{q}$ outputs the message $M$.

\end{enumerate}

$ $

We summarize BGV's modulus switch as follows:

\begin{tcolorbox}[title={\textbf{\tboxlabel{\ref*{subsec:bgv-modulus-switch}} BGV's Modulus Switch}}]

Suppose we have the current ciphertext modulus $q_l$ and new ciphertext modulus $\hat{q}$ where $q_l \equiv \hat{q} \equiv 1 \bmod t$ and $\hat{q} < q_l$. Therefore, $\hat{q}$ may or may not be one of the ciphertext moduli comprising a BGV ciphertext's multiplicative level moduli $q_0, q_1, \cdots, q_L$.

$ $

BGV's modulus switch from $q_l \rightarrow \hat{q}$ is equivalent to updating $(A, B) \bmod q_l$ to $(\hat A, \hat B) \bmod \hat{q}$ as follows:

$(A', B') = \left(\left\lceil\dfrac{\hat{q}}{q_l}\cdot A\right\rfloor, \left\lceil\dfrac{\hat{q}}{q_l}\cdot B\right\rfloor\right) \in \mathcal{R}_{\langle n, \hat{q} \rangle}^2$

$ $

$\epsilon'_A = \hat{q}\cdot A - q_l\cdot A'$ \textcolor{red}{ $\rhd$ where $\epsilon'_A \in \mathbb{Z}_{q_l}$}

$\epsilon'_B = \hat{q}\cdot B - q_l\cdot B'$ \textcolor{red}{ $\rhd$ where $\epsilon'_B \in \mathbb{Z}_{q_l}$}

$ $

$H_A = q_l^{-1}\cdot\epsilon'_A \bmod t$

$H_B = q_l^{-1}\cdot\epsilon'_B \bmod t$

$ $

$\hat{\textsf{ct}} = (\hat{A}, \hat{B}) = (A' + H_A, B' + H_B) \bmod \hat{q}$

$ $

BGV's modulus switch is used for ciphertext-to-ciphertext multiplication (will be covered in \autoref{subsec:bgv-mult-cipher}). Meanwhile, After BGV's modulus switch (i.e., the noise scaling factor), $\Delta = t$ stays the same as before. The secret key $S$ also stays the same as before. The noise gets scaled down roughly by $\dfrac{\hat{q}}{q_l}$, but this does not decrease the noise-to-ciphertext modulus ratio. The noise-to-ciphertext modulus ratio can be reduced by modulus bootstrapping (will be covered in \autoref{subsec:bgv-bootstrapping}). 

\end{tcolorbox}

\subsubsection{Difference between Modulus Switch and \textsf{ModDrop}}
\label{subsubsec:bgv-moddrop-vs-modswitch}

In the case of CKKS (\autoref{subsubsec:ckks-mult-cipher-rescale}), the difference between modulus switch and \textsf{ModDrop} is that the former scales down the plaintext's scaling factor by $\dfrac{q_l}{q_{l-1}} \approx \dfrac{1}{\Delta}$, whereas \textsf{ModDrop} does not affect the plaintext's scaling factor. 

Similarly, in the case of BGV, modulus switch (rescaling) and \textsf{ModDrop} from $q_l \rightarrow q_{l-1}$ both lower a BGV ciphertext's modulus from $q_l \rightarrow q_{l-1}$. However, the key difference is that rescaling also decreases the noise's scaling factor by $\dfrac{q_l}{q_{l-1}} \approx \dfrac{1}{\Delta}$, whereas \textsf{ModDrop} keeps the noise's scaling factor the same as it is. Therefore, rescaling is used only during ciphertext-to-ciphertext multiplication (to be explained in \autoref{subsec:bgv-mult-cipher}) when scaling down the noise's scaling factor in the intermediate ciphertext from $\Delta^2 \rightarrow \Delta$. Meanwhile, \textsf{ModDrop} is used to reduce the modulo computation time during an application's routine when it becomes certain that the ciphertext will not undergo any additional ciphertext-to-ciphertext multiplication (i.e., no need to further decrease the ciphertext's modulus). 

The main difference in modulus switch between CKKS and BGV is that the former decreases the plaintext's scaling factor by approximately $\dfrac{1}{\Delta}$, whereas the latter decreases the noise's scaling factor by approximately $\dfrac{1}{\Delta}$. 

\para{Source Code:} Examples of BGV modulus switch can be executed by running \href{https://github.com/fhetextbook/fhe-textbook/blob/main/source%20code/bgv.py}{\underline{this Python script}}.

\subsection{Ciphertext-to-Ciphertext Multiplication}
\label{subsec:bgv-mult-cipher}

\noindent \textbf{- Reference:} 
\href{https://www.inferati.com/blog/fhe-schemes-bgv}{Introduction to the BGV encryption scheme}

Since BGV uses a leveled ciphertext modulus chain like CKKS, BGV's ciphertext-to-ciphertext multiplication scheme is exactly the same as CKKS's scheme (Summary~\ref*{subsec:ckks-mult-cipher} in \autoref{subsec:ckks-mult-cipher}), except for the rescaling step which uses BGV's modulus switch (\autoref{subsec:bgv-modulus-switch}). 

\begin{tcolorbox}[title={\textbf{\tboxlabel{\ref*{subsec:bgv-mult-cipher}} BGV Ciphertext-to-Ciphertext Multiplication}}]

Suppose we have the following two RLWE ciphertexts:

$\textsf{RLWE}_{S, \sigma}(M^{\langle 1 \rangle} + \Delta E^{\langle 1 \rangle}) = (A^{\langle 1 \rangle}, B^{\langle 1 \rangle})$, \text{ } where $B^{\langle 1 \rangle} = -A^{\langle 1 \rangle} \cdot S +  M^{\langle 1 \rangle} + \Delta E^{\langle 1 \rangle}$

$\textsf{RLWE}_{S, \sigma}(M^{\langle 2 \rangle} + \Delta E^{\langle 2 \rangle}) = (A^{\langle 2 \rangle}, B^{\langle 2 \rangle})$, \text{ } where $B^{\langle 2 \rangle} = -A^{\langle 2 \rangle} \cdot S +  M^{\langle 2 \rangle} + \Delta E^{\langle 2 \rangle}$

$ $

Multiplication between these two ciphertexts is performed as follows:

$ $

\begin{enumerate}
\item \textbf{\underline{Basic Multiplication}}

Compute the following:

$ $

$D_0 = B^{\langle 1 \rangle}\cdot B^{\langle 2 \rangle}$

$D_1 = A^{\langle 1 \rangle}\cdot B^{\langle 2 \rangle} + A^{\langle 2 \rangle}\cdot B^{\langle 1 \rangle}$

$D_2 = A^{\langle 1 \rangle} \cdot A^{\langle 2 \rangle}$

$ $

The decryption relation satisfies: $M^{\langle 1 \rangle}M^{\langle 2 \rangle} + \Delta\cdot (M^{\langle 1 \rangle}E^{\langle 2 \rangle} + M^{\langle 2 \rangle}E^{\langle 1 \rangle}) + \Delta^2E^{\langle 1 \rangle}E^{\langle 2 \rangle}$

$ = \underbrace{B^{\langle 1 \rangle}\cdot B^{\langle 2 \rangle}}_{D_0}  + \underbrace{(B^{\langle 2 \rangle}\cdot A^{\langle 1 \rangle} + B^{\langle 1 \rangle}\cdot A^{\langle 2 \rangle})}_{D_1} \cdot S + \underbrace{(A^{\langle 1 \rangle} \cdot A^{\langle 2 \rangle})}_{D_2} \cdot \underbrace{S \cdot S}_{S^2}$

$= D_0 + D_1\cdot S + D_2 \cdot S^2$

$ $

\item \textbf{\underline{Relinearization}} 

$\textsf{RLWE}_{S, \sigma}\bm(M^{\langle 1 \rangle}M^{\langle 2 \rangle} + \Delta\cdot (M^{\langle 1 \rangle}E^{\langle 2 \rangle} + M^{\langle 2 \rangle}E^{\langle 1 \rangle}) + \Delta^2E^{\langle 1 \rangle}E^{\langle 2 \rangle}\bm) $

$= \textsf{RLWE}_{S, \sigma}\bm{(}\text{ }D_0 + D_1\cdot S + D_2\cdot S^2\text{ }\bm{)}$

$  \approx C_\alpha + C_\beta, \text{ where } \text{ } C_\alpha = (D_1, D_0), \text{ }\text{ }\text{ } C_\beta = \bm{\langle}  \text{ } \textsf{Decomp}^{\beta, l}(D_2), \textsf{RLev}_{S, \sigma}^{\beta, l}(S^2)  \text{ } \bm{\rangle}$ 

$ $

\item \textbf{\underline{(Optional) Rescaling}}

To suppress the noise component scaled by $\Delta^2$ and return it to a factor of $\Delta$, switch the ciphertext's modulo from $q \rightarrow \hat q$ by updating $(A, B)$ to $(\hat A, \hat B)$ according to BGV's modulus switch explained in Summary~\ref*{subsec:bgv-modulus-switch} (\autoref{subsec:bgv-modulus-switch}).

$ $

After the above update of $(A, B)$ to $(\hat A, \hat B)$, the noise scaling factor $\Delta = t$ and the plaintext $M$ stay the same, as we proved in \autoref{subsec:bgv-modulus-switch} that $((\hat A \cdot S + \hat B) \bmod \hat q) \bmod t = M$.

$ $

\end{enumerate}

\para{Swapping the Order of \textsf{Relinearization} and \textsf{Rescaling}: } The order of relinearization and rescaling is interchangeable. Running rescaling before relinearization reduces the size of the ciphertext modulus, and therefore the subsequent relinearization can be executed faster. 

\end{tcolorbox}

\para{Details of the Optional Rescaling:} Before rescaling, the contents of the ciphertext are $M^{\langle 1 \rangle}M^{\langle 2 \rangle} + \Delta\cdot (M^{\langle 1 \rangle}E^{\langle 2 \rangle} + M^{\langle 2 \rangle}E^{\langle 1 \rangle}) + \Delta^2E^{\langle 1 \rangle}E^{\langle 2 \rangle} + \epsilon$, where $\epsilon$ is a relinearization error. Therefore, after each ciphertext-to-ciphertext multiplication, the noise's scaling factor will become squared as $\Delta^2, \Delta^4, \Delta^8, \cdots$. To reduce such exponential noise growth rate, we can optionally rescale down the ciphertext by $w_l = \dfrac{q_l}{q_{l-1}} > \Delta$ at the end of each relinearization at multiplicative level $l$, which is the noise's growth rate (effectively keeping the noise scaling factor as $\Delta$). After rescaling, the ciphertext gets scaled down by $w_l$ and then added by a new noise $\epsilon_2$. Before the rescaling, the noise grew roughly by the factor of $\Delta = t$ (as the largest noise term is $\Delta^2 E^{\langle 1 \rangle} E^{\langle 2 \rangle}$), but the rescaling process reduces this growth rate by the factor of $w_l$ and then introduces a new constant noise $\epsilon_2$. Therefore, if $w_l$ is sufficiently bigger than $\Delta = t$, the resulting noise will decrease compared to both $\Delta E^{\langle 1 \rangle}$ and $\Delta E^{\langle 2 \rangle}$. Due to this reason, when we design the modulus chain of BGV, we require each $w_l$ to be sufficiently bigger than $\Delta = t$ to effectively reduce the noise growth rate upon each ciphertext-to-ciphertext multiplication (while ensuring the property that its reduction modulo $t$ gives the plaintext $M$ as explained in \autoref{subsec:bgv-modulus-switch}). Meanwhile, the constant noise term $\epsilon_2$ gets newly added upon each rescaling, but this term becomes part of the rescaled ciphertext, which will be later reduced by the factor of $w_{l-1}$ in the future rescaling. Therefore, BGV's rescaling upon ciphertext-to-ciphertext multiplication effectively suppresses the noise growth. 

On the other hand, the above design strategy of noise reduction is inapplicable to CKKS, because in CKKS, we use the scaling factor $\Delta$ to scale the message $M$ (not the noise $E$), and thus CKKS requires each $w_l \approx \Delta$ in order to preserve the plaintext's scaling factor $\Delta$ as the same value across ciphertext-to-ciphertext multiplications. Because of this difference in design, CKKS inevitably increases the noise after each ciphertext-to-ciphertext multiplication.

\subsection{Homomorphic Key Switching}
\label{subsec:bgv-key-switching}

BGV's homomorphic key switching scheme changes an RLWE ciphertext's secret key from $S$ to $S'$. This scheme is exactly the same as BFV's key switching scheme (Summary~\ref*{subsec:bfv-key-switching} in \autoref{subsec:bfv-key-switching}).

\begin{tcolorbox}[title={\textbf{\tboxlabel{\ref*{subsec:bgv-key-switching}} BGV's Key Switching}}]
$\textsf{RLWE}_{S',\sigma}(M + \Delta E) = (0, B) + \bm{\langle} \textsf{Decomp}^{\beta, l}(A), \text{ } \textsf{RLev}_{S', \sigma}^{\beta, l}(S) \bm{\rangle}$
\end{tcolorbox}

\subsection{Homomorphic Rotation of Input Vector Slots}
\label{subsec:bgv-rotation}

BGV's homomorphic rotation scheme of input vector slots is exactly the same as BFV's rotation scheme (Summary~\ref*{subsec:bfv-rotation} in \autoref{subsec:bfv-rotation}).

\begin{tcolorbox}[title={\textbf{\tboxlabel{\ref*{subsec:bgv-rotation}} BGV's Homomorphic Rotation of Input Vector Slots}}]

Suppose we have a BGV ciphertext and a key-switching key as follows:

\[
\textsf{RLWE}_{S, \sigma}(M + \Delta E) = (A, B), \quad \textsf{RLev}_{S, \sigma}^{\beta, l}(S(X^{J(h)}))
\]

Then, the procedure of rotating the first-half elements of the ciphertext's original input vector $\vec{v}$ by $h$ positions to the left (in a wrapping manner among them) and the second-half elements of $\vec{v}$ by $h$ positions to the left (in a wrapping manner among them) is as follows: 

\begin{enumerate}
\item Update $A(X)$, $B(X)$ to $A(X^{J(h)})$, $B(X^{J(h)})$. 
\item Perform the following key switching (refer to \autoref{subsec:bfv-key-switching}) from $S(X^{J(h)})$ to $S(X)$:

$\textsf{RLWE}_{S(X),\sigma}\bm{(} M(X^{J(h)}) + \Delta E(X^{J(h)})\bm{)} $

$= \bm{(} 0, B(X^{J(h)}) \bm{)} \text{ } + \text{ } \bm{\langle}  \textsf{Decomp}^{\beta, l}\bm{(}A(X^{J(h)})\bm{)}, \text{ } \textsf{RLev}_{S(X), \sigma}^{\beta, l}\bm{(}S(X^{J(h)})\bm{)} \bm{\rangle}$
\end{enumerate}

\end{tcolorbox}

\subsection{Modulus Bootstrapping}
\label{subsec:bgv-bootstrapping}

\noindent \textbf{- Reference:} 
\href{https://eprint.iacr.org/2022/1363.pdf}{Bootstrapping for BGV and BFV Revisited}~\cite{cryptoeprint:2022/1363}

BGV's bootstrapping shares some common aspects with both BFV and CKKS's bootstrapping. The goal of BGV's bootstrapping is the same as that of CKKS, but the internal technique is closer to that of BFV. Like CKKS, BGV's bootstrapping resets the depleted ciphertext modulus from $q_l \rightarrow q_L$ (strictly speaking, from $q_l \rightarrow q_{l'}$ such that $l < l' < L$ because the bootstrapping operations between step 2$\sim$6 have consumed multiplicative levels). This modulus transition effectively not only resets the multiplicative level but also reduces the noise-to-ciphertext modulus ratio. To achieve this goal, one might think that BGV's bootstrapping can take the same \textsf{ModRaise} approach used by CKKS's bootstrapping. However, this is not a directly applicable solution because CKKS uses the sine approximation technique to eliminate the $q_0$-multiple overflows after the mod-raise. On the other hand, BGV is an exact encryption scheme that does not allow the approximation of plaintext values. Therefore, BGV uses BFV's digit extraction approach to eliminate its modulus overflows. To use digit extraction, as in the case of BFV, BGV also has to modify the plaintext modulus to a specially prepared one, $p^\varepsilon$. To configure both the plaintext modulus and the ciphertext modulus to the desired values (i.e., $p^\varepsilon$ and $q_L$), BGV employs the homomorphic decryption technique, as BFV does.

The technical details of BGV's bootstrapping are as follows. 

$ $

Suppose that we have an RLWE ciphertext $(A, B)  = \textsf{RLWE}_{S, \sigma}(M + \Delta E) \bmod q_l$, where $A\cdot S + B = M + \Delta E$, \text{ } $\Delta = t = p$ (a prime), and $q_l$ is the ciphertext modulus of the current multiplicative level. 

$ $

\begin{enumerate}

\item \textbf{\underline{Modulus Switch} from \boldmath$q_l \rightarrow \hat{q}$:} BFV's bootstrapping initially switches the ciphertext modulus from $q \rightarrow p^{\varepsilon-1}$ where $q \gg p^\varepsilon > t = p$. On the other hand, BGV's bootstrapping switches the ciphertext modulus to $\hat{q}$ that is a special modulus satisfying the relation: $\hat{q} \equiv 1 \bmod p^\varepsilon$ and $\hat{q} > p^\varepsilon$ (where $p^\varepsilon$ will be explained in the next step). In order for a modulus switch from $q_l \rightarrow \hat{q}$ (i.e., the special modulus) to be possible, the prime factor(s) comprising $\hat{q}$ have to be congruent with $q_{0}, \ldots, q_L \bmod t$, so that we can do a modulus switch from $q_l\cdot \hat{q} \rightarrow \hat{q}$ (based on the technique learned in \autoref{subsec:bgv-modulus-switch}). Eventually, this step's modulus switch transforms the ciphertext $(A, B) \bmod q_l$ to $(\hat{A}, \hat{B}) \bmod \hat{q}$, during which the plaintext modulus (i.e., noise's scaling factor) stays the same. 

$ $

\item \textbf{\underline{Ciphertext Coefficient Multiplication by \boldmath$p^{\varepsilon-1}$}:} 
The constant $p^{\varepsilon-1}$ is multiplied to each coefficient of the ciphertext polynomials, updating the ciphertext to $p^{\varepsilon-1} \cdot (\hat{A}, \hat{B}) = (A', B') \bmod \hat{q}$, where $A' = p^{\varepsilon-1}\hat{A}$ and $B' = p^{\varepsilon-1}\hat{B}$. This operation updates the original decryption relation $\hat{A}\cdot S + \hat{B} = M + p E + K\hat{q}$ to $A'\cdot S + B' = p^{\varepsilon-1} M + p^\varepsilon E + K'\hat{q}$ (where $||K'||_{\infty} \leq n + 1$). Notice that the plaintext modulus (i.e., noise's scaling factor) has been changed from $p \rightarrow p^\varepsilon$. When choosing $\varepsilon$, BGV enforces the following additional constraint: $\hat{q} > p^\varepsilon$ and $\hat{q}  \equiv 1 \bmod p^\varepsilon$.

$ $

\item \textbf{\underline{\textsf{ModRaise}}:} We mod-raise $(\hat{A}, \hat{B}) \bmod \hat{q}$ to $(\hat{A}, \hat{B}) \bmod q_L$, where $\hat{q} \ll q_L$. The mod-raised ciphertext's decryption relation is as follows:

$\hat{A}\cdot S + \hat{B} = p^{\varepsilon-1}M + p^\varepsilon E + K'\hat{q} \bmod q_L$

$ $

Note that $K'\hat{q}$ is the $\hat{q}$-multiple overflow and does not get reduced modulo $q_L$, because $K'\hat{q} \ll q_L$. We saw the same situation in the CKKS bootstrapping's \textsf{ModRaise} (\autoref{subsubsec:ckks-bootstrapping-high-level}) which resets the ciphertext modulus from $q_0 \rightarrow q_L$ at the cost of incurring a $Kq_0$ overflow, which is to be removed by \textsf{EvalExp}'s homomorphic (approximate) sine graph evaluation (\autoref{subsubsec:ckks-bootstrapping-evalexp-details}). Likewise, BGV's mod-raised ciphertext $(\hat{A}, \hat{B}) \bmod q_L$ is $\textsf{RLWE}_{S, \sigma}(p^{\varepsilon-1}M + p^\varepsilon E + K'\hat{q}) \bmod q_L$, an encryption of $p^{\varepsilon-1}M + p^\varepsilon E + K'\hat{q}$. In the later step, we will use digit extraction to homomorphically eliminate $K'\hat{q}$ like we did in BFV's bootstrapping. The reason BGV's bootstrapping uses digit extraction instead of approximated sine evaluation is that BGV is an exact encryption scheme like BFV (not an approximate scheme like CKKS). 

$ $

\item \textbf{\textsf{\underline{CoeffToSlot}}:} This step works the same way as CKKS and BFV's \textsf{CoeffToSlot} step: move the coefficients of the polynomial $p^{\varepsilon-1} M + p^\varepsilon E + \hat{q}K'$ to the input vector slots of a new ciphertext. We denote polynomial $Z = p^{\varepsilon-1} M + p^\varepsilon E + \hat{q}K'$, and each $i$-th coefficient of $Z$ as $z_i$. For the \textsf{CoeffToSlot} step, we homomorphically compute $Z \cdot n^{-1} \cdot \hathat W \cdot I_n^R$. Then, each input vector slot of the resulting ciphertext ends up storing each $z_i$ of the polynomial $Z$. 

$ $

\item \textbf{\underline{Digit Extraction}:} At this point, each input vector slot contains each coefficient of $p^{\varepsilon-1} M + p^\varepsilon E + \hat{q}K'$, which is $p^{\varepsilon-1} m_i + p^\varepsilon e_i + \hat{q}k'_i$. Recall that we designed the lowest multiplicative level's ciphertext modulus $\hat{q}$ and the homomorphic multiplication factor $p^\varepsilon$ such that $\hat{q} \equiv 1 \bmod p^\varepsilon$, or $\hat{q} = k^{\langle \hat{q} \rangle } \cdot p^\varepsilon + 1$ for some positive integer $k^{\langle \hat{q} \rangle }$. Therefore, the following holds: 

$p^{\varepsilon-1} m_i + p^\varepsilon e_i + \hat{q}k'_i$

$ = p^{\varepsilon-1} m_i + p^\varepsilon e_i + k'_i\cdot(k^{\langle \hat{q} \rangle } \cdot p^\varepsilon + 1) $  \textcolor{red}{ $\rhd$ applying $\hat{q} = k^{\langle \hat{q} \rangle } \cdot p^\varepsilon + 1$}

$ = p^{\varepsilon-1} m_i + k'_i + p^\varepsilon  \cdot (e_i + k'_i\cdot k^{\langle \hat{q} \rangle }) $ \textcolor{red}{ $\rhd$ rearranging the terms}

$ = p^{\varepsilon-1} m_i + k'_i + p^\varepsilon \cdot k_i^{\langle \hat{q} + \varepsilon \rangle } $ \textcolor{red}{ $\rhd$ where $ k_i^{\langle \hat{q} + \varepsilon \rangle } = e_i + k'_i\cdot k^{\langle \hat{q} \rangle } $}

$ \equiv p^{\varepsilon-1} m_i + k'_i \pmod{p^\varepsilon}$

$ $

To eliminate $k'_i$ from the above (where $|k'_i| \leq n + 1$), we use the same digit extraction polynomial $G_{\varepsilon, v}$ as in the BFV bootstrapping (\autoref{subsubsec:bfv-bootstrapping-digit-extraction}) :

$z_i = d_0 + \left(\sum\limits_{j=\varepsilon'}^{\varepsilon-1} d_* p^j\right)$ \textcolor{red}{ $\rhd$ where $d_0 \in \mathbb{Z}_p$, \text{ } and $d_*$ can be any integer, \text{ } and $1 \leq \varepsilon' \leq \varepsilon$}

$F_{\varepsilon'}(z_i) \equiv d_0 \bmod p^{\varepsilon'+1}$

$G_{\varepsilon}(z_i) \equiv (z_i - \underbrace{F_{\varepsilon-1} \circ F_{\varepsilon-2} \circ \cdots \circ F_1}_{\varepsilon - 1 \text{ times}} (z_i)) \cdot |p^{-1}|_{q}$

$ $

We evaluate the digit extraction polynomial $G_{\varepsilon}$ for $\{\varepsilon, \varepsilon-1, \varepsilon-2, \cdots, 2\}$ recursively a total of $\varepsilon-1$ times. This operation finally zeros out and right-shifts the least significant (base-$p$) $\varepsilon-1$ digits of $z_i$ as follows:

$G_{2} \circ G_{3} \circ \cdots \circ G_{\varepsilon-1} \circ G_{\varepsilon} (z_i)$

$= m_i + k_i''p \bmod q_{l'}$

$ $

, where $k_i''p^\varepsilon$ is some multiple of $p^\varepsilon$ to account for the original $p^\varepsilon$-multiple overflow term plus additional $p^\varepsilon$-multiple overflows generated during the digit extraction. Note that the digit extraction step reduces the ciphertext modulus from $q_L \rightarrow q_{l'}$ (where $l'$ is an integer smaller than $L$), because the homomorphic evaluation of the polynomial $G_{\varepsilon}$ requires some ciphertext-to-ciphertext multiplications, which consume some multiplicative levels. The output of the digit extraction step is $m_i + k''_ip \bmod q_{l'}$ stored in each input vector slot. 

$ $

\para{\underline{Handling the Divisions by $p$}:} Unlike in BFV (\autoref{subsubsec:bfv-bootstrapping-digit-extraction}), we cannot use scaling factor re-interpretation because the scaling factor setup of BGV is different from that of BFV. In BGV's digit extraction, each round should explicitly homomorphically multiply the slot values by $|p^{-1}|_{q^{\langle i \rangle}}$, where $q^{\langle i \rangle}$ is the ciphertext modulus at the $i$-th round of digit extraction. Given $(A^{\langle g_1 \rangle}, B^{\langle g_1 \rangle}) \bmod q^{\langle 1 \rangle}$ is the ciphertext at the  1st round of digit extraction just before inverse-$p$ division, the actual division-by-$p$ is equivalent to performing the ciphertext-to-plaintext multiplication (\autoref{subsec:bfv-mult-plain}) as follows: 

$(A^{\langle g_1 \rangle}, B^{\langle g_1 \rangle}) \cdot \textsf{Encode}(p^{-1}) \bmod q^{\langle 1 \rangle}$

$= (A^{\langle g_1 \rangle}, B^{\langle g_1 \rangle}) \cdot p^{-1} \bmod q^{\langle 1 \rangle}$ \textcolor{red}{ $\rhd$ since $\textsf{Encode}(p^{-1}) = p^{-1} + 0X + 0X^2 + \cdots + 0X^{n-1}$ }

$= (p^{-1}A^{\langle g_1 \rangle}, p^{-1}B^{\langle g_1 \rangle}) \bmod q^{\langle 1 \rangle}$

$ $

Therefore, by multiplying the coefficients of two BGV ciphertext polynomials $A^{\langle g_1 \rangle}$ and $B^{\langle g_1 \rangle}$ by $|p^{-1}|_{q^{\langle 1 \rangle}}$, we obtain the following effect: 

$(|p^{-1}|_{q^{\langle 1 \rangle}} \cdot A^{\langle g_1 \rangle})\cdot S + (|p^{-1}|_{q^{\langle 1 \rangle}} \cdot B^{\langle g_1 \rangle}) \bmod q^{\langle 1 \rangle}$

$= |p^{-1}|_{q^{\langle 1 \rangle}} \cdot (p^{\varepsilon - 1}M + \left\lfloor{K'}\right\rfloor_p + K''p^{\varepsilon}) \bmod q^{\langle 1 \rangle}$

$= p^{\varepsilon - 2}M + \left\lfloor\dfrac{K'}{p}\right\rfloor + K''p^{\varepsilon-1} \bmod q^{\langle 1 \rangle} $

$ $

Verbally speaking, the above inverse-$p$ multiplication to all polynomial coefficients of $(A^{\langle g_1 \rangle}, B^{\langle g_2 \rangle})$ has the following two effects: (1) scales down the plaintext and the noise by $p$; and (2) reduces the modulus garbage term $K'$ to $\left\lfloor\dfrac{K'}{p}\right\rfloor$ (i.e., right-shift by 1 base-$p$ digit). 

$ $

BGV's digit extraction procedure is equivalent to recursively evaluating $G_{\varepsilon}$ with the input $z_i$ by decreasing $\varepsilon$ by 1 at each round (total $\varepsilon-1$ rounds) as follows:

\textbf{Input:} $p^{\varepsilon - 1}M + K' + K''p^{\varepsilon} \bmod \hat{q}$

\textbf{1st Round:} $G_{\varepsilon}(z_i) \xRightarrow{\text{effect}}  p^{\varepsilon - 2}M + \left\lfloor\dfrac{K'}{p}\right\rfloor + K''^{\langle 1 \rangle}p^{\varepsilon-1} \bmod q^{\langle 1 \rangle}$

\textbf{2nd Round:} $G_{\varepsilon-1} \circ G_{\varepsilon}(z_i) \xRightarrow{\text{effect}}  p^{\varepsilon - 3}M + \left\lfloor\dfrac{K'}{p^2}\right\rfloor + K''^{\langle 2 \rangle}p^{\varepsilon-2} \bmod q^{\langle 2 \rangle}$

\textbf{3rd Round:} $G_{\varepsilon-2} \circ G_{\varepsilon-1} \circ G_{\varepsilon}(z_i)  \xRightarrow{\text{effect}}  p^{\varepsilon - 4}M + \left\lfloor\dfrac{K'}{p^3}\right\rfloor + K''^{\langle 3 \rangle}p^{\varepsilon-3} \bmod q^{\langle 3 \rangle}$

$\vdots$

\textbf{$\bm{\varepsilon - 1}$-th Round:} $G_{2}\circ \cdots \circ G_{\varepsilon} (z_i) \xRightarrow{\text{effect}}  M + K''^{\langle \varepsilon - 1 \rangle}p \bmod q^{\langle \varepsilon - 1 \rangle}$

$ $

As shown above, the entire digit extraction procedure results in two effects on the values stored in the plaintext slots: (1) scales down the message $p^{\varepsilon - 1}M$ to $M$; and (2) zeros out the modulus garbage $K'$ (i.e., $\left\lfloor\dfrac{K'}{p^{\varepsilon - 1}}\right\rfloor = 0$). The reason why $K''$ gets updated to $K''^{\langle 1 \rangle}, K''^{\langle 2 \rangle}, \cdots, K''^{\langle \varepsilon -1 \rangle}$ across rounds is that each $i$-th round's evaluation of function $F_{\epsilon'}$ and $G_{\epsilon}$ is done modulo $p^{\varepsilon - i}$, which produces new $p^{\varepsilon - i}$-multiple overflow garbage values each time. 

$ $

Beneficially, the $\varepsilon - 1$ rounds of inverse-$p$ multiplications yields the effect of scaling down the plaintext $p^{\varepsilon-1}m_i \rightarrow m_i$ and the noise $k''p^{\varepsilon} \rightarrow k''p$, which has the desired noise scaling factor $p$ for standard BGV ciphertexts. 

$ $

\item \textbf{\textsf{\underline{SlotToCoeff}}:} This step works the same way as BFV's \textsf{SlotToCoeff} step: move $m_i + k''_ip$ stored in the input vector slots back to the polynomial coefficient positions by homomorphically multiplying with $\hathat W^*$. Upon completing this step, the ciphertext modulus is some value $q_{l'}$ smaller than $q_L$, because the homomorphic operation of step $4\sim6$ has consumed a few multiplicative levels. The resulting plaintext coefficients are $m_i + k'''_ip$, where $k''' > k''$ due to the additional noise generated by homomorphically running the \textsf{SlotToCoeff} step. The plaintext modulus (i.e., the noise scaling factor) and the noise scaling factor are $p$ at this point, the standard one for BGV ciphertexts. We let these polynomials be $M$ and $K'''$ each, and the output ciphertext is $\textsf{RLWE}_{S, \sigma}(M + K'''p) \bmod q_{l'}$.

\end{enumerate}

\subsubsection{Discussion} 

\para{The Reason for Modulus Switch from $q_l \rightarrow \hat{q}$}: BGV switches the modulus from $q_l \rightarrow \hat{q}$ to eliminate the $q_l$-multiple overflows during bootstrapping. After switching the modulus $q_l \rightarrow \hat{q}$ and then \textsf{ModRaise}, the encrypted plaintext gets the $K'\hat{q}$ overflow term, which can be reduced to $K'$ from the plaintext modulus's perspective due to the special property $\hat{q} \equiv 1 \bmod p^\varepsilon$ (where $p^\varepsilon$ is the plaintext modulus).

$ $

\para{The Choice of $\varepsilon$:} The larger $\varepsilon$ is, the greater the (base-$p$) digit-wise gap between $p^{\varepsilon-1}M$ and $K'$ becomes; thus, the less likely it is that the decryption would fail (i.e., fail to zero out $K'$). However, a larger $\varepsilon$ means the digit extraction operation would be more expensive. 

$ $

\para{Generalization to $\Delta = p^r$:} Like the case of BFV's bootstrapping (Summary~\ref*{subsubsec:bfv-bootstrapping-summary} in \autoref{subsubsec:bfv-bootstrapping-summary}), we can generalize the plaintext modulus (i.e., noise scaling factor) to $p^r$ where $p$ is a prime and $r$ can be any positive integer.

\subsubsection{Comparing the Bootstrapping in BFV, BGV, and CKKS} 
\label{subsubsec:bootstrapping-differences}

Both BFV and BGV's bootstrapping use digit extraction, but for different purposes. In BFV, digit extraction is used to eliminate the noise in the lower-bit area. In BGV, digit extraction is used to eliminate the modulus garbage values in the lower-bit area generated by \textsf{ModRaise} from $\hat{q}\rightarrow q_L$. In addition, at each round of BGV's digit extraction, we explicitly multiply the coefficients of the ciphertext polynomials by $|p^{-1}|_{q^{\langle i \rangle}}$. In BFV, this explicit inverse-$p$ multiplication is skipped and we only conceptually re-interpret the scaling factor. 

CKKS's bootstrapping does not involve digit extraction because it has no values to eliminate in the lower-bit area. Instead, regarding the modulus garbage value \textsf{ModRaise} from $q_0 \rightarrow q_L$, this is eliminated by \textsf{EvalExp}'s homomorphic evaluation of an approximate sine function. Although CKKS's bootstrapping resets its modulus to a large value, this operation does not decrease the noise-to-message ratio, and this ratio continuously increases over homomorphic operations. 

%In the case of BFV's bootstrapping, we need homomorphic decryption (\autoref{subsubsec:bfv-bootstrapping-homomorphic-decryption}) because we need to simultaneously change the ciphertext's plaintext scaling factor from $p^{\varepsilon-1} \rightarrow \left\lfloor\dfrac{q}{p}\right\rfloor$ and the ciphertext modulus from $p^\varepsilon \rightarrow q$.For the similar reason, in the case of BGV's bootstrapping, we used digit extraction to change the noise scaling factor from  $\hat{q} \rightarrow q_L$ ((or technically, $\hat{q} \rightarrow q_{l'}$ where $\hat{q} < q_{l'} < q_L$) while keeping the noise scaling factor (i.e., the plaintext modulus) $\Delta = p^\varepsilon$. On the other hand, in the case of CKKS's bootstrapping, \textsf{ModRaise} instead of homomorphic decryption is sufficient because we only need to change the ciphertext modulus from $q_0 \rightarrow q_L$ (or technically, $q_0 \rightarrow q_{l'}$ where $q_0 < q_{l'} < q_L$) while keeping the same plaintext scaling factor $\Delta \approx \dfrac{q_l}{q_{l-1}}$.  

\clearpage

\section{RNS-variant FHE Schemes}
\label{sec:rns}
The FHE parameters of BFV, BGV, or CKKS schemes, which are secure enough, sometimes require the ring size of polynomial coefficients to be 1000 bits or more, consuming significant computational resources for 64-bit CPU architectures. To make the computation efficient, we can alternatively represent the coefficients of ciphertext polynomials using the number residue system (RNS)~\autoref{subsec:crt-application}, which allows for modulo addition and multiplication of elements from a large ring (e.g., 1000 bits) by combining values computed in small rings (e.g., 32$\sim$64 bits), each of which compactly fits in 64 bit CPU registers. Modern BFV, BGV, and CKKS schemes adopt this RNS approach by default for the efficient computation of large values. These are called RNS-variant FHE schemes.  

While RNS can directly compute modulo addition and multiplication, it does not directly support other operations such as \textsf{ModRaise} or modulus switching, which are essential for all FHE schemes. This section explains how we can design such corner-case operations based on RNS to accomplish a complete design of RNS-based FHE schemes. Besides BFV, BGV, and CKKS, TFHE can also theoretically use RNS for representing its ciphertext coefficients. However, TFHE's practically used coefficient size is less than $2^{32}$ (or $2^{64}$), which compactly fits within 32-bit (or 64-bit) modern CPU registers. Therefore, TFHE does not need RNS. Thus, this section will focus on RNS-based operations for BFV, BGV, and CKKS. 

Particularly in this section, we assume that the modulo reduction $a \bmod q = |a|_q$ implicitly uses a centered (i.e., signed) residue representation (\autoref{subsec:modulo-centered}), whose modulo overflow \& underflow boundaries are $\dfrac{q}{2} - 1$ and $-\dfrac{q}{2}$, respectively. This assumption is necessary to eliminate a certain modulo reduction operation when designing \textsf{FastBconvEx} (\autoref{subsec:rns-fastbconvex}) by using the assumption of limiting the possible range of certain residue arithmetic, as discussed in \autoref{subsec:modulo-centered}. 

\begin{tcolorbox}[
    title = \textbf{Required Background},    % box title
    colback = white,    % light background; tweak to taste
    colframe = black,  % frame colour
    boxrule = 0.8pt,     % line thickness
    left = 1mm, right = 1mm, top = 1mm, bottom = 1mm % inner padding
]

\begin{itemize}
\item \autoref{sec:modulo}: \nameref{sec:modulo}
\item \autoref{sec:group}: \nameref{sec:group}
\item \autoref{sec:field}: \nameref{sec:field}
\item \autoref{sec:polynomial-ring}: \nameref{sec:polynomial-ring}
\item \autoref{sec:decomp}: \nameref{sec:decomp}
\item \autoref{sec:modulus-rescaling}: \nameref{sec:modulus-rescaling}
\item \autoref{sec:chinese-remainder}: \nameref{sec:chinese-remainder}
\item \autoref{sec:polynomial-interpolation}: \nameref{sec:polynomial-interpolation}
\item \autoref{sec:ntt}: \nameref{sec:ntt}
\item \autoref{sec:lattice}: \nameref{sec:lattice}
\item \autoref{sec:rlwe}: \nameref{sec:rlwe}
\item \autoref{sec:glwe}: \nameref{sec:glwe}
\item \autoref{sec:glwe-add-cipher}: \nameref{sec:glwe-add-cipher}
\item \autoref{sec:glwe-add-plain}: \nameref{sec:glwe-add-plain}
\item \autoref{sec:glwe-mult-plain}: \nameref{sec:glwe-mult-plain}
\item \autoref{subsec:modulus-switch-rlwe}: \nameref{subsec:modulus-switch-rlwe}
\item \autoref{sec:glwe-key-switching}: \nameref{sec:glwe-key-switching}
\item \autoref{sec:bfv}: \nameref{sec:bfv}
\item \autoref{sec:ckks}: \nameref{sec:ckks}
\item \autoref{sec:bgv}: \nameref{sec:bgv}
\end{itemize}
\end{tcolorbox}

\clearpage 

\subsection{Fast Base Conversion: \textsf{FastBConv}}
\label{subsec:rns-fastbconv}

\noindent \textbf{- Reference 1:} 
\href{https://eprint.iacr.org/2016/510}{A Full RNS Variant of FV-like Somewhat Homomorphic Encryption Schemes}~\cite{rns-bfv}

\noindent \textbf{- Reference 2:} 
\href{https://eprint.iacr.org/2022/657}{BASALISC: Programmable Hardware Accelerator for BGV FHE}~\cite{rns-bfv2}

$ $

Suppose we have $x \in \mathbb{Z}_q$ (where $q$ is a big modulus). Then, we can express $x$ by using RNS (\autoref{subsec:crt-application}) as $(x_1, x_2, \cdots, x_k)$, where each $x_i \in \mathbb{Z}_{q_i}$, $\prod\limits_{i=1}^k q_i = q$, and $\{q_1, q_2, \cdots, q_k\}$ are co-prime. In RNS, we define base conversion as an operation of converting the RNS residues $(x_1, x_2, \cdots, x_k) \in \mathbb{Z}_{q_1} \times \mathbb{Z}_{q_2} \times \cdots \times \mathbb{Z}_{q_k}$ into $(c_1, c_2, \cdots, c_k) \in \mathbb{Z}_{b_1} \times \mathbb{Z}_{b_2} \times \cdots \times \mathbb{Z}_{b_l}$, where $\{b_1, b_2, \cdots, b_l\}$ is a new base, and $\{q_1, q_2, \cdots, q_k\}$ and $\{b_1, b_2, \cdots, b_l\}$ are all co-prime. The relationship between $x$ and $c$ is: $c = |x|_b$ (where $x \in \mathbb{Z}_{q}$ and $c \in \mathbb{Z}_b$). The standard way of performing base conversion is assembling $(x_1, x_2, \cdots, x_k)$ into $x$ by computing $x = \sum\limits_{i=1}^k |x_i z_i|_{q_i} \cdot y_i \bmod q$ (where $y_i = \dfrac{q}{q_i} \text{ and } z_i = y_i^{-1} \bmod q_i$), and then computing $c_j \equiv x \bmod b_j$ for $j \in [1, l]$. However, this computation is slow if the modulus $q$ is large. To compute the base conversion \textit{fast}, we design the fast base conversion operation $\textsf{FastBConv}$ as follows:

\begin{tcolorbox}[title={\textbf{\tboxlabel{\ref*{subsec:rns-fastbconv}} Fast Base Conversion: \textsf{FastBConv}}}]

\textbf{\underline{Input}:} $(x_1, x_2, \cdots, x_k) \in \mathbb{Z}_{q_1} \times \mathbb{Z}_{q_2} \times \cdots \times \mathbb{Z}_{q_k}$ \textcolor{red}{ $\rhd$ which represents the big value $x \in \mathbb{Z}_q$, where $q = \prod\limits_{i=1}^kq_i$, and $\{q_1, q_2, \cdots, q_k\}$ are co-prime}

$ $

$\textsf{FastBConv}(x, q, b)= \textsf{FastBConv}(\{x_i\}_{i=1}^{k}, \{q_i\}_{i=1}^{k}, \{b_i\}_{i=1}^{l})$

$ = \left( \sum\limits_{i=1}^{k} |x_i \cdot z_i|_{q_i} \cdot y_i \bmod b_j \right)_{j \in [1,l]}$  

\textcolor{red}{ $\rhd$ where $y_i = \dfrac{q}{q_i} \text{, } z_i = y_i^{-1} \bmod q_i$, and $b = \prod\limits_{i=1}^lb_i$} 

$ $

$ = (c_1, c_2, \cdots, c_l) \in \mathbb{Z}_{b_1} \times \mathbb{Z}_{b_2} \times \cdots \times \mathbb{Z}_{b_l}$ \textcolor{red}{ $\rhd$ which represents the big value $c \in \mathbb{Z}_b$}

$ $

$ $

The input to this \textsf{FastBConv} function is a list of RNS residues $(x_1, x_2, \cdots, x_k)$ having the prime moduli $(q_1, q_2, \cdots, q_k)$ as the base. This RNS vector represents the big value:

$x = \left(\sum\limits_{i=1}^k x_i\cdot y_i \cdot z_i\right) \bmod q = \left(\sum\limits_{i=1}^k |x_i\cdot z_i|_{q_i} \cdot y_i\right) \bmod q$ \textcolor{red}{ $\rhd$ Theorem~\ref*{sec:chinese-remainder}.1}

$ $

The output of this \textsf{FastBConv} function is a list of RNS residues $(c_1, c_2, \cdots, c_l)$ having the prime moduli $(b_1, b_2, \cdots, b_l)$ as the base. This RNS vector represents the big value 

$c = \left(\sum\limits_{i=1}^l c_i\cdot y'_i \cdot z'_i\right) \bmod b = \left(\sum\limits_{i=1}^l |c_i\cdot z'_i|_{b_i} \cdot y'_i\right) \bmod b$ \textcolor{red}{ $\rhd$ where $y'_i = \dfrac{b}{b_i}$ and $z'_i = {y_{i}'}^{-1} \bmod b_i$}

$ $

The relationship between $c$ and $x$ is as follows: 

$c = x + uq \bmod b$ (where $u$ is an integer with the magnitude $|u| \leq \dfrac{k}{2} + 1$) 

\textcolor{red}{ $\rhd$ i.e. the fast-base-converted $c$ gets noise $|uq|_b$}

\end{tcolorbox}

\begin{myproof}

We will prove why a fast base conversion of $x$ into $c$ gets an additional noise $|u\cdot q|_b$ (where integer $|u| \leq \dfrac{k}{2}+1$) compared to a standard base conversion. If we did a standard (i.e., exact) base conversion of $x$ from base moduli $(q_1, \cdots, q_k)$ to $(b_1, \cdots, b_l)$, then we would compute the following:

$$
\left( \left( \sum\limits_{i=1}^{k} |x_i \cdot z_i|_{q_i} \cdot y_i \bmod q \right) \bmod b_j \right)_{j \in [1,l]} = \left( x \bmod b_j \right)_{j \in [1,l]}
$$

But $\textsf{FastBConv}$ omits the intermediate (big) reduction modulo $q$ and directly applies (small) reduction modulo $b_j$ for the sake of fast computation, so that our conversion process does not need to handle large values whose magnitude can be as large as $\pm\dfrac{q}{2}$. In this approach of fast base conversion, for each $i \in [1, k]$, the computation result of $|x_i \cdot z_i|_{q_i} \cdot y_i$ is some value between $\left[-\left\lceil\dfrac{q}{2}\right\rceil, \left\lfloor\dfrac{q}{2}\right\rfloor\right]$, because $|x_i \cdot z_i|_{q_i}$ is some integer between $\left[-\left\lceil\dfrac{q_i}{2}\right\rceil, \left\lfloor\dfrac{q_i}{2}\right\rfloor\right]$ and $y_i = \dfrac{q}{q_i}$. Therefore, $-\dfrac{q+1}{2} \leq |x_i \cdot z_i|_{q_i} \cdot y_i \leq \dfrac{q}{2}$. If we sum $k$ such values for $i \in [1, k]$, then the total sum $x' = \sum\limits_{i=1}^{k} |x_i \cdot z_i|_{q_i} \cdot y_i = x + u\cdot q$ (Summary~\ref*{sec:chinese-remainder} in \autoref{sec:chinese-remainder}) for some integer $u$ (where $u\cdot q$ represents the $q$-multiple overflows). And since we have shown that $-\dfrac{q+1}{2} \leq |x_i \cdot z_i|_{q_i} \cdot y_i \leq \dfrac{q}{2}$ for each $i \in [1, k]$, $uq$ has to be greater than $-k\cdot\dfrac{q+1}{2}$ and smaller than $k\cdot\dfrac{q}{2}$ (i.e., $u$ is an integer between $-\dfrac{k}{2} - 1 \leq u \leq \dfrac{k}{2}$). Therefore, $\sum\limits_{i=1}^{k} |x_i \cdot z_i|_{q_i} \cdot y_i$ can have maximum $-\left(\dfrac{k}{2} + 1\right)\cdot q$ underflows and $\dfrac{k}{2}\cdot q$ overflows. Thus, while standard (i.e., exact) base conversion computes each residue as $\hat{c}_j = \left(\sum\limits_{i=1}^{k} |x_i \cdot z_i|_{q_i} \cdot y_i \bmod q\right) \bmod b_j$ (i.e., $\hat{c}_j = x \bmod b_j$), fast (i.e., approximate) base conversion computes each residue as $c_j = \left(\sum\limits_{i=1}^{k} |x_i \cdot z_i|_{q_i} \cdot y_i\right) \bmod b_j$ (i.e., $c_j = x + uq \bmod b_j$, where integer $|u| \leq \dfrac{k}{2} + 1$). Notice that the residual difference (i.e., error) between each $\hat{c}_j$ and $c_j$ is $uq \bmod b_j$, and the collective noise generated by fast base conversion from $q \rightarrow  b$ is  $uq \bmod b$. Also, note that the RNS residue vector $(c_1, c_2, \cdots, c_l) \in \mathbb{Z}_{b_1} \times \mathbb{Z}_{b_2} \times \cdots \times \mathbb{Z}_{b_l}$ represents the big value $c = x + uq \bmod b$. 

Importantly, \textsf{FastBConv} does not guarantee the correctness of base conversion, because the $q$-multiple overflow would generate a non-negligible error. Yet, \textsf{FastBConv} is used as an essential building block for various RNS-based operations such as \textsf{ModRaise\textsubscript{RNS}} (\autoref{subsec:rns-modraise}) and \textsf{ModSwitch\textsubscript{RNS}} (\autoref{subsec:rns-modswitch}).  

\end{myproof}

\subsection{Small Montgomery Reduction Algorithm: \textsf{SmallMont}}
\label{subsec:rns-smallmont}

One problem of the \textsf{FastBConv} (i.e., the fast base conversion) operation is that it creates a non-negligible noise. Specifically, suppose we use \textsf{FastBConv} to convert the base of $x \in \mathbb{Z}_q$ (where $q = q_1 \cdot q_2 \cdot \cdots \cdot q_k$ moduli) into $c = x + uq \bmod b$ (where $b = b_1 \cdot b_2 \cdot \cdots \cdot b_l$ moduli),  where integer $|u| \leq \dfrac{k}{2}+1$. Then, the noise generated by this conversion is between $-\left(\dfrac{k}{2}+1\right)\cdot q \bmod b$ and $\left(\dfrac{k}{2}+1\right)\cdot q \bmod b$. To reduce this noise, we will explain the small Montgomery algorithm (\textsf{SmallMont}) which reduces the noise generated by fast base conversion from $uq$ to $u'q$, such that $u' \in \{-1, 0, 1\}$. The small Montgomery algorithm is designed as follows:

\begin{tcolorbox}[title={\textbf{\tboxlabel{\ref*{subsec:rns-smallmont}} Fast Modulo Reduction: \textsf{SmallMont}}}]

\textbf{\underline{Input}:} $c = (c_1, c_2, \cdots, c_l, c_{l+1}) \in \mathbb{Z}_{b_1} \times \mathbb{Z}_{b_2} \times \cdots \times \mathbb{Z}_{b_l} \times \mathbb{Z}_{b_\alpha}$ \textcolor{red}{ $\rhd$ $b_\alpha$ is a prime and co-prime to $b$, where $b = \prod\limits_{i=1}^lb_i$}

$ $

, where $c = \textsf{FastBConv}\bm(|b_\alpha \cdot x|_q, q, bb_\alpha\bm) = |b_\alpha \cdot x|_q + uq$ \textcolor{red}{ $\rhd$ where $x \in \mathbb{Z}_q$ and integer $|u| \leq \dfrac{k}{2}+1$}

$ $

\textbf{\underline{Main Steps}}

\textbf{$\textsf{SmallMont}(c, bb_\alpha, b_\alpha, q): $} 

\begin{enumerate}
\item $c' = |c \cdot q^{-1}|_{b_\alpha}$
\item For each $i \in [1, l]$, compute $r_i = \Big|(c_{b_i} - |q|_{b_i}\cdot c') \cdot b_\alpha^{-1} \Big|_{b_i}$
\end{enumerate}

$ $

\textbf{\underline{Output:}} $r = (\overbrace{r_1, r_2, \cdots, r_l}^{l}) \in \overbrace{\mathbb{Z}_{b_1} \times \mathbb{Z}_{b_2} \times \cdots \times \mathbb{Z}_{b_l}}^{l}$ \textcolor{red}{ $\rhd$ without $r_\alpha \in \mathbb{Z}_{b_\alpha}$}

$ $

The output satisfies the relation: $r = x + u'q \bmod b$ (where $u' \in \{-1, 0, 1\}$)

\end{tcolorbox}

\begin{myproof}

\begin{enumerate}
\item Given $c' = |c \cdot q^{-1}|_{b_\alpha}$, notice that $c - q \cdot c'$ is exactly divisible by $b_\alpha$ as shown below: 

$c - q \cdot c' \bmod b_\alpha$

$= c - q \cdot |c \cdot q^{-1}|_{b_\alpha} \bmod b_\alpha$ \textcolor{red}{ $\rhd$ substituting $c' = |c \cdot q^{-1}|_{b_\alpha}$}

$= c - c \bmod b_\alpha$ \textcolor{red}{ $\rhd$ by canceling out $|q|_{b_\alpha}$ and $|q|^{-1}_{b_\alpha}$}

$= 0 \bmod b_\alpha$

$ $

Since $c - q \cdot c' = 0 \bmod b_\alpha$, this implies that $c - q \cdot c'$ is a multiple of $b_\alpha$ (i.e., $c - q \cdot c'$ is exactly divisible by $b_\alpha$). This also implies that $\dfrac{c - q \cdot c'}{b_\alpha}$ is an integer. 

$ $

\item Given $c = |b_\alpha \cdot x|_q + uq \bmod b$ and $c' = |c \cdot q^{-1}|_{b_\alpha}$, we can express $\dfrac{c - q \cdot c'}{b_\alpha} \bmod b$ as follows:

%%%$\left|\dfrac{c - q \cdot c'}{b_\alpha}\right|_{bb_\alpha} = \left|\dfrac{c - q \cdot c'}{b_\alpha}\right|_{b}$ \textcolor{red}{ $\rhd$ since $c < bb_\alpha$, it is guaranteed that $\dfrac{c - q \cdot c'}{b_\alpha} < b$}
$\left|\dfrac{c - q \cdot c'}{b_\alpha}\right|_{b}$

$ = \left|\dfrac{c - q \cdot |c \cdot q^{-1}|_{b_\alpha}}{b_\alpha}\right|_{b}$ \textcolor{red}{ $\rhd$ by substituting $c' = |c \cdot q^{-1}|_{b_\alpha}$}

$ = \left|\dfrac{|b_\alpha \cdot x|_q + uq - q \cdot \Bigg|\Big||b_\alpha \cdot x|_q + uq\Big|_{b} \cdot q^{-1}\Bigg|_{b_\alpha}}{b_\alpha}\right|_{b}$ \textcolor{red}{ $\rhd$ by substituting $c = \Big||b_\alpha \cdot x|_q + uq\Big|_{b}$}

$ = \Bigg|\dfrac{b_\alpha\cdot x + vq + uq - q \cdot \Big||b_\alpha\cdot x + vq + uq|_{b} \cdot q^{-1}\Big|_{b_\alpha}}{b_\alpha}\Bigg|_{b}$ \textcolor{red}{ $\rhd$ by rewriting $|b_\alpha\cdot x|_{q}$ as $b_\alpha\cdot x + vq$ (where $v$ is some integer representing the $q$-overflows of $b_\alpha\cdot x$)}

$ $

$ = \Bigg|x + \dfrac{vq + uq - q \cdot \Big||b_\alpha\cdot x + vq + uq|_{b} \cdot q^{-1}\Big|_{b_\alpha}}{b_\alpha}\Bigg|_{b}$ \textcolor{red}{ $\rhd$ since $x = \dfrac{b_\alpha \cdot x}{b_\alpha}$}

$ = \Bigg|x + q\cdot \dfrac{v + u - \Big||b_\alpha\cdot x + vq + uq|_{b} \cdot q^{-1}\Big|_{b_\alpha}}{b_\alpha}\Bigg|_{b}$ \textcolor{red}{ $\rhd$ taking out the common multiple $q$}

$ $

The above computation result is guaranteed to be an integer (as we proved in the proof step 1). And $q$ and $b_\alpha$ are co-prime (by the input definition). This leads to the conclusion that 

$ \dfrac{v + u - \Big||b_\alpha\cdot x + vq + uq|_{b} \cdot q^{-1}\Big|_{b_\alpha}}{b_\alpha}$ is guaranteed to be an integer. Therefore, if we choose $b_\alpha$ (i.e., a prime and co-prime to both $q$ and $b$) as a sufficiently large value, then $ \dfrac{v + u - \Big||b_\alpha\cdot x + vq + uq|_{b} \cdot q^{-1}\Big|_{b_\alpha}}{b_\alpha}$ will converge to $\{-1, 0, 1\}$. This is because as $b_\alpha$ increases: (1) $v$ grows slower than $b_\alpha$ (since $|b_\alpha\cdot x|_q = b_\alpha\cdot x + vq$); (2) the magnitude of $u$ stays smaller than $\dfrac{k}{2} + 1$ (as integer $|u| \leq \dfrac{k}{2}+1$); and (3) $\Big||b_\alpha\cdot x + vq + uq|_{b} \cdot q^{-1}\Big|_{b_\alpha}$ is guaranteed to be an integer between $\left[-\dfrac{b_\alpha+1}{2}, \dfrac{b_\alpha}{2} - 1\right]$. In conclusion, if $b_\alpha$ is sufficiently large, then we get the following relation: 

$\left|\dfrac{c - q\cdot c'}{b_\alpha}\right|_{b} = x + u'q \bmod b$ \textcolor{red}{ $\rhd$ where $u' \in \{-1, 0, 1\}$}

$ $

Also, the following is true: 

$\dfrac{c - q\cdot c'}{b_\alpha} \bmod b = (c - q\cdot c')\cdot b_\alpha^{-1} \bmod b$ \textcolor{red}{ $\rhd$ because $b_\alpha$ divides $c - q\cdot c'$ and $b_\alpha$ is co-prime to $b$} 

$ $

\item It is possible to express the final output $x + u'q \bmod b$ as an RNS vector with the residues of the base moduli $(b_1, \cdots, b_l)$. For this, we convert $(c - q\cdot c')\cdot b_\alpha^{-1}$ into the RNS vector $(r_1, r_2, \cdots, r_l) \in \mathbb{Z}_{b_1} \times \mathbb{Z}_{b_2} \times \cdots \times \mathbb{Z}_{b_l}$ by computing the following for each $i \in [1, l]$:

$r_i = |(c - q\cdot c')\cdot b_\alpha^{-1}|_{b_i}$

$ = |(c_{b_i} - |q|_{b_i}\cdot c') \cdot b_\alpha^{-1} |_{b_i}$

\end{enumerate}
\end{myproof}

\subsubsection{Improving \textsf{FastBConv} by Using \textsf{SmallMont}}
\label{subsubsec:rns-smallmont-fastbconv}

Notice that by using \textsf{SmallMont} in Summary~\ref*{subsec:rns-smallmont}, the accuracy of the raw output of $\textsf{FastBConv}(x, q, b) = |x + uq|_b$ (where integer $|u| \leq \dfrac{k}{2}+1$) is improved to $|x + u'q|_b$ (where $u' \in \{-1, 0, 1\}$) as follows:

$\textsf{SmallMont}\bm(\textsf{FastBConv}\bm(|b_\alpha \cdot x|_q, q, bb_\alpha\bm), bb_\alpha, b_\alpha, q\bm) $

$\textsf{SmallMont}\bm(\Big||b_\alpha \cdot x|_q + uq\Big|_{bb_\alpha}, bb_\alpha, b_\alpha, q\bm) $

$= |x + u'q|_b$ \textcolor{red}{ $\rhd$ where $u' \in \{-1, 0, 1\}$}

$ $

\subsection{RNS-based ModRaise: \textsf{ModRaise\textsubscript{RNS}}}
\label{subsec:rns-modraise}

\noindent \textbf{- Reference:} 
\href{https://eprint.iacr.org/2018/931.pdf}{A Full RNS Variant of
Approximate Homomorphic Encryption}~\cite{rns-ckks}

$ $

\textsf{ModRaise} is an operation that raises a ciphertext's modulus from $q$ to $qb$ (where $q \ll qb$). We used \textsf{ModRaise} in BFV's ciphertext-to-ciphertext multiplication (Summary~\ref*{subsubsec:bfv-mult-cipher-summary} in \autoref{subsubsec:bfv-mult-cipher-summary}) and in CKKS's modulus bootstrapping (Summary~\ref*{subsubsec:ckks-bootstrapping-summary} in \autoref{subsubsec:ckks-bootstrapping-summary}). The RNS-based \textsf{ModRaise} operation is designed as follows:

\begin{tcolorbox}[title={\textbf{\tboxlabel{\ref*{subsec:rns-modraise}} \textsf{ModRaise\textsubscript{RNS}}}}]

\textbf{\underline{Input}:} $(x_1, x_2, \cdots, x_k) \in \mathbb{Z}_{q_1} \times \mathbb{Z}_{q_2} \times \cdots \times \mathbb{Z}_{q_k}$ \textcolor{red}{ $\rhd$ which represents the big value $x \in \mathbb{Z}_q$}

$ $

$\textsf{ModRaise\textsubscript{RNS}}({\{x_i\}_{i=1}}^k, q, qb)$ \textcolor{red}{ $\rhd$ where $q$ and $b$ are co-prime}

$= \textsf{FastBConv\textsubscript{RNS}}({\{x_i\}_{i=1}}^k, q, qb)$ 

$ = (x_1, x_2, \cdots, x_k, \textsf{FastBConv}(\{x_i\}_{i=1}^k, q, b))$ 

$ = (x_1, x_2, \cdots, x_k, c_1, c_2, \cdots, c_l) \in \mathbb{Z}_{q_1} \times \mathbb{Z}_{q_2} \times \cdots \times \mathbb{Z}_{q_k} \times \mathbb{Z}_{b_1} \times \mathbb{Z}_{b_2} \times \cdots \times \mathbb{Z}_{b_l}$

$ = (\chi_1, \chi_2, \cdots, \chi_{k+l}) \in \mathbb{Z}_{q_1} \times \mathbb{Z}_{q_2} \times \cdots \times \mathbb{Z}_{q_k} \times \mathbb{Z}_{b_1} \times \mathbb{Z}_{b_2} \times \cdots \times \mathbb{Z}_{b_l}$

\textcolor{red}{ $\rhd$ which represents the value $\chi \in \mathbb{Z}_{qb}$}

$ $

The relationship between $\chi$ and $x$ is as follows: 

$\chi \equiv x + u\cdot q \bmod qb$ \textcolor{red}{ $\rhd$ the noise generated by \textsf{ModRaise\textsubscript{RNS}} is $|uq|_{qb}$ (where integer $|u| \leq \dfrac{k}{2}+1$)}

$\chi \equiv x \bmod q$

\end{tcolorbox}

\begin{myproof}

In \autoref{subsec:rns-fastbconv}, we proved that $x' = \sum\limits_{i=1}^{k} |x_i \cdot z_i|_{q_i} \cdot y_i = x + u\cdot q$ (where integer $|u| \leq \dfrac{k}{2}+1$). Therefore, the following holds:

$x' \equiv x_i \bmod q_i$ for $i \in [1, k]$ \textcolor{red}{ $\rhd$ since $x' = x + u\cdot q \equiv x_i \bmod q_i$ (as $q_i$ divides $q$, thus $u \cdot q \equiv 0 \pmod{q_i}$, and $x \equiv x_i \bmod q_i$)}

$x' \equiv c_j \bmod b_j$ for $j \in [1, l]$ \textcolor{red}{ $\rhd$ where each $c_j = x + uq \bmod b_j$}

$ $

Therefore, $x' \bmod qb$ can be represented as the following RNS residues: 

$(x_1, x_2, \cdots, x_k, c_1, c_2, \cdots, c_l) \in \mathbb{Z}_{q_1} \times \mathbb{Z}_{q_2} \times \cdots \times \mathbb{Z}_{q_k} \times \mathbb{Z}_{b_1} \times \mathbb{Z}_{b_2} \times \cdots \times \mathbb{Z}_{b_l}$

$ = (x_1, x_2, \cdots, x_k, \textsf{FastBConv}(\{x_i\}_{i=1}^k, q, b)) \in \mathbb{Z}_{q_1} \times \mathbb{Z}_{q_2} \times \cdots \times \mathbb{Z}_{q_k} \times \mathbb{Z}_{b_1} \times \mathbb{Z}_{b_2} \times \cdots \times \mathbb{Z}_{b_l}$ 

$ $

Our ideal goal of mod-raising $x \in \mathbb{Z}_q$ from $q \rightarrow qb$ is to derive an RNS vector of $x \bmod qb$. However, the above RNS vector represents $x' \bmod qb$, where $x' = x + uq$ (with integer $|u| \leq \dfrac{k}{2}+1$). Therefore, we can interpret the above RNS vector as representing $x \bmod qb$ with the additional noise $|uq|_{qb}$.

\end{myproof}

\subsection{RNS-based ModDrop: \textsf{ModDrop\textsubscript{RNS}}}
\label{subsec:rns-moddrop}

\textsf{ModDrop} (\autoref{subsec:ckks-moddrop}, \autoref{subsec:bgv-moddrop}) is an operation of decreasing a ciphertext's modulus from $q \rightarrow q'$ (where $q'$ divides $q$) without affecting the plaintext's scaling factor(in the case of CKKS) or the noise's scaling factor (in the case of BGV). 

In an RNS-based ciphertext representation, \textsf{ModDrop} is equivalent to removing some of the base moduli in the ciphertext without affecting the scaling factor $\Delta$. This can be achieved by converting the ciphertext's base from $q$ to $\bar{q}$ where the base moduli set of $\bar{q}$ are a subset of that of $q$; that is, $\bar{q}$ divides $q$. Specifically, suppose that we have an input $(x_1, x_2, \cdots, x_k) \in \mathbb{Z}_{q_1} \times \mathbb{Z}_{q_2} \times \cdots \times \mathbb{Z}_{q_k}$, and a new subset base $\bar{q} = q_1 \cdot q_2 \cdot \cdots \cdot q_{k'}$, where $k' < k$. In this setup, the fast base conversion from $q \rightarrow \bar{q}$ is equivalent to simply extracting the input value's RNS residues associated with the base moduli $(q_1, q_2, \cdots, q_{k'})$. This is because of the following reasoning:

$\textsf{FastBConv}(\{x_i\}_{i=1}^{k}, q, \bar{q}) = \left( \sum\limits_{i=1}^{k} |x_i \cdot z_i|_{q_i} \cdot y_i \bmod q_j \right)_{j \in [1,k']}$  

$ = x + uq \bmod \bar{q}$ \textcolor{red}{ $\rhd$ Summary~\ref*{subsec:rns-fastbconv} in \autoref{subsec:rns-fastbconv}}

$ = x \bmod \bar{q} $ \textcolor{red}{ $\rhd$ $uq$ gets eliminated because $\bar{q}$ divides $uq$}

$ = (x_1, x_2, \cdots, x_{k'}) \in \mathbb{Z}_{q_1} \times \mathbb{Z}_{q_2} \times \cdots \times \mathbb{Z}_{q_{k'}}$

$ $

Notice that the above fast base conversion from $q \rightarrow \bar{q}$ (where $\bar{q}$ divides $q$) does not generate any noise. This is different from the case of fast base conversion from $q \rightarrow b$ (Summary~\ref*{subsec:rns-fastbconv} in \autoref{subsec:rns-fastbconv}) where $q$ and $b$ are co-prime, which generates the noise $|uq|_b$ (where integer $|u| \leq \dfrac{k}{2}+1$). 

The \textsf{ModDrop} operation is supported in all of BFV, BGV, and CKKS ciphertexts that are represented in RNS forms. However, note that \textsf{ModDrop} is possible only if the scaled plaintext (in the case of BFV and CKKS) or the scaled noise (in the case of BGV) does not exceed the ciphertext modulus after the mod-drop operation, because otherwise correct decryption is not possible. \textsf{ModDrop\textsubscript{RNS}} is summarized as follows:

\begin{tcolorbox}[title={\textbf{\tboxlabel{\ref*{subsec:rns-moddrop}} \textsf{ModDrop\textsubscript{RNS}}}}]

\textbf{\underline{Input}:} $(x_1, x_2, \cdots, x_k) \in \mathbb{Z}_{q_1} \times \mathbb{Z}_{q_2} \times \cdots \times \mathbb{Z}_{q_k}$ 

$ $

$\textsf{FastBConv}(\{x_i\}_{i=1}^{k}, q, \bar{q}) = \left( \sum\limits_{i=1}^{k} |x_i \cdot z_i|_{q_i} \cdot y_i \bmod q_j \right)_{j \in [1,k']}$  

\textcolor{red}{ $\rhd$ where $\bar{q}$ is a product of co-primes $q_1 \cdot q_2 \cdot \cdots \cdot q_{k'}$, and $\bar{q}$ divides $q$}

$ = (x_1, x_2, \cdots, x_{k'}) \in \mathbb{Z}_{q_1} \times \mathbb{Z}_{q_2} \times \cdots \times \mathbb{Z}_{q_{k'}}$ \textcolor{red}{ $\rhd$ no noise generated during the conversion}

\end{tcolorbox}

\subsection{RNS-based Modulus Switch: \textsf{ModSwitch\textsubscript{RNS}}}
\label{subsec:rns-modswitch}

Modulus switch is an operation of reducing a ciphertext's modulus from $q$ to $q'$ (where $q' < q$) and updating the target value from $x$ to $\left\lceil x\cdot \dfrac{q'}{q}\right\rfloor$. Modulus switch is used for lowering the multiplicative level of a ciphertext upon each ciphertext-to-ciphertext multiplication (in the case of BFV, CKKS, or BGV) or even upon each ciphertext-to-plaintext multiplication (in the case of CKKS). Upon each modulus switch from $q \rightarrow q'$ of a ciphertext, the scaling factor of the underlying plaintext in the ciphertext also gets reduced by the same proportion: $\dfrac{q'}{q}$. 

The modulus switch operation of an RNS-based ciphertext is denoted as \textsf{ModSwitch\textsubscript{RNS}}, which requires that the output base moduli are a subset of the input base moduli. In other words, like the case of \textsf{ModDrop\textsubscript{RNS}}, it only supports a modulus switch from $qb \rightarrow q$, where $q$ and $b$ are co-prime.

Suppose we have $(\chi_1, \chi_2, \cdots, \chi_{k+l}) \in \mathbb{Z}_{q_1} \times \mathbb{Z}_{q_2} \times \cdots \times \mathbb{Z}_{q_k} \times \mathbb{Z}_{b_1} \times \mathbb{Z}_{b_2} \times \cdots \times \mathbb{Z}_{b_l} $, which represents the value $\chi = \left(\sum\limits_{i=1}^{k}\left|\chi_i\cdot\left(\dfrac{qb}{q_i}\right)^{-1}\right|_{q_i}\cdot\dfrac{qb}{q_i}\right) + \left(\sum\limits_{j=k+1}^{k+l}\left|\chi_j\cdot\left(\dfrac{qb}{b_j}\right)^{-1}\right|_{b_j}\cdot\dfrac{qb}{b_j}\right) \bmod qb$. 

$ $

Given $\chi \in \mathbb{Z}_{qb}$, \textsf{ModSwitch\textsubscript{RNS}} from $qb \rightarrow q$ is an operation of updating $\chi \in \mathbb{Z}_{qb}$ to some $y \in \mathbb{Z}_q$ where $y \approx \left\lceil\dfrac{\chi}{b}\right\rfloor$. Unlike in regular modulus switch where we can directly arithmetically divide $\chi$ by $b$ and round it, an RNS vector is incompatible with direct arithmetic division on the residues. Therefore, our alternative strategy is to find some small value $\hat{\chi}$ such that $\chi \equiv \hat{\chi} \bmod b$. Once we find such $\hat\chi$, then $\chi - \hat{\chi} \bmod qb$ becomes divisible by $b$ (since their difference is some multiple of $b$), and thus we can compute $\dfrac{\chi - \hat{\chi}}{b} \approx \left\lceil\dfrac{\chi}{b}\right\rfloor$. Note that in this computation, the additionally introduced error of modulus switch caused by replacing $\chi$ with $\chi - \hat\chi$ is equivalent to: $\Bigg|\left\lceil\dfrac{\chi}{b}\right\rfloor - \dfrac{\chi - \hat\chi}{b}\Bigg| \approx \left\lceil\dfrac{\hat\chi}{b}\right\rfloor$. After the (exact) division of $\dfrac{\chi - \hat\chi}{b}$, we directly replace the modulus $qb$ with $q$. This direct replacement of modulus is arithmetically allowed because the computation result of $\dfrac{\chi - \hat{\chi}}{b}$ is guaranteed to be within $-\dfrac{q}{2}$ and $\dfrac{q}{2} -1$ (since $-\dfrac{qb}{2}\leq \chi \leq \dfrac{qb}{2}-1$). Therefore, we can derive the following formula:

$\dfrac{\chi - \hat{\chi}}{b} \bmod q = |b^{-1}|_{q}\cdot (\chi - \hat{\chi}) \bmod q$  %\textcolor{red}{ $\rhd$ since $\chi - \hat\chi$ is divisible by $b$, and $b$ has an inverse in $\mathbb{Z}_q$}

$ $

In the above relation, we can arithmetically replace $b$ with $|b^{-1}|_q$, because $\chi - \hat{\chi}$ is divisible by $b$ and $b$ is guaranteed to have an inverse modulo $q$ (since $b$ and $q$ are co-prime). Next, we can compute  $|b^{-1}|_{q}\cdot (\chi - \hat{\chi}) \bmod q$ based on their RNS residues as follows:

$|b^{-1}|_q\cdot (\chi - \hat{\chi}) \bmod q$

$= \bm(|b^{-1}|_{q_1} \cdot (\chi_1 - \hat\chi_1), \text{ } |b^{-1}|_{q_2} \cdot (\chi_2 - \hat\chi_2), \text{ } \cdots, \text{ } |b^{-1}|_{q_k} \cdot (\chi_k - \hat\chi_k)\bm) \in \mathbb{Z}_{q_1} \times \mathbb{Z}_{q_2} \times \cdots \times \mathbb{Z}_{q_k} $

$= (y_1, y_2, \cdots, y_k) $ \textcolor{red}{ $\rhd$ where each $y_i = |b^{-1}|_{q_i} \cdot (\chi_i - \hat\chi_i) \bmod q_i$}

$ $

Now, our task is to derive an expression for some small $\hat\chi$ such that $\chi - \hat\chi$ is divisible by $b$. We propose that $\hat\chi = |\chi|_b + ub$ for some small integer $|u| \leq \dfrac{l}{2}+1$. Then, notice that $\chi - \hat\chi$ is divisible by $b$ as follows:

$|\chi - \hat\chi|_b = \Big||\chi|_b - (|\chi|_b + ub)\Big|_b = |-ub|_b = 0$

$ $

Now, we will derive the RNS vector of $\hat\chi \bmod q = |\chi|_b + ub \bmod q$, which is to be plugged into $|b^{-1}|_q\cdot (\chi - \hat{\chi}) \bmod q$. First, we derive the RNS vector of $|\chi|_{b}$ as follows: 

$(|\chi|_{b_1}, |\chi|_{b_2}, \cdots, |\chi|_{b_l}) = (\chi_{k+1}, \chi_{k+2}, \cdots, \chi_{k+l}) \in \mathbb{Z}_{b_1} \times \mathbb{Z}_{b_2} \times \cdots \times \mathbb{Z}_{b_l}$

$ $

Next, we can compute its fast base conversion from $b \rightarrow q$ as follows:

$\textsf{FastBConv}(\{\chi_{k+i}\}_{i=1}^{l}, b, q)$

$= (\hat\chi_1, \hat\chi_2, \cdots, \hat\chi_{k}) \in \mathbb{Z}_{q_1} \times \mathbb{Z}_{q_2} \times \cdots \times \mathbb{Z}_{q_k}$

$ $ 

Now, notice that the above RNS residue vector $(\hat\chi_1, \hat\chi_2, \cdots, \hat\chi_k)$ represents the value $\hat\chi = |\chi|_b + u \cdot b \bmod q$ (where integer $|u| \leq \dfrac{l}{2}+1$), which is our desired formula for $\hat\chi$. Therefore, $\hat\chi = \textsf{FastBConv}(\{\chi_{k+i}\}_{i=1}^{l}, b, q)$.

Note that $\hat\chi \ll \dfrac{q}{2}-1$ and $-\dfrac{q}{2} \ll \hat\chi$, because $\Big||\chi|_b + u\cdot b\Big| < \left(\dfrac{l}{2} + 1\right)\cdot b + \dfrac{b}{2} \ll \dfrac{q}{2}$ (here we assume that $b \ll q$, as we assume the modulus switch operation is used to remove only a single prime factor from the large base $q$). Therefore, the magnitude of the error generated by computing $ b^{-1} \cdot (\chi - \hat\chi)$ is approximately $\left\lceil\dfrac{\hat\chi}{b}\right\rfloor <  \left\lceil\dfrac{\left(\dfrac{l}{2} + 1\right)\cdot b + \dfrac{b}{2}}{b}\right\rfloor = \left\lceil\dfrac{lb + 3b}{2b}\right\rfloor = \left\lceil\dfrac{l + 3}{2}\right\rfloor < \dfrac{l}{2} + 2$. 

%the relationship between $\$

$ $

We summarize the \textsf{ModSwitch\textsubscript{RNS}} operation as follows:

\begin{tcolorbox}[title={\textbf{\tboxlabel{\ref*{subsec:rns-modswitch}} \textsf{ModSwitch\textsubscript{RNS}}}}]

\textbf{\underline{Input}:} $(\chi_1, \chi_2, \cdots, \chi_{k+l}) \in \mathbb{Z}_{q_1} \times \mathbb{Z}_{q_2} \times \cdots \times \mathbb{Z}_{q_k} \times \mathbb{Z}_{b_1} \times \mathbb{Z}_{b_2} \times \cdots \times \mathbb{Z}_{b_l} $ 

\textcolor{red}{ $\rhd$ which represents $\chi = \left(\sum\limits_{i=1}^{k}\left|\chi_i\cdot\left(\dfrac{qb}{q_i}\right)^{-1}\right|_{q_i}\cdot\dfrac{qb}{q_i}\right) + \left(\sum\limits_{j=k+1}^{k+l}\left|\chi_j\cdot\left(\dfrac{qb}{b_j}\right)^{-1}\right|_{b_j}\cdot\dfrac{qb}{b_j}\right) \bmod qb$} 

$ $

\textbf{\underline{Notations}}

\begin{itemize}

\item The RNS vector $(\chi_{1}, \chi_{2}, \cdots, \chi_{k}) \in \mathbb{Z}_{q_1} \times \mathbb{Z}_{q_2} \times \cdots \times \mathbb{Z}_{q_k}$ represents the value: $|\chi|_q \in \mathbb{Z}_q$

\item $\textsf{FastBConv}(\{\chi_{k+1}\}_{i=1}^{l}, b, q) = (\hat{\chi}_1, \hat{\chi}_2, \cdots, \hat{\chi}_k) \in \mathbb{Z}_{q_1} \times \mathbb{Z}_{q_2} \times \cdots \times \mathbb{Z}_{q_k}$ 

, which represents the value $\hat{\chi} = |\chi|_b + ub \in \mathbb{Z}_q$ \textcolor{red}{ $\rhd$ where $|u| \leq \dfrac{l}{2}+1$}

\end{itemize}

$ $

\textbf{\underline{Main Steps}}

$\textsf{ModSwitch\textsubscript{RNS}}(\{\chi_i\}_{i=1}^{k+l}, qb, q)$

$= \{|b^{-1}|_{q_i}\cdot (\chi_i - \hat{\chi}_i) \bmod q_i\}_{i=1}^{k} \in \mathbb{Z}_{q_1} \times \mathbb{Z}_{q_2} \times \cdots \times \mathbb{Z}_{q_k}$ 

$ $

, whose RNS residue vector represents the value $|b^{-1}|_{q} \cdot (\chi - \hat{\chi}) \bmod q$. The magnitude of noise generated by \textsf{ModSwitch\textsubscript{RNS}} is roughly $\left\lceil\dfrac{\hat{\chi}}{b}\right\rfloor < \dfrac{l}{2} + 2$.

\end{tcolorbox}

\subsubsection{Comparing \textsf{ModSwitch\textsubscript{RNS}}, \textsf{ModRaise\textsubscript{RNS}}, and \textsf{ModDrop\textsubscript{RNS}}}

Given a big value $x \in \mathbb{Z}_q$ in an RNS vector, \textsf{ModSwitch\textsubscript{RNS}} reduces its modulus from $q \rightarrow  q'$ and explicitly decreases the modulo value $x$ by the proportion of $\dfrac{q'}{q}$ (i.e., it updates $x$ to $ \left\lceil x \cdot \dfrac{q'}{q}\right\rfloor$). On the other hand, \textsf{ModDrop\textsubscript{RNS}} from $q \rightarrow q'$ updates the modulo value from $x \rightarrow |x|_{q'}$ (where $q'$ divides $q$), which is different from decreasing $x$ by the proportion of $\dfrac{q'}{q}$ like modulus switch. \textsf{ModRaise\textsubscript{RNS}} from $q \rightarrow qb$ (where $q$ divides $qb$) increases the modulus without explicitly modifying the modulo value $x$, but it generates some $q$-overflow noise. \textsf{ModSwitch\textsubscript{RNS}} and \textsf{ModRaise\textsubscript{RNS}} generate some noise, whereas \textsf{ModDrop\textsubscript{RNS}} does not generate any noise.

\subsection{RNS-based Decryption}
\label{subsec:rns-dec}

This subsection will explain how to efficiently decrypt RNS-based ciphertexts for BFV, CKKS, and BGV.

\subsubsection{BFV Decryption: $\textsf{Dec}_{\textsf{RNS}}^{\textsf{BFV}}$}
\label{subsubsec:rns-dec-bfv}

Suppose we have a BFV ciphertext $(A, B)$ such that $B = A\cdot S + \Delta M + E$ (where $\Delta = \left\lfloor\dfrac{q}{t}\right\rfloor$). We decrypt the ciphertext as follows: $M = \left\lceil\dfrac{B - A\cdot S}{\Delta}\right\rfloor$ (Summary~\ref*{subsec:bfv-enc-dec} in \autoref{subsec:bfv-enc-dec}). However, RNS does not allow direct division and rounding. Therefore, we need to express this divide-and-round operation in terms of addition and multiplication.

Let's denote $\textsf{ct}(s) = \Delta m + e + kq$ (i.e., a decryption of ciphertext $\textsf{ct}$ without modulo-$q$ reduction). In this description, we will consider only a single set of coefficients $m$, $e$, and $k$ extracted from polynomials $M$, $E$, and $K$ for simplicity. 

As explained in \autoref{subsec:modulo-division}, modulo arithmetic does not support direct division. Meanwhile, the special relation $\dfrac{a}{b} \bmod p = a \cdot b^{-1} \bmod p$ holds if $b$ divides $a$ and an inverse of $b$ modulo $p$ exists (i.e., $b$ and $p$ are co-prime). Inspired by this, we can express the decrypted plaintext $m$ as follows: 

$m = \left\lceil\dfrac{|\textsf{ct}(s)|_q}{\Delta}\right\rfloor = \left\lfloor\dfrac{|\textsf{ct}(s)|_q}{\Delta}\right\rfloor + e_r$ \textcolor{red}{ $\rhd$ where $e_r \in [0, 1]$ is a rounding error}

$ = \left\lfloor |\textsf{ct}(s)|_q\cdot \dfrac{t}{q} \right\rfloor + e_r + e_d$  \textcolor{red}{ $\rhd$ where $e_d = \left\lfloor\dfrac{|\textsf{ct}(s)|_q}{\Delta}\right\rfloor - \left\lfloor |\textsf{ct}(s)|_q\cdot \dfrac{t}{q} \right\rfloor$ is a scaling error} 

$= \left\lfloor\dfrac{t\cdot |\textsf{ct}(s)|_q}{q}\right\rfloor + e_r + e_d$ 

$= \dfrac{t\cdot |\textsf{ct}(s)|_q - |t\cdot \textsf{ct}(s)|_q}{q} + e_r + e_d$ \textcolor{red}{ $\rhd$ where $|t\cdot \textsf{ct}(s)|_q \equiv t\cdot |\textsf{ct}(s)|_q \bmod q$, and therefore $t\cdot |\textsf{ct}(s)|_q - |t\cdot \textsf{ct}(s)|_q$ is divisible by $q$}

$ $

Now, we choose some prime number $\gamma$ which is co-prime to $t$ and $q$. Then, we derive the expression for $\gamma \cdot m$ as follows:

$\gamma \cdot m =  \gamma\cdot \left\lceil\dfrac{ |\textsf{ct}(s)|_q}{\Delta}\right\rfloor$

$ = \left\lceil\dfrac{\gamma\cdot |\textsf{ct}(s)|_q}{\Delta}\right\rfloor + e'_s $ \textcolor{red}{ $\rhd$ where $e'_s = \gamma\cdot \left\lceil\dfrac{ |\textsf{ct}(s)|_q}{\Delta}\right\rfloor - \left\lceil\dfrac{\gamma\cdot |\textsf{ct}(s)|_q}{\Delta}\right\rfloor$ is a multiplication error}

$ =  \left\lfloor\dfrac{\gamma \cdot |\textsf{ct}(s)|_q}{\Delta}\right\rfloor + e'_s + e'_r$ \textcolor{red}{ $\rhd$ where $e'_r \in [0, 1]$ is a rounding error}

$ =  \left\lfloor\gamma \cdot | \textsf{ct}(s)|_q\cdot \dfrac{t}{q} \right\rfloor + e'_s + e'_r + e'_d$  \textcolor{red}{ $\rhd$ where $e'_d =  \left(\left\lfloor\dfrac{\gamma \cdot |\textsf{ct}(s)|_q}{\Delta}\right\rfloor - \left\lfloor \gamma \cdot |\textsf{ct}(s)|_q\cdot \dfrac{t}{q} \right\rfloor\right)$ is a scaling error} 

$=  \left\lfloor\dfrac{\gamma \cdot t\cdot |\textsf{ct}(s)|_q}{q}\right\rfloor + e'_s + e'_r + e'_d$ 

$= \dfrac{\gamma \cdot t\cdot |\textsf{ct}(s)|_q -  |\gamma \cdot t\cdot \textsf{ct}(s)|_q}{q} + e'_s + e'_r + e'_d$ 

$ $

Next, we derive the expression for $|\gamma \cdot m|_{\gamma t}$ as follows:

$|\gamma \cdot m|_{\gamma t} = \left|\dfrac{\gamma\cdot t\cdot |\textsf{ct}(s)|_q - |\gamma\cdot t\cdot \textsf{ct}(s)|_q}{q} + e'_s + e'_r + e'_d\right|_{\gamma t}$

$= \left|\dfrac{\gamma\cdot t\cdot |\textsf{ct}(s)|_q -  |\gamma\cdot t\cdot \textsf{ct}(s)|_q}{q}\right|_{\gamma t} + |e'_s|_{\gamma t} + |e'_r|_{\gamma t} + |e'_d|_{\gamma t}$

$= \left|\dfrac{\gamma\cdot t\cdot |\textsf{ct}(s)|_q -  |\gamma\cdot t\cdot \textsf{ct}(s)|_q}{q}\right|_{\gamma t} + e'_s + e'_r + e'_d$ \textcolor{red}{ $\rhd$ assuming $|e'_s| \ll \dfrac{\gamma t}{2}$ and $|e'_r| \ll \dfrac{\gamma t}{2}$ and $|e'_d| \ll \dfrac{\gamma t}{2}$}

$= \Big|(\gamma\cdot t\cdot |\textsf{ct}(s)|_q -  |\gamma\cdot t\cdot \textsf{ct}(s)|_q)\cdot q^{-1}\Big|_{\gamma t} + e'_s + e'_r + e'_d$  \textcolor{red}{ $\rhd$ since $\gamma\cdot t\cdot |\textsf{ct}(s)|_q - |\gamma\cdot t\cdot \textsf{ct}(s)|_q$ is divisible by $q$, and $q$ is co-prime to $\gamma t$}

$ $

$= \left|-  |\gamma\cdot t\cdot \textsf{ct}(s)|_q\cdot q^{-1}\right|_{\gamma t} + e'_s + e'_r + e'_d$ \textcolor{red}{ $\rhd$ since $\gamma \cdot t\cdot |\textsf{ct}(s)|_q$ is a multiple of $\gamma t$}

$= \Big||\gamma\cdot t\cdot \textsf{ct}(s)|_q\Big|_{\gamma t} \cdot |-q^{-1}|_{\gamma t} + e'_s + e'_r + e'_d$

$ $

Given the above relation, notice that the computation result of $\textsf{FastBConv}\bm(\gamma\cdot t\cdot \textsf{ct}(s), q, \gamma t\bm) \cdot |-q^{-1}|_{\gamma t}$ can be expressed as follows: 

$\textsf{FastBConv}\bm(\gamma\cdot t\cdot \textsf{ct}(s), q, \gamma t\bm) \cdot |-q^{-1}|_{\gamma t}$

$ = \Big||\gamma t \cdot \textsf{ct}(s)|_q + uq\Big|_{\gamma t} \cdot |-q^{-1}|_{\gamma t} $ \textcolor{red}{ $\rhd$ where $|u| \leq \dfrac{k}{2}+1$ for the base moduli $q_1, q_2, \cdots, q_k$}

$ =  \Big||\gamma t \cdot \textsf{ct}(s)|_q \cdot |-q^{-1}|_{\gamma t}  - u\Big|_{\gamma t}$

$= |\gamma \cdot m|_{\gamma t}  - e'_s - e'_r - e'_d - u$ \textcolor{red}{ $\rhd$ as we previously showed that $|\gamma \cdot m|_{\gamma t} = \Big||\gamma\cdot t\cdot \textsf{ct}(s)|_q\Big|_{\gamma t} \cdot |-q^{-1}|_{\gamma t} +  e'_s + e'_r + e'_d$}

$= y$ \textcolor{red}{ $\rhd$ let's denote the above expression as $y$}

$ $

Then, if $e'_s, e'_r, e'_d, u$ are small enough such that $|e'_s + e'_r + e'_d + u| < \gamma$, then $|y|_{\gamma} = -e'_s -e'_r - e'_d - u$ (as $|\gamma \cdot m|_{\gamma t} \bmod \gamma = 0$ as a multiple of $\gamma$). Therefore, we can effectively remove the noise terms $e'_s, e'_r, e'_d, u$ and derive $m$ as follows:

$\bm{\Big|}(y - |y|_{\gamma}) \cdot |\gamma^{-1}|_t\bm{\Big|}_t$

$ = \bm{\Big|}|\gamma\cdot m|_{\gamma t} \cdot |\gamma^{-1}|_t\bm{\Big|}_t$

$ = \bm{\Big|}|\gamma\cdot m|_{t} \cdot |\gamma^{-1}|_t\bm{\Big|}_t$ \textcolor{red}{ $\rhd$ since $\Big||\gamma\cdot m|_{\gamma t}\Big|_t = |\gamma\cdot m|_{t}$}

$= |m|_t$

$ $

, which is the final decryption of \textsf{ct} we wanted to compute. Let's denote the RNS vector of $y$ as a $(y_{\gamma}, y_{t}) \in \mathbb{Z}_{\gamma} \times \mathbb{Z}_{t}$. Then, we can efficiently compute the term $\Big|(y - |y|_{\gamma}) \cdot |\gamma^{-1}|_t\Big|_t$ as follows:

$\Big|(y - |y|_{\gamma}) \cdot |\gamma^{-1}|_t\Big|_t$

$= \Bigg|\Big( \overbrace{\Big|y_{\gamma}\cdot t \cdot |t^{-1}|_{\gamma} + y_{t}\cdot \gamma \cdot |\gamma^{-1}|_{t} \Big|_{\gamma t}}^{y} - \overbrace{\Big|y_{\gamma}\cdot t \cdot |t^{-1}|_{\gamma} + y_{t}\cdot \gamma \cdot |\gamma^{-1}|_{t}\Big|_{\gamma}}^{|y|_{\gamma}}\Big) \cdot |\gamma^{-1}|_t\Bigg|_t$

$= \Bigg|\Big( \overbrace{\Big|y_{\gamma}\cdot t \cdot |t^{-1}|_{\gamma} + y_{t}\cdot \gamma \cdot |\gamma^{-1}|_{t} \Big|_{t}}^{y} - \overbrace{\Big|y_{\gamma}\cdot t \cdot |t^{-1}|_{\gamma} + y_{t}\cdot \gamma \cdot |\gamma^{-1}|_{t}\Big|_{\gamma}}^{|y|_{\gamma}}\Big) \cdot |\gamma^{-1}|_t\Bigg|_t$ \textcolor{red}{ $\rhd$ since $\Big||y|_{\gamma t}\Big|_t = |y|_t$}

$= \Bigg|\Big( \overbrace{\Big|y_{\gamma}\cdot t \cdot |t^{-1}|_{\gamma} + y_{t}\cdot \gamma \cdot |\gamma^{-1}|_{t} \Big|_t}^{y} - \overbrace{\Big|y_{\gamma}\cdot t \cdot |t^{-1}|_{\gamma}\Big|_{\gamma}}^{|y|_{\gamma}}\Big) \cdot |\gamma^{-1}|_t\Bigg|_t$ \textcolor{red}{ $\rhd$ since $y_{t}\cdot \gamma \cdot |\gamma^{-1}|_{t} \bmod \gamma = 0$}

$= \Bigg|\Big( \overbrace{\Big|y_{t}\cdot \gamma \cdot |\gamma^{-1}|_{t}\Big|_t}^{y} - \overbrace{\Big|y_{\gamma}\cdot t \cdot |t^{-1}|_{\gamma}\Big|_{\gamma}}^{|y|_{\gamma}}\Big) \cdot |\gamma^{-1}|_t\Bigg|_t$ \textcolor{red}{ $\rhd$ since $y_{\gamma}\cdot t \cdot |t^{-1}|_{\gamma} \bmod t = 0$}

$= \Big|(y_t - y_{\gamma}) \cdot |\gamma^{-1}|_t\Big|_t$ \textcolor{red}{ $\rhd$ since $|\gamma \cdot \gamma^{-1}|_t = 1$ and $|t \cdot t^{-1}|_{\gamma} = 1$}

$ $

We summarize $\textsf{Dec}_{\textsf{RNS}}^{\textsf{BFV}}$ as follows:

\begin{tcolorbox}[title={\textbf{\tboxlabel{\ref*{subsubsec:rns-dec-bfv}} $\textsf{Dec}_{\textsf{RNS}}^{\textsf{BFV}}$}}]

\textbf{\underline{Input}:} $\textsf{ct}(s) = \Delta m + e + kq$

\begin{enumerate}
\item Pick some prime number $\gamma$ which is co-prime to $t$ and $q$.

\item Compute $\textsf{FastBConv}(|\gamma \cdot t \cdot \textsf{ct}(s)|_q, q, \gamma t) \cdot |-q^{-1}|_{\gamma t}$

$ = (y_{\gamma}, y_{t}) \in \mathbb{Z}_{\gamma} \times \mathbb{Z}_{t}$

\item Compute $|m|_t = \Big|(y_{t} - y_{\gamma})\cdot |\gamma^{-1}|_t\Big|_t$

\end{enumerate}

\end{tcolorbox}

\subsubsection{CKKS and BGV Decryption}
\label{subsubsec:rns-dec-ckks-bgv}

CKKS and BGV ciphertexts can be decrypted efficiently by performing the \textsf{ModDrop} operation (Summary~\ref*{subsec:rns-moddrop} in \autoref{subsec:rns-moddrop}) to the lowest multiplicative level. After this, there remains only a single ciphertext modulus in the RNS base, so the regular decryption algorithm can be executed efficiently without any RNS components.

\subsection{BGV's RNS-based Modulus Switch: $\textsf{ModSwitch}_{\textsf{RNS}}^{\textsf{BGV}}$}
\label{subsec:rns-modswitch-bgv}

BGV's RNS-based modulus switch is not computed by using the standard  \textsf{ModSwitch\textsubscript{RNS}}, because BGV's original non-RNS modulus switch (Summary~\ref*{subsec:bgv-modulus-switch} in \autoref{subsec:bgv-modulus-switch}) is performed in a different manner than BFV or CKKS's non-RNS modulus switch (Summary~\ref*{subsec:modulus-switch-rlwe} in \autoref{subsec:modulus-switch-rlwe}). BGV's non-RNS modulus switch is computed as follows: 

$(A', B') = \left(\left\lceil\dfrac{\hat{q}}{q_l} A\right\rfloor, \left\lceil\dfrac{\hat{q}}{q_l}\cdot B\right\rfloor\right) \in \mathcal{R}_{\langle n, \hat{q} \rangle}^2$

$ $

$\epsilon'_A = \hat{q}\cdot A - q_l\cdot A'$ \textcolor{red}{ $\rhd$ where $\epsilon'_A \in \mathbb{Z}_{q_l}$}

$\epsilon'_B = \hat{q}\cdot B - q_l\cdot B'$ \textcolor{red}{ $\rhd$ where $\epsilon'_B \in \mathbb{Z}_{q_l}$}

$ $

$H_A = q_l^{-1}\cdot\epsilon'_A \bmod t$

$H_B = q_l^{-1}\cdot\epsilon'_B \bmod t$

$ $

$\hat{\textsf{ct}} = (\hat{A}, \hat{B}) = (A' + H_A, B' + H_B) \bmod \hat{q}$

$ $

Therefore, BGV's RNS-based modulus switch only needs to compute the above formulas for $\hat{A}$ and $\hat{B}$ based on RNS's $(+, \cdot)$ arithmetic. In the above computations, the only part that cannot be directly computed by RNS-based $(+, \cdot)$ operations is the rounding in $\left\lceil\dfrac{\hat{q}}{q_l} A\right\rfloor$ and $\left\lceil\dfrac{\hat{q}}{q_l}\cdot B\right\rfloor$. This rounding can be performed in RNS by using $\textsf{Dec}_{\textsf{RNS}}^{\textsf{BFV}}$, by setting $q = q_l$ and $t = \hat{q}$.

\subsection{Exact Fast Base Conversion: \textsf{FastBConvEx}}
\label{subsec:rns-fastbconvex}

\textsf{FastBConv} (\autoref{subsec:rns-fastbconv}) converts an input value $x$'s base moduli from $q \rightarrow b$, but generates a noise equivalent to $uq \bmod b$ where integer $|u| \leq \dfrac{k}{2}+1$. If we use \textsf{FastBConv} with \textsf{SmallMont} (\autoref{subsec:rns-smallmont}), we can reduce the generated noise from $uq$ to $u'q$ where $u' \in \{-1, 0, 1\}$. In this subsection, we introduce \textsf{FastBConvEx}, an algorithm for an exact fast base conversion that can eliminate the entire noise. However, using \textsf{FastBConvEx} has a restriction that the input value $x$ should be relatively much smaller than its modulus. This is different from the case of using \textsf{FastBConv} with \textsf{SmallMont} which has no restriction on the input $x$ (i.e., $x$ can be any value within its modulus range). \textsf{FastBConvEx} is designed as follows:

\begin{tcolorbox}[title={\textbf{\tboxlabel{\ref*{subsec:rns-fastbconvex}} Fast Exact Base Conversion: \textsf{FastBConvEx}}}]

\textbf{\underline{Input}:} $x = (x_1, x_2, \cdots, x_l, x_\alpha) \in \mathbb{Z}_{b_1} \times \mathbb{Z}_{b_2} \times \cdots \times \mathbb{Z}_{b_l} \times \mathbb{Z}_{b_\alpha}$ 

$ $

\textbf{\underline{Requirement}: } The size of $b_\alpha$ should be $b_{\alpha} \geq 2\cdot(l + \lambda)$, where $|x|_{b} = x + \mu \cdot b$, and $\mu \in [-\lambda, \lambda]$ (i.e., $\lambda$ and $-\lambda$ are the maximum and minimum possible values of $\mu$). 

\textcolor{red}{ $\rhd$ The constraint that $b_\alpha > \mu$ implies that the input $x$ should be much smaller than its modulus $bb_\alpha$ (i.e., $|x| \ll \dfrac{bb_\alpha}{2}$)}

$ $

\textbf{\underline{Main Steps}}

\begin{enumerate}
\item $\hat{x} = |x|_b = \textsf{ModDrop}(x, bb_\alpha, b)$ 

\item $x_\alpha = |x|_{b_\alpha} = \textsf{ModDrop}(x, bb_\alpha, b_\alpha)$

\item $\gamma = |(\textsf{FastBConv}(\hat{x}, b, b_\alpha) - x_\alpha)\cdot b^{-1}|_{b_\alpha}$

\item $\textsf{FastBConvEx}(x, bb_\alpha, q) = \Big|\textsf{FastBConv}(\hat{x}, b, q) - \gamma \cdot b\Big|_q = |x|_q$

\end{enumerate}

\end{tcolorbox}
                                        
We will prove why $\Big|\textsf{FastBConv}(\hat{x}, b, q) - \gamma \cdot b\Big|_q = |x|_q$. 

\begin{myproof}

\begin{enumerate}

\item $\textsf{FastBConv}(\hat{x}, b, b_\alpha) $%$ = \Bigg(\sum\limits_{i=1}^{l}|\hat{x}_i\cdot b_i \cdot b^{-1} |_{b_i} \cdot \dfrac{b}{b_i}\Bigg) \bmod b_\alpha$ 

$ = |\hat{x} + ub|_{b_\alpha}$ \textcolor{red}{ $\rhd$ where integer $|u| \leq \dfrac{l}{2}+1$}

$ = \Big||x|_{b} + ub\Big|_{b_\alpha}$ \textcolor{red}{ $\rhd$ since $|x|_{b} = \hat{x}$ by definition}

$ = |x + \mu b + ub|_{b_\alpha}$ \textcolor{red}{ $\rhd$ since $|x|_{b} = x + \mu b$ by definition}

$ $

\item $\gamma = |(\textsf{FastBConv}(\hat{x}, b, b_\alpha) - x_\alpha)\cdot b^{-1}|_{b_\alpha}$

$ = |(\textsf{FastBConv}(\hat{x}, b, b_\alpha) - x_\alpha - \mu b)\cdot b^{-1} + \mu|_{b_\alpha} $ \textcolor{red}{ $\rhd$ by adding $|(-\mu b + \mu b)\cdot b^{-1}|_{b_\alpha}$}

$ = |(x + \mu b + ub - x_\alpha - \mu b)\cdot b^{-1} + \mu|_{b_\alpha} $ \textcolor{red}{ $\rhd$ step 1 proved $ \textsf{FastBConv}(\hat{x}, b, b_\alpha) = |x + \mu b + u\cdot b|_{b_\alpha}$}

$ = |u + \mu|_{b_\alpha} $

$ = u + \mu $ \textcolor{red}{ $\rhd$ because $-\dfrac{b_\alpha}{2} \leq u + \mu \leq \dfrac{b_\alpha}{2} - 1$ (since $u + \mu < l + \lambda \leq \dfrac{b_\alpha}{2}$, and $-\dfrac{b_\alpha}{2} \leq -(l+\lambda) < u + \mu$)}

$ $

\item $\textsf{FastBConvEx}(x, b, q) = \Big|\textsf{FastBConv}(\hat{x}, b, q) - \gamma \cdot b)\Big|_q $

$ = \Big|\hat{x} + u b - \gamma \cdot b)\Big|_q $ \textcolor{red}{ $\rhd$ applying $\textsf{FastBConv}(\hat{x}, b, q) = \hat{x} + u b$ }

$ = \Big|(x + \mu b) + u b - \gamma \cdot b)\Big|_q $ \textcolor{red}{ $\rhd$ applying $\hat{x} = |x|_b = x + \mu b$}

$ = \Big|(x + \mu b) + u b - (u + \mu) \cdot b)\Big|_q $ \textcolor{red}{ $\rhd$ applying $\gamma = u + \mu$ from proof step 2}

$ = |x + \mu b + ub - ub - \mu b|_q $

$ = |x|_q $

\end{enumerate}

\end{myproof}

\para{Necessity of the Centered (i.e., Signed) Residue Representation:} In the proof step 2, we treated $|u + \mu|_{b_\alpha} = u + \mu$. To remove the modulo reduction operation, the canonical (i.e., unsigned) residue representation is inappropriate, because if $u + \mu$ becomes negative, then the residue will underflow and have to be wrapped around, which requires a modulo reduction operation. To prevent the occurrence of both overflow and underflow cases, we need the centered (i.e., signed) residue representation.

\subsection{Decomposing Multiplication: \textsf{DecompMult\textsubscript{RNS}}}
\label{subsec:rns-decompmult}

In FHE, gadget decomposition (\autoref{subsec:gadget-decomposition}) is used to compute ciphertext-to-plaintext multiplication with small noise. For example, BFV and CKKS's homomorphic key switching (Summary~\ref*{sec:glwe-key-switching} in \autoref{sec:glwe-key-switching}) uses gadget decomposition to compute $\textsf{RLWE}_{S',\sigma}(\Delta M) = B + A\cdot \textsf{RLWE}_{S', \sigma}(S)$ with small noise (where each coefficient of the polynomial $A$ can be any value within the range of the ciphertext modulus $q$). As another example, the relinearization process of ciphertext-to-ciphertext multiplication in BFV (Summary~\ref*{subsubsec:bfv-mult-cipher-summary} in \autoref{subsubsec:bfv-mult-cipher-summary}), CKKS (Summary~\ref*{subsubsec:ckks-mult-cipher-summary} in \autoref{subsubsec:ckks-mult-cipher-summary}), and BGV (Summary~\ref*{subsec:bgv-mult-cipher} in \autoref{subsec:bgv-mult-cipher}) uses gadget decomposition to derive the synthetic ciphertext $\textsf{RLWE}_{S', \sigma}(D_2\cdot S^2)$ when computing $\textsf{RLWE}_{S',\sigma}(\Delta^2 M^{\langle 1 \rangle} M^{\langle 2 \rangle}) = D_0 + D_1\cdot S + D_2\cdot S^2 = \textsf{ct}_\alpha + \textsf{ct}_\beta$, where $\textsf{ct}_\alpha = (D_0, D_1)$, $\textsf{ct}_\beta = \textsf{RLWE}_{S', \sigma}(D_2\cdot S^2)$, $D_0 = B_1B_2$, $D_1 = A_1B_2 + A_2B_1$, and $D_2 = A_1A_2$. Using gadget decomposition, we showed the following relations: 

$\textsf{RLWE}_{S', \sigma}(A\cdot S) = \bm{\langle} \textsf{Decomp}^{\beta, l}(A), \text{ } \textsf{RLev}_{S', \sigma}^{\beta, l}(S) \bm{\rangle}$ \textcolor{red}{ $\rhd$ used in key switching}

$\textsf{RLWE}_{S', \sigma}(D_2\cdot S^2) = \bm{\langle} \textsf{Decomp}^{\beta, l}(D_2), \text{ } \textsf{RLev}_{S, \sigma}^{\beta, l}( S^2) \bm{\rangle}$ \textcolor{red}{ $\rhd$ used in relinearization}

$ $

However, if we convert a value (e.g., $x$) into an RNS vector, then it cannot be directly expressed in a gadget-decomposed form based on the $\beta$ and $l$ parameters. Instead, given the relationship between the value $x$ and its RNS residues is $x = \sum\limits_{i=1}^k x \cdot \dfrac{q}{q_i} \cdot \left|\left(\dfrac{q}{q_i}^{-1}\right)\right|_{q_i} \bmod q$, we can treat each RNS residue as a gadget-decomposed element. For example, suppose our goal is to decompose $\textsf{RLWE}_{S', \sigma}( A \cdot S)$, where the RNS vector of $A = (A_1, A_2, \cdots, A_k)$ has base moduli $(q_1, q_2, \cdots, q_k)$. In other wods, the RNS representation of the polynomial $A$ is the set of polynomials $\{A_i\}_{i=1}^k$ such that $A_i = A \bmod q_i$. We can decompose $\textsf{RLWE}_{S', \sigma}( A \cdot S)$ as follows:

$\textsf{RLWE}_{S', \sigma}(A\cdot S) \bmod q$

$= \textsf{RLWE}_{S', \sigma}\left( S \cdot \left(A_{1}\dfrac{q}{q_1}\cdot\left|\left(\dfrac{q}{q_1}\right)^{-1}\right|_{q_1} + A_{2}\dfrac{q}{q_2}\cdot\left|\left(\dfrac{q}{q_2}\right)^{-1}\right|_{q_2} + \cdots + A_{k}\dfrac{q}{q_k}\cdot\left|\left(\dfrac{q}{q_k}\right)^{-1}\right|_{q_k}\right)\right)  \bmod q$ 

$= \textsf{RLWE}_{S', \sigma}\left(S \cdot A_{1}\dfrac{q}{q_1}\cdot\left|\left(\dfrac{q}{q_1}\right)^{-1}\right|_{q_1}\right) + \textsf{RLWE}_{S', \sigma}\left(S \cdot A_{2}\dfrac{q}{q_2}\cdot\left|\left(\dfrac{q}{q_2}\right)^{-1}\right|_{q_2}\right) + $

$\cdots + \textsf{RLWE}_{S', \sigma}\left(S \cdot A_{k}\dfrac{q}{q_k}\cdot\left|\left(\dfrac{q}{q_k}\right)^{-1}\right|_{q_k}\right) \bmod q$ 

$ $

$ $

$= A_{1}\cdot\textsf{RLWE}_{S', \sigma}\left(S \cdot \dfrac{q}{q_1}\cdot\left|\left(\dfrac{q}{q_1}\right)^{-1}\right|_{q_1}\right) + A_{2}\cdot\textsf{RLWE}_{S', \sigma}\left(S \cdot \dfrac{q}{q_2}\cdot\left|\left(\dfrac{q}{q_2}\right)^{-1}\right|_{q_2}\right) +$

$ \cdots + A_{k}\cdot\textsf{RLWE}_{S', \sigma}\left(S \cdot \dfrac{q}{q_k}\cdot\left|\left(\dfrac{q}{q_k}\right)^{-1}\right|_{q_k}\right) \bmod q$ 

$ $

$ $

$= \sum\limits_{i=1}^{k}\left(A_{i}\cdot \textsf{RLWE}_{S', \sigma}\left(S \cdot\dfrac{q}{q_i} \cdot \left| \left(\dfrac{q}{q_i}\right)^{-1}\right|_{q_i}\right)\right) \bmod q$

$ $

$ $

, where $\left\{\textsf{RLWE}_{S', \sigma}\left(S \cdot\dfrac{q}{q_i} \cdot \left| \left(\dfrac{q}{q_i}\right)^{-1}\right|_{q_i}\right)\right\}_{i=1}^{k}$ can be pre-generated as RNS key-switching keys.  

$ $

$ $

Applying the same reasoning as above, we can also derive the following for relinearization:

$\textsf{RLWE}_{S, \sigma}(D_2 \cdot S^2) = \sum\limits_{i=1}^{k}\Bigg( D_{2, i}\cdot \textsf{RLWE}_{S, \sigma}\left(S^2 \cdot\dfrac{q}{q_i} \cdot \left| \left(\dfrac{q}{q_i}\right)^{-1}\right|_{q_i}\right)\Bigg) \bmod q$

$ $

$ $

, where $\left\{\textsf{RLWE}_{S, \sigma}\left(S^2 \cdot\dfrac{q}{q_i} \cdot \left| \left(\dfrac{q}{q_i}\right)^{-1}\right|_{q_i}\right)\right\}_{i=1}^{k}$ can be pre-generated as relinearization keys.

$ $

$ $

RNS-based multiplication decomposition is summarized as follows:

\begin{tcolorbox}[title={\textbf{\tboxlabel{\ref*{subsec:rns-decompmult}} \textsf{DecompMult\textsubscript{RNS}}}}]

For key-switching:

$ $

\textbf{\underline{Input}:} $A = (A_1, A_2, \cdots, A_k) \in \mathcal{R}_{\langle n, q_1 \rangle} \times \mathcal{R}_{\langle n, q_2 \rangle} \times \cdots \times \mathcal{R}_{\langle n, q_k \rangle}$, 

\phantom{Input: } $S_{\langle S', \textsf{RNS}\rangle} = \left\{ \textsf{RLWE}_{S', \sigma}\left(S\cdot \dfrac{q}{q_i}\cdot\left|\left(\dfrac{q}{q_i}\right)^{-1}\right|_{q_i}\right) \right\}_{i=1}^{k}$ \textcolor{red}{ $\rhd$ key-switching keys}

$ $

$ $

$\textsf{DecompMult\textsubscript{RNS}}(A, S_{\langle S', \textsf{RNS}\rangle}) = \textsf{RLWE}_{S', \sigma}(A \cdot S) $

$= \sum\limits_{i=1}^{k}\Bigg(A_{i}\cdot \textsf{RLWE}_{S', \sigma}\left(S \cdot\dfrac{q}{q_i} \cdot \left| \left(\dfrac{q}{q_i}\right)^{-1}\right|_{q_i}\right)\Bigg) \bmod q$

$ $

\par\noindent\rule{\textwidth}{0.4pt}

For relinearization:

$ $

\textbf{\underline{Input}:} $D_2 = (D_{2,1}, D_{2,2}, \cdots, D_{2,k}) \in \mathcal{R}_{\langle n, q_1 \rangle} \times \mathcal{R}_{\langle n, q_2 \rangle} \times \cdots \times \mathcal{R}_{\langle n, q_k \rangle}$,

\phantom{Input: } $S_{\langle S, \textsf{RNS}\rangle}^2 = \left\{ \textsf{RLWE}_{S, \sigma}\left(S^2\cdot \dfrac{q}{q_i}\cdot\left|\left(\dfrac{q}{q_i}\right)^{-1}\right|_{q_i}\right) \right\}_{i=1}^{k}$ \textcolor{red}{ $\rhd$ relinearization keys}

$ $

$ $

$\textsf{DecompMult\textsubscript{RNS}}(D_2, S^2_{\langle S, \textsf{RNS}\rangle}) = \textsf{RLWE}_{S, \sigma}(D_2 \cdot S^2) $

$= \sum\limits_{i=1}^{k}\Bigg(D_{2,i}\cdot \textsf{RLWE}_{S, \sigma}\left(S^2 \cdot\dfrac{q}{q_i} \cdot \left| \left(\dfrac{q}{q_i}\right)^{-1}\right|_{q_i}\right)\Bigg) \bmod q$

\end{tcolorbox}

\subsection{Applying RNS Techniques to FHE Operations}
\label{subsec:rns-application}

This subsection will explain how the RNS primitives we have learned so far are used to handle FHE operations for RNS-based ciphertexts in BFV, CKKS, and BGV.

\subsubsection{Addition and Multiplication of Polynomials}
\label{subsubsec:rns-application-basic}

In BFV, CKKS, and BGV, ciphertext-to-plaintext addition, ciphertext-to-ciphertext addition, and ciphertext-to-plaintext multiplication are performed by only involving modulo additions and multiplications among polynomial coefficients. Therefore, we can represent each polynomial coefficient as an RNS residue vector and compute coefficient-wise additions and multiplications by using RNS-based addition and multiplication of residues, as explained in Summary~\ref*{subsec:crt-application} (\autoref{subsec:crt-application}). For example, suppose we have the following two polynomials $P^{\langle 1 \rangle}$ and $P^{\langle 2 \rangle}$:

$P^{\langle 1 \rangle} = \sum\limits_{a=0}^{n-1} c^{\langle 1 \rangle}_a \cdot X^a \in \mathcal{R}_{\langle n, q\rangle}$

$P^{\langle 2 \rangle} = \sum\limits_{b=0}^{n-1} c^{\langle 2 \rangle}_b \cdot X^b \in \mathcal{R}_{\langle n, q\rangle}$

$ $

In the RNS-variant FHE schemes, we express each coefficient of the polynomial as an RNS residue vector as follows:

$c^{\langle 1 \rangle}_a = (c^{\langle 1 \rangle}_{a,1}, c^{\langle 1 \rangle}_{a,2}, \cdots, c^{\langle 1 \rangle}_{a,k}) \in \mathbb{Z}_{q_1} \times \mathbb{Z}_{q_2} \times \cdots \times \mathbb{Z}_{q_k}$ \textcolor{red}{ $\rhd$ for $a \in [0, n - 1]$}

$c^{\langle 2 \rangle}_b = (c^{\langle 2 \rangle}_{b,1}, c^{\langle 2 \rangle}_{b,2}, \cdots, c^{\langle 2 \rangle}_{b,k}) \in \mathbb{Z}_{q_1} \times \mathbb{Z}_{q_2} \times \cdots \times \mathbb{Z}_{q_k}$ \textcolor{red}{ $\rhd$ for $b \in [0, n - 1]$}

$ $

Given the above RNS setup, when we add or multiply two polynomials, each coefficient-to-coefficient addition is computed as element-wise additions of two RNS residue vectors as follows:

$c^{\langle 1 \rangle}_a + c^{\langle 2 \rangle}_b \equiv \sum\limits_{i=1}^k (c^{\langle 1 \rangle}_{a,i} + c^{\langle 2 \rangle}_{b,i}) y_i z_i \bmod q $ \textcolor{red}{ $\rhd$ where $y_i = \dfrac{q}{q_i}$, $z_i = |y_i^{-1}|_{q_i}$}

$\Longleftrightarrow (c^{\langle 1 \rangle}_{a,1} + c^{\langle 2 \rangle}_{b,1}, \text{ } c^{\langle 1 \rangle}_{a,2} + c^{\langle 2 \rangle}_{b,2}, \cdots, \text{ } c^{\langle 1 \rangle}_{a,k} + c^{\langle 2 \rangle}_{b,k}) \in \mathbb{Z}_{q_1} \times \mathbb{Z}_{q_2} \times \cdots \times \mathbb{Z}_{q_k}$ 

$ $

Similarly, each coefficient-to-coefficient multiplication is computed as element-wise multiplications of two RNS residue vectors as follows:

$c^{\langle 1 \rangle}_a \cdot c^{\langle 2 \rangle}_b \equiv \sum\limits_{i=1}^k (c^{\langle 1 \rangle}_{a,i} \cdot c^{\langle 2 \rangle}_{b,i}) y_i z_i \bmod q $ 

$\Longleftrightarrow (c^{\langle 1 \rangle}_{a,1} \cdot c^{\langle 2 \rangle}_{b,1}, \text{ } c^{\langle 1 \rangle}_{a,2} \cdot c^{\langle 2 \rangle}_{b,2}, \cdots, \text{ } c^{\langle 1 \rangle}_{a,k} \cdot c^{\langle 2 \rangle}_{b,k}) \in \mathbb{Z}_{q_1} \times \mathbb{Z}_{q_2} \times \cdots \times \mathbb{Z}_{q_k}$

$ $

Using the above isomorphism, we can efficiently compute ciphertext-to-plaintext addition, ciphertext-to-ciphertext addition, and ciphertext-to-plaintext multiplication of big polynomial coefficients (e.g., 1000 bits big) based on small RNS residues (e.g., 30 bits each).

\subsubsection{Key Switching}
\label{subsubsec:rns-application-key-switching}

In BFV or CKKS, an RNS-based ciphertext's key-switching operation from $S \rightarrow S'$ is performed by computing the following formula in RNS vectors:

$\textsf{RLWE}_{S', \sigma}(\Delta M) = B + \bm{\langle} \textsf{Decomp}^{\beta, l}(A), \text{ } \textsf{RLev}_{S', \sigma}^{\beta, l}(S) \bm{\rangle}$

$ $

In the above formula, the computation of $\bm{\langle} \textsf{Decomp}^{\beta, l}(A), \text{ } \textsf{RLev}_{S', \sigma}^{\beta, l}(S) \bm{\rangle}$ can be performed by using $\textsf{DecompMult\textsubscript{RNS}}$ (Summary~\ref*{subsec:rns-decompmult} in \autoref{subsec:rns-decompmult}), after which $B$ can be added to it by using regular RNS-based addition. 

$ $

Similarly, in the case of the BGV, an RNS-based key-switching operation on a ciphertext from $S \rightarrow S'$ is performed by computing the following in RNS vectors:

$\textsf{RLWE}_{S', \sigma}(M) = B + \bm{\langle} \textsf{Decomp}^{\beta, l}(A), \text{ } \textsf{RLev}_{S', \sigma}^{\beta, l}(S) \bm{\rangle}$

\subsubsection{Input Slot Rotation}
\label{subsubsec:rns-rotation}

In BFV or CKKS, an RNS-based ciphertext's input slot rotation is performed by computing the following formulas in RNS vectors: 

$ $

\begin{enumerate}
\item $ \textsf{RLWE}_{S(X^{J(h)}), \sigma}\bm(\Delta M(X^{J(h)})\bm) = \bm(A(X^{J(h)})$, $B(X^{J(h)})\bm)$ \textcolor{red}{ $\rhd$ where $J(h) = 5^h \bmod 2n$}

\item Key-switch $\textsf{RLWE}_{S(X^{J(h)}), \sigma}\bm(\Delta M(X^{J(h)})\bm)$ to $\textsf{RLWE}_{S(X), \sigma}\bm(\Delta M(X^{J(h)})\bm)$
\end{enumerate}

$ $

Step 1 is equivalent to re-positioning the coefficients within each polynomial and flipping their signs whenever they cross the boundary of the $n$-th degree term. This step can be done with RNS-based coefficients by moving around each set of RNS residue vectors as a whole whenever the coefficient they represent is re-positioned to a new degree term, and flipping the signs of the residues in the same RNS vector altogether whenever their representing coefficient's sign is to be flipped. Step 2's RNS-based key switching can be done in the same way as explained in the previous subsection (\autoref{subsubsec:rns-application-key-switching}). 

$ $

Similarly, in BGV, an RNS-based ciphertext's input slot rotation is performed by computing the following formulas in RNS vectors: 

$ $

\begin{enumerate}
\item $ \textsf{RLWE}_{S(X^{J(h)}), \sigma}\bm(M(X^{J(h)})\bm) = \bm(A(X^{J(h)})$, $B(X^{J(h)})\bm)$ \textcolor{red}{ $\rhd$ where $J(h) = 5^h \bmod 2n$}

\item Key-switch $\textsf{RLWE}_{S(X^{J(h)}), \sigma}\bm(M(X^{J(h)})\bm)$ to $\textsf{RLWE}_{S(X), \sigma}\bm(M(X^{J(h)})\bm)$
\end{enumerate}

$ $

We can compute the above formulas in RNS by using the same strategy explained for BFV or CKKS.

\subsubsection{BFV's Ciphertext-to-Ciphertext Multiplication}
\label{subsubsec:rns-cipher-mult-bfv}

BFV's ciphertext-to-ciphertext multiplication (Summary~\ref*{subsubsec:bfv-mult-cipher-summary} in \autoref{subsubsec:bfv-mult-cipher-summary}) comprises \textsf{ModRaise} $\rightarrow$ polynomial multiplication $\rightarrow$ relinearization $\rightarrow$ rescaling, where the order of relinearization and rescaling can be swapped. In RNS-based ciphertext-to-ciphertext multiplication, we will swap the order of these two steps. The procedure is as follows: (1) \textsf{ModRaise\textsubscript{RNS}} from $q \rightarrow qb$; (2) polynomial multiplication; (3) constant multiplication by $t$; (4) \textsf{ModSwitch} from $qb \rightarrow b$; (5)
\textsf{FastBConvEx} from $b \rightarrow q$; and (6) relinearization. Among these, step $3 \sim 5$ corresponds to the rescaling operation. We will explain how each of these steps works. %Throughout the explanation, we denote the polynomial constant to apply the RNS operations (i.e., the coefficients $\{a_0, \cdots, a_{n-1}\}$, $\{b_0, \cdots, b_{n-1}\}$ of polynomial $A, B$ of a ciphertext) as $x$.  

$ $

\begin{enumerate}

\item \textbf{\underline{\textsf{ModRaise\textsubscript{RNS}}} from $q \rightarrow qbb_\alpha$:}

Let $b$ be a new RNS base where $b > \Delta$ so that $qb$ is large enough to prevent a multiplied scaled plaintext in ciphertexts (i.e., $\Delta^2M^{\langle 1 \rangle}M^{\langle 2 \rangle}$) from exceeding its allowed limit (Summary~\ref*{subsec:bfv-enc-dec} in \autoref{subsec:bfv-enc-dec}) during ciphertext-to-ciphertext multiplication. $b_\alpha$ is also added for exact fast base conversion to be performed later. 
Specifically, we mod-raise the modulus of each polynomial coefficient of ciphertexts $(A^{\langle 1 \rangle}, B^{\langle 1 \rangle})$ and $(A^{\langle 2 \rangle}, B^{\langle 2 \rangle})$ as follows:

$(\hat{A}^{\langle 1 \rangle}, \hat{B}^{\langle 1 \rangle}) = (A^{\langle 1 \rangle} + U_A^{\langle 1 \rangle}q, B^{\langle 1 \rangle} + U_B^{\langle 1 \rangle}q) \bmod qbb_\alpha$

$(\hat{A}^{\langle 2 \rangle}, \hat{B}^{\langle 2 \rangle}) =(A^{\langle 2 \rangle} + U_A^{\langle 2 \rangle}q, B^{\langle 2 \rangle} + U_B^{\langle 2 \rangle}q) \bmod qbb_\alpha$

$ $

, where each coefficient of $U_A^{\langle 1 \rangle}, U_B^{\langle 1 \rangle}, U_A^{\langle 2 \rangle}, U_B^{\langle 2 \rangle}$ is either $\{-1, 0, 1\}$. Decrypting these two (noisy) ciphertexts with the private key $S$ would give the following outputs:

$\hat{A}^{\langle 1 \rangle}\cdot S + \hat{B}^{\langle 1 \rangle} \bmod qbb_\alpha$

$ = (A^{\langle 1 \rangle} + U_A^{\langle 1 \rangle}q) \cdot S + (B^{\langle 1 \rangle} + U_B^{\langle 1 \rangle}q) \bmod qbb_\alpha$

$ = A^{\langle 1 \rangle}\cdot S + B^{\langle 1 \rangle} + U_A^{\langle 1 \rangle}q \cdot S + U_B^{\langle 1 \rangle}q \bmod qbb_\alpha$ 

$ = \Delta M^{\langle 1 \rangle} + E^{\langle 1 \rangle} + U_A^{\langle 1 \rangle}q \cdot S + U_B^{\langle 1 \rangle}q  + K^{\langle 1 \rangle} q  \bmod qbb_\alpha$ \textcolor{red}{ $\rhd$ where $ + K^{\langle 1 \rangle} q$ is the $q$-overflows of the decryption process}

$ $

$ \hat{A}^{\langle 2 \rangle}\cdot S + \hat{B}^{\langle 2 \rangle}  \bmod qbb_\alpha$ 

$ = (A^{\langle 2 \rangle} + U_A^{\langle 2 \rangle}q) \cdot S + (B^{\langle 2 \rangle} + U_B^{\langle 2 \rangle}q) \bmod qbb_\alpha$

$ = A^{\langle 2 \rangle}\cdot S + B^{\langle 2 \rangle} + U_A^{\langle 2 \rangle}q \cdot S + U_B^{\langle 2 \rangle}q  \bmod qbb_\alpha$ 

$ = \Delta M^{\langle 2 \rangle} + E^{\langle 2 \rangle} + U_A^{\langle 2 \rangle}q \cdot S + U_B^{\langle 2 \rangle}q + K^{\langle 2 \rangle} q  \bmod qbb_\alpha$

$ $

\item \textbf{\underline{Polynomial Multiplication:}}

 Compute $(\hat{B}^{\langle 1 \rangle}\hat{B}^{\langle 2 \rangle}, \hat{A}^{\langle 1 \rangle}\hat{B}^{\langle 2 \rangle} + \hat{A}^{\langle 2 \rangle}B^{\langle 1 \rangle}, \hat{A}^{\langle 1 \rangle}\hat{A}^{\langle 2 \rangle}) \bmod qbb_\alpha$, whose decryption relation is as follows:

$\hat{B}^{\langle 1 \rangle}\hat{B}^{\langle 2 \rangle} + (\hat{A}^{\langle 1 \rangle}\hat{B}^{\langle 2 \rangle} + \hat{A}^{\langle 2 \rangle}B^{\langle 1 \rangle})\cdot S + (\hat{A}^{\langle 1 \rangle}\hat{A}^{\langle 2 \rangle})\cdot S^2  \bmod qbb_\alpha$

$ = (\hat{A}^{\langle 1 \rangle}\cdot S + \hat{B}^{\langle 1 \rangle}) \cdot (\hat{A}^{\langle 2 \rangle}\cdot S + \hat{B}^{\langle 2 \rangle})  \bmod qbb_\alpha$

$ = (\Delta M^{\langle 1 \rangle} + E^{\langle 1 \rangle} + U_A^{\langle 1 \rangle}q \cdot S + U_B^{\langle 1 \rangle}q + K^{\langle 1 \rangle} q) \cdot ( \Delta M^{\langle 2 \rangle} + E^{\langle 2 \rangle} + U_A^{\langle 2 \rangle}q \cdot S + U_B^{\langle 2 \rangle}q + K^{\langle 2 \rangle} q)  \bmod qbb_\alpha$

$ $

\item \textbf{\underline{Constant Multiplication} by $\bm t$:} 

Step $3 \sim 5$ are equivalent to rescaling the plaintext's scaling factor from $\Delta^2 \rightarrow \Delta$ as well as switching the ciphertext's modulus from $qbb_\alpha$ to $q$.
In this step, we multiply $t$ to each coefficient of the resulting polynomials from the previous step as follows: 

$(t\cdot\hat{B}^{\langle 1 \rangle}\hat{B}^{\langle 2 \rangle}, \text{ } t\cdot\hat{A}^{\langle 1 \rangle}\hat{B}^{\langle 2 \rangle} + t\cdot\hat{A}^{\langle 2 \rangle}B^{\langle 1 \rangle}, \text{ } t\cdot\hat{A}^{\langle 1 \rangle}\hat{A}^{\langle 2 \rangle}) \bmod qbb_\alpha$

$ $

, which is equivalent to a ciphertext encrypting the following plaintext:

$ t\cdot(\Delta M^{\langle 1 \rangle} + E^{\langle 1 \rangle} + U_A^{\langle 1 \rangle}q \cdot S + U_B^{\langle 1 \rangle}q  + K^{\langle 1 \rangle} q)\cdot(\Delta M^{\langle 2 \rangle} + E^{\langle 2 \rangle} + U_A^{\langle 2 \rangle}q \cdot S + U_B^{\langle 2 \rangle}q  + K^{\langle 2 \rangle} q) \bmod qbb_\alpha$

$ $

\item \textbf{\underline{\textsf{ModSwitch\textsubscript{RNS}}} from $qbb_\alpha \rightarrow bb_\alpha$:}

We switch the modulus of the ciphertext from $qb$ to $b$ by using  \textsf{ModSwitch\textsubscript{RNS}} as follows:

$\left(\left\lceil\dfrac{t\cdot\hat{B}^{\langle 1 \rangle}\hat{B}^{\langle 2 \rangle}}{q}\right\rfloor, \left\lceil\dfrac{t\cdot\hat{A}^{\langle 1 \rangle}\hat{B}^{\langle 2 \rangle} + t\cdot\hat{A}^{\langle 2 \rangle}B^{\langle 1 \rangle}}{q}\right\rfloor, \left\lceil\dfrac{t\cdot\hat{A}^{\langle 1 \rangle}\hat{A}^{\langle 2 \rangle}}{q}\right\rfloor\right) \bmod bb_\alpha$

$ $

, which is (almost, considering the rounding error) equivalent to a ciphertext encrypting the following plaintext:

$ \left\lceil\dfrac{t\cdot(\Delta M^{\langle 1 \rangle} + E^{\langle 1 \rangle} + U_A^{\langle 1 \rangle}q \cdot S + U_B^{\langle 1 \rangle}q  + K^{\langle 1 \rangle} q)\cdot(\Delta M^{\langle 2 \rangle} + E^{\langle 2 \rangle} + U_A^{\langle 2 \rangle}q \cdot S + U_B^{\langle 2 \rangle}q  + K^{\langle 2 \rangle} q)}{q}\right\rfloor \bmod bb_\alpha$

$ $

$= \left\lceil (M^{\langle 1 \rangle} + \dfrac{t}{q}\cdot E^{\langle 1 \rangle} + U_A^{\langle 1 \rangle}t \cdot S + U_B^{\langle 1 \rangle}t + K^{\langle 1 \rangle} t + \epsilon_d)\cdot(\Delta M^{\langle 2 \rangle} + E^{\langle 2 \rangle} + U_A^{\langle 2 \rangle}q \cdot S + U_B^{\langle 2 \rangle}q + K^{\langle 2 \rangle} q)\right\rfloor \bmod bb_\alpha$ 

\textcolor{red}{ $\rhd$ where $\epsilon_d = \dfrac{t}{q}\cdot \Delta M^{\langle 1 \rangle} - M^{\langle 1 \rangle}$ is a rounding error caused by treating $\dfrac{q}{t} \approx \left\lfloor\dfrac{q}{t}\right\rfloor = \Delta$}

$ $

\item \textbf{\underline{\textsf{FastBConvEx\textsubscript{RNS}}} from $b \rightarrow q$:} 

We exactly convert the base of the ciphertext from $bb_\alpha \rightarrow q$ as follows:

$\left(\left\lceil\dfrac{t\cdot\hat{B}^{\langle 1 \rangle}\hat{B}^{\langle 2 \rangle}}{q}\right\rfloor, \left\lceil\dfrac{t\cdot\hat{A}^{\langle 1 \rangle}\hat{B}^{\langle 2 \rangle} + t\cdot\hat{A}^{\langle 2 \rangle}B^{\langle 1 \rangle}}{q}\right\rfloor, \left\lceil\dfrac{t\cdot\hat{A}^{\langle 1 \rangle}\hat{A}^{\langle 2 \rangle}}{q}\right\rfloor\right) \bmod q$

$ $

, which is equivalent to a ciphertext encrypting the following plaintext:

$= \left\lceil (M^{\langle 1 \rangle} + \dfrac{t}{q}\cdot E^{\langle 1 \rangle} + U_A^{\langle 1 \rangle}t \cdot S + U_B^{\langle 1 \rangle}t + K^{\langle 1 \rangle} t + \epsilon_d)\cdot(\Delta M^{\langle 2 \rangle} + E^{\langle 2 \rangle} + U_A^{\langle 2 \rangle}q \cdot S + U_B^{\langle 2 \rangle}q + K^{\langle 2 \rangle} q)\right\rfloor \bmod q$

$ = \Big\lceil \Delta M^{\langle 1 \rangle} M^{\langle 2 \rangle} +  \dfrac{t}{q}\cdot \Delta M^{\langle 2 \rangle} E^{\langle 1 \rangle} + U_A^{\langle 1 \rangle}t \Delta M^{\langle 2 \rangle} \cdot S + U_B^{\langle 1 \rangle}t\Delta M^{\langle 2\rangle}  + M^{\langle 1 \rangle} E^{\langle 2 \rangle} + \dfrac{t}{q}\cdot E^{\langle 1 \rangle} E^{\langle 2 \rangle} $

\text{ } \text{ } $+ U_A^{\langle 1 \rangle}  E^{\langle 2 \rangle}t\cdot S + U_B^{\langle 1 \rangle} E^{\langle 2 \rangle}t + K^{\langle 1 \rangle}t\Delta M^{\langle 2 \rangle} + K^{\langle 1 \rangle}t E^{\langle 2 \rangle} 
+ \epsilon_d\cdot(\Delta M^{\langle 2 \rangle} + E^{\langle 2 \rangle} + U_A^{\langle 2 \rangle}q \cdot S + U_B^{\langle 2 \rangle}q + K^{\langle 2 \rangle} q)\Big\rfloor \bmod q $ 

$ \approx \Delta M^{\langle 1 \rangle} M^{\langle 2 \rangle} \bmod q $ \textcolor{red}{ $\rhd$ all other terms are relatively much smaller than $\Delta M^{\langle 1 \rangle} M^{\langle 2 \rangle}$ in modulo $q$}

$ $

\item \textbf{\underline{Relinearization}:} 

Once we have derived the rescaled polynomial triple $(D'_0, D'_1, D'_2) \bmod q$, the final relinearization step is equivalent to deriving the synthetic ciphertexts $\textsf{ct}_\alpha$ and $\textsf{ct}_\beta$ and then computing $\textsf{ct}_\alpha + \textsf{ct}_\beta$. $\textsf{ct}_\alpha$ is simply $(D'_0, D'_1)$, and we can derive $\textsf{ct}_\beta = \textsf{RLWE}_{S, \sigma}(D_2 \cdot S^2)$ by using the \textsf{DecompMult\textsubscript{RNS}} operation (Summary~\ref*{subsec:rns-decompmult} in \autoref{subsec:rns-decompmult}). The final ciphertext-to-ciphertext addition of $\textsf{ct}_\alpha + \textsf{ct}_\beta$ can be performed by using regular RNS addition.

\end{enumerate}

\subsubsection{CKKS's Ciphertext-to-Ciphertext Multiplication}
\label{subsubsec:rns-cipher-mult-ckks}

CKKS's ciphertext-to-ciphertext multiplication (Summary~\ref*{subsubsec:ckks-mult-cipher-summary} in \autoref{subsubsec:ckks-mult-cipher-summary}) is almost the same as BFV's, except that CKKS does not need the \textsf{ModRaise} operation in the beginning (because each multiplicative level's modulus $q_l$ is large enough to hold a multiplied scaled plaintext $\Delta^2M^{\langle 1 \rangle}M^{\langle 2 \rangle}$). Therefore, CKKS's RNS-based multiplication is the same as BFV's except that it does not require step 1 (\textsf{ModRaise}), step 3 (constant multiplication by $t$), and step 5 (exact fast base conversion). Since a CKKS ciphertext's scaling factor $\Delta$ is approximately the same as the prime modulus factor of each multiplicative level, each ciphertext-to-ciphertext multiplication only needs to perform a 
modulus switch to a lower level.
%\textsf{ModSwitch\textsubscript{RNS}} to the lower level.

\subsubsection{BGV's Ciphertext-to-Ciphertext Multiplication}
\label{subsubsec:rns-cipher-mult-bgv}

BGV's ciphertext-to-ciphertext multiplication (Summary~\ref*{subsec:bgv-mult-cipher} in \autoref{subsec:bgv-mult-cipher}) is almost the same as CKKS's, except that BGV uses its own modulus switch ($\textsf{ModSwitch}_{\textsf{RNS}}^{\textsf{BGV}}$ as described in Summary~\ref*{subsec:bgv-modulus-switch} in \autoref{subsec:bgv-modulus-switch}) during the rescaling step. Therefore, BGV's RNS-based ciphertext-to-ciphertext multiplication is the same as CKKS's, except that \textsf{ModSwitch\textsubscript{RNS}} is replaced by $\textsf{ModSwitch}_{\textsf{RNS}}^{\textsf{BGV}}$.

\subsubsection{BFV's Bootstrapping}
\label{subsubsec:rns-bfv-bootstrapping}

BFV's original bootstrapping procedure (Summary~\ref*{subsubsec:bfv-bootstrapping-summary} in \autoref{subsubsec:bfv-bootstrapping-summary}) is as follows: (1) modulus switch from $q \rightarrow p^\varepsilon$; (2) homomorphic decryption; (3) \textsf{CoeffToSlot}; (4) \textsf{EvalExp}; (5) \textsf{SlotToCoeff}; and (6) re-interpretation. 

However, in RNS, we cannot mod-switch to $p^\varepsilon$ because RNS's base moduli have to be co-prime to each other, whereas the factors of $p^\varepsilon$ are not. To avoid this issue, RNS-based BFV's bootstrapping instead performs the following: (1) \textsf{ModRaise\textsubscript{RNS}} from $q \rightarrow qbb_\alpha$, where $bb_\alpha$ is an auxiliary base; (2) coefficient multiplication by $p^\varepsilon$; (3) \textsf{ModSwitch\textsubscript{RNS}} from $qbb_\alpha \rightarrow bb_\alpha$; (4) \textsf{FastBConvEx} from $bb_\alpha \rightarrow q$; (5) homomorphic decryption to adjust the scaling factor of the plaintext; (6) \textsf{CoeffToSlot}; (7) \textsf{EvalExp}; (8) \textsf{SlotToCoeff}; and (9) re-interpretation. The detailed procedure is described as follows:

$ $

\textbf{\underline{Input}:} The input BFV ciphertext to bootstrap is $(A, B) \bmod q$, which would decrypt to:  

$A\cdot S + B = \Delta M + E + Kq$ \textcolor{red}{ $\rhd$ where $\Delta = \dfrac{q}{p^r}$}

$ $

\begin{enumerate}

\item \textbf{\underline{\textsf{ModRaise\textsubscript{RNS}}} from $q \rightarrow qbb_\alpha$: } 

Mod-raise ciphertext $(A, B) \bmod q$ to $(A, B) \bmod qbb_\alpha$, which would decrypt to:  

$A\cdot S + B = \Delta M + E + Kq + Uq \pmod{ qbb_\alpha}$ 

\textcolor{red}{ $\rhd$ where $Uq$ is the \textsf{FastBConv} + \textsf{SmallMont} error, and $U$'s coefficients are either $\{-1, 0, 1\}$}

$ $

\item \textbf{\underline{Coefficient Multiplication} by $p^\varepsilon$: } 

Multiply the coefficients of $(A, B) \bmod q$ by $p^\varepsilon$ to update the ciphertext to  $(p^\varepsilon A,p^\varepsilon B) \bmod{ qbb_\alpha}$, which would decrypt to: 

$p^\varepsilon A\cdot S + p^\varepsilon B = \Delta p^\varepsilon M + p^\varepsilon E + p^\varepsilon Kq + p^\varepsilon Uq \pmod{qbb_\alpha}$

$ $

\item \textbf{\underline{\textsf{ModSwitch\textsubscript{RNS}}} from $qbb_\alpha \rightarrow bb_\alpha$: } 

Mod-switch the ciphertext $(p^\varepsilon A,p^\varepsilon B) \bmod pbb_\alpha$ to $\left(\left\lceil\dfrac{p^\varepsilon A}{q}\right\rfloor,\left\lceil\dfrac{p^\varepsilon B}{q}\right\rfloor\right) \pmod{bb_\alpha}$, which would decrypt to: 

$\left\lceil\dfrac{p^\varepsilon A}{q}\right\rfloor\cdot S + \left\lceil\dfrac{p^\varepsilon B}{q}\right\rfloor = \dfrac{\Delta p^\varepsilon M}{q} + \dfrac{p^\varepsilon E}{q} + \dfrac{p^\varepsilon Kq}{q} + \dfrac{p^\varepsilon Uq}{q} + \epsilon  \pmod {bb_\alpha}$ 

\textcolor{red}{ $\rhd$ $\epsilon$ is a small rounding error}

$ = p^{\varepsilon - r}M + \dfrac{p^\varepsilon E}{q} + p^\varepsilon K + p^\varepsilon U +\epsilon \pmod{ bb_\alpha}$ 

$ $

\item \textbf{\underline{\textsf{FastBConvEx}} from $bb_\alpha \rightarrow q$: } 

Exact fast base conversion of $(p^\varepsilon A, p^\varepsilon B) \bmod bb_\alpha$ to $(p^\varepsilon A, p^\varepsilon B) \bmod q$, which would decrypt to:

$p^{\varepsilon - r}M + \dfrac{p^\varepsilon E}{q} + p^\varepsilon K + p^\varepsilon U +\epsilon \bmod q$

$ $

\item \textbf{\underline{Homomorphic Decryption}:} 

Now, we have the ciphertext $(p^\varepsilon A, p^\varepsilon B) \bmod q = \textsf{RLWE}_{S, \sigma}\left(p^{\varepsilon - r}M + \dfrac{p^\varepsilon E}{q} + p^\varepsilon K + p^\varepsilon U +\epsilon\right) \bmod q$. We do homomorphic decryption by using the encrypted private key $\textsf{RLWE}_{S, \sigma}(\hat\Delta S)$, where $\hat\Delta = \left\lfloor\dfrac{q}{p^\varepsilon}\right\rfloor$. The output is $\textsf{RLWE}_{S, \sigma}\bm(\hat\Delta\cdot(p^{\varepsilon - r}M + \dfrac{p^\varepsilon E}{q} + p^\varepsilon K + p^\varepsilon U +\epsilon)\bm) \bmod q$. 

$ $

\item Perform \textsf{CoeffToSlot}, digit extraction, and \textsf{SlotToCoeff}. These operations can be performed by only regular RNS-based additions and multiplications. The final digit-extracted ciphertext is $\textsf{RLWE}_{S, \sigma}\bm(\hat\Delta\cdot(p^{\varepsilon - r}\cdot M +  Kp^\varepsilon +  Up^\varepsilon)\bm)$, where all noise values smaller than the (base-$p$) $(\varepsilon-r)$-th digits are eliminated. 

\begin{comment}
$ $

\item \textbf{\underline{Scaling Factor Re-interpretation:}} 

Theoretically re-interpret the ciphertext without any additional mathematical computation. The ciphertext $\textsf{RLWE}_{S, \sigma}\bm(\hat\Delta\cdot(p^{\varepsilon - r}\cdot M +  Kp^\varepsilon +  Up^\varepsilon)\bm)$ is mathematically the same as: 

$\textsf{RLWE}_{S, \sigma}\bm(\hat\Delta\cdot(p^{\varepsilon - r}\cdot M +  Kp^\varepsilon +  Up^\varepsilon)\bm) $

$= \textsf{RLWE}_{S, \sigma}\bm(\Delta M + (K + U)\cdot q\bm) $ \textcolor{red}{ $\rhd$ since $\Delta = \dfrac{q}{p^r}$, and $\hat\Delta = \dfrac{a}{p^\varepsilon}$}

$= \textsf{RLWE}_{S, \sigma}\bm(\Delta M\bm) \bmod q$.
\end{comment}

\end{enumerate}

\subsubsection{CKKS's Bootstrapping}
\label{subsubsec:rns-ckks-bootstrapping}

CKKS's original bootstrapping procedure (Summary~\ref*{subsubsec:ckks-bootstrapping-summary} in \autoref{subsubsec:ckks-bootstrapping-summary}) is as follows:
(1) \textsf{Modraise}; (2) homomorphic decryption; (3) \textsf{CoeffToSlot}; (4) \textsf{EvalExp}; (5) \textsf{SlotToCoeff}; (6) Re-interpretation. In the RNS-based CKKS bootstrapping, we perform \textsf{ModRaise\textsubscript{RNS}} at step 1, and all other steps are computed using regular RNS-based addition and multiplication operations. Step 1's \textsf{ModRaise\textsubscript{RNS}} operation generates a 
$u\cdot q_0$ noise (where $u \in \{-1, 0, 1\}$ using \textsf{SmallMont}) for each polynomial coefficient during \textsf{FastBConv}. Therefore, step 2's homomorphic decryption outputs $\Delta M + E + W\cdot q_0 + K\cdot q_0$, where $W\cdot q_0$ represents the aggregation of all coefficient noise terms that are multiplied by the $q_0$-overflow noises generated by \textsf{FastBConv} and \textsf{SmallMont}. The $W\cdot q_0 + K\cdot q_0$ term is eliminated by step 4's \textsf{EvalExp}, which performs approximated modulo reduction based on a sine-graph evaluation whose period is $q_0$.

\subsubsection{BGV's Bootstrapping}
\label{subsubsec:rns-bgv-bootstrapping}

BGV's original bootstrapping procedure (\autoref{subsec:bgv-bootstrapping}) is as follows:
(1) modulus switch from $q_l \rightarrow \hat{q}$; (2) ciphertext coefficient multiplication by $p^{\varepsilon - 1}$; (3) \textsf{ModRaise}; (4) \textsf{CoeffToSlot}; (5) digit extraction; (6) homomorphic multiplication by $|p^{-(\varepsilon-1)}|_{q^{l'}}$; (7) \textsf{SlotToCoeff}; (8) noise term re-interpretation. Given this procedure, the RNS-based bootstrapping steps are as follows.

$ $

Suppose the target BGV ciphertext to bootstrap is $(A, B) \bmod q_l$, where the plaintext modulus (i.e., noise scaling factor) is $p$. 

\begin{enumerate}

\item \textbf{\underline{$\textsf{ModSwitch}_{\textsf{RNS}}^{\textsf{BGV}}$}} from $q_l \rightarrow \hat{q}$ where $\hat{q}$ is a special modulus satisfying the following requirements: $\hat{q} \equiv 1 \bmod p^\varepsilon$,  $\hat{q}$ is a prime and is co-prime with $q_l$, and $p^{\varepsilon} < \hat{q} < q_l$.

$ $
 
\item \textbf{\underline{Constant multiplication}} by $p^{\varepsilon-1}$ to the coefficients of the ciphertext polynomials $(\hat{A}, \hat{B})$, which increases the underlying plaintext's noise scaling factor $\Delta$ and the plaintext modulus from $p \rightarrow p^{\varepsilon}$. This effectively updates the underlying plaintext to $p^{\varepsilon-1} M + p^{\varepsilon}E$. 

$ $

\item \textbf{\underline{\textsf{ModRaise\textsubscript{RNS}}}} from $\hat{q} \rightarrow q_L$, which generates additional noise $|u\cdot \hat{q}|_{q_L}$ (where $u \in \{-1, 0, 1\}$ using \textsf{SmallMont}). At this point, the ciphertext is $\textsf{RLWE}_{S, \sigma}(p^{\varepsilon-1} M + p^{\varepsilon}E + \hat{q}K) \bmod q_L$, whose underlying plaintext is:

$p^{\varepsilon-1} M + p^{\varepsilon}E + \hat{q}K$

$= p^{\varepsilon-1} M + K \bmod p^{\varepsilon}$ \textcolor{red}{ $\rhd$ since $\hat{q} \equiv 1 \bmod p^{\varepsilon}$}

\end{enumerate}

$ $

After the above steps, the remaining steps (i.e., \textsf{CoeffToSlot}, \textsf{EvalExp}, homomorphic multiplication by $|p^{-(\varepsilon-1)}|_{q^{l'}}$, \textsf{SlotToCoeff}, and re-interpretation) can be performed using regular RNS-based addition and multiplication operations.

\subsubsection{Noise Impact of RNS Operations}
\label{subsubsec:rns-noise}

When RNS techniques are used in FHE operations, the noise generated by \textsf{FastBConv}, \textsf{ModRaise\textsubscript{RNS}}, and \textsf{ModSwitch\textsubscript{RNS}} is directly added to each coefficient of the ciphertext polynomials $A$ and $B$. Since the decryption relation is $A\cdot S + B$, even the noise added to the coefficients of the polynomial $A$ is multiplied by a large factor due to the polynomial multiplication with $S$. Therefore, it is important to always ensure that the generated noise of each \textsf{FastBConvEx\textsubscript{RNS}} is reduced by using it with \textsf{SmallMont}. 

\subsubsection{Python Source Code of RNS Primitives}
\label{subsubsec:rns-source-code}

We provide a \href{https://github.com/fhetextbook/fhe-textbook/blob/main/source%20code/rns.py}{\underline{Python script}} implementing the following exemplary RNS primitives: \textsf{FastBConv}, 
\textsf{ModRaise\textsubscript{RNS}}, 
\textsf{ModDrop\textsubscript{RNS}}, 
\textsf{ModSwitch\textsubscript{RNS}},
\textsf{SmallMont}, and 
\textsf{FastBConvEx}.

\clearpage

\section{FHE Scheme Comparison and Summary}
%\subsection{Scheme Comparison}
%\label{subsec:scheme-comparison}

We summarize and compare TFHE, BFV, CKKS, and BGV as follows:

\begin{comment}
\clearpage

\end{comment}

\captionsetup{justification=raggedright,singlelinecheck=false,labelfont=bf}

\begin{table}[h]
\begin{tabular}{|c||l|}
\hline
&\textbf{Hard Problem Basis}\\\hline\hline
\textbf{TFHE}&LWE\\\hline
\textbf{BFV}&\\
\textbf{CKKS}&RLWE\\
\textbf{BGV}&\\\hline
\end{tabular}
\caption{\textbf{Hard Problem Basis}}
\end{table}

\begin{table}[h]
\begin{tabular}{|c||l|}
\hline
&\textbf{Unit Data Type}\\\hline\hline
\textbf{TFHE}&Vector\\\hline
\textbf{BFV}&\\
\textbf{CKKS}&Polynomial\\
\textbf{BGV}&\\\hline
\end{tabular}
\caption{\textbf{Unit Data Type}}
\end{table}

\begin{table}[h]
\begin{tabular}{|c||l|}
\hline
&\multicolumn{1}{c|}{\textbf{Plaintext}}\\\hline\hline
\textbf{TFHE}&Number $m \in \mathbb{Z}_t$ \text{ } \textcolor{red}{ $\rhd$ $t$ is a power of 2}\\\hline
\textbf{BFV}&Polynomial $M \in \mathbb{Z}_t[X]/X^n+1$ \text{ } \textcolor{red}{ $\rhd$ $t$ is a prime, and $n$ is a power of 2}\\\hline
\textbf{CKKS}&Polynomial $M \in \mathbb{R}[X]/X^n+1$ \text{ }\textcolor{red}{ $\rhd$ $n$ is a power of 2}\\\hline
\textbf{BGV}&Polynomial $M \in \mathbb{Z}_t[X]/X^n+1$ \text{ } \textcolor{red}{ $\rhd$ $t$ is a prime, and $n$ is a power of 2}\\\hline
\end{tabular}
\caption{\textbf{Plaintext}}
\end{table}

\begin{table}[h]
\begin{tabular}{|c||l|}
\hline
&\multicolumn{1}{c|}{\textbf{Secret Key}}\\\hline\hline
\textbf{TFHE}&Vector $\vec{s} \xleftarrow{\$} \mathbb{Z}_2^k$ sampled from $\{0, 1\}$ \textcolor{red}{ $\rhd$ $\$$ is a uniform random distribution}\\\hline
\textbf{BFV}&\\
\textbf{CKKS}&Polynomial $S \xleftarrow{\$} \mathbb{Z}_3[X]/X^n+1$, where $\mathbb{Z}_3 = \{-1, 0, 1\}$\\
\textbf{BGV}&\\\hline
\end{tabular}
\caption{\textbf{Secret Key}}
\end{table}

\begin{table}[h]
\begin{tabular}{|c||l|}
\hline
&\multicolumn{1}{c|}{\textbf{Ciphertext}}\\\hline\hline
\textbf{TFHE}&(Vector $\vec{a}$, Number $b$) = $(\vec{a} \xleftarrow{\$} \mathbb{Z}_q^k$, $\text{ } b \in \mathbb{Z}_q)$ \text{ } \textcolor{red}{ $\rhd$ $q \gg t$, and $t$ divides $q$}\\\hline
\textbf{BFV}&\\
\textbf{CKKS}&(Polynomial $A, B$) = $(A \xleftarrow{\$} \mathbb{Z}_q[X]/X^n+1$, \text{ }  $B \in \mathbb{Z}_q[X]/X^n+1)$ \text{ } \textcolor{red}{ $\rhd$ $q \gg t$}\\
\textbf{BGV}&\\\hline
\end{tabular}
\caption{\textbf{Ciphertext}}
\end{table}

\begin{table}[h]
\begin{tabular}{|c||l|}
\hline
&\multicolumn{1}{c|}{\textbf{Noise}}\\\hline\hline
\textbf{TFHE}&Number $e \xleftarrow{\chi} \mathbb{Z}_q$ \text{ } \textcolor{red}{ $\rhd$ $\chi$ is a Gaussian random distribution}\\\hline
\textbf{BFV}&\\
\textbf{CKKS}&Polynomial $E \xleftarrow{\chi} \mathbb{Z}_q[X]/X^n+1$\\
\textbf{BGV}&\\\hline
\end{tabular}
\caption{\textbf{Noise}}
\end{table}

\begin{table}[h]
\begin{tabular}{|c||l|}
\hline
&\multicolumn{1}{c|}{\textbf{Scaling Factor}}\\\hline\hline
\textbf{TFHE}&Used for $\Delta m$, \text{ } where $\Delta = \dfrac{q}{t}$ \text{ } \textcolor{red}{ $\rhd$ $t$ divides $q$}\\\hline
\textbf{BFV}&Used for $\Delta M$, \text{ } where $\Delta = \left\lfloor\dfrac{q}{t}\right\rfloor$  \text{ } \textcolor{red}{ $\rhd$ $t$ is a prime}\\\hline
\textbf{CKKS}&Used for $\Delta M$, \text{ } where $\Delta \cdot ||M||_\infty \ll q_0$\\
& \text{ } \textcolor{red}{ $\rhd$ $q_0$ is the lowest multiplicative level's ciphertext modulus}\\\hline
\textbf{BGV}&Used for $\Delta E$, \text{ } where $\Delta = t$ \text{ } \textcolor{red}{ $\rhd$$t$ is a prime}\\\hline
\end{tabular}
\caption{\textbf{Scaling Factor}}
\end{table}

\begin{table}[h]
\begin{tabular}{|c||l|}
\hline
&\multicolumn{1}{c|}{\textbf{Encryption}}\\\hline\hline
\textbf{TFHE}&$(\vec{a}, b)$ \text{ } where $\vec{a} \xleftarrow{\$} \mathbb{Z}_q^k$, $\text{ } b = \Delta m + e - a\cdot s \bmod q$, \text{ } $e \xleftarrow{\chi} \mathbb{Z}_q$\\
&\text{ } \text{ } \textcolor{red}{ $\rhd$ After using $e$ each time, throw it away}\\\hline
\textbf{BFV}&$(A, B)$ \text{ } where $A \xleftarrow{\$} \mathbb{Z}_q[X] / (X^n + 1)$, $\text{ } B = \Delta M + E - A\cdot S \bmod q$, \text{ } $E \xleftarrow{\chi} \mathbb{Z}_q[X]/(X^n+1)$\\
\textbf{CKKS}&\text{ } \text{ }\textcolor{red}{ $\rhd$ After using $E$ each time, throw it away}\\\hline
\textbf{BGV}&$(A, B)$ \text{ } where $A \xleftarrow{\$} \mathbb{Z}_q[X] / (X^n + 1)$, $\text{ } B = M + \Delta E - A\cdot S \bmod q$, \text{ } $E \xleftarrow{\chi} \mathbb{Z}_q[X]/(X^n+1)$\\
&\text{ } \text{ }\textcolor{red}{ $\rhd$ After using $E$ each time, throw it away}\\\hline
\end{tabular}
\caption{\textbf{Encryption}}
\end{table}

\begin{table}[h]
\begin{tabular}{|c||l|}
\hline
&\multicolumn{1}{c|}{\textbf{Cryptographic Relation}}\\\hline\hline
\textbf{TFHE}&$b + a\cdot s = \Delta m + e \pmod q$, \text{ } where $\Delta = \dfrac{q}{t}$  \text{ } \textcolor{red}{ $\rhd$ $t$ divides $q$}\\\hline
\textbf{BFV}&$B + A\cdot S = \Delta M + E \pmod q$, \text{ } where $\Delta = \left\lfloor\dfrac{q}{t}\right\rfloor$ \text{ } \textcolor{red}{ $\rhd$ $t$ is a prime}\\\hline
\textbf{CKKS}&$B + A\cdot S = \Delta M + E \pmod q$, \text{ } where $\Delta \cdot ||M||_\infty \ll q_0$\\
& \text{ } \textcolor{red}{ $\rhd$ $q_0$ is the lowest multiplicative level's ciphertext modulus}\\\hline
\textbf{BGV}&$B + A\cdot S = M + \Delta E \pmod q$, \text{ } where $\Delta = t$  \text{ } \textcolor{red}{ $\rhd$ $t$ is a prime}\\\hline
\end{tabular}
\caption{\textbf{Cryptographic Relation}}
\end{table}

\begin{table}[h]
\begin{tabular}{|c||l|}
\hline
&\multicolumn{1}{c|}{\textbf{Decryption Formula}}\\\hline\hline
\textbf{TFHE}&$m = \left\lceil\dfrac{(b + a\cdot s \bmod q)}{\Delta}\right\rfloor \bmod t$\\
&$\textcolor{white}{m} = \left\lceil\dfrac{(\Delta m + e)}{\Delta}\right\rfloor \bmod t$ \text{ } \textcolor{red}{ $\rhd$ $e$ gets eliminated if $e < \dfrac{\Delta}{2}$}\\\hline
\textbf{BFV}&$M = \left\lceil\dfrac{(B + A\cdot S \bmod q)}{\Delta}\right\rfloor  \bmod t$\\
&$\textcolor{white}{M} = \left\lceil\dfrac{(\Delta M + E)}{\Delta}\right\rfloor  \bmod t$ \text{ } \textcolor{red}{ $\rhd$ $E$ gets eliminated if $||E||_\infty < \dfrac{\Delta}{2}$}\\\hline
\textbf{CKKS}&$M = \left\lceil\dfrac{(B + A\cdot S \bmod q)}{\Delta}\right\rfloor_{\frac{1}{\Delta}}$\\
&$\textcolor{white}{M} = \left\lceil\dfrac{\Delta M + E}{\Delta}\right\rfloor_{\frac{1}{\Delta}}$ \text{ } \textcolor{red}{ $\rhd$ The final noise remains as $\dfrac{E}{\Delta}$ (increase $\Delta$ to reduce it)}\\\hline
\textbf{BGV}&$M = (B + A\cdot S \bmod q) \bmod t$\\
&$\textcolor{white}{M} = (M + \Delta E) \bmod t$ \text{ } \textcolor{red}{ $\rhd$ $E$ gets removed if $\Delta E < q$}\\\hline
\end{tabular}
\caption{\textbf{Decryption Formula}}
\end{table}

\begin{table}[h]
\begin{tabular}{|c||l|}
\hline
&\multicolumn{1}{c|}{\textbf{Ciphertext Modulus}}\\\hline\hline
\textbf{TFHE}&A single number $q$ \text{ } \textcolor{red}{ $\rhd$ $q \gg t$, and $t$ divides $q$}\\\hline
\textbf{BFV}&A single number $q$ \text{ } \textcolor{red}{ $\rhd$ $q \gg t$}\\\hline
\textbf{CKKS}&An $L$-multiplicative-level modulus chain $\{q_0, q_1, \cdots, q_L\}$\\
& \text{ } \textcolor{red}{ $\rhd$ each $q_i = \prod\limits_{j=0}^{l}w_j$, and each $w_j$ is a CRT modulus}\\
&\text{ } \text{ } \text{ } \textcolor{red}{ having the property: $w_0 \gg \Delta \cdot ||M||_{\infty}, \text{ } w_j \approx \Delta$ (for $1 \leq j \leq L$)}\\\hline
\textbf{BGV}&An $L$-multiplicative-level modulus chain $\{q_0, q_1, \cdots, q_L\}$\\
&\text{ } \textcolor{red}{ $\rhd$ each $q_i = \prod\limits_{j=0}^{l}w_j$, and each $w_j$ is a CRT modulus}\\
&\text{ } \text{ } \text{ } \textcolor{red}{ having the property: $w_0 \equiv w_1 \equiv \cdots \equiv w_L \bmod t$}\\\hline
\end{tabular}
\caption{\textbf{Ciphertext Modulus}}
\end{table}

\begin{table}[h]
\begin{tabular}{|c||l|}
\hline
&\multicolumn{1}{c|}{\textbf{Ciphertext-to-Ciphertext Addition}}\\\hline\hline
\textbf{TFHE}&- Ciphertext $\textsf{LWE}_{\vec{s}, \sigma}(\Delta m_1) = (\vec{a}_1, b_1) = (a_{1,0}, a_{1, 1}, \cdots, a_{1, k-1}, b_1) \bmod q$\\
&- Ciphertext $\textsf{LWE}_{\vec{s}, \sigma}(\Delta m_2) = (\vec{a}_2, b_2) = (a_{2,0}, a_{2, 1}, \cdots, a_{2, k-1}, b_2) \bmod q$\\
&$\textsf{LWE}_{\vec{s}, \sigma}(\Delta (m_1 + m_2)) = (\vec{a}_1+\vec{a}_2, b_1+b_2) \bmod q$\\\hline
\textbf{BFV}&- Ciphertext $\textsf{RLWE}_{S, \sigma}(\Delta M_1) = (A_1, B_1) \bmod q$\\
&- Ciphertext $\textsf{RLWE}_{S, \sigma}(\Delta M_2) = (A_2, B_2) \bmod q$\\
&$\textsf{RLWE}_{S, \sigma}(\Delta (M_1 + M_2)) =(A_1+A_2,B_1+B_2) \bmod q$\\\hline
\textbf{CKKS}&- Ciphertext $\textsf{RLWE}_{S, \sigma}(\Delta M_1) = (A_1, B_1) \bmod q_l$\\
&- Ciphertext $\textsf{RLWE}_{S, \sigma}(\Delta M_2) = (A_2, B_2) \bmod q_l$\\
&$\textsf{RLWE}_{S, \sigma}(\Delta (M_1 + M_2)) =(A_1+A_2,B_1+B_2) \bmod q_l$\\\hline
\textbf{BGV}&- Ciphertext $\textsf{RLWE}_{S, \sigma}(M_1) = (A_1, B_1) \bmod q_l$\\
&- Ciphertext $\textsf{RLWE}_{S, \sigma}(M_2) = (A_2, B_2) \bmod q_l$\\
&$\textsf{RLWE}_{S, \sigma}(M_1 + M_2) =(A_1+A_2,B_1+B_2) \bmod q_l$\\\hline
\end{tabular}
\caption{\textbf{Ciphertext-to-Ciphertext Addition}}
\end{table}

\begin{table}[h]
\begin{tabular}{|c||l|}
\hline
&\multicolumn{1}{c|}{\textbf{Ciphertext-to-Plaintext Addition}}\\\hline\hline
\textbf{TFHE}&- Ciphertext $\textsf{LWE}_{\vec{s}, \sigma}(\Delta m_1) = (\vec{a}_1, b_1) = (a_{1,0}, a_{1, 1}, \cdots, a_{1, k-1}, b_1) \bmod q$\\
&- Plaintext number $c \in \mathbb{Z}_t$\\
&$\textsf{LWE}_{\vec{s}, \sigma}(\Delta (m_1 + c)) = (\vec{a}_1, b_1+\Delta c) \bmod q$\\\hline
\textbf{BFV}&- Ciphertext $\textsf{RLWE}_{S, \sigma}(\Delta M_1) = (A_1, B_1) \bmod q$\\
&- Plaintext polynomial $C \in \mathbb{Z}_t[X]/(X^n + 1)$\\
&$\textsf{RLWE}_{S, \sigma}(\Delta (M_1 + C)) =(A_1,B_1+\Delta C) \bmod q$\\\hline
\textbf{CKKS}&- Ciphertext $\textsf{RLWE}_{S, \sigma}(\Delta M_1) = (A_1, B_1)  \bmod q_l$\\
&- Plaintext polynomial $C \in \mathbb{R}[X]/(X^n + 1)$\\
&$\textsf{RLWE}_{S, \sigma}(\Delta (M_1 + C)) =(A_1,B_1+\Delta C) \bmod q_l$\\\hline
\textbf{BGV}&- Ciphertext $\textsf{RLWE}_{S, \sigma}(M_1) = (A_1, B_1) \bmod q_l$\\
&- Plaintext polynomial $C \in \mathbb{Z}_t[X]/(X^n + 1)$\\
&$\textsf{RLWE}_{S, \sigma}(M_1 + C) =(A_1,B_1+C) \bmod q_l$\\\hline
\end{tabular}
\caption{\textbf{Ciphertext-to-Plaintext Addition}}
\end{table}

\begin{table}[h]
\begin{tabular}{|c||l|}
\hline
&\multicolumn{1}{c|}{\textbf{Ciphertext-to-Plaintext Multiplication}}\\\hline\hline
\textbf{TFHE}&- Ciphertext $\textsf{LWE}_{\vec{s}, \sigma}(\Delta m_1) = (\vec{a}_1, b_1) = (a_{1,0}, a_{1, 1}, \cdots, a_{1, k-1}, b_1) \bmod q$\\
&- Plaintext number $c \in \mathbb{Z}_t$\\
&$\textsf{LWE}_{\vec{s}, \sigma}(\Delta (m_1 \cdot c)) = (\vec{a}_1 \cdot c, \text{ } b_1 \cdot c) \bmod q$\\\hline
\textbf{BFV}&- Ciphertext $\textsf{RLWE}_{S, \sigma}(\Delta M_1) = (A_1, B_1) \bmod q$\\
&- Plaintext polynomial $C \in \mathbb{Z}_t[X]/(X^n + 1)$\\
&$\textsf{RLWE}_{S, \sigma}(\Delta (M_1 \cdot C)) =(A_1 \cdot C, \text{ } B_1 \cdot C)$\\\hline
\textbf{CKKS}&- Ciphertext $\textsf{RLWE}_{S, \sigma}(\Delta M_1) = (A_1, B_1)  \bmod q_l$\\
&- Plaintext polynomial $C \in \mathbb{R}[X]/(X^n + 1)$\\
&1. \underline{Basic Multiplication}\\
&\textcolor{white}{1. } $\textsf{ct} = \textsf{RLWE}_{S, \sigma}(\Delta^2 (M_1 \cdot C)) =(A_1 \cdot \Delta C, \text{ } B_1 \cdot \Delta  C)  \bmod q_l$\\
&2. \underline{Rescaling} by $\dfrac{1}{\Delta}$: $\left\lceil\dfrac{\textsf{ct}}{\Delta}\right\rfloor = \textsf{RLWE}_{S, \sigma}(\Delta M_1 C) \bmod q_{l-1}$\\\hline
\textbf{BGV}&- Ciphertext $\textsf{RLWE}_{S, \sigma}(M_1) = (A_1, B_1) \bmod q_l$\\
&- Plaintext polynomial $C \in \mathbb{Z}_t[X]/(X^n + 1)$\\
&$\textsf{RLWE}_{S, \sigma}(M_1 \cdot C) =(A_1 \cdot C, \text{ } B_1\cdot C)  \bmod q_l$\\\hline
\end{tabular}
\caption{\textbf{Ciphertext-to-Plaintext Multiplication}}
\end{table}

\begin{table}[h]
\begin{tabular}{|c||l|}
\hline
&\multicolumn{1}{c|}{\textbf{Ciphertext-to-Ciphertext Multiplication}}\\\hline\hline
\textbf{TFHE}&- Ciphertext $\textsf{LWE}_{\vec{s}, \sigma}(\Delta m_1) = (\vec{a}_1, b_1) = (a_{1,0}, a_{1, 1}, \cdots, a_{1, k-1}, b_1) \bmod q$\\
&- Ciphertext $\textsf{LWE}_{\vec{s}, \sigma}(\Delta m_2) = (\vec{a}_2, b_2) = (a_{2,0}, a_{2, 1}, \cdots, a_{2, k-1}, b_2)  \bmod q$\\
&1. \underline{Programmable Bootstrapping}:\\ 
&\textcolor{white}{1. } Convert $\textsf{LWE}_{\vec{s}, \sigma}(\Delta m_2)$ into $\textsf{GSW}_{\vec{s}, \sigma}^{\beta, l}(m_2)$.\\
&2. \underline{Homomorphic Multiplication}:\\
&\textcolor{white}{2. } Compute $\textsf{LWE}_{\vec{s}, \sigma}(\Delta m_1) \cdot \textsf{GSW}_{\vec{s}, \sigma}^{\beta, l}(m_2)$\\
&\text{ }$ = \sum\limits_{i=0}^{k-1} \langle \textsf{Decomp}^{\beta, l}(a_{1,i}),  \textsf{Lev}_{\vec{s}, \sigma}^{\beta, l}(-s_i\cdot m_2) \rangle + \langle \textsf{Decomp}^{\beta, l}(b_{1}),  \textsf{Lev}_{\vec{s}, \sigma}^{\beta, l}(m_2) \rangle$ \\
&\text{ } $ = \textsf{LWE}_{\vec{s}, \sigma}(\Delta m_1 m_2)$\\\hline
\textbf{BFV}&- Ciphertext $\textsf{RLWE}_{S, \sigma}(\Delta M_1) = (A_1, B_1)  \bmod q$\\
&- Ciphertext $\textsf{RLWE}_{S, \sigma}(\Delta M_2) = (A_2, B_2)  \bmod q$\\
&1. \underline{\textsf{ModRaise}} from $q$ to $Q = q \cdot \Delta$\\
&\textcolor{white}{1. } - Ciphertext $\textsf{RLWE}_{S, \sigma}(\Delta M_1) = (A_1, B_1)  \bmod Q$\\
&\textcolor{white}{1. } - Ciphertext $\textsf{RLWE}_{S, \sigma}(\Delta M_2) = (A_2, B_2)  \bmod Q$\\
&2. \underline{Polynomial Multiplication}: \\
& \textcolor{white}{1. } $(A_1A_2, \text{ } A_1B_2 + A_2B_1, \text{ } B_1B_2) \equiv (D_2, D_1, D_0) \pmod Q$\\
&3. \underline{Relinearization}: $\textsf{ct}_\alpha = (D
_1, D_0),$\text{ } $\textsf{ct}_\beta = \bm{\langle} \textsf{Decomp}^{\beta, l}(D_2), \text{ } \textsf{RLev}_{S, \sigma}^{\beta, l}( S^2) \bm{\rangle}$\\
&\textcolor{white}{3. Relinearization: } $\textsf{ct}_\alpha + \textsf{ct}_\beta = \textsf{ct}_{\alpha + \beta} = \textsf{RLWE}_{S, \sigma}(\Delta^2 M_1 M_2) \bmod Q$\\
&4. \underline{Rescaling} by $\dfrac{1}{\Delta}$: $\left\lceil\dfrac{\textsf{ct}_{\alpha+\beta}}{\Delta}\right\rfloor = \textsf{RLWE}_{S, \sigma}(\Delta M_1 M_2) \bmod q$\\\hline
\textbf{CKKS}&- Ciphertext $\textsf{RLWE}_{S, \sigma}(\Delta M_1) = (A_1, B_1)  \bmod q_l$\\
&- Ciphertext $\textsf{RLWE}_{S, \sigma}(\Delta M_2) = (A_2, B_2)  \bmod q_l$\\
&1. \underline{Polynomial Multiplication}: \\
& \textcolor{white}{1. } $(A_1A_2, \text{ } A_1B_2 + A_2B_1, \text{ } B_1B_2) \equiv (D_2, D_1, D_0) \pmod{q_l}$\\
&2. \underline{Relinearization}: $\textsf{ct}_\alpha = (D
_1, D_0),$\text{ } $\textsf{ct}_\beta = \bm{\langle} \textsf{Decomp}^{\beta, l}(D_2), \text{ } \textsf{RLev}_{S, \sigma}^{\beta, l}( S^2) \bm{\rangle}$\\
&\textcolor{white}{3. Relinearization: } $\textsf{ct}_\alpha + \textsf{ct}_\beta = \textsf{ct}_{\alpha + \beta} = \textsf{RLWE}_{S, \sigma}(\Delta^2 M_1 M_2) \bmod q_l$\\
&3. \underline{Rescaling} by $\dfrac{1}{\Delta}$: $\left\lceil\dfrac{\textsf{ct}_{\alpha+\beta}}{\Delta}\right\rfloor = \textsf{RLWE}_{S, \sigma}(\Delta M_1 M_2) \bmod q_{l-1}$\\\hline
\textbf{BGV}&- Ciphertext $\textsf{RLWE}_{S, \sigma}(M_1) = (A_1, B_1)  \bmod q_l$\\
&- Ciphertext $\textsf{RLWE}_{S, \sigma}( M_2) = (A_2, B_2)  \bmod q_l$\\
&1. \underline{Polynomial Multiplication}: \\
& \textcolor{white}{1. } $(A_1A_2, \text{ } A_1B_2 + A_2B_1, \text{ } B_1B_2) \equiv (D_2, D_1, D_0) \pmod{q_l}$\\
&2. \underline{Relinearization}: $\textsf{ct}_\alpha = (D
_1, D_0),$\text{ } $\textsf{ct}_\beta = \bm{\langle} \textsf{Decomp}^{\beta, l}(D_2), \text{ } \textsf{RLev}_{S, \sigma}^{\beta, l}( S^2) \bm{\rangle}$\\
&\textcolor{white}{3. Relinearization: } $\textsf{ct}_\alpha + \textsf{ct}_\beta = \textsf{ct}_{\alpha + \beta} = \textsf{RLWE}_{S, \sigma}( M_1 M_2) \bmod q_l$\\
&3. (Optional) \underline{Rescaling} by $\dfrac{1}{\Delta}$: $\left\lceil\dfrac{\textsf{ct}_{\alpha+\beta}}{\Delta}\right\rfloor_t = \textsf{RLWE}_{S, \sigma}(M_1 M_2) \bmod q_{l-1}$\\
&\textcolor{red}{ $\rhd$ $\lceil\rfloor_t$ means rounding to the nearest multiple of $t$}\\
&\textcolor{red}{ $\rhd$ The future noise growth rate gets reduced if the ciphertext is rescaled}\\\hline
\end{tabular}
\caption{\textbf{Ciphertext-to-Ciphertext Multiplication}}
\end{table}

\begin{table}[h]
\begin{tabular}{|c||l|}
\hline
&\multicolumn{1}{c|}{\textbf{Maximum Possible Multiplications (without Bootstrapping)}}\\\hline\hline
\textbf{TFHE}&Unlimited with programming bootstrapping (but not possible without it)\\\hline
\textbf{BFV}&Unlimited\\\hline
\textbf{CKKS}&As many times as the length of the modulus chain\\\hline
\textbf{BGV}&As many times as the length of the modulus chain\\\hline
\end{tabular}
\caption{\textbf{Maximum Possible Multiplications (without Bootstrapping)}}
\end{table}

\begin{table}[h]
\begin{tabular}{|c||l|}
\hline
&\multicolumn{1}{c|}{\textbf{Key Switching}}\\\hline\hline
\textbf{TFHE}& Key-switching from $\vec{s} \rightarrow \vec{s}_{'}$:\\
&$\textsf{LWE}_{\vec{s}_{'},\sigma}(\Delta m) = b + a\cdot \textsf{LWE}_{\vec{s}_{'}, \sigma}(s)$\\
&\textcolor{white}{$\textsf{LWE}_{\vec{s}_{'},\sigma}(\Delta m) $} $ = b + \bm{\langle} \textsf{Decomp}^{\beta, l}(\vec{a}), \text{ } \textsf{Lev}_{\vec{s}_{'}, \sigma}^{\beta, l}(\vec{s}) \bm{\rangle}$\\\hline
\textbf{BFV}& Key-switching from $S \rightarrow S'$:\\
\textbf{CKKS}&$\textsf{RLWE}_{S',\sigma}(\Delta M) = B + A\cdot \textsf{RLWE}_{S', \sigma}(S)$\\
&\textcolor{white}{$\textsf{RLWE}_{S',\sigma}(\Delta M) $} $ = B + \bm{\langle} \textsf{Decomp}^{\beta, l}(A), \text{ } \textsf{RLev}_{S', \sigma}^{\beta, l}(S) \bm{\rangle}$\\\hline
\textbf{BGV}& Key-switching from $S \rightarrow S'$:\\
&$\textsf{RLWE}_{S',\sigma}(M) = B + A\cdot \textsf{RLWE}_{S', \sigma}(S)$\\
&\textcolor{white}{$\textsf{RLWE}_{S',\sigma}(M) $} $ = B + \bm{\langle} \textsf{Decomp}^{\beta, l}(A), \text{ } \textsf{RLev}_{S', \sigma}^{\beta, l}(S) \bm{\rangle}$\\\hline
\end{tabular}
\caption{\textbf{Key Switching}}
\end{table}

\begin{table}[h]
\begin{tabular}{|c||l|}
\hline
&\multicolumn{1}{c|}{\textbf{Modulus Drop (\textsf{ModDrop})}}\\\hline\hline
\textbf{CKKS}&- Ciphertext with the multiplicative level $l$: $\textsf{RLWE}_{S, \sigma}(\Delta M) = (A, B)  \bmod q_l$\\
\textbf{BGV}& - Ciphertext with the multiplicative level $l-1$:\\ 
&\textcolor{white}{ - } $\textsf{RLWE}_{S, \sigma}(\Delta M) = (A', B') = (A \bmod q_{l-1}, B \bmod q_{l-1})$\\\hline
\end{tabular}
\caption{\textbf{Modulus Drop (\textsf{ModDrop})}}
\end{table}

\begin{table}[h]
\begin{tabular}{|c||l|}
\hline
&\multicolumn{1}{c|}{\textbf{Encoding and Decoding the Plaintext}}\\\hline\hline
\textbf{TFHE}&No need, because each plaintext is a single number\\\hline
\textbf{BFV}&Must convert the input slots into polynomial coefficients to support batch processing:\\
\textbf{CKKS}&- Encoding input slots $\vec{v}$ into polynomial coefficients: $\vec{m} = n^{-1}\cdot\vec{v}\cdot I_n^R \cdot \hathat W$\\
\textbf{BGV}&- Decoding polynomial coefficients $\vec{m}$ into input slots:  $\vec{v} = \vec{m} \cdot \hathat W^{*}$\\\hline
\end{tabular}
\caption{\textbf{Encoding and Decoding the Plaintext}}
\end{table}

\begin{table}[h]
\begin{tabular}{|c||l|}
\hline
&\multicolumn{1}{c|}{\textbf{Input Slot Rotation}}\\\hline\hline
\textbf{TFHE}&Not applicable, because its plaintext is a single number (i.e., a single slot)\\\hline
\textbf{BFV}&Given $\textsf{ct} = \textsf{RLWE}_{S(X), \sigma}\bm(\Delta M(X)\bm) = \bm(A(X), B(X)\bm)$,\\
&\text{ } to rotate the input slots by $h$ positions to the left:\\
\textbf{CKKS}&1. Update $\textsf{ct}$ to $ \textsf{RLWE}_{S(X^{J(h)}), \sigma}\bm(\Delta M(X^{J(h)})\bm) = \bm(A(X^{J(h)})$, $B(X^{J(h)})\bm)$\\
&\text{ } \text{ } (where $J(h) = 5^h \bmod 2n$)\\
&2. Key-switch $\textsf{RLWE}_{S(X^{J(h)}), \sigma}\bm(\Delta M(X^{J(h)})\bm)$ to $\textsf{RLWE}_{S(X), \sigma}\bm(\Delta M(X^{J(h)})\bm)$. \\\hline
\textbf{BGV}&Given $\textsf{ct} = \textsf{RLWE}_{S(X), \sigma}\bm(M(X)\bm) = \bm(A(X), B(X)\bm)$,\\
&\text{ } to rotate the input slots by $h$ positions to the left:\\
&1. Update $\textsf{ct}$ to $\textsf{RLWE}_{S(X^{J(h)}), \sigma}\bm(M(X^{J(h)})\bm) = \bm(A(X^{J(h)})$, $B(X^{J(h)})\bm)$\\
&2. Key-switch $\textsf{RLWE}_{S(X^{J(h)}), \sigma}\bm( M(X^{J(h)})\bm)$ to $\textsf{RLWE}_{S(X), \sigma}\bm( M(X^{J(h)})\bm)$. \\\hline
\end{tabular}
\caption{\textbf{Input Slot Rotation}}
\end{table}

\begin{table}[h]
\begin{tabular}{|c||l|}
\hline
&\multicolumn{1}{c|}{\textbf{Bootstrapping Goal}}\\\hline\hline
\textbf{TFHE}&To reset the noise.\\
\textbf{BFV}&\\\hline
\textbf{CKKS}&To reset the ciphertext modulus from $q_0 \rightarrow q_L$ (technically, to $q_{l'}$ where $q_0 < q_{l'} < q_L$).\\
\textbf{BGV}&\\\hline
\end{tabular}
\caption{\textbf{Bootstrapping Goal}}
\end{table}

\begin{table}[h]
\begin{tabular}{|c||l|}
\hline
&\multicolumn{1}{c|}{\textbf{Bootstrapping Details}}\\\hline\hline
\textbf{TFHE}&1. \underline{Modulus Switch} from $q \rightarrow 2n$ to convert $\textsf{LWE}_{\vec{s}, \sigma}(\Delta m) \rightarrow \textsf{LWE}_{\vec{s}, \sigma}(\hat\Delta m) = ( \vec{\hat a}, \hat b) \bmod 2n $,\\
&\text{ } \text{ } where $\hat\Delta = \dfrac{2n}{t}$. \text{ } \textcolor{red}{ $\rhd$ where $t$ divides $2n$}\\
&2. \underline{Blind Rotation}: Homomorphically rotate the GLWE-encrypted look-up table\\
&\text{ } \text{ }   polynomial $\textsf{GLWE}_{\vec{S},\sigma}(\Delta V)$ by $\Delta m + e$ positions to the left. This is done by\\
&\text{ } \text{ }  by homomorphically deriving $\textsf{GLWE}_{\vec{S}, \sigma}(\Delta V_k)$ as follows:\\
&\text{ } \text{ } $\textsf{GLWE}_{\vec{S}, \sigma}(\Delta V_0) = \textsf{GLWE}_{\vec{S},\sigma}(\Delta V) \cdot X^{-\hat b}$\\
&\text{ } \text{ } $\textsf{GLWE}_{\vec{S}, \sigma}(\Delta V_i) = \textsf{GLWE}_{\vec{S}, \sigma}(\Delta V_{i-1}) \cdot X^{\hat a_i s_{i-1}}$\\
&\text{ } \text{ } \text{ } $ = \textsf{GGSW}_{\vec{S},\sigma}^{\beta, l}(s_{i-1})\cdot \bm(\textsf{GLWE}_{\vec{S}, \sigma}(\Delta V_{i-1}) \cdot X^{\hat{a}_{i-1}} - \textsf{GLWE}_{\vec{S}, \sigma}(\Delta V_{i-1})\bm) + \textsf{GLWE}_{\vec{S}, \sigma}(\Delta V_{i-1})$\\
&3. \underline{Coefficient Extraction}: The rotated encrypted polynomial $V_k$'s constant term value is \\
&\text{ } \text{ }  $\Delta m$. Extract this encrypted constant term as $\textsf{LWE}_{\vec{s}, \sigma}(\Delta m)$ from $\textsf{GLWE}_{\vec{S}, \sigma}(\Delta V_k)$.\\\hline
\textbf{BFV}&1. \underline{Modulus Switch} from $q \rightarrow p^\varepsilon$ to convert $\textsf{RLWE}_{S, \sigma}(\Delta M) \rightarrow \textsf{RLWE}_{S, \sigma}(p^{\varepsilon-1} M) \bmod p^\varepsilon$\\
&2. \underline{Homomorphic Decryption}:\\
&\text{ } \text{ } $B + A\cdot \textsf{RLWE}_{S, \sigma}(\Delta' S) = \textsf{RLWE}_{S, \sigma}(\Delta' \cdot(p^{\varepsilon-1}M + E + Kp^\varepsilon)) \bmod q$, where $\Delta' = \left\lfloor\dfrac{q}{p^\varepsilon}\right\rfloor$\\
&3. \underline{\textsf{CoeffToSlot}}: Multiply to the ciphertext by $n^{-1}\cdot I_n^R \cdot \hathat W$ to move\\
& \text{ } \text{ } the plaintext coefficients of $p^{\varepsilon-1} M + E + Kp^\varepsilon$ to the input slots.\\
&4. \underline{Digit Extraction}: Given the digit extraction polynomial $G_{\varepsilon}(x)$,\\
&\text{ } \text{ }  homomorphically compute:\\
&\text{ } \text{ } $G_{2} \circ G_{3}\circ \cdots \circ G_{\varepsilon-1} (p^{\varepsilon-1} M + E + Kp^\varepsilon)$\\
&\text{ } \text{ }, and then the output $M + K^{\langle \varepsilon - 1 \rangle}p$ is stored in the plaintext slots.\\
&\text{ } \text{ } Use scaling factor re-interpretation to handle inverse-$p$ multiplications.\\
&5. \underline{\textsf{SlotToCoeff}}: Multiply to the ciphertext by $\hathat W^*$ to move $M + K^{\langle \varepsilon - 1 \rangle}p$ to the \\
&\text{ } \text{ } polynomial coefficient positions and get $\textsf{RLWE}_{S, \sigma}(\Delta'\cdot(M + K^{\langle \varepsilon - 1 \rangle}p)) \bmod q$\\
&\text{ } \text{ } $  = \textsf{RLWE}_{S, \sigma}(\Delta M) \bmod q$\\
\hline
\end{tabular}
\caption{\textbf{Bootstrapping Details: TFHE, BFV}}
\end{table}

\begin{table}[h]
\begin{tabular}{|c||l|}
\hline
&\multicolumn{1}{c|}{\textbf{Bootstrapping Details}}\\\hline\hline
\textbf{CKKS}&1. \underline{\textsf{ModRaise}:} View the ciphertext $(A, B) \bmod q_0$ as $(A, B) \bmod q_L$ \\
%&2. \underline{Homomorphic Decryption } (without scaling down by $\Delta$): \\
%&\text{ } \text{ } $B + A \cdot \textsf{RLWE}_{S, \sigma}(S) \bmod q_L = B + \langle \textsf{Decomp}^{\beta, l}(A), \textsf{RLev}_{S, \sigma}^{\beta, l}(S) \rangle$\\
%&$\text{ } \text{ } = \textsf{RLWE}_{S, \sigma}(\Delta M + E + Kq_0) \bmod q_L$\\
&2. \underline{\textsf{CoeffToSlot}}: Move the coefficients of $\Delta M + E + Kq_0$ to the input slots.\\
&3. \underline{\textsf{EvalExp}}: Homomorphically evaluate the polynomial approximation of\\
&\text{ } \text{ } the sine function with period $q_0$ at $\Delta M + E + Kq_0$,\\
& \text{ } \text{ } which outputs an encryption of $\Delta M + E$ in the plaintext slots.\\
&4. \underline{\textsf{SlotToCoeff}}: Move $\Delta M + E$ to the polynomial coefficient positions\\
&\text{ } \text{ } to get $\textsf{RLWE}_{S, \sigma}(\Delta M + E)$.\\\hline
\textbf{BGV}&1. \underline{Modulus Switch} from $q_{l} \rightarrow \hat{q}$ to convert\\
&\text{ } \text{ } $\textsf{RLWE}_{S, \sigma}(M) = (A, B) \bmod q_l \rightarrow \textsf{RLWE}_{S, \sigma}(M) = (\hat{A}, \hat{B}) \bmod \hat{q}$\\ 
&\text{ } \text{ }, where $\hat{q} \equiv 1 \bmod p^\varepsilon$\\
&2. \underline{Ciphertext Coefficient Multiplication by $p^{\varepsilon-1}$}:\\
& \text{ } \text{ } Compute $(p^{\varepsilon-1}\hat{A}, p^{\varepsilon-1}\hat{B}) = (A', B') \bmod \hat{q}$ (where $\hat{q} \equiv 1 \bmod p^\varepsilon$)\\
& \text{ } \text{ }, which the ciphertext $\textsf{RLWE}_{S, \sigma}(p^{\varepsilon-1}M) \bmod \hat{q}$ with noise $p^\varepsilon E$.\\
&3. \underline{\textsf{ModRaise}}: $(A', B') \bmod \hat{q} \rightarrow (A', B') \bmod q_L$\\
& \text{ } \text{ } , which is the ciphertext $\textsf{RLWE}_{S, \sigma}(p^{\varepsilon-1}M + p^\varepsilon E + K'\hat{q}) \bmod q_L$.\\
&4. \underline{\textsf{CoeffToSlot}}: Multiply to the ciphertext by $n^{-1}\cdot I_n^R \cdot \hathat W$ to move\\
& \text{ } \text{ } the plaintext coefficients of $p^{\varepsilon-1}M + p^\varepsilon E + K'\hat{q}$ to the input slots.\\
&5. \underline{Digit Extraction}: Given the digit extraction polynomial $G_{\varepsilon}(x)$, \\
&\text{ } \text{ }  homomorphically compute:\\
&\text{ } \text{ } $\bm(\text{ } G_{2} \circ G_{3}\circ \cdots \circ G_{\varepsilon} (p^{\varepsilon-1} M + p^\varepsilon E + K'\hat{q})\text{ } \bm)$\\
&\text{ } \text{ } , and then the output $M + K''p$ is stored in the plaintext slots.\\
&\text{ } \text{ } To handle inverse-$p$ multiplication in each $i$-th round, multiply $|p^{-1}|_{q^{\langle i \rangle}}$ to all\\
&\text{ } \text{ }  ciphertext polynomial coefficients to update their plaintext portion from\\
&\text{ } \text{ }  $Mp^{\varepsilon - i} + K'''p^{\varepsilon - i + 1} \pmod{q^{\langle i \rangle}}$ to $M^{\varepsilon - i - 1} + K^{\langle i \rangle}p^{\varepsilon - i} \pmod{q^{\langle i \rangle}}$\\
&\text{ } \text{ }   \textcolor{red}{ $\rhd$ where $q^{\langle i \rangle}$ is the ciphertext modulus at each specific round}\\
&6. \underline{\textsf{SlotToCoeff}}: Multiply to the ciphertext by $\hathat W^*$ to move $M + K^{\langle \varepsilon - 1 \rangle}p$ to the \\
&\text{ } \text{ } polynomial coefficient positions to get $\textsf{RLWE}_{S, \sigma}(M + K^{\langle \textit{last} \rangle}p)$\\
&\text{ } \text{ } $ = \textsf{RLWE}_{S, \sigma}(M + \Delta K^{\langle \textit{last} \rangle}) \bmod q_{l'}$ \textcolor{red}{ $\rhd$ where $\Delta = p$, and the final noise is $K^{\langle \textit{last} \rangle}$}\\\hline
\end{tabular}
\caption{\textbf{Bootstrapping Details: CKKS, BGV}}
\end{table}

\begin{table}[h]
\begin{tabular}{|c||l|}
\hline
&\multicolumn{1}{c|}{\textbf{Noise Management}}\\\hline\hline
\textbf{TFHE}&Their bootstrapping resets the noise.\\
\textbf{BFV}&\\\hline
\textbf{CKKS}&- The noise grows without stopping, because its bootstrapping resets only the\\
& modulus chain. To slow down the noise growth, we should increase the plaintext's\\
&  scaling factor $\Delta$.\\
&- CKKS's \textsf{EvalExp} cannot use the digit extraction polynomial to remove the noise, \\
&because CKKS's plaintext is not in a modulus ring, but is a real number.\\\hline
\textbf{BGV}&BGV's modulus switch has the special property of resetting the noise, and BGV's\\
& bootstrapping resets the modulus chain to enable indefinite modulus switches.\\\hline
\end{tabular}
\caption{\textbf{Noise Management}}
\end{table}

\clearpage

\section{Python Demo FHE Library}
We provide our Python Demo FHE library (TFHE, BFV, CKKS, BGV) for educational purposes. The source code and the manual are available at \href{https://github.com/fhetextbook/fhe-textbook}{\texttt{https://github.com/fhetextbook/fhe-textbook}}.

\clearpage

\bibliographystyle{unsrt}
\bibliography{z-bibfile}

%\clearpage
%\input{scratch-real}

\end{document}